# FACETS OF GRAVITY

A journey exploring spacetime from Planckian to
cosmological scales

Agustín Silva

**FACETS OF GRAVITY**

A journey exploring spacetime from Planckian to cosmological scales

Thesis, Radboud University Nijmegen

Radboud University

# FACETS OF GRAVITY

A journey exploring spacetime from Planckian to cosmological scales

Proefschrift ter verkrijging van de graad van doctor
aan de Radboud Universiteit Nijmegen
op gezag van de rector magnificus prof. dr. J.M. Sanders,
volgens besluit van het college voor promoties
in het openbaar te verdedigen op

donderdag 15 oktober 2026
om 16.30 uur precies

door

Agustín Silva

geboren op 5 juli 1996
te Santa Fe, Argentinië

**Promotoren**:

Prof. dr. Renate Loll

Prof. dr. Frank Saueressig

**Manuscriptcommissie**:

Prof. dr. Badri Krishnan

Prof. dr. Jan Ambjørn

Prof. dr. Astrid Eichhorn, Universität Heidelberg (Duitsland)

Dr. Dario Benedetti, CNRS (Frankrijk)

Dr. Tomislav Prokopec, Universiteit Utrecht

# Remarks on this extended version

This book is an extended version of my PhD thesis that was submitted to, reviewed by, and approved by the manuscript committee. The complete content of that original manuscript is preserved here unchanged in substance. In the time between submission and defence, I chose to add Part IV, *A personal detour towards a fundamental understanding of gravity*, which was not part of the manuscript assessed by the committee. Unfortunately, this extended version never became the definitive version of my thesis, since the extension required longer review times for my supervisors and the committee, and therefore my PhD thesis was shortened. Nevertheless, I wanted to include this extra part because it represents a particularly personal and important direction of my research: an attempt to step back from the usual question of how to quantize the metric and instead ask a more elementary one—what is it that we should regard as fundamental, and therefore quantize, in gravity? This line of thought has been one of the most inspiring and interesting outcomes of my PhD, at least for me, and has developed from a personal intuition into a concrete non-perturbative construction with first exploratory numerical results. Its inclusion here is therefore deliberately different in character from the rest of the thesis: it is not part of the work which the manuscript committee evaluated and approved the dissertation, but rather a record of a research direction that emerged from this journey and that I hope to continue exploring beyond it. Responsibility for the content of this added part is, of course, entirely my own.

# Preface

In scientific divulgation, the "quantization of gravity" is often invoked as the holy grail of physics, and the ultimate solution to extraordinarily ambitious problems. Sometimes it is a solution to the singularity problems inside black holes, or an explanation for the origin of the Universe, or a resolution to the information loss paradox, and more. These profound issues served as a seed that sparked my interest in this research area.

That seed kept growing with time, and a mere curiosity to understand spacetime at its most fundamental level became a necessity. This thesis is my attempt to satisfy that hunger for knowledge, and to clarify what it means to quantize gravity, as well as to assess what can be demanded and predicted from it. The chapters that follow are not organized as a linear derivation toward a single, final model that resolves all of the issues above. Instead, each part and chapter of this thesis should be read as a guided exploration of several complementary non-perturbative viewpoints. Each part is motivated by one or more questions, and these questions serve as genuine guidelines for investigating what a quantum theory of spacetime can, and cannot yet, provide.

A few remarks may help the reader navigate the material.

*Scope and philosophy:*

The selection of topics in this manuscript reflects a specific strategy: when standard tools are limited, one can still make progress by (i) constructing observables that remain meaningful in computational simulations of strongly fluctuating geometries, (ii) developing renormalization-group tools that connect microscopic definitions to macroscopic predictions, and (iii) questioning whether the variables we treat as fundamental are the right starting point in the first place. Parts II–IV implement these three elements in a deliberately multifaceted way.

*How to read this thesis:*

The parts are designed to be readable with minimal cross-dependence, with largely self-contained introductions and summaries for both individual chapters and each part as a whole. Part I sets the motivation and basic language. Part II develops geometric observables in a specific, fully non-perturbative path-integral setting. Part III shifts to the functional renor-malization group and focuses on predictivity and emergent infrared physics. Part IV is deliberately exploratory and addresses a more foundational choice: what we decide to quantize when we study spacetime.

Readers interested in what happens to spacetime near Planckian scales, how it looks, and its properties, may wish to begin with Part II. Readers primarily interested in phenomenology emerging from Planckian physics may wish to focus on Part III. Readers more versed in Quantum Gravity, and interested in foundational questions about what should be fundamentally quantized in gravity, may wish to focus on Part IV.

*Conventions:*

Whenever possible, standard conventions from quantum field theory and general relativity are used. Repeated indices are summed over using Einstein's summation convention. Greek indices typically denote spacetime components, while Latin indices are used for auxiliary or internal labels (e.g. lattice or theory-space indices), with exceptions noted locally. Unless stated otherwise, we work in units where the speed of light and the reduced Planck constant are equal to one ($c = \hbar = 1$). When working with Lorentzian signature, this thesis uses the so-called "mostly plus" signature for the spacetime metric, which in four dimensions is $(- + ++)$.

***Publication basis and authorship:***

This thesis is based on some of the work by the author and collaborators:

- Chapter 3 is based on: R. Loll, M. Schiffer, J. van der Duin and A. Silva, *"Quantum Gravity and Effective Topology"* (In: Eur. Phys. J. C, vol. 86, nr. 2, p. 102, (2026). doi:10.1140/$epjc/s$10052-026-15322-$x$) [1] and R. Loll, M. Schiffer, J. van der Duin and A. Silva, *"Exploring Quantum Spacetime with Topological Data Analysis"* (In: Phys. Lett. B, (2026), p. 140857. doi.org:10.1016/j.physletb.2026.140857) [2].

- Chapter 4 is based on: R. Loll and A. Silva, *"Measuring the homogeneity of the quantum universe"*, (In: Phys. Rev. D 107.8 (2023), p. 086013. doi:10.1103/PhysRevD.107.086013) [3].

- Chapter 5 is based on: J. van der Duin and A. Silva, *"Scalar curvature for metric spaces: Defining curvature for quantum gravity without coordinates"* (In: Phys. Rev. D 110.2 (2024), p. 026013. doi:10.1103/PhysRevD.110.026013) [4].

- Chapter 8 is based on: F. Saueressig and A. Silva, *"On harvesting physical predictions from asymptotically safe quantum field theories"* (In: Phys. Rev. D 110.8 (2024), p. 085005. doi:10.1103/PhysRevD.110.085005) [5] and S. Hennissen, F. Saueressig and A. Silva, *"Machine learning the predictions of Asymptotic Safety"* (To appear on ArXiv in 2026)[6].

- Chapter 9 is based on: A. Silva, *"Emergence of inflaton potential from asymptotically safe gravity"* (In: Phys. Lett. B 860 (2025), p. 139154. doi:10.1016/j. physletb.2024.139154) [7].

- Chapter 11 is based on: A. Silva, *"A pathway to non-perturbative Quantum Affine Gravity"* (In: Class. Quantum Grav. vol. 43, nr. 12, p. 125020, (2026). doi:10.1088/1361-6382/$ae7ada$) [8].

Throughout his PhD, the author has also contributed to other areas through scientific papers that are not included in this thesis. In the framework of black hole spacetimes, their thermodynamic and stability properties, the author published: C. Laporte and A. Silva, *"Universal black hole solutions for all $F(R)$ gravitational theories"* (In: Phys. Rev. D 111.4 (2025), p. 044020. doi:10.1103/PhysRevD.111.044020) [9].

# Acronyms

| | |
|---|---|
| AS | Asymptotic Safety |
| CDT | Causal Dynamical Triangulations |
| CM | Classical Mechanics |
| EAA | Effective Average Action |
| EDT | Euclidean Dynamical Triangulations |
| EFT | Effective Field Theory |
| EH | Einstein–Hilbert |
| FI | Flow Invariant |
| FINN | Flow Invariant Neural Network |
| FP | Fixed Point |
| FRG | Functional Renormalization Group |
| FRGE | Functional Renormalization Group Equation |
| GR | General Relativity |
| IR | Infrared |
| LHS | Left-Hand Side |
| LQAG | Lattice Quantum Affine Gravity |
| MCMC | Markov chain Monte Carlo |
| NGFP | Non-Gaussian Fixed Point |
| NN | Neural Network |
| PDE | Partial Differential Equation |
| PINN | Physics-Informed Neural Network |
| QCD | Quantum Chromodynamics |
| QFT | Quantum Field Theory |
| QG | Quantum Gravity |
| QM | Quantum Mechanics |
| RG | Renormalization Group |
| RHS | Right-Hand Side |
| SM | Standard Model |
| TDA | Topological Data Analysis |
| UV | Ultraviolet |

# Contents

# Part I

# The intersection of Quantum Mechanics and General Relativity

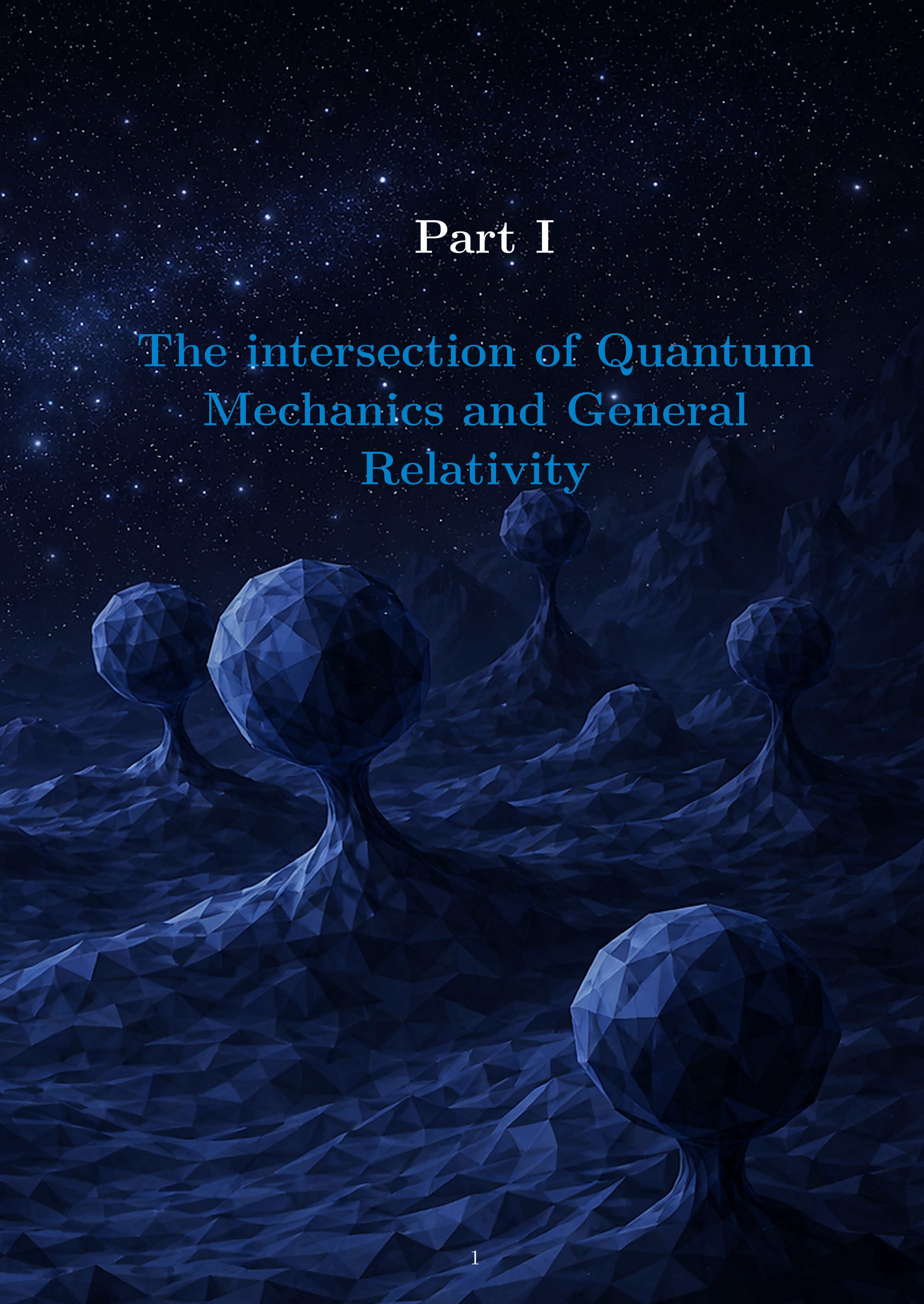

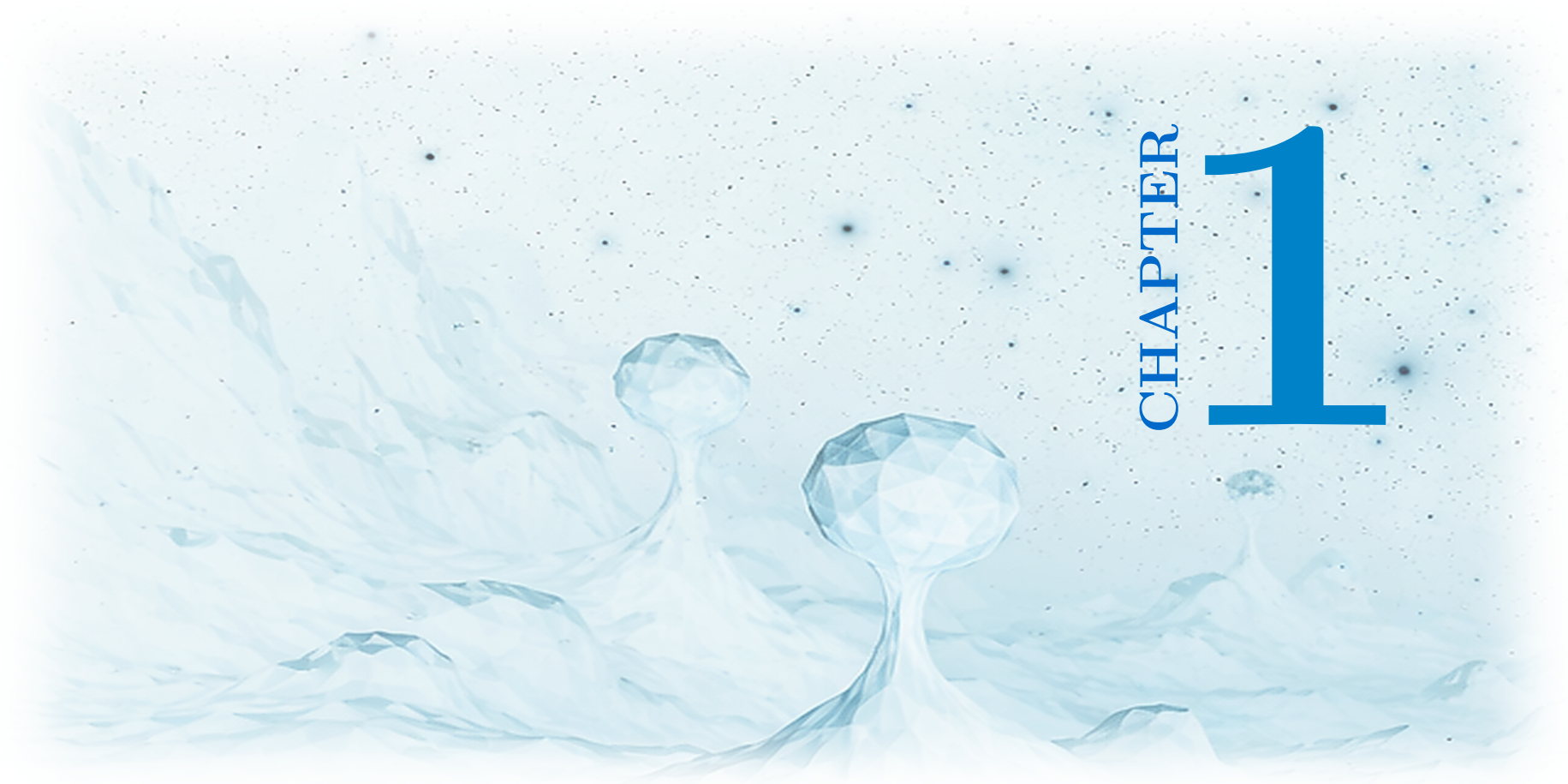

# General Introduction

Gravity is a name we physicists associate with a single fundamental interaction of our universe, but in practice, it plays many roles. Gravity is, in my personal opinion, the most fascinating of the known interactions of our universe. It shapes our reality and our perception of it in both direct and very subtle ways. Gravity is what grounds us on Earth. It is the quiet background that lets us speak of something like weight. It is the elegant orchestrator of planetary orbits that determine our days and nights. It is the mysterious artist who affects how we measure things like time. On astrophysical scales, it organizes the life of stars, galaxies, and the architecture of even more exotic objects, such as black holes. On cosmological scales, it becomes inseparable from the questions we ask about the origin, content, and fate of the Universe.

On the shortest scales, where the quantum nature of all interactions can no longer be ignored, we are forced into a more radical question: Is spacetime itself just a stage where all other interactions act, or is it a participant, an actor on its own, in the probabilistic nature of our universe?

As of now, we believe we understand to a great extent the classical description of gravity, which is currently modeled by Einstein's theory of General Relativity (GR). This model has been confirmed in every observation where it was tested. Fortunately, we live in times where technological progress is fast-paced, allowing scientists to push our frontier of knowledge and test GR in even more extreme cases, where deviations from it could appear, hinting at possible new physics. These are exciting times for the observation of strong gravitational regimes, providing us with windows to better understand black hole mergers via gravitational wave detection [10]. On top of that, current technological developments put us in the so-called era of "precision cosmology", where we can study snapshots of the dynamics of the early universe via telescopes so powerful that we can use to study in detail events that happened around 13 billion years ago [11].

So far, GR has been incredibly successful in explaining most of the observed gravitational phenomena on its own, with some exceptions, such as Dark Matter [12, 13], or the apparently a-causal correlations we observe in the primordial plasma that was our universe in its early stages [14–16]. While these two things can be solved without modifying our model for gravity, by proposing the existence of yet unobserved new particles or fields, the physics community still believes GR cannot be the final answer to our understanding of spacetime.

We have reasons to believe that this theory predicts its own demise with the appearance of solutions with curvature singularities, which the majority of the physics community does not believe are physical, such as black hole solutions, as well as at the beginning of the universe, where GR allows a singular, radiation-dominated beginning of time. These infinities are often quoted as a motivation to move away from the classical scales where GR works well and study the quantum realm, where it is expected that they will be resolved by a quantum theory of gravity.

There have been many claims of what a quantum theory of gravity should and should not do, going from merely solving singularities, to unifying all known interactions of our universe, explaining the origin of the arrow of time, solving information loss paradoxes, and the list goes on and on [17]. As well, due to its unresolved nature and the lack of experimental guidance, many claims have been made regarding what spacetime should or should not "look like" at very small scales: topology changes, microscopic wormholes, fundamental discreteness, baby universes branching from the fabric of spacetime, and more. Altogether, these claims contributed to an image of Quantum Gravity (QG) as a mysterious and exotic realm where seemingly anything can happen. A picture often invoked in scientific divulgation that reflects these ideas can be seen in Fig. 1.1.

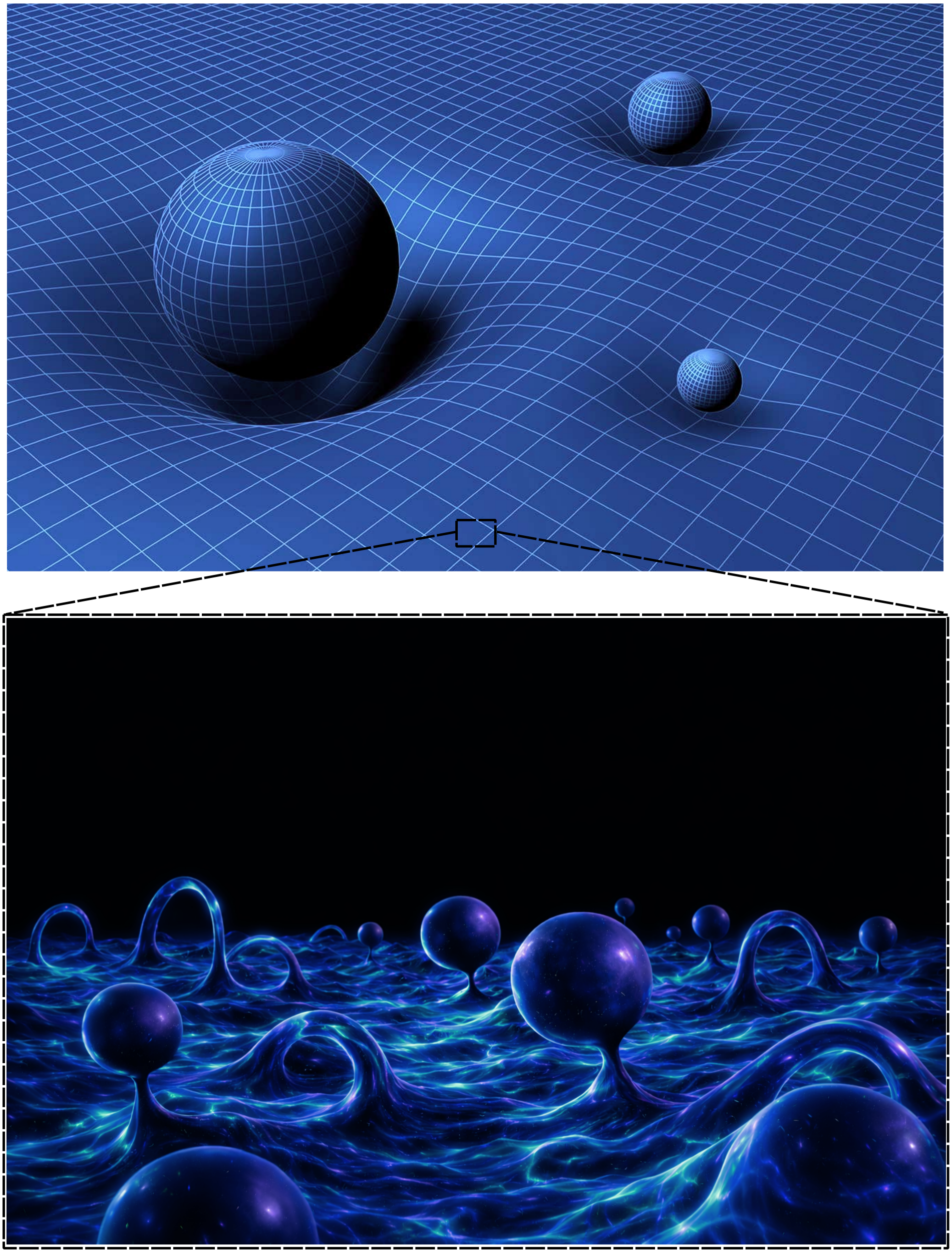

**Figure 1.1.:** Illustration of a zoom-in of the fabric of spacetime. From the top, the typical illustration of spacetime as a smooth curved surface due to the presence of matter, and on the lower part, the typical type of picture advocated when talking about what spacetime "should look like" at Planckian scales.

In this thesis, we will give a more concrete meaning to what we define as "quantum spacetime", with concrete mathematical formulations to explore high-energy regimes of the gravitational interaction, and we will set a more realistic picture of what Quantum Gravity is, rather than a "wonderland" where anything can happen and all problems of physics are solved.

In the theoretical physics community, fuzzy pictures such as Fig. 1.1 are less common, but the singularity-resolution narrative remains a frequently used headline motivation. However, treating singularity resolution as something to be "handed over" to the yet-to-be-found quantum theory of gravity strikes me as an overreach: before such a theory can address those problems, any candidate framework must first establish a consistent reconciliation of General Relativity and Quantum Mechanics, a task for which standard quantization strategies run into serious conceptual and technical difficulties, something that, for many decades, made physicists believe that Quantum Mechanics and General Relativity are seemingly incompatible.

In this thesis, we do not aim to specifically find a "quantum solution" to the singularity problems in gravity. Instead, we aim to get a better understanding of what happens to gravity on the smallest scales, where it is expected that the classical description of GR breaks down, and we will attempt to get a better grasp of what the outcomes of such a theory can be. Furthermore, we attempt to do so with a minimal amount of assumptions, and we aim to build such a theory using the known ingredients that work very well for constructing a quantum theory of all other known interactions in our Universe. Throughout the rest of this thesis, the reader will see how we explore a wide range of questions and topics within the rich realm of Quantum Gravity, which demands the development of new mathematical tools and ideas, and has been my working arena for the last four years.

Before going into the intricacies of how to quantize a classical theory, its obstacles in gravity, and the assumptions needed to bypass some of these problems, it is important to have in mind some of the basic ingredients to first study the classical features of a theory, and later the ones to perform its quantization, at least in a simple case where standard methods work. A simple example will help build the intuition needed for the understanding of the frameworks used in this thesis. Later, we will use the built knowledge from simpler scenarios as a basis to progress to the more complex situation of Quantum Gravity.

## 1.1 Least action principle and Einstein Field Equations

It is not often that physicists stumble upon deep and general unifying principles that can be used to describe a vast class of phenomena. In the case of Classical Mechanics (CM), the study of the dynamics of macroscopic objects, it was not until 1744, around 60 years after Isaac Newton published the foundations of Classical Mechanics, that Pierre Louis de Maupertuis noticed a profound underlying principle that would later be a pillar upon which our understanding of modern physics stands. This principle is commonly known as the *least action principle*. Maupertuis noticed that, based on the assumption of energy conservation of a system, its dynamics can be determined by finding the configuration that extremizes a quantity called *action*, typically denoted with an $S$. This principle was later generalized and formalized by Euler, Lagrange, and Hamilton, defining the action as the integral of a Lagrangian, typically denoted with an $L$. The Lagrangian is, in the simplest cases, just the difference between the kinetic energy of the system and the potential energy of the system, although, in more complicated cases like gravity, this splitting is not as evident.

This new formalism of finding a configuration of a system that extremizes the action addresses the following question. Given two configurations of the system $A$ and $B$ at two different times $t_A$ and $t_B$, which could be positions of a certain particle trajectory, or more complicated objects: What is the sequence of configurations, or path, that the system took to join $A$ and $B$ between times $t_A$ and $t_B$?

For concreteness, let us study the dynamics of a free particle moving in a single dimension in space, whose position as a function of time $t$ is determined by a function $q(t)$. In this toy model, we have that the Lagrangian is just the kinetic energy $L(q(t), \dot{q}(t)) = \frac{m}{2}\dot{q}^2(t)$, with $m$ being the mass of the particle. The least action principle yields the equation of motion $\frac{d^2q}{dt^2} = 0$, and its solution is a straight line $q_{\text{sol}}(t) = (q_B - q_A)\frac{t-t_A}{t_B-t_A} + q_A$ with $q_A$ and $q_B$ the two positions of the particle at the given times. The least action principle explores all possible configurations between $A$ and $B$, and finds the one that extremizes the action, such as depicted in Fig. 1.2.

This elegant idea of exploring different paths until finding the extremum of the action is the modern formulation of Newtonian mechanics, and the least action principle is a very practical way to find the equations of motion of a system. With this principle, one can explain why light rays follow the paths they do, and why large-scale objects do so as well, putting them together into the framework of Classical Mechanics [18].

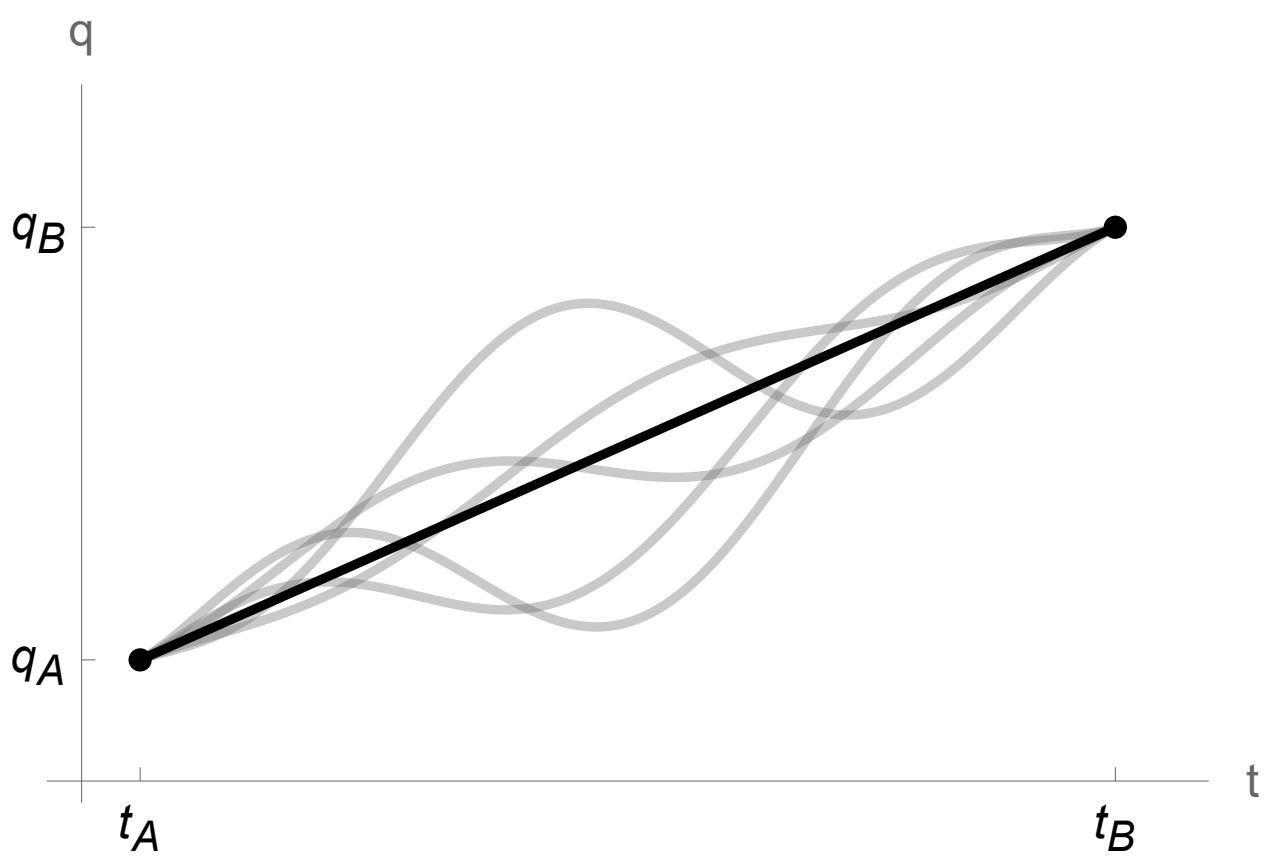

**Figure 1.2.:** Scheme of the least action principle for the one-dimensional free point particle. The solid black line represents the classical solution, the extremum of the action, and the gray lines represent other possible paths the system could take, that do not extremize the action.

This idea of "many possible paths" that yield different values of an action functional will come back in the next section, and is central to one of the approaches to Quantum Gravity we will study in this thesis.

This simple example helps us build intuition on how more complex physics works. In reality, many things require an infinite amount of degrees of freedom, and cannot be described by a single one, such as a point particle. A simple example are the vibrations of a drum, where the two dimensional surface of the drum is usually described by a continuum, rather than a point, and this description is given by a *field*, a function $\varphi(x)$ of the point $x$ on the drum, that describes the amplitude of its oscillation when it is played. The framework field describing a classical system is called *classical field theory*. General Relativity falls under this category.

GR was not built in a day. In 1905 Albert Einstein published, among four other seminal papers, one that set the foundation for how we understand gravity today [19]. Based on experimental evidence that light moved at the same speed, independently of the observer, he concluded that we should think of space and time as quantities that are relative to the observer. Time intervals and measured lengths were no longer absolute, but instead observer-dependent, and most importantly, they were modeled as one single entity called *spacetime*.

In 1915, Einstein published another seminal work where he proposed a radical new description of the gravitational interaction [20]. This extension of his previous work now proposes that spacetime is not only relative to observers, but is also shaped by the contents within it. Spacetime is modeled as a dynamical entity that bends in the presence of matter, and at the same time tells matter how to move in it. This description was made via a field, a tensor field, called the metric tensor $g_{\mu\nu}$. This metric tensor determines how we measure distances and time intervals, and characterizes the curvature properties of the fabric of spacetime. He determined the equations of motion for this tensor to be

$$R_{\mu\nu} - \frac{1}{2} R\, g_{\mu\nu} + \Lambda g_{\mu\nu} = 8\pi G\, T_{\mu\nu} \,, \tag{1.1}$$

where $G$ is Newton's gravitational constant, $\Lambda$ is the cosmological constant, $T_{\mu\nu}$ is the so-called stress energy tensor, which encodes the matter present in spacetime, and $R_{\mu\nu}$ and $R = g^{\mu\nu} R_{\mu\nu}$ are the Ricci tensor and Ricci scalar, two quantities that encode curvature properties of the spacetime characterized by the metric $g_{\mu\nu}$.

These equations can be derived from the least action principle by using the Einstein-Hilbert action including a cosmological constant [21]

$$S_{EH} = \frac{1}{16\pi G} \int d^4x \sqrt{|g|}\, \left(R - 2\Lambda\right) \,, \tag{1.2}$$

where $|g|$ denotes the absolute value of the metric determinant. The variation of the action with respect to the metric yields the left-hand side of the differential equations (1.1), while the right-hand side comes from the same variation but from the matter sources action (not included in (1.2)). Just like the particle could take many configurations but "chooses" the one that extremizes the action, so does the metric of spacetime with this action.

As mentioned before, this theory that models spacetime accurately describes observations at the classical level, or in other words, low-energy observations. As mentioned before, it is not believed that this macroscopic description is fundamental, and that there must be an underlying, more fundamental microscopic theory modelling spacetime at higher energies. In this thesis, we will explore realizations of a *Quantum Field Theory* of gravity for such a microscopic description. In the next few sections, we will give a more concrete notion of what low and high energy regimes mean in the context of this thesis.

## 1.2 Quantum Mechanics and the Path integral

The concept of an action had another large impact in physics almost two centuries later, on *Quantum Mechanics* (QM), where a whole framework to perform computations in QM was developed around the idea of multiple configurations that give different action values. It is beyond the scope of this thesis to introduce the reader to this fascinating physics area due to length constraints, and we will therefore take the leap from CM to QM without further details. It suffices to say this leap consists of more than 200 years of physics, and the interested reader can dive into the bibliography for a more detailed explanation [22].

There is no clear cut-off for distances below which CM stops working, and one must resort to QM, since this is highly system-dependent. Nevertheless, it suffices to say that below atomic scales, QM has proven incredibly successful in explaining experimental results, with over 100 years of successful predictions. In the case of gravity, the canonical distance and energy scales at which one expects to see quantum effects are

$$\ell_\mathrm{P} \equiv \sqrt{\frac{\hbar G}{c^3}} \simeq 1.616 \times 10^{-35}\ \mathrm{m} \qquad \text{(Planck length)}\,, \tag{1.3}$$

$$E_\mathrm{P} \equiv \sqrt{\frac{\hbar c^5}{G}} \simeq 1.22 \times 10^{19}\ \mathrm{GeV} \qquad \text{(Planck energy)}\,, \tag{1.4}$$

where $c$ is the speed of light and $\hbar$ is the reduced Planck's constant. To get an idea of how small the Planck length is, the Bohr radius of the hydrogen atom in its ground state is $a_0 \approx 5.29 \times 10^{-11}$ m, meaning that the scale at which QG effects are expected to be directly observable is approximately $10^{-25}$ times smaller than atomic scales. When we speak in this thesis about "Planckian scales", this is the order of magnitude of distances and energies we are referring to. This is one of the reasons why QG has been so elusive until now, since experiments cannot even come close to observing these scales, and we do not have any observational guidance on what the theory should be.

Therefore, if one wants to make progress in even defining such a theory, due to the lack of experimental guidance and evidence, I consider that the most reasonable path to take is the one that uses the fewest assumptions on what happens at those high energies. This is the philosophy followed in this thesis.

At small scales, below $a_0$, our universe is believed to be described by quantum (field) theories. At its core, QM assumes a fundamental uncertainty of the described systems rather than determinism, as in CM. Current interpretations of QM rely on probabilistic descriptions of physics, where outcomes are not determined by a set of initial conditions, but by a fundamental randomness of our universe, giving different results even if the initial conditions of the system are the same.

Thanks to the insights of Richard Feynman [23], we can construct both a visual intuition of this probabilistic nature, as well as a concrete framework to perform calculations on the probabilities for the outcomes of experiments. While in CM we determined a single configuration, or path, that joined two endpoints by using the principle of least action, in QM, thanks to Feynman, we can predict the outcome of experiments by assuming that a quantum system follows *all possible configurations* between two endpoints, with a certain probability, and where the action plays a central role.

It is important to remark that QM was designed to explain measurement results, given by the expectation values of certain quantities called *observables*. Observables in the context of QM are quantities that can be measured in experimental settings, such as the average position of a quantum particle. The physical results of the theory are the expectation values of these observables, and some quantities used to obtain those expectation values, such as the so-called *virtual paths* appearing in the path integral (see Fig. 1.3), can be regarded as mere tools to obtain predictions and have no physical meaning.

Going back to the example of the free point particle, following Feynman's recipe, the way one computes the expectation value of a certain quantity, say, the expected value of the position of the particle at a given time $t'$ between the endpoints $q_A$ and $q_B$ and times $t_A$ and $t_B$, denoted as $\langle q(t') \rangle$, requires

1.2

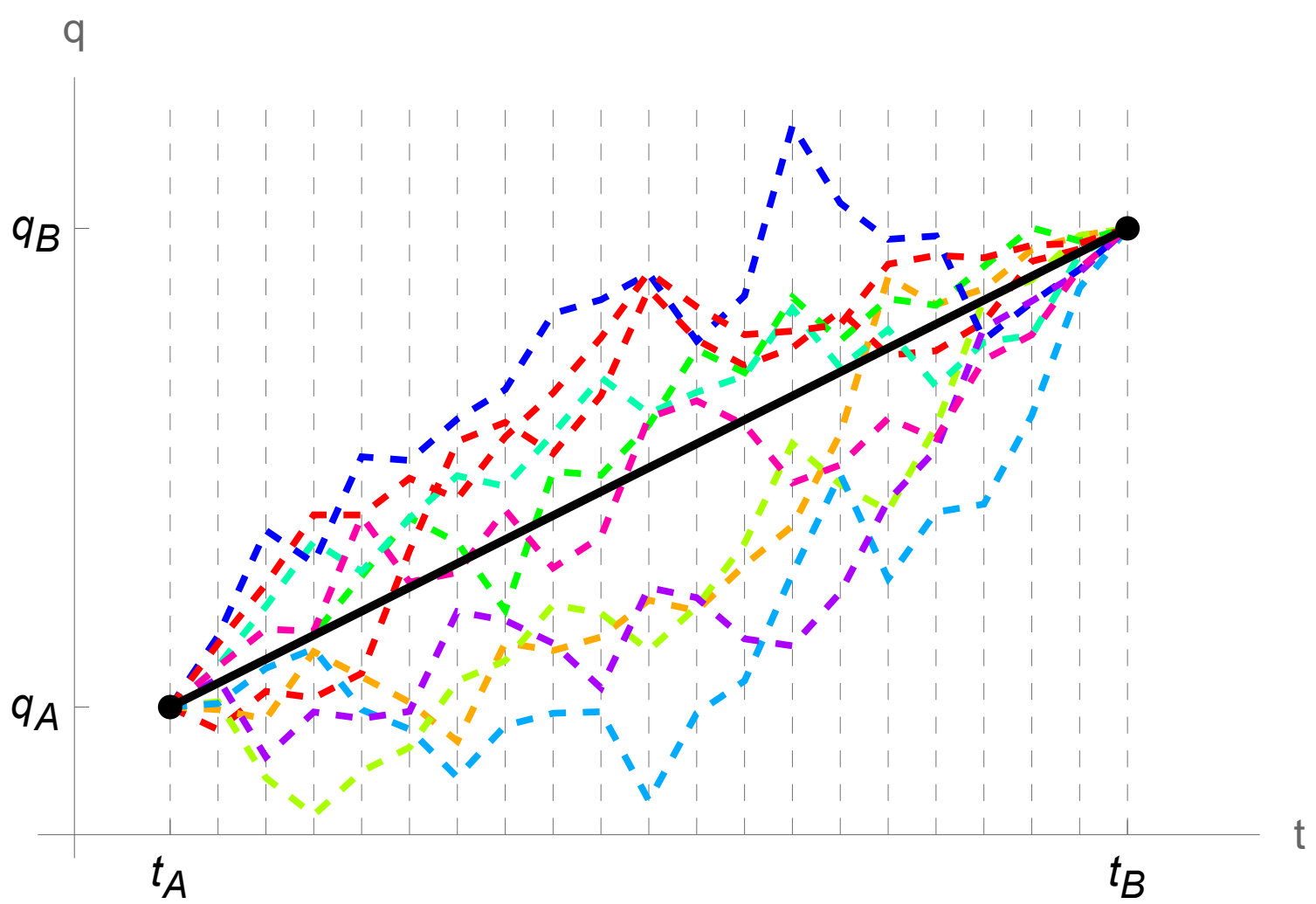

**Figure 1.3.:** Representation of the many paths that contribute to the path integral of a one-dimensional quantum point particle. The coloured dashed lines are configurations of the path integral, and the solid black line is the path resulting from the superposition of all virtual paths.

several steps. Schematically, the computation requires, first, discretizing the time interval between $t_A$ and $t_B$ in $N_T = \frac{t_B - t_A}{\Delta t}$ smaller time intervals of size $\Delta t$. Later, one discretizes the path between $q_A$ and $q_B$ as a piecewise linear path that joins points $q(t_n)$ at each of the times $t_n$ in the discretization of the interval. Finally, one assumes that at each of those discrete times, the coordinate of the particle $q(t_n)$ takes all possible values (in this case values in $(-\infty, \infty)$), and one integrates over all those possible values, weighting them with the exponential of the imaginary unit times the discretized action of the particle $S_{discrete}$, which can be obtained using finite differences. Each piecewise linear path that contributes to the integral is what we call a virtual path.

The expectation value $\langle q(t')\rangle$ takes the form of a limit, as $\Delta t \to 0$ of

$$\begin{aligned}\langle q(t')\rangle &\equiv \frac{1}{Z}\lim_{\Delta t\to 0}\prod_{n}^{N_T}\int_{-\infty}^{\infty} dq(t_n) e^{iS_{discrete}} q(t') \equiv \frac{1}{Z}\int \mathcal{D}q\, e^{i\,S} q(t') \\ &= \frac{1}{Z}\lim_{\Delta t\to 0}\prod_{n}^{N_T}\int_{-\infty}^{\infty} dq(t_n) e^{i\sum_j^{N_T}\Delta t\,\frac{m}{2}\left(\frac{q(t_j)-q(t_{j-1})}{\Delta t}\right)^2} q(t')\,, \end{aligned} \tag{1.5}$$

where the first line is the general definition of both the expectation value using the path integral, and the notation of the continuum limit of it, and the second line is a particular implementation for the point particle case.

Here, $Z$ is the so-called *quantum partition function* of the system, defined as

$$\begin{aligned} Z &\equiv \lim_{\Delta t\to 0}\prod_{n}^{N_T}\int_{-\infty}^{\infty} dq(t_n) e^{iS_{discrete}} = \int \mathcal{D}q\, e^{i\,S} \\ &= \lim_{\Delta t\to 0}\prod_{n}^{N_T}\int_{-\infty}^{\infty} dq(t_n) e^{i\sum_j^{N_T}\Delta t\,\frac{m}{2}\left(\frac{q(t_j)-q(t_{j-1})}{\Delta t}\right)^2}\,. \end{aligned} \tag{1.6}$$

A pictorial representation of this process can be seen in Fig. 1.3, where the black line represents the expected value of the position, and the piecewise flat coloured paths are the ones that contribute to the expectation value.

As mentioned before, the elements used to compute expectation values are not physical on their own; only expectation values are. This means that the piecewise flat paths that contribute to forming the expectation value are not physical. These paths are not paths a classical particle could even follow; in the limit where the path integral is defined, these paths are continuous but nowhere differentiable, unlike solutions to classical equations of motion. This computational framework is called the *path integral formalism*, and it extends beyond this simple case of a point particle to field theories. In particular, as we will see in this thesis, it can be used for General Relativity.

### 1.2.1 Quantum vs Statistical partition functions

A central object of the path integral formalism is the partition function. While formally divergent, because it is composed of infinitely many oscilla-tory exponentials, there are ways to make sense of these computations in physics. Central to the definition of the path integral was the discretization of the domain of the path integral configurations, in particular, the discretization of time. From this notion of time, one way to make sense of the divergent partition function $Z$ is to perform an analytic continuation of the time domain of the integration of the action to imaginary time [24, 25], by choosing a suitable closed contour in the complex domain, see Fig. 1.4 for an example.

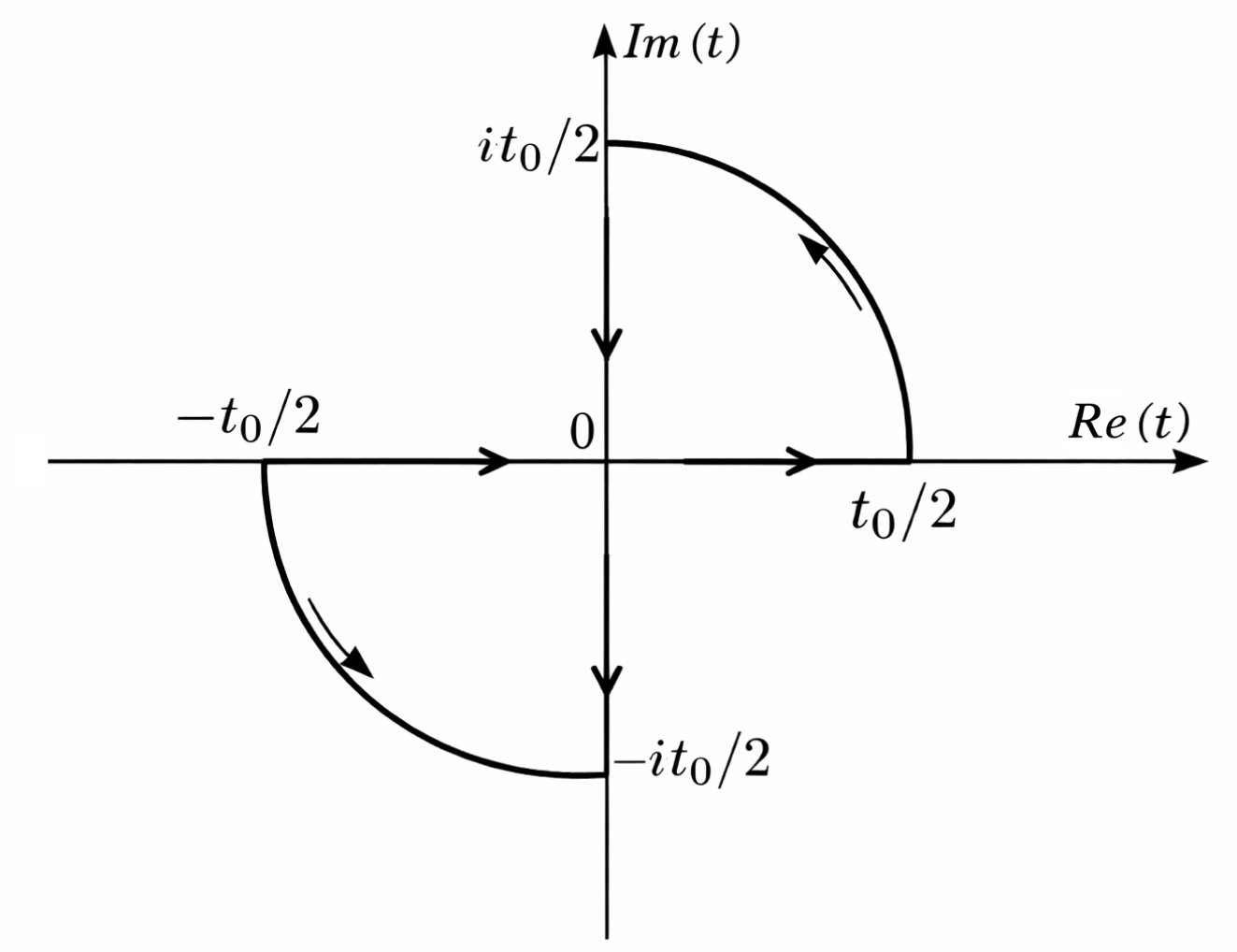

**Figure 1.4.:** Example of the contour in the complex plane, for an action being defined between times $-\frac{t_0}{2}$ and $\frac{t_0}{2}$, used for the analytic continuation of the action of a quantum system to define its partition function.

The intricacies and conditions upon which this works are not relevant at this stage, but the end result is that one can think of it as a "variable transformation" of the form $t \to -i\,\tau$, with $\tau$ a real parameter that ranges on the same interval as the original time coordinate.

The fact that in imaginary time the integrals defining the partition function are real makes them potentially amenable to computational methods, and this will be crucial for the computational simulations that we will use later on in this thesis. This imaginary time continuation makes the partition function $Z$ of the example in (1.6) convergent, since oscillatory integrals now become real Gaussian integrals. More concretely, what one does is to redefine the partition function as

$$Z = \lim_{\Delta t \to 0} \prod_{n}^{N_T} \int_{-\infty}^{\infty} dq(t_n) e^{i \sum_j^{N_T} \Delta t \, \frac{m}{2} \left( \frac{q(t_j) - q(t_{j-1})}{\Delta t} \right)^2}$$

$$\downarrow$$

$$Z \equiv \lim_{\Delta \tau \to 0} \prod_{n}^{N_T} \int_{-\infty}^{\infty} d\tilde{q}(\tau_n) e^{- \sum_j^{N_T} \Delta \tau \, \frac{m}{2} \left( \frac{\tilde{q}(\tau_j) - \tilde{q}(\tau_{j-1})}{\Delta \tau} \right)^2}, \qquad (1.7)$$

where $\tilde{q}(\tau)$ are new variables defined from the analytic continuation of the coordinates of the particle on imaginary time.

Making computations in imaginary time $\tau$ and later converting the results to real time by doing the reverse transformation $\tau \to i\,t$ is a way to make sense of the quantum partition function (1.6). This continuation to imaginary time is a standard practice for rendering some path integrals in quantum mechanics well-behaved, applicable to every interaction in the Standard Model of particle physics [26].

This imaginary time continuation of the partition function, often called "Wick rotation", has a very interesting interpretation. When going to imaginary time, the interval of integration of the new variable $\tau$ has the interpretation of the inverse of the temperature of the system, and the partition function in imaginary time can be thought of as a *statistical partition function* of the theory. The statistical analytical continuation of a relativistic quantum field theory is often called the "Euclidean version of the theory", and the resulting action in imaginary time, defined as minus the argument of the exponential in the partition function, is the "Euclidean action" ($S_E$).

While a study of this correspondence is beyond the scope of this thesis, it is important to keep in mind this distinction, as it will come again in other parts of the thesis: Computations done using imaginary time are statistical, not quantum, and to map them to a particular quantum theory computation, one must have a well defined mapping between the coordinates and field variables used in the statistical system and those that define an actual quantum theory, just like one has for the point particle example by doing $t \to -i\,\tau$ and $\tau \to i\,t$ [27].

Unfortunately, this simple trick of changing the time variable does not straightforwardly work for gravity. The issue when trying to use this imaginary time technique in gravity comes due to the fact that in gravity, time is not a fixed parameter associated with a particular background upon which one defines a quantum theory. In gravity, spacetime is quantum itself, and a definition of a notion of time is the first obstacle when trying to make a quantum theory. Unless one resorts to a particular background structure, defining an imaginary time continuation in gravity is a complicated issue.

We will see how this issue can be addressed in a particular computational approach that we will study in this thesis, where a more "geometric" analytic continuation achieves the same result. With these ingredients at hand, we are now in a position to discuss the main issues faced when attempting to construct a well-defined theory of Quantum Gravity, and how we will try and work around them in the next parts of this work.

### 1.2.2 Divergences, scale dependence and Renormalization

Quantum Mechanics deals with finitely many degrees of freedom, something that makes the definition of path integrals much easier than in Quantum Field Theory. QFT attempts to describe infinitely many degrees of freedom across arbitrarily many scales. This leads to the rather generic appearance of either infrared (IR) or ultraviolet (UV) divergences in the integrals involved in the path integrals. Furthermore, QFTs sometimes have the so-called *gauge symmetries*, field transformations that leave the theory invariant, and typically imply redundant field integration directions, introducing extra divergences in the path integrals of QFTs that one has to deal with. This means that even if one has a prescription like a Wick rotation to render a partition function better behaved, one still has to deal with those other divergences to render the computation of observables well-defined.

A central organizing principle that makes such computations meaningful is the framework of the *Renormalization Group* (RG) [28]. The basic idea is the following. One introduces a scale, for instance, a momentum cutoff, that regulates the theory and separates "short-distance" from "long-distance" contributions to the expectation values. One then allows the parameters of the theory, the couplings that appear in the bare action, to depend on this scale in such a way that physical quantities remain finite and stable when the regulator is removed or when the scale is changed. In other words, instead of thinking of the couplings in the action as fixed numbers, quantum field theory teaches us to think of them as *scale-dependent* quantities. This running with energy is not a technical detail: it is a characteristic and experimentally verified feature of quantum field theories.

Closely related to this is the notion of an *effective* description. The organizing principle mentioned before allows computing expectation values in a certain range of scales, and this allows for the definition of an "effective action", whose parameters encode the influence of the different scales that contributed to the expectation values. In this way, the dynamics of expectation values of fields can be captured by scale-dependent effective equations, derived using the least action principle from the effective action. The renormalization group is fundamentally a statement about how physics changes under coarse-graining.

A crucial feature a QFT must have to be considered a fundamental description at all energies is that its effective action must be finite at all scales. In particular, in this scenario, the scale-dependent couplings of the theory must be finite at high energies. The Standard Model is an example where this is not the case, due to the presence of Landau poles [29], and is thus expected to be replaced by another model at high energies. A possible mechanism to avoid those divergences at high energies is a quantum-scale invariant limit of the QFT, named Asymptotic Safety (AS). We will later come back to this mechanism in Part III, where scale dependence becomes the main handle to discuss high-energy theories of gravity, and possible infrared outcomes.

## 1.3 A quantum field theory of gravity

The standard computational engine of quantum field theory is perturbation theory: one expands around a solvable limit (a "free" theory), treats interactions as small corrections, and organizes observables in a series expansion in one or multiple perturbative couplings. The perturbative framework enables a correspondence between a particular expectation value and its corresponding diagrammatic representation, known as Feynman diagrams. Without diving into their technicalities, diagrams that do not contain loops in them are usually called tree diagrams, and the ones that do are called loop diagrams. The latter are, in general, divergent.

The expectation value of a certain quantity contains all possible diagrams of the theory, but in a perturbative theory, the more loops a diagram has, the higher its order in perturbation theory is, and so perturbative expansions are often called loop expansions of the theory. In Quantum Field Theories that are well behaved under perturbations, these divergences of the loop expansions are not a problem, because the divergences can be absorbed into a *finite* set of parameters already present in the bare theory: masses, couplings, and field normalizations.

1.3 Assuming that the fundamental field of the gravitational theory, the one to be quantized, is the metric, then one can try to proceed analogously to what is done with the interactions of the Standard Model, and construct a diagrammatic expansion of the quantum theory, using as a starting point the Einstein-Hilbert action (1.2). The way this is done is by splitting the full metric $g_{\mu\nu}$ into a fixed background $\bar{g}_{\mu\nu}$, plus a small fluctuation given by $h_{\mu\nu}$, which is the perturbation upon which one builds the perturbative expansion of the QFT

$$g_{\mu\nu} \;=\; \bar{g}_{\mu\nu} + \kappa\, h_{\mu\nu}\,, \qquad \kappa^2 = 32\pi G\,. \tag{1.8}$$

Using this splitting, one can expand the Einstein-Hilbert action in powers of $\kappa$, and compute expectation values order by order in $\kappa$. Without going into the technicalities of this approach, it suffices to say that one must, before performing computations of expectation values, deal with the gauge symmetry of this theory. In gravity, this gauge symmetry corresponds to *diffeomorphism transformations* of the manifold where the metric is defined. Eliminating this gauge symmetry, the perturbative approach leads to the conclusion that a quantum theory of GR has a fundamental quanta, a.k.a., the yet-unobserved particle called the "graviton", a spin two massless excitation of the gravitational field [30]. This is, schematically, the perturbative Quantum Gravity picture.

For many good reasons, it is canonical to use the Minkowski metric as a background, since that is the background used for quantizing all other interactions of the SM. The first problem one encounters is how to define the path integral, since, as we discussed before, in a real-time formalism, its partition function is ill-defined. Using the Minkowski background, one has a canonical time coordinate to use in perturbation theory (up to Lorentz transformations), and one can use the analytic continuation explained before, which works for the SM fields. Unfortunately, this yields a Euclidean action that is unbounded below, leading to the well-known "conformal factor problem" [31]. This makes the statistical partition function divergent as well, so this standard method proves useless for gravity.

Regardless of this, imaginary time continuations are not the only tool available for computing expectation values using the path integral, and other analytic continuation techniques allow one to control the infinities of the path integral and loop expansions. Unfortunately, when one attempts to quantize GR perturbatively around a fixed Minkowski background and computes loop corrections, the ultraviolet divergences do not remain confined to the Einstein-Hilbert form. In pure gravity, in the absence of a cosmological constant, the one-loop divergences vanish on-shell [32], but divergences appear once matter is included, and at two loops, pure gravity itself is divergent [33]. Either way, the divergences that arise cannot be absorbed solely by renormalizing Newton's constant $G$ and the cosmological constant $\Lambda$.

The remedy within perturbation theory is to enlarge the bare action by adding higher-derivative, diffeomorphism-invariant counterterms. A convenient way to parametrize this enlargement is to write the most general local action as an expansion in curvature invariants, for example

$$S[g] \sim \int d^4x \sqrt{|g|} \left[ \frac{1}{16\pi G} (R - 2\Lambda) + c_1 R^2 + c_2 R_{\mu\nu} R^{\mu\nu} + \cdots \right], \quad (1.9)$$

where $c_1, c_2, \ldots$ are independent couplings multiplying higher-curvature operators, and the ellipsis stands for further local curvature scalars. The important point is not simply that such terms can be written down, but that perturbation theory forces them upon us: once loop corrections are included, new divergent structures typically appear that require new counterterms, and hence new couplings, beyond those already present. As one proceeds to higher orders, the set of required operators grows rather than closing on a finite basis, which is the practical content of perturbative non-renormalizability.

This is the precise sense in which a perturbative approach to quantize General Relativity fails as a definition of a quantum field theory: the perturbative program does not yield a predictive framework with finitely many parameters.

These results led to a plethora of attempts to find a quantum theory for gravity, often by modifying at least one of the ingredients that we normally take for granted in quantum field theory: the choice of fundamental degrees of freedom, the continuum nature of spacetime, locality, or even the dimensionality and symmetry structure of the microscopic theory. The most prominent examples of attempts at such quantization are String Theory and Loop Quantum Gravity, but the landscape is broader. Without aiming at an exhaustive review, it is useful to keep in mind a rough map of directions, other than the ones pursued in this thesis, and the type of additional assumptions they commonly introduce.

A first large family of approaches changes the *fundamental* degrees of freedom. In String Theory, the point-particle picture is replaced by one-dimensional extended objects such as strings (or higher-dimensional generalizations), with consistency typically tied to additional symmetry and structural input, such as supersymmetry and extra dimensions [34, 35]. In Loop Quantum Gravity, one keeps the spirit of background independence but replaces the metric field by holonomies and fluxes, leading to spin network states and discrete spectra for geometric operators; its path integral counterpart is commonly formulated in terms of spin foams [36–39]. Closely related in spirit are Group Field Theory and tensor model approaches, where one builds quantum field theories on group manifolds whose quanta can be interpreted as "atoms" of geometry, and where continuum spacetime is expected to emerge collectively from many-body phases [40–42].

A second family keeps gravity as a theory of geometry but changes the *underlying structure* of spacetime. A prominent example is Causal set theory, which assumes that the fundamental structure is a locally finite partially ordered set encoding causal relations, a discrete spacetime rather than a differentiable manifold, with Lorentzian causality built in from the start [43–45].

A third family modifies the *principles* of the microscopic dynamics, either by assuming violations of Lorentz symmetry, or by proclaiming that gravity is an emergent phenomena guided by thermodynamics. Examples include scenarios where Lorentz invariance is not fundamental but emergent, and introduce anisotropic scaling between space and time at high energies, with the hope of improving ultraviolet behaviour by power counting [46, 47]. Closely related in motivation are many emergent gravity ideas, where gravity (and sometimes even spacetime) is not fundamental but arises as an effective, coarse-grained description of underlying quantum degrees of freedom, often guided by thermodynamic or information-theoretic principles [48–50].

A fourth family keeps the field content of gravity but modifies the microscopic dynamics through *extensions* of the gravitational action. From an EFT viewpoint, higher-curvature operators are expected to appear, and one can ask whether promoting them to the defining UV dynamics yields a consistent fundamental theory, still using standard perturbation theory. This leads to higher-derivative gravity, with its most well-known representative being quadratic gravity, improving the perturbative properties of the theory [51, 52]. Related non-local (often infinite-derivative) extensions improve the ultraviolet behaviour of quadratic gravity by softening propagators through form factors [53, 54].

The common denominator of the above list is not that any of these directions is “unnatural” or “wrong”, but that most of them seek ultraviolet control by adding new microscopic ingredients: extra new fundamental objects, extra symmetries, extra dimensions, extra propagating degrees of freedom, or discreteness assumptions. This is a perfectly legitimate strategy, and it has yielded profound insights and powerful mathematical frameworks. However, every single one of these approaches requires, to some degree, going beyond General Relativity, leading to a vast non-uniqueness to extend GR for its quantization. This generates the following prominent questions: Given that it is currently outside the reach of experimental physics to directly test any of these new ingredients introduced to construct a UV description of gravity, can we take a more conservative point of view? How far can one get while staying as close as possible to the standard ingredients of quantum field theory?

## 1.4 Non-perturbative methods as a key to Quantum Gravity

At this stage, it is useful to stress that the failure of perturbation theory is not unique to gravity, and it is not the only way to define a QFT of the same degrees of freedom. Even within the Standard Model, there are regimes where the perturbative expansion that works beautifully at high energies becomes useless. The most prominent example is Quantum Chromodynamics (QCD). At sufficiently high energies, the strong interaction becomes weakly coupled and perturbative methods give remarkably accurate predictions [55]. At low energies, however, the coupling grows, and the perturbative loop expansion loses predictive power [56]. In that regime, one does not conclude that the QFT of QCD is inconsistent; rather, one changes the methodology to make calculations.

1.4

A particularly successful strategy in QCD is to introduce an auxiliary regulator in the form of a lattice [57]. One replaces the continuum functional integral by a well-defined, regulated system with finitely many degrees of freedom, performs a Wick rotation to obtain a Euclidean/statistical partition function, and then evaluates expectation values numerically. The lattice is not assumed to be physical, but it is merely an auxiliary computational tool. The physics is extracted by searching for a quantum scale-invariant regime in which the lattice regulator can be removed, and by studying how observables depend on the scale at which the system is probed. Furthermore, non-perturbative renormalization group techniques allow us to study effective descriptions of the theory [58], connecting UV and IR through RG flows.

This logic provides a clean alternative path: when perturbation theory fails, one can still define the quantum theory through non-perturbative methods and use computational or RG techniques to explore its properties. This is the avenue pursued in this thesis for the quantization of gravity.

The idea to define a QFT of gravity is still that of a path integral, a sum over configurations of the field, in this case, all geometries. Schematically, one would like to define the formal expression

$$Z_{\text{grav}} \equiv \int \mathcal{D}[g_{\mu\nu}]\, e^{i\,S[g]}\,, \tag{1.10}$$

together with expectation values of observables $\mathcal{O}$

$$\langle\mathcal{O}\rangle \equiv \frac{1}{Z_{\text{grav}}} \int \mathcal{D}[g_{\mu\nu}]\, \mathcal{O}[g]\, e^{i\,S[g]}\,, \tag{1.11}$$

with $S[g]$ being the Einstein-Hilbert action (with the possible inclusion of extra terms), and where $\mathcal{D}[g_{\mu\nu}]$ denotes integration over all possible Lorentzian spacetime metrics, modulo the gauge symmetry of the theory, diffeomorphism transformations. As emphasized earlier, these expressions are formal: the measure is subtle, the diffeomorphism redundancy must be treated carefully, and the oscillatory weight makes the definition of these computations difficult to control. Part of the non-perturbative viewpoint taken in this thesis is then to proceed as in QCD: introduce a lattice regulator that turns the definition into a mathematically well-defined quantity, and use it to explore the theory beyond perturbation theory.

A central part of this process is again the role of a Wick rotation. In ordinary quantum field theory, Euclidean continuation provides a bridge between quantum amplitudes and statistical ensembles, replacing $e^{iS}$ by $e^{-S_E}$ and turning oscillatory amplitudes into well-behaved real weights. As we will see in the following part of this thesis, this can still be done for gravity, at least for the lattice approach studied in this thesis.

However, defining a regulated ensemble is only half of the problem. The other half is understanding how physics depends on scale, and whether the theory admits a meaningful ultraviolet quantum scale-invariant limit. In any regulated construction, such as a lattice implementation, there are small and large distance cutoffs. A continuum quantum field theory is obtained only if one can remove these cutoffs while keeping physical predictions finite, which typically requires approaching a scale-invariant regime, as we will discuss later on.

The renormalization group perspective introduced earlier provides a natural language to discuss precisely these questions, and it is not restricted to a particular implementation. Furthermore, its role becomes very important when one uses non-perturbative approaches. Computational lattice simulations are necessarily performed at finite lattice spacing and finite volume, and therefore are limited by computational resources. Thus, one typically has access to a finite window of scales, and extracting continuum physics requires controlling how observables change as the cutoff is varied. The renormalization group provides the conceptual bridge: it orders the scale dependence and allows one to map UV to IR physics.

From my perspective, this is also why it is valuable to keep lattice and renormalization group viewpoints in close contact. Lattice-inspired definitions provide a concrete non-perturbative arena where geometric observables can be measured directly, while renormalization group methods provide the language to connect scales and to interpret those measurements in terms of universal behaviour and effective dynamics. In the remainder of this thesis, these two perspectives will therefore be used as complementary tools to explore the same underlying question: how to define Quantum Gravity at the smallest scales, and how to understand its possible long-distance consequences.

## 1.5 Road map of this thesis

The parts of this thesis that follow pursue the goal of understanding Quantum Gravity from different angles, each one emphasizing a different question and a different set of tools.

Part II works with a lattice and statistical viewpoint, through a formalism called Causal Dynamical Triangulations (CDT). The driving questions of this part are: if we define gravity as a sum over geometries in a path integral, what do the typical quantum geometries look like, and how do we measure geometric properties in the high-energy regime of QG? The main emphasis is therefore on observables. We developed several concrete probes that let us address questions about emergence: Does an effective notion of topology stabilize under coarse-graining, or does this approach to QG predict a fractal universe? Do symmetries such as homogeneity appear at some scale, as we see in our universe? Can one construct curvature-related quantities that behave like meaningful geometric observables in a fluctuating quantum spacetime?

Part III switches viewpoint and asks a complementary question: Since lattice simulations have access to limited short-distance scales, how can we connect Planckian physics to infrared physics in a controlled way? Here, the renormalization group becomes the main language. Rather than sampling individual geometries, the focus is on the scale dependence of effective descriptions, and on the possibility that gravity can be ultraviolet complete through a non-trivial scaling regime, where we will work under the framework of Asymptotic Safety (AS). Within this framework, we develop tools aimed at making predictivity practical: We introduce and exploit RG flow-invariant structures that organize the space of renormalization group trajectories, and we present a computational strategy, including a machine-learning implementation, to construct predictions efficiently. We then move towards a scalar-tensor setting in which one can explore how infrared interaction potentials and cosmological dynamics may emerge from ultraviolet consistency conditions, illustrating how a quantum-gravity motivated framework can predict effective physics rather than merely accommodate it.

Part IV is my own personal detour towards Quantum Gravity, motivated by a question that sits on a lower level than the ones tackled in Parts II and III: What is it that we are choosing to quantize when we say "Quantum Gravity"? After exploring two metric-based non-perturbative viewpoints, I investigate a conservative alternative in which the quantized variable is not the metric but a more primitive one, the affine connection, and the metric (when it exists at all) is a derived, emergent object. I then take the next step and define an explicit lattice statistical partition function for the connection on a hypercubic lattice, together with an implementation of a Monte Carlo algorithm to explore its configurations, and I present the first exploratory numerical results. I finally describe the ongoing extension of this framework to a four-dimensional homogeneous and isotropic setting, where an emergent metric and genuinely geometric observables can be constructed.

Part V contains the conclusions that bring these threads together. They summarize what the lattice observables and the renormalization group tools teach us about possible quantum-spacetime behaviour, they emphasize the open problems that remain essential if one wants to connect a non-perturbative Planckian definition to testable infrared physics, and they discuss the opened new avenues and alternatives to understand gravity at its fundamental level.

*Join me in this journey exploring the many facets of spacetime from Planckian to cosmological scales.*

# Part II

# Quantum geometry as a gateway to Planckian universes

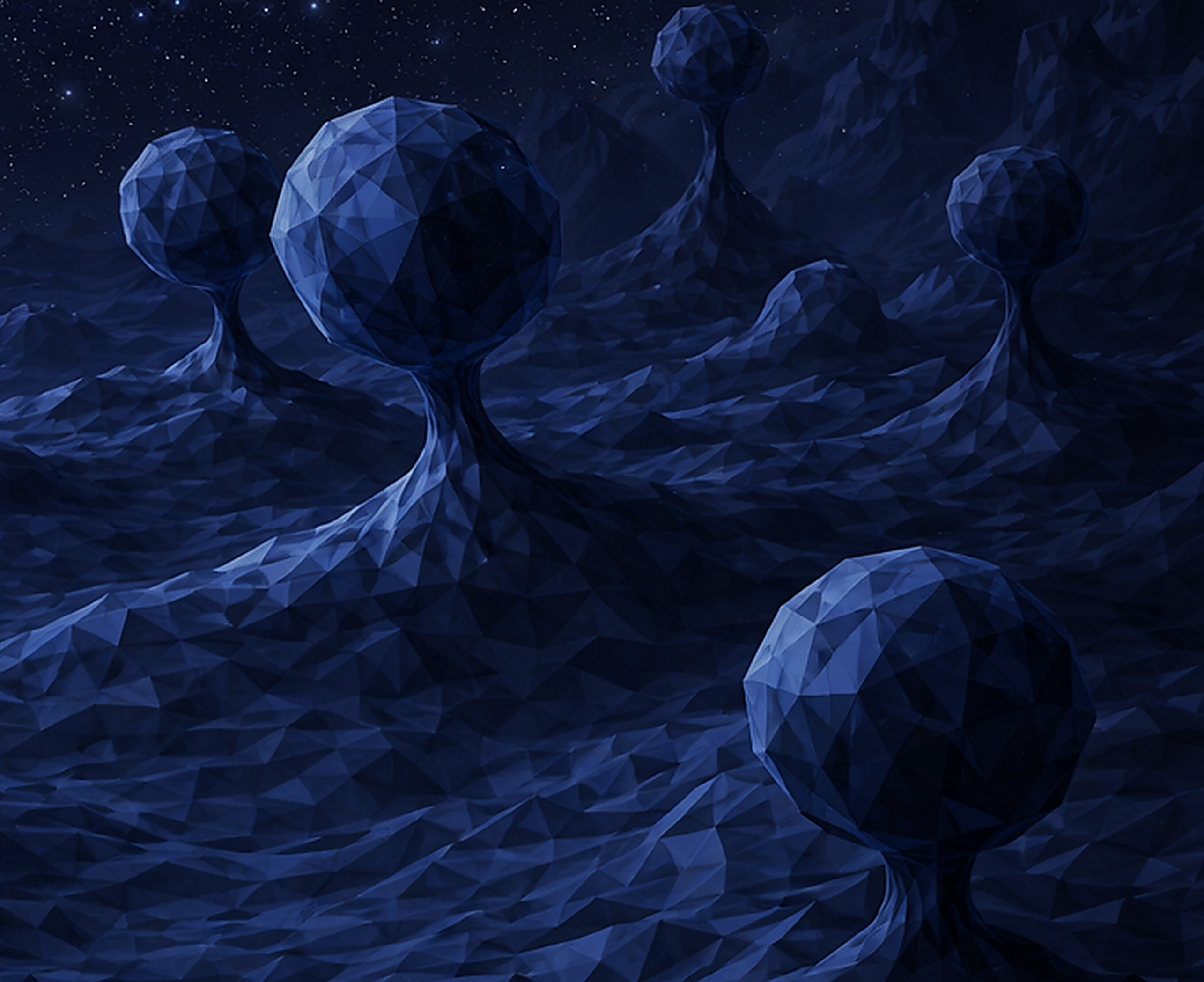

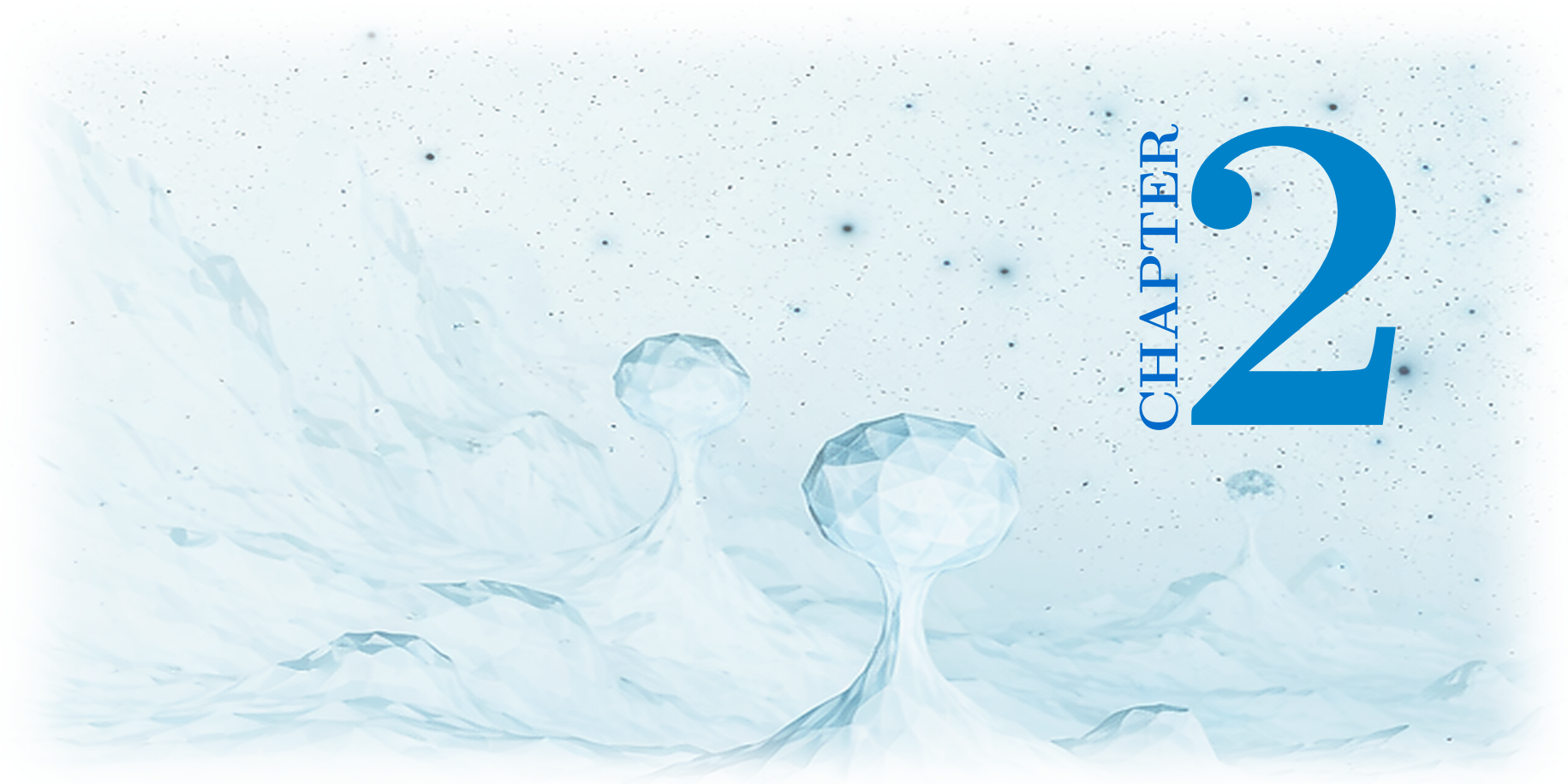

# CDT and the gravitational path integral

Before diving into the technicalities of Causal Dynamical Triangulations (CDT), I would like to first establish a very uncommon, but highly desirable, property of CDT within Quantum Gravity approaches: the ingredients of this approach are very simple! And this is what I would like to convey in the next paragraphs. To do so, I will first explain conceptually what CDT does, and later I will dive into some of its necessary technicalities, to provide the reader the possibility of a better understanding of this field.

Let us momentarily go back to the definition of the path integral for the free particle given in Part I. The configurations that contribute to the expectation values in the path integral formalism are those that are continuous, piecewise flat, and nowhere differentiable. If one is to attempt a numerical implementation of such a path integral, one option would be to approximate the sum over infinitely many configurations by some average over a large but finite subset of them, with a suitable finite discretization of their domain and extension in their target space. The next step in such a program will then be to reach the continuum limit of this approximation by suitably increasing the number of configurations while simultaneously taking into account and reducing the discretization artifacts.

This conservative approach is the one taken in Causal Dynamical Triangulations to define and regularize the gravitational path integral. CDT is the analogue of lattice QCD in the landscape of Quantum Gravity approaches. It defines and regularizes the abstract summation over all geometries in (1.10), with a concrete sum over all possible simplicial discretizations of spacetime which satisfy certain gluing rules [59–61]. This regularization is achieved via a discretization of the Einstein-Hilbert action using the so-called Regge calculus [62–64]. The used simplicial manifolds have a particular foliation structure, making them globally hyperbolic, and the formalism comes with a well-defined notion of Wick rotations of these discrete curved spacetimes [59, 65], where by discrete we mean that the manifolds are determined by an integer number of simplicial building blocks of certain types. Such a notion of Wick rotation for every configuration of the path integral is what allows this approach to use computational and statistical methods to perform explicit non-perturbative calculations, something not accessible in most other approaches.

The continuum limit of this regularization of the gravitational path integral can be obtained by suitably approaching a second-order phase transition in the parameter space of the theory, while simultaneously increasing the number of simplicial building blocks to infinity [66, 67].

These are all the ingredients that go into CDT. No need for the introduction of extra dimensions, unseen new symmetries, exotic particles, higher-order terms in the bare action, or anything else. Just lattice Quantum Field Theory as we know it.

Of course, the program is subject to one crucial assumption. To reach a continuum limit, there must be a second-order phase transition in parameter space, i.e., the numerical simulations should show a quantum scale-invariant UV regime for a suitable fine-tuning of the coupling constants of the theory. The existence of this phase transition is currently under investigation, but the latest numerical simulations suggest that there is at least one such transition line, so the approach looks promising [66, 68, 69].

To better understand this approach, let me explain in more detail how said simplicial regularization and foliation are implemented. To start with, the elementary building blocks of the path integral configurations of CDT are $d$-dimensional *simplices*, which are generalizations of triangles in $d$-dimensions. These $d$-simplices are pieces of flat $d$-dimensional Minkowski spacetime. The edges of such simplices are either spacelike, with squared length $\ell_s^2 = a^2$, or timelike with squared length $\ell_t^2 = -\alpha\, a^2$, where $a$ has length units and parameterizes the lattice spacing and $\alpha$ is a positive real. This asymmetry between space and time-like edges given by $\alpha$ is precisely what distinguishes the simplicial manifolds built from these blocks from being purely Euclidean [59].

To discretize the gravitational path integral, one first approximates spacetime with a simplicial manifold. It will be assumed that the geometries of the path integral configurations all accept a discrete foliation in terms of a timelike direction, and thus each type of $d$-simplex that constitutes the manifold can be uniquely determined by the number of its vertices that lie on the same spatial slice (slice of constant time). In the simplest case, and the only case we will explore in this thesis, that of two-dimensional spacetimes, this leaves only one possible building block, a triangle with two vertices on the same time slice, and the third one in a neighbouring slice, with two possible orientations, up or down, depicted in Fig. 2.1.

Once spacetime is discretized in terms of a simplicial manifold, one must also find the discrete expression for the Einstein-Hilbert action, including a cosmological constant $\Lambda$

$$S_{\text{EH}}[g_{\mu\nu}] = \frac{1}{16\pi G} \int_M d^d x \sqrt{|g|} \left(R - 2\Lambda\right) . \tag{2.1}$$

This we owe to Tullio Regge [62, 63]. The first term is proportional to the integral of the Ricci scalar over the entire spacetime, and the other is proportional to the volume of it. It is rather immediate to find a discrete analogue of the latter, since it would correspond to the total sum of the volumes of the building blocks of the simplicial manifold. The former is more complicated.

Since each simplicial complex is just a piece of Minkowski spacetime, its interior is flat. The only way curvature can arise is by the way these $d$-simplices $\sigma^{(d)}$ are glued. One of Regge's contributions was to note that this curvature was concentrated in the so-called *hinges*, denoted as $h$, which are co-dimension 2 sub-simplices defined by the intersection of $d$-dimensional simplices [63]. For example, in a two-dimensional manifold, the hinges are the vertices of the complex.

To define a quantity that represents the Ricci scalar under the integral sign, Regge made use of the concept of deficit angle [64], defined on hinges of the complex, and typically denoted as $\epsilon(h)$. To compute $\epsilon(h)$ one must determine all the incident $d$-simplices at a hinge $h$, meaning all those that share $h$, a set we will denote $I(h)$. Each incident simplex $\sigma_i^{(d)} \in I(h)$ will have an internal angle between its "faces", co-dimension 1 sub-simplices of

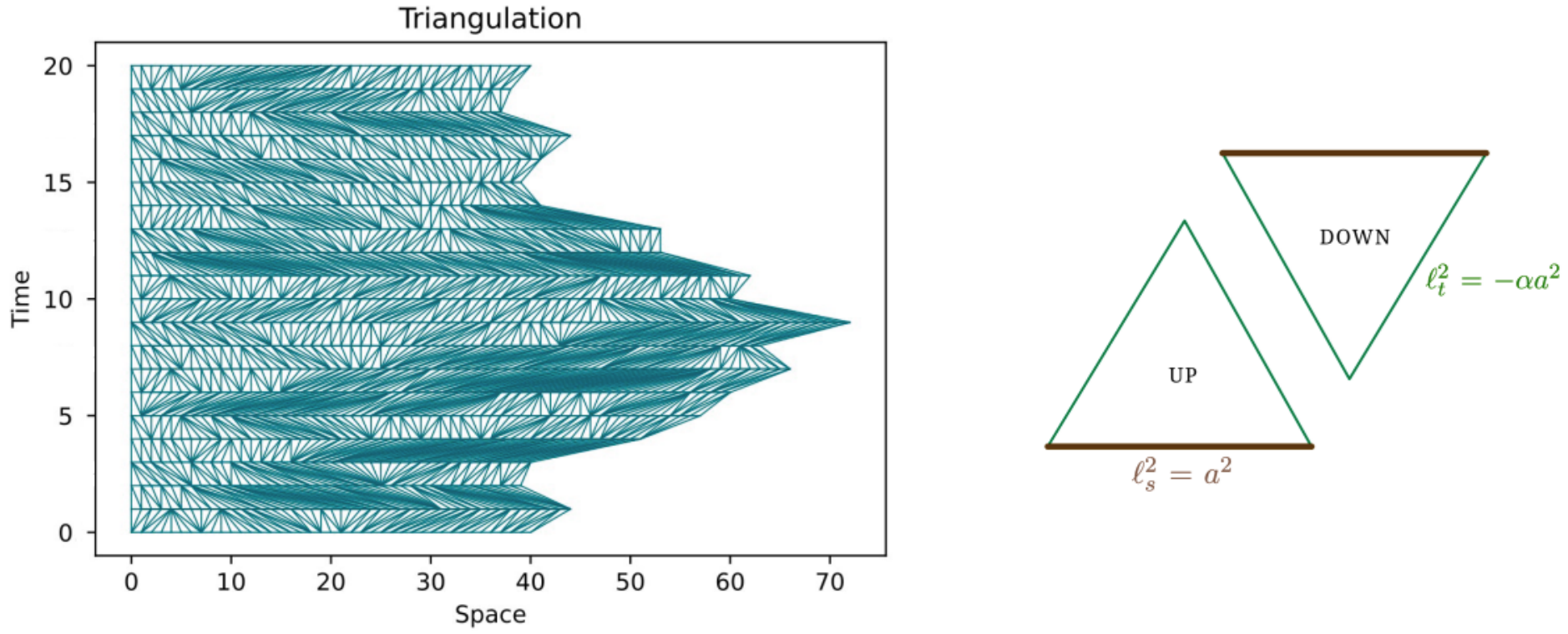

**Figure 2.1.:** On the left, one example of a 2D CDT path integral configuration with topology $S^1 \times S^1$ and represented on the plane by cutting the torus, $N_T = 20$ time slices, and a total volume of $N_2 = 2000$ triangles. Uppermost vertices are identified with the lowest ones, as well as left and right extreme vertices of each time slice. On the right, the two possible orientations of building blocks, UP and DOWN.

a $d$-simplex, which we will denote as $\delta(\sigma_i^{(d)})$. The deficit angle is defined as

$$\epsilon(h) = 2\pi - \sum_{\sigma_i^{(d)} \in I(h)} \delta(\sigma_i^{(d)}). \tag{2.2}$$

Local positive and negative Gaussian curvature are interpreted as $\epsilon > 0$ and $\epsilon < 0$ respectively, while $\epsilon = 0$ is interpreted as local flatness.

Regge proved that the sum $\sum_{h \in T} V(h)\,\epsilon(h)$, over all hinges in the simplicial manifold $T$, converges to the integral of the Ricci scalar over the entire manifold, in a suitable continuum limit of the simplicial manifold, thus, constructing a simplicial discretization of the integral of the Ricci scalar appearing in (2.1) [64].

Collecting these definitions, this discretized version of the Einstein-Hilbert action, called the *Regge action*, is

$$S_R[T] = \frac{1}{8\pi G} \sum_{h \in T} V(h)\,\epsilon(h) - \frac{\Lambda}{8\pi G} \sum_{\sigma^{(d)} \in T} V\left(\sigma^{(d)}\right), \tag{2.3}$$

where $T$ is the simplicial manifold, and the sums are over all hinges and all simplices on $T$. Now, this expression might not be very illuminating on its own, but each of the deficit angles and volumes has an expression in terms of the edge lengths of the simplices involved, and the number of incident sub-simplices at each hinge [63]. With all these tools at hand, we can take the leap and discretize and define the formal expression for the gravitational path integral as is done in the CDT approach

$$Z = \int \mathcal{D}[g_{\mu\nu}]\, e^{iS_{\mathrm{EH}}[g_{\mu\nu}]} \quad \to \quad Z = \sum_{T \in \mathcal{T}} \frac{1}{C_T}\, e^{iS_R[T]}\,, \tag{2.4}$$

where $\tilde{\mathcal{T}}$ is the set of all Lorentzian simplicial manifolds with fixed topology, and $C_T$ is the order of the automorphism group of $T$, whose role is to reduce the weight in the sum of discretizations with internal symmetries that make them equivalent [61].

A remarkable feature is that the resulting expression allows for an analytical continuation to $\alpha < 0$, effectively implementing a Wick rotation of the geometries. This rotation cancels the imaginary prefactor, leaving an exponential of minus a real-valued Euclidean Regge action $S_R^E$

$$Z = \sum_{T \in \tilde{\mathcal{T}}} \frac{1}{C_T}\, e^{iS_R[T]} \quad \to \quad Z = \sum_{T \in \mathcal{T}} \frac{1}{C_T}\, e^{-S_R^E[T]}\,, \tag{2.5}$$

with $\mathcal{T}$ the set of all Wick rotated simplicial manifolds with fixed topology. It is with the latter Euclidean version of the formalism that we can make use of numerical simulation tools [70]. This feature of a well-defined Wick rotation, with no known analogue in the continuum, is only present in CDT [65].

CDT explores the space of path integral configurations by keeping fixed the squared edge length scale $a^2$ of the simplices, and where the only degrees of freedom are the number of each type of simplices that appear in the simplicial manifold, and the way they are glued together, i.e., their connectivity. With this restriction, the bare action $S_R^E[T]$ has remarkably simple expressions in 2, 3, and 4 dimensions, being entirely determined by the total number of simplices of each type, and the values of the cosmological constant, Newtons constant, the asymmetry parameter $\alpha$, and the squared edge length scale $a^2$. So far, this formalism has produced astonishingly good results [59, 61].

For concreteness, in 2 dimensions, the term that corresponds to the scalar curvature integral in the bare action $S_R^E[T]$ is purely topological, and since topology is kept fixed, this term yields a constant that can be ignored. Removing the purely topological part, $S_R^E[T]$ takes the form

$$S_R^E[T] = \lambda N_2 \,, \tag{2.6}$$

where $\lambda = \frac{\Lambda a^2}{8\pi G}$ is a dimensionless bare parameter, and $N_2$ the total number of triangles in the manifold.

Taking a step back from the technicalities, it is important to summarize what we have achieved so far. With $S_R^E[T]$, there is a concrete definition of what the gravitational path integral over all geometries of fixed topology is. Furthermore, this was done over geometries that, before the Wick rotation, were the simplicial analogue of globally hyperbolic. This means that a notion of time is kept in the quantization and holds over all the path integral configurations [59]. Having a well-defined and computable notion of Euclidean time in all situations is necessary to map the statistical theory to a Quantum theory, as described in Part I.

On top of that, the formalism of Regge calculus, upon which CDT is built, does not require the use of coordinates to describe the path integral configurations. Since diffeomorphisms are only a gauge symmetry of smooth manifolds described using coordinates, the formulation of the theory without coordinates removes most of the redundancies associated with them. In the general setting of Regge calculus, one still has the freedom to change some edge lengths and leave the Regge action invariant, but this redundancy is eliminated in CDT, where all edges are of constant length, and thus, the issue of gauge fixing is avoided, together with all the technicalities that go with it. One aspect of this situation is that, since we are working with non-smooth manifolds, standard tensor calculus no longer works, and we are left with only two very fundamental tools: *lengths and volumes.*

Whatever spacetime property you want to unravel in these geometries, you have to find a way of doing that using only the connectivity structure of the manifold, the edge lengths, and the associated volumes of individual (sub)simplices. This forces us to develop new mathematics to explore geometrical properties that do not rely on the notion of tensors and calculus on pseudo-Riemannian manifolds. Furthermore, the geometries that appear in the non-perturbative path integral are not close to smooth classical manifolds with small fluctuations on top, but instead, they present fractal properties and very large fluctuations at short scales.

The current status of computational simulations of four-dimensional CDT is capable of studying geometries with an average linear extension of (up to) the order of 20 times the Planck length. This means that it provides a unique Planckian-sized window to study a highly non-perturbative sector of Quantum Gravity, where we can explore its properties and search for new physics [71, 72]. Unfortunately, larger sizes are currently unfeasible to simulate due to their high computational cost.

Despite the small sizes accessible in this approach, CDT has been able to provide new insights into the structure of spacetime at those scales. For example, it has produced strong evidence for the, now-conjectured universal property of dynamical reduction of the spectral dimension from 4 to 2 at small distances [68, 73], the emergence of a de Sitter-like ground state of the path integral, with positive average curvature and the effective topology of a sphere [59, 71, 74], and the existence of a rich phase diagram [66, 67], among many other features.

While great efforts were and are being made to reach the continuum limit of 4-dimensional CDT, in this thesis, we will only focus on two-dimensional quantum geometries, since our goal is to develop something we call "observables", which we will introduce in the next sections. Such observables can be used in $d$ dimensions, and to determine the properties of spacetime with very few accessible tools. Our main goal was to produce methodologies capable of efficiently extracting properties of spacetime, rather than focusing on the renormalization required to reach a continuum limit of the theory. Therefore, we will not dive deeply into the continuum limit of these discretizations, and instead, we will focus on their robustness and interpretations, and we will test such observables in the toy model of two-dimensional quantum geometries. We will see that even in such a simple setting, one encounters non-trivial results, with non-trivial difficulties and obstacles to overcome, that hopefully provide steps forward in the understanding of higher-dimensional spacetimes.

The details of the implementation of the computational simulations needed to perform calculations within CDT can be found in the Appendix A. A more technical reader can go to the Appendix and explore some of the intricacies of the implementations. An example of a 2-dimensional CDT path integral configuration is depicted in Fig. 2.1 on the plane, and a 3D embedding of another 2-dimensional CDT configuration in Fig. 2.2.

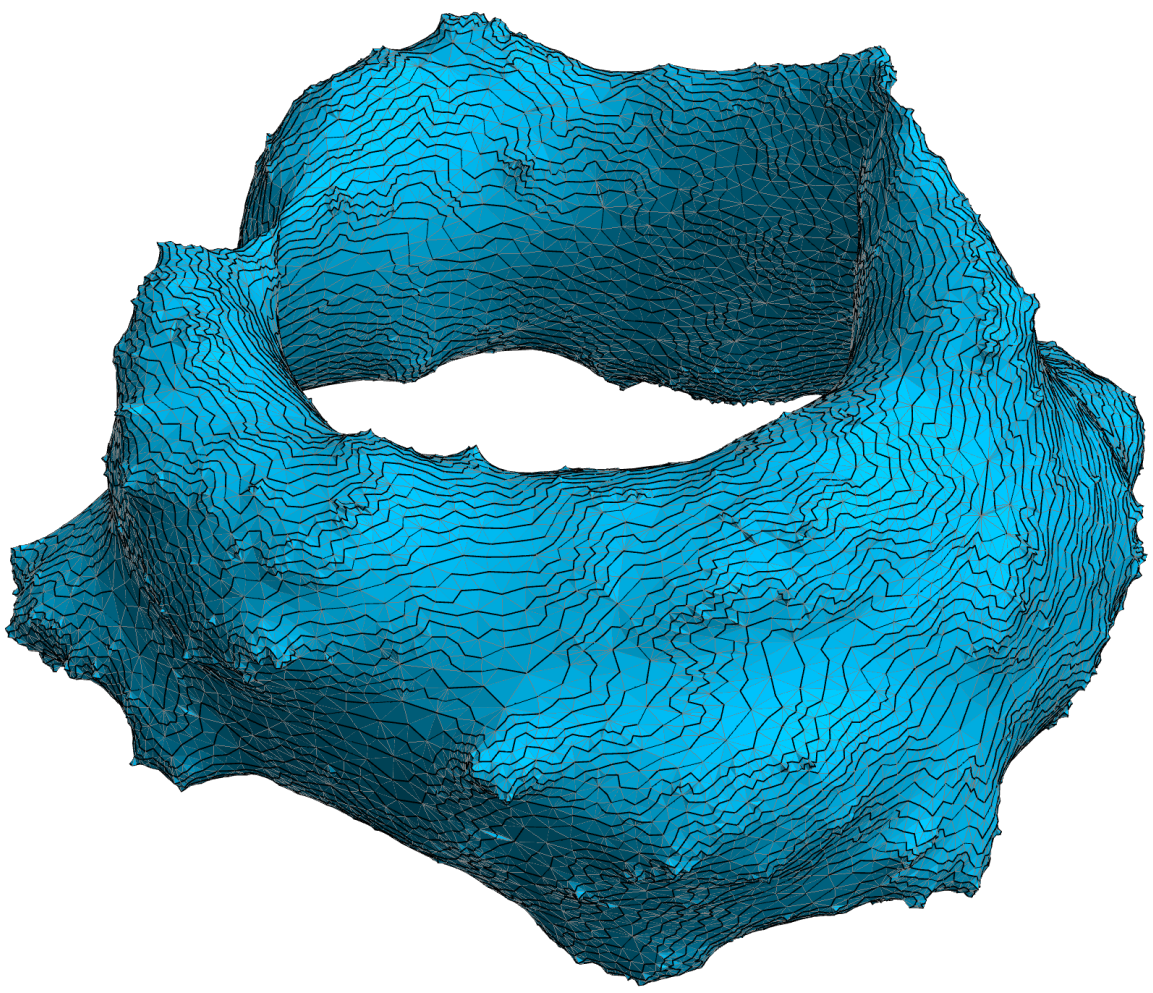

**Figure 2.2.:** Example of a 2D toroidal CDT configuration with $N_T = 63$ and $N_2 = 50000$. Thick black lines are fixed-time slices.

## 2.1 The need for observables

To extract geometrical information from the ensemble of CDT path integral configurations, one must resort to the accessible tools in those manifolds. As mentioned before, standard tensor calculus is not defined, and therefore, attempting to compute complicated tensors such as the Riemann tensor is not possible. Nevertheless, the connected nature of the simplicial manifolds, plus the constant edge lengths of the simplices, allows us to define lengths between different building blocks, as well as volumes of subsets of them.

On top of that, whatever quantity one defines to extract information about the geometries must have a meaningful continuum counterpart. This is because otherwise the expectation value of a quantity that only has a meaning in the discretized setting would not have one when taking the continuum limit of the theory. For example, computing the expectation value of the average number of edges in the manifolds would be meaningless in the continuum limit, since edges are just a mathematical tool to do computations.

Since there are no coordinates in each path integral configuration, and there is no canonical way to choose such a coordinate system, it is impossible to map a quantity evaluated at a given point of one configuration to the "corresponding point" of another configuration, and it is therefore impossible to compute an expectation value of such a quantity over all configurations. In this sense, any measured quantity in a given configuration must be "point-independent", for it to have a meaning across several configurations. In general, geometrical information is then extracted from non-local quantities, for example, by computing averages over the entire manifolds.

We call such quantities *Observables*, and will be generally denoted with the symbol $\mathcal{O}$. Observables are a crucial element of CDT, because CDT allows an explicit computation of such quantities in the non-perturbative regime of the theory, allowing one to determine geometric properties of spacetime at Planckian scales.

In general, the way one computes the expectation value of one such observable is by doing

$$\langle \mathcal{O} \rangle \;=\; \frac{1}{Z} \sum_{T \in \mathcal{T}} \frac{1}{C_T}\, \mathcal{O}[T]\, e^{-S_R^E[T]}\,. \tag{2.7}$$

A prototypical example of an observable is the Hausdorff dimension. For each configuration, one constructs geodesic balls $B_\delta(v)$ at each of the vertices $v$ of the manifold such that their distance is smaller than $\delta$, as $B_\delta(v) = \{v' \,/\, d(v, v') < \delta\}$. Then takes the average of the volumes of those balls over all vertices in the manifold, at each radius, obtaining $\bar{B}_\delta$. One repeats this for each configuration in the path integral, and from them computes an expectation of the volume profile as a function of the radius, usually denoted as $\langle \bar{B}_\delta \rangle$. By fitting a power law, such as $\langle \bar{B}_\delta \rangle \propto \delta^{d_h}$, the Hausdorff dimension $d_h$ of the emergent geometry can be determined.

Generally speaking, observables have been investigated using CDT lattice methods with two main motivations: firstly, to provide tests of the classical limit, by showing that their eigenvalues on sufficiently large scales are compatible with (semi-)classical expectations. Successful examples are the *shape* of the universe, i.e. its spatial three-volume $V_3$ as a function of proper time $\tau$, together with its quantum fluctuations $\delta V_3(\tau)$ [71, 72], and its average scalar curvature [75], which all turn out to match those of a de Sitter space. A second objective is to uncover new physics, in the form of genuine quantum signatures. The corresponding observables must of course be diffeomorphism-invariant and operationally well defined in the lattice framework, but by virtue of their nonperturbative nature, may not relate in any straightforward way to the geometric properties of classical spacetimes. These observables can take the form of scaling exponents, characterizing universal aspects of quantum spacetime in a Planckian regime. An example, as mentioned before, is the anomalous behaviour of the spectral dimension, which in CDT was discovered to be 2 (within error bars) near the Planck scale, instead of the classically expected value of 4, in what has since been conjectured to be a universal property of quantum gravity [73, 76].

In this thesis, we will develop observables with both motivations. Examples of questions that will be addressed in the following sections are: Is the emergent geometry highly fluctuating at all scales, or does it have some distance scale range where it resembles a piece of a smooth manifold in a suitable range? If so, does the emergent geometry possess some regime where a suitable statistical analogue notion of symmetries, like translation symmetry, makes any sense? If so, can one meaningfully say that in those scales the manifold resembles a piece of a known Riemannian manifold, like a Minkowski or de Sitter space?

Unfortunately, a full answer to these questions is out of reach, but it is remarkable that in CDT one has the necessary structure to even address them. The outcomes of our observables can be used as cross-checks for the necessary conditions to answer the previous questions with a yes. Since these questions are intrinsically associated with a certain range of scales, our observables are designed to probe geometries in a scale-dependent manner, allowing either finer or coarser-grained explorations of spacetime. The observables defined in the following chapters will be non-local in nature, since many of them involve averages over the entire configurations, as well as averages over some localized regions of them.

### 2.1.1 Emergence of Symmetries

As previously stated, one of the core results of four-dimensional CDT is the (apparent) emergence of a de Sitter-like universe, as suggested by its average spatial volume profile as a function of proper time. This is a global result, since it is derived from the spatial volume of the manifolds. It would be desirable to study the properties of such an emergent universe with more fine-grained quantities, capable of providing more detailed information apart from the global shape of the universe. In particular, we are interested in studying these properties with the motivation of understanding if this emergent spacetime from CDT resembles, in some meaningful way, the properties of our universe. For example, a non-perturbative analogue of the cosmological principle.

The definition of a symmetry in this non-perturbative setting, where group actions such as smooth translations and rotations are not defined, is challenging on its own. Nevertheless, we will see in the following sections that we can give a concrete meaning to suitable notions of symmetries, such as the "homogeneity" of these emergent spacetimes.

Throughout the rest of this part of the thesis, we will investigate particular observables designed to fine-grain the geometrical properties of the geometries of CDT. In particular, we will implement those observables in two-dimensional CDT, to test their robustness and understand their outcomes in these simplified $2D$ systems, which are closer to the scope of our intuition and provide us with a non-trivial testing ground.

For this, we will dive into topics such as topology to assess the fractal nature of these spacetimes, we will develop and use tools analogous to observational cosmology to determine statistical measures of emergent symmetries, and we will explore the technicalities of extended notions of curvature in spaces where not even the dimension is a given.

This is the organization of the following chapters. In chapter 3, we describe the notion of *effective topology*, how we use it to determine whether a quantum geometry resembles, or not, something like a smooth continuous manifold, and how to use it to determine new phenomena. In chapter 4, we explore a fundamental notion of *scale-dependent homogeneity*, and how we used it to study a statistical notion of translation symmetry in the quantum geometries we analyzed. In chapter 5, we explore a purely geometric construction of *generalized curvature observables* that we implement to characterize the emergent quantum geometries that come out of CDT. And at last, in 6, we conclude with a small summary of the achievements and possible new avenues to explore.

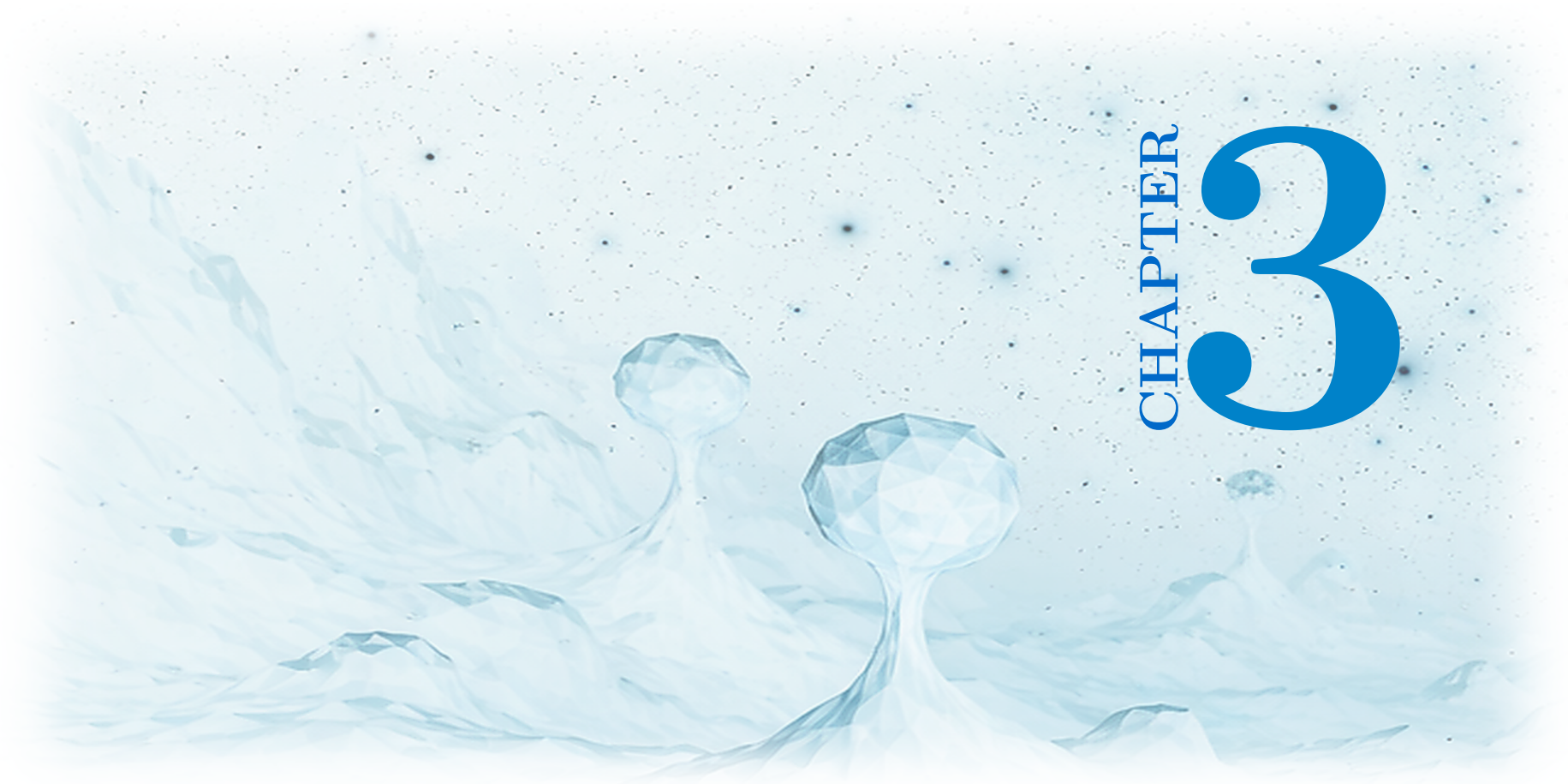

# Effective Topology

Within the realm of non-perturbative quantum gravity, where fluctuations are not small wiggles around a classical solution to Einstein's equations, it is difficult to establish properties of the underlying spacetimes. One particular task we are interested in is the determination of the emergence of symmetries at some coarse-grained distance scale. A type of obstacle one might encounter, at the most basic level of exploration of the spacetimes that come into place in CDT, is topological.

In CDT, where one is exploring the emergence of smooth spacetimes from a fundamental, non-perturbative ultraviolet theory, it is necessary to verify that some "exotic" possibilities do not occur at sufficiently large scales. For example, that the spacetimes that emerge from such a theory have a "foamy" nature, and possess wormholes or holes all over them.

While topology is fixed in CDT simulations, and therefore no such structures are allowed at the level of the construction of the geometries, at some scales topologically non-trivial structures might form. For example, there might be "necks" of a certain characteristic diameter. Such necks, when looked at with a certain lower resolution than that of the cut-off of the lattice, might look like "pinchings" of the geometry, i.e., places where a part of the geometry acquires non-manifold-like properties. Since this might only happen at a certain scale or resolution, we call them effective topology changes. While these "hand-waving" concepts might seem very unclear at this stage, we will properly define them later in this chapter.

In particular, in this chapter, we introduce a new methodology to characterize properties of quantum spacetime in a strongly quantum-fluctuating regime, using tools from topological data analysis. Starting from a microscopic quantum geometry, generated nonperturbatively in terms of dynamical triangulations (DT), we compute the Betti numbers of a sequence of coarse-grained versions of the geometry as a function of the coarse-graining scale, yielding a characteristic "topological fingerprint". We successfully implement this methodology in Lorentzian and Euclidean 2D quantum gravity, defined via lattice quantum gravity based on causal and Euclidean DT, yielding different results. This notion of effective topology will enable us to formulate necessary conditions for the recovery of spacetime symmetries in a classical limit.

In this chapter, we want to emphasize a motivation that relates to the classical limit, more precisely, the recovery of global, cosmological symmetries. This is particularly suggestive in view of the global de Sitter-like properties that have already been found in four dimensions and the fact that classical de Sitter space has a maximal number of isometries. It is not obvious that symmetries are or should be present in a Planckian regime, or can even be defined there in a operationally meaningful way. One may nevertheless expect that *at a sufficiently coarse-grained scale* some notion of *approximate* symmetry may apply. As discussed in [3] and the next chapter, and supported by a proof of principle, this requires the construction of diffeomorphism-invariant measures of homogeneity and isotropy, which is highly nontrivial. At the end of this chapter, we highlight the motivation for constructing these observables, that of finding new physical phenomena, using effective topology as a tool to quantitatively assess the "foaminess" of quantum geometry near the Planck scale, which is often conjectured to be some kind of *quantum spacetime foam* [77].

It is important to emphasize here that the presence of a symmetry, in the sense of an (approximate) invariance under a suitable notion of continuous translation or rotation, needs a suitable carrier space on which such transformations can be defined. This implies that at the coarse-graining scale considered, a quantum spacetime must be sufficiently similar to (a piece of) $\mathbb{R}^4$ in topological terms, without any topological obstructions in the form of holes, nontrivial connectivity (think wormholes) or parts of spacetime that become effectively disconnected. Investigating the effective topology of quantum geometry, as we will do below, allows us to formulate necessary criteria for being "sufficiently nice" to support notions of symmetry at a given scale. As we will see, already for the two-dimensional toy models of Lorentzian and Euclidean quantum gravity, the outcomes are very different.

The remainder of this chapter is organized as follows. We introduce the concept of effective topology in Sec. 3.1, and explain in Sec. 3.2 how methods from topological data analysis (TDA) can help in investigating observables related to this concept in the context of lattice quantum gravity based on dynamical triangulations. Sec. 3.3 contains a detailed description of how one proceeds in the particular case of two dimensions to generate a coarse-grained triangulation whose homology is then measured. This requires several steps: the selection of a coarse-grained vertex sample (Sec. 3.3.1), the construction of the associated Voronoi decomposition (Sec. 3.3.2) and subsequently its dual, coarse-grained Delaunay triangulation (Sec. 3.3.3). In Sec. 3.4 we describe and discuss the results of numerically measuring the expectation values of the Betti numbers for 2D Lorentzian and 2D Euclidean quantum gravity. Our conclusions and an outlook are contained in Sec. 3.6. Appendix B.1 contains additional technical details on how to construct the Voronoi decomposition, and Appendix B.2 describes how to locally adjust Delaunay triangulations to make them amenable to the computational library used to determine the Betti numbers.

## 3.1 Introducing effective topology

The key idea behind the new observables is to examine the connectivity properties of the quantum geometry – what we will call its *effective topology* or, more precisely, its *effective homology* – on small scales relative to their total linear extension, as a function of a linear coarse-graining scale $\delta$. Recall that the homology of a topological space keeps track of the number of its "holes" of various dimensions, where the dimension is defined as that of the hole's boundary.[1] For example, removing a two-dimensional disc from a two-dimensional sphere creates a one-dimensional hole, since the hole's boundary is a one-dimensional circle.

[1] Note that homology is also characterized by torsion coefficients. They are not computed by the open-source library we use, and will not be considered in this work.

3.1

Note that the topology of the spacetime configurations contributing to the lattice path integral is always fixed, usually to that of a sphere or torus, and not allowed to change during the Monte Carlo evolution.[1] However, due to the nonperturbative character of the gravitational dynamics, typical configurations in the continuum limit are highly nonclassical and nowhere differentiable, analogous to the nonclassical paths that support the Wiener measure of the quantum-mechanical path integral for a nonrelativistic particle. In particular, the microscopic building blocks can arrange themselves into geometries that are macroscopically indistinguishable from spaces with a *different* topology.

An example, in this case involving the global topology of spacetime, has been observed in 4D quantum gravity. The most common choice of lattice topology for this system is $S^1 \times S^3$, where for technical convenience the time direction has been compactified to a circle.[2] However, it turns out that in the so-called de Sitter phase the overall shape of the quantum geometry (the system's nonperturbative ground state) is driven dynamically to that of a four-sphere $S^4$ [79]! More specifically, a typical member of the ensemble of geometries consists of a thin "stalk", a proper-time interval during which the spatial three-volume $V_3(\tau)$ is close to the kinematically allowed minimum, whose spatial extension is of the order of the UV cut-off $a$ (the length of a lattice edge), and a spherical "blob", a complementary time interval during which the spatial universe is macroscopically extended, $V_3(\tau) \gg a^3$ (Fig. 3.1). From a macroscopic point of view (where $a \to 0$ in a continuum limit) the stalk has vanishing volume and can be neglected relative to the total volume of the blob, leading to an"effective" large-scale topology $S^4$.

We will look systematically for the possible presence of nontrivial effective topology at scale $\delta$ (with $\delta$ assumed small compared to the total system size), as a novel way to characterize the local nature of quantum geometry. Roughly speaking, this is meant to capture the topology "felt" by a probe of linear extension $\delta$ or a wavelike excitation of wave length $\delta$, and is related to the presence of holes of characteristic linear size $\leq \delta$. To determine the effective topology in practice, we apply a single coarse-graining step of magnitude $\delta$ to a given, typical path integral history.[1] This operation decimates the number of simplicial building blocks and generates a coarse-grained triangulation with edge length $\delta \geq 2$ in lattice units (i.e. setting $a = 1$), for a range of $\delta$.

[1] The gravitational path integral cannot be renormalized (unambiguously) by standard methods if a sum over topologies is included, due to the superexponential growth of the number of configurations as a function of the spacetime volume (see e.g. [17], Q14, for a discussion). The spaces of fixed topology we consider here are always compact without boundary.

[2] 4D CDT has also been investigated on a four-torus $T^4$. A discussion of how the choice of global topology may influence the nature of phase transitions can be found in [78].

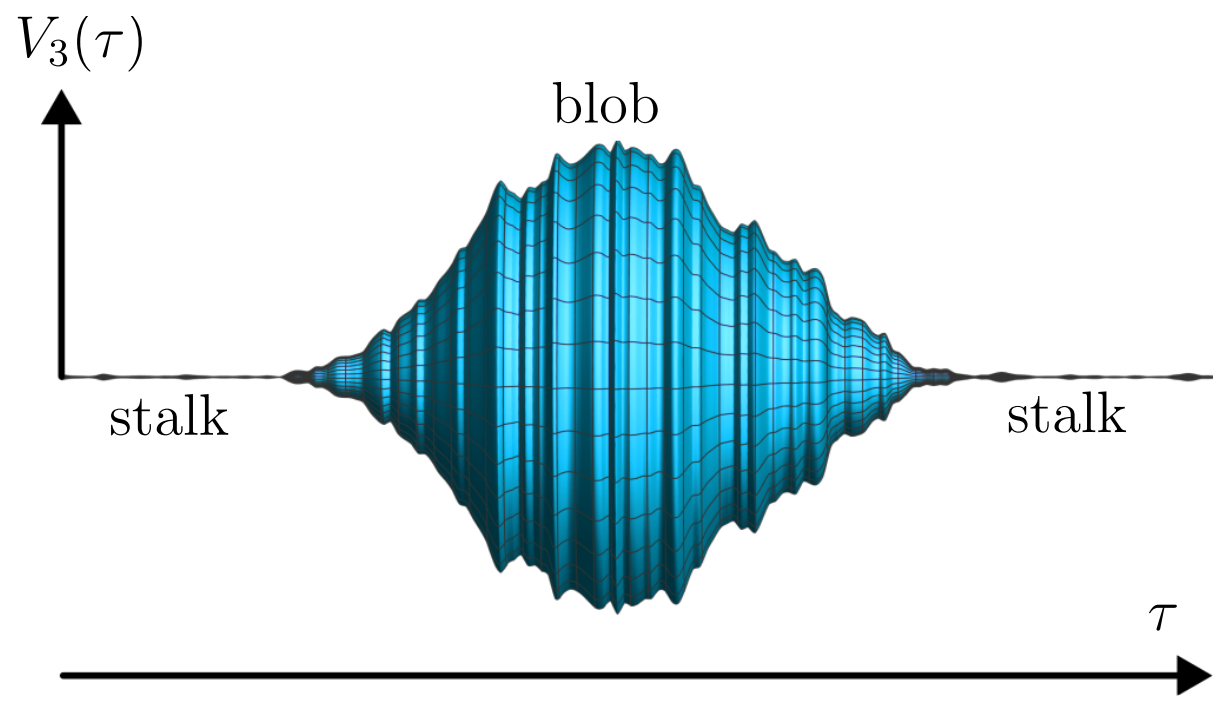

**Figure 3.1.:** Shape of a typical path integral configuration in 4D CDT lattice quantum gravity, illustrating the macroscopically emergent $S^4$-topology. The curve $V_3(\tau)$ has been made into a rotational body about the horizontal time axis $\tau$.

During the coarse-graining step we monitor for effective topology changes affecting the number of connected components, loops or higher-dimensional voids. For example, the original geometry may contain *necks*, which by definition are closed loops of edge length $\leq \delta$, whose length is minimal with respect to local deformations[2]. After the coarse-graining, these necks will typically be very thin or even pinch to a point, such that the two pieces of geometry on either side of the neck or pinching point can be regarded as effectively disconnected, corresponding to an effective topology change. As another example, consider a Swiss cheese-style quantum geometry. Depending on how the coarse-graining is set up, holes of linear size $\leq \delta$ may disappear in an effective sense at coarse-graining scale $\delta$.

[1] Recall that the configurations of (C)DT in $d$ dimensions are simplicial manifolds made of $d$-simplices (a 0-simplex is a vertex, a 1-simplex an edge, a 2-simplex a triangle, a 3-simplex a tetrahedron, etc.) with all edges of identical length, up to an overall multiplicative factor distinguishing space- from time-like link lengths.

[2] more precisely, local edge reroutings, since any path has to run along discrete lattice edges

3.1

## 3.2 Quantum geometry and homology

An attractive feature of this construction is the availability of powerful open-source software to compute the homology of large simplicial complexes, which includes the triangulations that are generated by our coarse-graining procedure. The larger context for these mathematical and numerical techniques is the field of topological data analysis (TDA), whose main objective is to characterize and analyze large sets of data in terms of a basic set of topological and geometrical properties that can be associated with them (see [80, 81] for motivation and introduction).

In a typical application of TDA, the input is a point cloud, i.e. a finite set of data points in a linear space $\mathbb{R}^n$, which may be of very high dimension, where each pair of data points has a mutual distance induced by the Euclidean metric on $\mathbb{R}^n$. The point cloud together with this distance matrix is then converted to a triangulated geometric object, more precisely, a simplicial complex [82]. The latter is defined as a set $\mathcal{K}$ of simplices such that any subsimplex ("face") of an element in $\mathcal{K}$ is again in $\mathcal{K}$, and the intersection of two elements $\sigma_1$, $\sigma_2$ of $\mathcal{K}$ is either empty or a subsimplex of both $\sigma_1$ and $\sigma_2$. There are several ways of obtaining such a complex, but a standard strategy is to draw balls of radius $\epsilon$ around the points in the cloud. Whenever $d+1$ of these $\epsilon$-balls have a nonempty intersection, the $d$-simplex spanned by their centre points is added to the simplicial complex, forming a so-called ech complex. In *persistent homology*, an important methodology in TDA, one studies the homology of the simplicial complexes associated with a given point cloud as a function of the spatial resolution $\epsilon$, and is particularly interested in the topological aspects that are stable as $\epsilon$ is varied (see [83] for a concise description).

We borrow some of these ideas and adapt and apply them to the quantum geometries generated in lattice quantum gravity. In the present work, we treat two toy models of quantum gravity in 2D explicitly, providing a proof of principle for the methodology. As far as we are aware, TDA tools have not been applied previously in quantum gravity, with the exception of [84], where they are employed in search of a geometrical spacetime interpretation of certain tensor decomposition data in the context of a canonical tensor model. In terms of ingredients, the present work also has some similarities with [85], which aimed to recover aspects of the homology of continuum spacetime from fundamentally discrete causal sets, invoking simplicial complexes at an intermediate stage of the computation.[1] In adjacent fields, persistent homology has been applied in string theory to study string compactification spaces and flux vacua (see [86] for a review), and in classical cosmology to describe the large-scale matter distribution of the universe (see e.g. [87] and references therein).

We will compute the expectation values of Betti numbers of coarse-grained quantum geometries as a function of the coarse-graining scale $\delta$, which plays the role of the resolution $\epsilon$ mentioned above. The Betti numbers are integers $\beta_k$ counting the number of $k$-dimensional holes of a simplicial complex $\mathcal{K}$ or, more formally, measuring the rank of its $k$th homology group $H_k(\mathcal{K})$ [88].

Our starting point are the triangulated configurations of the gravitational path integral on the lattice, given by what in the literature are variably called piecewise flat, simplicial or combinatorial $d$-manifolds [61, 89, 90], which in addition to their topological properties also carry metric properties, by virtue of length assignments to the edges of their constituent $d$-simplices.[2] By assumption, the interior of each $d$-simplex is flat, which entails that its metric properties are uniquely determined by its edge lengths. A simplicial manifold satisfies the condition that the *link* of each of its $i$-simplices, $i \in [0, 1, \ldots, d-1]$ is homeomorphic to the sphere $S^{d-i-1}$. [3] This is the simplicial analogue of the manifold condition that the neighbourhood of each point is homeomorphic to an open subset of $\mathbb{R}^d$. From the point of view of the underlying simplicial complex, it implies that every $i$-simplex is defined *uniquely* by a set of $i+1$ *distinct* vertices, which is also an input requirement for the "simplex tree" simplicial complex module of the GUDHI library [92] we used to determine the Betti numbers of the coarse-grained triangulations.

[1] We thank R. Sorkin for bringing this reference to our attention.

[2] Note that for the two-dimensional Euclidean dynamical triangulations considered in Sec. 3.4.2 below we used a slight generalization of combinatorial manifolds, the *restricted degenerate ensemble* in the classification of [91], where a pair of vertices can be connected by more than one edge.

[3] See [90] for a definition of "link" (not to be confused with link in the sense of edge) and further technical details, and [91] for a discussion and illustration of this and other, less stringent, regularity conditions on triangulations that have been used in 2D quantum gravity.

## 3.3 Coarse-graining triangulations

For a given triangulated configuration $T$ of the gravitational path integral, our coarse-graining procedure at resolution $\delta$ consists of three steps: (i) select a subset $\mathcal{S}_\delta$ of the set $V(T)$ of all vertices of $T$, such that the link distance between typical nearest neighbours – the number of edges in the shortest path connecting the two vertices – is of the order $\delta$, (ii) construct the Voronoi cells and Voronoi decomposition of $\mathcal{S}_\delta$, and (iii) construct the dual of the Voronoi decomposition, which is the searched-for coarse-grained Delaunay triangulation, whose Betti numbers are subsequently being determined. To illustrate the procedure and some of the subtleties involved, we will next describe these steps in detail for two-dimensional equilateral triangulations, which feature in the lattice formulations of both Euclidean quantum gravity and causal, Lorentzian quantum gravity after performing the Wick rotation. Elements of our construction below are similar to a coarse-graining proposed earlier in [93], albeit with a different motivation and application in mind.

### 3.3.1 Creating a vertex sample

Our aim is to construct a subset $\mathcal{S}_\delta$ of vertices which samples the vertex set $V(T)$ of the triangulation $T$ evenly at a given scale $\delta \in \mathbb{N}$. To achieve this, we use a construction loosely analogous to that of Poisson disk sampling on smooth manifolds [94, 95]. Defining an "open" geodesic ball of radius $\delta$ centred at a vertex $v$ by $B_\delta(v) := \{v' \in V(T) | d(v, v') < \delta\}$, where $d$ denotes the (integer-valued) link distance, we will require that (i) $\bigcup_{v \in \mathcal{S}} B_\delta(v) = V(T)$, i.e. each vertex of $T$ lies in at least one $\delta$-ball, and (ii) for any pair $v, v' \in \mathcal{S}_\delta$ with $v \neq v'$, $v' \notin B_\delta(v)$, i.e. no vertex from the sample $\mathcal{S}_\delta$ lies in the $\delta$-ball of another sample point, which also implies that $d(v, v') \geq \delta$. In addition, we define the $\delta$-annulus $A_\delta(v)$ of a vertex $v$ as the set $A_\delta(v) := \{v' \in V(T) | \delta \leq d(v, v') < 2\delta\}$.

To generate an evenly distributed vertex sample, we work with three dynamical sets[1]: the set $\mathcal{S}_\delta$, which at the end of the algorithm will be the searched-for sample set, the set $\mathcal{S}_{new} \subset \mathcal{S}_\delta$, which consists of sampled vertices whose annulus still needs to be explored, and the set $V_{cov} \subset V(T)$ of vertices that have already been covered by geodesic balls.

[1] By construction, the algorithm does not add vertices that were already present in $\mathcal{S}_\delta$ and $\mathcal{S}_{new}$; these sets may be implemented computationally as dynamical arrays. The set $V_{cov}$ can be implemented as a Boolean array.

At the outset, these lists are empty. After picking an initial vertex $v_0 \in V(T)$, the algorithm proceeds as follows:

1. Add $v_0$ to both $\mathcal{S}_\delta$ and $\mathcal{S}_{new}$ as a newly sampled vertex and add all vertices contained in the ball $B_\delta(v_0)$ around $v_0$ to $V_{cov}$, implying that those vertices have been covered.
2. Remove a randomly[1] chosen vertex $v$ from $\mathcal{S}_{new}$ and determine its annulus $A_\delta(v)$. Then
   A. select a random vertex $u$ from $A_\delta(v) \cap (V \backslash V_{cov})$, i.e. the part of the annulus that is not yet covered by geodesic balls;
   B. add $u$ to both $\mathcal{S}_\delta$ and $\mathcal{S}_{new}$ and add all vertices of the $\delta$-ball around $u$ to $V_{cov}$;
   C. if $A_\delta(v) \cap (V \backslash V_{cov})$ is nonempty, repeat from 2a.
3. If $\mathcal{S}_{new}$ is nonempty, repeat from 2.

Fig. 3.2 is a schematic, planar illustration of elements of this algorithm. Note that we only indicate the vertices that end up in the sample $\mathcal{S}_\delta$ (in green), and not any of the other vertices contained in the geodesic $\delta$-balls around them. In Fig. 3.2d, the combined red region represents the set $V_{cov}$ up to this stage, while the blue region contains the vertices that still need to be covered by the algorithm.

At the end of this process, we have created a vertex sample $\mathcal{S}_\delta$ which by construction satisfies condition (ii) from the beginning of Sec. 3.3.1. To see that it also satisfies (i), we can argue by contradiction. Assume that (i) does not hold and there therefore exists a vertex $w \notin V_{cov}$. This vertex must have a distance of at least $2\delta$ from all elements of $\mathcal{S}_\delta$, since the algorithm by construction covers all vertices inside this radius. Since $T$ by assumption is connected, there is a (possibly nonunique) vertex $p \in \mathcal{S}_\delta$ which has a finite, minimal link distance $d_{min} = d(p, w) \geq 2\delta$ to $w$. In other words, there is a (possibly nonunique) chain of $d_{min}$ consecutive edges between $p$ and $w$ that form a path of this minimal length. At $2\delta - 1$ steps from $p$ along this chain there is therefore a vertex $q$, which lies in the annulus $A_\delta(p)$, and has distance $d_{min} - 2\delta + 1$ from $w$. Although $q$ itself may not be in $\mathcal{S}_\delta$, it must lie in a $\delta$-ball of some other vertex $q' \in \mathcal{S}_\delta \cap A_\delta(p)$, since by construction the annulus $A_\delta(p)$ is completely covered by $\delta$-balls. From $d(q,q') \leq \delta - 1$ it then follows that $d(q',w) \leq d_{min} - \delta$, which implies that the sample vertex $q' \in \mathcal{S}_\delta$ is closer to $w$ than $p \in \mathcal{S}_\delta$, contradicting our original choice of $p$. Thus we have shown that condition (i) holds too. To summarize, we have shown that our construction delivers an evenly spread set of sample points.

[1] Here and elsewhere, "random" means according to a uniform distribution over all elements in the set.

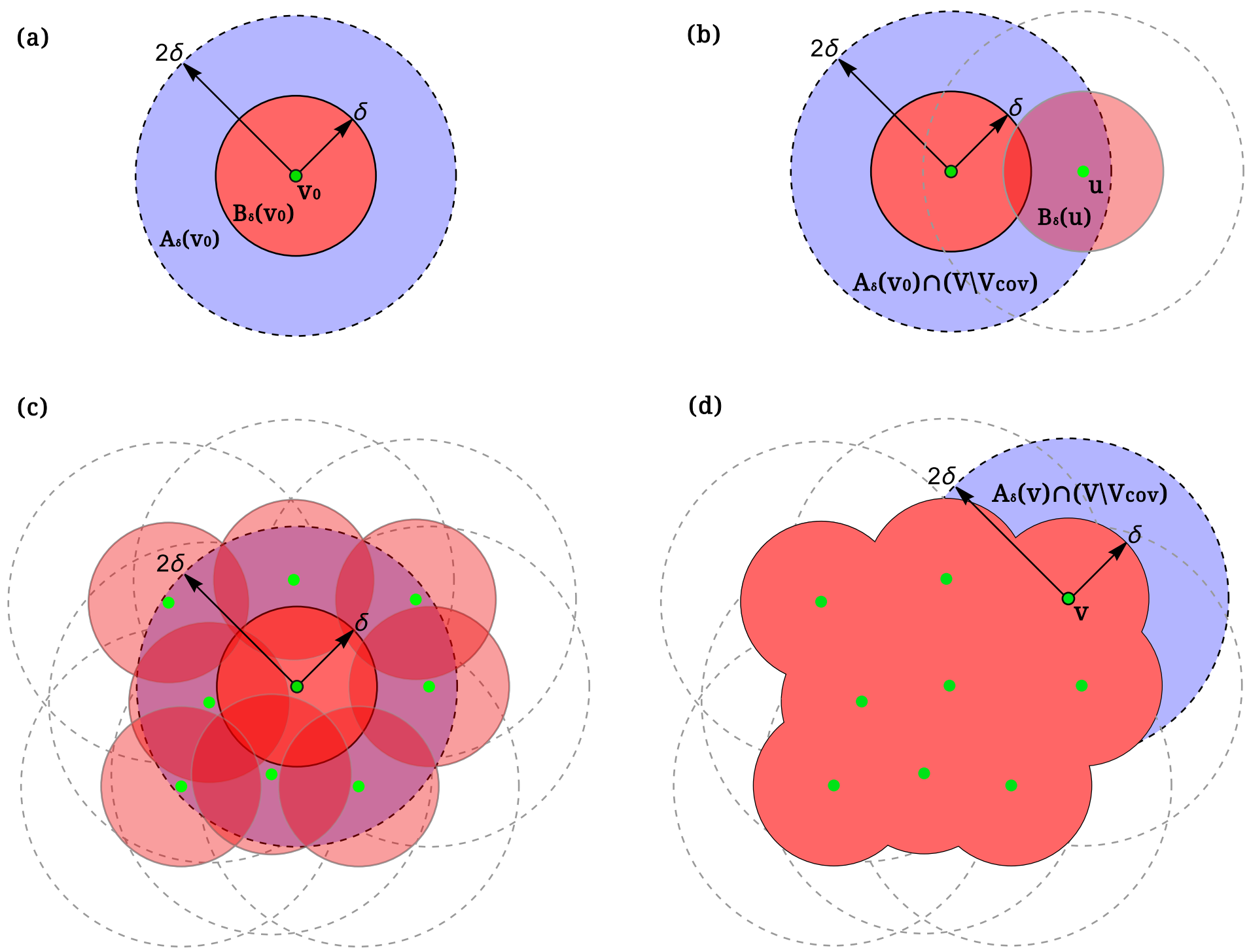

**Figure 3.2.:** Illustrating the covering algorithm: (a) initial vertex $v_0$, together with its $\delta$-ball $B_\delta(v_0)$ (red) and $\delta$-annulus $A_\delta(v_0)$ (blue); (b) random vertex $u$ from the part of the annulus not yet covered, together with its $\delta$-ball $B_\delta(u)$ (pink); (c) after several iterations, all vertices in $A_\delta(v_0)$ have been covered; (d) algorithm continues with a random vertex $v \in \mathcal{S}_{new}$ whose annulus is not yet covered.

### 3.3.2 Constructing Voronoi cells

Given a sample $\mathcal{S}_\delta$ for a given triangulation $T$, our next step is to partition the set $V(T)$ of all vertices into Voronoi cells, and subsequently extend this to a Voronoi decomposition of the entire two-dimensional triangulation. The Voronoi construction is most familiar as a prescription for decomposing ("tessellating") the Euclidean plane into $n$ cells associated with a set of seed points $\{p_i,\, i = 1, 2, \ldots n\}$, where the Voronoi cell associated with a given point $p_i$ is given by the set of all points closer to $p_i$ than to any other point $p_j$, $j \neq i$, or at an equal distance to such points. An analogous prescription can be used on a graph of the type considered here (see, e.g. [96]): the cell associated with a given vertex $v \in \mathcal{S}_\delta$ will by definition consist of all vertices closer to $v$ (in terms of the link distance) than to any other $v' \in \mathcal{S}_\delta$. If a vertex has the same distance to more than one seed vertex, a random assignment is made, as specified below.

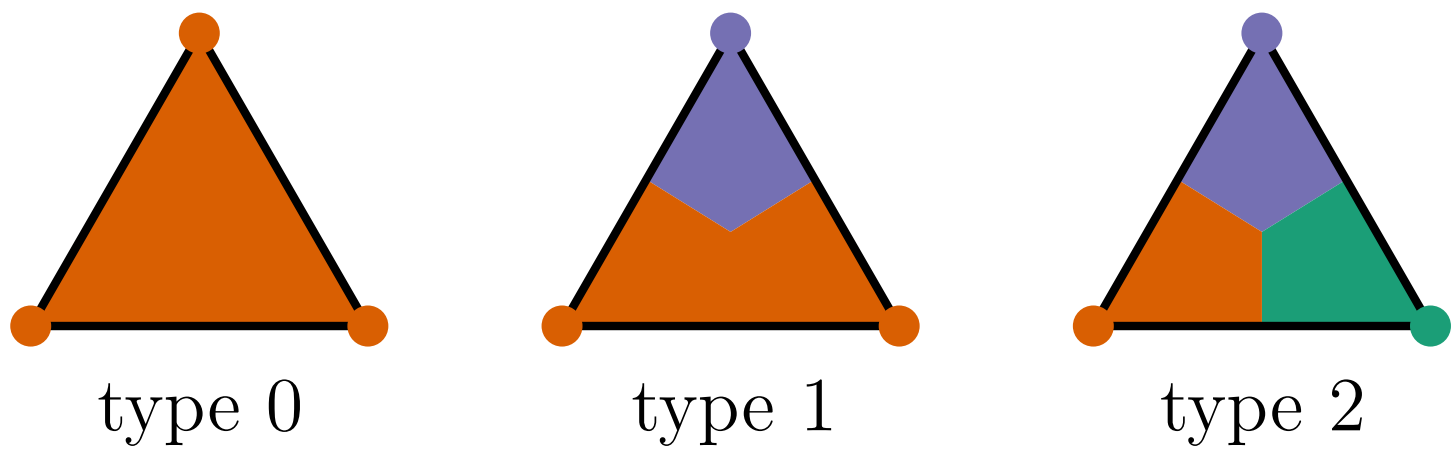

**Figure 3.3.:** Triangles can have three types of colouring, depending on the colouring of their vertices, generated during the decomposition of vertices into Voronoi cells.

Our seed points will be the vertices in $\mathcal{S}_\delta$, and the "cell" corresponding to a given seed vertex $v$ is found by performing a breadth-first search of its neighbouring vertices for increasing link distance $d = 1, 2, \ldots$ from $v$. Algorithmically, we perform the breadth-first search radially outward from all seeds simultaneously, in unit steps. This means that at the $d$th step we assign vertices to the cell of the nearest seed and make a random assignment (with uniform probability) in case there is more than one seed at distance $d$, before moving on to step $d + 1$. This guarantees that the Voronoi cells defined on the entire triangulation we will construct in the next step are connected.

We now extend the colouring to the triangles of $T$ by dividing each triangle into three "dual" area segments of the same colour as the associated corner vertex. Three types of triangle colouring are possible, depending on whether all three vertices have the same colour (type 0), exactly two vertices have the same colour (type 1), or all three have different colours (type 2), see Fig. 3.3. For type 1, the triangle contains a boundary between two (two-dimensional) Voronoi cells of different colour, while for type 2 it contains a point where three cells meet, as well as three pairwise boundaries. Fig. 3.4 shows a typical decomposition of a two-dimensional CDT configuration into such Voronoi cells, where we have associated one cell to each vertex of the evenly distributed sample set $\mathcal{S}_\delta$. In this example, each cell has the topology of a disc.

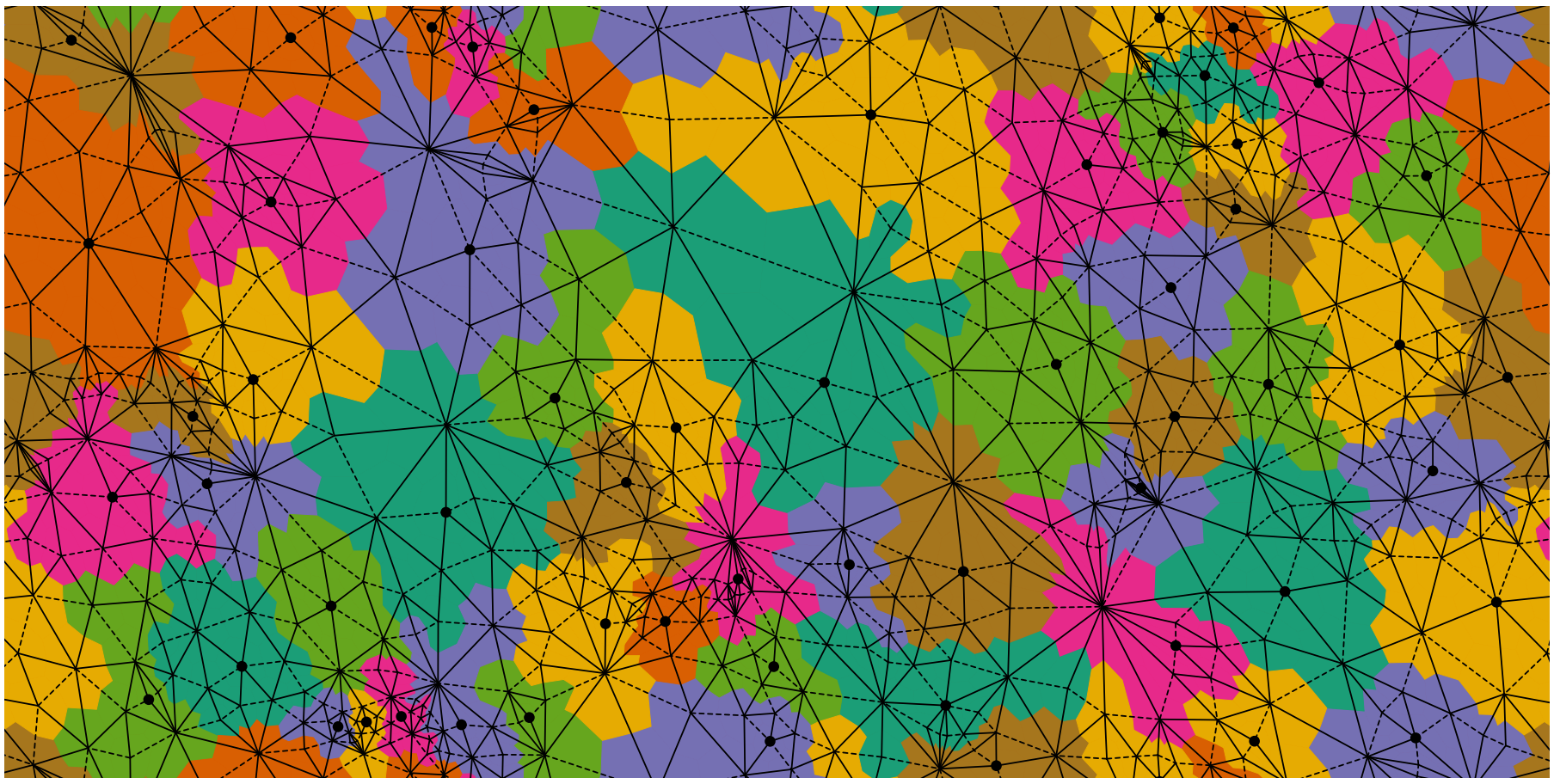

**Figure 3.4.:** Voronoi decomposition into cells of a 2D CDT configuration on a torus, for $\delta = 3$, using a Tutte or barycentric embedding, where each vertex is located at the barycentre of its neighbours. Dashed and solid lines represent space- and timelike edges respectively, and each black dot is a seed vertex for a cell of a given colour. Note the toroidal periodicity for opposite sides of the rectangle.

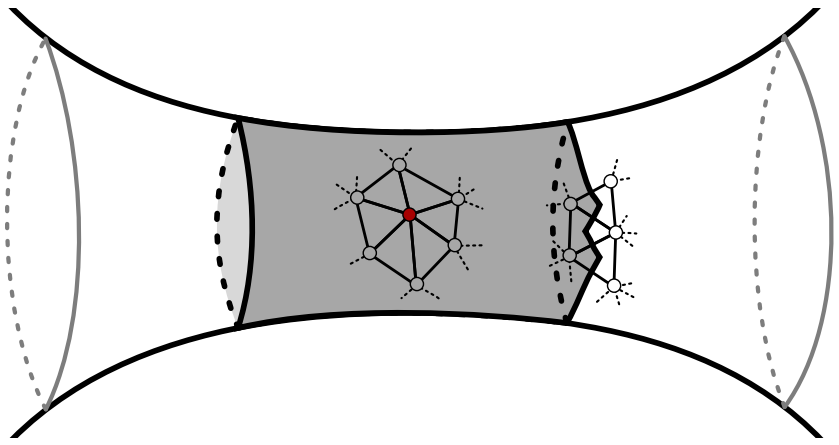

**Figure 3.5.:** Voronoi cell associated with a seed vertex (in red), with the topology of an annulus (shaded region), wrapping around a thin neck of the triangulation and shown in a three-dimensional embedding.

### Properties of Voronoi cells

As illustrated by Figs. 3.3 and 3.4, in our construction at most three Voronoi cells can meet in a point. Such a triple point is always located at the centre of a triangle, which can also be viewed as a vertex of the trivalent graph dual to the triangulation $T$. Likewise, the boundaries between neighbouring cells consist of edges of this dual graph. A sequence of dual boundary edges between two adjacent (along the boundary) triple points we will call a *boundary segment*.

The topology of a typical cell is that of a disc, but any connected subset of the original triangulated manifold can in principle occur. Since we will never consider a cell of maximal volume that covers the entire triangulation, a cell will always have one or more boundaries, each of which is topologically a circle. An important case that appears frequently when coarse-graining 2D Euclidean DT configurations is that of an annulus, i.e. a disc with a hole. This can happen when a cell wraps completely around a "thin neck" of the triangulation, as illustrated by Fig. 3.5. As we will see in Sec. 12, this is associated with a "pinching" of the coarse-grained triangulation.

Note also that two cells can meet along several, disconnected boundary segments, as illustrated by Fig. 3.6. This situation can occur when two cells wrap around a thin neck. It can also happen that an entire circular boundary consists of a single boundary segment only, without any triple points, for example, when a disc-shaped cell is surrounded by an annulus.

3.3

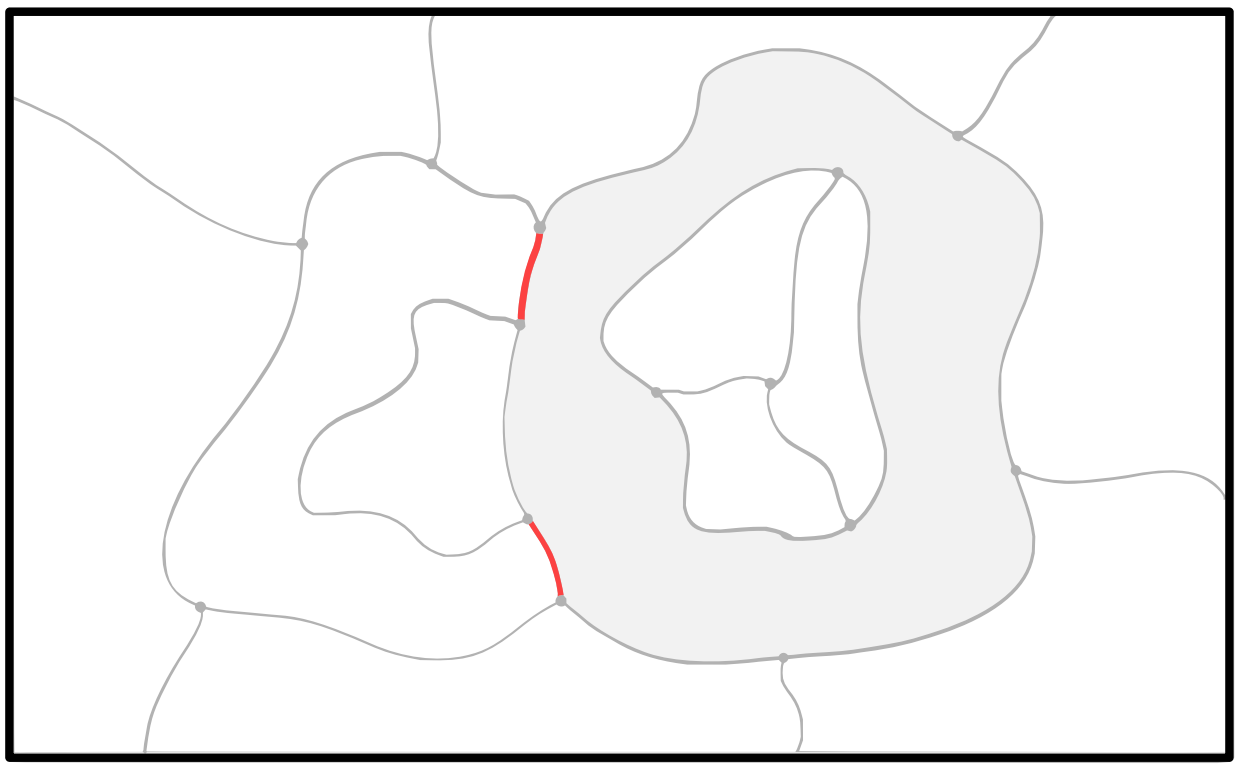

**Figure 3.6.:** An annulus-shaped (shaded gray) and a disc-shaped Voronoi cell meet along two disconnected boundary segments (in red). This schematic, planar drawing only shows boundary segments and triple points.

### 3.3.3 Constructing the dual Delaunay triangulation

Our next objective is to obtain a coarse-grained triangulation $T_\delta$, which is dual to the Voronoi decomposition described in Sec. 3.3.2, in a sense we will describe below. We are primarily interested in the effective topology of $T_\delta$, where features of up to a linear size $\delta$ are "disregarded". How precisely this is implemented is ultimately a matter of choice. Our general strategy will be to coarse-grain in a way that is maximally topology-preserving, and such that any localized topology changes that occur in this process are well controlled.[1] This requires a good understanding of the geometric and topological properties of a local neighbourhood of the triangulation, which is rather straightforward in dimension two, but more involved in higher dimensions.

The manner in which we will set up the coarse-graining leads to a slightly generalized notion of "triangulation" for $T_\delta$, compared to the one we started with for $T$. We will nevertheless call it a *Delaunay triangulation*[2], because it is inspired by a standard construction of the same name in the Euclidean plane (or, more generally, $\mathbb{R}^n$), with the following properties understood.

[1] Note in particular that for the choice $\delta = 1$ (no coarse-graining) the procedure of Secs. 3.3.1–3.3.3 leads to a dual Delaunay triangulation $T_1$ that reproduces the original triangulation $T$.

[2] also sometimes called a Delaunay complex

Firstly, $T_\delta$ will in general no longer be a topological manifold, in the sense that it will not look two-dimensional in every point. This happens when thin necks of the original triangulation pinch to a point or collapse to a sequence of one or more edges, which are not part of any triangle. In turn, these edge sequences can also branch into tree-like structures (see also Fig. 3.11 below). Secondly, even away from points which do not have a $\mathbb{R}^2$-like neighbourhood, the two-dimensional triangulation will in general not be a simplicial manifold – according to our definition in Sec. 3.2 above – because it can happen that two edges share the same endpoints, forming a closed loop of length 2. An explicit example will be presented in Sec. 12.

The coarse-grained Delaunay triangulation $T_\delta$ inherits the connectivity properties of its constituent vertices, edges and triangles from the Voronoi decomposition, and is subsequently endowed with metric properties by assigning a uniform length $\delta$ to all of its edges and by declaring its edges as straight and its triangles as flat. However, unlike reference [93], we will not be interested in the nontrivial curvature properties of $T_\delta$, but only in its effective topology. The first step to obtaining the Delaunay triangulation identifies subsets of the Voronoi decomposition as dual vertices, edges and triangles, keeping track of their neighbourhood relations:

(i) with each Voronoi cell we associate a “vertex” inside the cell. This vertex, dual to the Voronoi decomposition, can be thought of as midpoint of the cell, but its actual position is immaterial, since in this step we only keep track of the connectivity of the dual simplicial building blocks of what will become the Delaunay triangulation.

(ii) With each boundary segment shared by a pair of Voronoi cells we associate an “edge”, whose two endpoints are dual vertices as described in (i). We can construct this dual edge as a simple path that connects the two dual vertices and traverses the boundary segment at some (arbitrarily defined) midpoint. The details of this choice are again unimportant.

(iii) With each vertex of the Voronoi decomposition we associate a “triangle”, whose edges are the duals to the three boundary segments meeting at the vertex, as described in (ii).

Note that as subsets of the original triangulation $T$, the “edges” identified in (ii) are in general not straight, and the “triangles” in (iii) are not flat. The searched-for coarse-grained Delaunay triangulation $T_\delta$ is now *defined* as the piecewise flat space consisting of the abstract simplicial building blocks identified in steps (i)–(iii), together with their connectivities, where in addition we assign metric properties, namely that all edges are straight and of length $\delta$ and all triangles carry a flat Euclidean metric structure induced by their three boundary edges.

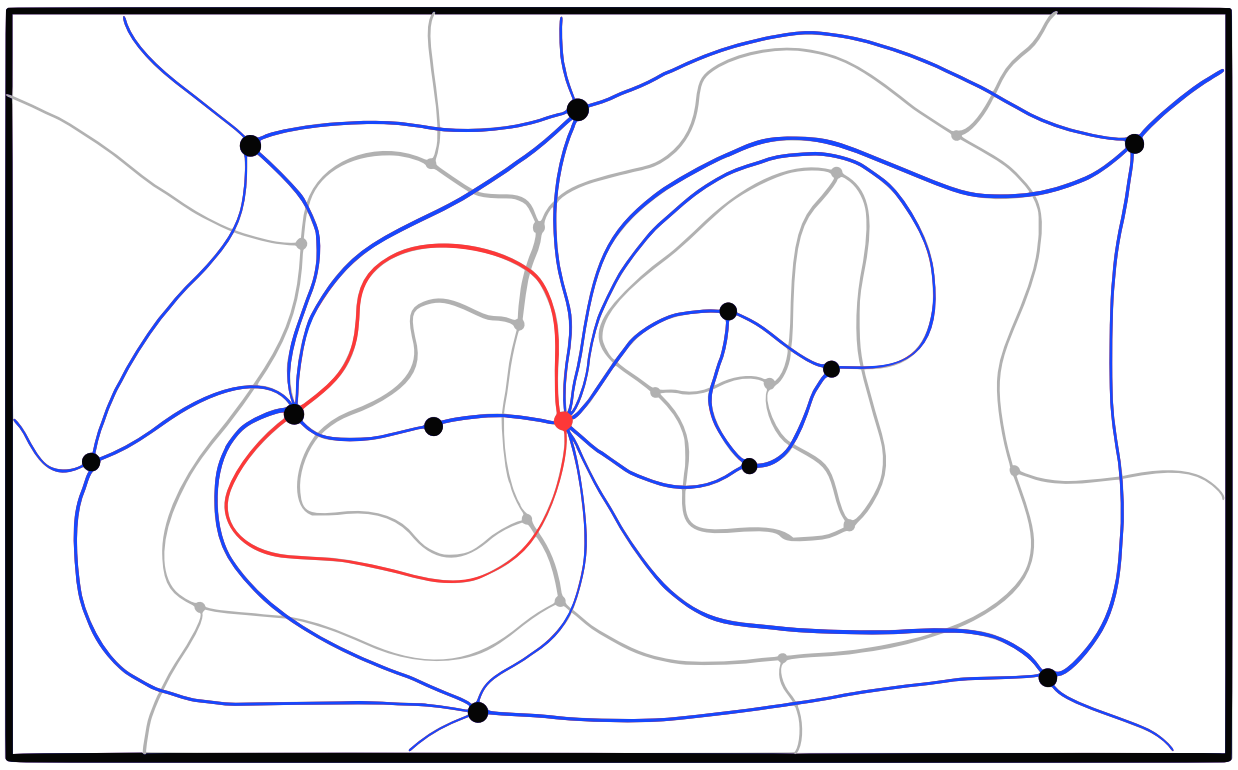

**Figure 3.7.:** A Voronoi decomposition (black) and its dual Delaunay triangulation (grey). Highlighted in red are a pinching vertex and a closed loop consisting of two edges, dual to the annulus-shaped Voronoi cell and the red boundary segments of Fig. 3.6 above.

This construction is illustrated by Fig. 3.7, which shows a piece of a Voronoi decomposition (the same as in Fig. 3.6 above) and its dual Delaunay triangulation in a schematic, planar representation. The dual vertices at the centre of Voronoi cells and the dual edges, connecting pairs of dual vertices, are drawn in grey. For better readability, we have not coloured the triangles of the Delaunay triangulation, which are dual to trivalent vertices of the Voronoi decomposition and whose boundaries are given by closed loops of three grey edges each.

### Properties of the Delaunay triangulation

Fig. 3.7 also illustrates some of the irregular features of the Delaunay triangulation $T_\delta$ we already mentioned above, and which would not occur for a Voronoi decomposition of the Euclidean plane. Firstly, there are the two cells, one with the topology of an annulus, the other one with the topology of a disc, which share two disconnected boundary segments.

According to prescription (ii), this leads to two distinct dual edges between the dual vertices associated with these two cells. They form a closed loop consisting of two edges in the Delaunay triangulation, thereby violating the simplicial manifold conditions.

Secondly, the vertex marked in red in Fig. 3.7, which is dual to the cell with the topology of an annulus, is what we call a pinching vertex. As was already mentioned in Sec. 3.1, it is associated with the presence of a neck in the original triangulation $T$, which in the coarse-grained Delaunay triangulation gets pinched down to a point. As a consequence, its neighbourhood is no longer homeomorphic to an open subset of $\mathbb{R}^2$, thereby violating the manifold condition.

For our purposes, the most important feature of the coarse-graining process is the pinching that occurs at the location of an annulus. Since the annulus, viewed as a cell of the Voronoi decomposition, by definition does not have any boundary segments in its interior, the dual Delaunay triangulation does not contain any edges that lie entirely inside the annulus. This implies that the coarse-grained triangulation is no longer a two-dimensional manifold at the pinching vertex.[1] Since the annulus separates the Voronoi cells into those lying in- and outside the annulus[2], the resulting Delaunay triangulation will in general have two parts located on either side of the pinching vertex, which are connected to each other only at the vertex.

In the simplest case, the neighbourhood of such a pinching point is homeomorphic to that of two two-dimensional cones meeting at their tips (Fig. 3.8). When the Voronoi cell has the topology of a disc with $h > 1$ holes instead of $h = 1$ (the annulus), the situation generalizes in the simplest case to that of a meeting point of the tips of $h + 1$ cones. Another type of generalization – beyond these simplest cases – occurs when the Voronoi decomposition contains two or more nested annuli. The shared boundary between two such annuli consists of a single boundary segment without any trivalent vertices located on it, which implies that the edge dual to this segment is not part of any triangle in $T_\delta$, but is instead what we will call a “loose” edge. The dual Delaunay triangulation has then not just a pinching

[1] For added clarity, one can envisage the pinching as a two-step process: (i) by hand, introduce a single dual edge that links the dual vertex inside the annulus to itself and winds around the annulus once; this preserves the manifold character of the resulting dual triangulation. (ii) Shrink the dual loop of length 1 to a point.

[2] This is not necessarily true when $T$ has a global topology with noncontractible loops, but will nevertheless lead to a local pinching.

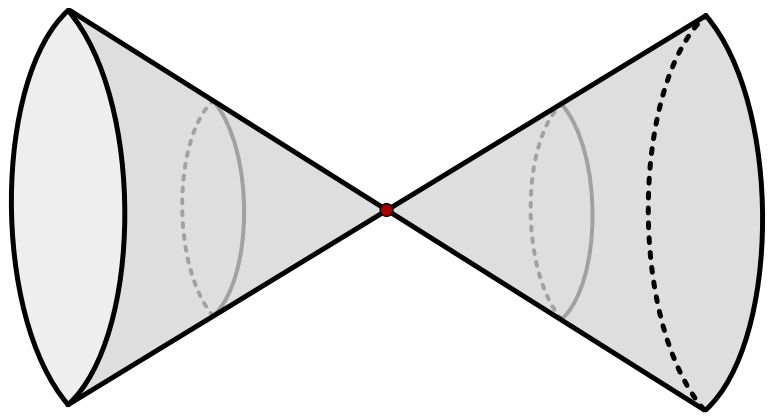

3.3

**Figure 3.8.:** Schematic depiction of a piece of coarse-grained Delaunay triangulation associated with a pinching vertex (red) at which two cones meet. It is dual to the Voronoi cell of Fig. 3.5, whose seed vertex can be identified with the pinching vertex here.

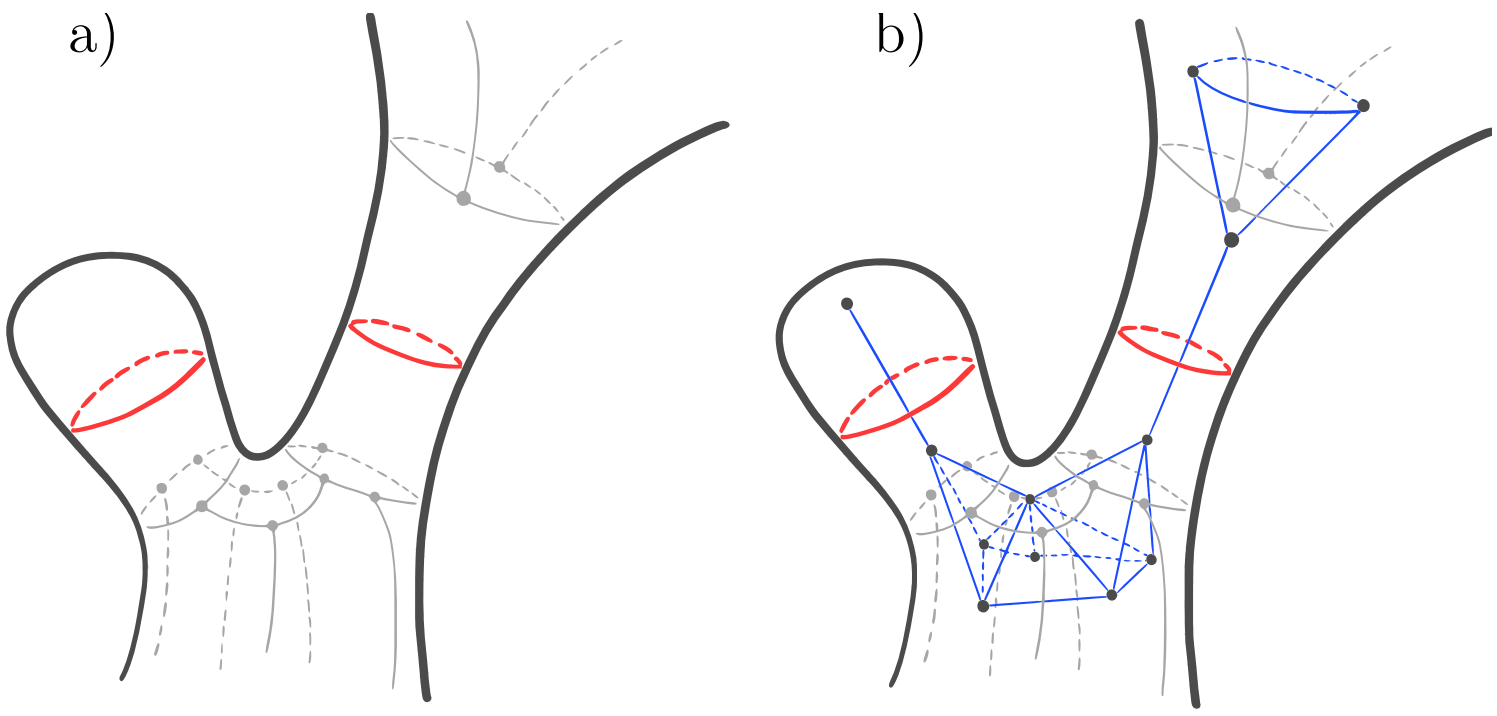

**Figure 3.9.:** (a): Schematic depiction of a Voronoi decomposition, containing two boundary segments without triple points marked in red. (b): ditto, with the Delaunay triangulation added (blue), containing two loose edges dual to the marked boundary segments.

vertex, but an entire “pinching edge”, along which the triangulation is locally one-dimensional. This is illustrated by Fig. 3.9a, which shows two instances of boundary segments without vertices (highlighted in red) of the Voronoi decomposition. The corresponding dual Delaunay triangulation is added in blue in Fig. 3.9b. The boundary segment between a disc-shaped cap and an annulus on the left branch of the geometry leads in the Delaunay triangulation to a loose edge ending in a single vertex, and the boundary segment between two annuli on the right branch leads to a loose edge connecting two different parts of the coarse-grained Delaunay triangulation.

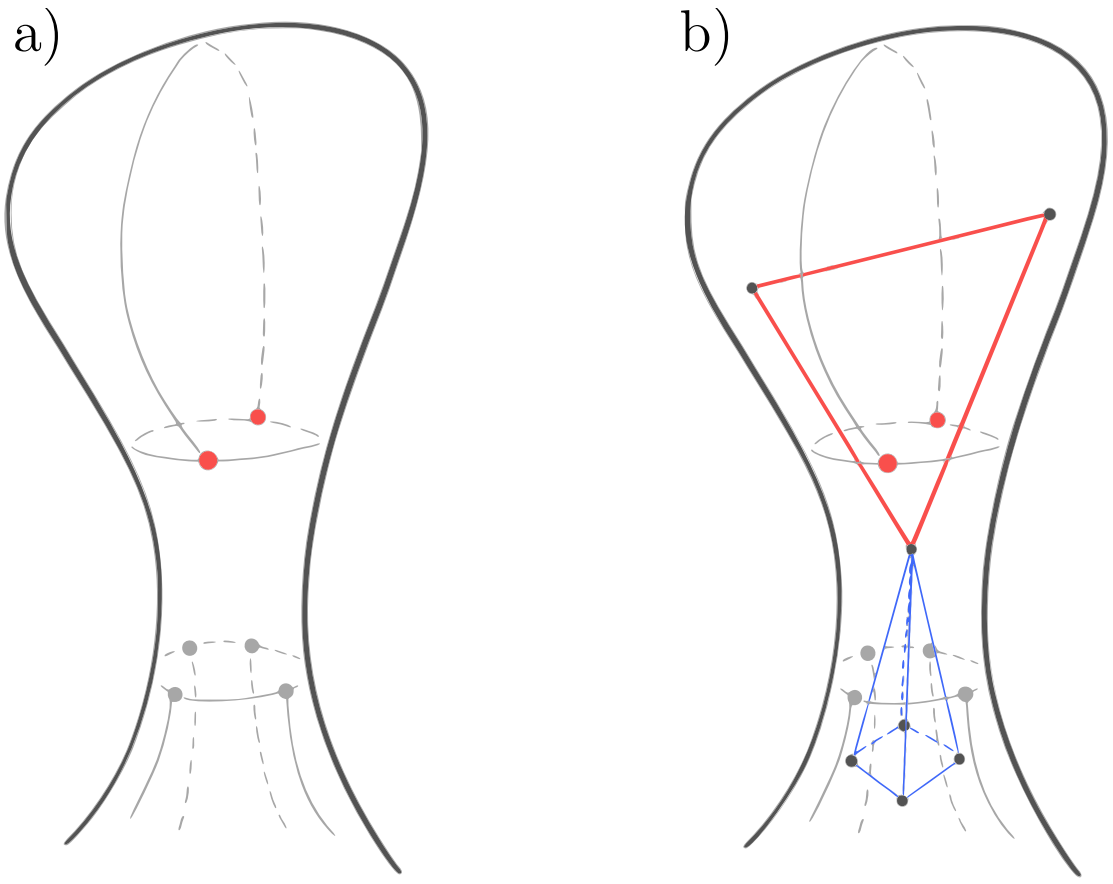

**Figure 3.10.:** (a): Voronoi decomposition where the same three Voronoi cells are shared by two distinct triple points (red). (b): In the dual Delaunay triangulation, this results in a pillow, consisting of two triangles (red) sharing the same edges and vertices.

Yet another irregular feature one can encounter during coarse-graining are so-called "pillows". By definition, a pillow – in this case of a Delaunay triangulation – consists of two distinct triangles, which share the same three (distinct) vertices and three (distinct) edges. This violates the combinatorial character of a simplicial manifold, since the triangles are not uniquely characterized in terms of their vertices. An example of how this can occur is illustrated by Fig. 3.10. It shows a Voronoi decomposition with the particular property that there are two distinct triple points (marked in red) at which the same three Voronoi cells meet; in the case at hand, these are two discs and one annulus (Fig. 3.10a). By virtue of our algorithm, this leads to two distinct Delaunay triangles forming a pillow, highlighted in red in Fig. 3.10b. In the example shown, it is connected to the remainder of the Delaunay triangulation by a pinching vertex, associated with the annulus of the Voronoi decomposition. The occurrence of this type of local structure is relevant for our analysis of the effective topology below, since a pillow is topologically a two-sphere and therefore contributes to the count of the Betti number $\beta_2$.

Note that by construction of the Delaunay triangulations $T_\delta$, no one-dimensional boundaries can appear during coarse-graining, and it also cannot happen that parts of $T_\delta$ become disconnected. Consequently, the overall topology of $T_\delta$ consists of two-dimensional spherical "bubbles" and generalized bubbles with more complicated shapes[1], connected to each other at pinching vertices or by (sequences of) loose edges, where multiple bubbles and/or loose edges can meet at a given pinching vertex, as illustrated by Fig. 3.11. The bubbles of minimal size are exactly the pillows we have just introduced.

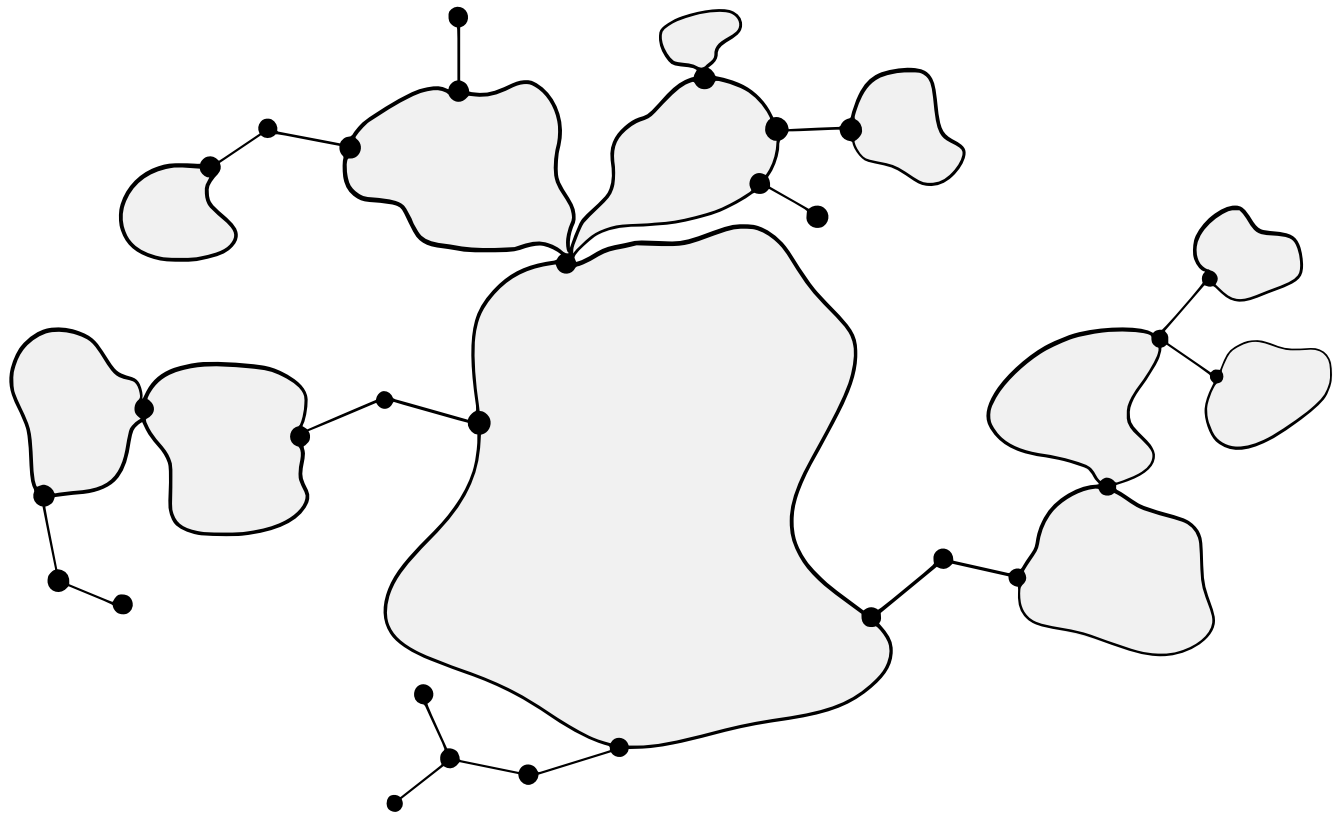

**Figure 3.11.:** Diagram illustrating the topological features that can occur in a coarse-grained Delaunay triangulation $T_\delta$ obtained from a triangulation $T$ of spherical topology. Each circle or "bubble" filled with gray lines represents a spherical triangulation of arbitrary size. Loose edges can connect between such bubbles or be attached to them as outgrowths.

## 3.4 Measuring the effective homology of quantum geometry

In the context of quantum gravity, the effective homology of space(-time) introduced above has the status of a diffeomorphism-invariant observable, as will be explained below. In what follows, we will measure this observable with the help of Monte Carlo simulations in two toy models of quantum gravity[2], defined nonperturbatively as continuum limits of gravitational

[1] e.g. enclosed by surfaces of higher-genus, depending on the topology of the original triangulations $T$

[2] The implementation for both models is open-source [97], with a repository available at `https://gitlab.com/dynamical-triangulation/dyntri-rs`.

path integrals regularized on dynamical lattices. We will treat the case of Lorentzian quantum gravity, formulated in terms of Causal Dynamical Triangulations (CDT), in Sec. 3.4.1 and that of Euclidean quantum gravity, formulated in terms of Euclidean Dynamical Triangulations (EDT), in Sec. 3.4.2. Among other things, this will help us to assess their suitability as carrier spaces of symmetry.

To set the context, let us recall the computation of Betti numbers of simplicial complexes, which we will use to analyze the Delaunay triangulations produced by the coarse-graining described in Sec. 3.3 (see e.g. reference [88]). It turns out to be a problem in linear algebra that is amenable to computation and scalable to large triangulations.

The setting is that of a $d$-dimensional simplicial complex $\mathcal{K}$, where by definition $d$ is the maximal dimension of any of its simplices. A *$p$-chain* $c$ is defined as the formal sum $c = \sum_i a_i \sigma_i^{(p)}$ of $p$-simplices $\sigma_i^{(p)} \in \mathcal{K}$, where for simplicity we use mod-2 coefficients $a_i$ which can take the values $a_i = 0, 1$. One can also define simplicial homology over larger finite coefficient fields, but for our present purposes this choice will not make a difference.[1] The $p$-chains form an abelian group $C_p(\mathcal{K})$ under addition. Next, we define a *boundary map* $\partial_p$, which maps $p$-chains to $(p-1)$-chains and on basis elements is defined as

$$\partial_p : C_p \to C_{p-1}, \quad \sigma^{(p)} = \{v_0, v_1, \ldots, v_p\} \mapsto \sum_{i=0}^{p} \{v_0, \ldots, \hat{v}_i, \ldots, v_p\}, \quad (3.1)$$

where we have represented the simplex $\sigma^{(p)}$ by the list of its $(p+1)$ vertices $v_i$, $\sigma^{(p)} = \{v_0, v_1, \ldots, v_p\}$, and the hat denotes the omission of the $i$th vertex. In geometric terms, eq. (3.1) says that the boundary of a simplex $\sigma^{(p)}$ is the sum of its $(p-1)$-dimensional faces. The boundary map allows us to define two subgroups of the group $C_p(\mathcal{K})$ of $p$-chains, namely (i) the group $Z_p(\mathcal{K})$ of *$p$-cycles* $c$ satisfying $\partial_p c = 0$, and (ii) the group $B_p(\mathcal{K})$ of *$p$-boundaries* $c$ satisfying $c = \partial_{p+1} d$ for some $(p+1)$-chain $d$. It then follows from the fundamental lemma of homology,

$$\partial_p \partial_{p+1}\, d = 0, \quad \forall d \in C_{p+1}(\mathcal{K}), \quad (3.2)$$

"the boundary of a boundary is zero", that the $p$-boundaries form a subgroup of the $p$-cycles, i.e. $B_p(\mathcal{K}) \subset Z_p(\mathcal{K})$.

[1] In the GUDHI implementation we have used the default field $\mathbb{F}_{11}$ instead of $\mathbb{F}_2$.

The $p$th homology group $H_p$ is then defined as the group of $p$-chains modulo the group of $p$-boundaries, $H_p(\mathcal{K}) := Z_p(\mathcal{K})/B_p(\mathcal{K})$. Since all of these groups are also vector spaces, this can be expressed equivalently as the quotient vector space

$$H_p(\mathcal{K}) = \text{kernel}(\partial_p)/\text{image}(\partial_{p+1}). \tag{3.3}$$

3.4 Its dimension,

$$\beta_p(\mathcal{K}) := \dim H_p(\mathcal{K}) = \dim Z_p(\mathcal{K}) - \dim B_p(\mathcal{K}), \tag{3.4}$$

is called the $p$th *Betti number*. In other words, the Betti number $\beta_p$ counts $p$-cycles that are *not* $p$-boundaries, which geometrically can be thought of as $p$-dimensional "holes". In two dimensions, the case at hand, $\beta_0$ counts the number of components, $\beta_1$ the number of loops (one-dimensional holes), and $\beta_2$ the number of two-dimensional holes. We also see that determining the Betti numbers of a simplicial complex involves the computation of the ranks of the linear maps $\partial_p$.

The Betti numbers are diffeomorphism-invariant quantities by construction, and therefore can be used as observables. For each coarse-grained simplicial manifold of characteristic scale $\delta$, we will measure the observables $\mathcal{O}_i[T] = \beta_i(\delta)$, and compute its expectation value following (2.7) over the ensemble of such configurations.

As already mentioned in Sec. 3.2, we have used the open-source C++ library GUDHI to compute the Betti numbers of the coarse-grained Delaunay triangulations $T_\delta$. To give an idea of its efficiency, it takes on the order of a second to compute the $\beta_i$ for a triangulation of size $100k$. However, there is one small step that must still be taken before this library can be used. Implicit in our definition (3.1) of the boundary map was the unique characterization of each $p$-simplex $\sigma^{(p)}$ in terms of its $p+1$ vertex labels. This vertex representation is required by GUDHI as an input format, but is in general not satisfied by our Delaunay triangulations, as we have already seen: the marked, closed loop consisting of two dual edges depicted in Fig. 3.7 is an explicit example, since both edges share the same vertices. This is merely a technical issue, since the homology of this generalized simplicial complex[1] is perfectly well defined. We resolve this by making small local adjustments to the Delaunay triangulation, without changing its homology, such that all of its edges and triangles are uniquely defined through their vertices. Details of this procedure can be found in Appendix B.2.

[1] Technically, it is an example of a $\Delta$-complex, which has a well-defined simplicial homology, see [82].

### 3.4.1 Lorentzian quantum gravity in D=2

We begin by evaluating the expectation values of Betti numbers in coarse-grained spacetimes of resolution $\delta$ in two-dimensional CDT. Recall that each CDT configuration is a sequence of one-dimensional spatial universes labeled by an integer time $t \in [0, 1, 2, \ldots, t_{\text{tot}}]$, where strips assembled from identical, triangular flat building blocks interpolate between adjacent spatial universes of variable size (Fig. 2.1). This stacked structure of the spacetimes is a lattice implementation of global hyperbolicity, a hallmark of the causal structure of CDT that is not present in the histories of Euclidean Dynamical Triangulations (EDT), which we will later explore. For convenience in the simulations, we cyclically identify the time direction, $t_{\text{tot}} \equiv 0$, and use compact spatial universes of $S^1$-topology, such that all configurations $T$ have the topology of a two-torus $T^2 \equiv S^1 \times S^1$. The geometric variable characterizing a spatial slice at integer $t$ is its one-dimensional volume, which in lattice units is given by the number $\ell(t)$ of its edges, where we demand $\ell(t) \geq 3$ to ensure that $T$ is a simplicial manifold.

We performed all our measurements of the simulations for ensembles of fixed total volume $N_2$, for eight equally spaced values $N_2 \in [50k, 400k]$. For each given $N_2$, we have performed measurements for various fixed time extensions $t_{\text{tot}}$. This changes the average spatial volume $\bar{\ell} = N_2/(2t_{\text{tot}})$, which can be interesting to exhibit the dependence of the Betti numbers on global properties of the underlying geometry. Following the considerations of [98], we have used 18 different ratios $r := t_{\text{tot}}^2/N_2$ for each $N_2$, evenly spaced in the range $r \in [0.08, 0.25]$. Whenever we study the volume-dependence of a given quantity, we will for simplicity focus on configurations with the intermediate value $r = 0.16$.

As already stated before, we use a Monte Carlo Markov chain (MCMC) algorithm[1] to generate sequences of independent CDT configurations, which allows us to approximate the expectation values $\langle \beta_i \rangle$ of the Betti numbers as the lattice volume is increased systematically. The typical number of measurements ranged from around $410k$ for the smallest volume $N_2 = 50k$ to around $40k$ for the largest volume $N_2 = 400k$.[1] We perform 200 sweeps between measurements, where each sweep consists of $N_2$ attempted Monte Carlo moves. Given the relatively high acceptance rate achieved in our simulations, this means that on average several Pachner moves are performed per simplex. We can therefore safely assume that our measurements are uncorrelated.

[1] For details on implementing MCMC for two-dimensional CDT, see e.g. [97, 99].

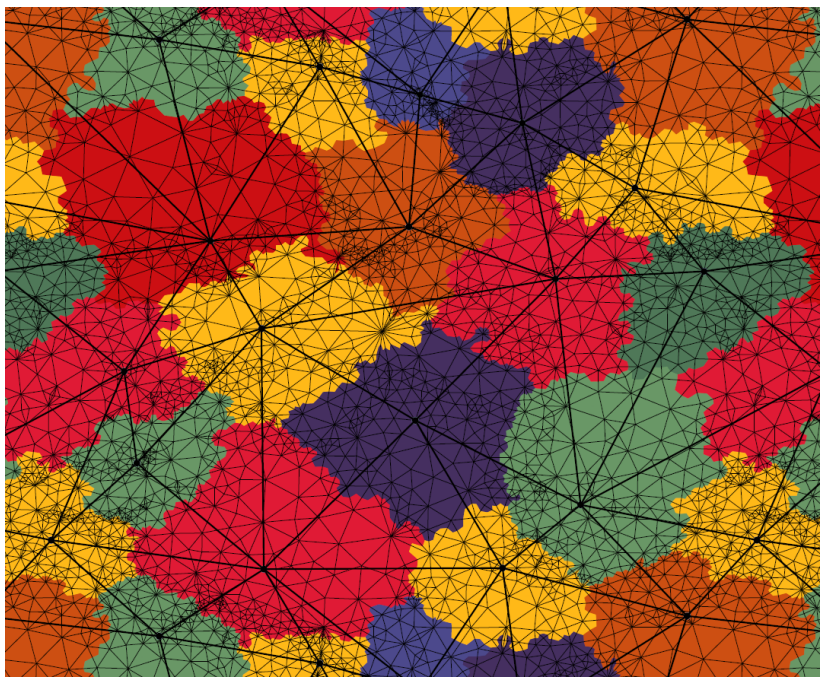

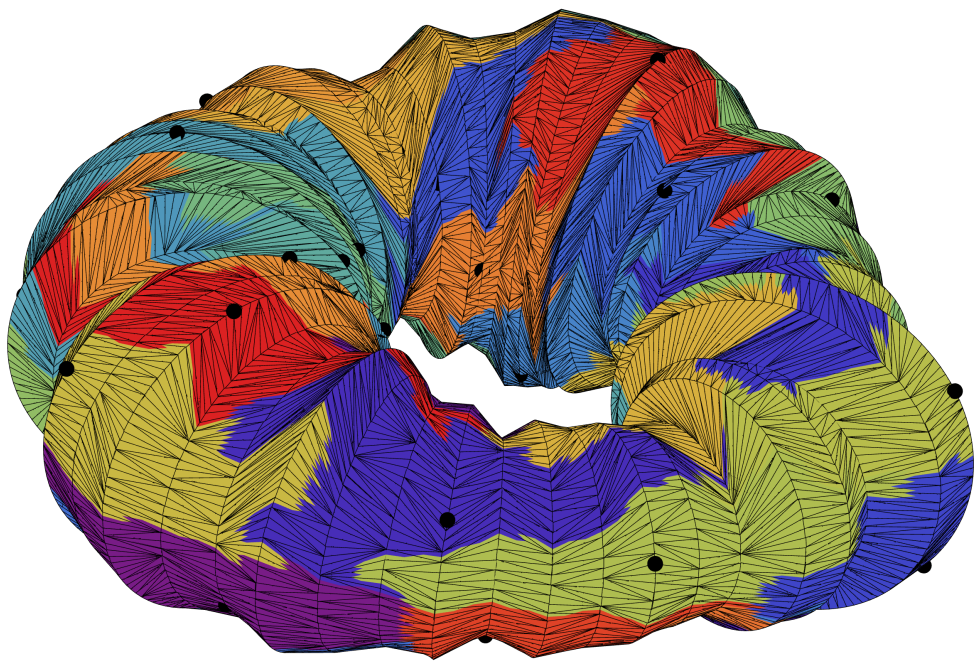

**Figure 3.12.:** Voronoi decompositions of typical CDT configurations on a two-torus at resolution scale $\delta = 8$, in a planar representation with $N_2 = 5.040$ and $t_{\text{tot}} = 35$ (left), and an embedding in $\mathbb{R}^3$ with $N_2 = 10.000$ and $t_{\text{tot}} = 52$ (right). Thin black lines are those of the original triangulation $T$. In the figure on the left, the edges of the dual Delaunay triangulation $T_\delta$ have been added as thick black lines.

Fig. 3.12 illustrates the nature of typical CDT configurations with their coloured Voronoi cells after coarse-graining with resolution $\delta = 8$, in a planar representation analogous to Fig. 3.4 (left), and superimposed on the original torus in a three-dimensional embedding (right). Each cell is associated with a vertex from the evenly spread sample $\mathcal{S}_8$ (indicated by a fat black dot), whose creation was described in Sec. 3.3.1. Although the original triangulation $T$ – still visible in both representations – is locally curved, the coarse-graining procedure contains random elements, and the representations are not strictly isometric, the resulting patterns of Voronoi cells are still fairly regular. All depicted cells have disc topology and a roundish shape, unlike what we will meet in Euclidean quantum gravity in Sec. 3.4.2 below.

[1] For the intermediate volumes $N_2 = 100k, 150k, 200k, 250k, 300k$ and $350k$, the typical number of measurements was on the order of $125k, 65k, 70k, 45k, 60k$ and $55k$ respectively.

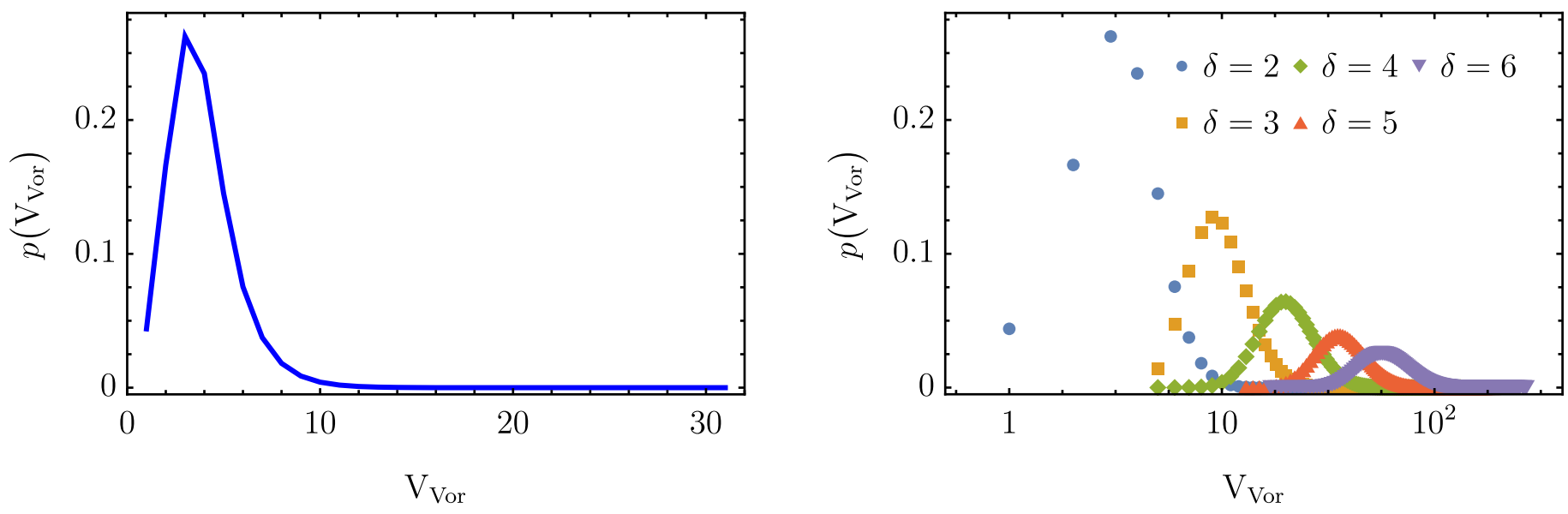

**Figure 3.13.:** Normalized distribution $p(V_{\text{Vor}})$ of the volume $V_{\text{Vor}}$ of individual Voronoi cells in units of vertices of the original triangulation $T$ for CDT configurations of volume $N_2 = 200k$ and time extension $t_{\text{tot}} = 179$, for $\delta = 2$ (left) and for the range $\delta \in [2, 6]$ (right). In this and other data plots below, the size of error bars is smaller than the dot size.

For a more detailed understanding of our coarse-graining procedure, we have monitored the volumes of the Voronoi cells at resolution $\delta$ in terms of the numbers of vertices of the triangulation $T$ they contain. As can be seen from Fig. 3.13, our algorithm has the property of creating cells of approximately equal size, with volume distributions that remain well peaked even for $\delta > 2$. Across a range of volumes $N_2$ and time extensions $t_{\text{tot}}$ we have investigated, these distributions are essentially unchanged.

We now turn our attention to the expectation values $\langle \beta_i(\delta) \rangle$ of the Betti numbers. Fig. 3.14 shows the measurement results for the CDT ensemble with volume $N_2 = 200k$ and time extension $t_{\text{tot}} = 179$, in the range $\delta \in [2, 82]$. The original triangulations $T$, which correspond to $\delta = 1$, i.e. the case without any coarse-graining, by construction have the topology of a torus, with $\beta_0 = 1$ counting its single connected component, $\beta_1 = 2$ the loops along the two directions of the torus, and $\beta_2 = 1$ the single two-dimensional hole or "cavity" enclosed by the torus. We see that throughout most or all of the observed $\delta$-range, the expectation values for the coarse-grained geometries $T_\delta$ reproduce these values within error bars. In other words, the effective homology coincides with the original homology

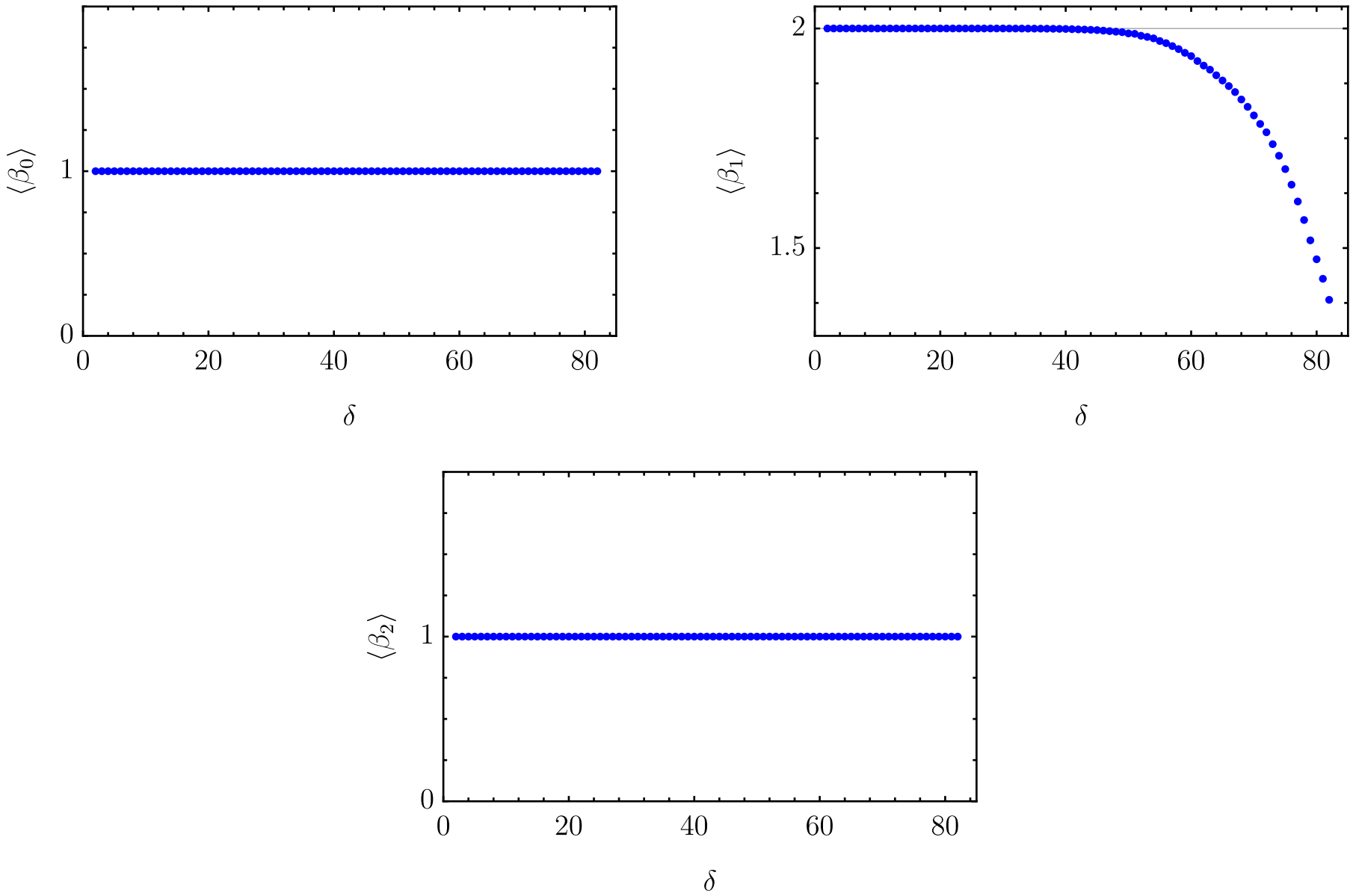

**Figure 3.14.:** Expectation values of the Betti numbers $\beta_0$, $\beta_1$ and $\beta_2$ in 2D Lorentzian quantum gravity with $N_2 = 200k$ and $t_{\text{tot}} = 179$, as a function of the resolution $\delta \in [2, 82]$. The straight line at $\langle \beta_1 \rangle = 2$ is included for comparison.

even for very large coarse-graining and in this sense does not reveal any new local structure.

Only when the resolution $\delta$ reaches a scale where the nontrivial global topology of the torus geometries comes into play, the average Betti numbers $\langle \beta_1 \rangle$ and $\langle \beta_2 \rangle$ can be affected. This happens because the diameter of a Voronoi cell can become large enough to wrap around one or both of the compact torus directions. This results in global pinchings of the geometry as described in Sec. 3.3.3, where the pinchings now take place at noncontractible loops of the original triangulation. By contrast, the expectation value $\langle \beta_0 \rangle$ cannot be affected, since by construction no part of the Delaunay triangulation detaches during a pinching.

The onset of these global features in the behaviour of the Betti numbers depends both on the time extension $t_{\text{tot}}$ and the spatial extension of the toroidal configurations. Because the size $\ell(t)$ of the spatial universe is subject to large quantum fluctuations (cf. Fig. 2.1), the global effect on observables of wrapping around the compact spatial direction becomes noticeable on length scales much smaller than the average spatial extension $\bar{\ell}$, as has been observed in previous studies of other observables in 2D CDT quantum gravity [98, 100]. This is also what we have found in the present study.

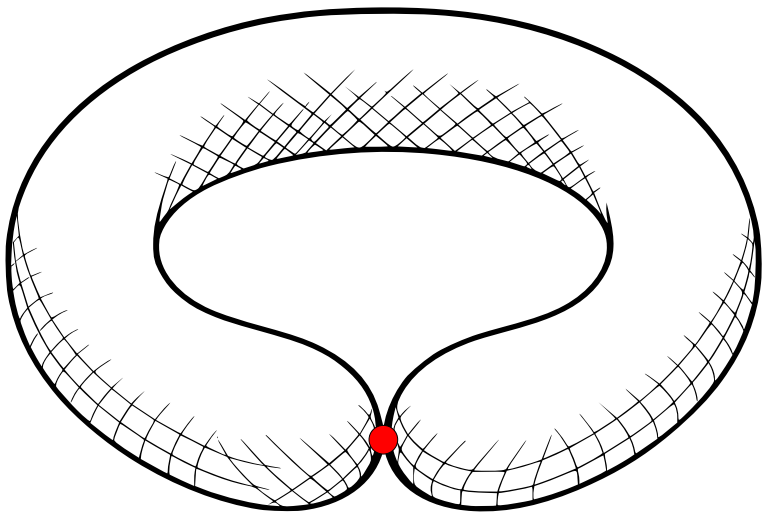

**Figure 3.15.:** The presence of a single global pinching vertex (red) lowers the Betti number $\beta_1$ of a toroidal Delaunay triangulation to 1.

For the parameter choices $N_2 = 200k$ and $t_{\text{tot}} = 179$ of the data plots shown in Fig. 3.14, a nontrivial wrapping of Voronoi cells in the time direction cannot occur, since the maximal diameter of such a cell is of the order of $2\delta$, which in the $\delta$-range probed is always smaller than $t_{\text{tot}}$. It implies that the deviation from constancy in $\langle\beta_1\rangle$ for $\delta \gtrsim 38$ observed in Fig. 3.14 is due to nontrivial global effects in the spatial direction. The leading contribution to the decrease of this expectation value is from configurations with a single global pinching vertex associated with contracting a spatial loop of winding number 1. Namely, for a given Delaunay triangulation $T_\delta$, $\beta_1$ drops from 2 to 1 in the presence of a single global pinching, since one of the noncontractible loops of the torus disappears, while $\beta_2$ remains unaffected, since there still is a two-dimensional cavity present. This is illustrated schematically by Fig. 3.15.

There are additional, smaller effects related to multiple pinchings, all of which occur with higher frequency as $\delta$ grows. For example, introducing a second pinching vertex in the configuration of Fig. 3.15, associated with the same direction but in a different location along the torus, will result in the appearance of two spherical cavities or "bubbles", with associated Betti number $\beta_2 = 2$. Correspondingly, additional pinchings can increase the number of bubbles and therefore $\beta_2$ even further. All of these lead to an increase in $\langle\beta_2\rangle$ as $\delta$ becomes larger. However, in the measurement of $\langle \beta_2 \rangle$ reported in Fig. 3.14 their occurrence is so rare that it does not lead to any appreciable deviation from constancy in the entire $\delta$-range considered.

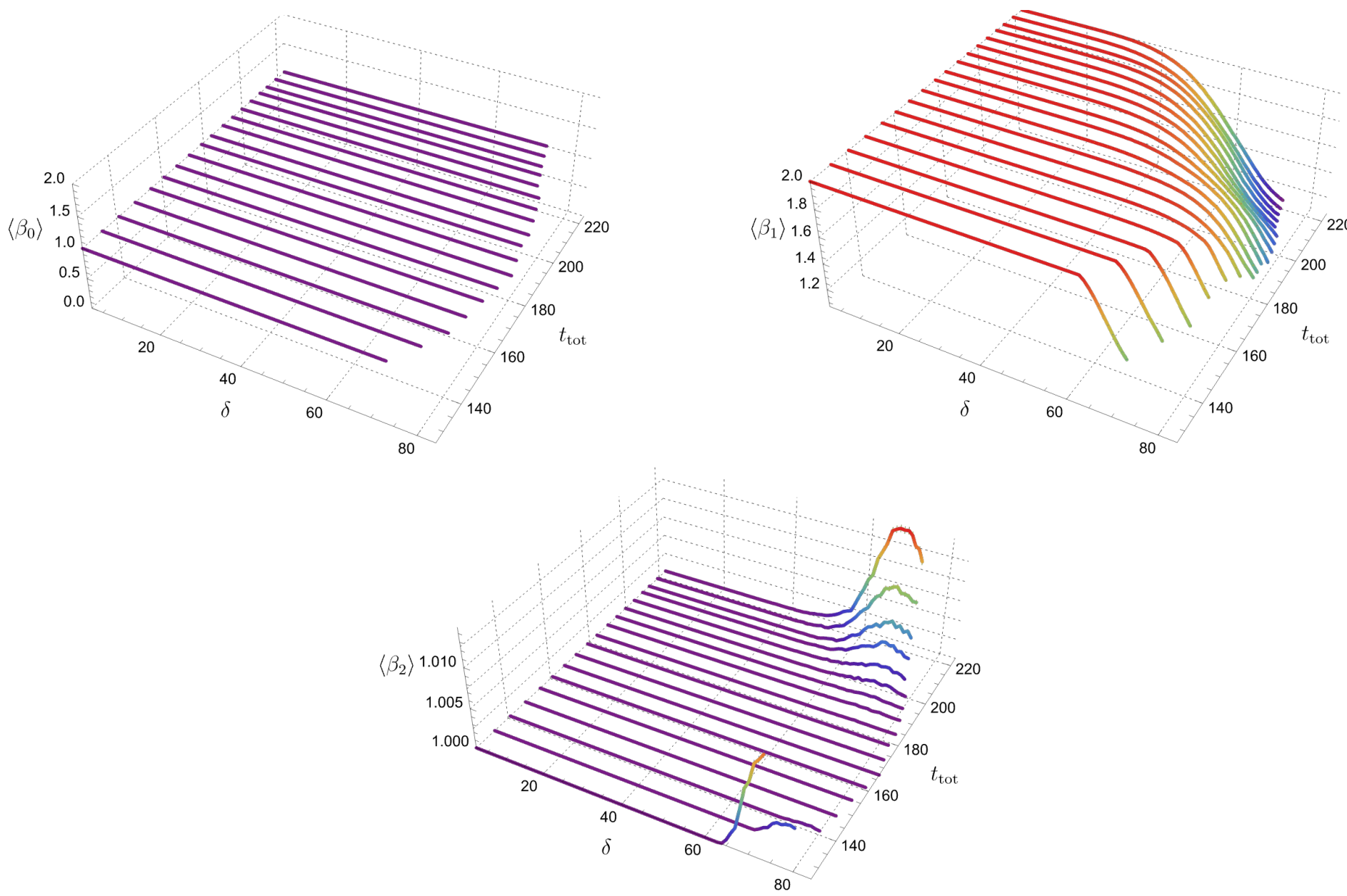

**Figure 3.16.:** Expectation values of the Betti numbers $\beta_0$, $\beta_1$ and $\beta_2$ in 2D Lorentzian quantum gravity with $N_2 = 200k$, as a function of the resolution $\delta$ and the time extension $t_{\text{tot}}$.

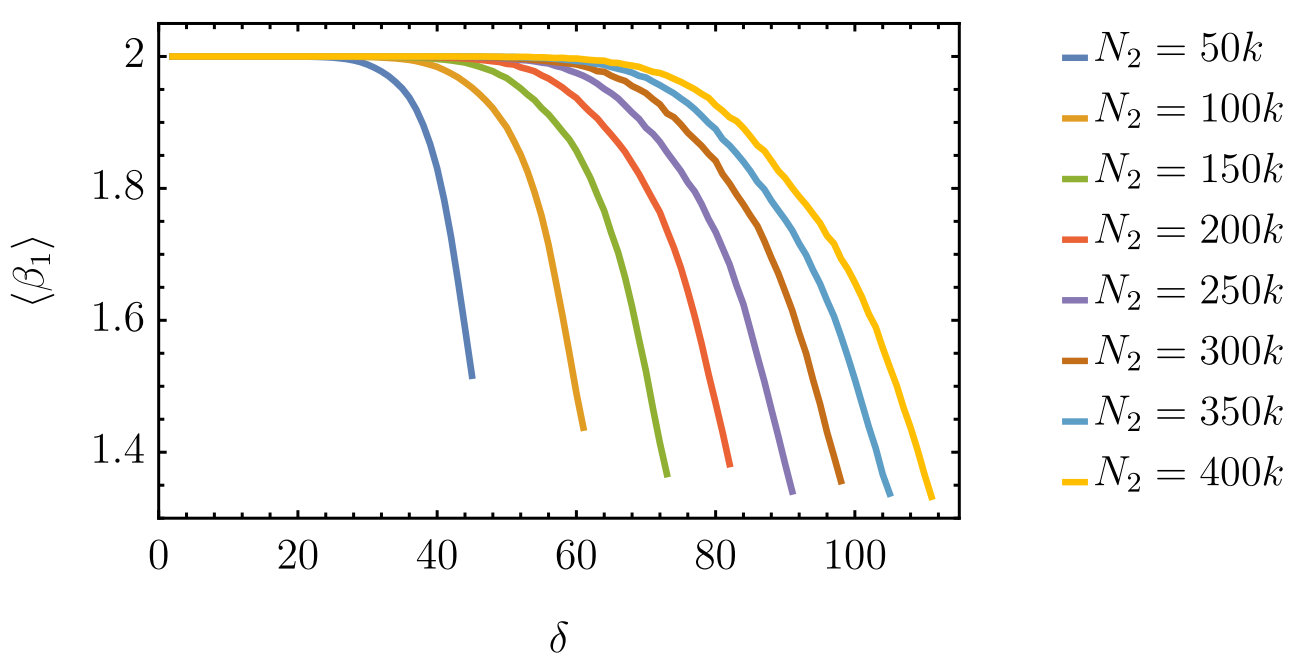

**Figure 3.17.:** Expectation value $\langle \beta_1 \rangle$ of the first Betti number in 2D Lorentzian quantum gravity as a function of the volume $N_2$, for a fixed ratio $r = 0.16$.

To study these effects more systematically, we have performed series of measurements of the Betti numbers at fixed volume $N_2 = 200k$, where in addition to $\delta$ we also vary the time extension, in the range $t_{\text{tot}} \in [126, 224]$, see Fig. 3.16. Note that for a given triangulation $T$, whenever the resulting Delaunay triangulation no longer contains any triangles we stop the coarse-graining and discard this configuration, independent of the values of its Betti numbers. It implies that the number of measurements we average over varies as a function of $\delta$. We stop plotting a curve for given $t_{\text{tot}}$ when more than 5% of measurements are discarded this way. We have verified on a test sample that this procedure captures the properties of the coarse-grained triangulations correctly, while allowing us to access sufficiently large $\delta$ to exhibit the nontrivial behaviour of the Betti numbers.

The expectation value of $\beta_0$, shown in Fig. 3.16, remains at the constant value 1 throughout, as anticipated. The location of the characteristic drop of the expectation value $\langle \beta_1(\delta) \rangle$ turns out to depend on the time extension $t_{\text{tot}}$. For small $t_{\text{tot}}$, it is primarily caused by pinchings along the time direction; these become rarer for increasing time extension, pushing the drop-off to larger values of $\delta$. Since a bigger $t_{\text{tot}}$ at constant two-volume implies a smaller spatial extension, pinchings along the spatial direction start appearing, leading to a monotonic decrease in the location of the drop-off beyond $t_{\text{tot}} \approx 155$. A complementary set of measurements of $\langle \beta_1 \rangle$, for increasing volume but a fixed ratio $r$ of the time and spatial extensions is given in Fig. 3.17. Again we observe that the $\delta$-interval where the expectation value is compatible with 2 grows as the volume becomes larger, roughly speaking $\propto \sqrt{N_2}$, as we have checked. The chosen ratio $r = 0.16$ is such that the effect comes from global pinchings of the spatial direction only. Lastly, as explained above, the Betti number $\beta_2$ is sensitive to the occurrence of multiple pinchings. Such pinchings along the time direction are in evidence in the measurements of its expectation value for small $t_{\text{tot}} \lesssim 148$ and large $\delta$, while the effects of multiple pinchings along the spatial direction start showing up in the plots for the largest time extensions $t_{\text{tot}} \gtrsim 200$ considered. Note that the absolute size of the change in $\langle \beta_2(\delta) \rangle$ away from the classical torus value is about two orders of magnitude smaller than that of $\langle \beta_1(\delta) \rangle$.

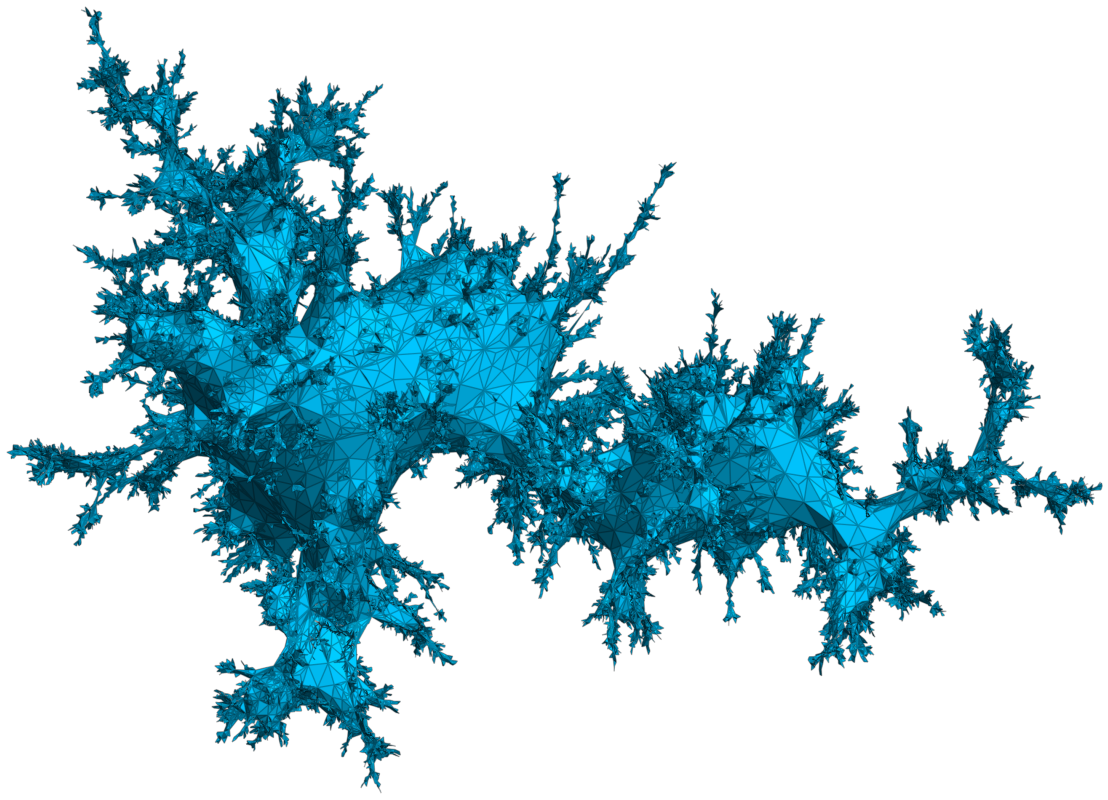

**Figure 3.18.:** Typical EDT configuration, with volume $N_2 = 100k$, depicting a Euclidean “spacetime” of spherical topology.

### 3.4.2 Euclidean quantum gravity in D=2

We turn next to the analysis of Euclidean quantum gravity in two dimensions (EDT). While it is not the main focus of this part of the thesis, unlike CDT, it is relevant to introduce some of its features, because in the following we will contrast the effective topology results obtained in CDT with those of EDT.

This toy model of quantum gravity is known to lie in a different universality class from the Lorentzian model considered in Sec. 3.4.1. The crucial difference is the absence of a causal structure for the histories that are summed over in the regularized Euclidean path integral

$$Z^{\mathrm{eu}}(\lambda) = \sum_T \frac{1}{C(T)} \, \mathrm{e}^{-\lambda N_2(T)}, \tag{3.5}$$

which otherwise is completely analogous to the Wick-rotated path integral (2.5) of CDT. As already mentioned in footnote 2, the sum is taken over EDT that are elements of a slightly generalized ensemble, compared to that of simplicial manifolds, which is known to not affect the continuum limit of the model. With these specifications observed, each triangulation $T$ in the sum is an arbitrary gluing of flat, equilateral triangles with the topology $S^2$ of a two-sphere. A typical EDT configuration is shown in Fig. 3.18, illustrating the well-known fractal nature of its quantum geometry.

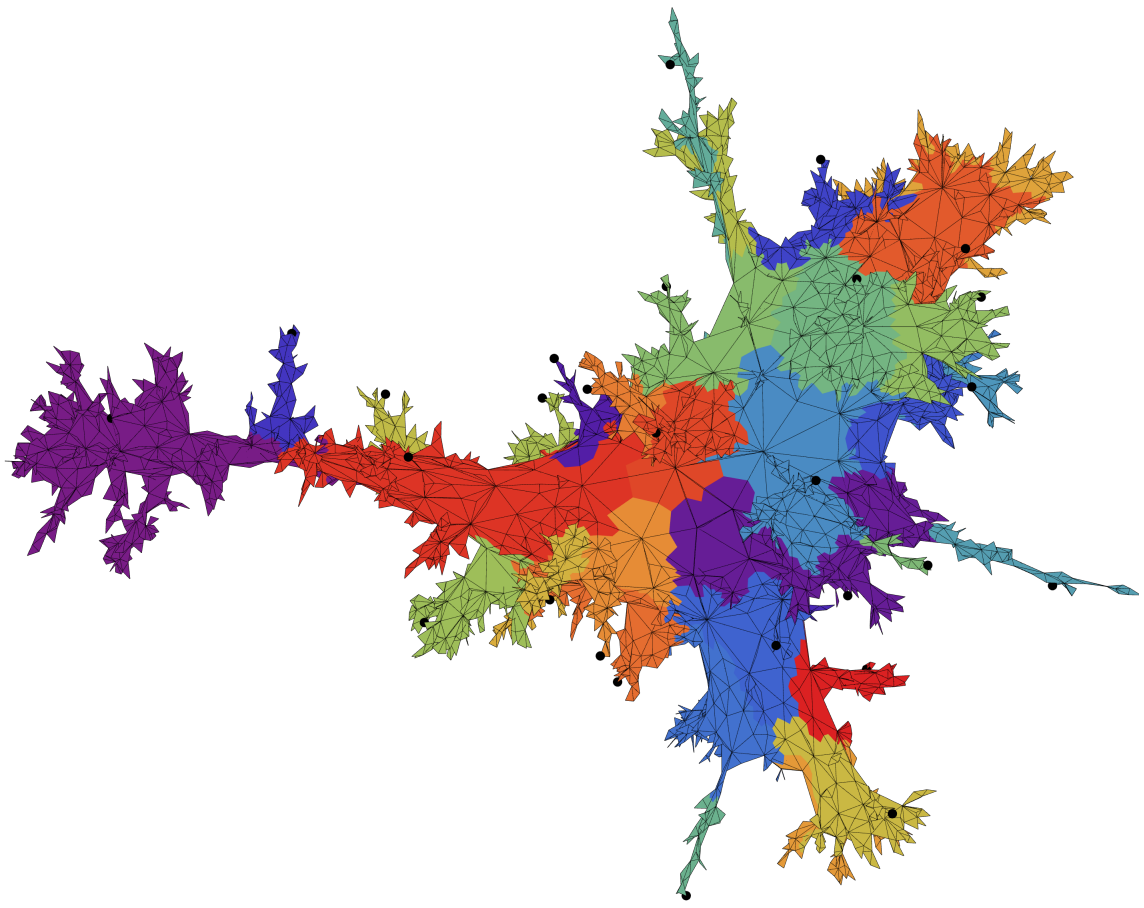

**Figure 3.19.:** Voronoi decomposition of a typical EDT configuration on a two-sphere at resolution scale $\delta = 8$, with $N_2 = 10k$. Thin black lines are those of the original triangulation $T$.

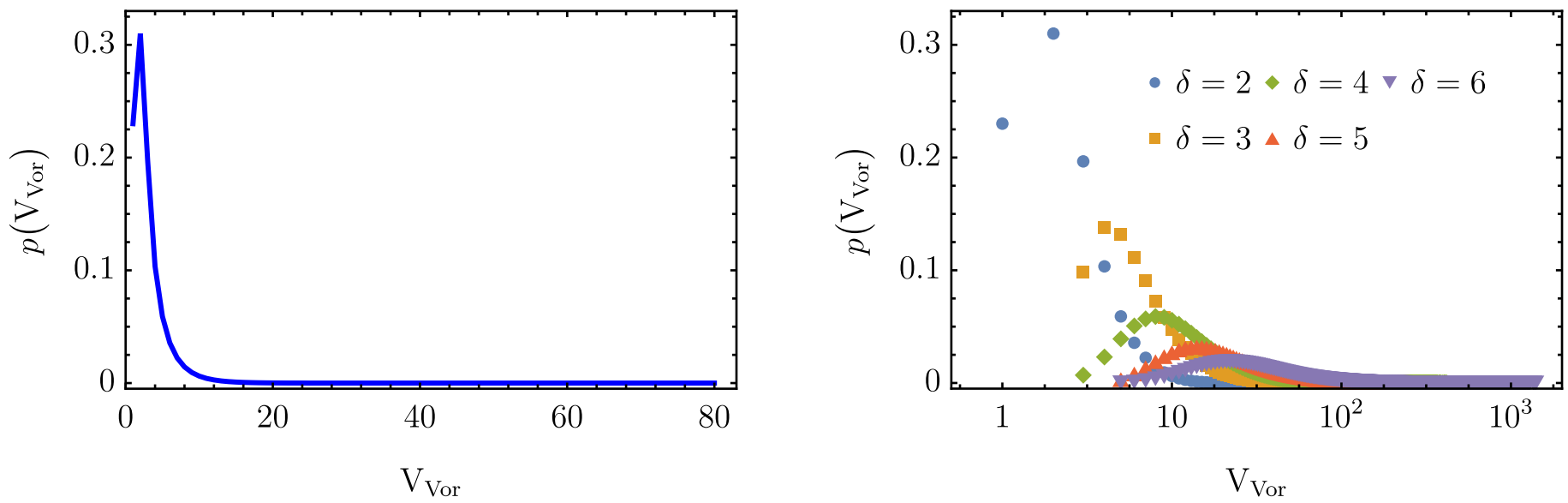

**Figure 3.20.:** Normalized distribution $p(V_{\mathrm{Vor}})$ of the volume $V_{\mathrm{Vor}}$ of individual Voronoi cells in units of vertices of the original triangulation $T$ for EDT configurations of volume $N_2 = 200k$, for $\delta = 2$ (left) and for the range $\delta \in [2, 6]$ (right).

We have again computed the expectation values of the Betti numbers, using the analogueof expression (2.7) for the Euclidean fixed-volume ensemble, for eight equally spaced values $N_2 \in [50k, 400k]$. For each volume, we have performed several hundred thousand measurements on independent configurations.[1] The EDT configurations are generated by a direct Monte

[1] For the volumes $N_2 = 50k$, $100k$, $150k$, $200k$, $250k$, $300k$, $350k$ and $400k$ we used $344k$, $322k$, $568k$, $552k$, $489k$, $516k$, $440k$ and $459k$ configurations.

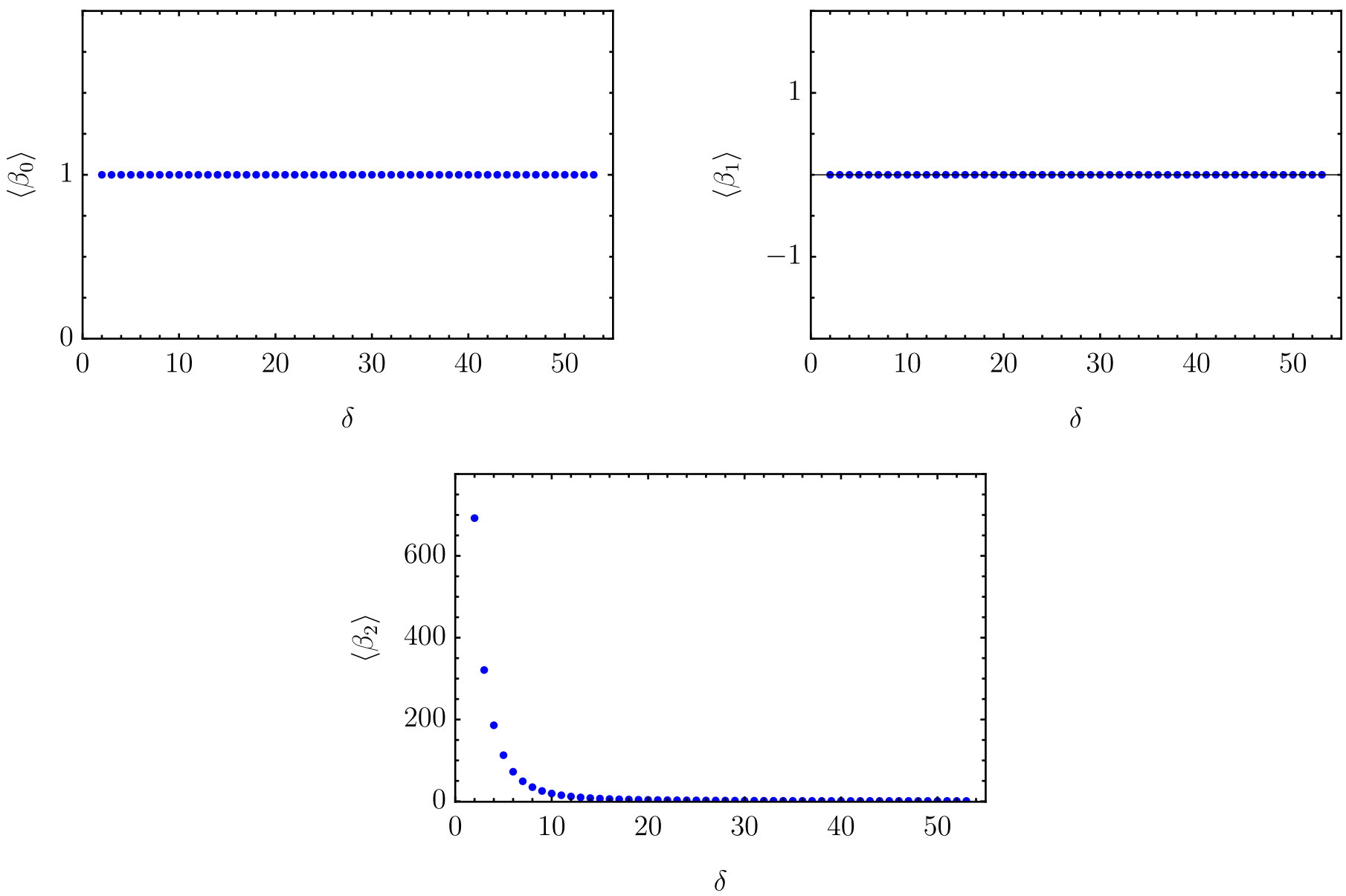

**Figure 3.21.:** Expectation values of the Betti numbers $\beta_0$, $\beta_1$ and $\beta_2$ in 2D Euclidean quantum gravity with $N_2 = 200k$, as a function of the resolution $\delta \in [2, 53]$.

Carlo sampling method, where each sampled configuration is automatically independent.[1] Fig. 3.19 shows a typical EDT configuration with coloured Voronoi cells after coarse-graining with resolution $\delta = 8$. Each cell is associated with a vertex from the evenly spread sample $\mathcal{S}_8$ (indicated by a fat black dot). Due to the fractal and "spiky" nature of the original triangulation $T$, the shapes of the cells are much less regular than in the Lorentzian case. Note also that the annuli discussed in Sec. 3.3.3 are present whenever a Voronoi cell of one colour wraps around an outgrowth that continues with a cell of another colour.

We have checked that our algorithm nevertheless distributes volume among the cells relatively evenly. This is illustrated by Fig. 3.20, which is the Euclidean analogue of Fig. 3.13 above, showing the distribution of volumes of the Voronoi cells at resolution $\delta$ in terms of the numbers of vertices of the triangulation $T$. The situation is qualitatively similar to that of the Lorentzian case, with the peaks of the distributions located at slightly lower values of the cell volume $V_{\text{Vor}}$, and somewhat broader distributions for the larger $\delta$-values.

[1] The direct sampling uses a bijection between tadpole-free triangulations and random Kreweras excursions[101], which can easily be sampled directly. The implementation is available at [97].

The measurements of the expectation value of the Betti numbers for the EDT ensemble with volume $N_2 = 200k$ in the range $\delta \in [2, 53]$ are displayed in Fig. 3.21. Before coarse-graining, the triangulations $T$ have the topology of a sphere, with $\beta_0 = 1$ counting its single connected component, $\beta_1 = 0$ reflecting the absence of noncontractible loops, and $\beta_2 = 1$ for the single two-dimensional hole enclosed by the sphere. Throughout the entire observed $\delta$-range, not only the expectation values $\langle\beta_0\rangle$ and $\langle\beta_1\rangle$ but the Betti numbers $\beta_0$ and $\beta_1$ for each individual coarse-grained geometry reproduce these values. The constancy of $\beta_0$ is unsurprising, since the algorithm we use preserves the connected character of the geometries. In the present case, since the geometries remain also simply connected, this also explains the constancy of $\beta_1$. However, pinchings can in principle happen, where the sphere after coarse-graining becomes a set of connected spherical bubbles, like the structure sketched in Fig. 3.11 above. This changes $\beta_2$, which counts the number of such bubbles.

The bottom plot of Fig. 3.21 for the expectation value of $\beta_2$ indicates that this is exactly what happens here and, unlike in the Lorentzian case, is a *local* phenomenon. Already for the smallest coarse-graining step $\delta = 2$, the expectation value of $\beta_2$ shoots up to a maximum, then decreases steeply as the resolution becomes larger, and asymptotes to 1 for the largest $\delta$-values considered.[1] A similar behaviour can be observed for a range of volumes $N_2 \in [50k, 400k]$, where correspondingly more bubbles appear, as shown in Fig. 3.22, left. The locality of this "bubble generation" is underscored by the fact that for $\delta = 2$, for which we have the best bubble statistics, the average Betti number $\langle\beta_2\rangle$ for large $N_2$ scales approximately linearly with the volume, as illustrated by Fig. 3.22, right.

The generation of these spherical bubbles is illustrated further by the sequence of Delaunay triangulations $T_\delta$, for $\delta = 2, 4, 6$ and 8, depicted in Fig. 3.23, which are coarse-grained versions of the triangulation of Fig. 3.18. At every iteration there is a "mother universe", drawn in white, which contains most of the volume, and smaller, coloured outgrowths or bubbles that are connected to the mother universe or to other bubbles by pinchings or loose edges. To convey a more quantitative understanding of the number and sizes of the bubbles, we present another diagrammatic representation of the same four configurations in Fig. 3.24. Individual spherical components or bubbles are represented by discs whose radius is proportional to $\log(\tilde{N}_2/2)$, where $\tilde{N}_2$ is the number of coarse-grained triangles in the bubble, which is also displayed alongside the disc. (The label for bubbles with $\tilde{N}_2 = 2$ is suppressed to avoid clutter.) A line is drawn between bubbles if they are connected by a pinching vertex or by one or more loose edges in the coarse-grained triangulation $T_\delta$.[1] All diagrams exhibit the presence of a mother universe at the centre, where the by far largest fraction of the volume is located, with a first generation of much smaller neighbouring bubbles, a much sparser second generation of small bubbles, and occasional instances of bubbles of a higher generation (in the examples shown, these are only present in the diagram for $\delta = 2$).

[1] Since we do not include configurations where all triangles have vanished, we always have $\beta_2 \geq 1$.

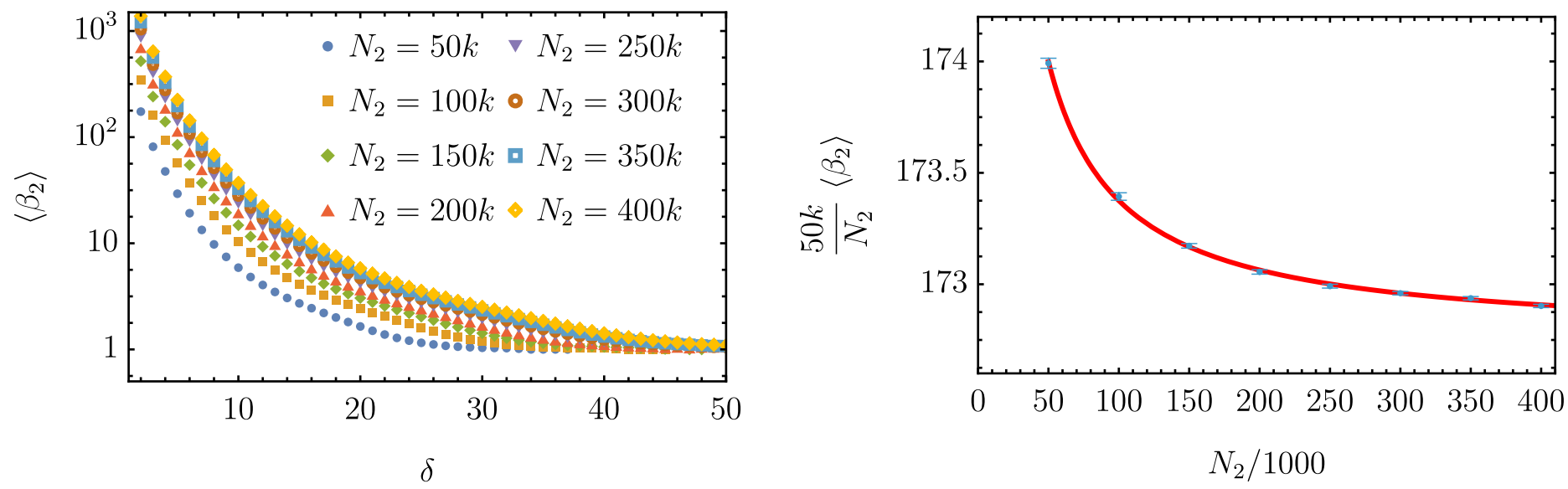

**Figure 3.22.:** Expectation value of the Betti number $\beta_2$ in 2D Euclidean quantum gravity as a function of the resolution $\delta$, for volumes $N_2 \in [50k, 400k]$ (left). The same expectation value, for $\delta = 2$, rescaled by the volume, and plotted as a function of the volume $N_2$. The red line is the best fit to $\frac{50k}{N_2}\langle\beta_2\rangle = A + B/N_2$, for fitting constants $A$ and $B$ (right).

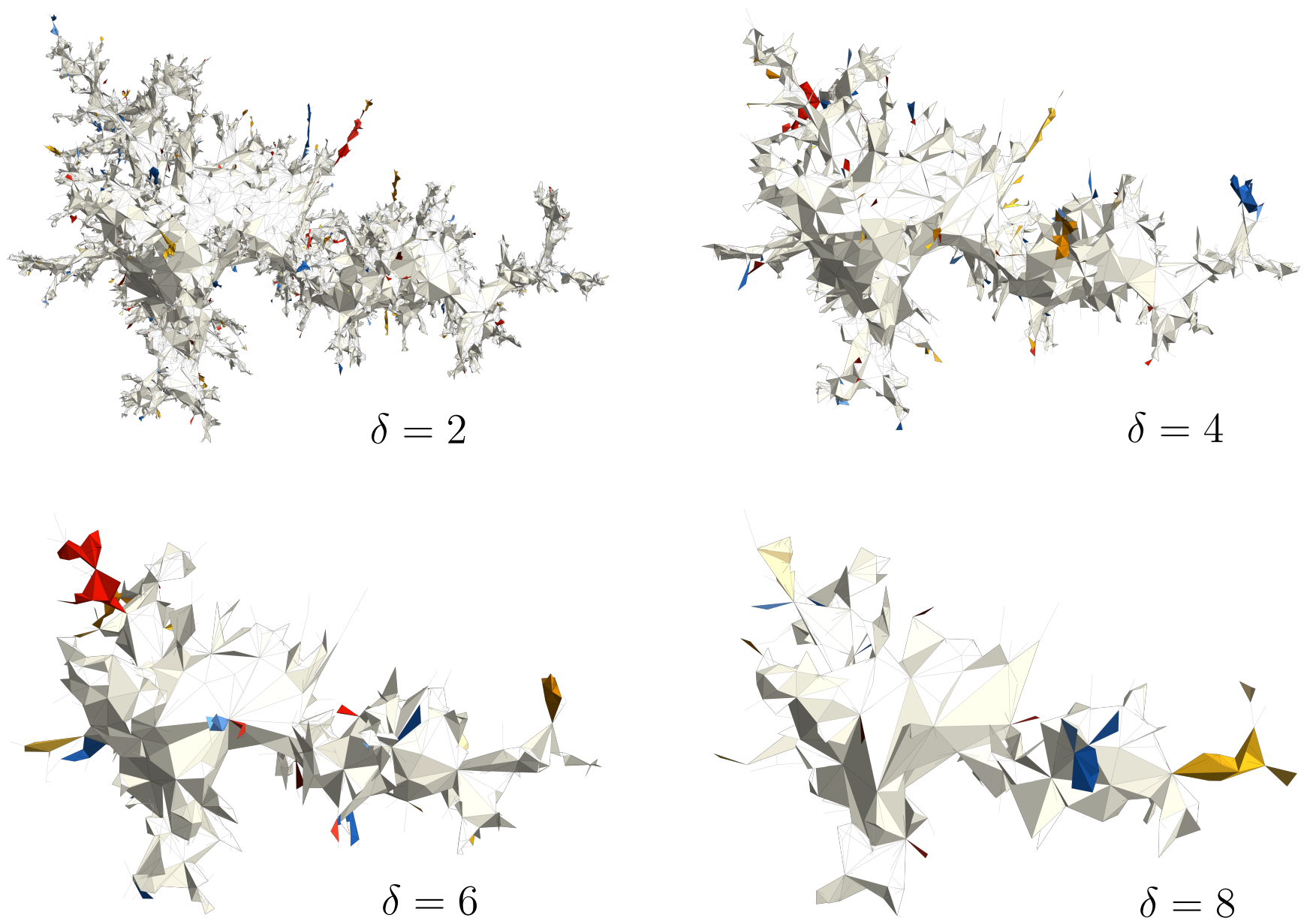

**Figure 3.23.:** Coarse-grainings of the spherical EDT configuration of Fig. 3.18 with initial volume $N_2 = 100k$, at resolution $\delta = 2, 4, 6$ and 8, showing a large mother universe (white) and bubbles that appear due to pinching (coloured).

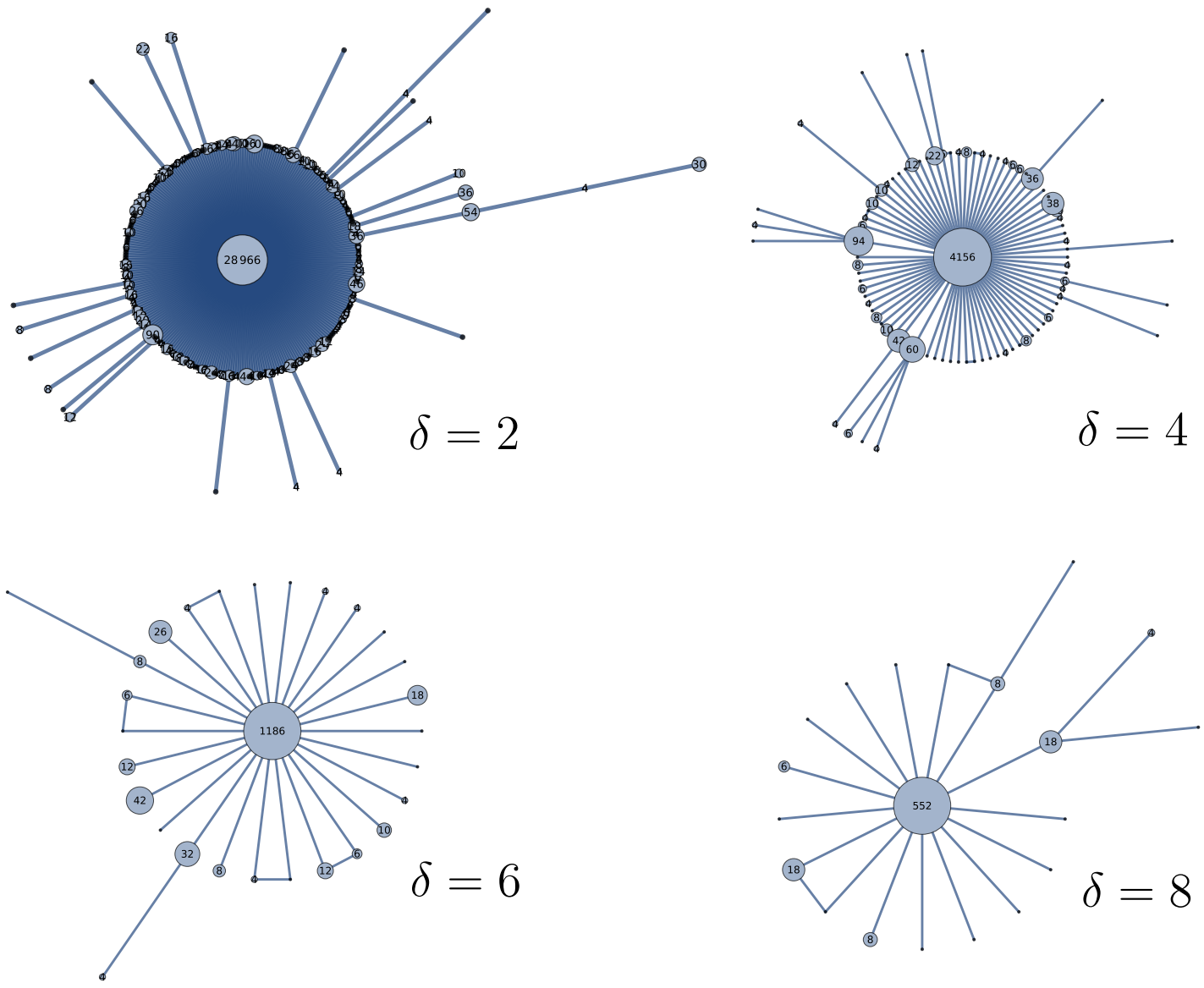

**Figure 3.24.:** Bubble diagrams of the four coarse-grained Delaunay triangulations depicted in Fig. 3.23, with discs associated to bubbles, and bubble volumes given in terms of coarse-grained triangle units, as described in the main text.

[1] Note that this leads to closed loops in the bubble diagram whenever there is a triple or higher-order meeting point between bubbles in $T_\delta$. In Fig. 3.24 this happens in the diagrams for $\delta = 6$ and $\delta = 8$. It does *not* imply that $T_\delta$ itself contains noncontractible loops (which it does not).

Unlike what we found in 2D Lorentzian quantum gravity, the nontrivial aspects of the effective homology in the Euclidean case are not just global, but also local[1], in the sense that the expectation value $\langle\beta_2(\delta)\rangle$ behaves nonclassically for all values of $\delta$. This behaviour is related to the well-known fractal structure of 2D Euclidean quantum gravity. It can be characterized by the presence of so-called minimal-neck baby universes ("minbus"), parts of the two-dimensional geometry that are connected to a mother universe by minimal necks, consisting of closed loops of three lattice links [102, 103]. Although our construction does not exactly identify such minbus, we show in the following section that the bubble structure we find is close enough to establish (for $\delta = 2$) a quantitative relation with the string susceptibility, a scaling parameter that in these references is extracted by measuring the statistical distribution of minbu sizes. Our bubble diagrams of Fig. 3.24 suggest that the fractal-like, hierarchical structure of a mother universe and subsequent generations of smaller bubbles is preserved during coarse-graining, at least at a qualitative level.

## 3.5 Recovering the string susceptibility

Our discussion of pinchings and thin necks is reminiscent of earlier work on 2D Euclidean quantum gravity involving so-called minimal-neck baby universes or "minbus" [102]. Minimal necks are thin necks of length 3 in an ensemble of equilateral triangulations that obey slightly stricter regularity conditions[2], but whose use in the path integral leads to the same continuum theory. Cutting open the triangulation along such a minimal neck will generically lead to a "mother universe", where most of the volume resides, and a much smaller minbu[3]. It has been shown that the distribution of minbu sizes $n$ for triangulations of fixed volume $N_2$ has a simple dependence on the so-called string susceptibility $\gamma_{\mathrm{str}}$ [102], and that measuring this distribution provides an efficient way to extract $\gamma_{\mathrm{str}}$ numerically [103].

[1] Strictly speaking, in 2D quantum gravity without topology change there is no distinction between local and global, since any length scale is set by a cosmological constant $\Lambda$, the continuum counterpart of the coupling $\lambda$ appearing in the bare actions in (2.1) and (3.5). However, in 2D CDT, by choosing suitable combinations of the two-volume $N_2$ and the time extension $t_{\mathrm{tot}}$, we can create a "bulk regime" largely free of the effects of global winding numbers, as we have seen.

[2] variably called combinatorial triangulations or simplicial manifolds, see e.g. [91]; for the triangulations used by us the minimal neck length is 2

[3] When the baby universe consists of a single triangle only, it is not counted as a minbu.

This scaling exponent governs the subleading behaviour of the partition function $\tilde{Z}(N_2)$ for fixed volume, which for large $N_2$ is known analytically to have the form

$$\tilde{Z}(N_2) \propto \mathrm{e}^{\lambda_0 N_2} N_2^{\gamma_{\mathrm{str}}-3}(1+\mathcal{O}(\tfrac{1}{N_2})) \tag{3.6}$$

for some (non-universal) constant $\lambda_0 > 0$. From this one can derive that the expectation value $\langle b(n)\rangle_{N_2}$ of the number of minbus with volume $0 \ll n < N_2/2$

$$\langle b(n)\rangle_{N_2} \propto n^{\gamma_{\mathrm{str}}-2}(N_2-n)^{\gamma_{\mathrm{str}}-2}. \tag{3.7}$$

The similarities with our construction raise the question of whether we can relate the behaviour of our bubbles in a quantitative way to the minbu analysis. This would require us to keep track of the volumes of the bubbles and consider their statistics. Despite the differences between the two set-ups, this turns out to be possible and leads to a surprisingly close match when extracting the scaling parameter analogous to $\gamma_{\mathrm{str}}$.

It is most natural to compare with the case where our necks have minimal length, i.e. $\delta = 2$, for which also the number of bubbles is maximal.[1] The potentially most relevant difference comes from the fact that bubbles and minbus are different objects. The standard minbu analysis for a given triangulation $T$ identifies all minimal necks and for each such neck records the volume of the smaller component (with $n < N_2/2$), regardless of whether it contains other minimal necks and associated minbus [103]. By contrast, for a given coarse-grained triangulation $T_2$, bubbles are mutually exclusive, and adding up their volumes – given in terms of coarse-grained triangle units – one obtains the total volume of $T_2$.

The reason why one can nevertheless expect a similar volume distribution is the typical baby universe structure of the triangulations under consideration, which we have investigated qualitatively in the previous sections [1]. It consists of a "mother bubble" of large volume, which via pinching vertices and loose edges is connected to a first generation of much smaller "baby bubbles", which in turn can have further bubbles as offspring, and so on. However, baby bubbles of second or higher generation are rare: less than 10% of first-generation bubbles have further offspring, and even if they do, their volumes tend to be small. It implies that many bubbles can be identified with minbus of the same or similar volume.

[1] The mismatch between $\ell_{\mathrm{min}} = 2$ and $\ell_{\mathrm{min}} = 3$ is a lattice artefact, which should not play much of a role.

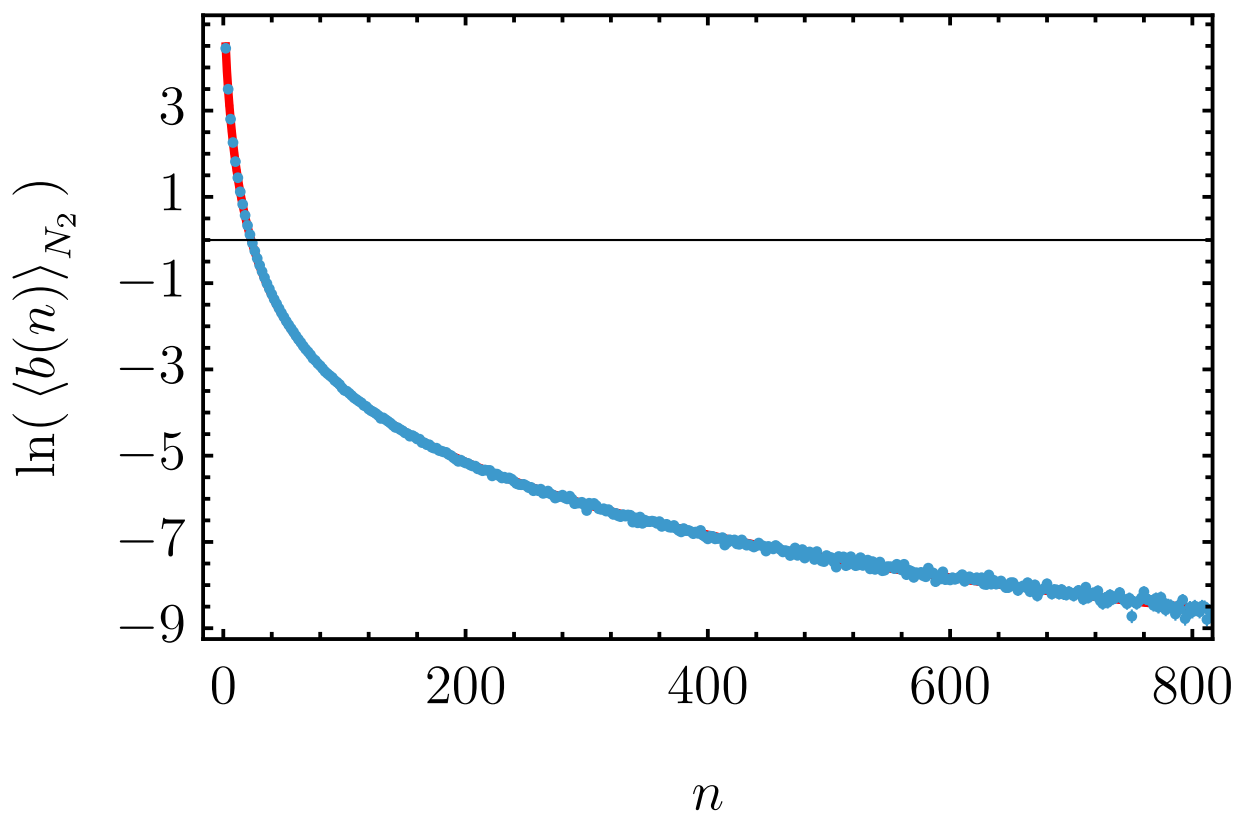

**Figure 3.25.:** Expectation value of the abundance of bubbles of volume $n$, for $\delta = 2$ and $N_2 = 50k$, together with a best fit for the string susceptibility.

We have measured the abundance $b(n)$ of bubbles of volume $n$ for several hundred thousand triangulations for each of the eight volumes $N_2 \in [50, 400k]$ considered and fitted the data to the functional form

$$\ln(\langle b(n)\rangle_{N_2}) = \alpha + (\gamma_{\text{str}} - 2)\ln\left(n(1 - \frac{n}{N_2^c})\right) + \frac{\kappa}{n}, \tag{3.8}$$

where $\alpha$ and $\kappa$ are fit parameters, $n$ and $N_2^c$ are measured in terms of coarse-grained volume units[1], $\kappa/n$ is a finite-size correction term, and a bubble size of 5% of the total volume $N_2^c$ was chosen as upper bound for the fitting window. For illustration, Fig. 3.25 shows the quantity (3.8) for $N_2 = 50k$; the fit quality for the other volumes is similar. The values for the fitted string susceptibility exponent decrease monotonically (within error bars) with the volume, ranging from $\gamma_{\text{str}} = -0.4861 \pm 0.0040$ for $N_2 = 50k$ to $\gamma_{\text{str}} = -0.5226 \pm 0.0056$ for $N_2 = 400k$. They are in good agreement with the known value $\gamma_{\text{str}} = -1/2$, especially considering the various mismatches with the minbu analysis in [103]. It shows that the bubble structure uncovered by effective homology is related to the universal properties of 2D Euclidean quantum gravity.

[1] We have verified that $N_2^c$ is roughly proportional to $N_2$, with a Gaussian spread.

## 3.6 Collecting the results of this chapter

By explicit construction, we have shown that the tool of *effective homology* can be used to characterize the quantum geometry in two-dimensional models of quantum gravity, defined nonperturbatively as continuum limits of dynamically triangulated lattice theories. The key idea is to construct a set of observables describing local metric properties of the quantum space(-time) by using the powerful machinery of TDA after applying a local coarse-graining algorithm to the path integral configurations. A coarse-graining of resolution $\delta$ produces a (generalized) triangulation with typical edge length $\delta$ (an integer in original lattice units): $\delta = 1$ reproduces the original triangulation, $\delta = 2$ results in a triangulation with edge length 2 (and correspondingly fewer triangles), and so forth. Although the original triangulations all have the same, fixed topology (of a torus in the Lorentzian and a sphere in the Euclidean case), this need not be the case after coarse-graining, since the latter by construction does not resolve substructures of linear size smaller than $\delta$.

The concrete observables we measured on lattices of volume $N_2 \leq 400k$ were the Betti numbers $\beta_i(\delta)$, $i = 0, 1, 2$, as a function of the coarse-graining scale. The only nontrivial behaviour of their expectation values we found in the Lorentzian model comes from large $\delta$, where the cells of the coarse-grained Voronoi decomposition become sufficiently large to completely wrap around one or both of the torus directions, leading to singular pinchings of the associated Delaunay triangulation that affect both $\langle\beta_1\rangle$ and $\langle\beta_2\rangle$ (Fig. 3.16). A similar pinching mechanism is also present in the Euclidean model, but already at a local scale, which affects the expectation value $\langle\beta_2\rangle$ already at the smallest nontrivial coarse-graining $\delta = 2$ (Fig. 3.21).

Returning to the theme of symmetry we introduced in Sec. 2.1.1, since the Lorentzian and Euclidean quantum gravity models in 2D do not have nontrivial classical limits, we cannot examine the issue of *recovering* any symmetries, but we can still ask whether their quantum geometries may support continuous isometries of some kind. For the Lorentzian model, our analysis has not found any obstructions, at least not of a local kind and when staying away from length scales where global pinchings can occur. By contrast, the bubble structure of Euclidean quantum gravity presents a clear obstacle to the existence of such symmetries. In the latter case, the nonclassical behavior of the Betti number $\beta_2$ revealed a simple kind of “foaminess” of its quantum geometry. We showed that this is related to the well-known fractal nature of the model by means of the string susceptibility, which was studied previously in terms of minimal-neck baby universes.

Our investigation demonstrates the feasibility of the concept of effective homology and its technical implementation, for nontrivial quantum gravity models of either spacetime signature in two dimensions. We have rediscovered the well-known fractal structure of Euclidean Liouville gravity, consisting of mother and baby universes, but otherwise have not discovered any new features. Given the well-documented and relatively straightforward properties of both the Lorentzian and Euclidean models, this was not to be expected either. However, our explicit set-up did not rely in an essential way on working in two dimensions, which opens the way to its application in four dimensions and Lorentzian signature, which is our main target and motivation.

The higher-dimensional case will be richer and inevitably more complex, with more freedom of how to handle topology changes during coarse-graining, generalizing the "pinching" that was the main feature in two dimensions. Any choices should be guided by what is expected to be important for phenomenologically interesting observables, including those related to matter coupling. Also, computational efficiency and lattice size, which hardly played a role in the current study, will become factors to be considered in $D > 2$. The 4D application will break genuinely new ground, given our current, limited understanding of its local quantum geometry [60, 104] and how to characterize it in terms of suitable observables. Is it really a quantum foam on short scales, and does it support local or global symmetries when coarse-grained suitably? – We look forward to investigating the physical case of 4D Lorentzian quantum gravity in the near future.

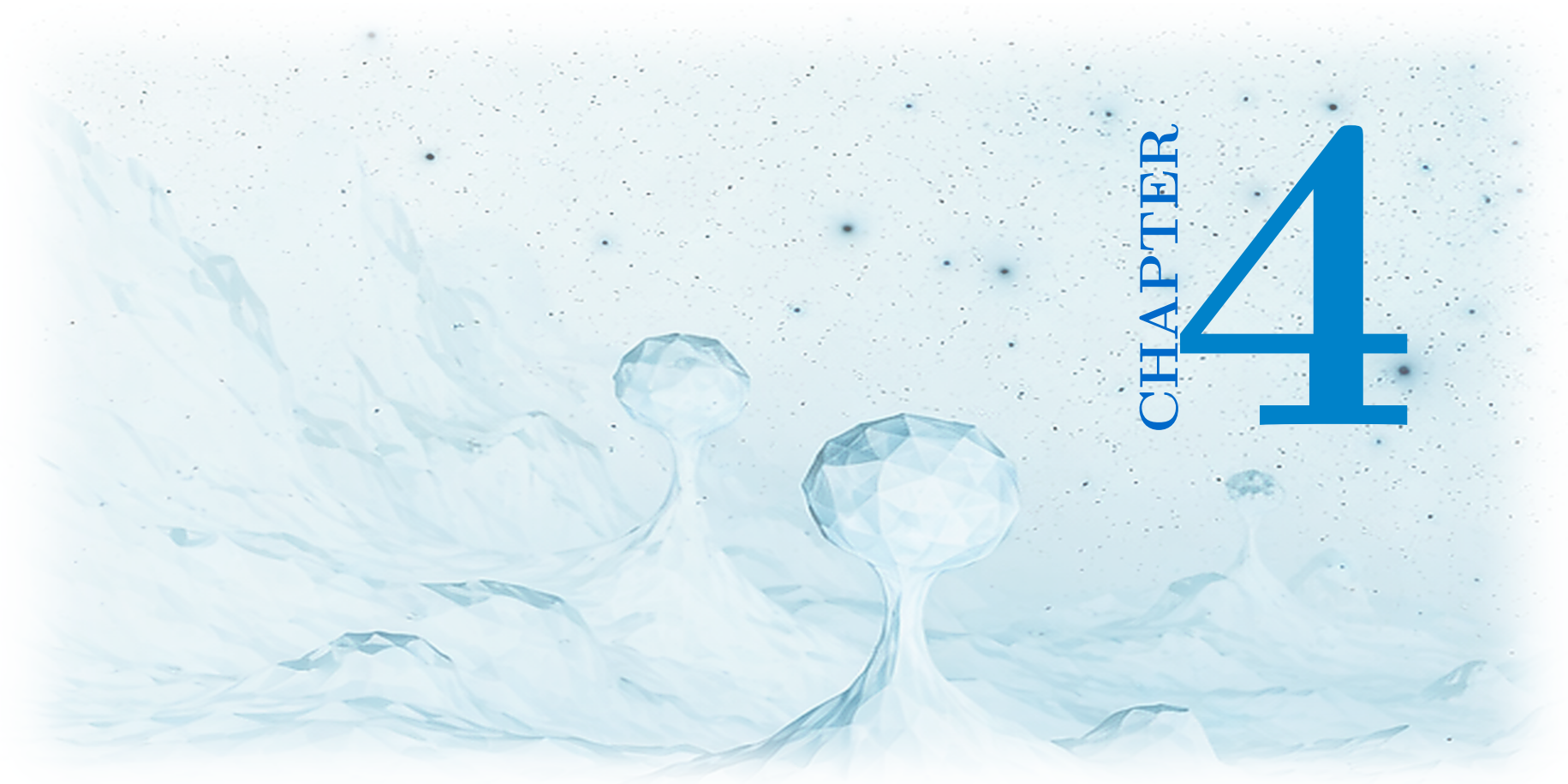

# Homogeneity

After the use of TDA, as described in the previous chapter, to determine if a quantum geometry satisfies the topological constraints necessary to embark on a more thorough investigation of the emergence of symmetries at some scale, we ask the obvious question: How can this be done when continuous symmetry groups, like translations or rotations, are not even defined in such spacetimes? This is the problem addressed in this chapter.

Another question we are particularly interested in, and that was one of the driving motivations for this chapter, is to what extent a given quantum spacetime can be related to a description of the very early universe. For instance, can we show that familiar properties of classical models of the pre-inflationary universe "emerge" from it on sufficiently large scales, or that it contains the quantum-gravitational seeds of structure formation?

In contrast to fully fledged quantum gravity, quantum cosmology aims to capture the quantum dynamics of the universe as a whole in terms of a small number of global variables, tantamount to making strong symmetry assumptions about the underlying spacetime [105]. In the most extreme case, when the spatial geometry is taken to be homogeneous and isotropic, the quantum dynamics of spacetime is reduced to the quantum mechanics of a single degree of freedom, the spatial volume (or scale factor). It would be interesting to understand whether some justification for such symmetry reductions can be derived from a theory of quantum gravity.

We are then tasked with the question: Are the spacetimes emerging from CDT homogeneous or not? The first natural question that will arise is “homogeneity with respect to which property?” Obviously, homogeneity depends on what property you are looking at.

With this motivation in mind, we will construct quantum observables below that enable us to quantitatively assess the degree of homogeneity of particular quasi-local properties of a quantum spacetime in a non-perturbative regime of full quantum gravity, focusing on the pure-gravity case. In addition to our theoretical considerations, we will present explicit computational results for a two-dimensional toy model of quantum gravity, to illustrate some of the technical and conceptual issues in constructing measures of homogeneity.

The aim of this chapter is to develop concrete tools to investigate the degree of homogeneity of a quantum spacetime, whose local structure may be foam-like or carry some other form of quantum excitations [106] (see also [77] for a recent review). As should be clear from our discussion above, the notions of scale and averaging will play important roles. Since, by assumption, we will work in a non-perturbative regime, we must make sure that any characterization of a quantum spacetime is well defined operationally. This implies that we define and evaluate suitable diffeomorphism-invariant observables, and extract the physical properties of quantum spacetime from their expectation values.

Following the initial question of “homogeneous with respect to what?”, we will study quantities associated with geodesic balls, which from a continuum point of view can be thought of as ball volume averages of particular local scalar properties of spacetime. Note that because of discretization effects, the analogues of local continuum quantities typically become quasi-local on the lattice, in the sense of being associated with a neighborhood of a lattice vertex. Given such a quasi-local quantity on a spacetime lattice and a prescription for how to average it over geodesic balls of size $\delta$, the associated quantum homogeneity measure will allow us to determine whether or not the quantity behaves approximately homogeneously on the scale $\delta$, in a way that will be spelled out in more detail below.

The general prescription we will use to characterize the symmetry properties of a non-perturbative quantum spacetime implements a more elementary idea than that of an isometry generated by a Killing vector field, namely that of “a quantum spacetime looking the same everywhere”. This idea is based on the intuitive notion of walking around a space, measuring a property, and if the space is homogeneous, then whatever property one is measuring is the same everywhere, at least at some scale. This is reminiscent of the so-called cosmological principle, which asserts that on suitably large scales the spatial universe looks the same to all observers or, in more technical terms, that it is statistically homogeneous and isotropic, motivating the use of Friedmann-Lemaître-Robertson-Walker (FLRW) models for their idealized, theoretical description.[1]

Regarding isotropy, one would require a notion of "looking the same in every direction", but due to the highly non-classical nature of the geometries contributing to CDT path integrals, it is not straightforward to define and parametrize (in terms of angular variables, say) what one means by a local "direction".[2] Even if one adopted a specific prescription, one would need a way to compare directions at different points, like parallel transport or distinguished global coordinates, which in a nonperturbative formulation are not readily available, as we explained earlier. Therefore, we will restrict ourselves to measures of homogeneity .

For the time being we will focus on scalar quantities, and will talk generically about their associated *measures of homogeneity*, which more precisely represent statistical measures of homogeneity. As discussed in the introduction of this part, sometimes it is more intuitive to work with continuous, Riemannian analogies, to understand why our definitions are diffeomorphism-invariant. So, in that spirit, let us discuss more precisely what we mean by statistical measures of homogeneity.

The main ingredient needed to specify such a measure of homogeneity on a given $D$-dimensional metric space $M$ of positive definite signature is a scalar quantity $Q(x, \delta)$ associated with a geodesic ball $B_x(\delta)$ of radius $\delta$ based at the point $x$, which has the interpretation of a quantity $Q$ averaged over the neighbourhood $B_x(\delta)$. In the simplest case, $Q(x, \delta)$ is the average of a local scalar $Q(x)$ over the ball $B_x(\delta)$, which in a continuum notation

[1] see [107, 108] for recent assessments of the observational evidence for the cosmological principle

[2] see [109], Sec. 4.3, for an illustration in the context of quantum Ricci curvature in two-dimensional Euclidean quantum gravity in terms of Dynamical Triangulations (DT)

reads

$$Q(x,\delta)=\frac{1}{vol(B_x(\delta))}\int\limits_{B_x(\delta)} d^D y\,\sqrt{\det g}\,Q(y)\,,$$

$$\text{with}\quad vol(B_x(\delta))=\int\limits_{B_x(\delta)} d^D y\,\sqrt{\det g}\,. \tag{4.1}$$

The explicit examples we discuss in Sec. 4.1.3 below are all of this type. We will comment in Sec. 4.2 on the nonuniqueness of this choice of averaging. It may happen that the quantities $Q(x)$ and $Q(x,\delta)$ have an additional dependence on some other scale $r$, which we will indicate by writing $Q(x;r)$ and $Q(x,\delta;r)$.

The intuitive interpretation of the quantity $Q(x,\delta)$ is that of "$Q(x)$ as seen with a lens of resolution $\delta$". In a sense, this comes from looking at a space with a somewhat blurry vision that cannot resolve details below a certain distance scale, and therefore one sees only the average of it. With this scale-dependent lens, we will define the following statistical measures of homogeneity.

The homogeneity measure of a quantity $Q(x,\delta)$ across $M$, as a function of the averaging scale $\delta$, will be assessed by comparing it to its average $\bar{Q}(\delta)$ on $M$

$$\bar{Q}[g](\delta)=\frac{1}{vol(M)}\int\limits_M d^D x\sqrt{\det g}\,Q(x,\delta)\,. \tag{4.2}$$

With these ingredients in hand, we can finally define the (absolute) homogeneity measure $\mathcal{H}_Q[g](\delta)$ of a scalar-valued quantity $Q(x,\delta)$ on a space with metric $g_{\mu\nu}$ by

$$\mathcal{H}_Q[g](\delta):=\sqrt{\frac{1}{vol(M)}\int\limits_M d^D x\sqrt{\det g}\Big(Q(x,\delta)-\bar{Q}(\delta)\Big)^2}\,. \tag{4.3}$$

If the averaged quantity $Q$ is homogeneous at resolution $\delta$ on $(M,g_{\mu\nu})$, we have $\mathcal{H}_Q[g](\delta)=0$.

The absolute homogeneity measure (4.3) is a natural choice in situations where absolute homogeneity in the sense of $\mathcal{H}_Q[g](\delta)=0$ can be attained exactly. However, this does not happen in the typical applications we have in mind, where the value zero will at most be approached in a limit or in an approximate sense. In these cases we will often be interested in quantifying the inhomogeneities, that is, the degree to which homogeneity is violated, relative to some other scale relevant to the physical situation at hand. One natural choice is the (absolute value of the) $\delta$-dependent sample average $\bar{Q}(\delta)$ of the quantity $Q(x,\delta)$. If it is non-vanishing, we can define a relative homogeneity measure $\mathcal{H}_Q^{rel}[g](\delta)$ by the dimensionless quotient

$$\mathcal{H}_Q^{rel}[g](\delta) := \mathcal{H}_Q[g](\delta)/|\bar{Q}[g](\delta)|, \qquad [\bar{Q}[g](\delta) \neq 0]. \tag{4.4}$$

Depending on the physical situation, other choices of reference scale may suggest themselves, which will then lead to alternative definitions of relative homogeneity measures. In the remainder of this chapter, we will stick with the definition (4.4).

Due to computational cost reasons, it is efficient to approximate the averages over the entire manifolds by averages over some "sufficiently representative" sample of points $S \in M$, to keep the computational effort manageable, and to limit the amount of double-counting (i.e. how many times local information at a point $y$ appears in averaged quantities associated with different geodesic balls $B_x(\delta)$). The discrete analogues of the previous averages are

$$\bar{Q}[g](\delta) \simeq \frac{1}{|S|}\sum_{x\in S} Q(x,\delta) \quad \text{and}$$
$$\mathcal{H}_Q[g](\delta) \simeq \sqrt{\frac{1}{|S|-1}\sum_{x\in S}\Big(Q(x,\delta)-\bar{Q}(\delta)\Big)^2}. \tag{4.5}$$

Here, $|S|$ denotes the number of points in $S$, and for later use, we have included the dependence on the metric $g$ (more precisely, on the diffeomorphism equivalence class of $g$) explicitly in the notation. The (non-unique) choice of a set $S$ should be such that $M$ is "well covered" by the associated geodesic balls $\{B_x(\delta)\,|\,x\in S\}$, in the sense that most points $y \in M$ should lie in exactly one ball, and the geometry of $M$ should be covered "evenly", without leaving large isolated regions uncovered. If the set $S$ is defined in geometric terms, without invoking any additional background structure that could bias the sampling, both of the measures in (4.5) are diffeomorphism-invariant classical observables, and suitable quantum implementations within the non-perturbative path integral, making them prime candidates for quantum observables. Explicit nontrivial examples will be discussed in Sec. 4.1. Of course, in the actual implementations on CDT, one works with discrete manifolds, so the previous approximations, averaging over a finite number of points, will come in handy in the next sections. One could adopt a more stringent mathematical characterization of such a covering or packing, as the one described in the previous section involving topological observables, but for the current purposes, the details of the prescription should be unimportant, as long as the linear size of $M$ is much larger than the ball radius $\delta$. This should be verified in any explicit implementation. A convenient choice of $S$ for geometries in two-dimensional CDT quantum gravity will be discussed in Sec. 4.1 below.

From the intuitive point of view, what this means is that one must first select some quantity of interest that reflects some aspect of the geometry, and later walk around the manifold, recording the average of that quantity on geodesic spheres around a sample of points in the manifold. Then compute the standard deviation of such averages to obtain how much the chosen quantity deviates from its mean value at a certain scale. This is very much like the $\frac{\delta T}{T} \approx 10^{-5}$ relative size fluctuations of temperature observed in the Cosmic Microwave Background. The obtained result from our measures is basically a number that determines the percentual deviations from perfect homogeneity, for each geometry.

Since our homogeneity measures refer to specific quasi-local or averaged quantities, we can only claim that the underlying spacetime is (approximately) homogeneous whenever (approximate) homogeneity has been shown for a complete set of such quantities. Classically, one might consider using a complete set of local curvature invariants (whose number depends on the dimension), but these tend not to have well-defined analogues in the non-perturbative quantum theory, where, anyway, few observables are available. When promoting the homogeneity measures to quantum observables below, we will therefore have little to say on how to define and construct a "complete set" of them. This being the case, one should be careful to refer to the homogeneity of the specific geometric quantity under consideration, rather than that of the quantum spacetime as such.

Our prescription is a notion of "coarse-graining" coming from a volume average of certain quasi-local geometric properties and their associated operators on a given quantum spacetime, in contrast to other types of notions of coarse-graining, like the one considered in renormalization group flows, or the one considered in the previous chapter 3. By using this volume average coarse-graining of quantities, we can probe its properties at various scales ("resolutions") of our choosing, which allows us to determine in quantitative terms whether a given property behaves (approximately) homogeneously in the range of physical scales considered. One could of course take a different approach, and instead of averaging a quantity over some geodesic ball, take the perspective of chapter 3 and measure the same quantity on the "coarse-grained geometry". These are two different notions, and will likely give different results. As mentioned in 3, the coarse-graining of a manifold on its own has many ambiguities, and therefore we will stick to the notion of coarse-graining introduced previously in this chapter, where the geometry is kept fixed, and one uses the covariant notion of geodesic balls on it to coarse-grain.

By working with an explicit averaging scale $\delta$, related to the resolution of a geodesic ball covering, we address the question of how (in)homogeneously an averaged quantity $Q(x,\delta;r)$ behaves when $\delta$ is varied while $r$ is kept fixed. In an explicit example of this kind discussed in Sec. 4.1.3 below, $Q(x,\delta;r)$ is the Hausdorff dimension extracted from measuring the volumes of spheres with a radius in the range $r\in[6,10]$, averaged over a geodesic neighborhood of radius $\delta$ based at the point $x$.

## 4.1 Implementation in $2D$ CDT

We are now in a position to demonstrate that the homogeneity measures (4.3) and (4.4) can be implemented in a nontrivial model of non-perturbative quantum gravity, namely in the two-dimensional gravitational path integral formulated in terms of CDT. As with the previous chapter, we will work in the well-studied toy model of 2D CDT, amenable to analytical methods (for a recent exposition, see [110] and references therein), which serves as a testing ground for new quantum gravity observables. We will use it here to implement and test the new quantum homogeneity observables numerically.

As before, we will adopt the standard choice of a $S^1\times S^1$-topology for the triangulated spacetimes, with compact spatial slices and a cyclically identified time direction for computational convenience, where once again there is a discrete proper-time parameter $t\in[0,1,2,\dots,t_{tot}]$, with times 0 and $t_{tot}$ identified, and where $\ell(t)\geq 3$ denotes the discrete length of the spatial universe at time $t$. For the simulations of this chapter, we also employ a Monte Carlo Markov chain algorithm to generate sequences of independent CDT geometries, allowing us to approximate the expectation values (2.7) by importance sampling of the full CDT ensemble.[1]

We perform simulations with ensembles of fixed volume $N_2$ and extract the continuum limit $N_2 \to \infty$ from a sequence of measurements on finite triangulations of increasing $N_2$, using finite-size scaling (see e.g. [111]). We work with CDT lattices in the volume range $N_2 \in [50.000, 350.000]$ and for time extensions $t_{tot} \in [62, 200]$, where as a result the average length $\bar{\ell}(t)$ of a spatial slice at integer time $t$ is of the order of 400–900. During data collection, we have performed $N_2 \times 1000$ attempted moves between measurements. Considering that the equilibrium acceptance ratios are around 10%, every site of the geometry was, on average, updated roughly 100 times between consecutive measurements.

In the context of the lattice calculations, $x$ will denote a vertex of a triangulation $T$, and $\delta$ (and other measures of linear distance) will be given in terms of the link distance. The (geodesic) link distance of two vertices $x$ and $y$ is defined as the number of links (edges) in the shortest path of contiguous links between $x$ and $y$. Given a vertex $x$, the sphere $\mathcal{S}_x(\delta)$ of radius $\delta$ at $x$ is defined as the set of all vertices at link distance $\delta$ from $x$. Similarly, the geodesic ball $B_x(\delta)$ of radius $\delta$ at $x$ – for which we will use the same notation as in the continuum – is given by the set of all vertices whose link distance to $x$ is smaller or equal to $\delta$. For a coarse-grained lattice quantity $Q(x,\delta)$, which is itself of the form of a geodesic-ball average of a local scalar $Q(x)$, the analogue of the relation (4.1) reads

$$Q(x,\delta) = \frac{1}{vol(B_x(\delta))} \sum_{y \in B_x(\delta)} Q(y), \quad \text{with} \quad vol(B_x(\delta)) = \sum_{y \in B_x(\delta)} 1. \tag{4.6}$$

Before presenting our numerical results, we still need to specify how we choose the set $S$ of points that will serve as centres of the sphere covering in the lattice versions of the homogeneity measures (4.3) and (4.4), namely

$$\mathcal{H}_Q[T](\delta) := \sqrt{\frac{1}{|S|-1} \sum_{x \in S} \Big( Q(x,\delta) - \bar{Q}[T](\delta) \Big)^2}, \quad \text{with} \quad \bar{Q}[T](\delta) = \frac{1}{|S|} \sum_{x \in S} Q(x,\delta), \tag{4.7}$$

[1] The observables implementation code used in this chapter was based on J. Brunekreef's CDT code. We thank him for sharing his work with us.

and

$$\mathcal{H}_Q^{rel}[T](\delta)=\mathcal{H}_Q[T](\delta)/|\bar{Q}[T](\delta)|, \qquad [\bar{Q}[T](\delta)\neq 0]. \qquad (4.8)$$

To generate a vertex sampling $S$ of resolution $\delta$ on a given CDT configuration $T$, we start by constructing an (initial) sphere covering based on a vertex set $S^{ini}$ of $T$. (This is not yet the final sphere covering that will be used to measure inhomogeneities). To obtain the sphere covering, we make use of the product structure of CDT configurations. We randomly pick a spatial slice $t_0$ and on it a vertex $x_0$, and construct a sphere $\mathcal{S}_{x_0}(\delta)$ of radius $\delta$ at $x_0$. Next, moving a distance $2\delta$ along vertices in the same slice, we pick a following vertex $x_1$ and construct its $\delta$-sphere, which "touches" the first sphere at time $t_0$. We continue moving along the slice in steps of length $2\delta$, constructing further spheres around vertices $x_2$, $x_3, \dots$, until no more spheres can be fitted without creating overlaps. Having thus constructed a sphere covering of a strip of time extension $\Delta t=2\delta$, we repeat the same construction, using as a starting vertex a randomly chosen vertex on the slice with label $t_0+2\delta$. After covering the corresponding strip $\Delta t=2\delta$ by spheres, we progress in the time direction in steps of $2\delta$ until no more strips can be fitted in without creating nontrivial overlaps between $\delta$-spheres (for illustration, see Fig. 4.1, left).

The sphere covering associated with the resulting finite set of vertices $S^{ini}=\{x_0,x_1,x_2,\dots\}$ will in general have some gaps of linear size $\leq 2\delta$, but this should not matter much for triangulations that are large compared to the scale $\delta$. The reason why we nevertheless do not use this sphere covering directly is a bias that could potentially arise from the alignment of sphere centres along the selected spatial slices. Recall that the main dynamics of the CDT model is given by large fluctuations of the spatial volume $\ell(t)$ as a function of time (Fig. 4.1, right). If a quantity whose homogeneity is being assessed correlates with the size of spatial slices, in particular when these are small, our initial selection $S^{ini}$ of sphere centres may not be sufficiently random. The selection principle we have adopted to avoid any such effects consists in picking one vertex from each of the spheres of the initial sphere covering and choose it as the centre of a sphere in a new sphere covering. The new set $S$ of circle centres is then spread roughly evenly over different spatial slices, at the price of creating some overlaps between pairs of spheres, and therefore a slightly less complete covering of the total set of vertices in $T$. We have compared averaged observables for sphere coverings based on both prescriptions $S$ and $S^{ini}$ for selected volumes, but have not found any differences. All measurements reported below are based on choosing sets of type $S$.

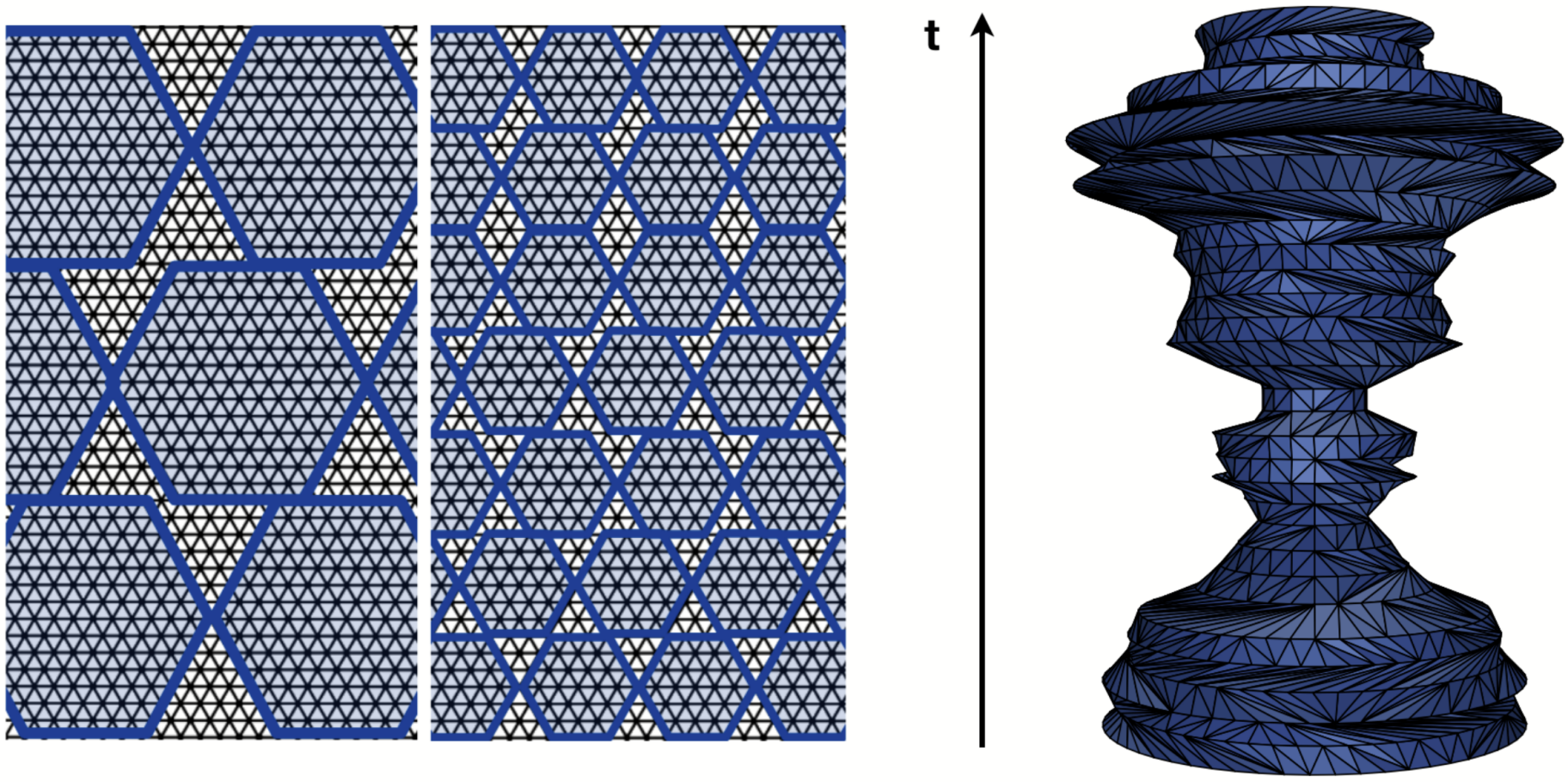

**Figure 4.1.:** Pieces of an initial sphere packing as described in the text, for resolution $\delta = 6$ and $\delta = 3$ (left). For ease of illustration, the packing has been implemented on a regular, everywhere flat CDT configuration. Generic CDT geometries carry local curvature and have spatial slices whose size $\ell(t)$ fluctuates strongly as a function of the time $t$, as illustrated by the sample configuration on the right.

In what follows, we will evaluate the expectation value $\langle \mathcal{H}_Q^{rel}(\delta) \rangle$ of the *relative* homogeneity measure (4.8) of a quantity $Q$ in the quantum theory. We will use a straightforward implementation of the observable $\mathcal{O}[T] = \mathcal{H}_Q^{rel}[T](\delta)$ on each two-dimensional piecewise flat CDT configuration $T$, where the set $S$ of vertices is selected as just described. When measuring and interpreting homogeneity observables we must take into account the inherent limitations of the lattice regularization. Firstly, to avoid finite-size effects due to the torus topology of the CDT configurations, we should only consider $\delta \ll L$, where $L$ is the linear size of a configuration $T$. In the case at hand this is given by the time extension $t_{\text{tot}}$ or the average spatial extension $\bar{\ell} = N_2/(2t_{\text{tot}})$, whichever is smaller, where it should be kept in mind that the spatial length $\ell(t)$ fluctuates strongly as a function of $t$.[1] At any rate, nothing interesting can be learned when the resolution $\delta$ approaches the linear size of the spacetime, since homogeneity is realized trivially there. In practice, we worked with an upper cutoff $\delta < t_{tot}/4$. Secondly, results for small $\delta$ suffer from discretization artefacts and therefore should not be considered in any continuum interpretation. Previous investigations of geometric observables on two-dimensional triangulations indicate that these discretization effects are definitely present on distance scales $\delta \leq 5$ in terms of link units [112]. In the explicit measurements discussed below, we find that this region is even larger, probably due to the complex and somewhat extended nature of our construction. Lastly, since we are interested in the homogeneity or otherwise of properties of the *continuum* theory, we study the expectation values of the homogeneity measures for increasing lattice volumes and search for convergent behaviour as a function of a physical, rescaled averaging scale.

[1] A quantitative analysis of the relation between the average and the minimal value of $\ell$ for the same CDT ensemble was made in [100].

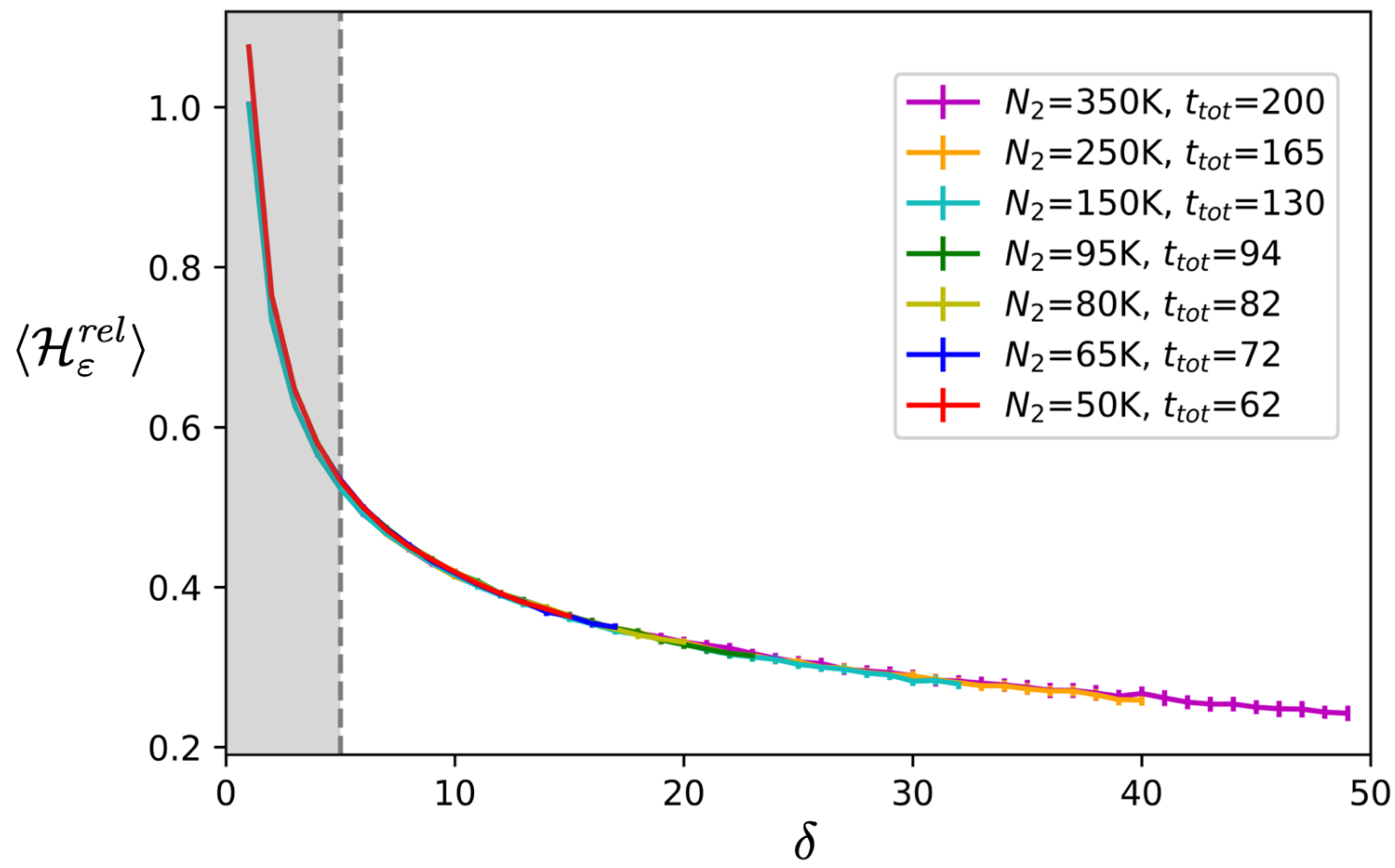

**Figure 4.2.:** Expectation value $\langle \mathcal{H}_\varepsilon^{rel}(\delta) \rangle$ of the relative homogeneity measure associated with the averaged deficit angle $\varepsilon$, as a function of the averaging scale $\delta$ in lattice units, for volumes in the range $N_2 \in [50k, 350k]$ and time extensions $t_{tot} \in [62, 200]$, including statistical error bars. The shaded area $\delta \leq 5$ indicates the region dominated by discretization artefacts.

### 4.1.1 Homogeneity of the deficit angle

4.1 We start by considering the deficit angle, $Q(x)=\varepsilon(x)$, which is a direct measure of the local curvature. For each fixed scale $\delta\in[1,49]$ we computed for each randomly sampled configuration $T$ of a CDT ensemble at fixed volume $N_2$ and time extension $t_{tot}$ the averaged deficit angle $\varepsilon(x,\delta)$ for geodesic balls of radius $\delta$ associated with a sphere covering. The expectation value $\langle\mathcal{H}^{rel}_{\varepsilon}(\delta)\rangle$ in the quantum ensemble for a range of different volumes and as a function of the link distance is shown in Fig. 4.2. Data points at integer values of $\delta$ have been joined up by straight line elements to guide the eye. Note that the curves end at different values of $\delta$, due to our choice $\delta<t_{tot}/4$. The initial region $\delta\leq 5$ is expected to be dominated by discretization artefacts. Our primary interest are the data points with $\delta>5$, which within error bars are seen to agree well for different volumes $N_2$ within their common $\delta$-ranges. For all lattice volumes, the homogeneity measure of the deficit angle is seen to decrease monotonically over the observed range, corresponding to an increase in homogeneity as a function of the averaging scale $\delta$, as one may have expected. However, since the relative homogeneity does not even fall to 24% (its value at $\delta=49$ is 0.242), this particular observable does not behave approximately homogeneously in the observed $\delta$-range.

To understand its behaviour in the continuum limit $N_2\to\infty$, we performed a finite-size scaling analysis by rescaling the lattice distances $\delta$ by the square root of the volume (Fig. 4.3, left), and plotted the data as a function of the rescaled, physical distance $\tilde{\delta}:=\delta/\sqrt{N_2}$.[1] The lack of overlap of the curves illustrates that the observable $\mathcal{H}^{rel}_{\varepsilon}$ is subject to strong short-distance lattice artefacts, extending also beyond $\delta=5$. Nevertheless, the presence of finite-size scaling can be exhibited by comparing the curves for sufficiently large $\delta$ and fixed physical distance $\tilde{\delta}$. When comparing data points at $\tilde{\delta}=0.065$, at the end of the range for which we have measurements from all volumes, we observe signs of convergence as a function of the volume $N_2$, as illustrated in Fig. 4.3, right. The interesting question is whether the homogeneity measure for sufficiently small $\tilde{\delta}\ll\tilde{L}$, where $\tilde{L}$ denotes the (rescaled) linear size of the spacetime, converges to a constant function in the continuum limit, and what the value of this constant is. We have fitted the seven data points at $\tilde{\delta}=0.065$ to functions $F(N_2)=c_0+c_1 N_2^{-\nu}$, for selected $\nu$ with $0<\nu\leq 1$. The fit shown is for $F(N_2)\approx 0.100 + 4.00 N_2^{-1/4}$. The fit for $\nu=1/2$ is of comparable quality and yields $c_0\approx 0.205$. Both of these would imply significant levels of inhomogeneity (of 10% and 20%) in the continuum limit. However, it should be kept in mind that the limited volume range of our data set does not allow us to discriminate between these different fall-off behaviours or to verify our simple ansatz for $F(N_2)$.

[1] We assume that the physical two-volume $V_2=a^2N_2$ is kept fixed as we scale $N_2\to\infty$, where $a$ denotes the dimensionful length of a lattice edge, such that $a\propto 1/\sqrt{N_2}$.

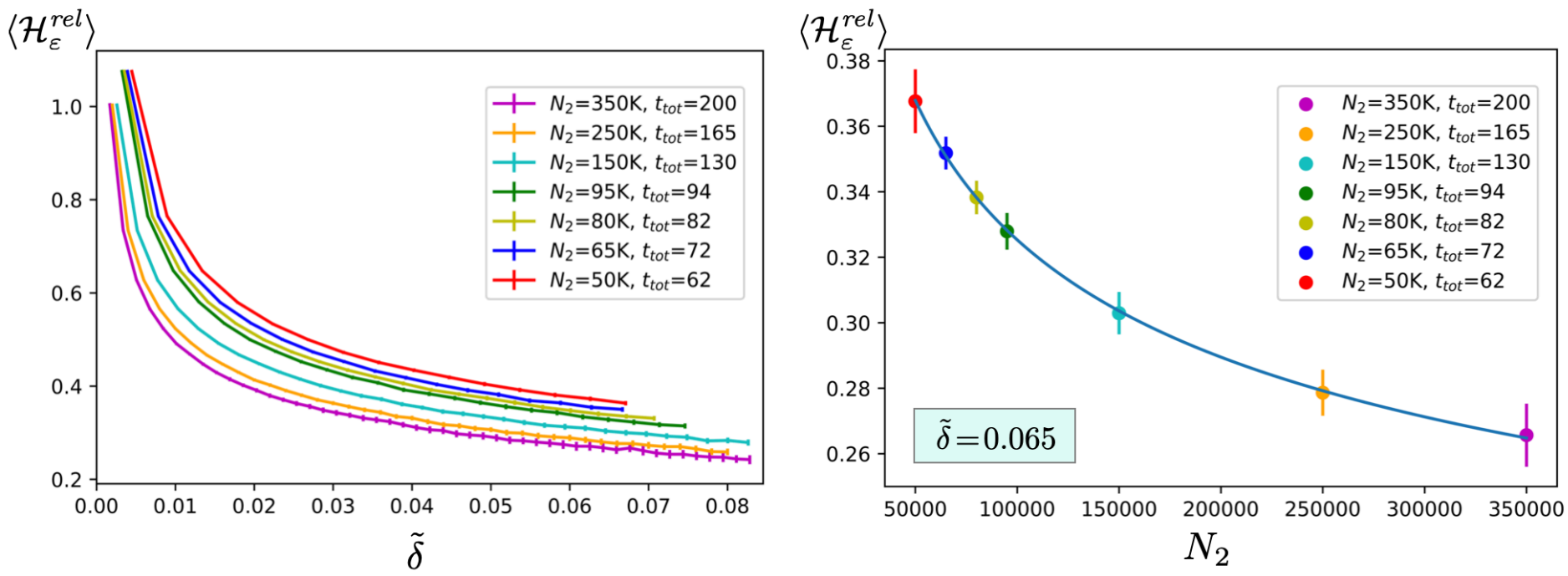

**Figure 4.3.:** Expectation value $\langle\mathcal{H}^{rel}_{\varepsilon}\rangle$ of the relative homogeneity measure associated with the averaged deficit angle $\varepsilon$, as a function of the rescaled averaging scale $\tilde{\delta}$ (left). Expectation value $\langle\mathcal{H}^{rel}_{\varepsilon}\rangle$ at fixed $\tilde{\delta}=0.065$, as a function of $N_2\in[50k, 350k]$, together with a curve fit (right).

We also do not have stringent theoretical arguments for how $\langle\mathcal{H}^{rel}_{\varepsilon}(\delta)\rangle$ should behave in the limit. Regarding the denominator of this expression according to (4.4), we have measured how the geodesic-ball average of the deficit angle behaves for $\delta\leq 100$, and found excellent agreement with a scaling $\propto 1/\delta$ toward zero from below. On the other hand, *if* the original distribution of the values $\varepsilon(x)$ on a given configuration $T$ could be treated as independent random events (which is strictly speaking not true, but may still be a good approximation), it would follow from the central limit theorem that samples of size $n$ approach a normal distribution with standard deviation $\propto 1/\sqrt{n}$. In the case at hand this would translate to a behaviour $\propto 1/\delta$, since the Hausdorff dimension of CDT configurations in the continuum limit is known to be $d_H=2$ [113, 114]. If this scaling scenario for the standard deviation was realized, it would mean that the relative homogeneity measure $\langle\mathcal{H}^{rel}_{\varepsilon}(\delta)\rangle$ for large $\delta$ approaches a constant, which is consistent with what we found in our measurements. However, without a more detailed understanding of the behaviour of the variance for large $\delta$, we do not know which value (including zero) this constant may have.

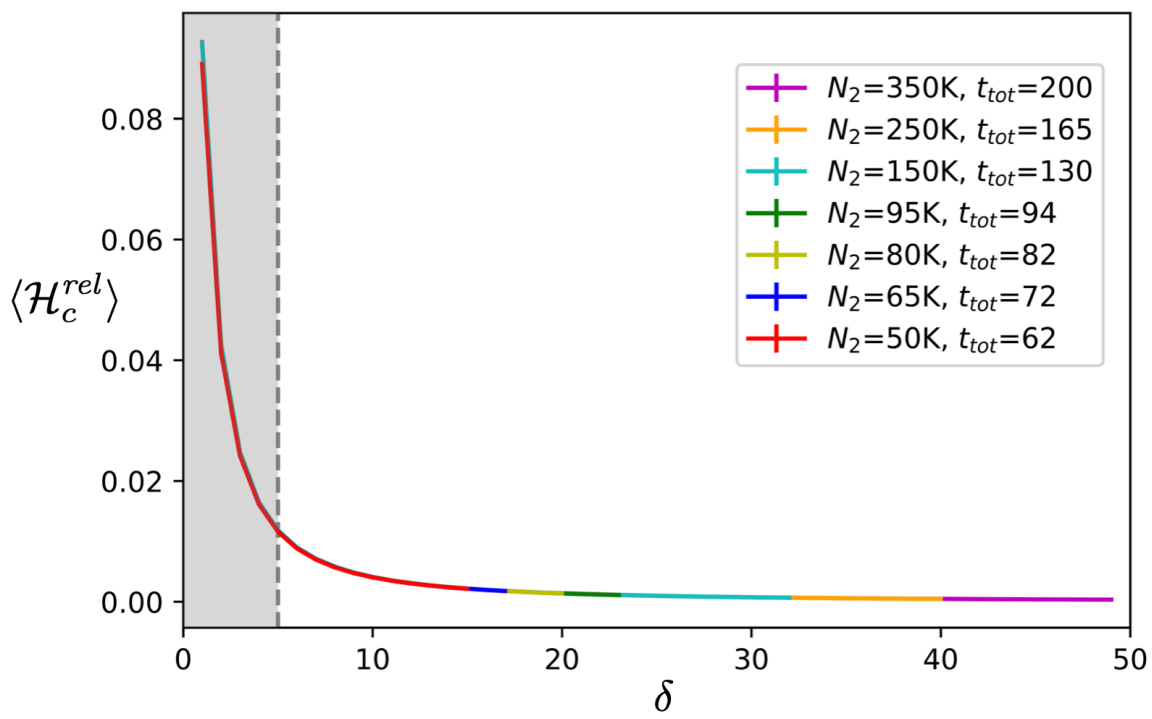

**Figure 4.4.:** Expectation value $\langle \mathcal{H}_c^{rel}(\delta) \rangle$ of the relative homogeneity measure associated with the averaged coordination number $c$, as a function of the averaging scale $\delta$ in lattice units, for volumes in the range $N_2 \in [50k, 350k]$ and time extensions $t_{tot} \in [62, 200]$. Error bars are too small to be shown. The shaded area $\delta \leq 5$ indicates the region dominated by discretization artefacts.

### 4.1.2 Homogeneity of the coordination number

Looking instead at the relative homogeneity measure of the coordination number, $Q(x) = c(x)$, avoids the potential problem of dividing by an average that is close to zero, since the geodesic-ball average of $c(x)$ will approach 6 for large $\delta$, by virtue of the global topological constraint. Repeating the same analysis as before yields the expectation values shown in Fig. 4.4. Their absolute scale is of course smaller than that associated with the deficit angle of Fig. 4.2, since we are dividing by a larger average. From the point of view of continuum physics, this absolute scale is really immaterial, since the coordination number as such does not have a continuum counterpart. Let us nevertheless proceed and accept the geodesic-ball average of $c(x)$ as an appropriate reference scale for use in the homogeneity measure. Since this average is well approximated by a nonvanishing constant $\approx 6$, what we are looking at here is – up to this overall constant – the absolute homogeneity measure (4.7) of the coordination number. Fig. 4.4 shows that there is a very close overlap between the data for different volumes, and a rather fast convergence toward a small expectation value as $\delta$ increases. We do not show the curves as a function of the rescaled distance $\tilde{\delta}$, which look similar to those for the deficit angle, with slightly less severe discretization effects. More interesting is the comparison of the expectation values at fixed $\tilde{\delta} = 0.065$ (Fig. 4.5, left), which indicates convergence to a small value near zero. A fit to $F(N_2) \approx 2.46 * 10^{-4} + 103 N_2^{-1}$ works well in this data range, subject to the same provisos as our earlier discussion of the deficit angle. We thus conclude from our measurements that in the $\delta$-range considered this particular observable attains a very high degree of relative homogeneity, in the sense of expectation values.

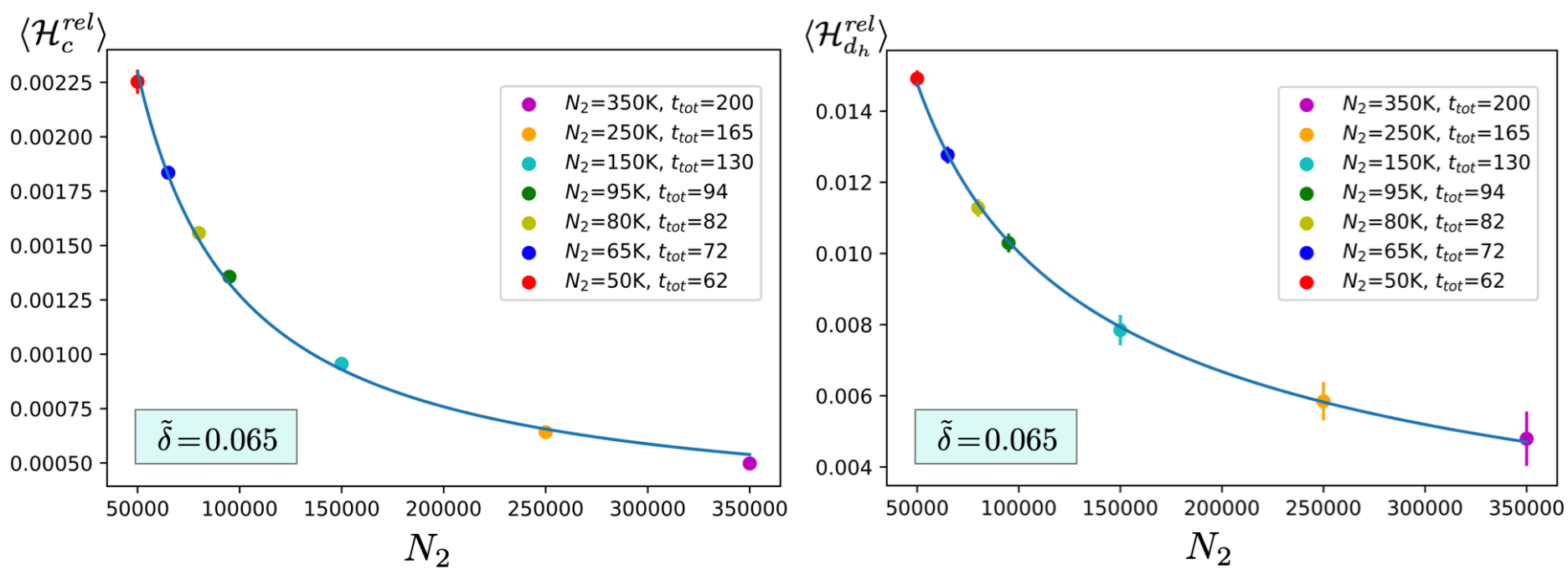

**Figure 4.5.:** Expectation value $\langle \mathcal{H}^{rel} \rangle$ of the relative homogeneity measures associated with the averaged coordination number $c$ (left) and the averaged Hausdorff dimension $d_h$ (right) at fixed $\tilde{\delta} = 0.065$, as a function of $N_2 \in [50k, 350k]$, together with curve fits.

A lesson we have learned so far is that the behaviour of the relative homogeneity depends on the quantity whose homogeneity is being assessed. It is ultimately a physics question which quantities can be meaningfully considered from a continuum point of view. Computational feasibility is also important, because of the dense sampling we perform of individual path integral configurations. A tentative interpretation for the behaviour of the expectation value $\langle \mathcal{H}^{rel}_c(\delta) \rangle$ associated with the coordination number is that the only inhomogeneities observed are confined to a region near the cut-off scale (which anyway we discard because of discretization artefacts), outside of which the homogeneity is almost perfect. This is the behaviour one would generically expect from any local, randomly distributed lattice property. In the case at hand, the homogeneity of the coordination number should probably be seen as a consequence of the unphysical nature of this local lattice property, rather than an indication that the quantum geometry of two-dimensional CDT is homogeneous. At any rate, we have already seen evidence that the behaviour of the relative homogeneity measure $\langle \mathcal{H}^{rel}_\varepsilon(\delta) \rangle$ associated with the deficit angle $\varepsilon$, a quantity more closely related to curvature, may be less trivial. Our measurements on lattice volumes up to $350k$ are compatible with a behaviour where the size of the fluctuations stays large compared to the mean value $\bar{\varepsilon}$, although larger lattices will be needed to corroborate this.

### 4.1.3 Homogeneity of the Hausdorff dimension

The next quantity whose homogeneity measure we will consider is not related to curvature and is given by the (local) Hausdorff dimension $d_h$. By definition, $d_h$ characterizes the growth of small geodesic spheres $\mathcal{S}_x(r)$ as a function of their radius $r$. In two-dimensional CDT quantum gravity, the Hausdorff dimension is known to be $d_h = 2$ analytically [113, 114], and has also been measured previously in numerical simulations [115, 116]. On a given triangulation $T$ of the CDT ensemble, the Hausdorff dimension $d_h(x)$ at a vertex $x \in T$ is extracted from the leading power law behaviour in $r$ of the sphere volumes $vol(\mathcal{S}_x(r))$, where $vol(\mathcal{S}_x(r))$ counts the number of vertices at link distance $r$ from $x$. We have used $r \in [6, 10]$ as a fitting range for the radius, just beyond the region of lattice artifacts. Our fitting function was of the form

$$vol(\mathcal{S}_x(r)) = c(x)\, r^{d_h(x)-1}, \tag{4.9}$$

for a local multiplicative constant $c$. It should be kept in mind that our primary aim is to establish homogeneity or otherwise of the *averaged* Hausdorff dimension, rather than a precision measurement of the Hausdorff dimension per se. Due to the rather complex determination of $d_h(x)$, which involves geometric data from an entire neighbourhood of radius 10 around $x$, it is rather costly computationally to extract the corresponding averaged quantities $d_h(x, \delta)$ according to the prescription (4.6), especially for large geodesic balls $\mathcal{B}_x(\delta)$. To keep the computational effort manageable, we opted for the relatively short fitting interval [6,10] and the simple fit function (4.9). We did not include an additive shift of $r$ in the fit, which is sometimes employed for optimization, after running some checks to make sure that the effect on our results is small.

Another feature worth noting is that because of the quasi-local nature of $d_h(x)$, the geodesic-ball averaged quantities $d_h(x, \delta)$ depend also on the geometry beyond $\delta$, more precisely, on data up to distance $\delta + 10$ from the central vertex $x$. This should be contrasted with the deficit angle and coordination number, whose homogeneity we discussed above, which are genuinely local lattice quantities. The quasi-local character of the Hausdorff dimension $d_h(x)$ is a characteristic feature of quantities with a more direct continuum interpretation; other examples are the spectral dimension and the quantum Ricci curvature. While the quasi-local nature should make little difference in any continuum limit $N_2 \to \infty$, how it affects our considerations of lattice versus physical effects at finite volume $N_2$ is a priori less clear. A possible effect of the quasi-locality could be that the curves of the homogeneity measures are shifted to some larger, "effective" $\delta$, compared to those for more local quantities. On the other hand, since the averaging over geodesic balls now effectively contains some overlap between neighbouring balls, this may lead to an increase in the apparent homogeneity.

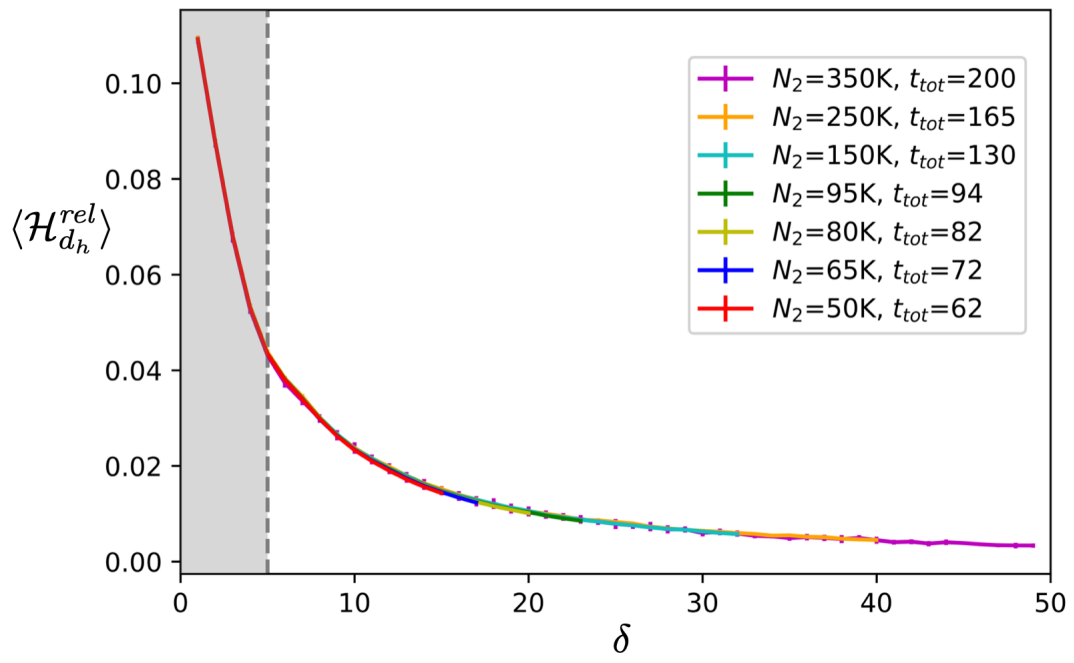

**Figure 4.6.:** Expectation value $\langle \mathcal{H}^{rel}_{d_h} \rangle$ of the relative homogeneity measure associated with the averaged Hausdorff dimension $d_h(\delta)$, as a function of the averaging scale $\delta$ in lattice units, for volumes in the range $N_2 \in [50k, 350k]$ and time extensions $t_{tot} \in [62, 200]$. The shaded area $\delta \leq 5$ indicates the region dominated by discretization artefacts.

Our measurement data for the expectation value of the relative homogeneity measure are shown in Fig. 4.6. We again observe a monotonic decrease as a function of the averaging scale $\delta$, like for the measures we investigated earlier. Since the Hausdorff dimension is a genuinely physical quantity, the data can perhaps best be compared to those of the homogeneity measure of the deficit angle. Compared to the latter, they indicate both a much greater degree of homogeneity and a faster fall-off for increasing $\delta$ in the case of the Hausdorff dimension. The spread of the curves as a function of the rescaled length scale $\tilde{\delta}$ is similar to that shown for the deficit angle in Fig. 4.3, left, but even bigger, indicating that the $\tilde{\delta}$-region with appreciable discretization effects is even larger. This is compatible with the expectation that the quasi-locality of the Hausdorff dimension leads to a larger "effective" $\delta$ than in the case of the more local deficit angle. The scaling with volume of the data at $\tilde{\delta}=0.065$ (Fig. 4.5, right) is within error bars compatible with an asymptotic approach to zero, using a fit with $\nu=1/2$, which leads to $F(N_2)\approx -1.42*10^{-3}+3.62N_2^{-1/2}$. Also the choice $\nu=1/4$ gives a fit of decent quality and yields $c_0\approx -0.012$. The difference between these $c_0$-values and the fact that both are negative illustrates the presence of systematic errors significantly larger than the statistical error bars, which are on the order of only 2%. We conclude that the Hausdorff dimension shows a highly homogeneous behaviour on CDT geometries, as quantified by the expectation value of our relative homogeneity measure (4.8). Taking into account the uncertainties present the outcome appears to be compatible with $\langle \mathcal{H}^{rel}_{d_h}\rangle(\tilde{\delta})=0$ in the continuum limit.

4.1

### 4.1.4 Some remarks about the constructed measures of homogeneity

Let us summarize what we have learned from implementing the relative quantum homogeneity measures in two-dimensional CDT quantum gravity. We performed Monte Carlo simulations for three different quantities and for finite system sizes up to $N_2=350.000$ triangular building blocks. We have focused on intermediate averaging scales $\delta$ above a lower cutoff $\delta=5$ and much smaller than the total linear system size, where a subsequent finite-size scaling analysis showed that the region where the homogeneity observables are affected by discretization artefacts extends beyond $\delta=5$. Taking this into account when interpreting the data, we can draw some interesting conclusions.

All three homogeneity measures are monotonically decreasing in $\delta$, but the details and their interpretation are different. The relative homogeneity measure of the coordination number $c$ falls off rapidly in $\delta$ to a value compatible with zero, and for fixed physical scale $\tilde{\delta}$ approaches this limit with a rate roughly proportional to $1/N_2$. However, we attributed this "perfect" statistically homogeneous behaviour to the unphysical nature of the coordination number, which does not have a direct continuum counterpart, rather than to the homogeneity of the underlying quantum geometry. The latter conclusion was supported by the fact that the relative homogeneity measure of the deficit angle $\varepsilon$, which does have a direct correspondence with Gaussian curvature, shows a different behaviour, with a slower fall-off in $\delta$. More importantly, our best fits from a finite-size scaling suggest an approach proportional to a fractional inverse power $N^{-\nu}$, with $\nu \sim 1/4 ... 1/2$, to a nonvanishing expectation value $\langle \mathcal{H}_{\varepsilon}^{rel} \rangle$ on the order of 10-20%, indicating a residual statistical inhomogeneity in the continuum limit. Lastly, the relative homogeneity measure of the Hausdorff dimension $d_h$ behaves rather similar to that of the deficit angle, including the rate of convergence to its continuum value. However, unlike what happens in the case of the deficit angle, this continuum value is zero within measuring accuracy. Because of the physical nature of the Hausdorff dimension, we interpret this as a genuine continuum result. The fact that the dimension is more homogeneous than the curvature may be attributable to the fact that the former is a more robust "pregeometric" property, while the inhomogeneity of the curvature may be a consequence of the pure quantum nature of two-dimensional quantum gravity, which does not have a classical limit. In particular, there is no nontrivial classical solution that the quantum geometry could converge to in an $\hbar \to 0$ limit.

Lastly, let us comment on the relation of our construction with a previous attempt to introduce homogeneity measures in CDT quantum gravity [117]. Its author works with variances that superficially resemble our absolute homogeneity measure (4.3), but without introducing a separate averaging scale (our "$\delta$"). Instead, the notion of scale is associated with the scale-dependence of the quantity whose variance is being investigated, which in our construction corresponds to the possible presence of another (linear) scale $r$, as mentioned just below eq. (4.1). There may of course be good reasons for doing so, but the resulting scale-dependent measures contain primarily information about the $r$-dependence of the underlying quantity (and its variance) rather than about the degree of homogeneity as a function of averaging. As we will illustrate by an explicit example in the Appendix C, the dependencies on $\delta$ and $r$ are in general different. One difficulty with interpreting $r$-dependent results is that the functional $r$-dependence of a generic quasi-local property of the quantum geometry in a non-perturbative approach like CDT is not known a priori.[1] If it were, it could be taken into account to try and extract the associated inhomogeneities from the variances proposed in [117].

[1] The two observables whose variance is studied in [117], the spectral dimension as a function of the diffusion time $\sigma$ and the volume of a geodesic ball of radius $r$, are cases in point.

## 4.2 Collecting the results of this chapter

Motivated by the search for quantitative measures of the degree of homogeneity of a quantum geometry at a given coarse-graining scale, we have introduced statistical homogeneity measures associated with particular quasi-local properties of quantum spacetime. By construction, these measures are very nonlocal and characterize the underlying quantum geometry in its entirety, through the statistics of the volume-averaged properties associated with a geodesic ball covering. This is reminiscent of the nonlocal character of the global symmetry transformations that are used to characterize a smooth classical geometry as homogeneous and isotropic.

Our application of the relative quantum homogeneity measures to two-dimensional quantum gravity led to interesting and apparently sensible results, demonstrating the viability of this new class of nonperturbative quantum observables. It also highlighted the dependence of the homogeneity or otherwise on the local property under consideration, and the importance of distinguishing between properties with a well-defined continuum interpretation and general discretization-dependent lattice properties. We do not claim uniqueness of our construction, and some of its elements can clearly be varied. The choice of how a geometry is covered by or divided into averaging regions is one example, although we do not expect it to influence results substantially, as long as one makes sure that the prescription is truly geometric and does not introduce an unwanted coordinate dependence or other biases.

However, there is another and more subtle issue that could influence the outcome of our homogeneity analysis, namely the choice of what we mean by "averaging". For good reasons, which were explained in Sec. 4 above, we have presently confined ourselves to studying local or quasi-local scalar quantities $Q(x)$. Although it is natural and simple to construct a coarse-grained quantity $Q(x,\delta)$ by taking a volume average of $Q(x)$ over a geodesic ball (or some other local region of linear size $\delta$), as we did in eq. (4.1), this is not necessarily a unique or the physically most appropriate choice. This has to do with the complex nature of local geometry, which in a continuum setting is captured by the tensorial character of the metric and curvature. For instance, a scalar may have a complicated, composite dependence on local curvature tensors and their covariant derivatives, as long as we make sure that all indices are contracted. However, an averaging of the individual components of a scalar $Q(x)$, however we define it, may not commute with the composition of its various components, leading to an ambiguity in what we may mean by "the averaged scalar $Q(x,\delta)$".

Related issues appear also for more general, nonsmooth geometries. For instance, the way in which we determined the quasi-local Hausdorff dimension $d_h(x)$ from the metric properties of a piecewise flat geometry in Sec. 4.1.3 involved a rather complex operation of fitting to sphere volumes centred at $x$, according to eq. (4.9). One way to determine an averaged $d_h(x,\delta)$ is to average $d_h(x)$ over the $\delta$-ball, as we did above, but one could also first average the sphere volumes over the $\delta$-ball and then extract an averaged Hausdorff dimension by fitting to these averaged volumes, which turns out to lead to a different result, with larger inhomogeneity. Whether a given way of averaging is appropriate is ultimately a question of how one operationally probes physics at the scale $\delta$, an issue that goes far beyond the scope of this paper. The only points we want to highlight here is that sticking to scalars is not necessarily a panacea, and that the way in which we implement averaging can in principle affect the outcomes of quantum homogeneity measures.

As already discussed in the introduction, a primary motivation for the present work and an important application is to assess the homogeneity properties of the de Sitter-like ground state of CDT quantum gravity in four dimensions. To extract reliable data in a sufficiently large $\delta$-interval will present computational challenges and may only be feasible for selected local properties. Whether approximately homogeneous behavior is present and whether its onset can be observed in a regime that is accessible computationally will have to be investigated. An application of the quantum homogeneity measures to other formulations of quantum gravity and other "quantum spacetimes", obtained dynamically or otherwise, should in principle be possible, given that our construction only relies on elementary metric ingredients. Of course, any actual implementation will hinge on the ability to perform (numerical) computations in a near-Planckian regime and the availability of well-defined local quantities whose quantum homogeneity measures can be evaluated.

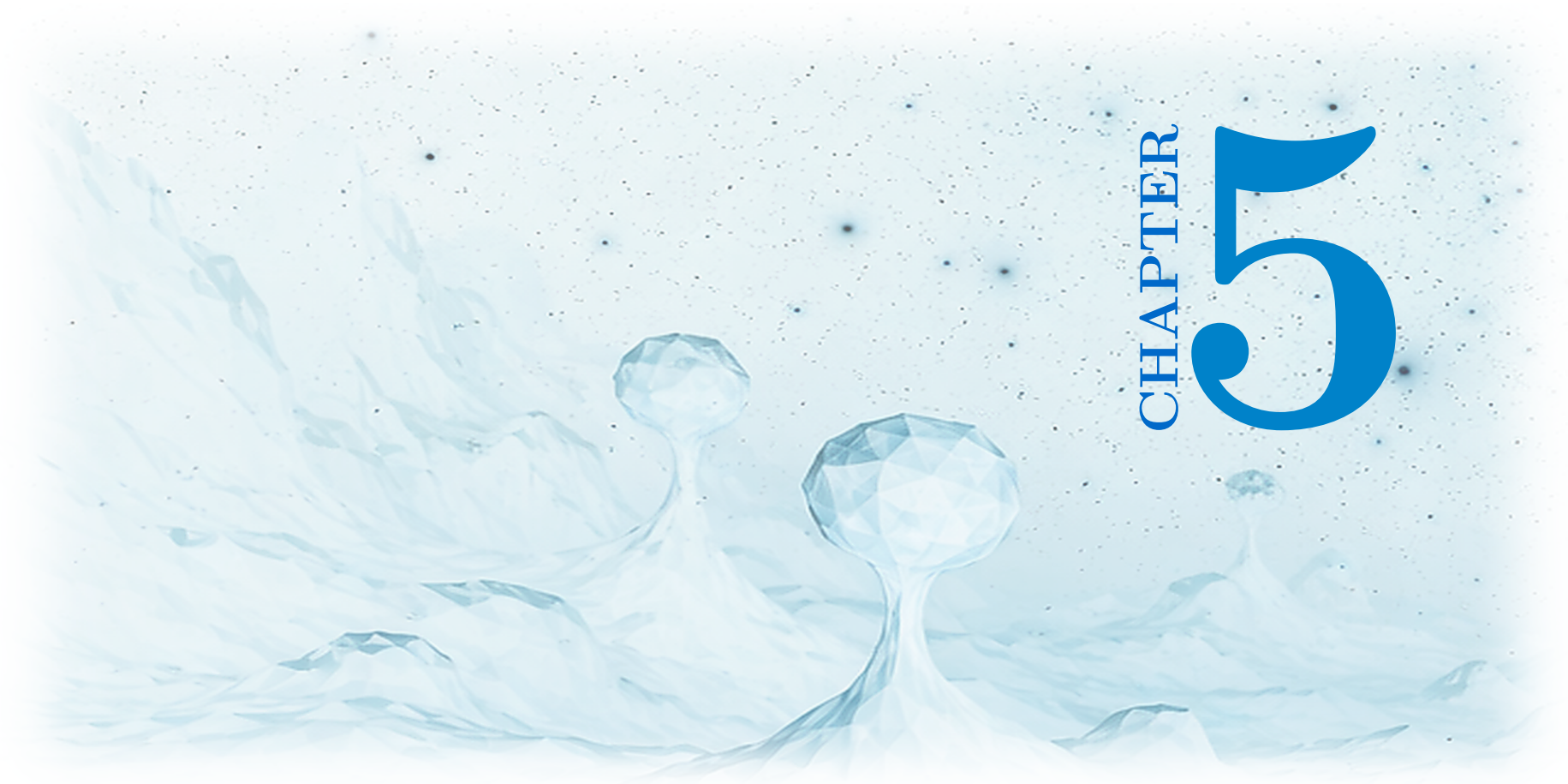

# Curvature as a geometric observable

The observables developed in the previous two chapters of this part were designed with the intention of establishing the emergence of symmetries of suitable geometrical properties of quantum geometries. The topological observables are a $0th$-order check that one must perform on a quantum geometry to see if its possibly "foamy" structure disappears at some scale, giving a topologically trivial spacetime at larger scales. If the quantum spacetime satisfies this condition, one can study its statistical inhomogeneities with the observables developed in the previous chapter.

Due to the non-perturbative nature of the studied geometries, it is difficult to come up with quantities whose continuum limit is interesting to study. In the previous chapter, we discussed the inhomogeneities of simply accessible quantities, like coordination numbers, deficit angles, and some slightly more sophisticated quantities like a local notion of the Hausdorff dimension. It would of course be more interesting to study the properties of a more familiar property of geometries: its curvature.

While the deficit angle in two dimensions is directly related to curvature and easily computable, it suffers from divergences when taking the continuum limit in higher dimensions. Since the goal of this thesis is to develop tools that could be applied in more general settings than the toy model of 2D CDT, we will attempt to define efficient, easily accessible, coarse-grained notions of local scalar curvature. More formally, in this chapter, we define coarse-grained notions of scalar curvature in metric spaces endowed with a volume measure or a random walk. In particular, our notions are well-suited for discrete spaces, like simplicial manifolds or graphs.

The notion of curvature is fundamental to the discussion of spacetime in the framework of General Relativity. To study the geometry of the spacetimes appearing in non-perturbative quantum gravity, one can expect curvature to be an important quantity to understand [118]. However, studying curvature in the discrete setting of simplicial manifolds is challenging, because the metric of the simplicial manifolds is not smooth. To be able to study curvature in discrete approaches to quantum gravity, like CDT, different methods have to be considered. In this chapter, we introduce generalizations of the Ricci scalar curvature based on the volume of spheres and the return probability of a random walker, that are especially suited for numerical computation. These generalized definitions of scalar curvature are motivated by their use in CDT, but can also be applied to other discrete settings, like a graph.

Previously, several notions of generalized curvature in simplicial manifolds have been proposed. A particular one of them is the notion of curvature proposed by Ollivier [119]. There, the author defined the Ricci curvature tensor by using the transport distance between two geodesic spheres. In this chapter, we apply Ollivier's idea of defining curvature indirectly, not from taking derivatives of the metric tensor, but instead, by computing well-defined quantities in metric measure spaces, that depend on the Ricci scalar in a Riemannian manifold. In [112], an adaptation of Ollivier's idea is made and is successfully applied to CDT, obtaining promising results. They do this by making a generalization of the Ricci tensor, called by the authors the Quantum Ricci Curvature (QRC), which is proportional to the Ricci curvature when applied to a Riemannian manifold. The QRC offers a computationally feasible, but expensive calculation in metric measure spaces; it requires the computation of the distance between all pairs of points of two geodesic spheres.

The idea behind the notions of curvature presented in [119] and [112] is to compute a quantity that depends on the Ricci curvature in a Riemannian manifold, and use this quantity to define curvature in a more general setting. These are particular cases of the general idea of: instead of defining curvature directly, by taking derivatives of the metric tensor, it is possible to define curvature indirectly, by computing quantities that depend on the curvature in a Riemannian manifold, that only rely on a distance function and some additional structure, like a volume measure or a diffusion process. This indirect/implicit definition of curvature is the method we will use in the following to establish some new ways to determine the scalar curvature of metric spaces.

Our method to define scalar curvature is similar in nature to the QRC, so, it is worth mentioning some of its features. More specifically, the QRC is defined through the average sphere distance $\bar{d}\left(S_p^\delta, S_{p'}^\delta\right)$, between a pair of spheres with radius $\delta$, centred at points $p$ and $p'$, which are themselves separated by geodesic distance $\delta$. The average sphere distance $\bar{d}\left(S_p^\delta, S_{p'}^\delta\right)$ is the average distance between pairs of points, one in each sphere [112].

In general, the outcome of the calculation of the average sphere distance is normalized by the scale $\delta$, and is parameterized as

$$\frac{\bar{d}\left(S_p^\delta, S_{p'}^\delta\right)}{\delta} := C\big(1 - K(p, p', \delta)\big). \tag{5.1}$$

Here the quantity $K(p, p', \delta)$ defines the *quantum Ricci curvature*, and it captures any deviation from constancy as a function of $\delta$. The prefactor $C$ is a positive real value, defined by $C := \lim_{\delta\to 0} \bar{d}\left(S_p^\delta, S_{p'}^\delta\right)/\delta$. This prefactor can be shown to be point independent when considering $\bar{d}\left(S_p^\delta, S_{p'}^\delta\right)$ on a smooth Riemannian manifold, only depending on the dimension of the manifold. However, when measured on a lattice, the limit of the definition has to be substituted with an appropriate construction on the lattice (see [112, 120] for details). On some regular lattices, the prefactor $C$ has been found to be a constant depending on the choice of lattice [112]. On a general lattice $C$ could in principle depend on the points $p$ and $p'$, but previous investigations [112] have demonstrated that some meaningful results can be obtained in more general lattices assuming $C$ to be point independent. The QRC $K(p, p', \delta)$ defines a notion of (coarse-grained) curvature, including directional information. It can be shown that in a Riemannian manifold it can be expanded in powers of curvature invariants, containing at the lowest order the Ricci tensor in the direction joining the two points $p, p'$ [112]. The point-pair $p, p'$ plays the role of the vector indicating this direction. Any deviation from $C$ in (5.1) can be interpreted as the existence of non-zero curvature of the space. Since it is out of the scope of this work to fully introduce the QRC, we refer the reader to the recent review [118].

Crucially, this way of capturing curvature has the disadvantage that the prefactor $C$ is in general unknown. This is specifically an issue in cases where analytical calculations of the QRC are out of reach, and one has to rely on numerical methods. To determine the value of $C$ in the QRC, one takes the smallest $\delta$ values accessible in the measurements, in CDT this is typically around $\delta = 5$, where one expects lattice artefacts to disappear for $5 \leq \delta$. But, if there are other agents influencing the numerical result, like discretization artefacts or some other noise, one would never know if the resulting number is close or far away from the actual value of $C$, if there is any, since these artefacts usually dominate the small $\delta$ region, and therefore it is hard to be conclusive in the results. Furthermore, there has been no clean way of disentangling the prefactor $C$ from the quantum Ricci curvature $K$, at least in CDT applications, so one only has access to the average sphere distance measurements.

The motivation for this discussion are correlation functions, which are fundamental objects of quantum field theory. As such, it is interesting to study these for curvature in the case of quantum gravity.

Using the average sphere distance (5.1) to compute the correlation functions has the problem that the correlations of the prefactor $C$ are in general mixed with the correlations of the QRC, and makes it more difficult to interpret the correlations. We will later explain how our definitions of curvature somehow bypass this problem by eliminating the presence of any point-dependent multiplicative prefactor, providing definitions of curvature where correlations of the obtained quantities might be easier to interpret than those of using (5.1).

The structure of this chapter is the following. In section 5.1 we construct two methods to define generalized notions of scalar curvature from scratch: one based on sphere volumes and one based on return probabilities. In section 5.2 we generalize this construction of a scalar curvature definition, and show how these methodologies still work for more general settings. In section 5.3 we implement our definitions for triangulations of the sphere and the plane and discuss the applicability of the different methods to extract curvature. In section 5.4 we apply our generalized scalar curvatures to the quantum geometries of 2D CDT. Finally, in section 5.5 we present our conclusions and discuss them.

## 5.1 Generalized scalar curvatures

We begin with an introduction to our generalized scalar curvatures by explicitly constructing them from scratch for two different types of spaces: metric measure spaces, and metric spaces with a random walk.

To this end, we start by introducing two scale-dependent quantities, which can be computed in a metric space, endowed with a measure or a random walk. The important property is that these quantities give the Ricci scalar curvature at the lowest order of a curvature expansion, if computed in a Riemannian manifold. We will call these quantities Generalized Scalar Curvatures.

First we give an example for a metric space endowed with a volume measure. Recall, a Metric Measure Space (MMS) [121], denoted by $(\mathcal{X}, d, \mu)$, is a mathematical space $\mathcal{X}$ that has a distance function $d$ and a measure $\mu$. A sphere is defined as $\mathcal{S}_p^r = \{p' \in \mathcal{X} : d(p,p') = r\}$, with its volume given by $\left\|\mathcal{S}_p^r\right\| \doteq \mu(\mathcal{S}_p^r)$. Consider the case where $\mathcal{X}$ is a $D$-dimensional Riemannian manifold, $d$ is the geodesic distance function, and $\mu$ is the invariant integration measure. Then, $\left\|\mathcal{S}_p^r\right\|$ has the expansion in $r$ [122]

$$\left\|\mathcal{S}_p^r\right\| = \frac{D\,\pi^{\frac{D}{2}}}{\Gamma\left(\frac{D+2}{2}\right)} r^{D-1}\left(1 - \frac{\mathcal{R}_p}{6D} r^2 + \mathcal{O}(r^4)\right), \tag{5.2}$$

where $\mathcal{R}_p$ is the Ricci scalar[1] at the point $p$. This expansion can be used to define the scalar curvature as the first scale-dependent correction, in this case radius-dependent. To construct general definitions of curvature, we manipulate the sphere volume, and define the following (scale-dependent) Generalized Scalar Curvatures $(\mathcal{SC})$[2]

$$\begin{aligned}
\mathcal{SC}_1\left(\left\|\mathcal{S}_p^r\right\|\right) &:= r\frac{\partial}{\partial r}\left(\frac{\partial \log \left\|\mathcal{S}_p^r\right\|}{\partial \log r}\right), \quad \textit{and} \\
\mathcal{SC}_2\left(\left\|\mathcal{S}_p^r\right\|\right) &:= \frac{1}{\left(\int_0^r dx\, x\,\right)} \int_0^r dx\, x \left(\frac{\partial \log \left\|\mathcal{S}_p^r\right\|}{\partial \log r} - \frac{\partial \log \left\|\mathcal{S}_p^x\right\|}{\partial \log x}\right).
\end{aligned} \tag{5.3}$$

[1] We denote the Ricci scalar with $\mathcal{R}$ instead of the commonly used $R$ to avoid confusion with the sphere radius $R$, which will be frequently used later.

[2] We will differentiate between the different methods to extract the scalar curvature with subscripts $\{1, 2\}$. We will use $\mathcal{SC}_i$ to refer to either of them.

In a Riemannian manifold, these two quantities can be expanded at small distance scales, obtaining

$$\mathcal{SC}_1(\|\mathcal{S}_p^r\|) = -\frac{2\mathcal{R}_p}{3D}r^2 + \mathcal{O}(r^4) \qquad \mathcal{SC}_2(\|\mathcal{S}_p^r\|) = -\frac{\mathcal{R}_p}{6D}r^2 + \mathcal{O}(r^4). \quad (5.4)$$

Notice that they recover the Ricci scalar in the limit $r \to 0$, up to a factor $-\frac{2}{3D}r^2$ and $-\frac{1}{6D}r^2$ respectively. This is why we call $\mathcal{SC}_1(\|\mathcal{S}_p^r\|)$ and $\mathcal{SC}_2(\|\mathcal{S}_p^r\|)$ generalized scalar curvatures.

Using these manipulations, we can generalize the notion of scalar curvature to any metric measure space, and $\mathcal{SC}_1(\|\mathcal{S}_p^r\|)$ or $\mathcal{SC}_2(\|\mathcal{S}_p^r\|)$ can be used as a measure of the coarse-grained scalar curvature at that point. Of course, for large values of $r$ the interpretation of the coarse-grained curvatures $\mathcal{SC}_1(\|\mathcal{S}_p^r\|)$ or $\mathcal{SC}_2(\|\mathcal{S}_p^r\|)$ can be more difficult due to possible higher-order curvature corrections. However, in an appropriate short-scale range the interpretation is straightforward. Note that, in cases where $r$ is a discrete variable and/or $\|\mathcal{S}_p^r\|$ is not defined for every continuum value of $r$, a smooth continuation is understood in the definitions (5.4). Discrete implementations of the derivatives and integrals are needed in explicit calculations. We discuss such implementations for the discrete examples we consider in the following sections.

Let us use the previous technique to define curvature from a different perspective. One can use similar methods to obtain a scalar curvature definition using the return probability density of a diffusion process. Diffusion processes can be constructed from Markov chains, where the probability of the next event in the chain depends only on the current event of the chain. More specifically, one can model a diffusion process on a metric space by defining a Random Walk in it [123]. The probability density of the random walker to move from point to point, is equivalent to the probability density of a particle that is being diffused on the space itself [124].

More formally, a Metric Space with a Random Walk (RWMS), denoted by $(\mathcal{X}, d, \{X_p\})$, is a mathematical space $\mathcal{X}$ that has a distance function $d$ and a set of random variables $\{X_p\}$ for each point $p$ in $\mathcal{X}$, where each random variable $X_p$ has to be integrable, and have a finite first moment [119, 125].

Since our main goal is to apply our techniques to discrete spaces, we will restrict our discussion to discrete Random Walks. In this scenario, each $X_p$ is a random variable describing the displacement from point $p$ to the next point in the random walk. For example, in a square 2D lattice of unit edge length, and for a random walker moving only to nearest neighbours of the lattice, if $p = \{0,0\}$, then $X_p \in \{\{1,0\}, \{0,1\}, \{-1,0\}, \{0,-1\}\}$. If the random walk is non-biased, then the probability distribution of $X_p$ is uniform over the points $\{\{1,0\}, \{0,1\}, \{-1,0\}, \{0,-1\}\}$.

Going back to a general discrete case, we define the "walking time" $n$, as the number of steps taken for the random walker since its starting point. In this setting, we will denote the probability of the random walker to go from $p$ to $p'$ as $P^n_{p,p'}$, and the probability that the walker returns to $p$ after time $n$, called the return probability, simply as $P^n_p = P^n_{p,p}$.

We would like to relate this to continuum manifolds, in order to define some notion of curvature. A formal identification of $P^n_{p,p'}$ with a discrete version of the Heat Kernel $K(p,p';\tau)$ is given in [124] (see Chapter 1). For this identification to hold, the relation between the number of steps of the random walker $n$ and the diffusion time $\tau$ of the Heat Kernel has to be linear, i.e. $n \propto \tau$. The proportionality constant has to be determined in each setting, and will be of importance in this work.

In particular, one can identify the return probability $P^n_p$ with the diagonal part of the Heat Kernel, by doing

$$P^n_p = K(p,p;\tau). \tag{5.5}$$

This identification has been widely used to study discrete spaces with unknown properties, such as their spectral dimension, see [126–131] for details and examples.

We use this correspondence between a random walk and a diffusion process to construct a definition of scalar curvature in a RWMS. Note that we will not assume that every metric space with a random walk satisfies this identity, and we only use it to show that our calculations recover the Ricci scalar in the case of a Riemannian manifold, with the previous diffusion conditions satisfied.

For this construction, note that in a D-dimensional Riemannian manifold, the heat kernel $K(p,p;\tau)$ has the well-known expansion [132]

$$K(p,p;\tau) = \frac{1}{(4\pi\tau)^{\frac{D}{2}}}\left(1 + \frac{\mathcal{R}_p}{6}\tau + \mathcal{O}(\tau^2)\right). \tag{5.6}$$

To this expression, we apply several operations to extract the scalar curvature, analogous to the construction used for the sphere volume. With this in mind, we define the Generalized Scalar Curvatures in a RWMS as

5.1

$$\mathcal{SC}_1(P_p^n) := n\frac{\partial}{\partial n}\left(\frac{\partial \log P_p^n}{\partial \log n}\right), \textit{and}$$
$$\mathcal{SC}_2(P_p^n) := \frac{1}{\left(\int_0^n dx\, x\right)}\int_0^n dx\, x\left(\frac{\partial \log P_p^n}{\partial \log n} - \frac{\partial \log P_p^x}{\partial \log x}\right). \tag{5.7}$$

In the cases where $n$ is a discrete variable, one is to understand this definition as an analytical continuation in $n$. We will return to the discretization of these definitions in the following sections. Note that on a Riemannian manifold, and using (5.5), (5.7) takes the form

$$\mathcal{SC}_1(P_p^n) = \frac{\mathcal{R}_p}{6}\tau + \mathcal{O}(\tau^2) \qquad \mathcal{SC}_2(P_p^n) = \frac{\mathcal{R}_p}{18}\tau + \mathcal{O}(\tau^2) \tag{5.8}$$

So we see that $\mathcal{SC}_1(P_p^n)$ and $\mathcal{SC}_2(P_p^n)$ recover the Ricci scalar in the limit of $n \to 0$, up to a factor $\frac{\tau}{6}$ and $\frac{\tau}{18}$ respectively.

Summarizing, the defined coarse-grained scalar curvatures $\mathcal{SC}_1(P_p^n)$, $\mathcal{SC}_1(||\mathcal{S}_p^r||)$, $\mathcal{SC}_2(P_p^n)$ and $\mathcal{SC}_2(||\mathcal{S}_p^r||)$ recover the Ricci scalar for a Riemannian manifold at small scales. In a general metric space, we can use these quantities to define a scale-dependent curvature scalar. Note that they are not fundamentally different, but just different methodologies to compute the same curvature scalar in the limit of the distance scales going to 0. For example, these scalar curvatures can be used in a graph where one uses the link distance as distance function, and the number of nodes in a set as the volume measure. In this space, sphere volumes can be measured with discrete radius $r$ and can be used to compute the scalar curvature at every point. Additionally, a random walk can be implemented to determine scalar curvature equivalently but based on a different computation. An advantage of these definitions, is that they are scale dependent. $\mathcal{SC}_1(||\mathcal{S}_p^r||)$ and $\mathcal{SC}_2(||\mathcal{S}_p^r||)$ depend on the radius of a sphere and $\mathcal{SC}_1(P_p^n)$ and $\mathcal{SC}_2(P_p^n)$ depend on the diffusion step of the random walker. This scale dependence can be used to renormalize the curvature by taking the continuum limit of the discrete theory, as is done in lattice theories of quantum gravity like CDT. In CDT the path integral over the gravitational degrees of freedom is regularized as a sum over discrete spacetimes, and a continuum limit is taken by rescaling all distances and taking the infinite-volume limit. Our definitions are perfectly suited for these discrete spacetimes, as there is no notion of tensor calculus defined on them.

Notice that if one wants to extract the precise dimensionful value of the scalar curvature $\mathcal{R}_p$ from (5.7), one has to know the exact linear relation between $n$ and $\tau$. To the best of our knowledge, this relation is not known in a general situation. Despite that, one can approximate this relation by the well-known, and widely used [125] identity obtained for a Gaussian random walk in a Euclidean space

$$\mathrm{Var}\,(S_n) = \sigma^2 n = 2D\eta\tau, \tag{5.9}$$

where $\sigma^2 = Var(X_p)$ is the variance of the random variables describing each step in the random walk, $S_n = \sum_i^n X_i$ is the random variable given by the sum of the steps up to time $n$, $\mathrm{Var}\,(S_n)$ is its variance, $D$ is the topological dimension of the space, and $\eta$ is the diffusion constant of the process. Using this identification, one can obtain a numerical value for $\mathcal{R}_p$ from (5.7).

Even though our implementations for a random walk will not be in a Euclidean space, and do not use Gaussian variables, we will still use identification (5.9) to determine dimensionful numerical values of the curvature, $R_p$. We will return to this in the next sections.

## 5.2 General scale-dependent quantities

Based on the constructions presented in the previous section, in this section we generalize the two methodologies ($\mathcal{SC}_1$ and $\mathcal{SC}_2$) to extract the Ricci scalar from a more general class of scale-dependent quantities, which, if computed in a Riemannian Manifold, contain curvature-dependent corrections. We will show how $\mathcal{SC}_1$ and $\mathcal{SC}_2$ still work for this more general class of quantities, even in the case where the parameters describing these quantities are random variables, and if one uses discrete notions of derivatives and integrals.

Specifically, we consider the class of quantities $\mathcal{Q}_p^s$, that take the functional form

$$\mathcal{Q}_p^s = C_p s^{\alpha_p}(1 + \mathcal{K}_p(s)), \tag{5.10}$$

where $p$ is a point on the space, $s$ is a distance[1] and $C_p, \alpha_p \in \mathbb{R}$ are scale independent, but potentially point dependent. In the cases of the quantities $\mathcal{Q}_p^s$ we will look into, if evaluated in a Riemannian manifold, a non-zero $\mathcal{K}_p(s)$ implies non-zero curvature of the manifold. Examples of such quantities $\mathcal{Q}_p^s$ are: the volume of a geodesic ball or sphere, the average distance between point pairs in a geodesic ball or sphere, or the trace of the heat kernel of the covariant Laplacian. In these cases, the scale $s$ is given by the sphere/ball radius $r$, or by the diffusion time $\tau$. For all of these quantities $\mathcal{K}_p(s) = 0$ in a flat manifold, but in a general Riemannian manifold this is not necessarily the case.

The reason for considering these types of quantities is that our main goal is to apply our methods $\mathcal{SC}_1$ and $\mathcal{SC}_2$ to the quantum geometries that appear in CDT. Since these geometries are irregular simplicial manifolds, and fractal in nature, one cannot expect that at every point $p$ volumes and return probabilities (such as $\left\|\mathcal{S}_p^r\right\|$ and $P_p^n$) behave in the same manner. This is why we generalized the quantities to be studied to have the form (5.10), allowing for point-dependent $C_p, \alpha_p \in \mathbb{R}$. In the following sections of this work, we will assume that $\left\|\mathcal{S}_p^r\right\|$ and $P_p^n$ can be cast within the form (5.10), to be able to interpret the measured $\mathcal{SC}_i$.

The idea is to extract $\mathcal{K}_p(s)$ from quantities of this type, without having to worry about what $C_p$ and $\alpha_p$ are, and use our methods to define a notion of coarse-grained scalar curvature that is still valid in the geometries obtained in CDT.

Following the previous section, we compute each Generalized Scalar Curvature applied to $Q_p^s$

$$\begin{aligned} \mathcal{SC}_1(Q_p^s) &:= s\,\frac{\mathrm{d}}{\mathrm{d}s}\left(\frac{\mathrm{d}\log Q_p^s}{\mathrm{d}\log s}\right), \quad \textit{and} \\ \mathcal{SC}_2(Q_p^s) &:= \frac{1}{\left(\int_0^s dx\, x\right)} \int_0^s dx\, x \left(\frac{\mathrm{d}\log Q_p^s}{\mathrm{d}\log s} - \frac{\mathrm{d}\log Q_p^x}{\mathrm{d}\log x}\right), \end{aligned} \tag{5.11}$$

keeping the point $p$ fixed.

Since $\log Q_p^s = \log C_p + \alpha_p \log s + \log\left(1 + \mathcal{K}_p(s)\right)$, one can show that due to the distributive properties of the derivative and the integral, that hold also in an appropriate discrete implementation, $\alpha_p$ and $C_p$ cancel out from the expressions of $\mathcal{SC}_1$ and $\mathcal{SC}_2$, and any deviation from 0 is associated to the presence of curvature.

[1] For example, this distance can be the geodesic radius of a sphere $r$, or the diffusion time $\tau$ of a diffusion process as we used in the previous section.

It is essential to compute (5.11) at a fixed point $p$, otherwise the cancellation of the (possibly random) variables $C_p$ and $\alpha_p$ does not happen. In fact, if one does this calculation after an average of $Q_p^s$ at different points, one cannot guarantee that the cancellation occurs, and a bias due to the possible correlation of the point-dependent variables would appear. In what follows in this chapter, we will compute these quantities at each point of the spaces being studied. The discrete implementations will be discussed in the following sections.

The interpretation of the presence of curvature in $\mathcal{SC}_1(Q_p^s)$ and $\mathcal{SC}_2(Q_p^s)$ applies to the quantities discussed in the previous section, $\left\|\mathcal{S}_p^r\right\|$ and $P_p^n$. In case we work in a metric measure space using the sphere volume, we use $\left\|\mathcal{S}_p^r\right\|$ as $\mathcal{Q}_p^s$ where $r$ is used as $s$. In the case of a metric space endowed with a random walk, we use $P_p^n$ as $\mathcal{Q}_p^s$ where $\tau = \frac{n\sigma^2}{2D\eta}$ is used as $s$, assuming (5.9) to hold.

Using the generalized scalar curvatures based on these quantities, we can define generalized notions of scalar curvature, $\mathcal{SC}_i$, on a metric space. And, we can interpret deviations from 0 as the presence of curvature[1].

The advantage of using $\left\|\mathcal{S}_p^r\right\|$ and $P_p^n$ to define coarse-grained curvatures, is that the continuum expressions in a Riemannian manifold, or at least expansions at small distance scales, are known. So positive or negative deviations from 0 at small distance scales can be directly related to the *sign* of the scalar curvature of the manifold.

Furthermore, it is useful to compare $\mathcal{SC}_i$ for all length scales $s$, not just in the limit of small scales, to the results in a constantly curved Riemannian manifold. For a $D$-dimensional constantly curved Riemannian manifold, this can be done for the quantity $\left\|\mathcal{S}_p^r\right\|$. For the return probability this is unfortunately impossible, because so far only asymptotic expansions are known for the heat kernel. The full continuum expression for each $\mathcal{SC}_i(\left\|\mathcal{S}_p^r\right\|)$ on a $D$-dimensional sphere and a hyperbolic $D$-space can be found in appendix D. Both expressions depend on 2 parameters, the dimension $D$ and the radius of the space $R$.

[1] After statistical noise and lattice artifacts are taken into account.

By comparing the results of computing $\mathcal{SC}_i(||\mathcal{S}_p^r||)$ in a given space with the results of appendix D, one can determine the resemblance to a sphere or a hyperbolic space. Furthermore, if there is a strong resemblance between the measured $\mathcal{SC}_i(||\mathcal{S}_p^r||)$ and the plot D.1 for a constantly curved Riemannian manifold, an effective radius can be assigned, by fitting the measured $\mathcal{SC}_i(||\mathcal{S}_p^r||)$ to the formulas given in the appendix.

## 5.3 Implementations and examples of use

The intended use of the defined $\mathcal{SC}_i$ is to evaluate scalar curvature in discrete metric spaces. In this section we define a discretization of (5.11) and evaluate it on triangulations of two-dimensional spaces with constant curvature, the two-sphere and the two-dimensional plane. These evaluations serve as a consistency check for the $\mathcal{SC}_i$, by comparing the measured curvature of the triangulations based on $\mathcal{SC}_i$, with the known curvature of the smooth two-dimensional space the triangulations are approximating. This allows us to evaluate the discretization artefacts and other limitations of our definition in a controlled setting.

### 5.3.1 Triangulation construction

To construct triangulations of the plane and the sphere, we will make use of so-called Delaunay triangulations of the smooth geometries. In order to do so, we consider an embedding of the two-sphere and the plane in a three-dimensional Euclidean space, and we will restrict ourselves to discussing Delaunay triangulations in this setting. Delaunay triangulations of a surface embedded in Euclidean space, are triangulations of a sample set of points contained in the surface, that have the property that no point in the sample is contained inside the circumcircle of any triangle of the triangulation [133]. In this work, we construct the sample set of points by using Poisson Disk sampled points [94]. Note that to create a triangulation of the plane we sample points in a square region of the plane, where we periodically tile these square regions to fill the plane, associating opposing sides. The distance function used to create the sample of points is the induced metric on the surfaces (plane and sphere) from the embedding in Euclidean space. Denoting the spaces being triangulated (plane and sphere) by $M$, we will denote a Delaunay triangulation of it as $\mathcal{D}_{\mathcal{T}}(M)$. The resulting Delaunay triangulation $\mathcal{D}_{\mathcal{T}}(M)$ is a random geometry, in the sense that its vertices are randomly sampled.

In each of the Delaunay triangulations we construct the metric spaces $(P(\mathcal{D}_{\mathcal{T}}(M)), d_l)$ and $(F(\mathcal{D}_{\mathcal{T}}(M)), d_d)$. Here $P(\mathcal{D}_{\mathcal{T}}(M))$ denotes the set of points constituting the vertices of the triangulation, together with the link distance $d_l$ as the distance function, which is the minimal number of links between two vertices in the triangulation; very similar to the construction used in [112]. $F(\mathcal{D}_{\mathcal{T}}(M))$ denotes the set of points constituting the triangles (dual vertices) of the triangulation, together with the dual distance $d_d$ as the distance function, which is given by the minimal number of edges traversed from triangle to triangle. Note that $F(\mathcal{D}_{\mathcal{T}}(M)$ can be seen as the dual of the Delaunay triangulation, a structure which is also known as a Voronoi tessellation. Additionally, we endow $(P(\mathcal{D}_{\mathcal{T}}(M)), d_l)$ with a Hausdorff measure $\mu_h$, to make it a metric measure space. This Hausdorff measure is given by $\mu_h(A) = \#A$ for any subset $A \subseteq P(\mathcal{D}_{\mathcal{T}}(M))$, where $\#A$ is the number of elements in the set. Also, we endow $(F(\mathcal{D}_{\mathcal{T}}(M)), d_d)$ with a random walk $\{X_i\}$ for each triangle $i$ in $F(\mathcal{D}_{\mathcal{T}}(M))$.

The reason for using the link (or dual) distance instead of the canonical distance function of the original smooth metric space, is to test the capability of our curvature definitions to recover geometrical properties of the underlying triangulated space without using the metric tensor of the underlying continuum surface. The motivation for doing this is twofold. First, we want to test the robustness of our curvature measures with respect to discretization artefacts. Replacing the continuum distance function from the embedding with a lattice distance function gives a discrete approximation of the continuum distance, thus serving as a test ground for discretization artefacts. Second, our main goal is to apply these curvature definitions to the CDT quantum geometries, where one does not even know if there is an underlying continuum manifold being approximated. In this setting one relies solely on the discrete setting of the geometries in the path integral. So, using the link distance in $P(\mathcal{D}_{\mathcal{T}}(M))$ also works as a preamble and test ground for implementing the generalized curvatures in those quantum geometries. In this controlled Delaunay triangulation setting, we expect both the link distance and dual distance to converge, up to proportionality factors, to the geodesic distance of the surfaces the triangulations are based on, in the limit of an infinite number of triangles. So we expect our generalized scalar curvature definitions to recover the Ricci scalar curvature of the underlying spaces in the infinite triangle limit. To give an idea of the type of triangulation one obtains, we show examples of Delaunay triangulations of a plane and sphere in Fig. 5.1; note that these triangulations contain only a small number of triangles compared to the Delaunay triangulations used in our measurements.

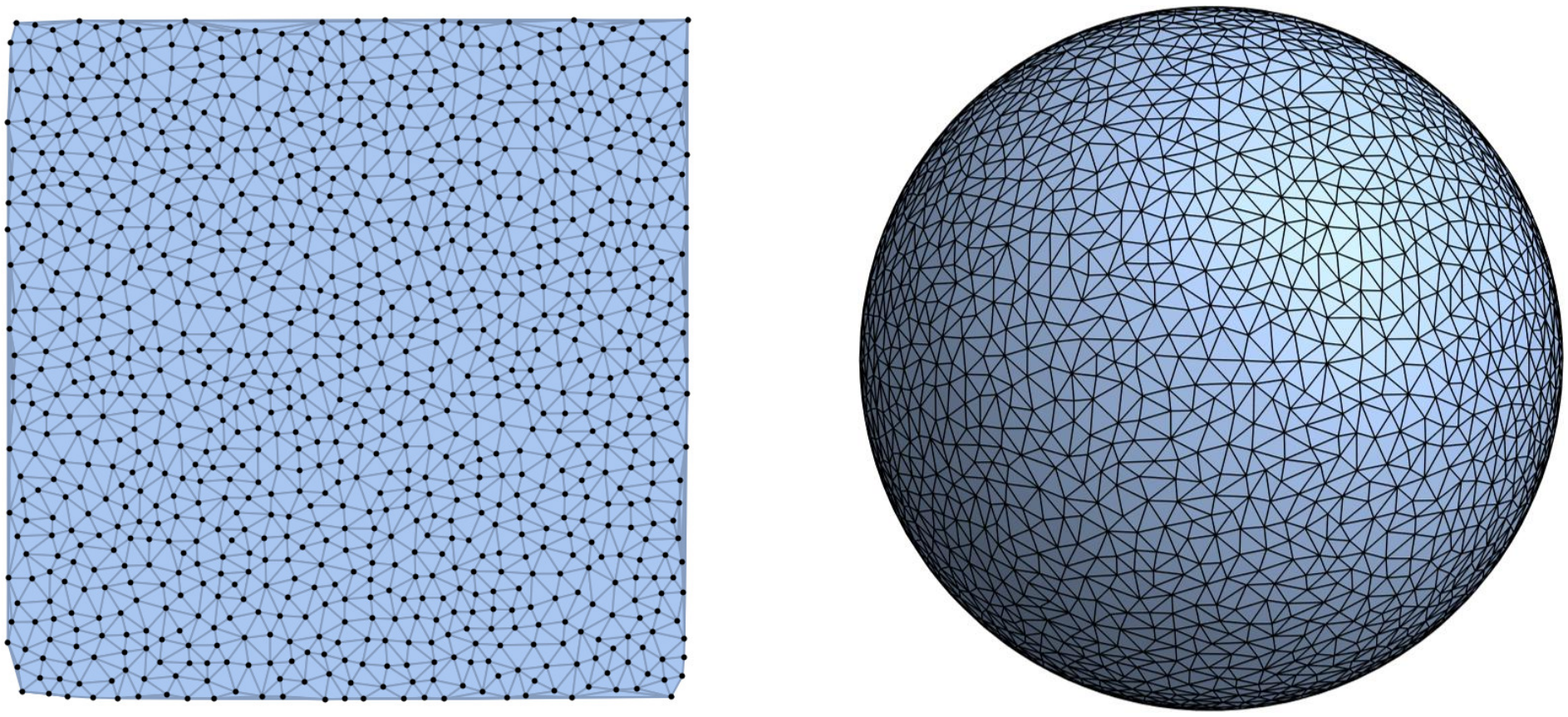

**(a)** Poisson Disk Sampled Plane **(b)** Poisson Disk Sampled Sphere

**Figure 5.1.:** Delaunay triangulations of a plane and a sphere, making use of Poisson disk sampling of points on the surface of smooth two-dimensional surfaces as described in the text. Note that we only show a Delaunay triangulation of a square region of the plane in this figure, but for the measurements we tile the square region to cover the plane.

We are interested in studying discrete spaces that are large enough to see curvature effects before reaching finite-size effects. To this end, we studied Delaunay triangulations with close to 150 k vertices. In the case of the Delaunay triangulation of the sphere, we can make a rough estimation of the effective radius in terms of the link length. When using the link distance as distance function means setting the length of the links to be 1, giving a triangle area of $A_\triangle = \sqrt{3}/4$. With a total of $N_2 = 300\,\text{k}$ triangles this gives a total triangulation area (two-volume) of $A_\triangle N_2$. Comparing this to the area of a continuum sphere $4\pi R^2$, we obtain a rough estimate of the effective radius of $R_{\text{eff}} = \sqrt{A_\triangle N_2/4\pi} \approx 102$. Note that we use the Poisson disk sampling such that all links will already have very similar lengths, so the scaling to link length 1 does not change the geometry strongly. Still, the triangulation after rescaling the link lengths will resemble something more like a "fuzzy" plane or sphere, with bumps and fluctuations around the actual smooth underlying sphere. Since to use this formula, one assumes that all these fluctuations created by the "fuzziness" are at the same distance $R_{\text{eff}}$ from the origin, but giving a larger area, this will cause the estimate of the effective radius to be an overestimate of the actual radius of the underlying continuum sphere. We expect our coarse-grained notions of scalar curvature to be capable of washing this "fuzziness" out, and provide a more accurate (smaller) effective radius of the underlying space.

### 5.3.2 Discrete $\mathcal{SC}$s

In order to measure the $\mathcal{SC}_i$ in discrete metric spaces we need to discretize the definitions (5.11). For this work we ended up using the following discretizations, where now $Q_p^s$ is defined on a discrete scale $s = 0, 1, 2, \ldots, s_{\text{max}}$, where $s_{\text{max}}$ represents the last $s$ for which $Q_p^s$ is measured. We discretize the logarithmic derivative using the simple forward finite-difference,

$$L_p^s := \frac{\log\left(Q_p^{s+1}/Q_p^s\right)}{\log((s+1)/s)}. \tag{5.12}$$

Also using the forward finite-difference for the second derivative and the cumulative sum for the integral, gives the discretized definitions

$$\mathcal{SC}_1(Q_p^s) = s\left(L_p^{s+1} - L_p^s\right), \tag{5.13}$$

$$\mathcal{SC}_2(Q_p^s) = \frac{\sum_{k=0}^{s} k\left(L_p^s - L_p^k\right)}{\sum_{k=0}^{s} k}. \tag{5.14}$$

Note that (5.12) is not well-defined at $s = 0$. However, we are not interested in the scalar curvature at very small discrete length scales, because at these scales the behaviour is dominated by discretization effects, as will be seen later. Hence, we simply leave $L_p^s$, and thus $\mathcal{SC}_1(Q_p^s)$ and $\mathcal{SC}_2(Q_p^s)$, undefined at $s = 0$, and we do not consider them in the results. Note that this may seem to have an effect on (5.14) for all $s$, but this is not the case as the factor $k$ makes sure that $L_p^k$ does not contribute to the sum at $k = 0$.

In practice, we find that taking the second derivative (more accurately, the finite difference implementation in (5.13)), results in the statistical fluctuations becoming very large. So, instead of taking a difference of $L_p^s$ directly we can take the difference of some moving (weighted) average, more

specifically we use

$$L_p^s \longmapsto \bar{L}_p^s = \sum_{k=-m}^{+m} w_k \, L_p^{s+k}, \tag{5.15}$$

where $m$ denotes the maximum number of points away from $s$ we take our average over, making it an average over $2m+1$ points. Additionally, we
5.3 use weights $w_k$, which are properly normalized, $\sum_{k=-m}^{m} w_k = 1$. Note that (5.15) is not defined for $s$ close to 0 or $s_{\text{max}}$. Since small scales $s$ are not interesting due to discretization artefacts, and we will explore a sufficiently large range of scales $s$, we leave the first and last points undefined and ignore them in our analysis.

### 5.3.3 Discrete diffusion process

Before we present the results of the discretized scalar curvatures (5.13) and (5.14) on the Delaunay triangulations of a plane and sphere, we explain the discrete implementation of the return probability $P_p^n$ measurements. First, as mentioned before, if we want to determine precise dimensionful values for the scalar curvature, and not only its signature, we need an identification between the random walk time and the corresponding diffusion time. In the case of a discrete geometry, a standard implementation of a random walk is to start at a point in the space, and move to a random neighbour of that point. The neighbour is chosen uniformly from all direct neighbours. In the case of an irregular discrete space, a node might not have the same number of neighbours all other vertices. Therefore, the probability distribution of the random variables determining the random walk is different for each point.

To get closer to a situation that best resembles the conditions that allow the identification (5.9), we implement the random walk in such a way that each step is equally distributed. This is because, if the variables are not identically distributed, the variance computed in (5.9) is not necessarily equal to a factor $\sigma^2$ times the number of steps taken. In the case of not equally distributed random variables describing each step, relation (5.9) could be regarded as an effective relation, where the proportionality factor $\sigma^2$ would be some effective variance of each step. In this situation, one would have to determine this relation by other means.

Because of this, we will restrict ourselves to a case where we can compute $\sigma^2$ exactly. This is the case for the metric space whose construction is based on the dual of the Delaunay triangulations, meaning the space based on the triangles with the dual distance. In the dual graph of the triangulation, each node will have the same number of nearest neighbours, and thus, the random variables describing each step are equal to a uniform distribution among the three nearest neighbours of each node. For a visualization of this effect on a Delaunay triangulation of a sphere and its dual graph, see Fig. 5.2. If the triangulation is equilateral, $(\sigma^2)$ can be computed exactly, as is done later.

Even though the random variables in this discrete setting are not Gaussian but uniform, we appeal to the Central Limit Theorem and assume that at some large value of $n$ the identification (5.9) will hold approximately. Therefore, we will use identification (5.9) for our diffusion processes in the dual lattice, even though we are not exactly in the situation where this identification is exact. We remark that an implementation on the vertices of the triangulation is also possible if one is interested in detecting the presence of curvature using our prescriptions. However, we wish to make use of a precise identification between diffusion steps and diffusion time, to obtain predictions for the dimensionful scalar curvature in the discrete spaces considered here.

Since in the Delaunay triangulation the edge length was set to 1 everywhere, the edge length in the dual graph is going to be $\frac{1}{\sqrt{3}}$ so in the dual we will also have the same "fuzziness" as in the Delaunay Triangulation. From now on, when we speak of the random walk in the Delaunay triangulations, we will be speaking of a random walk on the dual graph of that Delaunay triangulation, after rescaling all the edges to length 1.

The discrete diffusion process we use is implemented as follows. Let $P(p', p; n)$ be the probability of a random walker moving from point $p$ to $p'$ on the dual graph of the triangulation, in $n$ steps. We start in our diffusion process with $P(p', p; 0) = \delta_{p,p'}$, where $\delta$ is the Kronecker delta symbol; this signifies that the walk starts at point $p$ and is normalized such that $\sum_{p'} P(p', p; n) = 1$ for all $n$; the diffusion process is then implemented using the evolution equation [127, 134]

$$P(p', p; n+1) = (1-\eta)P(p', p; n) + \frac{\eta}{3} \sum_{q \sim p'} P(q, p; n), \tag{5.16}$$

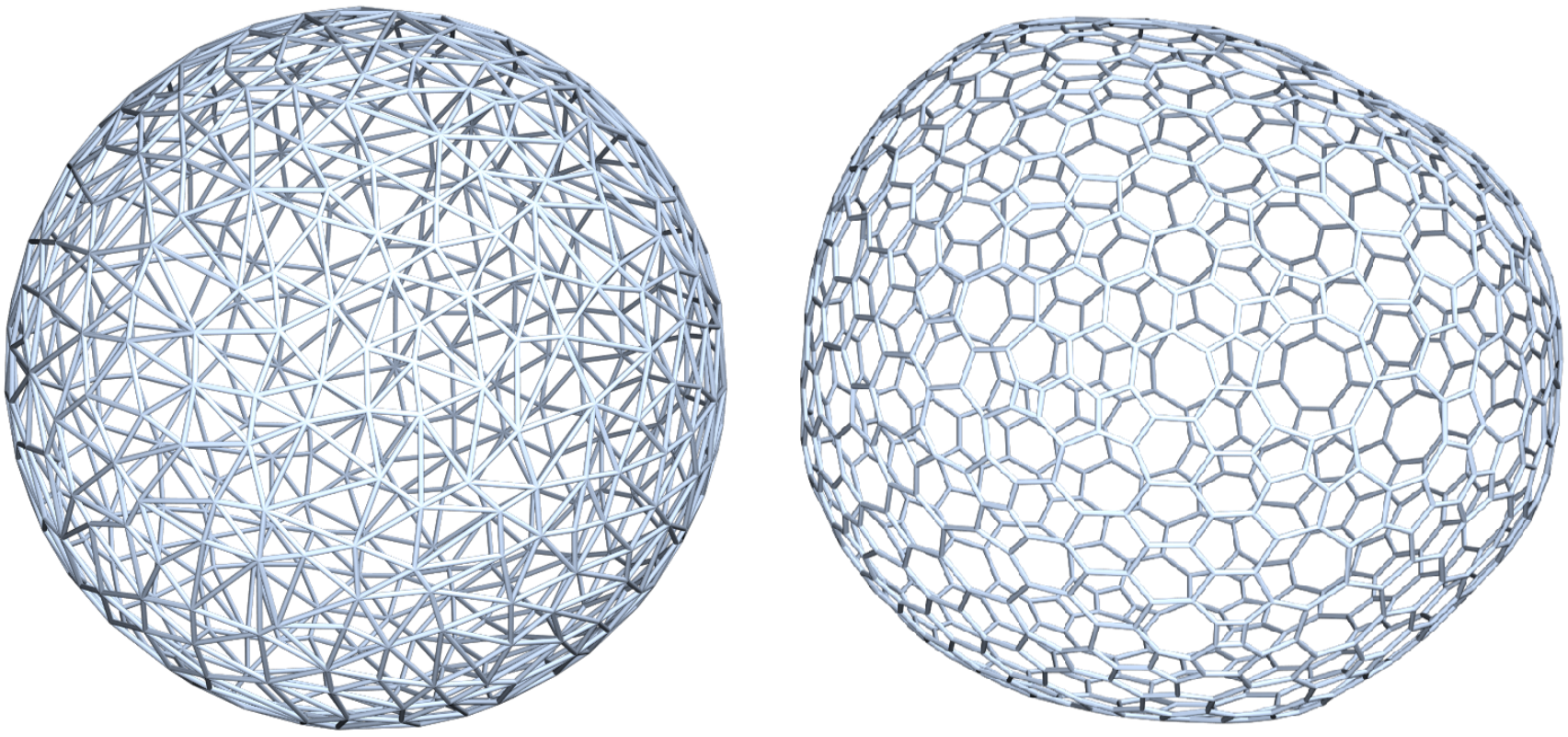

**Figure 5.2.:** On the left, a Delaunay triangulation of a sphere constructed from a sample of 500 points on a continuum sphere. On the right, the associated dual graph of the triangulation. Notice that each node in the dual graph has coordination number 3, even though the vertex graph of the triangulation does not have a fixed coordination number.

where $q \sim p'$ denotes the set of all direct neighbours of $p'$ in the graph. Also, $\eta \in (0, 1]$ takes the role of a discrete diffusion constant, which can be set smaller than 1 to mitigate the large differences between the return probability for odd and even $n$ for small $n$; we use $\eta = 0.9$ in our measurements. The return probability is determined by measuring $P_p^n = P(p, p; n)$ after each diffusion step.

### 5.3.4 Discrete scalar curvature

Using the discretizations (5.13) and (5.14), with the discrete sphere volume $\|S_p^r\|$ and return probability $P_p^n$ we are able to measure the scalar curvature in the Delaunay triangulation of the plane and sphere.

#### Sphere volume

The obtained results for $Q_p^s \to \|S_p^s\|$ with $s \to r$, are presented in Fig. 5.3 for both geometries. Note that the results presented here are a sample average of $\mathcal{SC}_i(\|S_p^r\|)$ for $10^5$ uniformly sampled origins $p$ on the Delaunay geometry. Additionally, $\mathcal{SC}_1(\|S_p^r\|)$ is computed using a moving average

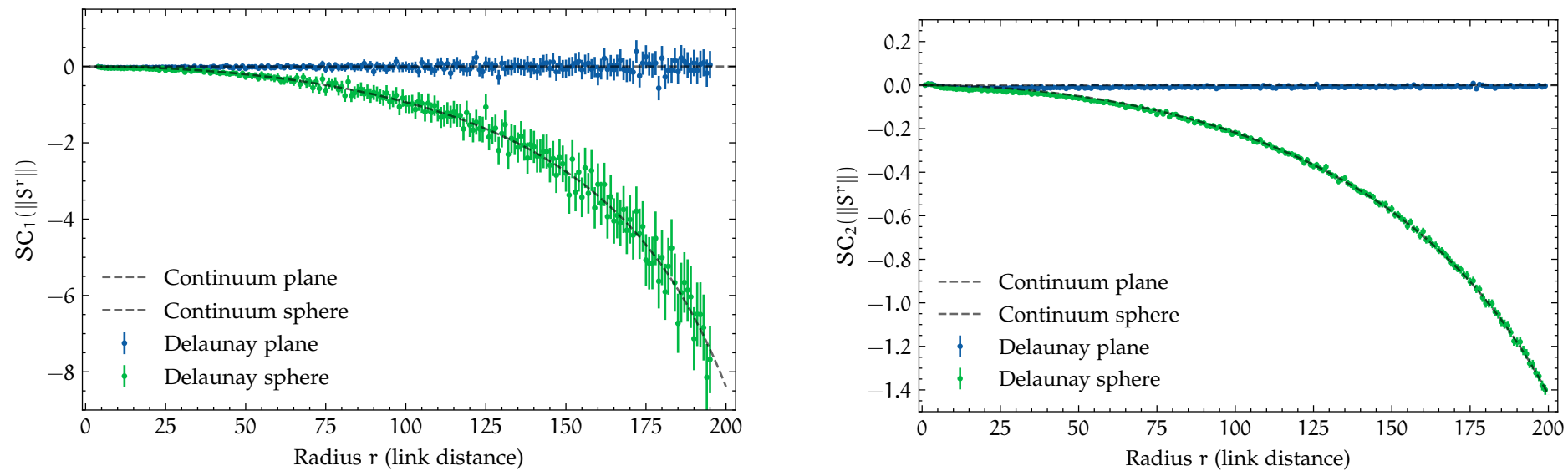

**Figure 5.3.:** Scalar curvatures $\mathcal{SC}_1(\|S^r\|)$ and $\mathcal{SC}_2(\|S^r\|)$ for the sphere volume for the Delaunay sphere and plane. These results are sample averages based on $10^5$ sampled origins in each geometry. Included are the continuum curves of the scalar curvatures of the plane and the two-sphere, the latter of which is fitted to the Delaunay sphere results.

as described by (5.15) with $m = 5$, in this case using uniform weights $w_k = 1/(2m+1)$.

From these results it is evident that $\mathcal{SC}_1(\|S^r\|)$ is more sensitive to discretization artefacts. This is due to the use of the second derivative, and precisely why a moving average as given by (5.15) is used, but even with this moving average the results are significantly more noisy than for $\mathcal{SC}_2(\|S^r\|)$. However, $\mathcal{SC}_2(\|S^r\|)$ shows lattice artefacts at small radii $r$, whereas $\mathcal{SC}_1(\|S^r\|)$ is virtually unaffected by these deviations from 0. The presence of lattice artefacts at larger scales $r$ is caused by the integral in the definition (5.14) of $\mathcal{SC}_2(\|S^r_p\|)$, which causes the lattice artefacts present in $S^r_p$ at small $r$ to have significant contributions at larger $r$. Finally, we note that in the presented $r$ range the lowest-order contribution, which is second-order, is clearly visible, and for large $r$ higher-order corrections are also present. This was determined by comparing to quadratic fits. As a consequence, one has to be careful when trying to extract curvature based on the lowest-order correction, as will be discussed later.

To compare these results of the Delaunay triangulations to the continuum geometries they are based on, we compare them to the continuum expectations of the scalar curvatures. In Fig. 5.3 we include a fit of $\mathcal{SC}_i(\|S^r_p\|)$ in a continuum sphere, as well as the identically 0 line of a continuum plane, both displayed by a dashed line. These are one-parameter fits as given in appendix D ($D = 2$), based on the radius $R$ of the two-sphere. From this fitting procedure we obtain the estimated two-sphere radii $R = 92.2 \pm 0.3$ and $R = 92.5 \pm 0.1$ based on $\mathcal{SC}_1(\|S^r\|)$ and $\mathcal{SC}_2(\|S^r\|)$ respectively. It is important to remark that since the fitting functions used do not have the initial over- or undershoot present in the numerical results, the fitting itself, and therefore the estimated $R$, is more accurate at larger scales. We note that these estimates of the radius are compatible with each other, indicating that $\mathcal{SC}_1(\|S^r\|)$ and $\mathcal{SC}_2(\|S^r\|)$ are expected to yield the same scalar curvature in the continuum limit.

The intended use of the scalar curvatures $\mathcal{SC}_i$ is to measure curvature using only (reasonably) small scales, not necessarily by fitting $\mathcal{SC}_i$ to an analytical expectation. As such we analyse the small-scale behaviour of the measured scalar curvatures to obtain an estimation of the two-sphere radius $R$. To define curvature using the knowledge of the continuum expansion of $\mathcal{SC}_i$, we have to be sure that we are not in a range of lattice artefacts. For $\mathcal{SC}_1(\|S^r\|)$ and $\mathcal{SC}_2(\|S^r\|)$ we observe lattice artefacts to be negligible for $r > 25$. Based on the continuum formulas for $\mathcal{SC}_i(\left\|S^r_p\right\|)$ in appendix D, we find that for $\frac{r}{R} \leq \frac{3}{4}$, a quadratic approximation of the expansion deviates less than 3% from its actual value. The expressions for each $\mathcal{SC}_i(\left\|S^r_p\right\|)$ of this relative error can be found in the appendix as well. So, it can be used as an approximation to compute the curvature in this range. In this range we estimate the scalar curvature $\mathcal{R}$, as

$$\mathcal{R}_p \simeq -\frac{3D}{2}\frac{\mathcal{SC}_1(\left\|S^r_p\right\|)}{r^2} + \mathcal{O}(r^2) \qquad \text{and} \qquad \mathcal{R}_p \simeq -6D\frac{\mathcal{SC}_2(\left\|S^r_p\right\|)}{r^2} + \mathcal{O}(r^2) \tag{5.17}$$

We apply this approximation around $r = 75$ for both $\mathcal{SC}_1(\left\|S^r_p\right\|)$, using $D = 2$. To avoid the large statistical error of a measurement at a single radius we take an estimate over the range $r \in [70, 79]$ for each $r$ and average the resulting estimates. With the obtained estimate of the Ricci scalar $\mathcal{R}_p$, one can estimate the more intuitive effective radius $R_{\text{eff}}$ defined through $\mathcal{R}_p := \frac{D(D-1)}{R^2_{\text{eff}}}$, which yields an estimation for the size of the approximate sphere being triangulated. We find the effective radii to be $R_{\text{eff}} = 89 \pm 9$ and $R_{\text{eff}} = 87 \pm 5$ based on $\mathcal{SC}_1(\|S^r\|)$ and $\mathcal{SC}_2(\|S^r\|)$ respectively. The errors represent the 68%-confidence interval based on the propagated errors of the $\mathcal{SC}_i$ together with the systematic error of the truncation to second order, based on the results in appendix D.

These estimates of the sphere radius based on the first-order correction are found to be compatible with each other, and almost compatible with the estimates based on a full fitting. However, the estimate based on $\mathcal{SC}_2(\|S^r\|)$ is found to be relatively low compared to the more accurate fitting estimates. This may be understood by the still significant lattice artefacts around $r = 75$, despite our expectations. It is possible that the integral in the definition of $\mathcal{SC}_2(\|S^r\|)$ causes the lattice artefact to still have a significant contribution around $r = 75$. But at this radius the higher-order corrections also have a significant contribution (around 3%), so it seems that $\mathcal{SC}_2(\|S^r\|)$ is not ideal to estimate curvature based on a single value in this case, because of the relatively large lattice artefacts. Finally, note that all the effective radius estimates are significantly smaller than the naive estimate of the radius of $R_{\text{eff}} \approx 102$ based on a total volume comparison, as was expected given that our definitions coarse-grain the fluctuations induced by setting all the link lengths to 1. Thus, we find the effective sphere the Delaunay triangulation describes to be slightly smaller than the underlying smooth sphere, as if the triangulation is slightly crumpled up.

We find that the best of the two prescriptions is $\mathcal{SC}_2(\|S_p^r\|)$, because of its robustness against noise and still reasonably fast convergence towards the continuum fit of the Delaunay triangulation. The fact that $\mathcal{SC}_2(\|S_p^r\|)$ only contains one derivative, makes it more robust at larger discrete scales, which is important if one desires to explore the continuum limit of a discrete geometry. By comparing the numerical results with the continuum formulas given in appendix D, we find that $\mathcal{SC}_2(\|S_p^r\|)$ is more suitable to estimate the curvature at larger scales, and not very reliable for estimating curvature at short scales due to the presence of small but still significant lattice artefacts.

### Return probability

The obtained results for $Q_p^s \to P_p^n$ with the diffusion time identified $s \to \tau = \frac{n\sigma^2}{2D\eta}$, are presented in Fig. 5.4 for both geometries. Again note that the results presented here are a sample average of $\mathcal{SC}_i(P_p^n)$ for 50 uniformly sampled origins $p$ on the Delaunay geometry. Additionally, $\mathcal{SC}_1(P_p^n)$ is computed using a moving average as described by (5.15) with $m = 72$, in this case using Gaussian weights $w_k = \exp\left(\frac{k^2}{2(m/4)^2}\right)$.

We studied diffusion steps up to values 2 orders of magnitude larger than the lattice distance used with the sphere surface. Nonetheless, within the measured range, our results only show the lowest-order correction to flatness, which in this case is linear. Therefore, it seems that the diffusion process explores distance scales much smaller than those probed for the sphere volumes in Fig. 5.4, where higher-order corrections are already observed at the end of the plots. Because of the already large number of diffusion steps it became computationally unfeasible to study distance scales equivalent with those of the sphere volume measurements. However, this is no problem as both methods can be used complementary to study different scale ranges in more detail in the same metric space. This indicates that $\mathcal{SC}_i(P_p^n)$ is ideal to define the scalar curvature at short scales, since the higher-order contributions are negligible until large diffusion times.

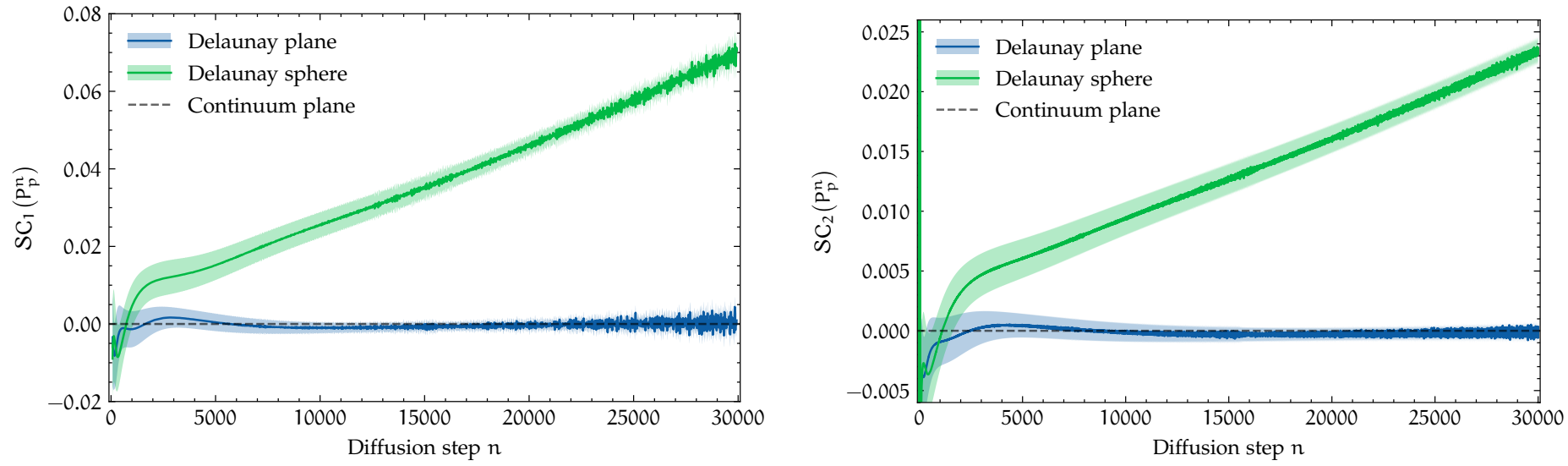

**Figure 5.4.:** Scalar curvatures $\mathcal{SC}_1(P^n)$ and $\mathcal{SC}_2(P^n)$ for the return probability on the Delaunay sphere and plane. These results are sample averages based on 50 sampled origins in each geometry. Included are the continuum curves of the zero scalar curvature of the plane.

Given that there is no closed formula for the return probability on a sphere, only asymptotic expansions, we do not have a closed analytical expression of $\mathcal{SC}_i(P_p^n)$ to fit to. Nevertheless, we can still estimate the curvature and the radius of the underlying sphere based on the first-order correction,

$$\mathcal{R}_p \simeq \frac{12\eta D}{\sigma^2}\frac{\mathcal{SC}_1((P_p^n)^{-1})}{n} + \mathcal{O}(n) \quad \text{and} \quad \mathcal{R}_p \simeq \frac{36\eta D}{\sigma^2}\frac{\mathcal{SC}_2((P_p^n)^{-1})}{n} + \mathcal{O}(n) \tag{5.18}$$

We apply this approximation to the range $n \in [6000, 10000]$ for both $\mathcal{SC}_i(P_p^n)$, where $D = 2$, because this appears to be a range of $n$ where lattice artefacts can be neglected, and higher-order correction are still negligible. For this estimate the value of the standard deviation $\sigma^2$ in (5.9) is necessary.

In the dual of a triangulation, this value can be computed exactly, to be[1] $\sigma^2 = \frac{1}{3}$. We then find the effective radii to be $R_{\text{eff}} = 96 \pm 4$ and $R_{\text{eff}} = 91 \pm 4$ based on $\mathcal{SC}_1(P^n)$ and $\mathcal{SC}_2(P^n)$. Where the errors again represent the 68%-confidence interval based on the propagated errors of the $\mathcal{SC}_i$.

The estimates of the effective radius made here based on the return probability turn out to be consistent with each other and with the results we obtained using the sphere volumes, further solidifying $\mathcal{SC}_1$ and $\mathcal{SC}_2$ as a consistent method of measuring curvature.

With the return probability $P_p^n$ we think that $\mathcal{SC}_2(P_p^n)$ is again the best prescription to obtain the curvature of the space. There is no significant difference with $\mathcal{SC}_1(P_p^n)$ regarding behaviour at large scales, but we think it is best to use $\mathcal{SC}_2(P_p^n)$ because it requires no moving average to obtain reasonable results, unlike $\mathcal{SC}_1(P_p^n)$ due to the presence of a second derivative.

Comparing our results based on the sphere volume with those based on the return probability, $\mathcal{SC}_2(P_p^n)$ seems to be more suitable to estimate curvature at short scales than both $\mathcal{SC}_i(\|S_p^r\|)$. This is because we observe the higher-order corrections to be significantly smaller on a similar distance scale. So ideally, used together, $\mathcal{SC}_i(P_p^n)$ will probe short-scale curvature, and $\mathcal{SC}_i(\|S_p^r\|)$ will probe larger distance scales, giving a larger overall range of scales at which scalar curvature can be measured in a metric space.

One last remark is in order. We expect that all our estimations of the scalar curvature will match exactly in the limit of vanishing lattice spacing. However, we did not perform a study of finite-size scaling of our results at different triangulation volumes, so we do not have a good understanding of how the errors will change with increasing volume, as we can only conjecture they will agree in the limit. Such a scaling analysis would be very useful in understanding the discretization effects of the presented curvature quantities, and is highly recommended being used in follow-up research.

In the following section, we will use these methods to determine the scalar curvature of the quantum geometries that appear in the path integral of 2D Causal Dynamical Triangulations. Since these geometries are also triangulations, we use the same prescriptions as the ones used for the Delaunay triangulations in this section.

[1] This can be computed as the standard deviation of a uniform distribution that can only take 3 possible values, moving at points at distance $\frac{1}{\sqrt{3}}$ from the centre.

5.4

## 5.4 Quantum scalar curvature

As discussed in the introduction, the main motivation for our generalized definitions of curvature is its use in a quantum geometry setting, in particular that of CDT. Here, there is no notion of tensor calculus, but we can define distances and a diffusion process. In order to create valid observables for quantum geometries, they must be diffeomorphism-invariant. We will define an average quantum scalar curvature observable, for which we will remove the point dependence of our curvature definitions by taking a manifold average. To that end, we define the *Quantum Scalar Curvature* $\mathcal{QSC}_i(\mathcal{Q}^s)$ based on an $s$ scale-dependent local observable $\mathcal{Q}^s_p$ to be:

$$\mathcal{QSC}_i(\mathcal{Q}^s) = \frac{\int \mathrm{d}^D p \sqrt{|g|}\, \mathcal{SC}_i(\mathcal{Q}^s_p)}{\int \mathrm{d}^D p \sqrt{|g|}}, \tag{5.19}$$

where the manifold average is taken over manifold $M$ with metric $g$, where $|g|$ denotes the metric determinant. Its obvious discretization used in CDT implementations is

$$\mathcal{QSC}_i(\mathcal{Q}^s) = \frac{1}{|T|} \sum_{p \in T} \mathcal{SC}_i(\mathcal{Q}^s_p) \tag{5.20}$$

where $|T|$ is the total number of points $p$ in the vertex (or dual) graph of the triangulation $T$.

We thus define the observables $\mathcal{O}_i[T] = \mathcal{QSC}_i(\mathcal{Q}^s)$ using (5.20). The expectation value of (5.20) is computed as an average over the ensemble of simplicial manifolds that come out of the simulations of CDT, as shown in equation (2.7). As in the previous chapters, instead of considering the ensemble of all triangulations, we consider the ensemble of triangulations with a fixed total volume $N(T) = N_2$, because of how convenient it is to work with computationally, making it the standard choice for numerical CDT [61]. Furthermore, in this chapter, we only consider triangulations with a toroidal topology without boundaries. Another illustration of what a typical CDT configuration looks like can be seen in Fig. 5.5.

We will estimate the quantum scalar curvature by sampling triangulations from the ensemble using Markov chain Monte Carlo methods. It is possible to compute the triangulation average of $\mathcal{SC}_i(\mathcal{Q}^s_p)$ in each triangulation, but computationally rather expensive. So, instead, we estimate the triangulation average with a sample average, based on a uniform sample of points. We tested that this approximation did not make any significant modifications to our results, supporting the results of the previous sections about the homogeneity of the geometries explored. To decrease the effect caused by the compactness of the geometry, something that was of importance in previous investigations of the QRC [100, 112], we perform the actual measurements on an unwrapped version of the simulation toroidal triangulations. Specifically, we cut the triangulations open along two closed non-contractible loops and tile the unfolded triangulation to create an infinite periodic triangulation of planar topology; for more details, see [99].

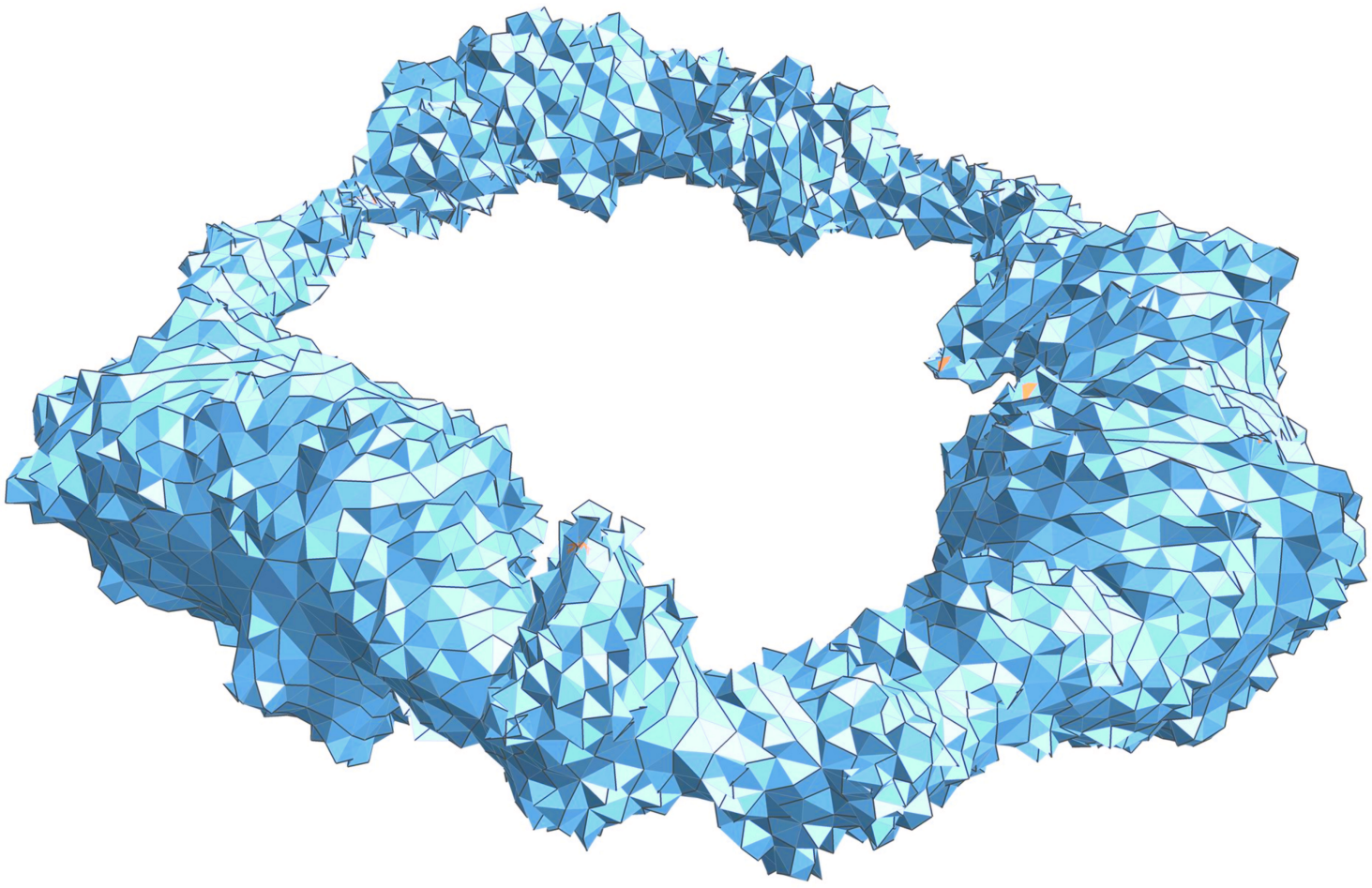

**Figure 5.5.:** Visualization of a typical triangulation of 2D CDT with $N_2 = 10\,k$ triangles and $\tau = 123$ time-slices, based on a spring embedding. Spatial slices of constant time are marked by dark lines.

Identical to the implementation in the Delaunay triangulations, the sphere volume measurements are implemented by counting the number of vertices in the vertex graph at a given link distance $r$ from the origin. The computation of the return probability of a random walker is also implemented identically to the Delaunay triangulations. We use the dual graph of the CDT triangulations and perform the random walk by updating the probability of the random walk with (5.16). The return probability at a point $p$ after $n$ steps in a configuration $T$, is computed by recording the value of $P_T(p, p; n)$ after each diffusion step in (5.16).

### 5.4.1 $\mathcal{QSC}$ results

5.4

Using the described methods we measured the quantum scalar curvatures for sphere volumes $\|S_p^r\|$ in CDT. We performed measurements on 5000 triangulations to approximate the ensemble average and approximated the manifold average by a sample average over 500 points. The resulting quantum scalar curvatures are displayed in Fig. 5.6. Note that just like was done for the Delaunay triangulations, $\mathcal{QSC}_1(\|S^r\|)$ is computed using a moving average as described by (5.15) with $m = 5$, using uniform weights $w_k = 1/(2m+1)$.

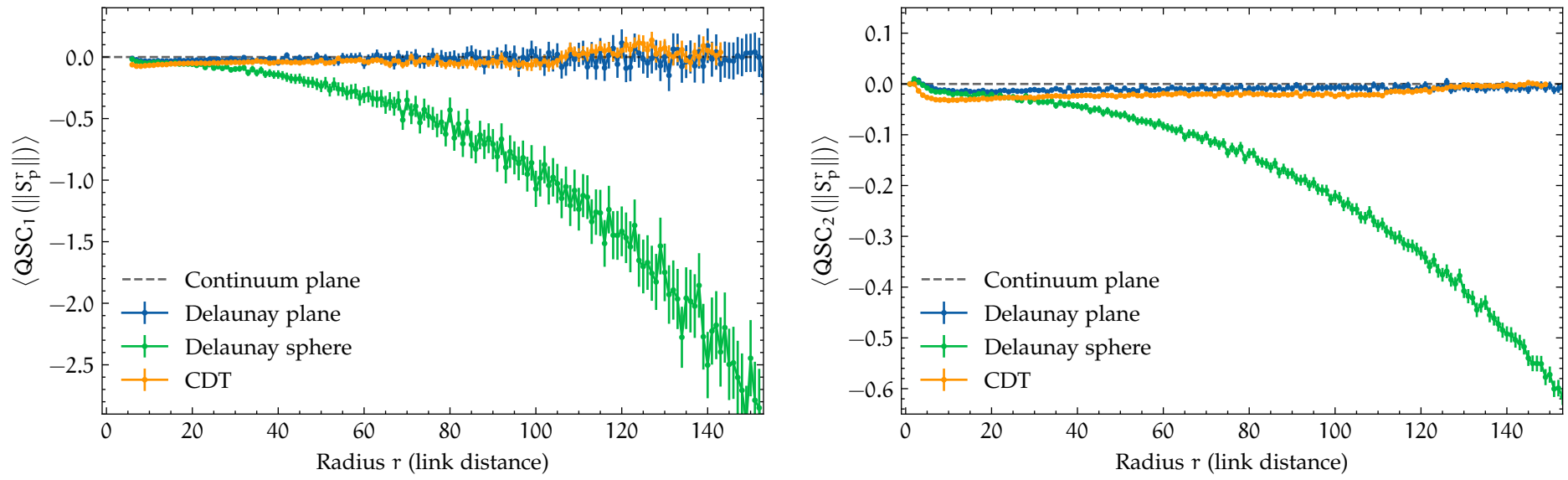

**Figure 5.6.:** The quantum scalar curvatures of the sphere volume $\langle \mathcal{QSC}_i(\|S_p^r\|)\rangle$ on 2D CDT with $N_2 = 300\,\mathrm{k}$ triangles and $\tau = 219$ time-slices. As a comparison the generalized scalar curvature results $\mathcal{SC}_i(\|S_p^r\|)$ of the Delaunay triangulations as already presented in Fig. 5.3 are also included with the same number of triangles, 300 k. The error bars represent 95%-confidence intervals; for CDT this is based on the fluctuations in $\mathcal{QSC}_i$ over 5000 CDT configurations.

From Fig. 5.6 we see some interesting features of the resulting quantum scalar curvatures. Both $\langle \mathcal{QSC}_1(\|S_p^r\|)\rangle$ and $\langle \mathcal{QSC}_2(\|S_p^r\|)\rangle$ exhibit a plateau, meaning that they are compatible with a flat curvature. We also observe that the deviation from 0 at small $r$ is much more significant than for the Delaunay triangulation, indicating stronger lattice artefacts. This is to be expected as the geometries of 2D CDT exhibit a lattice structure that is significantly less representative of a smooth geometry than the Delaunay triangulations. One may also notice that $\langle \mathcal{QSC}_2(\|S_p^r\|)\rangle$ is smaller than 0 for most of the measured range of $r$, and it seems to show a slow increase toward 0 for larger $r$. First, note that radii $r > 100$ will be subject to finite-size effects, that is to say that these radii are large enough to cover a two-volume of the order of the total volume of the geometry and can thus have behaviour that need not be reflective of the continuum $r \to 0$ behaviour of interest. Thus, we should not consider this region to be of interest when considering 'quasi-local' coarse-grained curvatures. Also, this deviation from 0 for CDT is significantly smaller than that of the Delaunay triangulation of a constant curvature sphere with the same number of triangles. For short scales ($r < 25$) we expect, just as with the Delaunay triangulations, that the discretizations of derivatives and especially integrals, have significant deviations from the continuum implementations, so we don't take into account those points. In conclusion, both $\mathcal{QSC}_i(\|S^r\|)$ indicate that, at least within the measured scales, the ground state of the 2D CDT path integral exhibits the properties of a space with zero average curvature, when extracting curvature using the geodesic sphere surface.

We want to stress that these results are not trivial and are not to be expected a priori. The non-triviality of curvature observables was measured and noticed, for example, in 2D CDT in [100]. There could be several reasons for this. The generalized scalar curvatures we are considering here are coarse-grained quantities in the sense that they depend on the triangulation in the neighborhood of the point $p$, with a size associated with the scale $s$ (in our cases $r$ or $n$). Because of this coarse-graining, the obtained results are not necessarily the same as for the Ricci scalar, which is a truly local quantity [118]. Even if the average of the Ricci scalar is zero on the whole manifold, since all our ensemble configurations have toroidal topology, higher-order scalar curvature invariants might not. The surprising finding in our work is that at least up to the measured scales, all of them seem to be zero. In general, the properties of such a coarse-grained curvature is non-trivial. These properties cannot be directly derived from the topological properties of the ensemble constituents, and can thus exhibit a non-trivial curvature average. If one is measuring a local quantity like the Regge curvature, the result of its average over the manifold is trivial in two dimensions, because it is fixed by the topology by the Gauss-Bonnet theorem [61], and is 0 for the toroidal topology we use here. Nevertheless, higher-order curvature invariants contribute to our coarse-grained observables, and non-trivial behaviour is to be expected.

Similarly, we performed measurements of the quantum scalar curvature based on the return probability $P_p^n$ in CDT, measured using the previously described methods. The ensemble average of the observable is estimated using 1370 triangulations with $N_2 = 300\,\mathrm{k}$ triangles and $\tau = 219$ time-slices.

The manifold average $\mathcal{QSC}_i(P^n)$ is approximated by a single sample of $\mathcal{SC}_i(P_p^n)$ at a point $p$ in each triangulation. The resulting $\mathcal{QSC}_i(P_p^n)$ are presented in Fig. 5.7. Again identically to the Delaunay triangulations $\mathcal{QSC}_1(\|P^n\|)$ is computed using a moving average as described by (5.15) with $m = 72$, using Gaussian weights $w_k = \exp\left(\frac{k^2}{2(m/4)^2}\right)$.

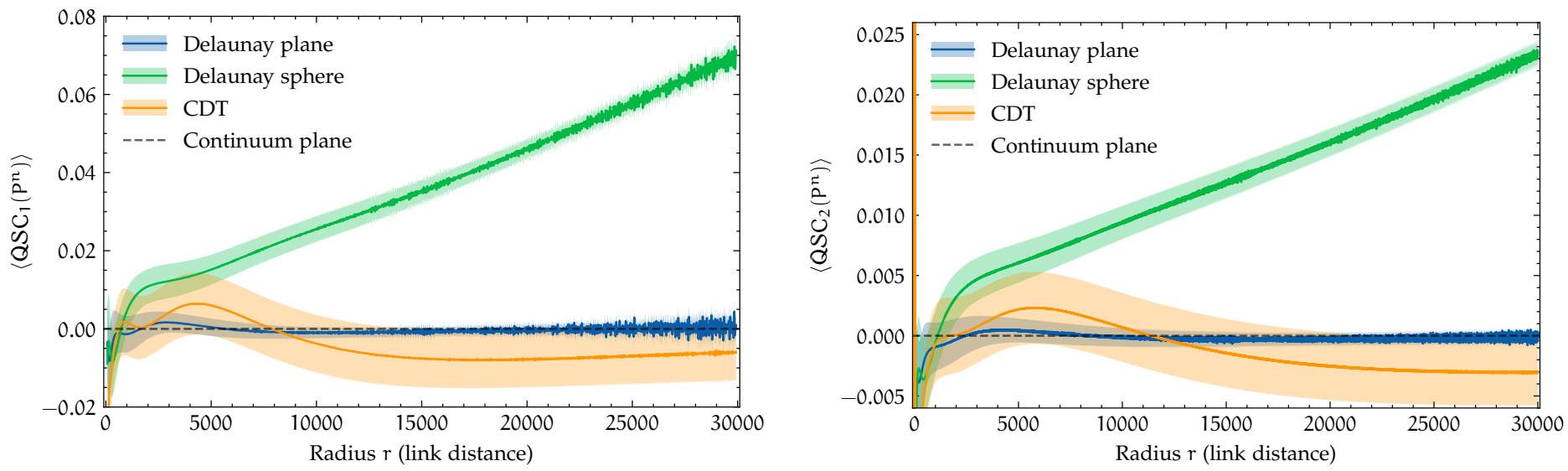

**Figure 5.7.:** The quantum scalar curvatures of the return probability $\langle \mathcal{QSC}_i(P^n)\rangle$ on 2D CDT with $N_2 = 300\,\mathrm{k}$ triangles and $\tau = 219$ time-slices. As a comparison, the generalized scalar curvature results $\mathcal{SC}_i(\|S_p^r\|)$ of the Delaunay triangulations as already presented in Fig. 5.4 are also included with the same number of triangles, 300 k, as well as the continuum result for generalized scalar curvature in a plane. The error bands represent 95%-confidence intervals; for CDT this is based on the fluctuations in $\mathcal{QSC}_i$ over 1370 CDT configurations, and for the Delaunay triangulations this is based on the fluctuation between 50 different points on the triangulation.

Similar to the sphere volume measurements we see that after a region of lattice artefacts both $\langle \mathcal{QSC}_i(P^n)\rangle$ are compatible with 0 average curvature within statistical error, again indicating a flat geometry. Note that the errors appear to be very large compared to the smoothness of the curve. This is the case because the scalar curvature $\mathcal{SC}_i(\|P_p^n\|)$ at a single point $p$ is already relatively smooth, however the $\mathcal{QSC}_i(\|P^n\|)$ varies strongly from configuration to configuration in the CDT path integral, yielding a smooth average with large errors. Nevertheless, we also observe that there are large overshoots for both $\mathcal{GSC}_i(P^n)$ at $n < 1000$. In particular, the deviations are of the size of that of the Delaunay triangulation of a sphere with the same number of triangles, meaning that the lattice artefacts of this order are large. Despite this, at $n > 5000$ these effects seem to be washed out, and results compatible with flatness were found. However, we acknowledge that the statistical error is large, so our results are also compatible with a non-zero curvature of CDT, albeit significantly smaller than that of the Delaunay triangulation with the same number of triangles.

Summarizing, we conclude that all our definitions of scalar curvature indicate that the path integral ground state of 2D CDT has average curvature equal to 0. This is different from the findings in [100], where the coarse-grained notion of curvature called Quantum Ricci curvature (QRC) is used. Here the QRC measured in 2D CDT seemed to be non-constant, which they acknowledge to be difficult to interpret, but may seem to indicate negative curvature. Although they also note that this curvature may vanish for scales larger than the ones studied in [100]. Despite the remarkable precision in their measurements, the QRC still has the problem of the unknown leading prefactor multiplying QRC, as mentioned in the introduction. The problem in interpreting the results of the QRC, is that even if a non-constant region of the scale-dependent QRC is observed that seems to flatten out, as it appeared to do in [100], it cannot be known how far away from the asymptote one is. Therefore, the actual numerical value of the QRC is meaningless unless one knows the multiplying prefactor, and because of that, one cannot know if the results are actually converging to it or diverging from it. We believe to have solved this issue with our generalized curvature scalars, by eliminating this unknown leading constants and powers. This allows us to measure scalar curvature as a deviation from 0, a known reference value. This makes the interpretation of the results easier than in the presence of unknown leading powers or multiplying constants, since any deviation from 0 can be directly interpreted as the presence of curvature.

## 5.5 Collecting the results of this chapter

In this chapter, we defined and implemented generalized notions of scalar curvature for metric spaces endowed with a measure and/or a random walk, with the aim of providing computationally feasible methods of determining curvature that can be implemented in the non-perturbative approach to quantum gravity called Causal Dynamical Triangulations. The definitions are successfully implemented in 2D CDT. This is done without the need to define or compute intermediate, more complex, tensorial quantities like the Riemann or Ricci tensor. We do not assume the existence of tensor calculus at any stage in our definitions of curvature in these spaces. Moreover, our constructions do not assume that the considered spaces are smooth in any sense. Indeed, they are designed to be used in discrete and fractal

5.5

spaces, like the ones appearing in CDT. Furthermore, our definitions are intentionally coordinate free by construction, since this is situation that is encountered in the geometries of CDT.

For the construction of these generalized notions of scalar curvature $\mathcal{SC}$, we use quantities that can be computed both in a Riemannian manifold, and in a metric space with a measure and/or a random walk. In particular, we use the volume of a sphere $\|S_p^r\|$ with a radius $r$, centred at a point $p$, and we use the return probability $P_p^n$ of a random walker from a point $p$ after $n$ steps. In a Riemannian manifold, both of these quantities define the Ricci scalar curvature by the leading-order scale-dependent correction from their calculation in Euclidean space. Summarizing, the procedure used to define some $\mathcal{SC}$ is the following: 1) Find a scale-dependent quantity that can be computed both in a Riemannian manifold and in a metric space with a measure and/or a random walk, and it should have the functional form (5.10). The quantities we consider here, the sphere volume $\|S_p^r\|$ and the return probability $P_p^n$, are of this form. 2) Apply a combination of integrals and derivatives to this scale-dependent quantity, to extract the leading-order curvature correction in a Riemannian manifold. 3) Replicate these calculations on the metric space in question, and use the result as a definition of scalar curvature, associated with this quantity.

We introduce two different methods $\mathcal{SC}_1$ and $\mathcal{SC}_2$ given in (5.11), that involve different combinations of integrals and derivatives. Their expansions in a Riemannian manifold, using $\|S_p^r\|$ and $P_p^n$, are given in (5.4) and (5.8) respectively. The reason for introducing two methods is that, in a discrete setting, different methods introduce different undesired numerical effects like noise and lattice artefacts. The methods are compared in order to find the one that has the most desirable features, which can be different in a different setting. We tested these definitions on random geometries generated by constructing Delaunay triangulations of a sphere and a periodic plane, and we find that applying these prescriptions to determine the generalized scalar curvature of the corresponding Delaunay triangulations, gives consistent results with the Ricci scalar of the underlying Riemannian manifolds. Hence, we conclude that this is a confirmation that our generalized notions of curvature can indeed be applied reliably to discrete metric spaces, and that their result can be interpreted as a notion of scalar curvature.

To determine the curvature of the ground state of the path integral in 2D Causal Dynamical Triangulations, we also measured the generalized notions of scalar curvature in this setting. In the quantum gravity setting we extend our definitions to a diffeomorphism-invariant quantum observable $\mathcal{QSC}$, by averaging over all points in the geometry (5.19). We estimated the expectation value of the $\mathcal{QSC}_i$ associated with $\|S_p^r\|$ and $P_p^n$, and determined that the expectation value of all curvature definitions are compatible with flatness; the results are given Fig. 5.6 and 5.7. This seems to suggest that the ground state of the 2D CDT path integral is flat. We cannot make a clear physical interpretation of this result because 2D CDT has no classical limit. In fact, the system is quantum at all scales, and to the best of our knowledge, there is no analogue in our universe that can be modeled with the quantum geometries obtained in this setting. Since 2D CDT is a toy model for the more complex quantum geometries that emerge in 4D CDT, we consider the successful application of our methods in 2D as a first step to implement them in the physically interesting setting of 4D CDT.

We emphasize that all our generalized scalar curvatures are equivalent in the sense that they all define the Ricci scalar curvature at small scales. There is no fundamental difference between them, only technical differences. In the discrete cases we considered the different methods do show different discretization effects, and we find that one of our definitions of curvature is better than the other. Based on our findings on the Delaunay triangulation we conclude that the best definition of curvature is $\mathcal{SC}_2(\|S_p^r\|)$ for larger distance scales. Additionally, we conclude that $\mathcal{SC}_2(P_p^n)$ provides a good prescription to obtain the curvature of the space on shorter scales, as the higher-order corrections are significantly smaller. Given that our definitions are scale-dependent, in the sense of defining a notion of curvature at a given distance scale in the space, they can be renormalized. This is extremely useful when the continuum limit of a discrete space is desired, as is the situation of CDT. In that situation, one studies the scaling of the results when changing the size of the universe itself, after performing calculations on a fixed-size discrete triangulation. This was not done in our calculations since we were virtually using an infinite size 2D universe, by using a periodically identified section of the plane. Therefore, our results are already approximately equal to the ones in an infinite size 2D CDT universe, and there was no need of rescaling. Nevertheless, it would be interesting to study the dependence of our results with the size of the system in question. As a matter of fact, even in the case of a Delaunay triangulation of a 2-sphere, one could study the convergence of the estimated scalar curvature using the different methods proposed here. We expect that at least in a classical space like the Delaunay triangulations of smooth manifolds, the numerical differences are just a discretization artefact, and once the limit of vanishing lattice spacing is taken, all prescriptions converge to the same Ricci scalar curvature value. But, we acknowledge that this convergence should in principle be studied to verify this claim, and a study of convergence is strongly recommended in a follow-up study.

Aside from our particular implementations and test spaces, the defined $\mathcal{SC}_i$ are ideal for the calculation of $n$-point functions. The reason is that they explicitly define scalar curvature, without additive or multiplicative unknown constants. Furthermore, our definitions allow us to determine more than just the presence or the absence of curvature on a space. Since there are no unknown multiplicative constants in our observables, we can extract the scalar curvature of the given space, under some approximations explained in the previous sections. This was done for the Delaunay triangulation of the sphere, obtaining results compatible with the underlying space being approximated by the triangulation. In future research, it would be interesting to apply these definitions to determine $n$-point functions of the scalar curvature in CDT, obtaining a non-perturbative result for the correlation function of the scalar curvature in quantum gravity. Furthermore, it would be interesting to extend the results of the previous chapter to study the homogeneity properties of CDT with respect to these generalized curvatures. This is left for future work.

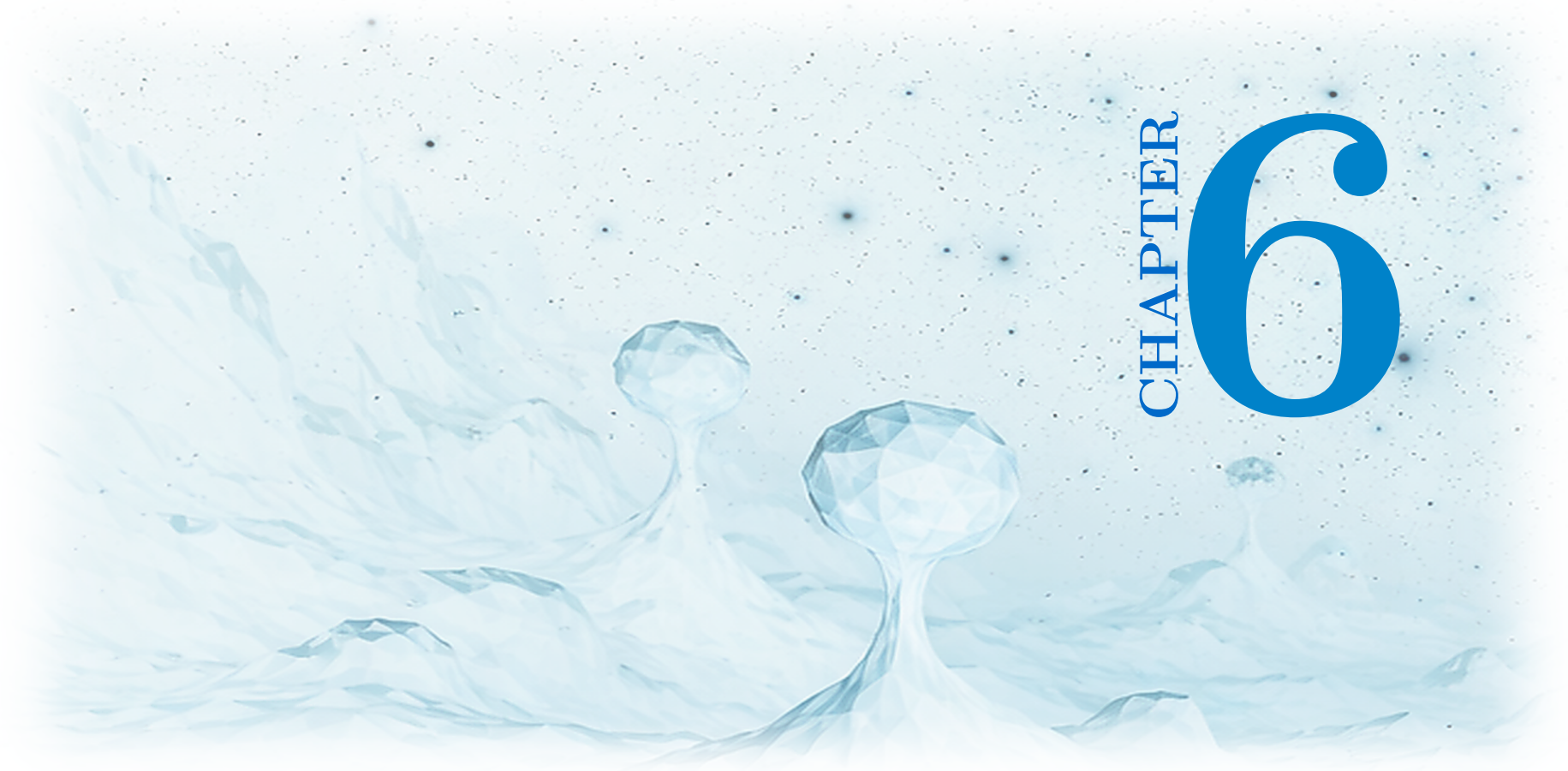

# Summary part II

The central objective of this part of the thesis was to construct and implement observables that make it possible to study the *emergence of symmetries* and geometric regularities in quantum spacetimes generated by Causal Dynamical Triangulations (CDT). Given that CDT configurations lack coordinates, tensor fields, or continuous group actions, traditional definitions of homogeneity, isotropy, or curvature cannot be applied directly. Therefore, we developed a hierarchy of scale-dependent, diffeomorphism-invariant observables; each designed to overcome one of these limitations and to build a bridge between CDT's discrete microscopic dynamics and the smooth, symmetric behavior expected in a semiclassical limit.

**Chapter 3 - Effective Topology.** We introduced a quantitative notion of *effective topology* based on tools from *Topological Data Analysis*. By coarse-graining the simplicial geometries at varying resolutions and computing their Betti numbers, we extracted a "topological fingerprint" that characterizes how the connectivity of quantum spacetime evolves with scale. In two dimensions, this method distinguishes between Lorentzian and Euclidean dynamical triangulations and identifies the presence or suppression of "baby universes" and local necks. Its outcome helps distinguish whether a quantum geometry can effectively behave as a smooth manifold at macroscopic scales, a prerequisite for any subsequent discussion of geometric symmetry. Besides its conceptual novelty, this chapter demonstrated the first systematic use of ideas inspired by persistent homology in CDT, thereby providing a scalable, coordinate-free diagnostic of the foaminess of spacetime and a new cross-check for semiclassical consistency.

**Chapter 4 - Homogeneity.** Building upon the topological characterization, we defined *scale-dependent measures of homogeneity* for discrete manifolds. Using local quantities accessible in CDT, like coordination numbers, deficit angles, and local Hausdorff dimensions, we constructed relative statistical fluctuations that quantify how homogeneous a configuration is at a given probing scale. This allowed us to identify the distances at which analogue concepts of translation symmetry could effectively emerge, such as large-scale homogeneity and isotropy. The framework also highlights the ambiguities in defining homogeneity in the absence of coordinates, and proposes canonical, computationally efficient observables that can later be generalized to higher dimensions. The results indicate that 2D Lorentzian CDT geometries display homogeneous behavior beyond a certain scale, strengthening the picture of a program for the emergence of a de Sitter-like ground state in 4D CDT.

**Chapter 5 - Curvature as a Geometric Observable.** Going beyond topological or statistical measures, we introduced a family of generalized scalar curvatures suitable for discrete and metric-measure spaces, to fine-grain our knowledge of these quantum geometries. Two independent constructions were developed: one based on volume growth of geodesic balls, and another on diffusion and return probabilities of random walks. These definitions reproduce the correct continuum limits for smooth manifolds while remaining numerically stable on triangulated geometries. Applied to CDT data, they reveal a near-flat average curvature, in contrast with previous research, and provide a more direct way to compute curvature correlators, something previously limited by the arbitrariness and computational cost of Regge calculus and other measures of curvature. These curvature observables thus supply a coordinate-free probe of the effective geometric nature of CDT, opening the door to more efficient studies of curvature correlation functions in CDT.

**Synthesis and a bridge to the next chapter.** Together, these three chapters established a toolkit of observables capable of probing topology, homogeneity, and curvature in fully non-perturbative quantum geometries. They collectively hint that, at least in 2D, CDT configurations satisfy the necessary conditions for approximating smooth manifolds and that geometric symmetries can indeed *emerge* at large scales without being imposed. The methodologies, TDA-based topology, statistical homogeneity measures, and generalized curvature, constitute complementary layers of description that can be extended to four dimensions. Beyond improving the numerical and conceptual status of CDT, these tools enhance the precision of correlation measurements and enable new cross-checks of continuum behavior.

Finally, while CDT provides a well-defined ultraviolet completion of the gravitational path integral, its direct phenomenological output remains limited because of the size of the simulated spacetimes, at least so far. While it would be desirable to have both the specific UV completion of quantum gravity, and simultaneously a scale window of observable phenomenology, this is unfortunately not the case with CDT.

This is the reason why in the next part of the thesis, we turn to the *Functional Renormalization Group (Asymptotic Safety)* program, which offers the means to connect UV completions of gravity, like CDT, and their microscopic dynamics to continuum effective field theories, effectively bridging the UV with the IR, and allowing us to test observable phenomenology.

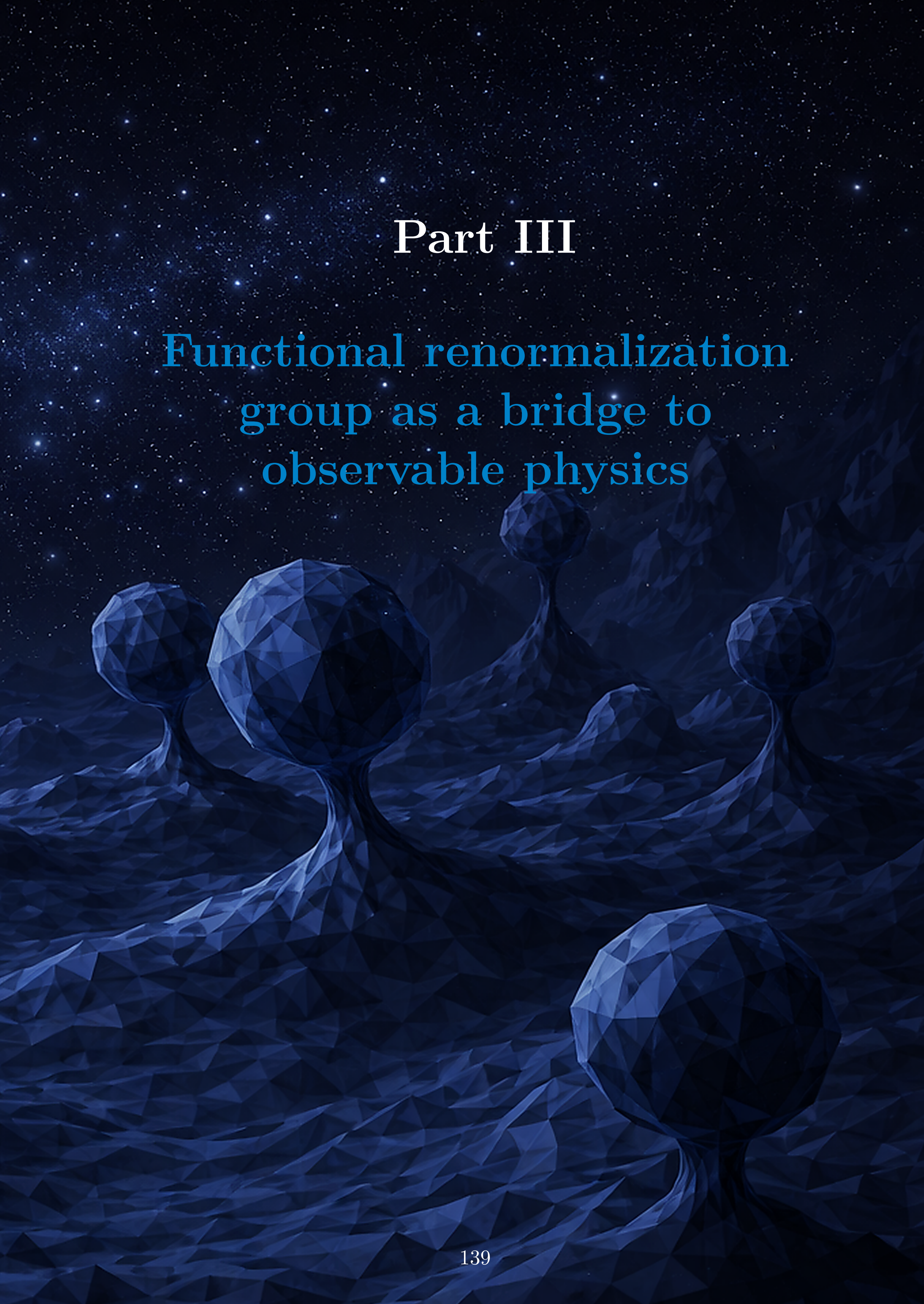

# Part III

# Functional renormalization group as a bridge to observable physics

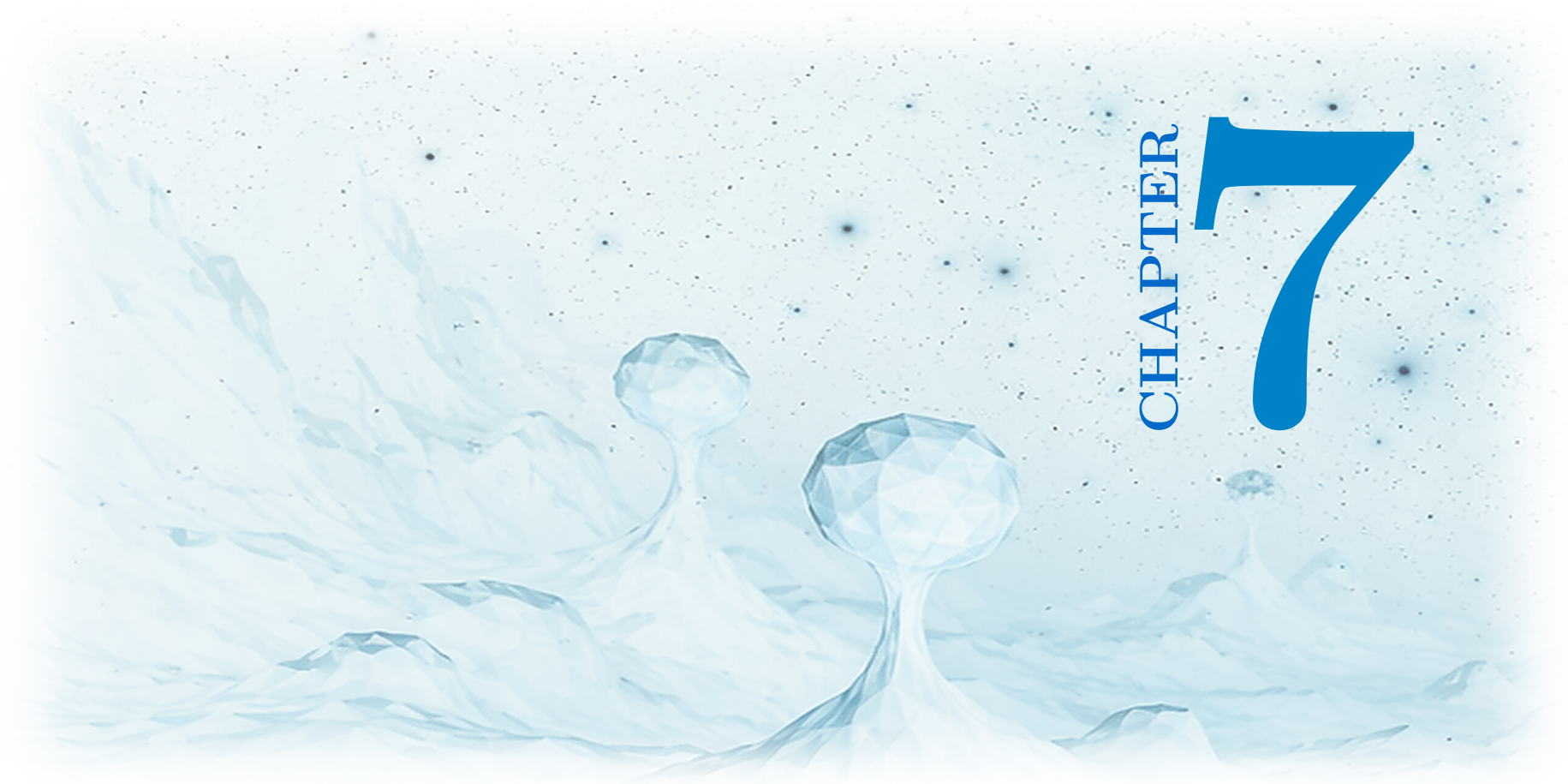

# Effective Field Theory and the FRGE

In Part I we mentioned how the notion of a quantum scale-invariant limit of a Quantum Field Theory (QFT) is a mechanism to achieve the well-posedness of the theory at high energies. In the case of quantum gravity, this mechanism will play an essential role in this part of the thesis. This mechanism of UV quantum scale invariance of any QFT is called *Asymptotic Safety*. While we already provided a specific candidate for the realization of such UV completion of gravity, called Causal Dynamical Triangulations, we also established that in that approach, the exploration of observable phenomenology is currently out of scope. In this part of the thesis, we will explore other non-perturbative techniques that allow us to explore the space of all possible Asymptotically Safe QFTs of gravity, regardless of the specific path integral formulation of such a theory. These techniques fall under the name of "Functional Renormalization Group" (FRG) techniques.

In the previous part of the thesis, we literally plunged into individual quantum geometries of CDT. In this part, the picture is quite different. Instead of exploring individual path integral geometries, we will explore an abstract space of possible Quantum Field Theories, whose coordinates are couplings determining the QFT, and where trajectories are renormalization group flows. The FRG will be our map and compass in this landscape.

We will explore the theory space of UV completions of gravity, and furthermore, we will use FRG methods to extract phenomenology of these

7.0

UV completions, in settings that can rule out entire classes of theories, such as inflationary cosmology. We will come back to these in the following chapters.

The power of the techniques used in this part of the thesis is remarkable. With the methods we will explore here, we will be able to build entire theories from just a few parameters determined by the UV completion of the QFT. The immense predictive power coming from the Asymptotic Safety scenario will allow us to construct QFTs parameterized by just a few parameters determined by the UV completion of the theories. These few parameters will determine the values of infinitely many couplings of the theory. The most tantalizing idea within this framework, is that a realization of the Asymptotic Safety hypothesis provides a highly predictive setting for gravity and the Standard Model.

The Asymptotic Safety scenario, in which gravity is described by a quantum field theory whose renormalization group flow approaches a non-trivial ultraviolet fixed point, has been deeply studied over the last three decades. The outcome of such exploration has been extremely rich and varied, covering topics in physics from the Standard Model (and beyond), to cosmology, black holes, and extensions to General relativity and their phenomenology. In this sense, substantial evidence for such UV completion of gravity via a fixed point has been accumulated in FRG studies [135–157], and references therein.

Before diving into the phenomenological aspects of these UV completions of gravity, it is important to understand the underlying mathematical structure behind them, and how the FRG can give us access to study the theory space in question from which phenomenological predictions are made.

The framework of QFT allows the construction of effective theories that describe the classical dynamics of the expectation values of the quantum fields. This effective dynamics is modeled by the so-called *Effective Action*. To construct the effective action and understand what it means, one has to go through the partition function. In this chapter, and also in the rest of this part of the thesis, we will work in Euclidean signature, and we will not be concerned about the notion of time; therefore, we will only focus on Euclidean QFTs.

At this point, it is useful to make explicit how this Euclidean choice should be read. As discussed in Part I, working in Euclidean signature turns the path-integral weight into a real factor and thereby places us, technically, in the realm of statistical field theory. The FRG flow we will study should therefore be interpreted as a *coarse-graining flow* for Euclidean effective descriptions of a field theory of the metric. As explained before, in gravity, this distinction is highly important because the naive Wick-rotation logic is obstructed by the absence of a preferred time variable and by the dynamical nature of the metric itself.

In this part of the thesis, we will not attempt to settle these conceptual issues. Instead, we adopt the standard working assumption behind most of the Asymptotic Safety evidence accumulated so far: Euclidean FRG computations can reliably capture universal properties of Quantum Gravity, and these properties can then be used to constrain infrared physics. This assumption is backed up by compelling works on the lines of Lorentzian FRG computations, which suggest that the so far obtained Euclidean results are at least qualitatively reliable even without a specific notion of time [149, 153–158]. Having established this disclaimer, let us dive into the (Euclidean) effective action.

Let us start with a certain QFT, whose field content is denoted by $\varphi$, and its Euclidean action is denoted by $S^E[\varphi]$. In this thesis, we will use the so-called *background field method*, which corresponds to splitting the field variables into a known (arbitrary) background field $\bar{\phi}$ and a fluctuation field $\hat{\varphi}$ such that $\varphi = \bar{\phi} + \hat{\varphi}$. The quantum theory then corresponds to integrating its Euclidean partition function with a source $J$, over the fluctuation field $\hat{\varphi}$. This integration is given by

$$Z[J, \bar{\phi}] = \int \mathcal{D}\hat{\varphi}\, e^{-S^E[\varphi] + \int d^D x\, J(x)\hat{\varphi}(x)}.. \tag{7.1}$$

From this partition function, one can compute every correlation function of the theory, by taking functional derivatives with respect to the source

$$\langle \hat{\varphi}(x_1) \cdots \hat{\varphi}(x_n) \rangle = \frac{1}{Z[J, \bar{\phi}]} \frac{\delta^n Z[J, \bar{\phi}]}{\delta J(x_1) \cdots \delta J(x_n)} \bigg|_{J=0}. \tag{7.2}$$

One can define another object that encodes the effective dynamics of the source $J$, by taking the logarithm of the partition function $W[J, \bar{\phi}] \equiv \ln Z[J, \bar{\phi}]$. This quantity $W[J, \bar{\phi}]$ generates the connected correlation functions, from which all physical quantities of the QFT are derived. Applying a Legendre transformation to $W[J, \bar{\phi}]$, one obtains the effective description of the expectation value of the field

$$\Gamma[\phi,\bar{\phi}] \equiv \sup_J \left\{ \int d^D x \, J(x)\phi(x) - W[J,\bar{\phi}] \right\} \tag{7.3}$$

$$\text{with} \quad \phi(x) \equiv \langle \hat{\varphi}(x) \rangle_J = \frac{\delta W[J,\bar{\phi}]}{\delta J(x)}\,, \tag{7.4}$$

where we defined the effective action $\Gamma[\phi,\bar{\phi}]$.

The effective action contains the effective description of the expectation value of the quantum field, and from it one can obtain its classical equations of motion by simply computing $\frac{\delta\Gamma[\phi,\bar{\phi}]}{\delta\phi(x)} = 0$. This object encodes all the information of the quantum theory, and therefore so do the resulting equations of motion. $\Gamma[\phi,\bar{\phi}]$ is potentially composed of an infinite number of terms, with an infinite number of associated couplings, all of which respect the symmetries of the original bare action $S^E[\varphi]$. This object is constructed by carefully integrating the QFT, by whatever means, such as perturbations or lattice methods, and is the quantity of interest in this part of the thesis.

Typically, a QFT is operationally defined in a certain range of energies, or equivalently, momentum scales. This is because most theories exhibit either IR or UV divergences, as mentioned in Chapter I, which leads to an energy dependence of the effective action.

The energy dependence of the effective action, or, as we will use equivalently in this thesis, the momentum dependence, is the result of a very complex chain of mathematical intricacies, such as the introduction of momentum cut-off regulators in the calculations, which leads to an energy dependence of the action to avoid physical results having a dependence on mathematical artifacts. The study of the momentum dependence of the effective action is called Renormalization Group. The functional dependence of the effective action on momenta forms a curve in theory space, and this curve is called an RG trajectory. The change of the effective action at different energies is often called RG flow. Since a detailed explanation of this is out of the scope of this thesis, we refer the reader to [159] and references therein.

The momentum dependence of the theory is encoded in the renormalized couplings of the theory. There are several techniques to determine the functional dependence of $\Gamma[\phi,\bar{\phi}]$ with respect to the fields and the energy. For example, in perturbation theory, one could use Feynman diagrams to construct the effective action in powers of some small couplings of the bare theory, assuming that contributions of the fluctuation field $\hat{\varphi}$ are small compared to the background $\bar{\phi}$. Since in this thesis we are focused on the non-perturbative aspects of a QFT of gravity, we will not follow this path. Instead, we will use a non-perturbative tool called (in this thesis) the *Functional Renormalization Group Equation* (FRGE).

The FRGE is a non-perturbative equation that determines the RG flow of an IR-regulated generalization of the effective action called the *Effective Average Action* (EAA), denoted by $\Gamma_k[\phi, \bar{\phi}]$, or simply by $\Gamma_k$, which depends on the IR regulator momentum scale $k$. The EAA contains the information of modes of the QFT with momentum $k^2 \leq p^2$, and can be interpreted as an interpolation between the physics described by UV modes ($k \to \infty$), and the full effective action, containing the information about all modes of the QFT when $k \to 0$. We will return to the interpretation of this object later on.

A derivation of the FRGE will not be presented here, and we refer the reader to [160]. The FRGE takes the form

$$k\,\partial_k \Gamma_k = \frac{1}{2}\mathrm{STr}\left[\left(\frac{\delta^2\Gamma_k}{\delta\phi^2} + R_k\right)^{-1} k\,\partial_k R_k\right], \tag{7.5}$$

where STr denotes the functional trace, which traces over all modes and indices of the fields that determine the QFT, and $\frac{\delta^2\Gamma_k}{\delta\phi^2}$ is the second functional derivative of the EAA with respect to the field variable $\phi$, at fixed background $\bar{\phi}$. The quantity denoted by $R_k$ is a momentum-dependent ($p$) IR regulator that satisfies

$$R_k(p^2) \sim k^2, \quad \text{for } p^2 \ll k^2, \qquad R_k(p^2) \sim 0, \quad \text{for } p^2 \gg k^2., \tag{7.6}$$

effectively giving a mass to low-squared-momentum modes of the QFT. A typical example of such a regulator is

$$R_k(p^2) = (k^2 - p^2)\,\Theta(k^2 - p^2), \tag{7.7}$$

with $\Theta(k^2 - p^2)$ being the Heaviside step function.

While the derivation of the FRGE makes use of a certain path integral formulation, the equation itself is independent of the bare action used to derive it. This makes it ideal for agnostic exploration of theory spaces, without particular operational implementations for deriving $\Gamma_k[\phi, \bar{\phi}]$ [161]. With this perspective, (7.5) can be used as a stand-alone equation, and therefore as a tool agnostic of a particular path integral prescription.

By using the FRGE as a stand-alone equation, in this thesis, we will explore possible solutions to it of the form

$$\Gamma_k[\phi, \bar{\phi}] = \sum_i \bar{g}_i(k)\mathcal{O}_i[\phi, \bar{\phi}] = \sum_i g_i(k)\, k^{-d_i}\, \mathcal{O}_i[\phi, \bar{\phi}]\,, \tag{7.8}$$

where the dimensionless couplings $g_i(k) \equiv \bar{g}_i(k)k^{d_i}$ were introduced, using the mass dimension of their associated operators $d_i$, and where the sum is over a, possibly infinite, number of operators $\mathcal{O}_i[\phi, \bar{\phi}]$, engineered such that they satisfy a certain set of symmetries. For example, diffeomorphism invariance, in the case of gravity. Notice that in (7.8), the $k$ dependence of the couplings was explicitly shown for clarity, but from now onward, this dependence will not be shown.

At finite non-zero values, associating a physical meaning to $\Gamma_k$ is at best ambiguous in QFT with multiple intrinsic mass scales, since the running with the regulator scale $k$ is not the same as the running with a physical energy scale [162], and therefore, it is safest to interpret the EAA $\Gamma_k$ merely as an interpolation function between two physically interesting limits, $k = \infty$ (if that limit is well defined), and the $k = 0$ limit. Despite the interpolative nature of the EAA, the $k = 0$ limit is still the full effective action of the theory (whichever underlying specific path integration creates it), containing all the information about all the modes of the underlying QFT. This is because, at this scale, the mathematical artifacts used to define the FRGE disappear, and one recovers a physical quantity $\Gamma_{full} = \lim_{k\to 0} \Gamma_k$. In particular, at $k = 0$ it contains all the information about the underlying QFT [161]. Obtaining the object $\Gamma_{full}$ is going to be the central goal of this part of the thesis. This limit of $k \to 0$ is notoriously difficult to approach, at least with the current methods in place, so we will come back to this in the coming chapters.

The Asymptotic Safety program in gravity, explores the theory space of gravitational actions using the FRGE, with the hope of finding a suitable UV completion, i.e., the program searches for scaling solutions, *Fixed Points*

(FP) of the *beta functions* of the couplings ($\beta_{g_i}$) that satisfy

$$\beta_{g_i} \equiv k\,\partial_k g_i \quad \longrightarrow \quad \text{FP} = \{g_{i*}\} \text{ such that } \quad \beta_{g_i}|_{\text{FP}} = 0 \tag{7.9}$$

in the limit of $k \to \infty$, where the values for the couplings $\{g_{i*}\}$ are all finite. This is a very particular limit of flow, and its interpretation is that of a scale-invariant limit. Exactly at this point, increasing or decreasing the energy does not change the theory at all. Since energy is associated with an inverse distance scale, one can intuitively think of this as a point in theory space where zooming in or out in spacetime does not affect the effective description of it. One could zoom in and out, but the effective dynamics would be the same.

In particular, if such a limit is found, one has found a UV starting point for solving the FRGE. Such a UV limit is called an "ultra-violet completion". For that limit to be predictive, one requires it to have a finite amount of *Relevant Directions*. These relevant directions count the number of couplings one has to fine-tune, such that the theory approaches this limit, and they are defined as the perturbations of the FRGE that have negative eigenvalues. In particular, in the more intuitive case where one is working with a finite number $N$ of couplings $g_i$, relevant directions are defined as the left eigenvectors $V^\alpha$ of the *Stability Matrix* $B_{i\,j}$ with positive *Critical Exponents* $\theta_\alpha$, which are defined as minus the eigenvalues of those eigenvectors, as depicted in (7.10)

$$B_{i\,j} \equiv \left.\frac{\partial \beta_{g_j}}{\partial g_i}\right|_{FP} \qquad \longrightarrow \qquad \sum_i V_i^\alpha\, B_{i\,j} = -\theta_\alpha\, V_j^\alpha \tag{7.10}$$

where the $V^\alpha$ are vectors in theory space $\mathcal{T}$, which, in the case of finitely many couplings, is isomorphic to $\mathbb{R}^N$. A relevant direction $V^\alpha$ has an associated positive critical exponent $\theta_\alpha > 0$. From now on, we will call a FP that has non-trivial critical exponents and a finite amount of relevant directions a Non-Gaussian Fixed Point (NGFP).

With these directions, one can span the asymptotically safe theory space near the FP, meaning all the points that will approach the FP in the $k \to \infty$ limit, by approximating near the FP

$$g_i \approx g_{i*} + \sum_\alpha C_\alpha\, V_i^\alpha \left(\frac{k_0}{k}\right)^{\theta_\alpha}, \tag{7.11}$$

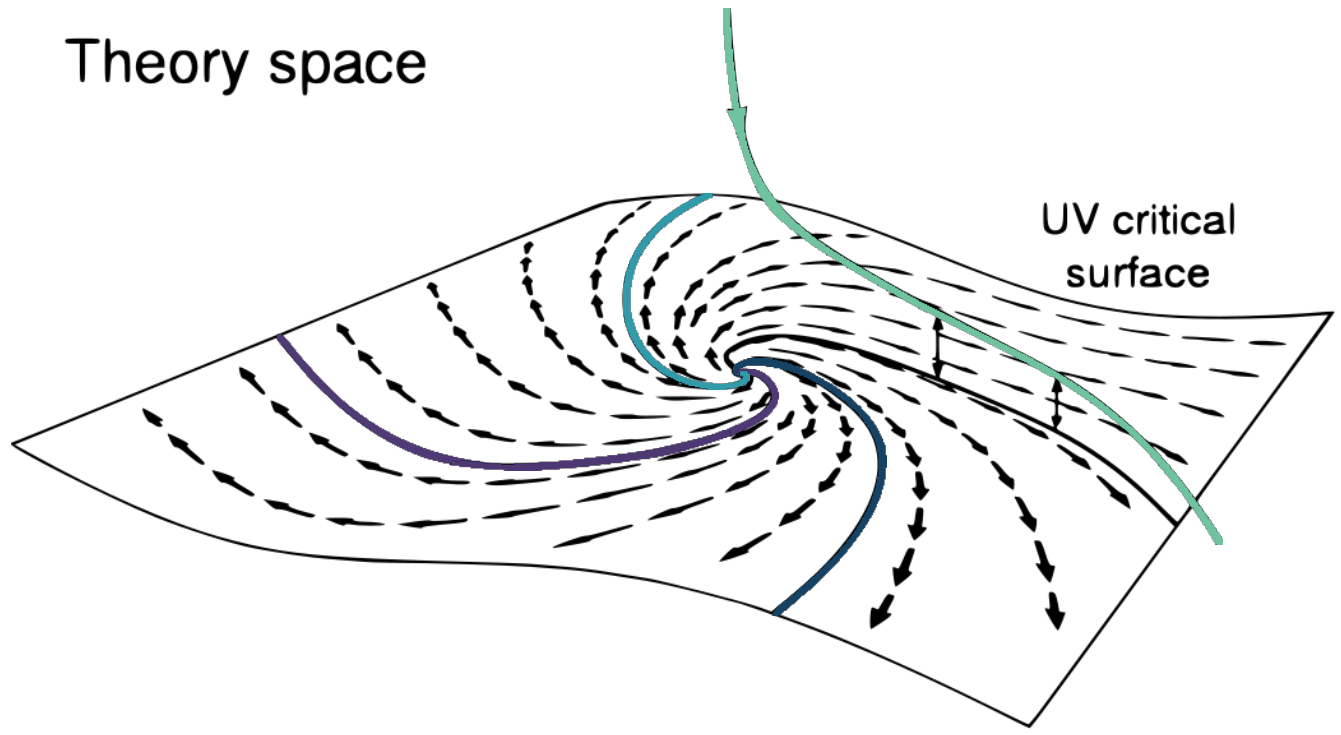

**Figure 7.1.:** Representation of $\mathcal{S}_{UV}$. The arrows are the vector field given by the beta functions of the theory, pointing towards the IR. The light blue, dark blue, purple and black solid lines, are different theories flowing towards the central FP, and are therefore contained in $\mathcal{S}_{UV}$. The light green trajectory is not contained in $\mathcal{S}_{UV}$, and therefore diverges from the FP when approaching the UV.

where there is an arbitrary reference scale $k_0 \gg 0$, and arbitrary linear combination constants $C_\alpha$. The predictivity of an Asymptotically Safe QFT relies on it having a finite number of relevant directions, so that there is a finite number of free parameters one would have to tune in theory space, to find a renormalization group trajectory that would reach the UV NGFP. In such a case, since the relevant directions are linear combinations of couplings, they are then parameterized by a finite number of relevant couplings (this is a choice one can make among the parameters that appear in each $V^\alpha$). Tuning these free parameters close to the FP is equivalent to selecting an RG trajectory that reaches the FP in the UV. This trajectory only required a finite amount of fine-tuning, and its endpoint at $k = 0$ only depends on this finite number of parameters; therefore, the theory obtained from the values of the couplings at $k = 0$ has a finite number of free parameters that created it.

This is the predictive power of Asymptotic Safety. If one finds a UV-complete EAA, with a NGFP, then the obtained $\Gamma_{full}$ at $k = 0$ is determined by a finite number of free parameters. All irrelevant couplings are then functions of the relevant ones. There is a small caveat in this affirmation, since the FRGE is a functional differential equation, and thus, boundary conditions also contribute to the parameterization of $\Gamma_{full}$, but we will return to this in the last chapter of this part of the thesis.

The subset of theory space that is parameterized by these relevant directions is called the "UV Critical Surface", and it will be denoted by $\mathcal{S}_{UV}$. It contains all trajectories that in the limit $k \to \infty$ arrive at the NGFP. Trajectories that are not contained in this surface will diverge from the NGFP in the limit $k \to \infty$, and are thus not UV complete. The $\mathcal{S}_{UV}$ is the region of theory space $\mathcal{T}$ that is asymptotically safe, parameterized by a finite number of couplings. A pictorial representation of $\mathcal{S}_{UV}$ can be seen in Fig. 7.1. This surface will be used to reach the $k = 0$ limit of gravitational theories, and will be a central part of the next chapter.

## 7.1 Background approximation and truncations

While the formal expression given in the FRGE is exact and works without an explicit path integral formulation that constructs the EAA, it is a functional equation for the EAA, which, in principle, has infinitely many terms. To explore the theory space of possible UV-complete EAA, in practice, one must resort to certain approximations. In this thesis, we will work with two main approximation schemes, one called the *background approximation*, and another extra approximation method called *truncation* [159].

The background approximation evaluates the FRGE at zeroth order in the mean fluctuation field $\phi$. Importantly, this projection is performed *after* constructing the Hessian entering the right-hand side of the flow equation. Thus, one first computes $\frac{\delta^2 \Gamma_k[\phi, \bar{\phi}]}{\delta \phi^2}$, taking two functional derivatives with respect to the mean fluctuation field at fixed background, and only afterwards evaluating the result at $\phi = 0$. In gauge theories, and in gravity in particular, the gauge-fixing, ghost, and regulator sectors retain a separate dependence on background and fluctuation fields during this calculation. For the linear split used here, setting $\phi = 0$ implies that the expectation value of the total field coincides with the background field. This is a standard and convenient approximation, since it simplifies further computations significantly. While there is no a priori reason for doing this from an agnostic point of view of effective field theory, more sophisticated techniques that take into account this fluctuation field [163], suggest that the background approximation gives results that are consistent with the existence of a UV FP for gravity, therefore making the approximation a good toy model to test new tools, as we will do in this thesis. In what follows, we denote this projected, zero-fluctuation sector simply by

$$\Gamma_k[0,\bar{\phi}] = \Gamma_k[\bar{\phi}] = \sum_i \bar{g}_i \mathcal{O}_i[\bar{\phi}] = \sum_i g_i\, k^{-d_i}\, \mathcal{O}_i[\bar{\phi}]\,, \tag{7.12}$$

already stating that we will use the background approximation of the FRGE. This approximation should therefore be understood as the projected zero-fluctuation sector used to parameterize the flow, and not as implying that the Hessian in the FRGE is obtained by taking derivatives with respect to $\bar{\phi}$.

On top of this approximation, in practice, one must resort to an extra approximation, called truncation. A truncation is a projection of the FRGE into a controlled, known, basis of operators. What is done, is to choose a basis of operators that belong to the theory space of the studied QFT, express the EAA as a linear combination of those operators, and project the FRGE into that basis. These truncations are chosen based on expectations of what the Effective Action of a certain QFT would be. For example, in the case of Gravity, the EAA is typically chosen to contain the same operators as the Einstein-Hilbert action plus other terms that are allowed by diffeomorphism invariance, such as higher curvature terms, or interactions with other matter fields that are expected to appear in the effective action of gravity. Notice that a projection is different from expansion, in the sense that none of the terms in the EAA are assumed to be small.

There are two types of truncations: *non-functional truncations*, and *functional truncations*. The first case is the simplest, so we will start with it. A non-functional truncation is one that only involves a finite linear combination of operators whose dependence with respect to the field is known. For example, in the gravitational case, one could do the so-called Einstein-Hilbert truncation where

$$\Gamma_k[g] = \int d^4x\,\sqrt{g}\,\frac{R-2\,\bar{\Lambda}}{16\,\pi\,\bar{G}} = \int d^4x\,\sqrt{g}\,\frac{k^2\,R-2\,k^4\,\Lambda}{16\,\pi\,G}\,, \tag{7.13}$$

where the Ricci scalar $R$ is computed in the background field, and the $k$ dependence is encoded in the couplings $G$ and $\Lambda$. Here, the only unknowns are the $k$ dependence of the couplings, not the operators themselves. This was the first truncation used to test for AS in gravity, giving rise to the now well-established *Reuter fixed point*. An example of the RG flow of this truncation is shown in Fig. 7.2, where the beta functions were taken from [164].

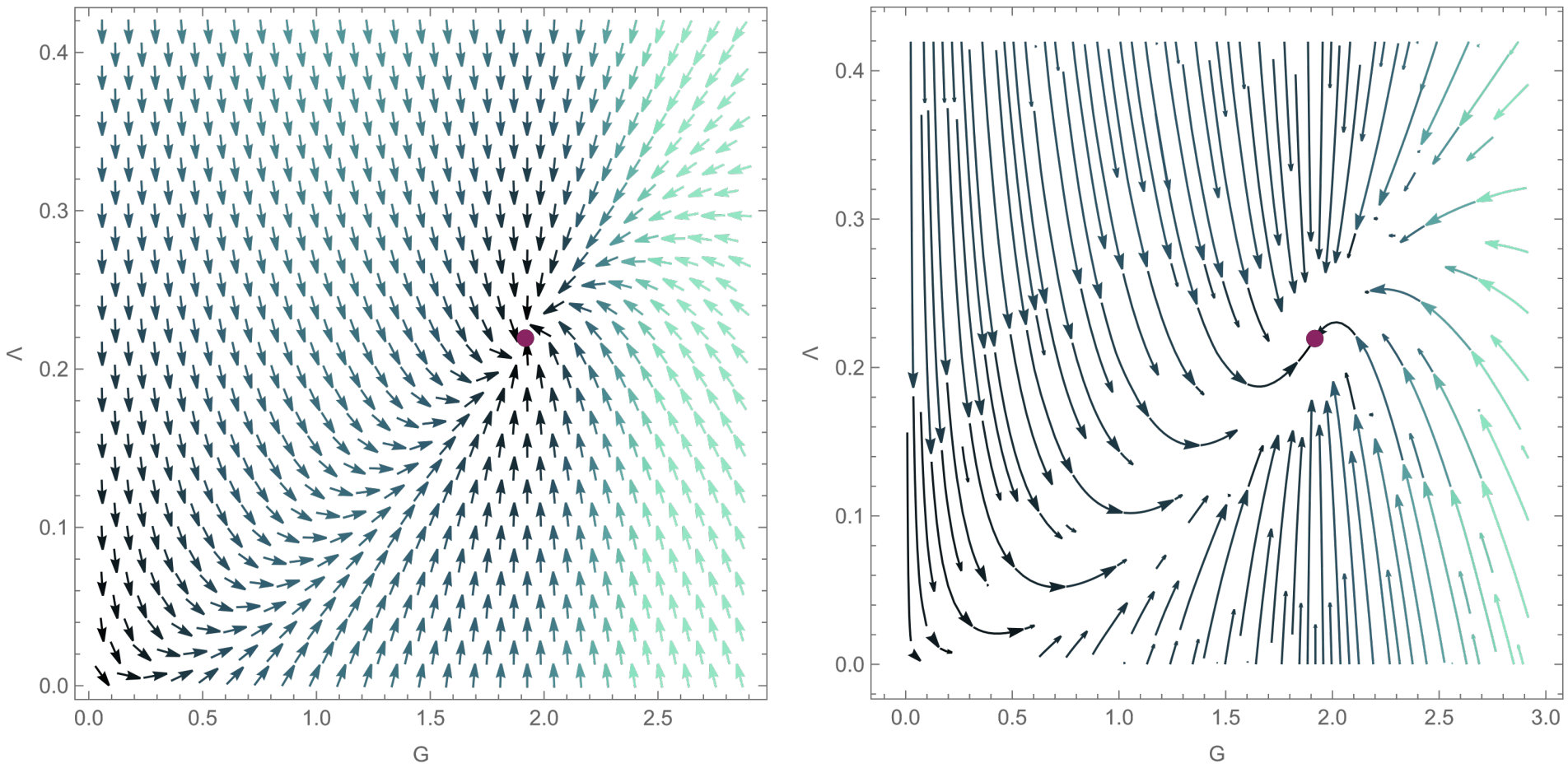

**Figure 7.2.:** RG flow of the Einstein-Hilbert truncation. The left image is a vector field plot corresponding to the beta functions of Newton's coupling and the cosmological constant's beta functions. The right image shows the integral lines of that same vector field. Notice that the arrows point towards the UV, as opposed to Fig. 7.1, ending at the attractive FP marked with a dot. The beta functions were taken from [164].

Of course, more sophisticated truncations have been studied, establishing that the Reuter fixed point prevails regardless of the approximations or technical artifact used, such as regulator choice, gauge choice, and background, thus giving compelling evidence for the existence of a UV completion of gravity [135–157].

In this thesis, we will also explore other kinds of truncations, called *functional truncations*. These types of truncations are built on unspecified operator dependence of the EAA. For example, one could use a functional truncation for pure gravity of the form

$$\Gamma_k[g] = \int dx^4 \sqrt{g}\, F_k(R)\,, \tag{7.14}$$

called $F(R)$ truncations. Here, one assumes that the EAA is an unknown scalar function of the Ricci scalar, whose dependence on $R$ and $k$ will be given by solving the FRGE. These truncations are more general than the finite truncations and make direct use of the power of the FRGE to treat functional dependence non-perturbatively. We will come back to these types of truncations in the next sections.

There are some caveats to these approximations. Every calculation requires a gauge fixing for the fluctuation field, and therefore the addition of ghost and gauge fixing terms in the EAA. On top of that, the projection of the FRGE on a particular operator, requires the choice of particular backgrounds that isolate certain terms, allowing the extraction of the beta functions of their associated couplings. In addition, each particular calculation requires the choice of a specific IR regulator. While background and truncation approximations are very useful for computational purposes, approximations of the FRGE lead to the dependence of the results on the mathematical artifacts used for performing the calculations.

The aim of this part of the thesis is to develop new tools to make predictions from an UV completion of gravity, based on the assumption that the RG flow of the system was computed by other means a priori. It is out of the scope of this thesis to assess the dependence of our results on the previously mentioned mathematical artifacts. Therefore, we will not treat these very important topics, and we will work on top of previously computed RG flows for UV-complete theories.

This is the organization of the following chapters. In chapter 8, we construct novel techniques to map the UV Critical Surface of QFTs within finite truncations. For this, we will have to submerge into the use of Neural Networks, and into the depths of the interplay of gravity and scalar matter, for extracting phenomenologically interesting predictions. In chapter 9, we develop new techniques to extract predictions from functional truncations of the FRGE, and in particular, we will explore how AS can provide models for the early universe expansion, compatible with observations. We will explore inflationary cosmology and the beautiful story told by the Cosmic Microwave Background. And at last, in chapter 10 we conclude with a small summary of the achievements and possible new avenues to explore.

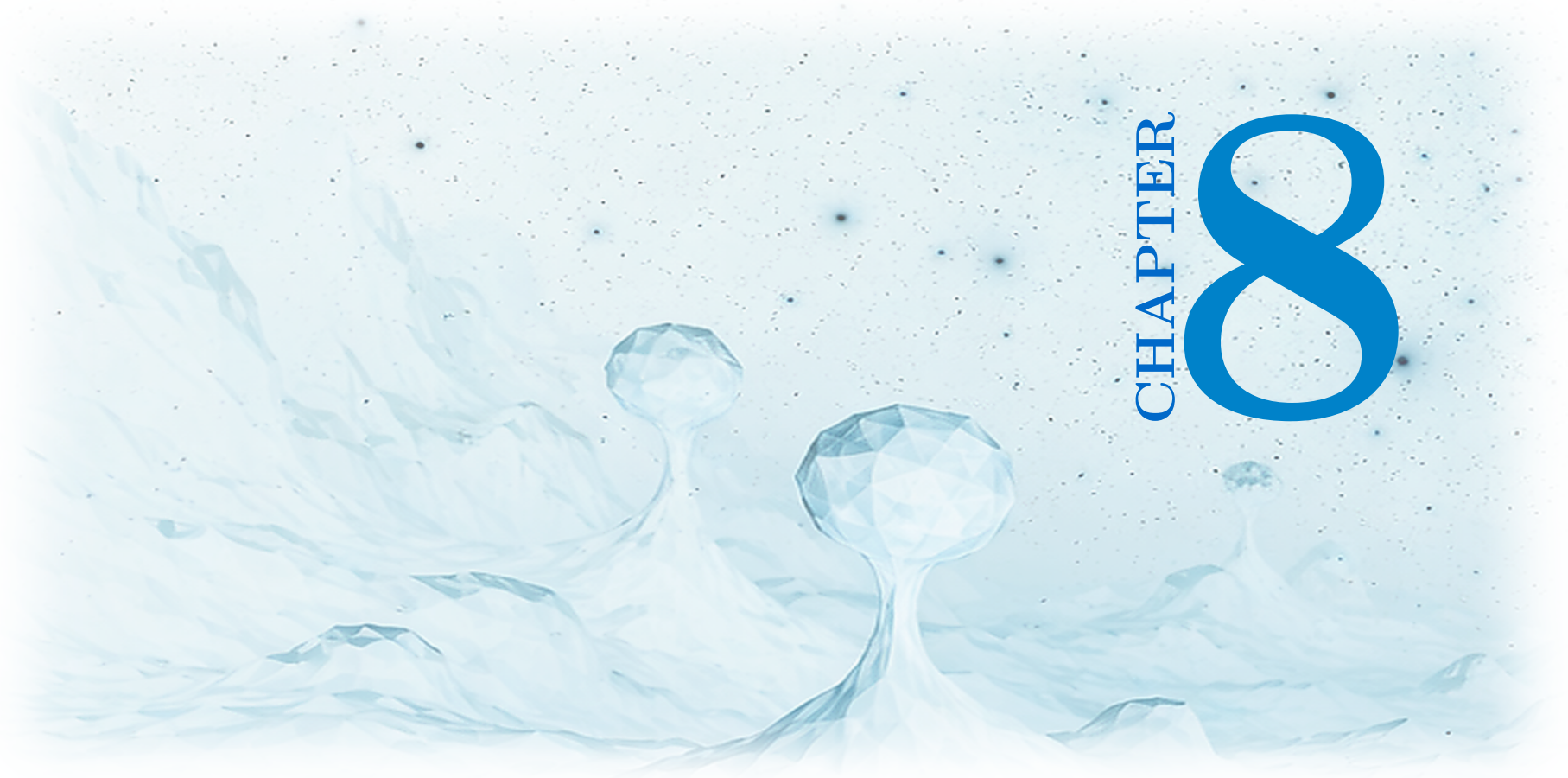

# Flow Invariants to make IR predictions

To better understand how the techniques developed in this part of the thesis apply to QG, we will take a small detour to put the developed methods and the motivation behind them into a larger setting.

Our theoretical understanding of elementary particles and their interactions rests on the standard model of particle physics. This highly successful model is capable of explaining the outcome of particle physics experiments based on approximately 20 free parameters, which are fixed from experimental inputs [165]. The modern understanding of the theory is that it is an effective field theory: while it works very well at energy scales probed by the most energetic collider experiments, it is expected to break down once the energy scales of the physical processes exceed a certain threshold [29]. Moreover, the model is incomplete in the sense that gravitational effects are not taken into account.

It is then a thought-provoking idea that this model, and its extension by including the gravitational interactions, can be obtained from a more fundamental theory where just a few parameters at high energies determine the entirety of the effective field theory. As explained before, within the realm of relativistic quantum field theory, such a situation may be realized through the asymptotic safety mechanism, using the FRGE. In the AS scenario, the enhanced symmetry, called quantum scale invariance, provides significant predictive power, potentially forecasting and postdicting free parameters of the standard model effective field theory, including the Higgs mass and quark masses [166, 167]. It may also lead to predictions testable within the context of inflationary cosmology [7, 168]. These prospects provide a strong incentive to develop efficient computational tools to map out the predictions cast by asymptotic safety and test their robustness with respect to approximations made in their derivation.

Before explaining the core ideas behind this chapter, it is important to set the stage. In all the cases analyzed in this chapter, the domain where we will apply our methods is a compact subset, denoted as $\Omega$, of a truncated theory space $\mathcal{T} \sim \mathbb{R}^N$ for some finite integer $N$, where the explored domains take the form of products of compact intervals, i.e. hypercubes, and our expositions and developed techniques will be restricted to such subspaces. Throughout this chapter, we will work under the assumption that there is a UV-critical surface $\mathcal{S}_{UV}$ of dimension $M$ embedded in $\mathbb{R}^N$, and we will attempt to find it. While our main goal is to apply these ideas to renormalization group flows of QFTs, it is easier to introduce the developed techniques in simpler settings, as we will do below. With these ideas in mind, we are now in the position of introducing the notion of what we call in this thesis a *Flow Invariant* (FI), a central quantity of this chapter.

Given a vector field $\beta : \Omega \to \mathbb{R}^N$ a *Generating Function* (of the field $\beta$) is a function $F : \Omega \to R$ which satisfies

$$\beta \cdot \nabla F = 0\,, \tag{8.1}$$

everywhere in $\Omega$, where $\cdot$ is just the Euclidean dot product in $\mathbb{R}^N$, and $\nabla$ is the gradient operator. This implies that $F$ is a flow invariant, i.e., is a constant along the flow lines of $\beta$.

The idea is that if one wants to find a $M$-dimensional subspace $\mathcal{S}$ in $\Omega$, that is tangent to the flow of $\beta$, one can determine it from the intersection of the zeroth level sets, a.k.a. kernels, of $N - M$ solutions $F_i$ with $n = 1, \ldots, N - M$ of (8.1), obtaining

$$\mathcal{S} = \left\{x \in \{\{F_1^{-1}(0)\} \cap \{F_2^{-1}(0)\} \cap \ldots \cap \{F_{N-M}^{-1}(0)\}\right\}. \tag{8.2}$$

This set of generating functions must be independent, in the sense that their gradients must be linearly independent in $\Omega$. Of course, not every vector field $\beta$ admits such a set of independent generating functions $\{F_i\,, n = 1, \ldots, N - M\}$ on any given domain.

At least in the studied domains, the vector fields with which we worked have no intersecting flow lines, are at least $C^1$ smooth, and thus one can guarantee that local solutions $F_i$ satisfying the criteria we spelled out before exist. This can be seen by simply choosing a set of coordinates locally orthogonal to the vector field at a point.

Nevertheless, whether such solutions $F_i$ can be found in a larger domain cannot be assured for any vector field $\beta$, and this is something that we assumed to be true in the studied examples, and we found no obstacles to doing so. This is a strong assumption, but we will see that at least in the studied cases, this assumption leads to correct numerical results. A proof of the existence of such a set of generating functions is out of the scope of this thesis.

Going back to the course of the explanation of the idea, under these assumptions, the kernel of a generating function contains all integral curves of the field $\beta$. This is the reason why FIs are ideal candidates for finding "surfaces" that contain all possible flow trajectories, called like that in this part of the thesis, even if they have a dimension different from two. We will explore how these flow invariants can be constructed with a mix of analytical and numerical techniques, and harness their full power later on when applying our developed formalism to vector fields $\beta$ that describe RG flows.

The set of solutions of (8.1) is degenerate, since any monotonic $C^1$ function $\tilde{F}(F)$ is also a solution. If it satisfies $\tilde{F}(0) = 0$, then this new solution also shares the same kernel. Fortunately, this non-uniqueness can be eliminated via a Taylor expansion of (8.1) close to a certain point, by fixing the lowest order coefficients of a Taylor expansion of $F$ at that point to values that solve the equation at that order. This requires $\beta$ to be sufficiently smooth at the point of expansion. In this thesis, we will work with cases where this is always true.

There is arguably no better starting point to explain a concept in physics, and to have a visual picture of this method, than the harmonic oscillator. Let us consider the simple classical 1-dimensional harmonic oscillator. This physical system is described by the Hamiltonian

$$\mathcal{H}(q,p) = \frac{p^2}{2m} + \frac{m\,\omega^2}{2} q^2 \,, \tag{8.3}$$

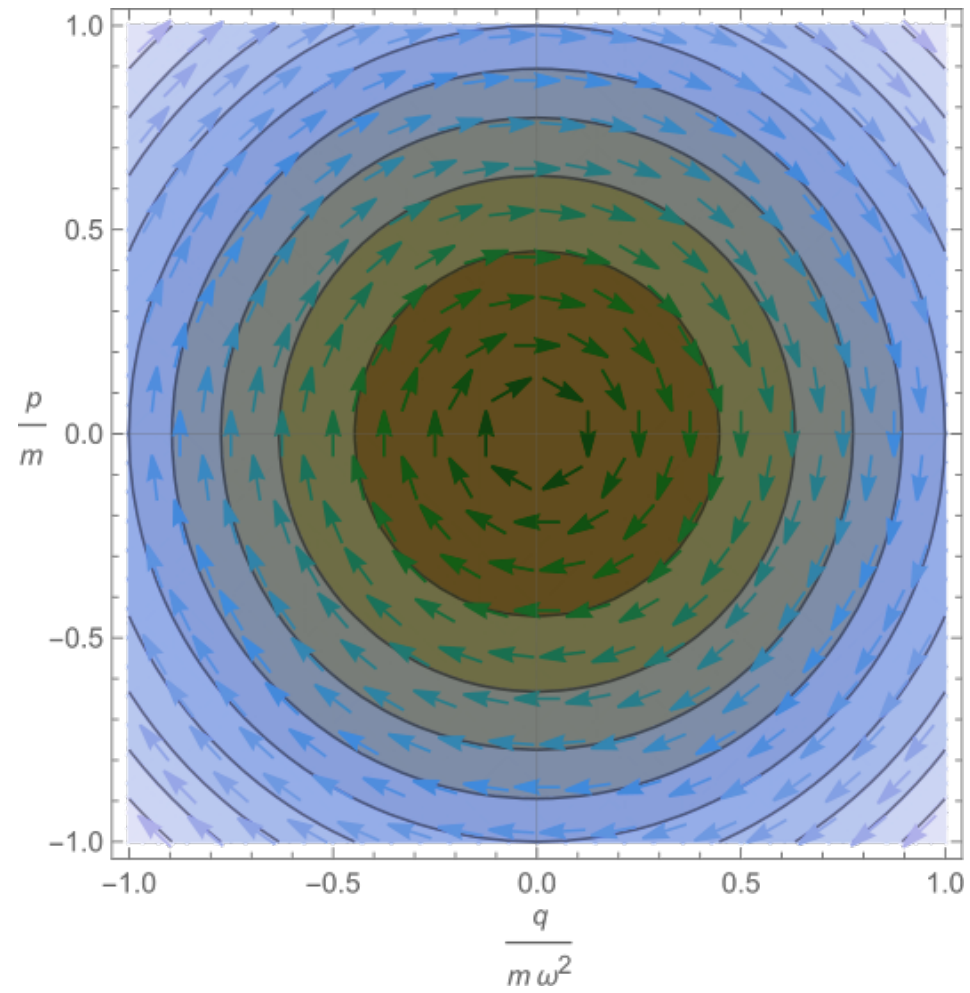

**Figure 8.1.:** Representation of the Hamiltonian vector field $\beta_{\mathcal{H}}$, together with the kernel of different $F_{\mathcal{H}}$, each associated with different initial points in phase space.

where $m$ is the mass and $\omega$ its characteristic frequency. The equations of motion of this system are given by

$$\frac{dq}{dt} = \frac{\partial \mathcal{H}}{\partial p} \qquad \frac{dp}{dt} = -\frac{\partial \mathcal{H}}{\partial q} \qquad \Longrightarrow \qquad \beta_{\mathcal{H}} \equiv \{\frac{\partial \mathcal{H}}{\partial p}, -\frac{\partial \mathcal{H}}{\partial q}\}, \tag{8.4}$$

where we defined the Hamiltonian vector field $\beta_{\mathcal{H}}$ to which all trajectories in phase space $\{q(t), p(t)\}$ of the system are tangent. Given a point $\{q(t_0), p(t_0)\} = \{q_0, p_0\}$ which is assumed to be contained in an underlying tangent surface to $\beta_{\mathcal{H}}$, which in this case should just be a one dimensional curve, one proceeds as follows. The first step, is to Taylor expand (8.1) around $p_0$, until the first non-trivial coefficient appears. The equation allows for infinitely many of those coefficients, as its solution space is non-degenerate, but once one chooses the first coefficients that solve the first order in the expansion, the rest are determined in terms of those.

In this case, the linear term is already non-trivial. Choosing the coefficients so that the result has energy units, making the interpretation of the generating function more evident, one can prove that an exact solution to (8.1) that contains trajectories that pass through the given initial point, is simply a quadratic function

$$F_{\mathcal{H}}(q,p) = \frac{p_0}{m}(p-p_0) + m\omega^2 q_0(q-q_0) + \frac{(p-p_0)^2}{2m} + \frac{m\omega^2}{2}(q-q_0)^2\,, \tag{8.5}$$

which, after rearranging, one can see that it is nothing else but the energy conservation equation

$$F_{\mathcal{H}}(q,p) = \left(\frac{p^2}{2m} + \frac{m\,\omega^2}{2}q^2\right) - \left(\frac{p_0^2}{2m} + \frac{m\,\omega^2}{2}q_0^2\right)\,. \tag{8.6}$$

When solving for its kernel, $F_{\mathcal{H}}(q,p) = 0$ gives us the subset of phase space where the trajectory passing through the initial point $\{q_0,\, p_0\}$ will lie in. A scheme of this is shown in Fig. 8.1. In this simple setting, one can say that this is a rather obvious flow invariant, since energy was known to be a conserved quantity from the start, and the level sets of the generating function (8.5) are nothing more than a rephrasing of that statement. This trivial case, where the flow invariant could be found with a mere Taylor expansion up to second order in the variables, serves as the basis to explore more complex systems.

In more complex systems, this is not as trivial, but the logic of how we approach numerical solutions of the generating functions is very similar. Suppose one is given an initial point $p_0$ which is assumed to be contained in some underlying surface tangent to a vector field $\beta$. Then, one can attempt a Taylor expansion around this point of the equation (8.1). This expansion gives the first non-trivial coefficients, and from it, one can determine the tangent surface near $p_0$. In all the settings studied in this thesis, the linearized equation is already non-trivial, allowing us to determine a normal vector to the surface at $p_0$.

Another example of a vector field $\beta$, that helps with a better visual intuition on the type of complex situations that this method can handle, is depicted in the left image of Fig. 8.2. The red dot is a given point $p_0$, which is assumed to be contained in some underlying surface, and the tangent plane to the surface is depicted in red, as well as the normal vector to that plane. The normal vector and the tangent plane are determined from a Taylor expansion of (8.1) using this particular vector field. As mentioned earlier, fixing the first coefficients of the expansion removes the non-uniqueness of the solution and provides a starting point for finding a solution of (8.1) in a certain domain. In the following sections, we will explain numerical techniques to address this problem and to progress, from the simple and reduced initial information of a point and a tangent plane, towards a solution of the generating function valid away from the starting point, and to reveal an underlying tangent surface to the given vector field. In the right image of Fig. 8.2, one can see the kernel of the obtained solution $F$, where the tangent surface is depicted in blue.

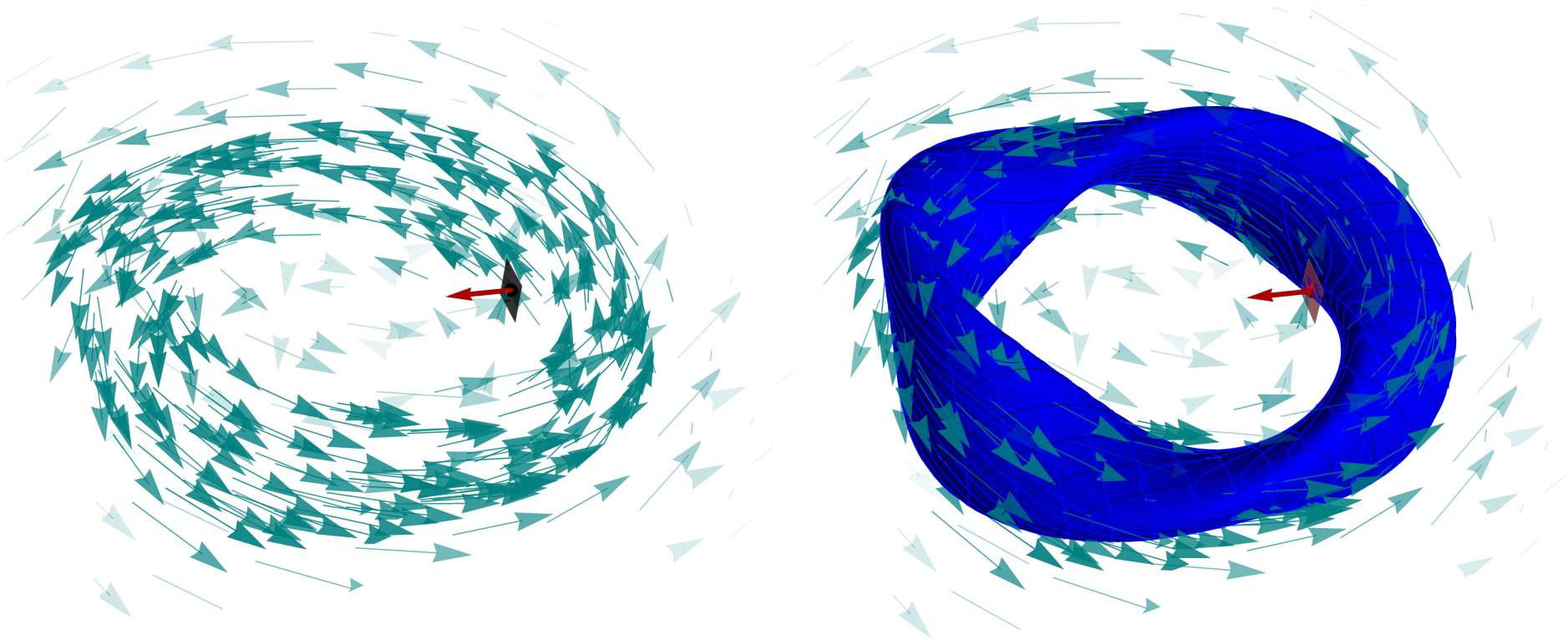

**Figure 8.2.:** Example of a given vector field $\beta$. In red, a point $p_0$ and a tangent plane to the underlying tangent surface containing that point. In blue, the obtained flow invariant surface using the generating function method.

Notice that, in the general case, one could attempt an *explicit* parameterization, using a single valued smooth function, of these tangent surfaces, instead of using the implicit parameterization given by the generating functions, but the very simple example of the harmonic oscillator already shows a situation where an explicit parameterization of the surface with a single-valued smooth function is not enough, since the surface is a circle. It is expected that more complicated situations, where the vector field has a more complicated functional dependence in the studied domain, also do not allow for explicit parameterizations, and the implicit methods for finding these surfaces are the right way to go to avoid problems, such as global topology.

In a more generic case, one cannot expect to always be able to find analytic solutions for $F$, and one must resort to numerical methods. In the following sections, we will explain the technical developments used for finding solutions to the generating functions, and apply this formalism to the $\beta$ fields that represent the RG flow of asymptotically safe QFTs, to make predictions using the FRGE.

## 8.1 General set-up for FRG applications and toy model example

In this section, we will explain the general setup needed to use our numerical techniques to find numerical solutions of the "master equation" (8.1). We will explain some numerically convenient parameterizations of theory space that we use in our implementations. We will later apply the flow invariant methods to make IR predictions of a toy model QFT: a 3-dimensional interacting real scalar field.

Suppose one is given a finite truncation of a QFT, with an EAA determined by a finite polynomial of operators

$$\Gamma_k[\bar{\phi}] = \sum_{i=1}^{N} \bar{g}_i \mathcal{O}_i[\bar{\phi}] = \sum_{i=1}^{N} g_i\, k^{-d_i}\, \mathcal{O}_i[\bar{\phi}]\,, \tag{8.7}$$

where $N$ is some positive integer, and where, as mentioned before, the $k$ dependence of the couplings will be dropped from now onward for simplicity. For our method to be applicable, we will assume that the beta functions for this truncation were already previously computed, and will be taken as an input for our program. Also, we will assume that

1. The beta functions admit a NGFP. Within the truncation under consideration, the projection of the NGFP is a saddle point, i.e., it comes with both UV-attractive and UV-repulsive eigen directions. When assessing the stability of results under a systematic increase of the truncation size $N$, the properties of the fixed point, including its critical exponents, show convergence.

2. RG trajectories can actually reach the limit $k = 0$. This entails that the beta functions have a well-defined $k \to 0$-limit and there are no singularities which lead to a termination of the trajectories at a finite value $k$.[1]

[1] A prototypical example of such a singularity is the $\lambda = 1/2$-line found in the Einstein-Hilbert truncation [136].

These conditions are not guaranteed a priori and must be checked on a case-by-case basis. In $F(R)$-type truncations, one typically satisfies 1) but faces problems when sending $k \to 0$. Conversely, not all fixed points encountered in scalar or scalar-tensor theories satisfy criterion 1), at least at the level of approximation that we checked. The explicit examples discussed in this chapter have been selected on the basis that they satisfy the criteria spelled out above.

To be able to use numerical methods to find solutions for (8.1), it is necessary to go through a number of steps. Since the method developed here is computational, one needs to study a compact domain of $\mathcal{T}$. To compactify the domain of study of $\mathcal{S}_{UV}$ and to have a better grasp of the domain where it lies in $\mathcal{T}$, it is particularly useful to perform the following steps.

First, a convenient reparameterization of the couplings of the action is in place, namely, the use of a single dimensionful coupling to make all other parameters of the theory dimensionless. The procedure works as follows

- Identify a relevant coupling $\tilde{g}_{n_0}$, for a given $n_0$, with a non zero canonical mass dimension $d_{n_0}$. If the coupling has $d_{n_0} > 0$, redefine it as the inverse, so that $d_{n_0} < 0$. This coupling should be a coupling whose sign does not change for any physical RG trajectory, for example, a mass-squared coupling. If the coupling is always negative, just redefine it as minus itself. Since the dimensionful theory obtained when removing the regulator scale must be finite, and therefore so must $\lim_{k\to 0} \tilde{g}_{n_0}$, the dimensionless version of this coupling, $g_{n_0} = \tilde{g}_{n_0} k^{-d_{n_0}}$, must reach the value 0 when $k \to 0$, setting the $k = 0$ limit at $g_{n_0} = 0$.
- Use $\tilde{g}_{n_0}$ to make all other couplings of the theory dimensionless. For example, say one has another coupling $\tilde{g}_m$ with mass dimension $d_m$, then define the coupling $\lambda_m$ via $\tilde{g}_m = \lambda_m \tilde{g}_{n_0}^{-\frac{d_m}{d_{n_0}}}$. The coupling $\lambda_m$ is now dimensionless without the need for any rescaling with $k$.

These two previous steps allow us to study compact domains in the theory space. There are two reasons for this. The first one, is that in the presence of a NGFP, then $\lim_{k\to\infty} |g_{n_0}| < \infty$. This allows us to only explore the compact domain $g_{n_0} \in [0, g_{n_0 *}]$, since we know that it will contain the UV and IR limits of the theory. In a sense, the selected coupling $g_{n_0}$ will play the role of a RG "clock", which we will use to determine $\mathcal{S}_{UV}$ both in the UV and the IR.

The second reason is that, since in this parameterization all other couplings $\lambda_m$ are dimensionless without the need for any rescaling with $k$, all one needs is an estimate of the ranges where those couplings range between the UV and the IR to have an idea of what is the region of theory space one needs to explore to determine the desired portion of $\mathcal{S}_{UV}$.

In an ideal scenario, external constraints would indicate the phenomenologically interesting sector of theory space, such as experimental evidence. In this chapter, we will focus on testing if the method works for simple, controlled QFTs where their RG flow is sufficiently simple to test this method, and therefore, we will not use experimental evidence as a guide to where to implement the method, but instead, we will use methods that are known to work in these settings.

To find a region of theory space to test our method, and where we know that at least a portion of $\mathcal{S}_{UV}$ must be contained, we use the shooting method. The idea is to use the shooting method with varied initial conditions near the NGFP to find integral lines of the RG flow. From these lines, one can estimate the coupling ranges $[\lambda_{m_{min}}, \lambda_{m_{max}}]$ that contain the NGFP, and at the same time contain IR endpoints of RG flow trajectories.

After these parameterizations and interval estimations, one has all the necessary ingredients to use our method.

### 8.1.1 The generating function approach to $\mathbf{S_{UV}}$

The asymptotic safety hypothesis states that any asymptotically safe RG trajectory must be part of the UV-critical surface of the underlying NGFP. On this basis, one can predict the value of (irrelevant) couplings based on the coordinates spanning $\mathcal{S}_{\text{UV}}$. In [5], we developed an efficient methodology to explicitly construct the embedding of $\mathcal{S}_{\text{UV}}$ into a (truncation of) the theory space $\mathcal{T}$. Since this construction forms the starting point of our algorithmic treatment, we briefly explain its core ideas, adding some minor modifications.

We start from a $N$-dimensional truncation of $\mathcal{T}$, parameterized by the dimensionless couplings

$$\{ g_n \}, \qquad n = 1, \cdots, N\,. \tag{8.8}$$

8.1

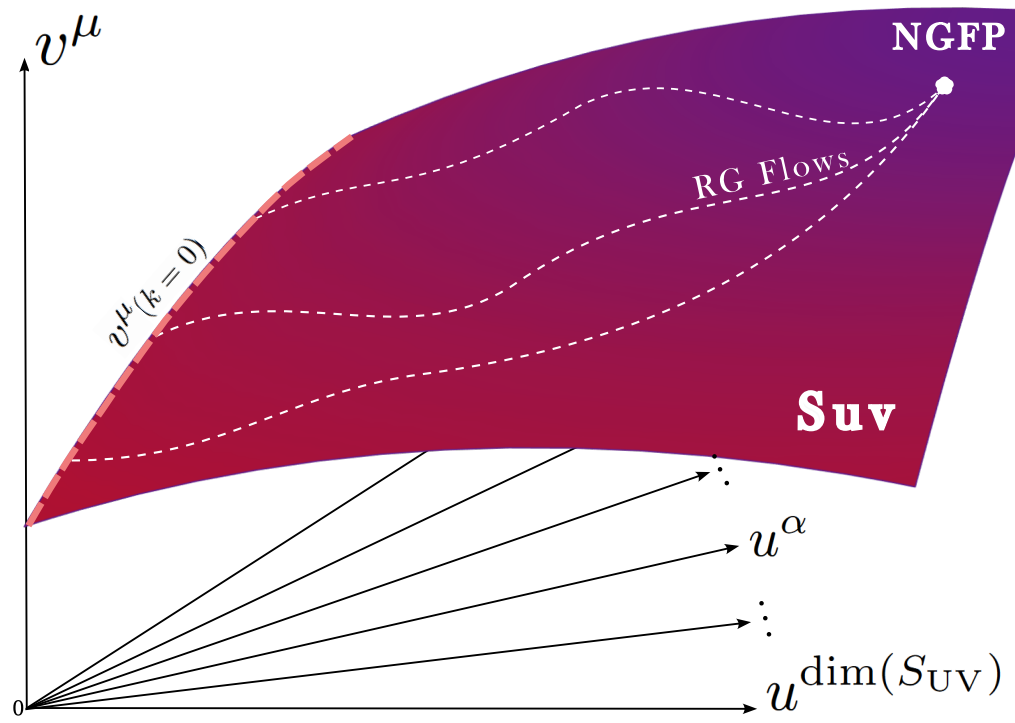

**Figure 8.3.:** Illustration of the embedding of $\mathcal{S}_{\rm UV}$ in theory space $\mathcal{T}$. We use coordinates $\{u_1, \cdots, u_{dim(\mathcal{S}_{\rm UV})}\}$ on $S_{\rm UV}$. The embedding of the surface then gives the values of $v_\mu$ as a function of these coordinates. In terms of physics predictions, we are interested in evaluating these relations at the endpoint of the RG flow, $k = 0$, where all quantum fluctuations are taken into account.

We assume that the corresponding beta functions $\beta_{g_n}$ admit a NGFP $\{g_{n*}\}$. Furthermore, we assume that the fixed point gives rise to a UV-critical surface with dimension $\dim(\mathcal{S}_{\rm UV}) < N$. This entails that $N - \dim(\mathcal{S}_{\rm UV})$ parameters can be predicted based on asymptotic safety. In order to harvest these predictions, we decompose the couplings $g^n$ according to

$$\{ g_n \} \mapsto \{ u_\alpha \, , \, v_\mu \} \, , \tag{8.9}$$

where $\alpha = 1, \cdots, \dim(\mathcal{S}_{\rm UV})$ and $\mu = 1, \cdots, N - \dim(\mathcal{S}_{\rm UV})$. The $u_\alpha$ provide coordinates on $\mathcal{S}_{UV}$, and are called *relevant couplings*, while $v_\mu$ encode the predictions of asymptotic safety, and are called *irrelevant couplings*. On the UV-critical hypersurface, we then have that

$$v_\mu|_{S_{\rm UV}} = v_\mu(u_\alpha) \, . \tag{8.10}$$

The UV-critical surface $\mathcal{S}_{UV}$ of the fixed point is then the set of points in $\mathcal{T}$ spanned by all RG trajectories which hit the fixed point in this limit, see Fig. 8.3 for an illustration.

Next, we focus on the embedding of $\mathcal{S}_{\rm UV}$ in $\mathcal{T}$. The embedding can be described by a set of functions $\tilde{F}_\mu : \mathbb{R}^N \to \mathbb{R}$ with $\mu = 1, \cdots, N - \dim(\mathcal{S}_{\rm UV})$, that encode $\mathcal{S}_{\rm UV}$ implicitly, which we choose so that their zero level set contains $\mathcal{S}_{\rm UV}$. More explicitly, these functions satisfy

$$\tilde{F}_\mu(g)\Big|_{S_{\rm UV}} = 0 \,, \quad \mu = 1, \cdots, N - \dim(\mathcal{S}_{\rm UV})\,. \tag{8.11}$$

Solving these relations for the couplings $v_\mu$ yields the explicit form (8.10).

One can see that, on the UV critical surface, they satisfy $k\frac{d\tilde{F}_\mu}{dk}\Big|_{S_{\rm UV}} = 0$. Since the existence of $S_{\rm UV}$ is guaranteed by the analysis of the stability matrix, as done in (7.10), then the existence of these functions is as well. And thus, the fact that they are constant on the UV critical surface implies

$$\sum_{n=1}^{N} \left( \frac{\partial \tilde{F}_\mu}{\partial g_n} \beta_{g_n} \right)\Bigg|_{S_{\rm UV}} = 0 \,, \quad \mu = 1, \cdots, N - \dim(\mathcal{S}_{\rm UV})\,. \tag{8.12}$$

Given a theory space domain $\Omega = [g_{1_{min}}, g_{1_{max}}] \times \cdots \times [g_{N_{min}}, g_{N_{max}}]$, we do not know $\mathcal{S}_{\rm UV}$ a priori. To construct a numerical technique that can explore theory space we will, as mentioned before, make a stronger hypothesis, and we will assume that there exists a set of functions (Generating Functions) $F_\mu : \mathbb{R}^N \to \mathbb{R}$ with $\mu = 1, \cdots, N - \dim(\mathcal{S}_{\rm UV})$ such that

$$\sum_{n=1}^{N} \frac{\partial F_\mu}{\partial g_n} \beta_{g_n} = 0 \quad \forall\, g \in \Omega\,, \tag{8.13}$$

with $\Omega$ some custom, compact hypercube in theory space, that is assumed to contain at least a portion of $S_{\rm UV}$. This means that we are assuming the existence of FIs in a chosen domain $\Omega$, where we expect $S_{\rm UV}$ to be contained. As explained before, this is not guaranteed for a general vector field $\beta$, but this assumption simplifies significantly the numerical exploration of theory space because these equations constitute a decoupled set of linear, partial differential equations for the generating functions. This assumption will allow us to use efficient tools to determine, in the cases where a solution set to (8.13) can be found, the underlying $S_{\rm UV}$.

Finding numerical approximations of solutions of (8.13), to unravel the underlying $S_{\rm UV}$ and map the predictions (8.10) by extracting the kernel of the flow invariants $\{\, F_\mu \,\}$ requires several steps. These equations are to be solved using suitable initial conditions. The initial data is obtained from linearizing the beta functions close to the NGFP.

The linearization is obtained by Taylor expanding (8.13) near the NGFP $g_*$, within the space spanned by the relevant directions around the NGFP, i.e., the tangent space of $\mathcal{S}_{UV}$. An element of this tangent plane is a linear combination of the relevant directions $V_\nu$, with some arbitrary real coefficients $c_\nu$, and components given by $R_n = \sum_{\nu=1}^{dim(\mathcal{S}_{UV})} c_\nu \, V_{\nu\,,n}$. At linear order, one finds that the expansion of (8.13) leads to

$$\sum_{j=1}^{N} \left( \sum_{\nu=1}^{dim(\mathcal{S}_{UV})} \theta_\nu \, c_\nu \, V_{\nu\,,j} \right) n_{j\,,\mu} = 0 \qquad \forall \;\; c_\nu \, \in \mathbb{R}, \tag{8.14}$$

where we introduced the vectors $n_{j\,,\mu} \equiv \left. \frac{\partial F_\mu}{\partial g_j} \right|_{g=g_*}$, where $j$ denotes the component of the vector, and $\mu$ labels it with a corresponding generating function. This means that the vectors $n_{j\,,\mu}$ form a basis of the null space of the relevant directions $V_{i,\nu}$. Carrying out a Gram-Schmidt process, one can find $n_{j\,,\mu}$ with $\mu = 1, \cdots, N - dim\,(\mathcal{S}_{UV})$, which correspond to the normal vectors to $\mathcal{S}_{UV}$ at the NGFP. The result should be a set of linearly independent vectors $\{\{V_{i,\nu}\}\,, \{n_{j,\mu}\}\}$ that span $\mathbb{R}^N$, at the NGFP.

By means of this expansion, one can say that the Generating Functions have a linear expansion near the NGFP of the form

$$F^*_\mu(g) \equiv \sum_{i=1}^{N} n_{i\,,\mu} \, (g_i - g_{i_*}) \;, \tag{8.15}$$

where we defined the notation $F^*_\mu$ for the linearized generating functions.

As mentioned before, the set of solutions of (8.13) is degenerate, since any monotonic $C^1$ function $\tilde{F}(F)$ that satisfies $\tilde{F}(0) = 0$ is also a solution that gives the same kernel. Fortunately, this non-uniqueness is eliminated when imposing the linear behaviour (8.14) close to the NGFP, by fixing $\{n_{i\,,\mu}\}$.

The independence of the equations appearing in (8.13) is something we will later on use to our advantage, when approximating solutions with independent numerical methods for each generating function, a feature that allows a major speed up compared to another method that has to solve coupled equations determining $\mathcal{S}_{UV}$.

Taking a step back from the technicalities, it is useful to highlight again the generality of the approach we are taking. Given the RG flow vector field $\beta$ of a truncated QFT, possessing a fixed point with a finite number of UV attractive directions, we attempt to determine a surface containing all possible flow trajectories that reach that fixed point, by constructing conserved quantities of the flow in a certain domain. We do this by examining the intersection of the kernels of flow invariant quantities, our Generating Functions. If the assumption of the existence of such conserved quantities is correct on a given domain, this general strategy allows us to map the Asymptotically Safe theory space of a QFT with a set of independent equations.

Finding $\mathcal{S}_{UV}$ in a given domain, with a continuous approach like this one, has, to the best of our knowledge, not been pursued so far. This is because previous methods could only extract individual trajectories, for example, finding individual flow lines by integrating the system of PDEs given by the set of beta functions. Despite how far individual trajectories have taken us, a shooting method requires determining the right initial conditions to find a phenomenologically viable trajectory. This is computationally expensive, as we will discuss later on, in cases where the truncation order $N$ and the dimension of $\mathcal{S}_{UV}$ are significantly larger than 1. Our techniques potentially allow mapping all possible flows at once, providing a computational engine capable of searching phenomenological outcomes of UV-complete QFTs, which we hope can be used as a powerful complement to other techniques.

### 8.1.2 Toy model example: Scalar field in three dimensions

One of the core strengths of the master equation (8.13) is its linear nature. This linearity of the master equation makes the use of pseudo-spectral methods and collocation techniques [169, 170] an efficient tool for obtaining solutions.[1] For fixed-sized truncations, or very simple, increasingly large truncations, this approach is particularly simple to implement and sufficient as a first approach to test the method within the FRG. In the following sections, we will discuss other methods that are more suitable for our settings.

Given a truncated theory space with $N$ couplings, the basic idea of the pseudo-spectral method is to select a dense set of functions $\{\psi_i : \mathbb{R}^N \to \mathbb{R}\}$, with $i = 1, \ldots, N_p + N$ for a fixed number $N_p$ of what we will call later as

[1] For earlier applications of pseudo-spectral methods in the context of the functional renormalization group see [157, 171–173].

8.1

"collocation points". Typical examples used in practice include Chebyshev polynomials [174] or radial basis functions [175]. The generating functions are then approximated by expanding in this basis

$$F_\mu(g) \simeq \sum_{i=1}^{N_p+N} p_{\mu,i}\, \psi_i(g)\,, \tag{8.16}$$

i.e., the functions are determined by the $(N_p + N) \times (N - \dim(\mathcal{S}_{\rm UV}))$ free parameters $p_{\mu,i}$. Since (8.13) contains derivatives of $F_\mu(g)$ only, it determines solutions up to an additive constant. Following (8.15), we fix this freedom by demanding that the first derivatives of $F_\mu$ evaluated at the fixed point are taken to agree with the linearized approximation

$$\left.\frac{\partial F_\mu}{\partial g_n}\right|_{g_{n*}} = \frac{\partial F^*_\mu}{\partial g_n}\,, \tag{8.17}$$

giving $N$ derivative conditions for each $F_\mu$. This enforces that at least one of the expansion coefficients in (8.16) must take non-vanishing values, thereby eliminating the trivial solution.

Subsequently, one chooses $N_p$ collocation points in a domain of theory space where one expects the UV critical surface to be contained, points which will be used as anchors for a numerical approximation of the solution. It is natural to take the NGFP as one of these points and construct a grid covering the parameter space of interest. Substituting the expansion (8.16) into (8.13) and evaluating the resulting expressions at the collocation points then leads to a system of algebraic equations which determines the coefficients $p_{\mu,i}$. Since at the NGFP the beta functions are zero, the supplementary equation one uses at this point is

$$F_\mu(g_\star) = 0\,. \tag{8.18}$$

The predictions $v_\mu$ compatible with asymptotic safety are then found by evaluating the approximate solution at $\{\lim_{k\to 0} u_\alpha\}$.

*Systematic improvements.* In the simplest case, one may arrange the collocation points $(g_{n_{c,i}})$ in a regular lattice and expand the generating functions in terms of Multivariate Cauchy Distributions (MCDs)

$$\psi_i(g) = \left(1 + \sum_n \frac{(g_n - g_{n_{c,i}})^2}{\sigma^2}\right)^{-1} . \tag{8.19}$$

Here $\sigma$ is a smoothness parameter whose typical value is chosen to be of the order of the average separation between the collocation points. Since the substitution of (8.16) into the system (8.13) results in a linear equation, taking one collocation point for each MCD leads to a unique solution for the parameters $p_{\mu,i}$.

Depending on the concrete application, this "basic" setup may be improved by working with fewer parameters than collocation points, and/or with base functions more suitable for describing sharp edges, periodicity, or optimized convergence [176]. Moreover, the positioning of the collocation points may be optimized near interesting regions.

We illustrate the working of our general algorithm based on a scalar field theory in a three-dimensional Euclidean spacetime. The RG flow possesses a NGFP, the Wilson-Fisher fixed point. This fixed point is a saddle point in theory space, and we will therefore use it as a test bed for our method.

Working with the Wilson-Fisher fixed point comes with the advantage that it is already seen in simple approximations. For our purpose, it suffices to work with the local potential approximation (LPA) [160]. In this case, the effective average action is approximated by

$$\Gamma_k[\phi] \simeq \int d^3x \left[\frac{1}{2}(\partial\phi)^2 + U_k[\phi]\right] , \tag{8.20}$$

where $\phi(x)$ is a real scalar field with mass-dimension $[\phi] = \frac{1}{2}$. Implementing our convention on the couplings, we expand the potential according to

$$U_k[\phi] = \frac{1}{4!\,\tilde{u}}\phi^4 + \sum_{0<n\,,\,n\neq 2}^{N_{max}} \frac{v_n}{(2n)!}\,(\tilde{u})^{n-3}\,(\phi^2)^n\,. \tag{8.21}$$

This singles out the coupling $\tilde{u}$ (defined to have a negative mass dimension) associated with the $\phi^4$-term as the free parameter allowed by asymptotic safety. The IR-value of the dimensionless couplings $v_n = (v_1, v_3, v_4, \cdots)$ will be predicted based on the UV-critical surface of the NGFP.

**Table 8.1.:** Position of the Wilson-Fisher fixed point (LPA, eq. (8.21)).

| $N_{max}$ | $u_*$ | $v_{1*}$ | $v_{3*}$ | $v_{4*}$ | $v_{5*}$ |
|---|---|---|---|---|---|
| 4 | 0.129 | −0.0013 | | | |
| 6 | 0.081 | −0.00092 | 541 | | |
| 8 | 0.071 | −0.00088 | 889 | 1167462 | |
| 10 | 0.068 | −0.00087 | 1021 | 1786707 | 3427010600 |

**Table 8.2.:** Stability coefficients of the Wilson-Fisher fixed point.

| $N_{max}$ | $\theta_1$ | $\tilde{\theta}_1$ | $\tilde{\theta}_2$ | $\tilde{\theta}_3$ | $\tilde{\theta}_4$ |
|---|---|---|---|---|---|
| 4 | 1.84 | −1.18 | | | |
| 6 | 1.69 | −1.13 | −12.20 | | |
| 8 | 1.58 | −0.96 | −8.82 | −32.52 | |
| 10 | 1.54 | −0.78 | −6.52 | −22.00 | −60.30 |

In Tables 8.1, 8.2 we report the NGFP properties, and in Table 8.3 we report the $k = 0$ results up to order $N_{max} = 10$.[1] For simplicity, we limit our exposition to the simplest case where $N_{max} = 2$, i.e., we are dealing with two couplings $(u, v = v_1)$.

We start from the beta functions given in [164]. Adapting to our parameterization of the potential and setting $N_{max} = 2$ and $d = 3$, we have

$$\begin{aligned} \beta_u &= u - \frac{u^6}{\pi^2\,(u^2+v)^3}, \\ \beta_v &= -\frac{u^5}{6\pi^2\,(u^2+v)^2} - \frac{2\,v\,u^5}{\pi^2\,(u^2+v)^3}. \end{aligned} \tag{8.22}$$

The first term in $\beta_v$ is associated with the scale-dependence of $v$. The second term is proportional to the quantum part of $\beta_u$ and captures the contribution of an anomalous scaling dimension associated with $\tilde{u}_k$.

[1] These results serve as an illustration of the method and should not be mistaken as an attempt of a precision computation. The latter may be achieved by extending the approximation along the lines [177, 178].

**Table 8.3.:** Asymptotic safety predictions for $\lim_{k\to 0} v_n$.

| $N_{max}$ | $v_1\|_{k=0}$ | $v_3\|_{k=0}$ | $v_4\|_{k=0}$ | $v_5\|_{k=0}$ |
|---|---|---|---|---|
| 4 | 0.0048 | | | |
| 6 | 0.0017 | 1043 | | |
| 8 | 0.0013 | 1581 | 1652938 | |
| 10 | 0.0012 | 1738 | 2169522 | −3615823144 |

The NGFP of the system is located at $(u_*, v_*) = (\frac{2197}{1728\pi^2}, -\frac{371293}{2985984\pi^4})$. The stability coefficients and eigenvectors entering into the linearized solution (8.14) are

$$\begin{aligned} \theta_1 &= \frac{1}{6}(2+\sqrt{82})\,, \qquad & V_1 &= \left\{1\,,\ \frac{864\left(\sqrt{82}-11\right)\pi^2}{2197}\right\}\,, \\ \tilde{\theta}_1 &= \frac{1}{6}(2-\sqrt{82})\,, \qquad & \tilde{V}_1 &= \left\{1\,,\ -\frac{864\left(\sqrt{82}+11\right)\pi^2}{2197}\right\}\,. \end{aligned} \tag{8.23}$$

Thus we are dealing with a saddle-point with one UV-attractive and one UV-repulsive eigendirection. The UV-critical surface is one-dimensional and we expect a prediction for $v$ from asymptotic safety. Following the procedure leading to (8.15), the generating function for the UV-critical surface is

$$F^*(u,v) = -u + \frac{864\left(\sqrt{82}-11\right)\pi^2 v}{2197} + \frac{169\left(15+\sqrt{82}\right)}{3456\pi^2}\,, \tag{8.24}$$

where we have chosen the normalization of the normal vector $n_\mu$ such that the coefficient multiplying $u$ is set to one. Based on this input we compute the prediction for $v$ using 1) the traditional shooting method, 2) a solution of the system (8.13) based on series expansions of orders 10 to 15, and 3) a spectral solution evaluating the system (8.13) for a linear combination of 500 MCDs on a regular lattice centered on the NGFP, covering the domain shown in Fig. 8.4. All methods yield

$$v_{|k=0} = \lim_{k\to 0} v \simeq 4.8\times 10^{-3}\,, \tag{8.25}$$

when $N_{max} = 2$, with the numerical difference appearing in the third relevant digit. Fig. 8.4 illustrates the structures underlying this prediction.

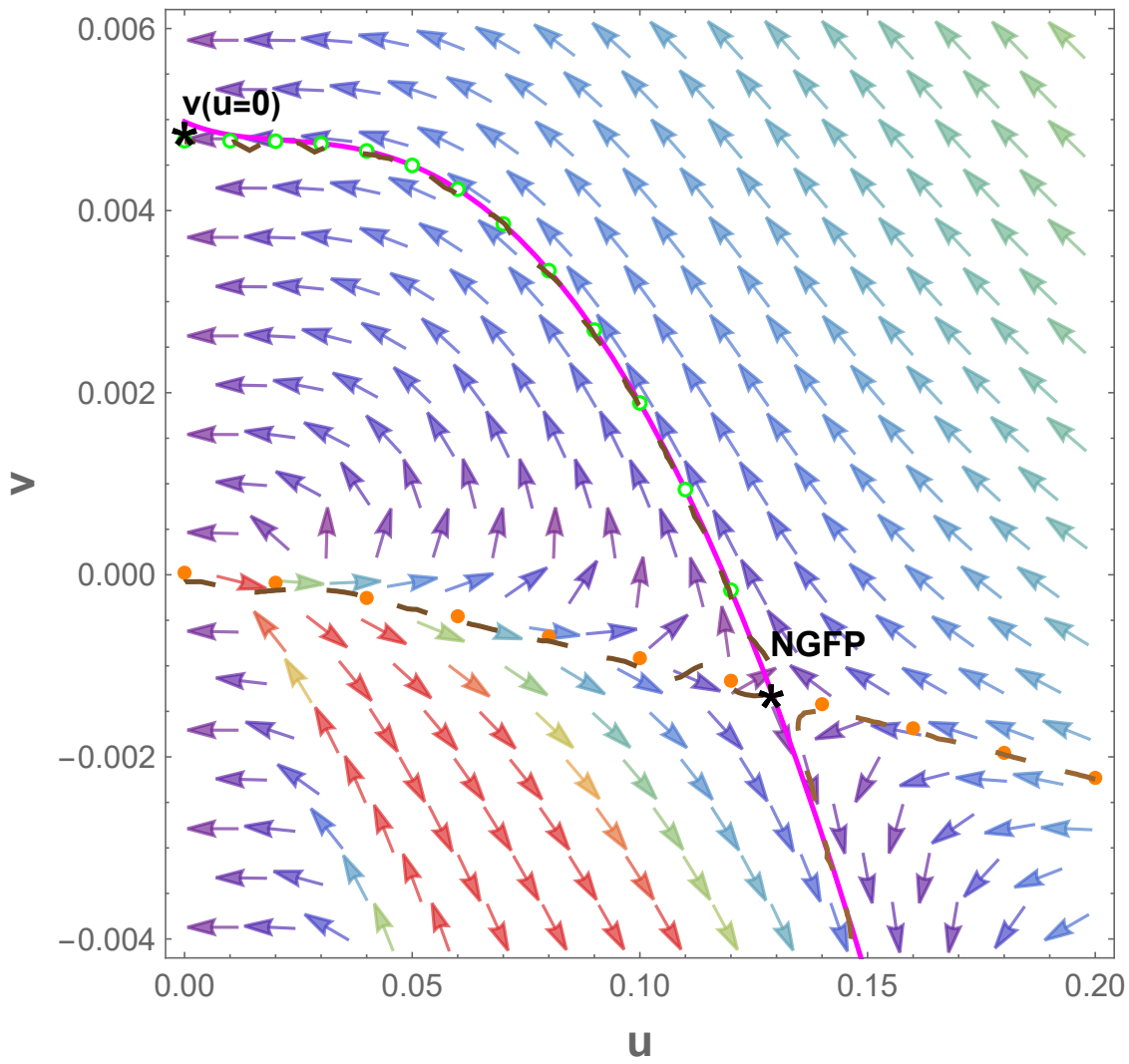

**Figure 8.4.:** Phase diagram generated by the beta functions (8.22) with arrows pointing towards lower values of $k$. The NGFP and the endpoint of $\mathcal{S}_{\mathrm{UV}}$ giving rise to the prediction (8.25) are marked by $*$. The UV-critical surface obtained from the shooting method is marked by hollow green dots, the solution of the numerical generating functions by brown dashes, and the one of the Taylor expansion of the generating functions by a solid magenta line. For completeness, we added the RG trajectories generated by $\tilde{V}_1$ as solid orange dots.

This example shows that the FI approach to find $\mathcal{S}_{UV}$ worked in this simple scalar field toy model, and gave us the confidence to push forward with this approach on more complicated settings, as we will see in the next section. While this simple approach of using fixed collocation points proved useful for this case, matching two other approximation schemes, we can do better. One of the disadvantages of this simple approach of using collocation points and a deterministic grid for approximating the Generating Functions, is that for each new truncation order $N$, one has to find a completely new approximation for the flow invariants. This seems inefficient in the sense that truncations of the lowest order are expected to give results that are reasonably close to a truncation of slightly higher order; otherwise, the truncation method itself is not useful at all. If one wants to use this information, one would have to switch to methods that are capable of *learning* from the previous truncations, so that one does not need to start from scratch every time one adds a new coupling to the truncation. With this in mind, we now move on to the next section, where we will explore modern and efficient techniques to approximate solutions of PDEs.

## 8.2 Machine Learning to unravel theory space

The previously used approximation of the network using fixed collocation points and pseudo-spectral methods involved using exactly as many free parameters as points where the master equation (8.13) was imposed. On top of that, every iteration of the method with truncations of a different order requires starting from scratch.

In the following sections, we will consider a more scalable method that uses optimization and statistical sampling, rather than exact solvers at fixed points. We will resort to *Neural Networks* (NN) as approximants to our generating functions. One can intuitively think of a neural network as a function with some free parameters. Within this thesis, the *training* of a NN is then just a parameter optimization, or simply a best fit, that tunes the parameters to optimize a certain condition. In the following sections, we will describe in more detail the structure of these NN, and how we use specific loss functions to address our problem of finding the underlying $\mathcal{S}_{UV}$ of an asymptotically safe QFT. We will briefly describe how these NN are trained, and why they are efficient approximators, making them suitable for our problem.

### 8.2.1 Neural networks in a nutshell

The problem of finding the generating functions as presented in section 8.1.1 can be treated by a variety of algorithms, and in particular, solving the master equation (8.13) can be viewed computationally as solving an optimization problem. The goal of any optimization algorithm is to optimize (minimize or maximize) one or multiple *loss functions*. A loss function scores a proposed solution to the optimization problem, evaluating one of the characteristics that the solution is supposed to encode.

The method we employed is the training of a *physics-informed neural network* (PINN). PINNs are a subclass of models within the larger class of *(artificial) neural networks*. Neural networks are a type of universal function approximators. They can, in theory, be used to model any continuous function $f : \mathbb{R}^n \mapsto \mathbb{R}^m$, whose behaviour is determined by the constraint that some loss function $\ell[f]$ is minimized, to any desired degree of accuracy [179]. This result is known as the *universal approximation theorem* [180]. If multiple loss functions are used, they can be combined linearly, using weighing factors, into one master loss function. A PINN is merely a neural network (NN) for which at least one of the loss functions is a partial differential equation (PDE) [181]. The term "physics-informed" stems from the idea that raw data from nature can often be hard to learn for a neural network, so introducing the correct PDE expected to describe the data will inform the network of the relevant physics [182]. A good recent overview of PINNs is found in [183].

### Architecture of a neural network

The used neural networks in this work are directed, weighted graphs consisting of layers of nodes, called *neurons*, and weighted edges (sometimes called *synapses*) connecting neurons of subsequent layers. A NN made to model a function $f : \mathbb{R}^n \mapsto \mathbb{R}^m$ will have an *input layer* consisting of $n$ neurons and an *output layer* consisting of $m$ neurons. In between can be any number of *hidden layers* that can each have any number of neurons. The number of layers is called the *depth* of the NN, and the maximum number of neurons in any layer is called the *width* of the NN [184, 185]. The general architecture of a neural network is illustrated in Fig. 8.5, using the notation developed in the next paragraph.

Let us denote the $i$'th neuron of the $j$'th layer as $N_j^i$. $N_j^i$ is connected to $N_{j+1}^k$ by a weighted, directed edge $E_j^{ik}$ that has weight $W_j^{ik} \in \mathbb{R}$. There are only edges between layer $j$ and $j+1$, not between $j$ and $j+2$ or any deeper layer. The neuron $N_j^i$ receives a real number signal from neurons in the previous layers. This signal is processed in the neuron using an *activation function* $\sigma_{(j)} : \mathbb{R} \mapsto \mathbb{R}$, after which the signal is sent to all neurons in the next layer. There are multiple choices of activation functions, but the important thing is that they are almost always non-linear, which allows even a small network to encode complex behaviour [186, 187]. The signal passes through the weighted edges $E_j^{ik}$, connecting to all neurons $N_{j+1}^k$. The signal in each edge is multiplied by the weight $W_j^{ik}$ of the edge. The value entering neuron $N_{j+1}^k$ is $\sum_i W_j^{ik} \sigma(N_j^i) + B_{j+1}^k$, where $B_j^k \in \mathbb{R}$ is called the *bias*. A bias is introduced to effectively shift the activation function, which may be critical for successful learning [184, 188].

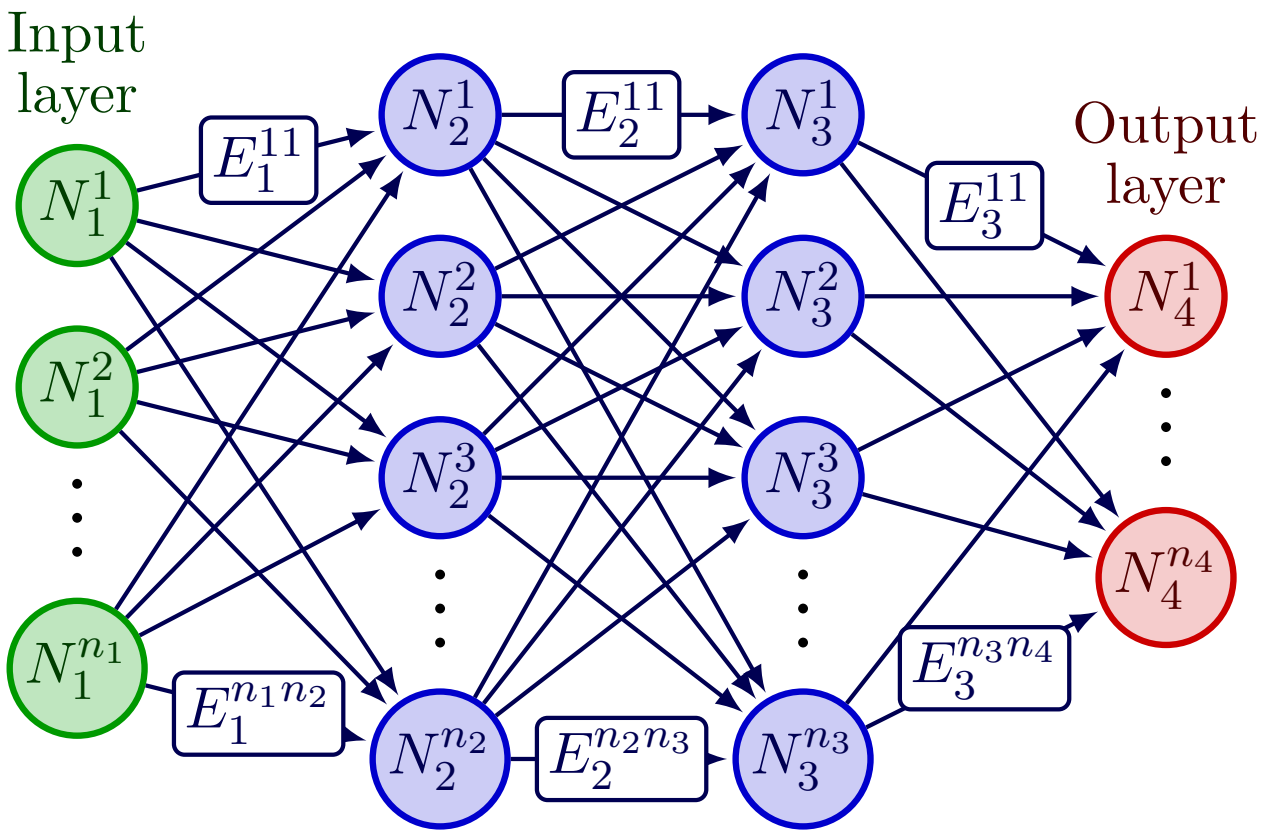

**Figure 8.5.:** An illustrative depiction of the architecture of a neural network with depth 4. The signal flows from the input layer to the output layer through the hidden layers. Each edge $E_j^{ik}$ has a weight $W_j^{ik}$ associated with it. Each node $N_j^i$ has a bias $B_j^i$ that is added to the incoming signal. Each layer has an activation function $\sigma_{(j)}$ that endows the network with the ability to model non-linear behaviour.

The computation of data flowing through the network is done efficiently through matrix multiplication [184]. If layer $j$ contains $n_j$ neurons then we can represent the signal entering it by an $n_j$-dimensional vector $N_{(j)} = [N^1_{(j)}, \cdots, N^{n_j}_{(j)}]^T$. The signal exiting is $\Sigma_{(j)} := \sigma_{(j)}(N_{(j)})$, where we understand the activation function to have been applied to each neuron. We can represent the weights connecting layer $j$ and $j+1$ by an $n_j \times n_{j+1}$-dimensional matrix $W_{(j)}$. The bias added is represented by an $n_{j+1}$-dimensional vector $B_{(j)}$. The signal entering layer $j+1$ is given by the $n_{j+1}$-dimensional vector $N_{(j+1)} = W_{(j)} \cdot \Sigma_{(j)} + B_{(j)}$. This signal then passes through the activation function and gets sent to the next node, where the cycle repeats.

The choice in depth and width of the network is highly problem-specific [189, 190]. Deep networks allow the representation of hierarchical structures, which are present in non-linear systems [191]. This is certainly required for finding $\mathcal{S}_{UV}$ with our method. The width of the network helps capture smooth, global features and is important for networks with multiple outputs [192]. This is also important for our problem, as we expect to at least see some global behaviour. The number of parameters ($\#P$) in a general network that maps $F : \mathbb{R}^N \mapsto \mathbb{R}$ with depth $D$, weights $W_j$, and biases $B_j$

is

$$\#P = \sum_{j=1}^{D-1} \left(|W_{(j)}| + |B_{(j)}|\right) , \tag{8.26}$$

with $|W_{(j)}|$ and $|B_{(j)}|$ denote the number of components of each matrix and vector.

The neural network can be trained by tuning the weights and biases. To train a NN, one has to gather or construct a set of input data that can be fed through the network. The optimization process is iterative. At each iteration, one feeds a set of input data through the network. The input and output of the network are then used to calculate a loss value using the loss function, which the training should optimize/minimize. This optimization is done through a process called *backpropagation*. Backpropagation is essentially the calculation of the gradient of the loss function with respect to the weights and biases. This calculation is done by applying the chain rule layer by layer, working from the output layer to the input layer. Working backward through the network avoids redundant calculations of intermediate terms in the chain rule [193].

With the gradients available, the weights can be adjusted using gradient descent. In the simplest form, this means taking a small step in weight-space in the opposite direction of the gradient of the loss function. This completes one iteration of the learning procedure. Iterating this procedure over the optimization domain leads to the minimization of the loss function and the convergence to the desired solution [193, 194].

### 8.2.2 The Flow Invariant Neural Network

We are now in a position to use NNs tailored for our purpose: finding $\mathcal{S}_{UV}$. To model the generating functions, we developed the *Flow Invariant Neural Network* (FINN). The FINN builds on top of two pillars. A standard PINN loss function, that solves the PDE (8.13) via the loss function (8.27)

$$\ell_M[F] := \sum_{\mu} \frac{1}{(||\beta|| + ||\nabla F_\mu|| + \epsilon)} \left( \sum_n \frac{\partial F_\mu}{\partial g_n} \beta_{g_n} \right)^2 , \tag{8.27}$$

where $||\beta|| \equiv \sqrt{\sum_n (\beta_{g_n})^2}$, $||\nabla F_\mu|| \equiv \sqrt{\sum_n (\frac{\partial F_\mu}{\partial g_n})^2}$ and where a small auxiliary parameter $\epsilon = 10^{-10}$ was introduced to avoid divisions by zero during training. The reason for dividing the loss function by these norms is to avoid the problem of different orders of magnitude of the values taken by the beta functions and the FINN gradient itself. The second pillar consists of the addition of a second loss function (8.28), thus approximating the linear behaviour of the flow invariants

$$\ell_{FP}[F] := (F_\mu - F^*_\mu)^2. \tag{8.28}$$

Loss function $\ell_M$ is imposed over the entire training domain, as discussed at the beginning of this chapter, while $\ell_{FP}$ is imposed in an infinitesimal region around the fixed point, which in the studied case of this chapter is on a ball of radius $\delta = 10^{-4}$ around the NGFP. Both loss functions must be simultaneously minimized.

The computational advantage present in our setting with respect to standard solvers of PDEs, is that the system (8.13) is composed of independent generating functions. Thus, each generating function is approximated by a separate FINN, instead of solving the system with a global PINN that approximates all generating functions at once. The uncoupled nature of our system spanning $\mathcal{S}_{UV}$ is a major advantage, allowing a parallel training of each FINN.

From an architecture and modeling point of view, splitting the network forces the modeling of the independence of the generating functions, instead of relying on this feature to emerge from training. We effectively introduce an *inductive bias*. Introducing an inductive bias into the model by forcing known features is always preferred over trying to have them emerge during training [195–197].

### Purpose, efficiency, and use of this method

We do not claim that our method outperforms the shooting method in all scenarios, nor that it is a replacement for it. In fact, given the fact that the FI method relies on the assumption of the existence of invariant quantities outside of a local domain, an assumption that was not mathematically proven, we think of this method as work in progress, where some of its elements must be better understood, even though it gives promising results on the studied settings.

Therefore, we believe our developed techniques should be used in a complementary manner to more established methods, such as shooting, to cross-check the predictions obtained from one and the other. There is, however, one particular situation where the use of our method can prove more useful than the shooting method. Suppose one is given an experimental range of allowed values for $N$ couplings of an EFT, and one is tasked with determining whether those coupling values are compatible with the Asymptotic Safety hypothesis.

For simplicity of this argument, we can think of the given range of couplings as a product of intervals, one associated with each coupling, forming a "box" in theory space. Assuming that one can capture the RG flow of such an EFT using a truncation containing those $N$ couplings, and the FRGE, we could proceed as follows.

By using the shooting method, one has two main options. The first option is to select a sufficiently well-spread sample of points in that box, say, a uniform grid with spacing $\epsilon$ of $(\frac{1}{\epsilon})^N$ points in that box. Then, one integrates the $N$ beta functions of that truncated EFT towards the UV and determines if some of the points (approximately) reach or not the NGFP that generates a UV-completion of the theory. Unfortunately, the computational cost of this increases with the power of $N$, a very inefficient method.

The second option, is to start near the NGFP in the UV, and perform a series of integrations towards the IR of the beta functions using points contained in $S_{\text{UV}}$. If one does not know how the flow will affect the trajectories, a possible avenue is to construct a regular grid with spacing $\delta$ of $(\frac{1}{\delta})^{\dim(\mathcal{S}_{UV})}$ points. Unfortunately, the RG flow is not necessarily divergence-free, and samples of points that evenly cover a region in the UV can end up being collimated/concentrated at lower energies, where the sampled trajectories might not end up passing through the experimentally constrained box. That could lead to the conclusion that a certain IR region is asymptotically safe, even if it is not, just because of a wrong initial sampling. Also, it could require iterations of different samplings of initial shooting points near the UV, an inefficient process, and highly difficult to automate.

Our FINNs method allows us to automate this process in a single iteration. With our method, all one needs to do is construct a region of theory space that contains the given experimental box, as well as the UV-completion of the theory space, train the FINNs in that region, and determine if there is a common kernel or not within it. The structure and size of the NN will probably also scale with $N$, but we expect its width and depth to grow at most linearly with $N$. The cost of training a NN in a given domain is highly optimized, and with a network with width and depth proportional to $N$, the cost can at most be proportional to $\mathcal{O}(S\,B\,N^3)$, where $S$ is the number of training steps, $B$ is the size of batch of points in the domain used for training. In typical PINNs, the random sample of points in the domain used for training in each step allows one to use fewer points than with a standard deterministic fit function. This allows for the conservative assumption that $S$ and $B$ grow as a polynomial in $N$, not as some power of it, making the total computational cost of finding FINNs polynomial in $N$.

Of course, one still has to find a common kernel of the FINNs. If one wants to cover the UV critical surface domain uniformly, one can construct a grid of spacing $\epsilon$ of $(\frac{1}{\epsilon})^{\dim(\mathcal{S}_{\text{UV}})}$ points in the relevant couplings domain. This means that the total computational cost of the FINNs method is, conservatively speaking, at most polynomial with the size $N$, and proportional to $(\frac{1}{\epsilon})^{\dim(\mathcal{S}_{\text{UV}})}$. In this particular scenario, where one is interested in a verification of external coupling constraints versus the AS hypothesis, our method could overcome standard shooting techniques for large truncation sizes $N$.

These reasons make FINNs an interesting methods, but there is still the issue of the initial assumption of the existence of such flow invariants. Nevertheless, that assumption worked well in the toy model scalar field example, and as we will see below, it will as well with more complicated models. This justifies the exploration of the FINN's method, viewing it as a powerful complementary tool to shooting method techniques, and we leave the study of the FI existence assumption for future work.

## 8.3 4-dimensional scalar-tensor as a test model

Scalar-tensor theories provide one of the most conservative and, at the same time, most versatile extensions of General Relativity (GR). In their simplest form, they supplement the metric $g_{\mu\nu}$ with a real scalar field $\tilde{\varphi}$ that couples non-minimally to the Ricci scalar $\tilde{R}$. The prototype is the Brans-Dicke theory, where the effective Planck mass is promoted to a dynamical field and gravity is mediated by both the metric and a scalar degree of freedom [198]. Modern scalar-tensor models generalize this idea by allowing for arbitrary functions of $\tilde{\varphi}$ in the non-minimal coupling and in the scalar potential, and in the most general case also for derivative self-interactions of the scalar, as captured by the Horndeski class [199–201]. Within this broad family, the theories considered in this chapter belong to the Jordan-frame subclass with canonical kinetic term and non-minimal coupling $f(\tilde{\varphi})\,\tilde{R}$.

At the classical level, the scalar-tensor theories we will look at are of the form

$$\Gamma = \int d^4\tilde{x}\,\sqrt{|g|}\,\left[\tilde{f}(\tilde{\varphi})\,\tilde{R} - \tilde{v}(\tilde{\varphi}) - \frac{1}{2}\,\partial_\mu\tilde{\varphi}\,\partial^\mu\tilde{\varphi}\right], \tag{8.29}$$

where $\tilde{f}$ and $\tilde{v}$ are dimensionful functions of $\tilde{\varphi}$, and where the tildes are used to distinguish the elements of this expression from dimensionless quantities defined later on. For $\tilde{f} = \frac{1}{16\pi\tilde{G}} = const.$ one recovers minimally coupled scalar field theory on a GR background. Allowing for a non-trivial $\tilde{f}(\tilde{\varphi})$ opens up a rich phenomenology while staying within second-order equations of motion and avoiding extra propagating degrees of freedom beyond the scalar and the graviton.

A key structural property of this class is its relation to $F(R)$ theories at the classical level. For a wide range of functions $\tilde{f}$ and $\tilde{v}$, a Legendre transform followed by a conformal rescaling of the metric maps the Jordan-frame action (8.29) to an Einstein-frame action with canonical Einstein-Hilbert term and a scalar potential $V_{\text{eff}}(\phi)$ determined by the original functions [202–205]. Conversely, many $F(R)$ models can be rewritten as scalar-tensor theories with a specific choice of $\tilde{f}$ and $\tilde{v}$. This duality is not just a formal curiosity: it implies that constraints on scalar-tensor models immediately inform the viability of a large class of modified gravity theories, and vice versa.

From a phenomenological perspective, scalar-tensor theories have been extensively used in both early- and late-time cosmology. On the early-universe side, non-minimal couplings naturally arise in inflationary model building: the classic examples include non-minimally coupled inflaton models and Higgs inflation, where the Standard Model Higgs doublet plays the role of the inflaton thanks to a sizable coupling $\xi\,\tilde{\varphi}^2\tilde{R}$ [206, 207]. Through the scalar-tensor / $F(R)$ correspondence, Starobinskys $R^2$ inflation model also fits into this framework [205, 208]. On the late-time side, scalar-tensor models provide paradigmatic realizations of dynamical dark energy and modified gravity on cosmological scales, including quintessence and more general Horndeski-type dark energy scenarios [209–211]. Solar-system tests and, more recently, the measurement of the gravitational-wave speed from binary neutron-star mergers have placed strong constraints on the allowed scalar-tensor interactions, carving out an experimentally viable subspace of the broader theory space [212, 213].

In the context of Asymptotic Safety, scalar-tensor theories occupy a special role as one of the simplest settings where matter and gravity are genuinely intertwined, yet the number of degrees of freedom remains minimal. Non-perturbative functional RG studies have shown that truncations of the form (8.29) can exhibit non-Gaussian fixed points with a finite number of relevant directions, suggesting that Asymptotically Safe UV completions of scalar-tensor gravity exist [214–217]. At the same time, the functional freedom in $\tilde{f}$ and $\tilde{v}$ makes it possible to connect these UV completions to phenomenologically interesting cosmological models, including inflationary scenarios and late-time acceleration, once the RG flow is continued towards the IR [141, 218, 219].

In this chapter, we will not attempt an exhaustive survey of scalar-tensor phenomenology. Instead, we focus on the concrete Euclidean scalar-tensor truncation of the form (8.29), which is sufficiently general to capture non-trivial interplays between the scalar sector and the gravitational coupling while still being amenable to a detailed functional RG analysis. The model provides an ideal testbed for our Flow Invariant Neural Network (FINN) machinery: it has a well-understood non-Gaussian fixed point with a small number of relevant directions [217], and the resulting IR effective action can in principle be confronted with cosmological data, as illustrated for closely related setups in [7]. Our goal in the following is therefore twofold: first, to demonstrate that FINNs can reliably reconstruct the UV-critical surface of such scalar-tensor systems, and second, to prepare the ground for future applications where the Asymptotically Safe scalar-tensor dynamics could be matched directly to inflationary and dark-energy observables.

As a first application of the FINNs framework developed above, we will therefore consider the models given by the Euclidean, scale-dependent version of (8.29). The Euclidean effective average action of this class of theories takes the general form

$$\Gamma_k = \int d^4\tilde{x}\sqrt{|g|}(-\tilde{f}(k,\tilde{\varphi})\tilde{R} + \tilde{v}(k,\tilde{\varphi}) + \frac{1}{2}\partial_\mu\tilde{\varphi}\partial^\mu\tilde{\varphi})\,, \tag{8.30}$$

where $\tilde{f}$ and $\tilde{v}$ are undetermined, dimensionful, even functions of the scalar field, $g$ is the metric determinant, and $R$ is the Ricci scalar.

In this paper, we will use as a starting point the RG flow equations derived [217], and the derivation of the NGFP properties in this section will closely follow the one given there. Since it is outside the scope of this thesis to derive RG flow equations, and as stated before, in the FINNs setting, we assume flows as an input for our program, we will just use their derived equations as a starting point. Using the dimensionless variables

$$\varphi \equiv \frac{\tilde{\varphi}}{k} \qquad f \equiv \frac{\tilde{f}}{k^2} \qquad v \equiv \frac{\tilde{v}}{k^4}\,, \tag{8.31}$$

in [217] they obtained the RG flow equations

$$\dot{v} = -4\,v + \phi\,v' + 3\,c_4 - c_4\,\frac{f\,v''}{3f'^2 + f\,(1+v'')} \tag{8.32}$$

$$\dot{f} = -2\,f + \phi\,f' + \frac{15}{4}c_4 - 4\,c_4\,\frac{3f'^2 + f}{12\,(\,3f'^2 + f\,(1+v'')\,)} - 2c_4\,\frac{\big(3f'^2 + f\big)\big(f\,f'' + f'^2\big)}{\big(3f'^2 + f\,(1+v'')\big)^2}\,, \tag{8.33}$$

with $c_4^{-1} = (4\pi)^{4/2}\Gamma\big(\frac{4}{2}+1\big)$, and where the $\dot{}$ denotes a derivative with respect to the logarithm of RG flow scale $k$, and a $'$ denotes a derivative with respect to the field variable. The field and scale dependences were removed for simplicity.

In these equations, the flow is computed at constant $\varphi$, for all values of $k$, as is customary in the Asymptotic Safety community.

From these choices, one can determine the existence of a UV fixed point of the form

$$v_* = \frac{5}{128\pi^2} \approx 0.003958\,, \tag{8.34}$$

$$f_* = \frac{157}{2304\pi^2} \approx 0.006904\,, \tag{8.35}$$

which, after an analytical study of perturbations of the flow equations, one finds only 4 relevant perturbation directions $r_i = (\delta v, \delta f)_i$, with corresponding critical exponents $\theta_i$ given by

$$\begin{aligned}
\theta_1 &= 4, & r_1 &= (\delta v, \delta f)_1 = (1, 0)\,, \\
\theta_2 &= \frac{314}{123}, & r_2 &= (\delta v, \delta f)_2 = \left(-\frac{72}{89}, 1\right)\,, \\
\theta_3 &= 2, & r_3 &= (\delta v, \delta f)_3 = \left(\phi^2 - \frac{29}{544\pi^2}, \frac{157}{3264\pi^2}\right)\,, \\
\theta_4 &= \frac{68}{123}, & r_4 &= (\delta v, \delta f)_4 = \left(\frac{56913}{3094352\pi^2} - \frac{72}{89}\phi^2, -\frac{17741}{350304\pi^2} + \phi^2\right)\,.
\end{aligned} \tag{8.36}$$

Notice that $r_1$ corresponds to the cosmological constant term, which fortunately decouples from the flow of the rest of the couplings. This leaves us with 3 parameters that encode $\mathcal{S}_{UV}$. Notably, this NGFP is stable under polynomial truncation expansions, and therefore, this setting satisfies both necessary conditions (1) and (2) stated at the beginning of Sec. 8.1.

Following the steps depicted in sec. 8.2.1, we will study a *finite* polynomial truncation in the scalar field of (8.30), assuming $v$ and $f$ even functions of the field. We will explore the results for a truncation of the form

$$\begin{aligned}
v_T(t, \varphi) &= \frac{1}{(16\pi G)^2}\left(\lambda + 10^2 v_2 (16\pi G\,\varphi^2) + v_4 (16\pi G\,\varphi^2)^2\right)\,, \\
f_T(t, \varphi) &= \frac{1}{16\pi G}\left(1 + \frac{f_2}{10}(16\pi G\,\varphi^2) + \frac{f_4}{10^3}(16\pi G\,\varphi^2)^2\right)\,,
\end{aligned} \tag{8.37}$$

where $T$ stands for truncation, and $t$ is the RG time, defined as $dt \equiv dk/k$.

To obtain the beta functions of the truncated potentials, one replaces them in (8.33) and (8.32), and expands and truncates the equations. Matching the coefficients of the same order in powers of the scalar field on both sides of the resulting truncated equation, one extracts the non-perturbative beta functions. The results are not shown here due to their complexity. Within this truncation, the relevant directions are associated with the couplings $\lambda$, $G$, $f_2$, and $v_2$. Due to the decoupling of $\lambda$, this means that the irrelevant couplings $f_4|_{\mathcal{S}_{UV}} = f_4(G, f_2, v_2)$ and $v_4|_{\mathcal{S}_{UV}} = v_4(G, f_2, v_2)$, and in particular, in the IR limit, they satisfy $\lim_{k\to 0} f_4|_{\mathcal{S}_{UV}} = f_4(0, f_2, v_2)$ and $\lim_{k\to 0} v_4|_{\mathcal{S}_{UV}} = v_4(0, f_2, v_2)$, meaning that we can plot them as surfaces embedded in 3D.

To obtain numerical results for the desired $\mathcal{S}_{UV}$, we implemented a Python script that trains 2 independent FINNs, one for each of the irrelevant couplings of this truncation. We established the structure of the FINNs, having a fixed width $W = 10$ for the hidden layers, depth $D = 10$. Due to our lack of knowledge of what the structure should be in advance, we explored different ranges of $D$ and $W$ between 5 and 15, and we did not notice a substantial difference between them, especially when $D$, $W > 7$. Since we never changed $N$, we cannot make statements on how $D$ and $W$ should scale with $N$. It is left for future research to determine how different truncation sizes affect the preferred architecture of the used NNs.

Even at low truncation order, the beta functions can already grow to be extremely complex for symbolic and even numeric methods to evaluate. This slows down computations or can even prevent them from being carried out entirely. To address this issue, we Taylor expand the beta functions for small values of $f_2$, $f_4$ and $v_4$. The reason for only expanding in these couplings is twofold. The first one, is that $G$ acts as an RG flow clock, so if we want to keep information from both the UV and IR, we must not assume it is close to zero. The second is that $v_2$ appears in the denominators of the beta functions and acts as a regulator for their numerical value. We expanded these beta functions up to $4th$ order in these couplings, and fed this expansion to the algorithm. This expansion simplifies the beta functions' input part significantly, as the expanded beta functions are simpler symbolic expressions, and therefore way easier to deal with in computation and numerical evaluations. We checked that the numeric error we accumulate by doing this is insignificant compared to the precision of the algorithm.

Because of the used expansion, we had to constrain our analysis to the domain $G \in [0, G_\star]$, $f_2 \in [0, \frac{1}{2}]$, $v_2 \in [0, 4]$, $f_4 \in [-\frac{1}{2}, 0]$, and $v_4 \in [-\frac{1}{2}, 0]$, keeping the expanded couplings smaller than 1, where numerical results are reliable after the Taylor expansion. Within this domain, the training of each FINN consisted of a run of 10000 gradient descent iteration steps, with a learning rate of $\ell_r = 10^{-3}$ (the gradient descent step size), scheduled to decrease by a factor of 2 every time the loss function decreased one order of magnitude. In each run, a random sample of 2000 points was used in the selected domain, and given as input to the FINNs for training. This random selection of input points for the FINNs is a common practice when solving PDEs, since it avoids overfitting.

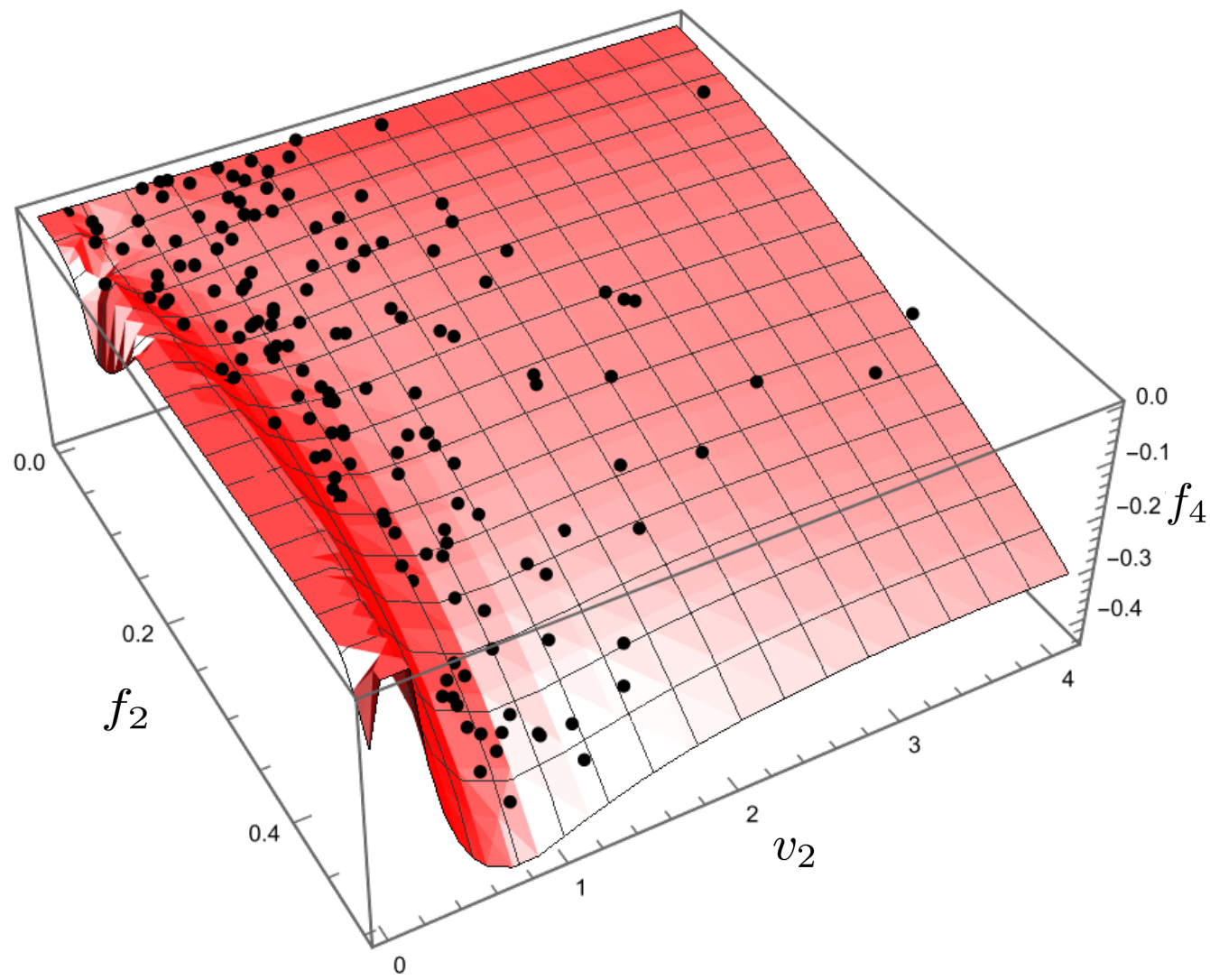

**Figure 8.6.:** $k = 0$ limit of the $\mathcal{S}_{UV}$ determining the predicted functional dependence of $f_4$ in terms of the other relevant couplings $v_2$ and $f_2$. The dots correspond to the end points of RG trajectories with random seeds close to the FP.

From the resulting FINNs, we extracted the common kernel by minimizing the function

$$||F_{FINNs}||^2 \equiv (F_1[G, f_2, v_2, f_4, v_4])^2 + (F_2[G, f_2, v_2, f_4, v_4])^2 \tag{8.38}$$

over the optimization domain. The result of this minimization are the implicit functions $f_4|_{\mathcal{S}_{UV}} = f_4[G, f_2, v_2]$ and $v_4|_{\mathcal{S}_{UV}} = v_4[G, f_2, v_2]$ depicted in figures 8.6 and 8.7.

In figures 8.6 and 8.7, one can also see the end points of 150 RG flow trajectories, whose starting points were randomly selected from a close neighborhood of the NGFP (8.35), spanned by the relevant directions (8.36). Our results show that within the studied domain, the results from the shooting method $(v_{4_{shoot}}, f_{4_{shoot}})$ and the FINNs results $(v_{4_{FINN}}, f_{4_{FINN}})$ satisfy $\frac{|v_{4_{shoot}} - v_{4_{FINN}}|}{|v_{4_{shoot}}|}, \frac{|f_{4_{shoot}} - f_{4_{FINN}}|}{|f_{4_{shoot}}|} < 0.01$, meaning they differ by less than 1% for each point studied. We take these results as a good sign that our method delivers precisely what we were after.

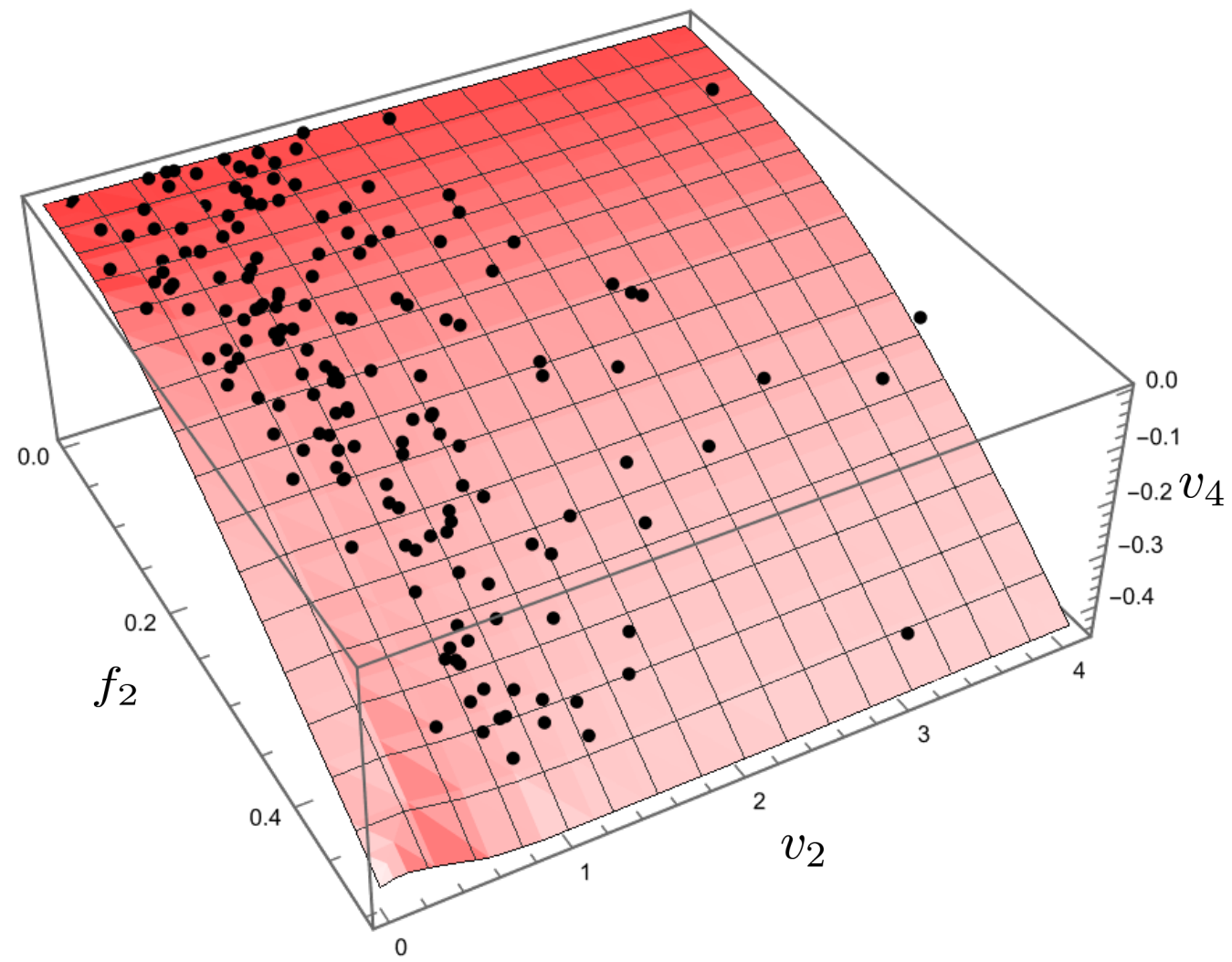

**Figure 8.7.:** $k = 0$ limit of the $\mathcal{S}_{UV}$ determining the predicted functional dependence of $v_4$ in terms of the other relevant couplings $v_2$ and $f_2$. The dots correspond to the end points of RG trajectories with random seeds close to the FP.

It is remarkable that even without a careful study of the architecture of the FINNs, and a lack of experimentation regarding the training schedulers, and the optimization methods, a simple feed-forward NN, using standard gradient descent methods with a fixed learning rate, is capable of reproducing these results with such accuracy. We consider this a highly positive feature of our approach. The roughly implemented code finishes the optimization cycle in approximately 2 minutes, but without a truncation increasing analysis, it is impossible to say how these computation times would scale. Despite the fact that we did not explore how this computational cost increases with $N$, our method has the advantage that NN can be trained with distinct inputs. In our setting, the NNs were initialized with random Gaussian weights in every layer. Nevertheless, at a given $N + 1$ truncation order, one could feed as an initial guess the values of the NN from the previously trained FINNs at order $N$, most likely improving convergence time significantly.

While it would be desirable to explore a theory space domain where these couplings could give rise to observable phenomenology, such as inflationary cosmology, at this stage, we are only able to reconstruct $\mathcal{S}_{UV}$ in small domains, and for small truncations. A comparison with other FRGE computations that attempt a reconstruction of the theory in the $k \to 0$ limit using different methods and different fixed points is precisely the kind of robustness checks the Asymptotic Safety approach needs. For example, if one were able to extend the domain of our FINNs in this setting, up to phenomenologically interesting ones, one could compare the obtained results with the ones of [7]. In [7], this same model was numerically studied in its functional form, i.e., without the need of a polynomial truncation, but using as a starting point another fixed point of the RG flow equations (8.33) and (8.32). Since it is expected that very different theories emerge from different NGFP, it would be desirable to use our techniques to rule out certain NGFPs from their emergent EFTs at $k = 0$. We believe our methods pave the way for an efficient and systematic exploration of the landscape of Asymptotically Safe gravity.

## 8.4 Collecting the results of this chapter

In this chapter, we developed a general and versatile framework for extracting infrared predictions from renormalization group flows by exploiting the geometric structure encoded in their flow invariants. The central idea is that the UV-critical surface of an asymptotically safe fixed point can be reconstructed from the kernel of a set of generating functions, whose level sets capture all RG trajectories attracted to the fixed point in the ultraviolet. The master equation (8.1), expressing the constancy of flow invariants along the vector field, provides the unifying ingredient behind this construction.

Starting from simple systems, including the harmonic oscillator and a three-dimensional toy model, we illustrated how flow invariants can be obtained analytically in favorable situations, and we described how to proceed to more complicated scenarios. These examples served as prototypes for the functional RG applications developed later in the chapter. There, the generating-function formalism was adapted to RG flows in truncated theory spaces, leading to a set of linear PDEs (8.13) whose solutions, when they exist, encode the embedding of the UV-critical surface $\mathcal{S}_{UV}$. Previous approaches typically constructed individual RG trajectories by shooting methods, becoming increasingly inefficient as the dimension of $\mathcal{S}_{UV}$ grows. In contrast, the flow-invariant strategy captures the entire family of admissible trajectories at once and has the potential of being more scalable than traditional methods.

We demonstrated the viability of this approach using both pseudo-spectral approximations and machine learning tools. For polynomial truncations of the three-dimensional scalar theory, the master-equation method reproduced the outcomes of conventional trajectory-based analyses with excellent accuracy, confirming its robustness and efficiency. This laid the foundation for the second half of the chapter, where we introduced Flow-Invariant Neural Networks (FINNs), a physics-informed neural network architecture tailored to learning the generating functions themselves. The FINNs exploit the linear structure of the master equation, incorporate the correct linearization at the fixed point, and leverage the independence of the flow invariants to train separate networks in parallel.

As a proof of concept, the FINN framework was applied to a four-dimensional scalar-tensor theory. In this system, the non-Gaussian fixed point possesses four relevant directions, and the simple truncation (8.37) thereby predicts two irrelevant couplings, as functions of the relevant ones. Training independent FINNs for each generating function, and minimizing their joint kernel (8.38), we reconstructed the UV-critical surface over a compact region of theory space. The resulting predictions match the end-points of RG trajectories obtained by the shooting method at the percent level, while requiring only modest network architectures and minimal hyperparameter tuning. This agreement provides a strong validation of the flow-invariant method and of the FINNs that implement it.

Beyond the specific models studied here, the general framework developed in this chapter provides a new computational paradigm for exploring asymptotically safe theory spaces. The master equation reformulation recasts the search for UV-complete trajectories into the problem of solving a PDE whose solutions potentially encode the full predictive subspace of the theory. The FINN architecture then provides a scalable and flexible solver for these PDEs, with the potential of handling higher-dimensional truncations where traditional numerical integration would be impractical. In particular, it offers a machine-learning tool to explore and map the UV-critical surface and its IR projections, so that one can systematically characterize which effective theories admit a UV completion, and which lie outside the asymptotically safe landscape.

Looking forward, several natural extensions suggest themselves. The first and most obvious is an investigation of the generality of the existence of FIs. In the studied cases, the assumption of their existence proved useful, and we were able to find numerical solutions, but so far, we have no proof of existence in other settings. If this is not the case, our method, as it is defined now, would not work. We hope to return to this topic in the future. On the methodological side, more advanced neural architectures–attention mechanisms, adaptive–depth networks, or reinforcement–learned sampling strategies could enhance accuracy or extend the accessible domain within theory space. On the physics side, applying the FINN framework to larger truncations of gravity–matter systems, including SM and beyond-SM interactions, may enable precision studies of asymptotically safe predictions for Higgs physics, neutrino masses, gauge sectors, or inflationary observables [139–152]. In the scalar-tensor context, extending the reconstructed domain toward phenomenologically relevant regions would allow direct comparison with functional RG analyses such as the one we will see in the next chapter, and ultimately confront asymptotically safe predictions with cosmological data.

In summary, the present work establishes machine learning, specifically physics-informed neural networks, as a new, potentially general-purpose methodology to chart the high-energy domain of quantum field theory in the EFT setting, and bridge it with low-energy phenomenology. By combining theoretical structure with computational power, this framework, if applicable in other models apart from the ones studied here, opens a door to systematic, automated, and potentially scalable explorations of the Asymptotic Safety paradigm and its extensions. Its future development promises to deepen our understanding of the ultraviolet completion of field theories and their relation to the fundamental principles underlying quantum gravity.

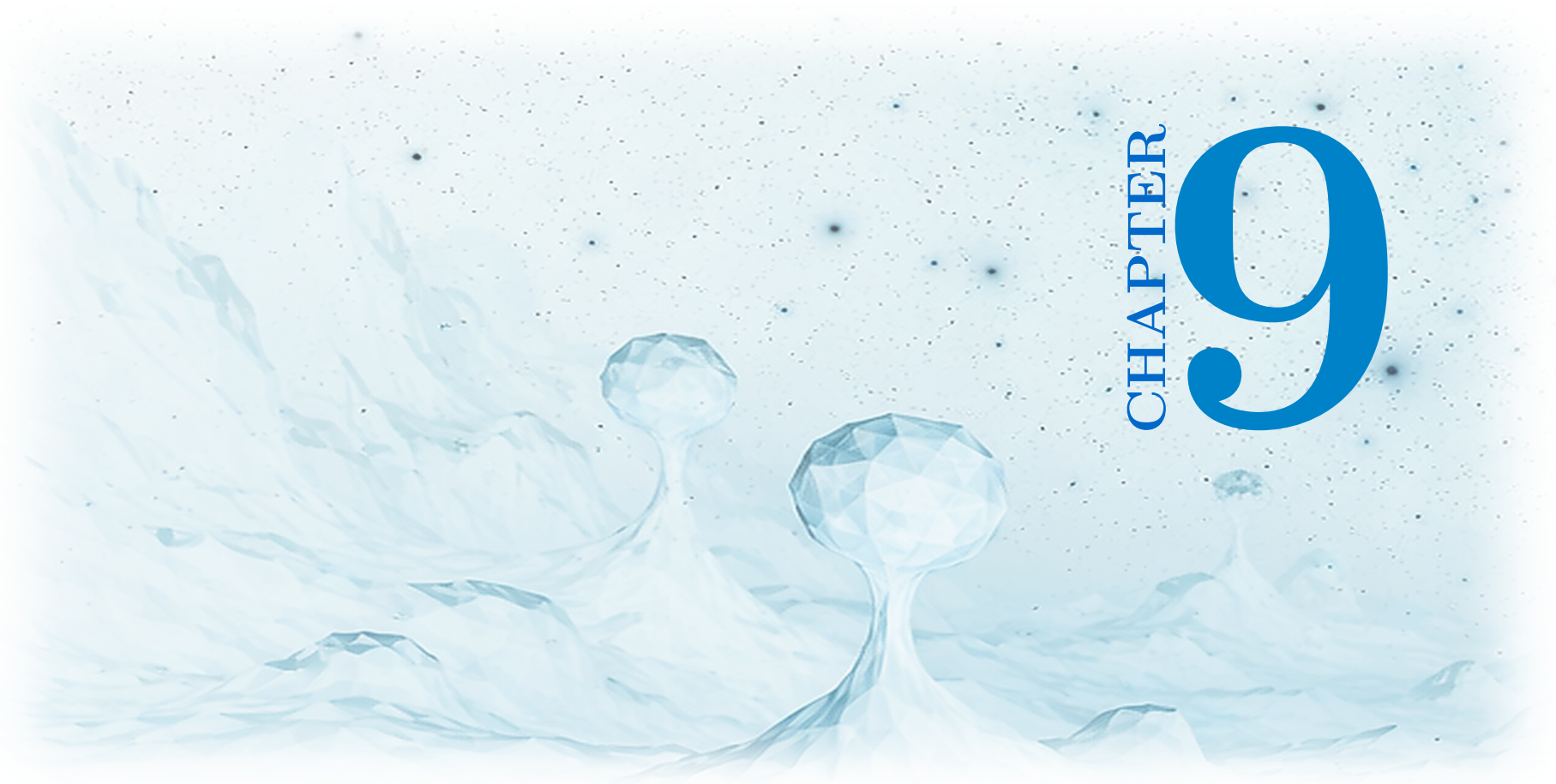

# Inflaton emergence from Asymptotic Safety

We have reached the exciting point of the thesis where we attempt to connect high-energy results directly with observations of our universe. While in the previous chapters, we have developed a toolbox to connect the functional renormalization group to low-energy physics, unfortunately, the methods developed there are not yet at the stage where one can explore phenomenologically interesting regions of theory space. In this chapter, we will explore various numerical methods, particularly those tailored for scalar-tensor theories, and examine their applications in early universe cosmology.

In chapter 7, we introduced the effective average action and the Wetterich equation as our main non-perturbative tool, and we discussed how Asymptotic Safety can provide ultraviolet (UV) completions of quantum field theories, including gravity, in terms of non-Gaussian fixed points and their UV-critical surfaces. We highlighted that in practice, one has to resort to approximations of the infinite-dimensional theory space, most notably the background-field approximation and the use of truncations of the effective average action. In this chapter, we take a complementary step and explore the predictive power of *functional truncations.* Instead of specifying the effective average action in terms of finitely many couplings, we keep the full functional dependence of certain building blocks with a prescribed functional form.

Functional truncations have already played a role in this thesis. In chapter 8, we used a scalar-tensor system as a testbed for the FINN framework, but there we projected the functional freedom in $\tilde{f}_k$ and $\tilde{v}_k$ onto a finite set of couplings by means of low-order field expansions. In the present chapter, we go one step further and work with the full functional flow equations for a scalar-tensor model.

Exactly as before, we will work on scalar-tensor models to define a functional truncation (8.30). In the previous chapter, we used this functional truncation as a basis, to later obtain the finite truncation used in the approach developed there. Here, we will allow the effective potentials of the scalar field appearing in (8.30) to remain arbitrary, without assuming a power expansion of the potentials. This truncation still restricts the full theory space, but it retains an infinite number of operators.

The predictive power of functional truncations is closely tied to the choice of boundary conditions. The functional flow equations are typically third or second-order partial differential equations in the field variable and first order in the RG scale. As a consequence, their solutions are not uniquely fixed by the requirement of asymptotic safety alone. One has to supplement the fixed-point data at high scales with additional boundary conditions encoding physical information, such as the measured values of Newton's constant and the cosmological constant, or regularity requirements in field space. In the scalar-tensor setting, these boundary conditions partially determine how the scalar field sector departs from its UV behaviour and how non-trivial interactions are generated along the flow. In this way, the combination of an asymptotically safe fixed point and carefully chosen boundary data selects a *functional trajectory* in the space of actions, from which entire effective potentials can be read off at low scales.

The main goal of this chapter is to show that such a functional trajectory can be connected to cosmological inflation. By integrating the scalar-tensor RG flow from the UV fixed point down to IR scales, a surprising yet encouraging thing happens for the AS program. By fine-tuning just 2 free parameters, we obtain an emergent inflaton potential and a prediction for the initial value of the inflaton field. Remarkably, the resulting slow-roll model of inflation is compatible with current observational constraints. An important feature of this result is that tuning only 2 parameters of the action, the functional potential that emerges is purely determined by the RG flow, rather than being chosen *ad hoc* to satisfy observational constraints.

Before delving into the details of the scalar-tensor model and its RG integration, it is useful to briefly review standard single-field inflationary cosmology and to clarify how the effective potentials obtained in this chapter are confronted with cosmological data. The next section is devoted to this recap.

## 9.1 Recap of single-field inflationary cosmology

In this section, we motivate why inflation is introduced in cosmology by looking at the cosmic microwave background (CMB) and the horizon problem, and then review the basic ingredients of single-field inflation in the Einstein frame, and how it is confronted with observations (see, e.g., [220–222] for textbook and review treatments).

The CMB is the relic radiation from the epoch of recombination, when the universe was about 380 000 years old and photons last scattered off free electrons. It provides us with a snapshot of the universe at that time, in the form of a nearly perfect black-body spectrum at an average temperature $T_0 \simeq 2.7$ K with tiny temperature anisotropies with a size of the order $10^{-5}$ [223, 224]. A full-sky map of these anisotropies, such as the one measured by WMAP and Planck, is shown schematically in Fig. 9.1: each direction on the sky corresponds to a point on the surface of last scattering, and the colour encodes the fractional temperature fluctuation $\Delta T/T$ [225, 226].

Several robust features emerge from such maps. First, the average temperature is extremely uniform: directions in the sky that correspond to regions of the last-scattering surface separated by tens of degrees have almost the same temperature. Second, the small residual anisotropies are highly correlated: the two-point correlation function of temperature fluctuations is non-zero over a wide range of angular scales, and its structure encodes acoustic oscillations in the pre-recombination plasma [224]. Observationally, we are thus confronted with a pattern of strong correlations on very large angular scales.

Something worth highlighting is that *correlation does not necessarily imply causation.* One could imagine that the initial conditions of the universe are specified in some acausal way so as to produce a correlated pattern on the CMB sky without any dynamical mechanism relating distant regions. However, if we interpret the correlations as *caused* by ordinary local physics,

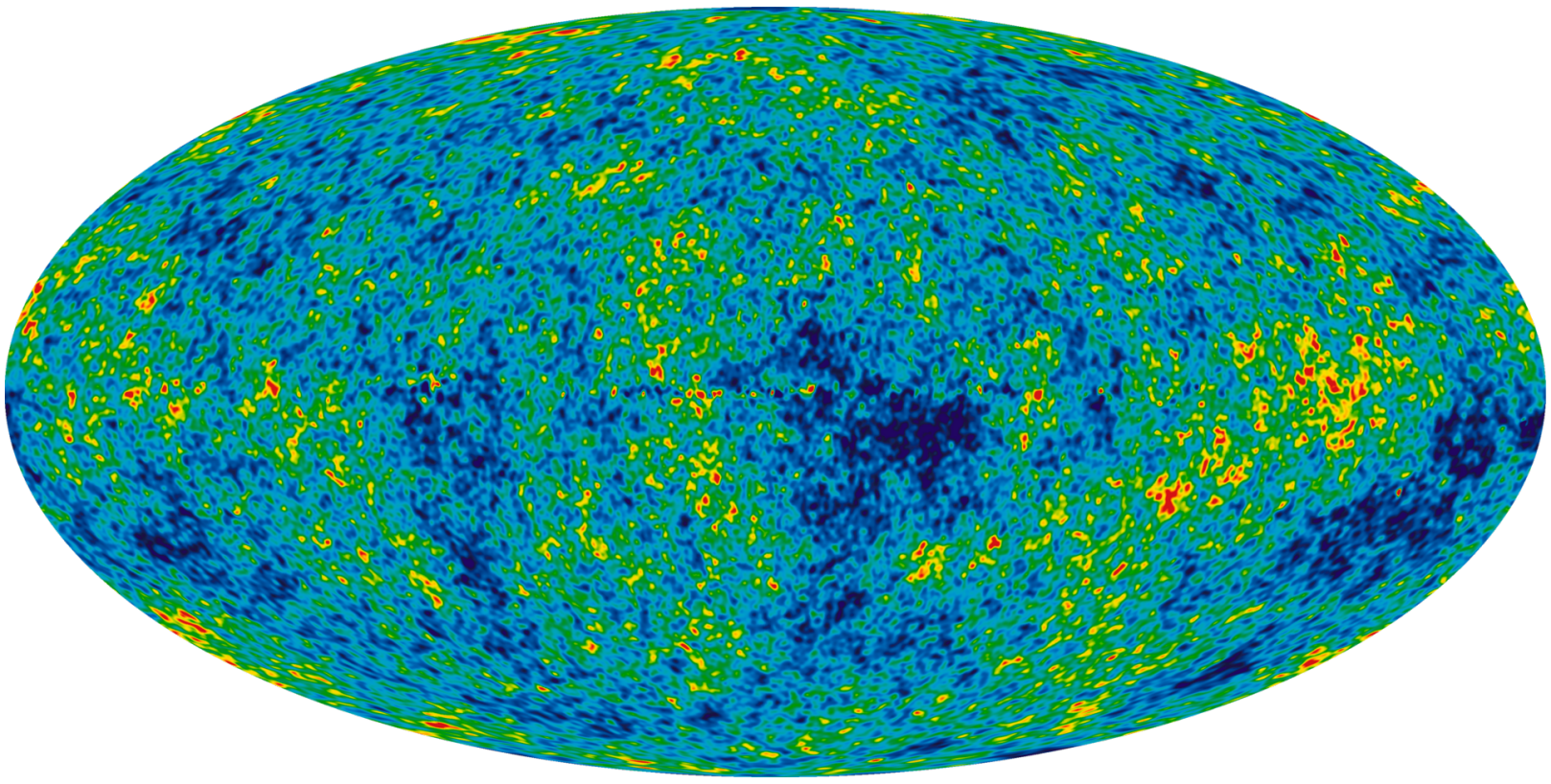

9.1

**Figure 9.1.:** Full-sky map of CMB temperature anisotropies $\Delta T/T$ at the $10^{-5}$ level, obtained by experiments such as WMAP and Planck [225, 226]. Hot and cold spots correspond to tiny density inhomogeneities on the surface of last scattering, which later seed the formation of large-scale structure.

for instance, by the thermalization processes and plasma oscillations that precede recombination, then the regions that we see as correlated today must have been in mutual causal contact at some time in the past. In other words, if we take seriously the idea that the similarity of temperatures across the sky is the result of local processes, then the past light cones of the corresponding points on the last-scattering surface must have overlapped at some point in the past.

Based on the CMB observations, the way standard early-universe cosmology is modeled is by assuming that the cosmological principle holds almost exactly: on sufficiently large scales, the universe is assumed to be spatially homogeneous and isotropic. The “almost” will be explained in the following sections. This ansatz leads to the Friedmann–Lemaître–Robertson–Walker (FLRW) class of geometries to model spacetime, whose metric can be written as

$$ds^2 = -dt^2 + a(t)^2 d\vec{x}^2 \tag{9.1}$$

for a spatially flat universe, with a single function $a(t)$, the scale factor, encoding the expansion. The matter content is then described by coarse-grained fluids (radiation, non-relativistic matter, vacuum energy), and Einsteins equations reduce to a set of ordinary differential equations for $a(t)$ [221, 227].

Within the hot Big Bang theory for early universe cosmology, which uses the Λ-CDM model, the matter content included in it gives a decelerated expansion of the universe at early times [227, 228]. In such a decelerating universe the comoving Hubble radius $(aH)^{-1}$ grows monotonically with time. As a consequence, at the time of recombination [229], the surface of last scattering (what we see in the CMB) is divided into many causally disconnected patches: the comoving distance that light could have traveled since the Big Bang is much smaller than the comoving size of the region that we observe today. This situation is often summarized by saying that the CMB sky consists of many regions that have never had the opportunity to come into causal contact, and yet they display almost identical temperatures and a correlated pattern of fluctuations. This is the *horizon problem* [14, 220].

As physicists, we often try to pinpoint the underlying causes for things to be the way they are, and nowadays there is almost a general consensus among the physics community that the correlations seen in the CMB have causation behind them. Working under this assumption, the horizon problem is not just an initial condition problem, but an actual inconsistency in the standard Big Bang story. *Inflation* offers a simple dynamical way to restore causal contact. It posits an early epoch of accelerated expansion, driven in the simplest case by the potential energy of a scalar field, a.k.a., the *inflaton field*. During inflation, the scale factor $a(t)$ grows (almost) exponentially while the Hubble parameter $H$ is approximately constant. As a result, the comoving Hubble radius $(aH)^{-1}$ *decreases* with time, and scalar field fluctuation modes whose scales are well inside the Hubble radius at the beginning of inflation are stretched to scales far beyond it. In the conformal diagram of Fig. 9.2, the effect is that the past light cones of all points in the observable universe intersect a common region during the inflationary epoch: the entire last-scattering surface descends from a single causally connected patch [14, 15, 222, 230].

In this picture, the correlations observed in the CMB arise in two steps. First, microphysical processes in a small, causally connected patch drive the state of the primordial plasma towards homogeneity and generate quantum fluctuations of the inflaton and the metric [231–234]. Second, inflation stretches this patch to cosmological scales, converting the micro-scopic fluctuations into classical, super-Hubble perturbations that later seed the structures we observe today. The qualitative evolution from an almost homogeneous, slightly perturbed early universe to the rich large-scale structure of galaxies and clusters is illustrated in Fig. 9.3.

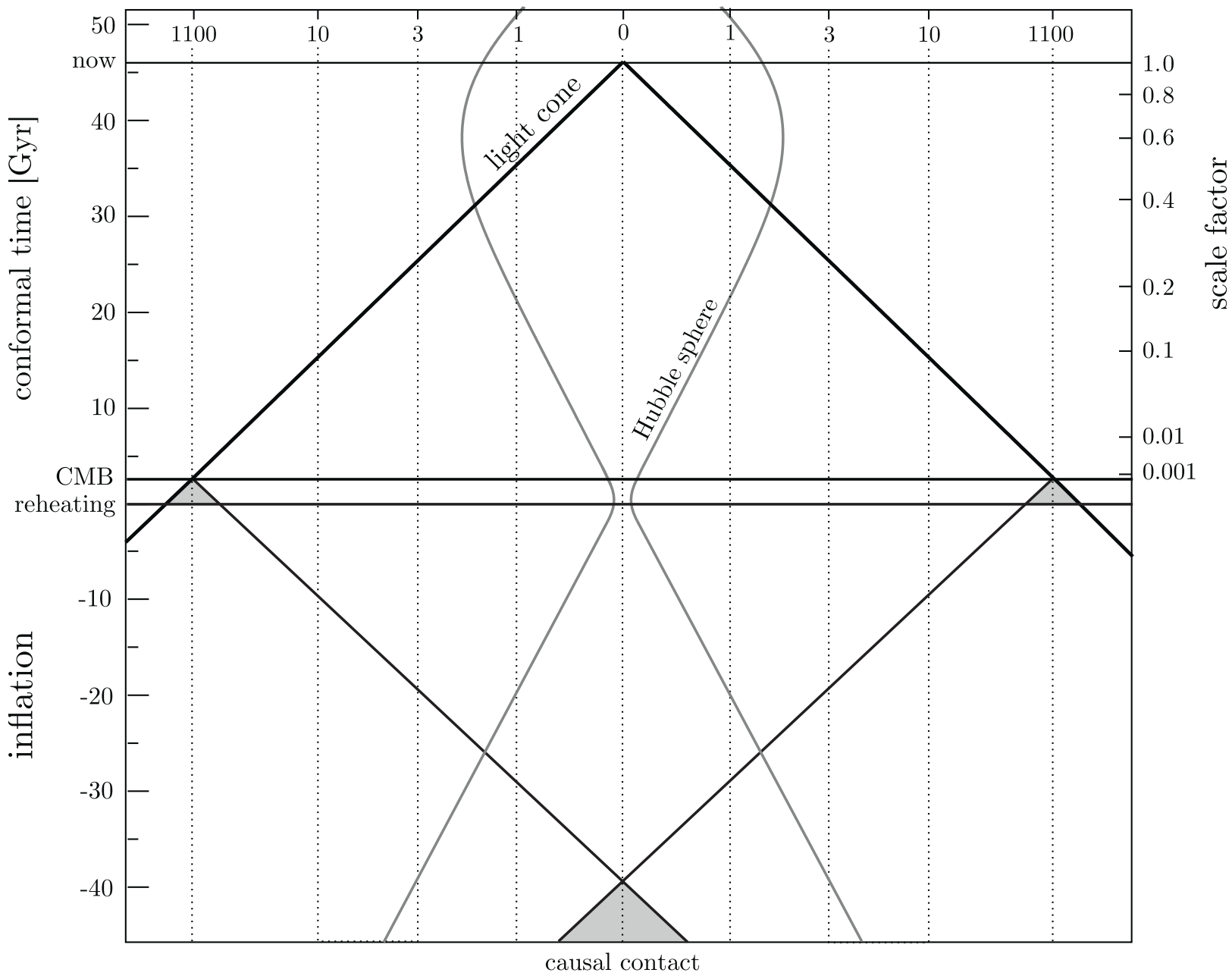

**Figure 9.2.:** Conformal-time diagram of the horizon problem and its resolution by inflation. In a decelerating universe (upper triangle) the comoving Hubble radius grows and the past light cones of points on the last-scattering surface do not overlap. With an early phase of inflation (lower triangle) the comoving Hubble radius shrinks during inflation, so that the entire region corresponding to the observable CMB was once contained in a single causal patch [14, 15, 230].

In the remainder of this section, we will describe the simplest realization of this idea in terms of a single scalar field minimally coupled to Einstein gravity, which will provide the bridge between the effective potentials emerging from the functional RG analysis and the cosmological observables [220, 221].

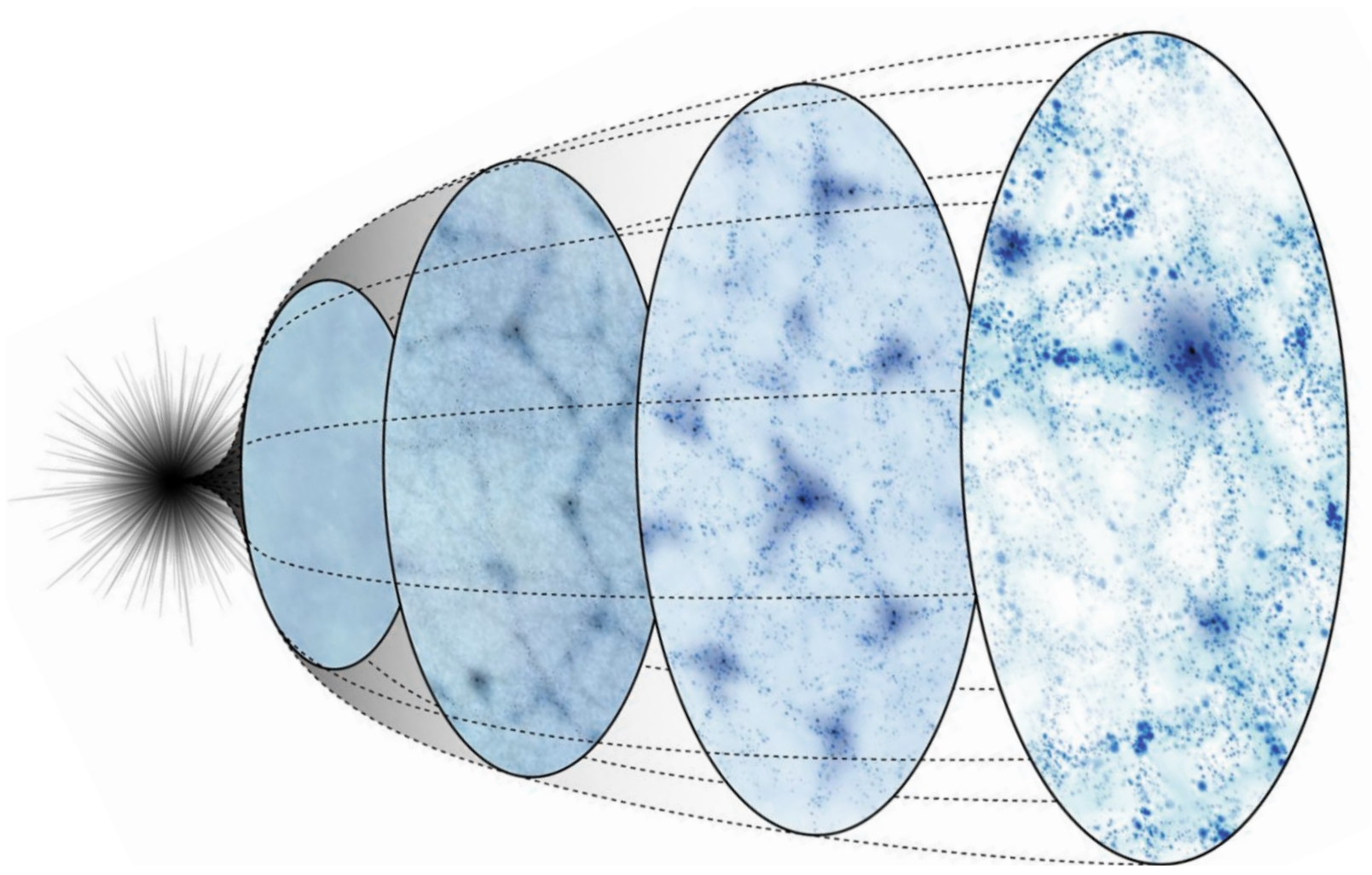

**Figure 9.3.:** Schematic evolution of cosmic structure: tiny initial inhomogeneities, generated and stretched to large scales during inflation, grow under gravitational instability into the web of large-scale structures observed at late times [220, 221].

### 9.1.1 Background dynamics

On top of the homogeneous and isotropic FLRW background introduced above, inflation is modeled, in the Einstein frame, by a single scalar field $\phi$ minimally coupled to gravity with an action (in Lorentzian signature) of the form

$$S_{\text{inf}} = \int d^4x\sqrt{-g}\left[\frac{M_{\text{P}l}^2}{2}\,R - \frac{1}{2}\,\partial_\mu\phi\,\partial^\mu\phi - V_{\text{eff}}(\phi)\right], \tag{9.2}$$

where $M_{\text{P}l} = \frac{1}{\sqrt{8\pi G}}$ is the reduced Planck mass and $V_{\text{eff}}(\phi)$ is the inflaton potential. Assuming a homogeneous inflaton $\phi = \phi(t)$, the Einstein equations reduce to the Friedmann equations

$$H^2 = \frac{1}{3M_{\text{P}l}^2}\left(\frac{1}{2}\dot{\phi}^2 + V_{\text{eff}}(\phi)\right), \tag{9.3}$$

$$\dot{H} = -\frac{1}{2M_{\text{P}l}^2}\,\dot{\phi}^2\,, \tag{9.4}$$

where $H \equiv \dot{a}/a$ is the Hubble parameter and dots denote time derivatives. The inflaton equation of motion is the Klein–Gordon equation in the expanding background,

$$\ddot{\phi} + 3H\dot{\phi} + V'_{\text{eff}}(\phi) = 0\,, \tag{9.5}$$

where primes denote derivatives with respect to $\phi$ [220, 221].

9.1

Cosmic acceleration occurs when the scale factor satisfies $\ddot{a} > 0$. Using the Friedmann equations, this condition can be written as

$$\frac{\ddot{a}}{a} = H^2(1 - \epsilon_H) > 0\,, \tag{9.6}$$

where we introduced the Hubble slow-roll parameter

$$\epsilon_H \equiv -\frac{\dot{H}}{H^2} = \frac{\dot{\phi}^2}{2M_{\text{P}l}^2 H^2}\,. \tag{9.7}$$

Inflation corresponds to the regime

$$\epsilon_H < 1\,. \tag{9.8}$$

In typical models, inflation ends when $\epsilon_H$ grows to unity. The total amount of inflation is conveniently quantified by the number of e-folds

$$N(t) \equiv \ln\frac{a(t_{\text{end}})}{a(t)} = \int_t^{t_{\text{end}}} H(t')\,dt'\,, \tag{9.9}$$

measured between a given time $t$ and the end of inflation at $t_{end}$.

### 9.1.2 Slow-roll regime and potential slow-roll parameters

A large class of single-field models admits a slow-roll regime, in which the inflaton potential energy dominates over its kinetic energy, and the field evolution is effectively dictated by the potential. The slow-roll approximation is defined by

$$\dot{\phi}^2 \ll V_{\text{eff}}(\phi)\,, \qquad |\ddot{\phi}| \ll 3H|\dot{\phi}|\,. \tag{9.10}$$

In this limit the background equations simplify to

$$H^2 \simeq \frac{V_{\text{eff}}(\phi)}{3M_{\text{Pl}}^2}\,, \qquad\qquad 3H\dot{\phi} \simeq -V'_{\text{eff}}(\phi)\,. \tag{9.11}$$

It is often convenient to parameterize slow roll in terms of the *potential* slow-roll parameters

$$\epsilon_V(\phi) \equiv \frac{M_{\text{Pl}}^2}{2}\left(\frac{V'_{\text{eff}}(\phi)}{V_{\text{eff}}(\phi)}\right)^2, \qquad\qquad \eta_V(\phi) \equiv M_{\text{Pl}}^2\frac{V''_{\text{eff}}(\phi)}{V_{\text{eff}}(\phi)}\,. \tag{9.12}$$

In the slow-roll regime one has $\epsilon_H \simeq \epsilon_V$, and inflation requires $\epsilon_V \ll 1$ and $|\eta_V| \ll 1$. The end of inflation is typically defined by the condition

$$\epsilon_V(\phi_{end}) \simeq 1\,. \tag{9.13}$$

The number of e-folds between a field value $\phi_*$ and the end of inflation can be expressed in terms of the potential as

$$N(\phi_*) \simeq \frac{1}{M_{\text{Pl}}^2}\int_{\phi_{end}}^{\phi_*} d\varphi\,\frac{V_{\text{eff}}(\varphi)}{V'_{\text{eff}}(\varphi)}\,. \tag{9.14}$$

In particular, the field value $\phi_*$ associated with the CMB pivot scale is defined implicitly by $N(\phi_*) \simeq 60$, which is approximately the number of e-folds needed to make the light cones of the observed correlated points in the CMB be causally connected [220, 227].

### 9.1.3 Primordial perturbations and observables

Quantum fluctuations of the inflaton and the metric during inflation generate a nearly scale-invariant spectrum of scalar (curvature) perturbations and tensor (gravitational-wave) perturbations [231–234]. In single-field slow-roll models the power spectrum of scalar perturbations at horizon exit for modes with momentum $k = aH$, after identifying the Hubble factor with the scalar field via the classical equations of motion, is, to leading order

$$\mathcal{P}_\zeta(k) \simeq \frac{1}{24\pi^2 M_{\text{Pl}}^4}\frac{V_{\text{eff}}(\phi)}{\epsilon_V(\phi)}\,, \tag{9.15}$$

while the tensor spectrum is

$$\mathcal{P}_h(k) \simeq \frac{2}{3\pi^2}\frac{V_{\text{eff}}(\phi)}{M_{\text{Pl}}^4}\,. \tag{9.16}$$

The scalar spectral index $n_s$ and the tensor-to-scalar ratio $r$ evaluated at the pivot scale are given by

$$n_s - 1 \equiv \left.\frac{d\ln\mathcal{P}_\zeta}{d\ln k}\right|_{k_*} \simeq -6\epsilon_V(\phi_*) + 2\eta_V(\phi_*)\,, \tag{9.17}$$

$$r \equiv \left.\frac{\mathcal{P}_h}{\mathcal{P}_\zeta}\right|_{k_*} \simeq 16\,\epsilon_V(\phi_*)\,. \tag{9.18}$$

Current cosmic microwave background measurements indicate a slightly red-tilted scalar spectrum with $n_s \simeq 0.965$ and place upper bounds on the tensor-to-scalar ratio at the few-percent level [228, 235]. These constraints select a relatively narrow band of viable slow-roll potentials, with a clear preference for concave potentials ($V''_{\text{eff}}(\phi_*) < 0$) over simple convex monomials [228, 235].

In the context of this chapter, the relations (9.12)–(9.18) provide the bridge between the functional RG analysis and cosmological observations. Once the scalar-tensor RG flow has been integrated down to low scales, we will identify the canonically normalized inflaton $\phi$ and its potential $V_{\text{eff}}(\phi)$, and then evaluate the slow-roll parameters, the number of e-folds, and the observables $n_s$ and $r$ [220, 221, 233].

## 9.2 Emergence of inflaton potential from asymptotically safe gravity

In this section we return to the scalar–tensor truncation already used in chapter 8 as a testbed for the FINN framework. Since it will be the central object of our analysis, we recall it here for convenience. In this section, we now return to the FRGE setting, where we will use the EAA and background formalism to connect RG flows with observable physics. Our starting point is the (Euclidean) effective average action

$$\Gamma_k = \int d^4\tilde{x}\,\sqrt{|g|}\left(-\tilde{f}(k,\tilde{\varphi})\,\tilde{R} + \tilde{v}(k,\tilde{\varphi}) + \frac{1}{2}\,\partial_\mu\tilde{\varphi}\,\partial^\mu\tilde{\varphi}\right), \tag{9.19}$$

where $\tilde{\varphi}$ is a real scalar field, $\tilde{f}(k,\tilde{\varphi})$ plays the role of a non-minimal coupling to gravity, $\tilde{v}(k,\tilde{\varphi})$ is a scalar potential, and $k$ is the RG momentum scale that appeared in previous sections of this part of the thesis. Both functions are assumed to be even in $\tilde{\varphi}$, and we work with a standard Einstein-Hilbert kinetic term for the metric and a canonical kinetic term for the scalar. This is the same functional truncation as in eq. (8.30); we reproduce it here explicitly so that the discussion in this section is self-contained.

As in the original functional RG analysis of this system, it is convenient to work with dimensionless quantities. For later reference we therefore reintroduce the dimensionless field and potentials

$$\varphi \equiv \frac{\tilde{\varphi}}{k}, \qquad f(k,\varphi) \equiv \frac{\tilde{f}(k,\tilde{\varphi})}{k^2}, \qquad v(k,\varphi) \equiv \frac{\tilde{v}(k,\tilde{\varphi})}{k^4}, \tag{9.20}$$

where $k$ is the RG scale. In the following, all $k$-derivatives are understood at fixed $\varphi$.

Within this truncation and for a specific choice of background, gauge and regulator (the same scheme used in Ref. [217]), the Wetterich equation reduces to a coupled system of non-linear partial differential equations for the dimensionless functions $f(k,\varphi)$ and $v(k,\varphi)$. The explicit flow equations were already presented in chapter 8, but in this chapter, we will use approximated and simpler flow equations derived from those, which read

$$\dot{v} = \varphi\, v' - 4\, v + \frac{1}{16\pi^2} + \frac{3\, f'^2 + f}{32\pi^2\, (3f'^2 + f\,(v''+1))} \tag{9.21}$$

$$\dot{f} = \varphi\, f' - 2\, f + \frac{37}{384\pi^2} + \frac{f\,\left((3f'^2+f)\,(-3f''+3v''+1) + 2f\,v''^2\right)}{96\pi^2\,(3f'^2 + f\,(v''+1))^2} \tag{9.22}$$

where $\dot{}$ denotes a derivative with respect to the logarithm of $k$ and $'$ a derivative with respect to $\varphi$. For simplicity, the dependence on $k$ and $\varphi$ was left out.

The flow equations (9.21) and (9.22) are closely related to the equations underlying the finite truncations used in chapter 8. Equations (9.21) and (9.22) are an approximation of equations (8.33) and (8.32), where one neglects the $k$ derivatives on the RHS of the FRGE. This approximation is called “1-loop approximation of the FRGE” in [217], because it neglects scale dependence on the trace part of the FRGE, which in some situa-tions resembles the results obtained using 1-loop perturbative computations. While this is not the case in gravity, this approximation is simple enough to test new methods, but at the same time it retains the physics of a UV-completion and its predictive power. The fact that in this approximation the only scale derivative appears on the LHS, and is not entangled in a complicated manner on the RHS, makes it more amenable for standard numerical integration solvers. Because of these arguments, we will use these approximated equations, instead of the full equations (8.33) and (8.32) to perform the calculations of this section.

As discussed previously, these equations are second-order in the field variable, and first-order in the RG scale variable. To find a specific numerical solution to these equations, one must impose boundary conditions in both variables. We will tackle this specific problem in the coming sections.

### 9.2.1 Fixed-point structure and convenient parameterizations

The first step before diving into constructing predictions from a given RG flow is to assess its fixed-point structure. The flow equations (9.21) and (9.22) admit a non-Gaussian fixed point in the gravitational sector, with constant, *shift-symmetric* ($\varphi \to \varphi + c$) potentials in the scalar field sector. As shown in Ref. [217], one finds a fixed point

$$f_* = \frac{41}{768\pi^2}\,, \qquad v_* = \frac{3}{128\pi^2}\,, \tag{9.23}$$

which is compatible with the Asymptotic Safety hypothesis for gravity and will serve as the ultraviolet anchor of our RG trajectories.

In order to assess the predictivity of the UV completion (9.23) of (9.19), one must determine the number of parameters of the theory that are a priori undetermined by the RG flow. In other words, determining the stability properties of the found NGFP and the number of relevant directions. In the case of the model (9.19), to determine its relevant directions, the authors of [217] perturbed the flow equations (8.32) and (8.33) at linear order in deviations from the NGFP using the ansatz $f \simeq f_* + \delta f\,(\frac{k0}{k})^\theta$ and $v \simeq v_* + \delta v\,(\frac{k0}{k})^\theta$, where $k0$ is a finite value for the coarse-graining scale, and $\theta$ is the so-called critical exponent. The authors of [217] determined 2 directions that were relevant,

$$(\delta f, \delta v)_1 = (0,1),\ \theta_1 = 4\,, \qquad (\delta f, \delta v)_2 = (1,0),\ \theta_2 = 2\,, \tag{9.24}$$

and all other directions were found to be marginal or irrelevant.

Since we are working with a functional truncation, if one finds a boundary condition that is compatible with the NGFP, but that breaks shift-symmetry away from it, one can generate non-trivial interactions in the scalar field sector. One can think of selecting boundary conditions as particular functional trajectories. For a reader more used to working with finite truncations, and following the lines of the previous chapter: selecting a particular boundary condition is conceptually similar to selecting a starting point for an RG flow trajectory close to the NGFP and within the $\mathcal{S}_{UV}$.

The typical discussion within the Asymptotic Safety framework restricts its arguments to finite truncations, and thus the argument for the predictive power of AS relies on "only relevant directions parameterize the entire theory". This is strictly speaking not true. The FRGE is a functional PDE, and thus, boundary conditions are also an important factor when addressing the predictive power of a UV completion. Once one chooses particular boundary conditions that are compatible with the NGFP, then the rest of the theory is purely determined in terms of the relevant couplings. Following the logic developed in the previous chapter, given a choice of boundary conditions, one can then parameterize the entire theory in terms of those relevant couplings, with an implicit dependence in terms of the boundary conditions.

Having said that, let us explore how to connect AS to cosmology. To connect this scalar-tensor system to inflation, it is useful to separate the purely gravitational sector from the field-dependent part, and to adopt variables which remain meaningful when we flow from the UV to the IR. In the limit where the regulator scale disappears, $k \to 0$, the effective action must still be well defined and finite. Working directly with the dimensionless field $\varphi = \tilde{\varphi}/k$ can be problematic in the limit $k \to 0$, since $\varphi \to \infty$ at a finite value of the physical dimensionful field $\tilde{\varphi}$. To avoid such $\infty \cdot 0$ limits, we follow Ref. [7] and reparameterize the potentials in terms of

$$f(k, \frac{\phi}{g_k}) \equiv \mathtt{f}(k, \phi) \equiv \frac{1 + \mathsf{f}\,(g_k, \phi^2)}{16\pi\, g_k^2}\,,$$
$$v(k, \frac{\phi}{g_k}) \equiv \mathsf{v}(k, \phi) \equiv \frac{2\lambda_k + \mathsf{v}(g_k, \phi^2)}{\left(16\pi\, g_k^2\right)^2}\,, \tag{9.25}$$

where

$$g_k^2 \equiv G_k \equiv \tilde{G}_k k^2\,, \qquad \lambda_k \equiv 16\pi \tilde{G}_k \tilde{\Lambda}_k\,, \qquad \phi \equiv g_k\, \varphi = \sqrt{\tilde{G}_k}\, \tilde{\varphi} = \sqrt{G_k}\, \varphi\,. \tag{9.26}$$

Here $\tilde{G}_k$ and $\tilde{\Lambda}_k$ denote the dimensionful running Newton coupling and cosmological constant, respectively. The dimensionless coupling $g_k$ is therefore the square root of the dimensionless Newton constant, $\lambda_k$ is a dimensionless parameter constructed from Newton's coupling and the cosmological constant, and $\phi$ is the scalar field measured in scale-dependent Planck units. For this splitting, we assume that $\mathsf{f}(g_k, 0) = \mathsf{v}(g_k, 0) = 0$, thus at vanishing field we recover the gravitational action.

Notice that $\phi$ is dimensionless by construction, without the need for any rescaling with $k$, making it very useful for numerical simulations, where it is more convenient to never introduce dimensionful scales. In the $k \to 0$ limit, its interpretation is simply that of the field in standard Planck mass units.

The functions $\mathsf{f}(g_k, \phi^2)$ and $\mathsf{v}(g_k, \phi^2)$ encode the genuine scalar-field dependence beyond the pure Einstein-Hilbert sector. Notice that we already put an explicit dependence on the relevant parameter $g_k$, since we will integrate flows from the NGFP (9.23). The reason $\lambda$ does not appear is because just like in the previous chapter, it decouples from the RG flow of the rest of the parameters, leaving only $g_k$ as the relevant parameter encoding the interactions. The dependence on the boundary conditions is not given explicitly, but we will return to this later on.

Just like in chapter 8, the coupling $g_k$ will be used as an "RG clock", interpreting the RG scale limits as $k \to 0 \;\sim\; g = 0$ and $k \to \infty \;\sim\; g = g_\star$. The fixed point (9.23) can be re-expressed in these variables. In this parameterization, we have the following reformulation of the NGFP

$$\begin{aligned}
&\lim_{k\to\infty} g_k \equiv g_* = \sqrt{\frac{48\pi}{41}} \simeq 1.92\,, \\
&\lim_{k\to\infty} \lambda_k \equiv \lambda_* = \sqrt{\frac{995328\pi^3}{68921}} \simeq 21.2\,, \\
&\lim_{k\to\infty} \{\mathsf{f}(g_k,\phi^2),\, \mathsf{v}(g_k,\phi^2)\} \equiv \{\mathsf{f}_*,\, \mathsf{v}_*\} = \{0,\, 0\}\,,
\end{aligned} \tag{9.27}$$

that shows explicitly that the field dependent parts vanish in the UV. Therefore, in the UV limit $k \to \infty$ the asymptotically safe trajectories we are interested in will approach

$$g_k \to g_*\,, \qquad \lambda_k \to \lambda_*\,, \qquad \mathsf{f}(g_k,\phi^2) \to 0\,, \qquad \mathsf{v}(g_k,\phi^2) \to 0\,, \tag{9.28}$$

so that the action becomes shift-symmetric in the scalar field and reduces to the fixed-point form (9.23).

The main advantage of the parametrization (9.25)–(9.26) is that it cleanly separates the purely gravitational couplings $(g_k, \lambda_k)$ from the scalar-sector dimensionless functions $\mathsf{f}$ and $\mathsf{v}$, which will play the role of the unknown "effective potentials" to be solved for. Their dimensionless nature without the need to introduce dimensionful quantities makes them ideal for numerical simulations.

To obtain explicit flow equations for $\mathsf{f}(g,\phi^2)$ and $\mathsf{v}(g,\phi^2)$, one combines the splitting (9.25) with a change of variables from $(k,\varphi)$ to $(k,\phi)$, where we now want to keep fixed the field variable $\phi$, so that we can meaningfully interpret results at finite values of the field at $k = 0$. Denoting by $\mathtt{f}(k,\phi)$ and $\mathtt{v}(k,\phi)$ the intermediate functions obtained by rewriting $f(k,\varphi)$ and $v(k,\varphi)$ in terms of $(k,\phi)$, a straightforward application of the chain rule yields the following transformation rules. These transformations should be applied to (9.22) and (9.21) to obtain the flow equations in terms of a constant $\phi$ field variable

$$\begin{aligned} k\, f^{(1,0)}(k,\varphi)\Big|_{\varphi} &\rightarrow \frac{\beta_g}{g_k}\,\phi\,\partial_\phi \mathtt{f}(k,\phi) + k\,\partial_k \mathtt{f}(k,\phi)\,, \\ f^{(0,n)}(k,\varphi)\Big|_{\varphi} &\rightarrow g_k^n\,\partial_\phi^n \mathtt{f}(k,\phi)\,, \\ k\, v^{(1,0)}(k,\varphi)\Big|_{\varphi} &\rightarrow \frac{\beta_g}{g_k}\,\phi\,\partial_\phi \mathtt{v}(k,\phi) + k\,\partial_k \mathtt{v}(k,\phi)\,, \\ v^{(0,n)}(k,\varphi)\Big|_{\varphi} &\rightarrow g_k^n\,\partial_\phi^n \mathtt{v}(k,\phi)\,, \end{aligned} \tag{9.29}$$

with $n = 0, 1, 2$ and $\beta_g \equiv k\,\partial_k g_k$. The field variable transforms as $\varphi \rightarrow \phi/g_k$. Finally, expressing $\mathtt{f}$ and $\mathtt{v}$ in terms of $\mathsf{f}$ and $\mathsf{v}$ via (9.25), and inserting (9.25) into (9.21)–(9.22), one arrives at coupled flow equations for the unknown functions $\mathsf{f}$ and $\mathsf{v}$ as functions of $(g, \phi^2)$. The explicit form of these transformed flow equations is lengthy and complicated, and adds no value to the current discussion. If the reader wants to see them, they can go to [7], and the attached Mathematica notebook [236]. In the main text, we will focus on their qualitative structure, the associated boundary value problem, and the emergent inflationary dynamics.

As stated before, the main goal of this separation is to allow us to encapsulate all irrelevant directions inside $f$ and $v$, and we would like to make predictions for all those. In particular, we want to determine the functional form of $\lim_{k\to 0}\{\mathsf{f}(g_k,\phi^2), \mathsf{v}(g_k,\phi^2)\} = \{\mathsf{f}(0,\phi^2), \mathsf{v}(0,\phi^2)\}$. This is what we are after.

## 9.3 Numerical solution of RG flow equations

The resulting flow equations for $\mathsf{f}(g,\phi^2)$ and $\mathsf{v}(g,\phi^2)$ are given by previously mentioned transformations of (9.21) and (9.22), and they are coupled, non-linear, second-order differential equations. So far, the only constraints to be compatible with the NGFP (9.27) are that $\mathsf{f}(g_*,\phi^2) = \mathsf{v}(g_*,\phi^2) = 0$, and due to the splitting used, we have $\mathsf{f}(g,0) = \mathsf{v}(g,0) = 0$. In order to find the number of free parameters that will be undetermined by the RG flow, and to see if there are any extra degrees of freedom, it is not sufficient to look only at the stability properties around the fixed point (see chapter 7), but one must also look at the differential order of the equation determining the RG flow. In our case, since the flow equations are second order, but we only have the conditions $\mathsf{f}(g_*, \phi^2) = \mathsf{v}(g_*, \phi^2) = 0$ and $\mathsf{f}(g, 0) = \mathsf{v}(g, 0) = 0$, there is still one extra boundary condition to be fixed. This can add extra degrees of freedom to the solutions, apart from the relevant directions around the fixed point.

To close the system of equations, we will impose boundary conditions by using the derivatives $\mathsf{f}^{(0,1)}(g, 0) \equiv f_1(g)$ and $\mathsf{v}^{(0,1)}(g, 0) \equiv v_1(g)$. Akin to an initial value problem, we will use conditions at a fixed value of the field $\phi = 0$ and $g = g_\star$ as initial conditions to find numerical solutions.

A priori, one does not know which functions $f_1(g)$ and $v_1(g)$ are consistent with the equations determining $\mathsf{f}(g, \phi^2)$ and $\mathsf{v}(g, \phi^2)$, and would give rise to solutions. In order to find candidates for these boundary conditions, we expand the flow equations (9.21) and (9.22) in their $\phi$ field representation, near $\phi = 0$ up to order $\phi^2$, assuming $\mathsf{f}(g, \phi^2) \simeq f_1(g)\phi^2$ and $\mathsf{v}(g, \phi^2) \simeq v_1(g)\phi^2$. To determine $v_1(g)$ and $f_1(g)$ analytically, one can assume $|f_1(g)|$, $|v_1(g)| < 1$, and expand to the lowest non-trivial order in $f_1$ and $v_1$. This would give us an idea of what kind of boundary conditions are compatible with the PDEs. The resulting expanded equations are

$$\begin{aligned} 82g\, v_1(g) + \left(48\pi - 41g^2\right) v_1'(g) = 0\,, \\ 12g\, f_1(g)^2 + \left(96\pi^2 - 82\pi g^2\right) f_1'(g) = 0\,. \end{aligned} \tag{9.30}$$

For this case, one can find analytic solutions, and they are

$$\mathsf{v}^{(0,1)}(g, 0) = v_1(g) = \frac{41\left(\frac{48\pi}{41} - g^2\right) m_0^v}{48\pi}\,, \tag{9.31}$$

and

$$\mathsf{f}^{(0,1)}(g, 0) = f_1(g) = \frac{m_0^f}{1 - \frac{3}{41\pi}\, m_0^f\, \log\left(\frac{48\pi - 41g^2}{48\pi}\right)}\,, \tag{9.32}$$

where $m_0^v$, $m_0^f$ are free parameters, that, due to the assumptions (expansions) we made to arrive at these equations, should be smaller than 1. Nevertheless, this procedure was merely to get an idea of acceptable boundary conditions to the system, and does not guarantee that these boundary conditions would give rise to solutions of the system. One can explore all kinds of values for $m_0^v$, $m_0^f$, even larger than 1, until finding solutions of the system. These boundary conditions are compatible with the NGFP, since they satisfy $\lim_{g\to g_*} v_1(g) = \lim_{g\to g_*} f_1(g) = 0$. These two free parameters $m_0^f$ and $m_0^v$ determine the "infrared mass couplings" of both potentials $\mathsf{f}$ and $\mathsf{v}$, since they satisfy $\lim_{g\to 0} f_1(g) = m_0^f$ and $\lim_{g\to 0} v_1(g) = m_0^v$.

This is a clear example of how boundary conditions also determine the predictivity of the FRGE. It is not only up to the stability properties of the NGFP and the number of relevant directions in a truncation to establish the number of free parameters that encode the full theory. The functional nature of the system has extra degrees of freedom given by the order of the PDEs. In this case, when using these boundary conditions, we have 2 extra free parameters, $m_0^v$, $m_0^f$, that come from boundary conditions compatible with the NGFP, and that would not have been seen in a truncation analysis of the stability properties of the NGFP.

Notice that this method to determine the boundary conditions is not unique, and one can come up with other boundary conditions that are well motivated and are still compatible with the NGFP and the flow equations themselves. In this chapter, we will restrict ourselves to these boundary conditions, but modifications of those can be found in the supplementary Mathematica notebook [236].

Using these results, one can close the system of equations with the boundary conditions (9.31) and (9.32), plus $\mathsf{f}(g_*, \phi^2) = \mathsf{v}(g_*, \phi^2) = \mathsf{f}(g, 0) = \mathsf{v}(g, 0) = 0$. The numerical solution implementation can be found in the supplementary material [236], where we use standard numerical solvers in Mathematica to find a solution for equations (9.21) and (9.22) in their $\phi$ field representation on the range $\phi \in [0, 10]$.

The numerical solutions for the particular choice $m_0^f = \mathsf{f}^{(0,1)}(0,0) = 0.42$ and $m_0^v = \mathsf{v}^{(0,1)}(0,0) = 2.75 \times 10^{-10}$, can be seen in Fig. 9.4. These values are fine-tuned in order to get the desired results in the next section, and different values would change the outcome. In a way, these two degrees of freedom span the set of possible theories emerging from the NGFP.

In Fig. 9.4 we can see that the shift-symmetry of the NGFP is broken when $g < g_*$, i.e. when the coarse graining scale $k$ is less than infinity. These types of solutions are only allowed thanks to the symmetry-breaking boundary conditions, without which the potentials $\mathsf{f}$ and $\mathsf{v}$ will remain 0 throughout the entire flow, preserving the shift-symmetry of the NGFP.

The interaction potentials obtained at $k = 0$, $\mathsf{f}(0, \phi^2)$ and $\mathsf{v}(0, \phi^2)$ can be extracted from Fig. 9.4. Their functional dependence is depicted in Fig. 9.5. One can see that the predicted potentials $\mathsf{f}(0, \phi^2)$ and $\mathsf{v}(0, \phi^2)$ have non-trivial dependence on the scalar field in the studied range. These results are subject to a particular choice of the free parameters, and different choices give different results. At least for the analysed values, one can see that the effect of changing the parameters $m_0^f$ and $m_0^v$ is an overall multiplication factor, and does not change the shape of the potentials (see appendix E).

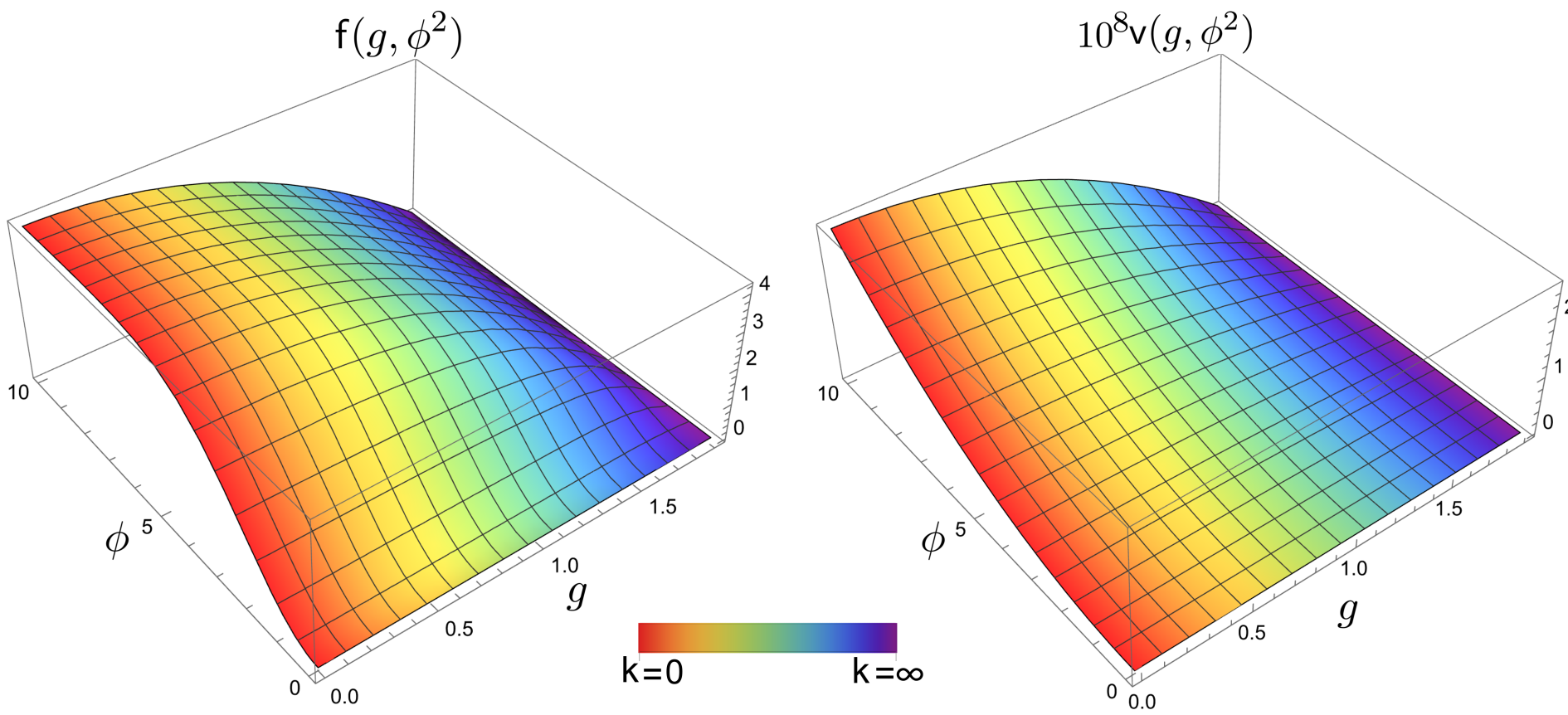

**Figure 9.4.:** Numerical solutions for the functions $\mathsf{f}(g,\phi^2)$ and $\mathsf{v}(g,\phi^2)$ in the domain $(g,\phi) \in [0,\sqrt{\frac{48\pi}{41}}] \times [0,10]$ using boundary conditions (9.31) and (9.32), fixing $m_0^f = 0.42$ and $m_0^v = 2.75 \times 10^{-10}$. The predicted potentials can be extracted from the intersection of the surfaces with the plane $g = 0$. The colour coding is a linear function of $g$ that interpolates between red and purple, meant to represent small and high values of the original coarse-graining scale $k$.

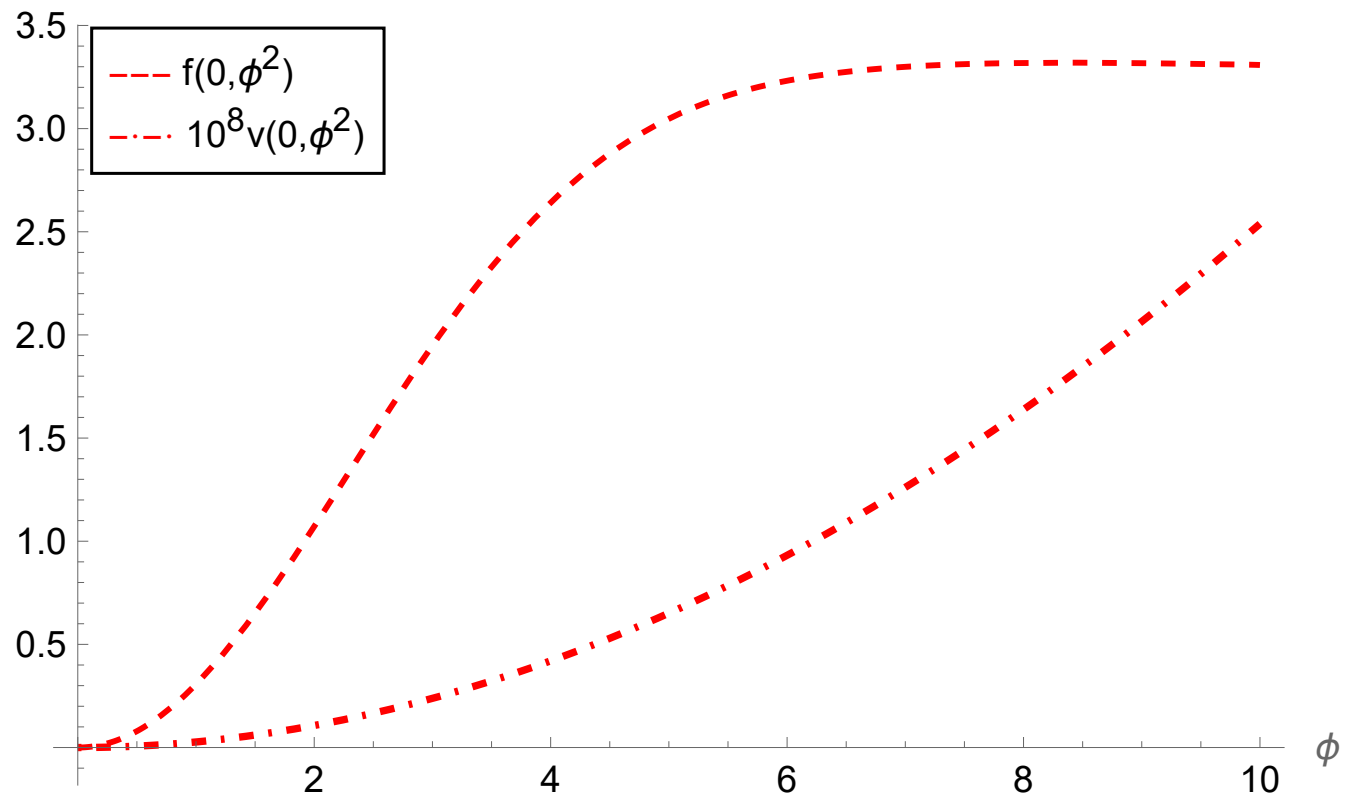

**Figure 9.5.:** Numerical result for the predicted functional dependence of $\mathsf{f}$ an $\mathsf{v}$ at vanishing regulator ($k = 0$), in the range $\phi \in [0,10]$, for the choices $m_0^f = 0.42$ and $m_0^v = 2.75 \times 10^{-10}$.

Now, we are interested in the phenomenology of the predicted potentials. We found that this is easier to understand after a change of variables in the effective action, as we will show in the next section.

## 9.4 Emergence of Inflaton Potential

The phenomenological consequences of the potentials obtained in Fig. 9.5, can be easily interpreted after a conformal change of variables. Specifically, the scalar-tensor model given in (8.30) is expressed in the so-called "Jordan frame", but the same action can be transformed to the "Einstein frame" (E) [237, 238]. To change to such frame, one performs a variable transformation

$$g^E_{\mu\nu} \equiv (1 + \mathsf{f}(g, \phi^2))g_{\mu\nu} \tag{9.33}$$

such that the action (8.30) is turned into

$$\Gamma_k = \int d^4x \sqrt{|g^E|}(-\frac{R^E}{16\pi g^2} + \frac{V_{\text{eff}}(g, \phi(\sigma))}{(16\pi g^2)^2} + \frac{\partial_\mu\sigma\partial^\mu\sigma}{2\, g^2}), \tag{9.34}$$

where

$$V_{\text{eff}}(g, \phi(\sigma)) \equiv \frac{2\lambda + \mathsf{v}(g, \phi^2(\sigma))}{(1 + \mathsf{f}(g, \phi^2(\sigma))^2}, \tag{9.35}$$

and where the new scalar field satisfies

$$(\frac{\partial\,\sigma}{\partial\,\phi})^2 = \frac{1}{(1 + \mathsf{f}(g, \phi^2))} + \frac{3\,(\mathsf{f}'(g, \phi^2))^2}{(1 + \mathsf{f}(g, \phi^2))^2}. \tag{9.36}$$

As we said before, we are interested in the phenomenology of the effective potentials once all quantum fluctuations are integrated out. This is obtained by looking at $V_{\text{eff}}(0, \phi(\sigma))$. In Fig. 9.6 one can see the resulting effective potential as a function of the variable $\phi$. In this result, we approximated the relevant parameter $\lambda_{k=0} \simeq 10^{-120} \simeq 0$, since we found it made no difference in the numerical predictions. Presenting the results as a function of $\phi$ is merely for simplicity, and the phenomenology is equivalent for the potential as a function of $\sigma$. If one wants to see the shape of the potential as a function of $\sigma$, one must solve (9.36) numerically, at $g = 0$. This can be found in the supplementary material [236], and the appendix E.

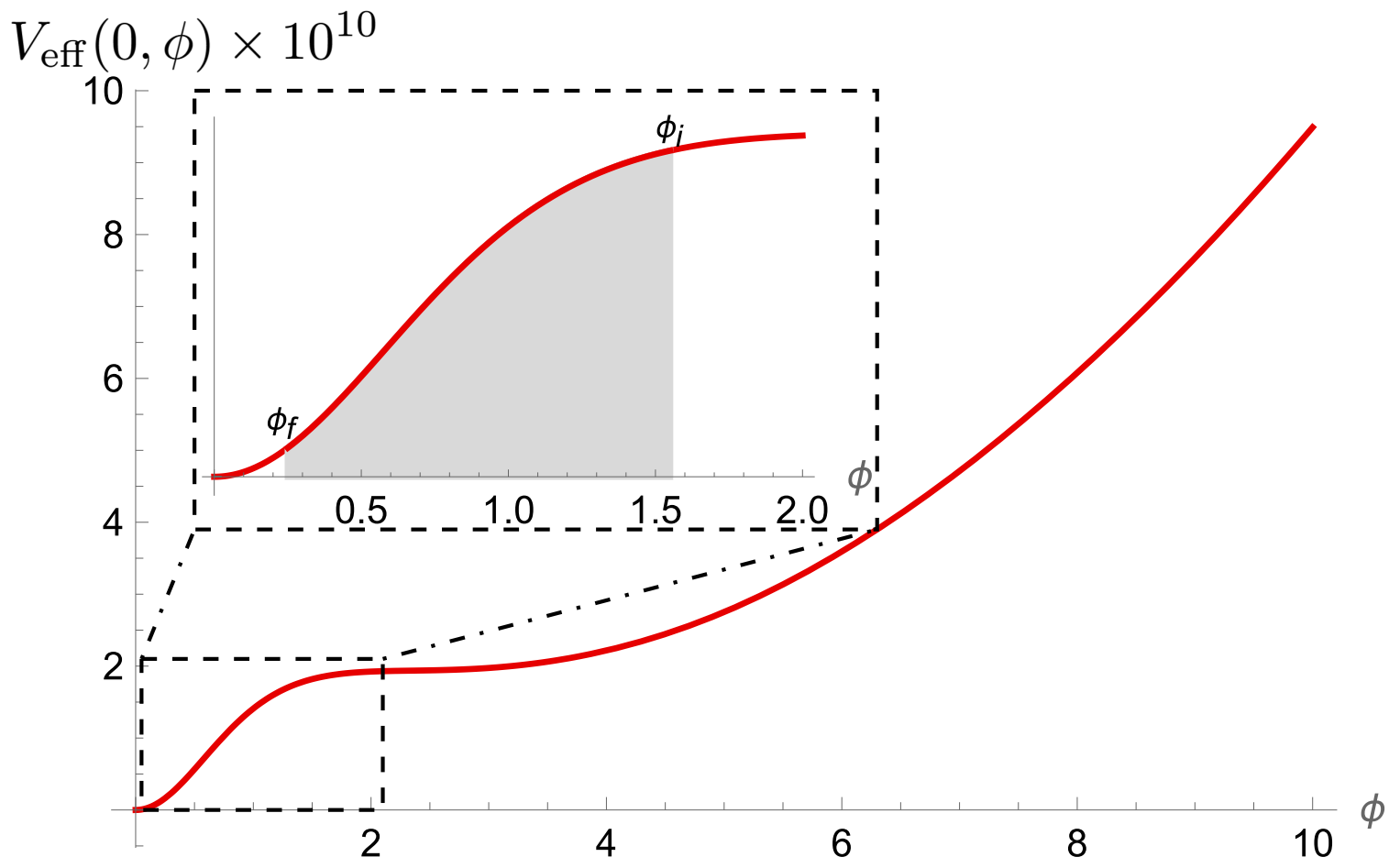

**Figure 9.6.:** Obtained effective potential for the inflaton field as a function of $\phi$, in the range $\phi \in [0, 10]$. The zoomed in region represents the region where a period of cosmological inflation, compatible with observations, can occur. The initial and final values of the field for generating such period are $\phi_i = 1.55$ and $\phi_f = 0.25$.

As one can immediately see in Fig. 9.6, the potential presents an approximate plateau at values of $\phi$ between 1.5 and 3, which, as mentioned before, can be interpreted equivalently to values of the original scalar field $\tilde{\varphi}$ between 1.5 and 3 times the Planck mass.

This led us to compute inflationary observables associated with a field starting at some value of such a plateau. In the way the potential $V_{\text{eff}}$ is defined, the slow-roll observables [11] take the form

$$\begin{aligned}
\epsilon &= \frac{1}{8\pi}\frac{1}{2}\left(\frac{V_{\text{eff}}'}{V_{\text{eff}}}\right)^2 & \eta &= \frac{1}{8\pi}\left(\frac{V_{\text{eff}}''}{V_{\text{eff}}}\right) \\
n_s &= 1 - 6\,\epsilon + 2\,\eta & r &= 16\,\epsilon \\
A_s &= \frac{1}{4}\frac{V_{\text{eff}}}{24\pi^2\epsilon} & N_{ef} &= 8\pi\int_{\sigma_i}^{\sigma_f}\frac{d\,\sigma}{\sqrt{2\,\epsilon}},
\end{aligned} \tag{9.37}$$

where the prime denotes a derivative with respect to $\sigma$, $n_s$ is the spectral index, $r$ is the scalar to tensor ratio, $A_s$ is the amplitude of scalar fluctuations and $N_{ef}$ is the number of e-folds between the initial value $\sigma_i$ and the final $\sigma_f$. The parameters $\epsilon$ and $\eta$ are the previously introduced slow-roll parameters. As mentioned before, the condition for slow-roll inflation to happen is that they are sufficiently small ($|\epsilon|$, $|\eta| << 1$) for a certain range of values of the scalar field. Here we will compute everything in terms of the field $\phi$, using the chain rule and the relation between $\sigma$ and $\phi$ given by (9.36).

If we assume the start (of the end) of inflation is at $\phi_i = 1.55$, and determine the final of inflation when one of the slow roll parameters becomes 1 (in our case it was $\epsilon$), it gives us the value for the field $\phi_f = 0.25$. For that interval, one can say that the field slowly rolls on the potential, and one has the following results for the inflationary observables

$$\begin{aligned} n_s &\simeq 0.965 \qquad & r &\simeq 0.005 \\ A_s &\simeq 2.06 \times 10^{-9} \qquad & N_{ef} &\simeq 66\,, \end{aligned} \tag{9.38}$$

where $n_s$, $r$ and $A_s$ were computed at $\phi_i$, and $N_{ef}$ between $\phi_i$ and $\phi_f$. These values are in agreement with current observations made by [239].

At this stage, it is important to emphasise that these results were obtained on the basis of a small amount of fine-tuning. Only the numerical value of two mass couplings ($m_0^f$, $m_0^v$) had to be fixed to obtain this result. The non-trivial field dependence of the obtained effective potential is the result of the interplay between the gravitational degrees of freedom and the scalar degrees of freedom throughout the renormalization group flow between the NGFP and the infrared. To put it in other words, its shape is a consequence of the quantization of gravity plus the scalar degree of freedom, and its UV completion.

We have achieved what we claimed at the beginning of this work, and obtained, from a UV completion of gravity, an effective potential for a scalar degree of freedom that is consistent with a period of cosmological inflation. This potential is derived, not constructed, from fundamental physics, using only Wilsonian renormalization group techniques.

Up to now, every result was independent of any identification of the coarse-graining scale with a physical scale, such as the curvature of spacetime, or with the age of the universe, as is often done in the so-called "RG improvement" procedures. All our results were obtained exactly in the limit of vanishing regulator. Nevertheless, one can take one more step. In the following, we will accept that one can relate the renormalization group scale $k$ with some physical scale, such that there is an identification between $k$ and physical time. This identification should be such that large values of $k$ are associated with early times of the universe, and small values of $k$ with later times. We do not need any particular identification, since our discussion does not require it.

If one accepts that there is such identification, an argument that has been largely used [141, 150, 240–244], but also very criticized [143, 245], one can provide an explanation for the initial value of the inflaton field near the end of inflation. In Fig. 9.7, one can see snapshots of the RG flow of the effective potential $V_{\rm eff}$ as a function of $\phi$, in the studied range. Interestingly, the potential flows from the scale-invariant, and shift symmetric NGFP, where it takes the value $\lim_{k\to\infty} V_{\rm eff} = 2\lambda_*$, given by (9.27), to the inflaton potential shown in Fig. 9.6.

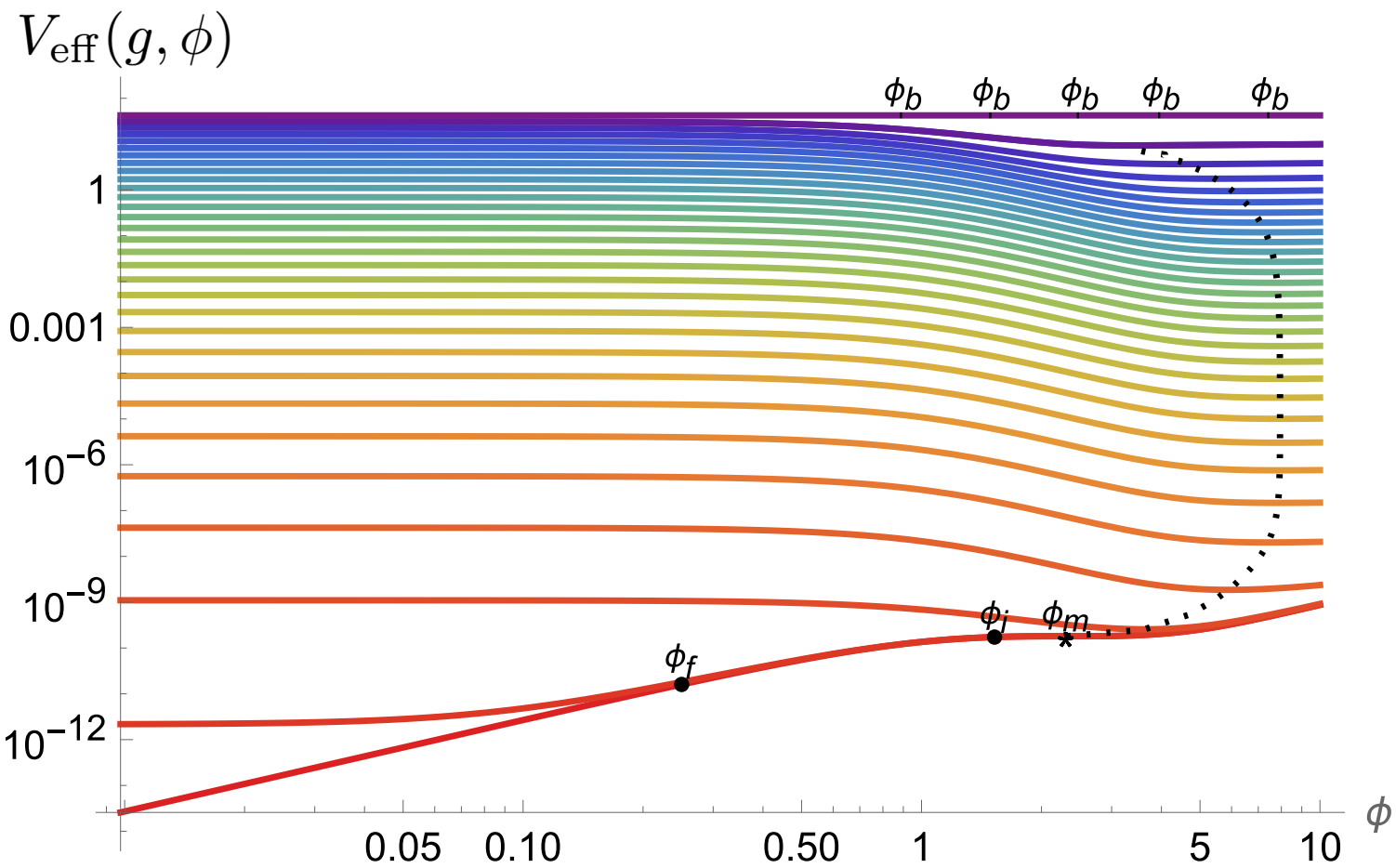

**Figure 9.7.:** Snapshots of the RG evolution of $V_{\rm eff}$, from the UV to the IR in logarithmic scale. The adiabatic evolution of the universe is shown by the dashed line, where it follows a stable minimum from the UV, that becomes unstable near $k \simeq 0$, around $\phi_m \simeq 2.3$.

A tantalizing idea, extracted from the RG evolution of $V_{\rm eff}$, is that if the very early universe was effectively described by an action close to (8.30) near the NGFP, the initial conditions for the inflaton field do not have to be fine-tuned. This is because the NGFP is scale invariant, so the action at that fixed point will be independent of the value of the scalar field, and therefore the universe can have arbitrary initial conditions ($\phi_b$) for the field $\phi$ at its beginning. Furthermore, for values of $k < \infty$, $V_{\text{eff}}$ develops a stable minimum as a function of the scalar field $\phi$. If the universe begins near the NGFP, and is allowed to evolve adiabatically, the scalar field values will follow that minimum. This minimum becomes unstable near $k \simeq 0$, around $\phi_m \simeq 2.3$. This value is slightly above $\phi_i$, so that the field can roll down to $\phi_i$, where we start seeing the primordial fluctuations. Between $\phi_m$ and $\phi_i$ the universe experiences more than 70 e-folds, thus this region is not observable. An adiabatic evolution of the universe, with respect to the RG flow, will follow this trajectory and might explain the initial conditions of the inflaton near the end of inflation, without the need to fine tune its initial value, since it can start at an arbitrary value in the UV ($\phi_b$).

In Fig. 9.7, snapshots of such evolution are presented, where the dashed line represents the values of the scalar field if the universe evolves adiabatically, with an arbitrary starting value of the scalar field $\phi_b$ at the beginning of the universe. Once all quantum fluctuations are integrated out, the stable minimum becomes unstable, and the end of inflation begins, giving rise to the observed fluctuations.

These and similar arguments have been used in several works before to describe the phenomenology of the RG flow, in particular when using RG improvements techniques, but also in other approaches, like using scale invariant solutions of the Wetterich equation [168, 246–248]. The underlying justification for the assumption that the universe at its beginning is described by the highest energy degrees of freedom of the path integral ($\Gamma_k$ for large $k$) relies on the interpretation of $\Gamma_k$. Since at large values of $k$, $\Gamma_k$ contains information only of the highest energy modes of the theory, it is reasonable to believe that a denser universe in its beginnings will be mostly described by those degrees of freedom. As the universe evolves in time, it expands and becomes less dense, thus requiring the inclusion of lower energy modes for an accurate effective description ($\Gamma_k$ for lower values of $k$). We would like to emphasise that we did not use any particular identification of $k$ with a physical quantity, and if one desires to study details of the time evolution of the universe using Fig. 9.7, one should stick to a particular identification, such as $k \propto \frac{1}{t}$ or something more sophisticated.

## 9.5 Collecting the results of this chapter

In this chapter we combined asymptotically safe gravity with early-universe cosmology and followed a full functional renormalization group trajectory all the way from a non-Gaussian fixed point in the ultraviolet to an inflaton potential in the infrared. In contrast to the finite truncations used in the previous chapter, we kept the full field dependence of the scalar-tensor potentials $\tilde{f}_k$ and $\tilde{v}_k$ and treated their RG evolution as a coupled functional boundary value problem. The central message is that a concrete, phenomenologically viable slow-roll model of inflation can *emerge* from an asymptotically safe scalar-tensor theory once the fixed point and a small set of boundary data are specified.

Technically, the starting point was the scalar-tensor truncation (9.19) equipped with the simplified one-loop-type flow equations (9.21) and (9.22). We showed that these equations admit a shift-symmetric non-Gaussian fixed point (9.23) with a finite number of relevant directions in the gravitational sector. To make the functional problem numerically tractable and physically transparent, we reparameterized the theory in terms of the dimensionless Planck-unit field $\phi$ and the running couplings $g_k$ and $\lambda_k$, splitting off the pure Einstein-Hilbert part and encoding all scalar self-interactions in two dimensionless functions $\mathsf{f}(g, \phi^2)$ and $\mathsf{v}(g, \phi^2)$, cf. (9.25)-(9.26). In this parametrization, $g$ plays the role of a RG clock, compactifying the flow to the finite interval $g \in [0, g_*]$ and encoding both the UV and IR limits.

A key conceptual point of this chapter is that, for functional truncations, the predictivity of asymptotic safety is controlled not only by the number of relevant directions at the fixed point but also by the differential order of the flow equations and the associated boundary conditions. The second-order nature of the PDEs for $\mathsf{f}$ and $\mathsf{v}$ implies that, in addition to the UV conditions at the fixed point and regularity at $\phi = 0$, one extra piece of functional data must be specified. By expanding the flow near $\phi = 0$ and solving for consistent boundary functions, we identified a two-parameter family of admissible conditions, conveniently encoded in the infrared mass couplings $m_0^f$ and $m_0^v$ that determine $\mathsf{f}^{(0,1)}(g, 0)$ and $\mathsf{v}^{(0,1)}(g, 0)$ through (9.32) and (9.31). These parameters span a continuum of asymptotically safe scalar-tensor EFTs, all of which share the same UV completion but differ in their low-energy phenomenology.

Using these boundary conditions, we numerically solved the coupled, non-linear flow equations for $\mathsf{f}(g, \phi^2)$ and $\mathsf{v}(g, \phi^2)$ on the domain $(g, \phi) \in [0, g_*] \times [0, 10]$. The resulting scaling surfaces are shown in Fig. 9.4. They interpolate between a shift-symmetric, flat configuration at the fixed point and fully developed scalar self-interactions in the IR, demonstrating explicitly how shift symmetry is broken dynamically along the flow once one leaves the UV. The predicted interaction potentials at vanishing regulator, $\mathsf{f}(0, \phi^2)$ and $\mathsf{v}(0, \phi^2)$, extracted from the $g = 0$ slice of these surfaces and displayed in Fig. 9.5, exhibit non-trivial but smooth field dependence up to field values of 10 times the reduced Planck mass. For the ranges explored, varying $m_0^f$ and $m_0^v$ mainly rescales the overall height of the potentials, while their qualitative shape remains robust.

9.5

To connect these results to cosmology, we transformed the Jordan-frame action to the Einstein frame via a conformal rescaling and a field redefinition, arriving at an action of the form (9.34) with a canonically normalized scalar field $\sigma$ and an effective inflaton potential $V_{eff}(g, \phi(\sigma))$ given in (9.35)-(9.36). Evaluated at $g = 0$, this yields an emergent single-field inflaton potential $V_{eff}(0, \phi)$, shown in Fig. 9.6. For the fine-tuned choice $m_0^f = 0.42$ and $m_0^v = 2.75 \times 10^{-10}$, this potential develops a broad, almost plateau-like region for $\phi$ of order a few Planck masses. Using standard slow-roll expressions (9.37) and identifying the observable window with the field range $\phi \in [\phi_f, \phi_i] = [0.25, 1.55]$, we found $n_s \simeq 0.965$, $r \simeq 0.005$, $A_s \simeq 2.06 \times 10^{-9}$, $N_{ef} \simeq 66$, compatible with current CMB bounds [239]. Crucially, the functional dependence of $V_{eff}$ with the field is not postulated by hand, but dictated by the RG flow from the UV fixed point, up to the two mass parameters $m_0^f$ and $m_0^v$.

Beyond reproducing a viable slow-roll potential, the RG evolution of $V_{eff}(g, \phi)$ suggests a compelling explanation for the initial conditions of inflation. As illustrated in Fig. 9.7, the UV fixed point is scale invariant and shift symmetric, so the early universe can start with essentially arbitrary values of the scalar field. Along the RG flow, the effective potential develops a moving minimum that adiabatically drags the field towards a particular trajectory as $g$ decreases. Near $g \simeq 0$, this minimum turns into an inflection point at $\phi_m \simeq 2.3$, slightly above the value $\phi_i$ required to generate more than 60 e-folds of inflation. In an adiabatic picture in which the cosmological evolution tracks the RG flow, this mechanism naturally prepares the inflaton near the appropriate part of the potential without an extra fine-tuning of its initial field value; the only tunings are the two mass parameters that label the functional trajectory itself.

The analysis presented in this chapter is subject to several approximations and therefore opens many avenues for future work. Our flow equations were derived in Euclidean signature, for a specific gauge choice and regulator, and within a one-loop-type approximation to the full functional RG. It will be important to test the robustness of the emergent inflaton potential, and of the inflationary predictions it implies, under changes of gauge, regulator, truncation scheme, and, ultimately, using Lorentzian flows. Nevertheless, the central result stands: a concrete asymptotically safe scalar-tensor model can generate, through its RG flow, a realistic inflaton potential and a plausible dynamical origin for its initial conditions, with only two phenomenological parameters left to be fixed. In this sense, the chapter provides a concrete example of how Asymptotic Safety can bridge the gap between UV quantum gravity and observational cosmology, and sets the stage for extending this bridge to more general matter sectors and late-time cosmological observables in more complex truncations.

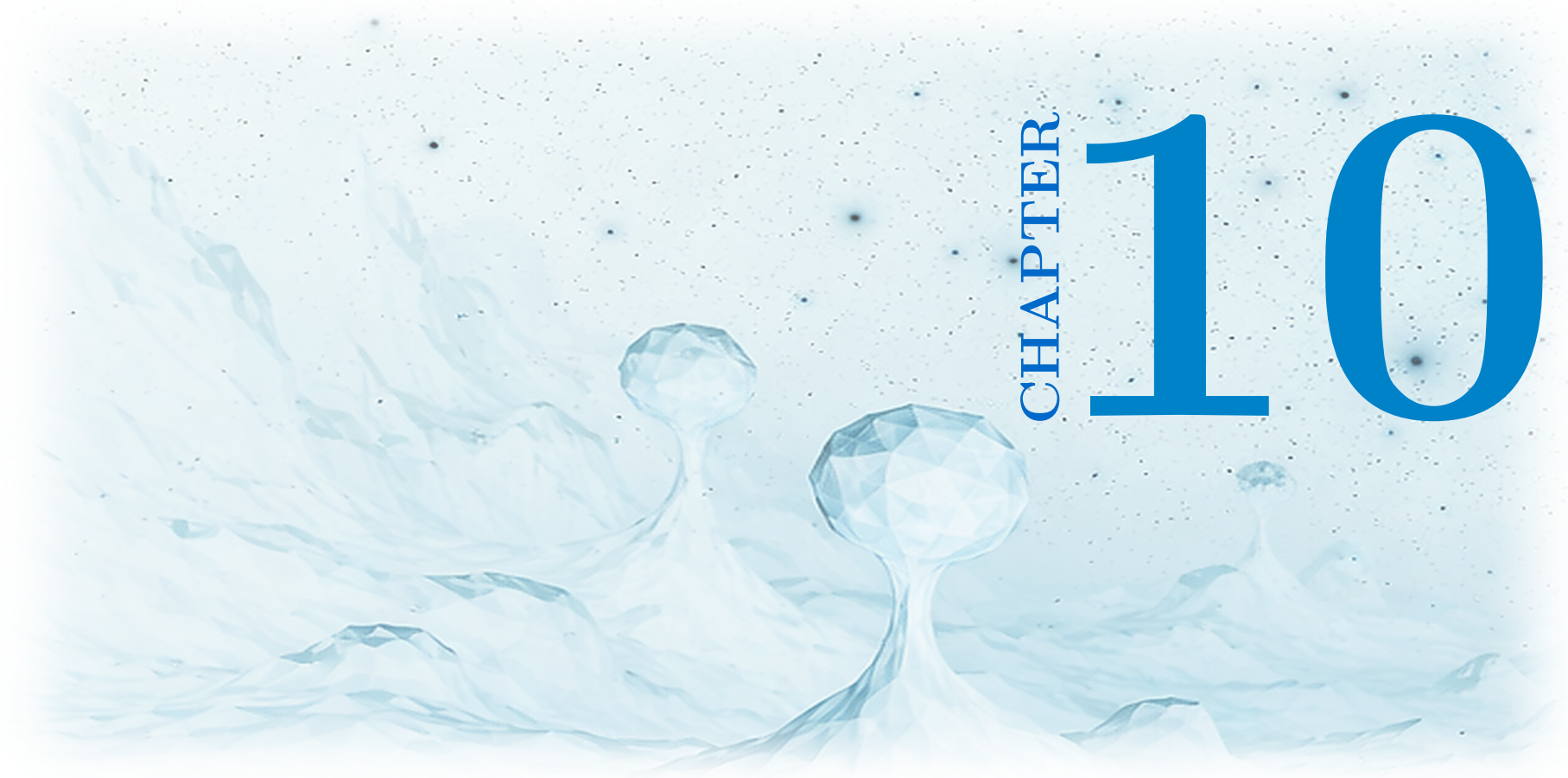

# Summary part III

The central objective of this part of the thesis was to understand how the Functional Renormalization Group (FRG) and the Asymptotic Safety program can be used not only to define ultraviolet (UV) complete theories of gravity, but also to make concrete predictions for infrared (IR) observables. Rather than treating the FRG as a purely formal tool, the chapters in this part developed practical frameworks to (i) reconstruct the UV-critical surface of non-Gaussian fixed points, (ii) parameterize entire renormalization group (RG) trajectories in terms of a small set of relevant data, and (iii) derive effective scalar potentials with direct applications to cosmology. In this way, we explored the use of the Asymptotic Safety hypothesis not just as a consistency condition in the UV, but as a mechanism that constrains the space of emergent effective field theories (EFTs) at low energies.

**Chapter 8 - Flow Invariants to make IR predictions.** We introduced the notion of *flow invariants*, defined by what we call *generating functions*, as a way to characterize the UV-critical surface of a non-Gaussian fixed point in a given truncation. Starting from a finite set of couplings and their beta functions, we showed that one can construct functions $F_\mu(g^n)$ that remain constant along RG trajectories attracted to the fixed point. These generating functions satisfy a linear “master equation” and their level sets $F_\mu = 0$, define subsets of theory space that contain the UV-critical surface.

The UV-critical surface is obtained as the intersection of these hypersurfaces, and the irrelevant couplings become implicit functions of the relevant ones.

This construction makes the predictive content of Asymptotic Safety fully explicit: once the values of the relevant couplings are fixed at some scale, the flow invariants determine the values of all irrelevant couplings along the entire trajectory, including in the IR. We illustrated this in a three-dimensional scalar model, where the UV-critical surface of the Wilson-Fisher fixed point was reconstructed and used to predict IR couplings. The comparison with standard trajectory-by-trajectory shooting methods showed agreement at the percent level, demonstrating that flow invariants provide a robust and efficient way to encode the geometry of theory space and the constraints imposed by the fixed point.

Building on the analytical framework of flow invariants, we then introduced the *Flow Invariant Neural Network* (FINN), a physics-informed neural network architecture designed to solve the master equations for the generating functions directly. The key observation is that each generating function satisfies a linear partial differential equation and different $F_\mu$ do not couple to each other. This makes it natural to represent each invariant by an independent neural network, trained with a loss function that enforces the master equation in the bulk of theory space and the correct linear behaviour near the fixed point.

In contrast to traditional approaches, which reconstruct the UV-critical surface by integrating individual RG trajectories, FINNs learn the geometry of the surface as a global object. We validated this strategy in an asymptotically safe scalar-tensor model, where the UV-critical surface obtained from FINNs reproduces the one obtained from standard methods with percent-level precision. Importantly, the method scales naturally to higher-dimensional truncations, where trajectory-based techniques become computationally expensive. Thus, FINNs provide a practical, flexible, and systematically improvable way to explore the structure of theory space and to map the allowed region of couplings compatible with a given UV completion.

**Chapter 9 - Inflaton emergence from Asymptotic Safety.** The final chapter of this part moved beyond polynomial truncations and addressed *functional truncations* in scalar-tensor gravity, in which the full field dependence of the non-minimal coupling $f_k(\tilde{\varphi})$ and scalar potential $v_k(\tilde{\varphi})$ is retained. Within a simplified but non-trivial approximation to the FRG flow, we derived functional RG equations for these potentials and reparametrized them in terms of variables adapted to numerical computations: a dimensionless Newton coupling $g_k$, a cosmological constant parameter $\lambda_k$, and two dimensionless functions $\mathsf{f}(g_k, \phi^2)$ and $\mathsf{v}(g_k, \phi^2)$, where $\phi$ is the scalar field measured in (scale-dependent) Planck units. In these variables, the UV fixed point corresponds to a scale-invariant, shift-symmetric configuration, while any non-trivial scalar dynamics emerges along the flow towards the IR.

A central conceptual point of this chapter is that the predictive power of functional truncations is controlled not only by the number of relevant directions at the fixed point, but also by the differential order of the functional flow equations and the associated boundary conditions. For the scalar-tensor model studied here, compatibility with the fixed point and regularity at vanishing field still leave room for two additional parameters, encoded in the first derivatives of $\mathsf{f}$ and $\mathsf{v}$ at $\phi = 0$. These parameters act as IR mass-like couplings for the scalar sector. Once they are specified, the full functional trajectory is uniquely determined, and the complete field-dependent potentials at $k = 0$ follow as predictions of the RG flow. This provides a concrete example of how functional truncations extend the usual Asymptotic Safety picture, adding boundary conditions as extra undetermined parameters, while keeping the predictive structure under control.

Transforming the Jordan-frame action to the Einstein frame, we extracted an effective scalar potential $V_{eff}(\phi)$ that exhibits a broad plateau at field values of order the Planck scale. For a fine-tuning of only the two IR mass parameters, this plateau supports a phase of slow-roll inflation whose observables (spectral index, tensor-to-scalar ratio, scalar amplitude, and number of e-folds) agree with current CMB constraints. Crucially, in this setup, the inflaton potential is not chosen by hand, but *emerges* from an asymptotically safe scalar-tensor theory, given a fixed point, a set of functional flow equations, and a small amount of boundary data.

By tracking the full RG evolution of the Einstein-frame potential, we further uncovered a dynamical scenario for the initial conditions of inflation. Near the UV fixed point, the action is approximately scale invariant and insensitive to the initial value of the scalar field. As the RG scale is lowered, the effective potential develops a moving minimum, and under the assumption of an adiabatic identification between RG scale and physical evolution, the scalar field tends to follow this minimum. This mechanism naturally biases the field into the right region of the potential by the time inflation becomes observationally relevant, thereby alleviating the usual fine-tuning problem of initial conditions in bottom-up inflationary model building.

Taken together, the chapters in this part show how FRG methods and Asymptotic Safety can be turned into quantitative tools that link UV fixed points to IR phenomenology. Flow invariants provide a geometric description of UV-critical surfaces and make use of a small set of relevant directions to determine an infinite tower of irrelevant couplings. FINNs extend this idea into a practical numerical framework, making it possible to reconstruct critical surfaces in higher-dimensional truncations where analytic methods are no longer feasible. Functional scalar-tensor flows demonstrate that, once boundary conditions are specified, entire effective potentials can be derived instead of postulated, and that these potentials can have rich cosmological implications, including realistic scenarios for inflation and its initial conditions.

Conceptually, this part complements the CDT-based results of Part II. While CDT provides a fully non-perturbative realization of the gravitational path integral, the FRG and Asymptotic Safety framework offers a continuum, effective field-theoretic language in which UV completions can be connected to observational data. In Asymptotic Safety, the emphasis is on constraining the space of admissible EFTs from above: only those lying on the UV-critical surface of a suitable fixed point, and consistent with phenomenology, can be regarded as viable low-energy limits of quantum gravity.

This perspective becomes particularly relevant in the context of the “zoo” of modified gravity theories that populate the literature, many of which are constructed in a bottom-up fashion with loosely motivated additional parameters. The methods developed in this part provide a template for how such theories should be judged: not merely by their ability to reproduce a given set of observations, but by whether they can be embedded into a controlled RG flow stemming from a sensible UV completion.

# Part IV

# A personal detour towards a fundamental understanding of gravity

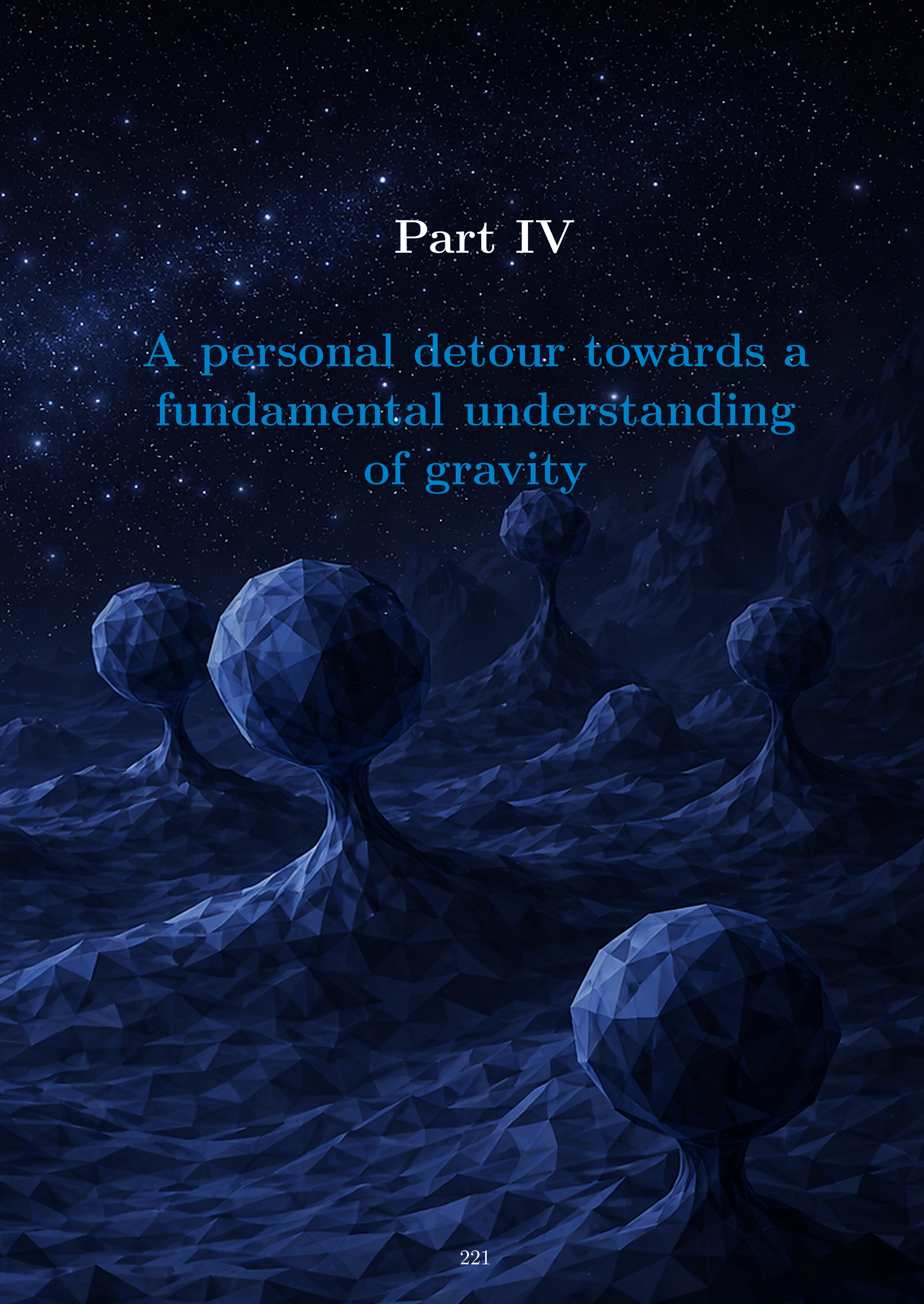

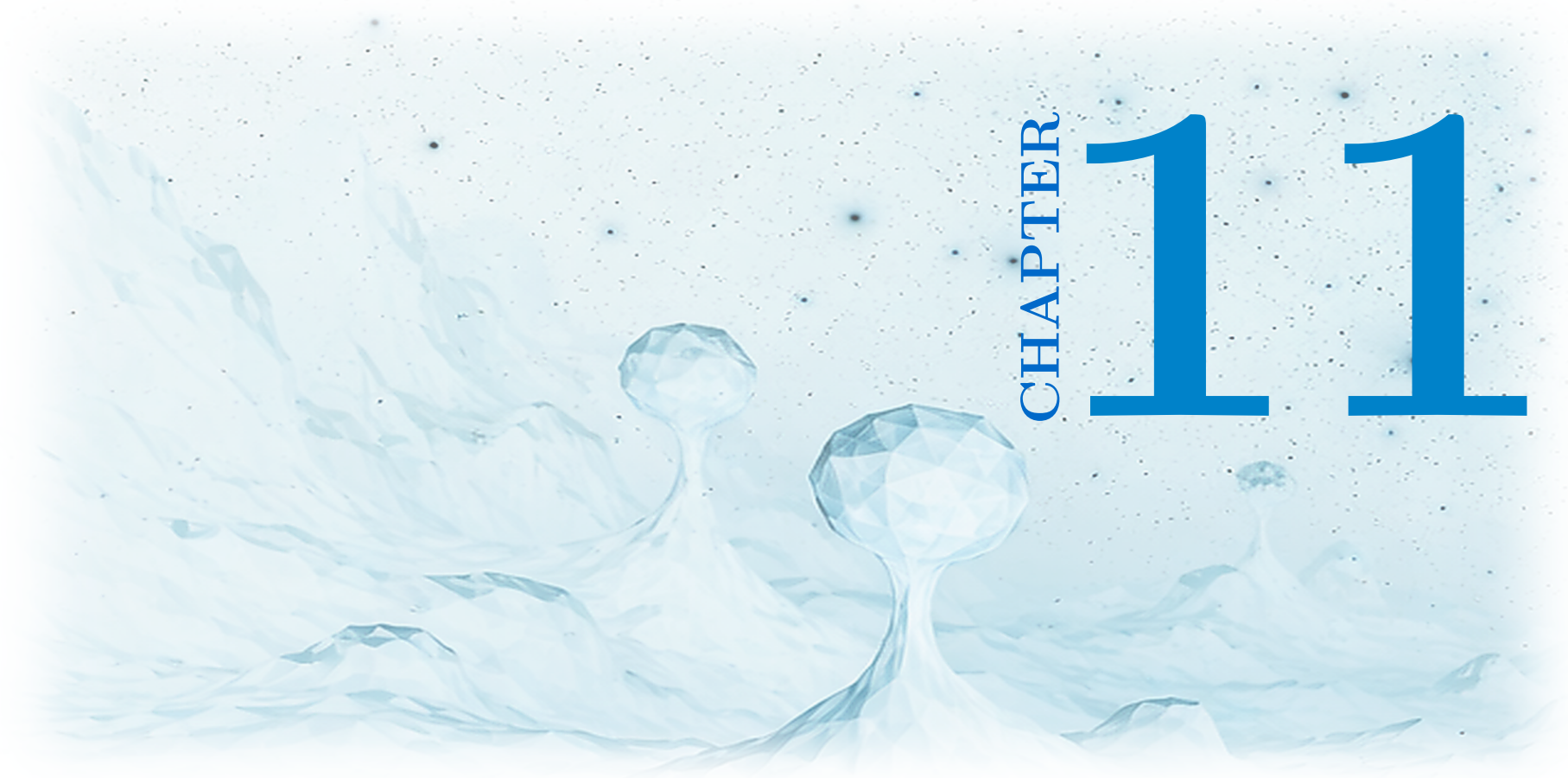

# Lattice Quantum Affine Gravity

Up to this point of the thesis, the reader has seen two complementary ways of thinking about quantum gravity without leaning on perturbation theory. On the one hand, lattice-inspired approaches provide a concrete arena where the elusive “sum over geometries” can be given a concrete meaning and explored numerically. On the other hand, renormalization group ideas provide a language in which ultraviolet completion can be phrased as the existence of a suitable scaling regime, and in which one can ask how Planckian physics may constrain the space of effective theories in the infrared.

In both viewpoints, a recurring theme is that quantum gravity is studied from a non-perturbative point of view, and each approach has access to some particular aspects of the quantum nature of gravity. In the context of CDT, one has a Planckian-sized window where it is possible to analyze quantum geometries that contribute to the path integral, providing great intuition of what a spacetime “looks like” at those scales. Despite many of its positive features, in its formalism, there is no room for smooth differential calculus, the familiar language of tensors, where General Relativity thrives. In the FRG context, the tools from smooth tensor calculus of (pseudo-)Riemannian manifolds are available, making it possible to compute more familiar observables, such as the ones computed in the previous part of this thesis. Nevertheless, despite its good features, the FRG explores theory space in an “agnostic manner”, in the sense that it does not rely on a specific path integral formulation, and thus, has no access to explore geometrical properties of quantum geometries like CDT does.

With that context in mind, this final block is where I allow myself to take a more personal detour. I will try to combine the appealing features from both approaches explored in the previous parts of this thesis, their non-perturbative nature, access to smooth tensor calculus, and access to quantum geometries from a specific path integral realization. Instead of asking only *how* to quantize General Relativity as it is usually written, I want to also ask *what do we choose to quantize?*.

The starting point of the two approaches explored in this thesis, CDT and Asymptotic Safety using the FRG, is the metric $g_{\mu\nu}$. Here I will explore a different possibility: a purely affine description, which is classically equivalent to GR under mild conditions, but in which the fundamental variable is the affine connection. Here, the metric, when it exists at all, is an emergent, derived object.

11.0 The main novelty of this part of the thesis is to bring this purely affine formulation into a concrete non-perturbative setting by constructing an explicit lattice statistical model for the connection, together with a working Monte Carlo implementation, made publicly available.

The purely affine viewpoint goes back to Eddington's proposal of an action built from the determinant of the (symmetric) Ricci tensor [249], Einstein's theory of the affine field [250], and Schrödinger's extensive study of affine theories [251]. In the modern literature, purely affine and metric-affine formulations have been analyzed in detail, both at the classical and quantum level, giving birth to Eddington-type and Ricci-based theories such as Eddington-inspired Born-Infeld gravity (EiBI), and other extensions to General Relativity (GR) [252–271]. These models show that determinant-type actions can be classically equivalent to GR in suitable regimes and parameter ranges, as we will later re-derive for clarity.

Beyond this classical equivalence, there is also an aesthetic motivation for taking the connection as fundamental. All Standard Model interactions are mediated by excitations of gauge connections (photons, $W^{\pm}$, $Z$, gluons). From this perspective, it is natural to ask whether gravity might also be described in terms of excitations of a connection rather than of a metric field. In a purely affine formulation, a "graviton" would be replaced, at least heuristically, by an excitation of the affine connection itself, a "connecton".

The framework explored here is a step towards making such a picture precise in a non-perturbative setting.

Our starting point is the Palatini formulation of GR in $D$ spacetime dimensions [272], where we re-derive a purely affine theory that is equivalent to GR. For $D > 2$, and under mild assumptions such as having an arbitrarily small, non-vanishing positive cosmological constant, the metric can be eliminated algebraically in terms of the symmetric part of the Ricci tensor (and the matter sources if present). This yields a purely affine action of Eddington-type that is classically equivalent to GR. Thus, via this algebraic mechanism, the affine and metric descriptions carry the same local propagating degrees of freedom. A crucial property for our purposes is that, once the algebraic solution for the metric is inserted back into the action, the resulting Lagrangian density has a *definite sign*, determined by the sign of the cosmological constant. For a positive cosmological constant, the action is bounded below and strictly positive. This contrasts sharply with the usual metric formulation given by the Einstein-Hilbert action, whose unboundedness due to the conformal mode is a long-standing obstacle to defining a gravitational path integral. *The bounded nature of the purely affine formulation is the core motivation for this work*, because it allows one to define a non-perturbative statistical weight directly from the action.

In this section, we construct a non-perturbative, lattice formulation of the purely affine Eddington-type action and study it as a statistical field theory of the affine connection. We discretize spacetime on a fixed hypercubic lattice, approximate the relevant quantities by finite differences (and, as an alternative, via gravitational Wilson loops [56]), and define a statistical partition function with Boltzmann weight $\exp(-S)$, where $S$ is the lattice version of the derived purely affine action. Because the action has a definite sign, this defines a well-posed Gibbs measure over affine connections, provided we specify the ensemble and handle redundant degrees of freedom, as we will do later on. The lattice explicitly breaks diffeomorphism invariance, alleviating some technical problems, but creating others, such as the need to study the recovery of diffeomorphism invariance in a suitable lattice-spacing-to-zero limit.

The main results can be summarized as follows:

1. On a hypercubic lattice, we formulate torsionful and torsionless ensembles of affine connections, and construct the corresponding statistical partition functions.

2. We implement a local Metropolis algorithm for the affine connection and provide a publicly available C++ code [273] capable of simulating these ensembles in arbitrary dimension $D$. As a first numerical test, we perform Monte Carlo simulations in $D = 2$ of the torsionless ensemble on lattices of three different sizes, defining and computing the expectation values of two different diffeomorphism-invariant quantities that characterize different volume measures of the emergent geometries. We find that the resulting statistical system is well-behaved: the distribution of the measured quantities depends smoothly on the parameters of the theory, with no signs of instabilities or pathologies in the explored regime.

The scope of the original work presented in the first sections of this chapter is to take the first steps into this new research avenue and to establish a concrete formulation and computational baseline from which more refined ensembles and detailed studies can be developed. We will later use precisely this baseline to explore a first four-dimensional, symmetry-reduced ensemble in which an emergent metric can be reconstructed configuration by configuration.

We provide specific ensembles to define the partition function, as well as a code to simulate the systems in $D$ dimensions. In the unrestricted two-dimensional lattice simulations, we do not yet constrain these ensembles so that specific metric properties emerge from them, as we will discuss later on. Since these physical constraints are not imposed in that first numerical study, we initially restrict the analysis to the simplest possible scenario. The limitations encountered there will motivate the four-dimensional construction developed at the end of the chapter, where fixed signature and metric reconstruction are treated explicitly. As well, we do *not* claim to have constructed a Lorentzian, unitary quantum theory of gravity, as is done in CDT. We work entirely within a statistical framework and do not address the existence of a Wick rotation.

Treating gravity as a statistical field theory is standard practice in the early stages of development of an approach, and has not prevented the community from exploring such models even when the precise relation to a Lorentzian, unitary theory is not fully under control. As we have seen in the previous parts of this thesis, a prominent example is the Asymptotic Safety scenario, in which gravity is described by a quantum field theory whose renormalization group flow approaches a non-trivial ultraviolet fixed point.

While recently there has been progress towards Lorentzian calculations [149, 153–157], in the last three decades most progress has been achieved using FRG computations that do not assume a particular Wick rotation of the effective field theories studied, nor a fixed signature for the quantized metrics. While some computations do fix the signature of the full metric, most split the metric into background and fluctuation and only fix the signature of the background, but not of the full metric. In this sense, while substantial evidence for such a fixed point has been accumulated in FRG studies of "quantum Einstein gravity" and extensions [135–152], the setting where many of those results are derived can be qualified as purely statistical. In particular, all the studies of scalar-tensor theories performed in Part III of this thesis are of that nature.

On the lattice, a similar path has been pursued. Early work constructing gravitational ensembles includes studies of simplicial geometries within Euclidean Dynamical Triangulations [274–278], which we already explored in Part II of this thesis. While EDT is a statistical framework with no clear notion of a time direction and no explicit implementation of a Wick rotation, its exploration led to more refined ensembles with a notion of proper-time foliation and a concrete Wick rotation, i.e., the framework we used the most to explore quantum gravity in Part II, Causal Dynamical Triangulations [61, 69, 71, 73, 78, 279–285].

Just like in CDT, a scale-invariant limit in parameter space would also be required to define a continuum limit of the statistical model studied in this part of the thesis. While some contact between affine gravity and FRG studies exists to date, such as [286–289], this has not been analyzed to the extent that metric formulations have been, and to the best of my knowledge, a lattice counterpart such as the one studied in this section has never been explored.

In that sense, our goal here is modest: to show that the purely affine Eddington-type action admits a concrete, statistical, non-perturbative lattice realization, with controllable Monte Carlo dynamics, and to provide a computational framework from which more refined ensembles and detailed studies can be developed. This work should not be viewed as a complete realization of the vast amount of work that this new line of research opens, but as a solid starting point.

The rest of this part of the thesis is organized as follows. In Sec. 11.1 we review the Palatini formulation, adding simple matter content for generality, and re-derive the purely affine Eddington-type action that is classically equivalent to GR, including a discussion of the conditions for the existence of an emergent metric and the special role of two dimensions. In Sec. 11.2 we discuss the lattice approximation for the partition function of this affine theory, and define the torsionful and torsionless statistical ensembles, including a discussion of gauge fixing and path integral measure definitions. In Sec. 11.3 we test the developed code in the simplest toy model accessible within Monte Carlo simulations, that of a two-dimensional manifold, and provide concrete diffeomorphism-invariant quantities that can be computed in the simulations, as well as specific numerical results. We conclude in Sec. 12 with a summary and an outlook on extensions to higher dimensions, improved discretizations, and the question of continuum and Lorentzian limits.

## 11.1 Affine formulation of General Relativity

11.1

As stated before, we will explore the first steps for quantizing gravity under the assumption that the metric is not fundamental, but instead the connection is. It has long been known that GR can be written in a purely affine form, as Eddington already observed almost a century ago [249]. In this section, we briefly review this construction in a form adapted to our purposes, emphasizing the conditions under which it is equivalent to the usual metric formulation.

### 11.1.1 Palatini formulation and two algebraic routes

We start from the Einstein-Hilbert action in Palatini form, and for generality of the discussion, we include a minimally coupled real scalar field with self-interaction potential $V(\phi)$, in $D$ spacetime dimensions,

$$S[g,\Gamma,\phi] = \frac{1}{16\pi G}\int d^Dx\,\sqrt{|g|}\,g^{\mu\nu}R_{\mu\nu}(\Gamma) + \int d^Dx\,\sqrt{|g|}\,\Big(-\tfrac{1}{2}g^{\mu\nu}\,\partial_\mu\phi\,\partial_\nu\phi - V(\phi)\Big), \tag{11.1}$$

where $R_{\mu\nu}(\Gamma)$ is the Ricci tensor of an *a priori* independent affine connection $\Gamma^\lambda{}_{\mu\nu}$.

Varying with respect to $\Gamma^{\lambda}{}_{\mu\nu}$ and restricting for simplicity to torsionless connections, one finds

$$\nabla_{\lambda}\big(\sqrt{|g|}\, g^{\mu\nu}\big) - \delta^{\nu}{}_{\lambda}\, \nabla_{\sigma}\big(\sqrt{|g|}\, g^{\mu\sigma}\big) = 0, \tag{11.2}$$

which implies metric compatibility,

$$\nabla_{\lambda} g_{\mu\nu} = 0, \tag{11.3}$$

and hence that $\Gamma$ is the Levi-Civita connection of $g_{\mu\nu}$. Substituting this solution back into (11.1) yields the usual metric formulation of GR with a scalar field.

For our purposes, it is useful to emphasize that the Palatini system admits *two* algebraic routes:

- the standard one, just recalled, in which one solves the connection equation $\delta_{\Gamma} S = 0$ and eliminates $\Gamma$ in favour of $g_{\mu\nu}$;
- an alternative route, in which one treats the metric as an auxiliary field and solves the metric equation $\delta_g S = 0$ algebraically in terms of the affine curvature and matter sources, eliminating $g_{\mu\nu}$ in favour of $R_{(\mu\nu)}(\Gamma)$.

In the second case, the affine connection is taken as the fundamental variable, and the metric, when it exists, becomes a derived quantity.

11.1

### 11.1.2 Eliminating the metric

Varying (11.1) with respect to $g_{\mu\nu}$ gives

$$\begin{aligned} R_{(\mu\nu)}(\Gamma) - \tfrac{1}{2} g_{\mu\nu} g^{\alpha\beta} R_{(\alpha\beta)}(\Gamma) \;&=\; 8\pi G\, T_{\mu\nu}, \\ \text{with}\;\; T_{\mu\nu} = \partial_\mu\phi\,\partial_\nu\phi &- g_{\mu\nu}\Big(\tfrac{1}{2} g^{\rho\sigma}\partial_\rho\phi\,\partial_\sigma\phi + V(\phi)\Big), \end{aligned} \tag{11.4}$$

where parentheses denote symmetrization. In the Palatini setting $R_{(\mu\nu)}(\Gamma)$ is built purely from the affine connection, and (11.4) is an *algebraic* equation for the metric.

A convenient way to see how $g_{\mu\nu}$ can be solved for is to move the $\partial_\mu\phi\,\partial_\nu\phi$ term to the left-hand side and collect everything proportional to $g_{\mu\nu}$ on the right-hand side. Defining

$$P_{\mu\nu} \;\equiv\; R_{(\mu\nu)}(\Gamma) - 8\pi G\, \partial_\mu\phi\, \partial_\nu\phi, \tag{11.5}$$

one finds that (11.4) is equivalent to a proportionality relation of the form $P_{\mu\nu} = C\, g_{\mu\nu}$, where the scalar coefficient $C$ depends only on the trace of the equation. Taking the trace and using $D \neq 2$ to avoid degeneracy, the trace algebra fixes this coefficient uniquely to $C = \frac{16\pi G}{D-2}\, V(\phi)$. Therefore, for $D > 2$ and for potentials $V(\phi)$ that do not vanish, $g_{\mu\nu}$ is proportional to $P_{\mu\nu}$ and can be reconstructed by inversion.[1] A convenient way to write the result is

$$g_{\mu\nu} \;=\; \frac{D-2}{16\pi G\, V(\phi)} \Big( R_{(\mu\nu)}(\Gamma) - 8\pi G\, \partial_\mu\phi\, \partial_\nu\phi \Big), \tag{11.6}$$

which is only valid when the RHS is invertible.

This expression makes it clear that an emergent metric can be reconstructed provided the following conditions hold:

11.1

1. $D \neq 2$, so that the trace of (11.4) does not degenerate;
2. the scalar potential $V(\phi)$ does not vanish in the region of field space explored by the theory, so that the overall scale of $g_{\mu\nu}$ is well defined;
3. the symmetric tensor $P_{\mu\nu} = R_{(\mu\nu)}(\Gamma) - 8\pi G\, \partial_\mu\phi\, \partial_\nu\phi$ is non-degenerate, so that the resulting metric is invertible.

If any of these conditions fails, the algebraic relation cannot be inverted and the metric description breaks down. In particular, the $D \to 2$ limit of (11.6) is singular, and no metric can be defined in that case, even though the purely affine action itself has a smooth $D \to 2$ limit, as we will see later.

[1] See e.g. [253, 255, 256, 258] for more detailed discussions in the purely affine and metric-affine literature.

### 11.1.3 Purely affine action and sign properties

Substituting (11.6) back into the action (11.1), one obtains a purely affine action depending only on $\Gamma$ and $\phi$,

$$S_{\rm aff}[\Gamma,\phi] \;=\; \beta_D \int d^Dx\, \sqrt{\left|\det\Big(R_{(\mu\nu)} - 8\pi G\,\partial_\mu\phi\,\partial_\nu\phi\Big)\right|}\;|V(\phi)|^{-\frac{D}{2}}\,V(\phi)\,, \tag{11.7}$$

with $\beta_D = \frac{(D-1)(D-2)^{\frac{D-2}{2}}}{(16\pi G)^{\frac{D}{2}}}$, and where we dropped explicit dependence on the coordinates and on $\Gamma$ for compactness. In the pure gravity case, obtained by setting $\phi = 0$ and $V(\phi) = \frac{2\Lambda}{16\pi G}$, this reduces to Eddington's purely affine action

$$S_{\rm Edd}[\Gamma] \;=\; \frac{(D-1)(D-2)^{\frac{D-2}{2}}\,2\Lambda}{(16\pi G)^{\frac{D+2}{2}}}\left|\frac{2\Lambda}{16\pi G}\right|^{-\frac{D}{2}}\int d^Dx\,\sqrt{|\det R_{(\mu\nu)}(x)|}. \tag{11.8}$$

As in the original purely affine literature, this action is classically equivalent to GR in $D > 2$ under the conditions listed above: we have only used the metric equations of motion algebraically, without introducing new dynamics.

The dependence on $V(\phi)$ (or on the cosmological constant in the pure gravity case) is central. For the metric to be well defined along the lines of (11.6)-(11.7), the potential must not cross zero in the region of field space explored by the theory. If one writes $V(\phi) = \frac{2\Lambda}{16\pi G} + V_{\rm int}(\phi)$ with an interaction $V_{\rm int}(\phi) \geq 0$, then a negative cosmological constant $\Lambda < 0$ generically implies the existence of field values for which $V(\phi) = 0$, where the expression for $g_{\mu\nu}$ diverges. This signals that the purely affine description in terms of a regular emergent metric breaks down when $V(\phi)$ changes sign, suggesting that a positive cosmological constant is the most interesting parameter range to explore in this approach.

For our purposes, the crucial feature of (11.8) is that for a positive cosmological constant $\Lambda > 0$ the action has a definite sign and is bounded below: the overall prefactor in (11.8) is positive and the integrand involves $\sqrt{|\det R_{(\mu\nu)}(x)|}$, which is non-negative. This contrasts sharply with the usual Einstein-Hilbert action formulated in terms of the metric, whose

unboundedness due to the conformal mode is a long-standing obstacle to defining the gravitational path integral.

In the rest of this work, we take the purely affine action (11.8) as our starting point and treat the affine connection as the fundamental field in a statistical theory. In particular, in the next section, we discretize (11.8) on a hypercubic lattice and use it to define a Gibbs ensemble of affine connections with Boltzmann weight $\exp(-S_{\text{Edd}})$.

## 11.2 Partition function on a hypercubic lattice

To define a non-perturbative path integral for the purely affine theory, we discretize the manifold where the connection is defined using a fixed, periodic, hypercubic lattice and we approximate the Eddington action (11.8) in this lattice. The lattice is introduced purely as an ultraviolet regulator: it explicitly breaks diffeomorphism invariance and is not dynamical, in contrast to Causal Dynamical Triangulations [61, 281]. The fundamental dynamical variables on the lattice are the affine connection components ${\Gamma^\lambda}_{\mu\nu}(x)$, treated as real fields on the sites of the lattice.

We work on a $D$-dimensional manifold $M$ and discretize a coordinate patch into a hypercubic lattice with spacing $\varepsilon$ and periodic boundary conditions. At each lattice site $x$ we identify a tangent-space basis with the coordinate basis aligned along the links of the lattice. All indices $(\mu, \nu, \dots)$ refer to this fixed frame, and repeated indices are summed over unless explicitly stated otherwise.

Given an affine connection ${\Gamma^\alpha}_{\beta\mu}(x)$ on the lattice site $x$, we can approximate derivatives by finite differences. For any lattice field $f(x)$ (with arbitrary index structure) we define the central finite difference $(\Delta^c_\mu f)(x) \equiv \frac{f(x+\varepsilon\hat{\mu})-f(x-\varepsilon\hat{\mu})}{2\varepsilon}$ where $\hat{\mu}$ is the unit vector in direction $\mu$. Applying this to the connection components gives a straightforward lattice approximation of the continuum expression for the Riemann tensor

$$
\begin{aligned}
{R^{\text{lat}\,\alpha}}_{\beta\mu\nu}(x) \;\equiv\; & \Delta^c_\mu {\Gamma^\alpha}_{\beta\nu}(x) - \Delta^c_\nu {\Gamma^\alpha}_{\beta\mu}(x) \\
& + {\Gamma^\alpha}_{\kappa\mu}(x)\,{\Gamma^\kappa}_{\beta\nu}(x) - {\Gamma^\alpha}_{\kappa\nu}(x)\,{\Gamma^\kappa}_{\beta\mu}(x)\,,
\end{aligned}
\tag{11.9}
$$

which reduces to the continuum formula in the limit $\varepsilon \to 0$.

The corresponding lattice Ricci tensor is obtained by contraction:

$$R^{\rm lat}_{\rho\nu}(x) \;\equiv\; R^{{\rm lat}\,\alpha}{}_{\rho\alpha\nu}(x)\,, \tag{11.10}$$

and when needed, we take its symmetric part $R^{\rm lat}_{(\mu\nu)}(x)$.

This finite-difference Ricci tensor is the basic approximation used in the main text of this work. In Appendix F.0.2 we describe an alternative discretization in which the curvature is reconstructed from gravitational Wilson loops on plaquettes. Both discretizations are expected to define the same continuum theory, differing only by lattice artifacts.

### 11.2.1 Lattice Eddington action and statistical partition function

We now discretize the purely affine Eddington action (11.8) by replacing the continuum Ricci tensor by its lattice counterpart and the integral by a sum over sites. In the pure gravity case, with $\Lambda > 0$, this gives

$$S^{\rm lat}_{\rm Edd}(\Gamma) \;=\; \frac{(D-1)(D-2)^{\frac{D-2}{2}}}{(16\pi G)^{\frac{D}{2}}}\left(\frac{2\Lambda}{16\pi G}\right)^{\frac{2-D}{2}} \sum_x \varepsilon^D \sqrt{\left|\,\det R^{\rm lat}_{(\mu\nu)}(x)\right|}\,. \tag{11.11}$$

At the classical level, this approximation has a continuum limit $\varepsilon \to 0$ that recovers the Eddington action,

$$\lim_{\varepsilon\to 0} S^{\rm lat}_{\rm Edd}(\Gamma) \;=\; S_{\rm Edd}(\Gamma). \tag{11.12}$$

Because the continuum action has a definite sign for $\Lambda > 0$, the lattice action (11.11) is non-negative configuration by configuration. Furthermore, for large (algebraic) values of the connection components, due to the determinant structure, the action grows at least linearly with it; therefore, large values of the connection components are exponentially damped. This encourages us to define a statistical partition function in complete analogy with ordinary lattice field theory,

$$Z_{\rm stat} \;=\; \int \mathcal{D}\Gamma\;\exp\!\big[-S^{\rm lat}_{\rm Edd}(\Gamma)\big]\,, \tag{11.13}$$

where $\mathcal{D}\Gamma$ denotes an appropriate product of Lebesgue measures over the connection components at each lattice site $\mathcal{D}\Gamma = \prod_{x,\lambda,\mu,\nu} d\Gamma^{\lambda}{}_{\mu\nu}(x) \times$ (Gauge/Constraints), subject to some constraints for different ensembles, to be specified below.

This construction is deliberately modest: we simply treat $S_{\text{Edd}}^{\text{lat}}$ as an energy functional in a classical statistical mechanics model and interpret (11.13) as a Gibbs measure over affine connections. We do *not* derive this weight from a concrete Wick rotation of a Lorentzian path integral, nor do we assume that such a Wick rotation necessarily exists in the usual sense. As discussed in the introduction, this purely statistical point of view is common in non-perturbative approaches to quantum gravity, as it is out of the scope of this work to go beyond that.

### 11.2.2 Ensembles and path-integral measure

The path integral (11.13) is an integral over the affine connection at each lattice site. There are two natural choices for the integration domain:

- a *torsionful* ensemble, in which all $D^3$ components $\Gamma^{\lambda}{}_{\mu\nu}(x)$ are integrated over;
- a *torsionless* ensemble, in which one restricts from the outset to symmetric connections $\Gamma^{\lambda}{}_{(\mu\nu)}(x)$ and enforces $\Gamma^{\lambda}{}_{[\mu\nu]}(x) = 0$ at all sites.

These correspond to two different choices of configuration space for the statistical system; in both cases the connection is the only dynamical field.

Independent of this choice, the lattice already breaks continuum diffeomorphism invariance. However, there remain internal redundancies associated with the connection field itself. One fundamental issue, which will be addressed later in this part, is that in the non-perturbative formulation we are trying to construct, the action (11.8) and its lattice version, just like any scalar built from many degrees of freedom, is not injective. This means that all manifolds that have the same value for (11.8) may be extremely different configurations. Depending on which observable one is computing, these "flat directions" of theory space, i.e., configurations that leave the action invariant, can either stall Monte Carlo simulations or even render some observables ill-defined, which can be a problem for the definition of the whole program. In this section, that will not be a problem since we will only look at observables where these flat directions do not change their value and factor out, but we will revisit this issue when describing our first attempts to go to four-dimensional quantum gravity within this formalism.

Going back to the redundancies relevant to this section, in the torsionful case there is a projective symmetry; in the torsionless case one must account for the Jacobian associated with restricting to symmetric connections. We now discuss these issues in turn and show that in both cases the corresponding Jacobians are field-independent, so no non-trivial ghost determinants arise.

### Torsionful ensemble

In the torsionful ensemble, the affine connection is not assumed to be symmetric in its lower indices. In this case, the symmetric part of the Ricci tensor is invariant under the projective transformation [290]

$$\Gamma^{\alpha}{}_{\beta\mu}(x) \;\longrightarrow\; \Gamma^{\alpha}{}_{\beta\mu}(x) + \delta^{\alpha}{}_{\beta}\,\eta_{\mu}(x)\,, \tag{11.14}$$

where $\eta_{\mu}(x)$ is an arbitrary vector field. This is a local gauge redundancy of the purely affine theory [255, 256] that must be fixed in the path integral.

A convenient gauge choice is

$$\Gamma^{\alpha}{}_{\alpha\mu}(x) = 0\,, \tag{11.15}$$

which fixes the arbitrary projective degree of freedom at each site. The transformation (11.14) is a translation in field space,

$$\frac{\delta\Gamma'^{\alpha}{}_{\beta\mu}(x)}{\delta\Gamma^{\rho}{}_{\sigma\nu}(y)} = \delta^{\alpha}{}_{\rho}\,\delta^{\sigma}{}_{\beta}\,\delta^{\nu}{}_{\mu}\,\delta(x,y), \tag{11.16}$$

so the Jacobian of the transformation in the measure $\mathcal{D}\Gamma$ is unity. The associated Faddeev-Popov determinant for the gauge condition (11.15) is also a field-independent constant and can be absorbed into the normalization of $Z$.

One may therefore write the torsionful partition function either with an explicit gauge constraint,

$$Z_{\rm tf} \;=\; \int \prod_{x,\lambda,\mu,\nu} d\Gamma^{\lambda}{}_{\mu\nu}(x)\,\prod_{x,\sigma}\delta\big(\Gamma^{\rho}{}_{\rho\sigma}(x)\big)\; e^{-S^{\rm lat}_{\rm Edd}(\Gamma)}\,, \tag{11.17}$$

or, equivalently, with a Gaussian gauge-fixing term in the action,

$$Z_{\rm tf}(\alpha_{\rm gf}) \;=\; N(\alpha_{\rm gf}) \int \prod_{x,\lambda,\mu,\nu} d\Gamma^{\lambda}{}_{\mu\nu}(x)\ \exp\Big[-S^{\rm lat}_{\rm Edd}(\Gamma) - \sum_{x,\sigma} \frac{\big(\Gamma^{\rho}{}_{\rho\sigma}(x)\big)^2}{2\,\alpha_{\rm gf}}\Big], \tag{11.18}$$

where $\alpha_{\rm gf}$ is a gauge-fixing parameter and $N(\alpha_{\rm gf})$ is an irrelevant normalization constant. Both forms are equivalent at the level of expectation values. In practice, one can either work with a strict traceless constraint such as (11.17), or work directly with (11.18), letting the gauge-fixing term suppress excursions away from (11.15).

### Torsionless ensemble

In the torsionless ensemble, the lower indices of the connection are constrained to be symmetric from the outset:

11.2

$$\Gamma^{\alpha}{}_{\beta\mu}(x) = \Gamma^{\alpha}{}_{\mu\beta}(x)\,. \tag{11.19}$$

There is no residual projective symmetry of the form (11.14) within this restricted space of connections, since such transformations remove the connection from the torsionless constraint. However, one must check the Jacobian associated with restricting the functional integration to the symmetric sector.

A convenient way to see that this Jacobian is constant is to change variables from the full connection to its symmetric and antisymmetric parts,

$$S^{\alpha}{}_{\beta\mu} \equiv \Gamma^{\alpha}{}_{(\beta\mu)}, \qquad A^{\alpha}{}_{\beta\mu} \equiv \Gamma^{\alpha}{}_{[\beta\mu]}. \tag{11.20}$$

For each fixed index $\alpha$ and each unordered pair $\{\beta,\mu\}$, the transformation to these new variables has constant determinant

$$\begin{pmatrix} S^{\alpha}{}_{\beta\mu} \\ A^{\alpha}{}_{\beta\mu} \end{pmatrix} = \frac{1}{2}\begin{pmatrix} 1 & 1 \\ 1 & -1 \end{pmatrix}\begin{pmatrix} \Gamma^{\alpha}{}_{\beta\mu} \\ \Gamma^{\alpha}{}_{\mu\beta} \end{pmatrix} \qquad \to \qquad \det\left(\frac{1}{2}\begin{pmatrix} 1 & 1 \\ 1 & -1 \end{pmatrix}\right) = -\frac{1}{2}, \tag{11.21}$$

so the Jacobian of the change of variables is a constant factor that can be absorbed into the overall normalization of the path integral.

Restricting to $A^{\alpha}{}_{\beta\mu} = 0$ is then equivalent to inserting a product of delta functions in the measure. The torsionless partition function can be written as

$$Z_{\text{tl}} = \int \prod_{x,\lambda} \prod_{\mu \leq \nu} d\Gamma^{\lambda}{}_{(\mu\nu)}(x) \prod_{x,\lambda} \prod_{\beta<\gamma} d\Gamma^{\lambda}{}_{[\beta\gamma]}(x) \prod_{x,\lambda,\beta<\gamma} \delta\big(\Gamma^{\lambda}{}_{[\beta\gamma]}(x)\big)\, e^{-S^{\text{lat}}_{\text{Edd}}(\Gamma)} \,. \tag{11.22}$$

Here $\prod_{\mu \leq \nu}$ avoids double counting of symmetric pairs. Equivalently, one can integrate only over the symmetric components and drop the delta functions, since the antisymmetric components are fixed to zero.

In the numerical implementation described in Sec. 11.3 we will show the numerical results of working with the torsionless ensemble, enforcing the symmetry $\Gamma^{\lambda}{}_{\mu\nu} = \Gamma^{\lambda}{}_{\nu\mu}$ in each local update. The torsionful ensemble and its projective gauge fixing are discussed here mainly to show that both natural choices of configuration space lead to simple, local measures with trivial Jacobians.

### 11.2.3 Size of theory space and comparison to metric variables

It is useful to quantify the size of the configuration space of the purely affine theory and compare it with the familiar metric formulation.

In $D$ dimensions, a general affine connection $\Gamma^{\lambda}{}_{\mu\nu}$ has $D^3$ real components at each spacetime point. In the torsionless ensemble we impose symmetry in the lower indices, $\Gamma^{\lambda}{}_{\mu\nu} = \Gamma^{\lambda}{}_{\nu\mu}$, so the number of independent components per point is

$$N_{\Gamma}^{(\text{torsionless})} = D \times \frac{D(D+1)}{2} = \frac{D^2(D+1)}{2}. \tag{11.23}$$

In the torsionful ensemble, we keep all $D^3$ components, but as mentioned before, there is a projective gauge symmetry which, when fixed, removes $D$ gauge directions per point, leaving

$$N_{\Gamma}^{(\text{torsionful, gauge-fixed})} = D^3 - D \tag{11.24}$$

independent components per point after projective gauge fixing.

For comparison, the metric tensor $g_{\mu\nu}$ has $N_g = D(D+1)/2$ components. There is a substantial difference in the number of components to be integrated over in both cases.

This shows that, at fixed dimension $D$, the purely affine theory requires the exploration of a much larger local theory space than the metric formulation. In four dimensions, for example, the metric has ten components per point, while the torsionless affine connection has forty, and the torsionful, projectively gauge-fixed connection has sixty. This increase in the number of local integration variables is the main price to pay for taking the connection as fundamental and avoiding coordinate gauge fixing at the microscopic level, and it is one of the reasons why numerical simulations in the affine formulation are computationally expensive. The tradeoff of the affine approach is the boundedness of the action, which allows a path integral formulation, as discussed before.

### 11.2.4 Lattice symmetries

While the periodic hypercubic lattice breaks diffeomorphism invariance, there are still symmetries of the discretized system that we do not fix. The periodic lattice admits two discrete symmetries: 90-degree rotations and translations along lattice edges. Since the lattice is finite and periodic, both of these symmetries have a finite volume and leave both the integration measure and the discretized action invariant. Therefore, not fixing them cannot lead to divergences or runaway behaviour in the simulations.

At the same time, the existence of these symmetries implies that no lattice vertex or lattice direction is physically distinguished. In particular, local quantities evaluated at a specific vertex label (or in a specific coordinate direction) are not, by themselves, symmetry-invariant observables; their expectation values must be independent of the chosen vertex label and direction. Equivalently, one may promote such local quantities to symmetry-invariant observables by averaging them over the lattice within each configuration.

Furthermore, Eddington's action is scale invariant, and so is the discretized version. This dilation symmetry will be fixed by choosing a particular lattice spacing, as we will see below.

We will come back to the definition of particular observables later on.

## 11.3 Computational implementation

In this section we describe the numerical implementation of the lattice theory defined above. We have developed a C++ code [273] that simulates the purely affine Eddington action on a hypercubic lattice for different choices of ensemble and curvature discretization, in arbitrary dimension $D \geq 2$. All numerical results shown in this work are obtained with this code.

### 11.3.1 Ensembles and discretizations

The code is capable of realizing both statistical ensembles introduced in Sec. 11.2.2, defined by the lattice Eddington action $S^{\rm lat}_{\rm Edd}(\Gamma)$ and the Boltzmann weight $\exp\left[-S^{\rm lat}_{\rm Edd}\right]$:

- A *torsionful ensemble*, in which all components $\Gamma^{\lambda}{}_{\mu\nu}(x)$ are integrated over, subject only to the projective gauge condition (11.15). The lower indices of the connection are not constrained to be symmetric.
- A *torsionless ensemble*, in which one restricts from the outset to symmetric connections $\Gamma^{\lambda}{}_{(\mu\nu)}(x)$ and enforces $\Gamma^{\lambda}{}_{[\mu\nu]}(x) = 0$ at all sites, as in Eq. (11.22). In this case there is no residual projective symmetry in the integration domain.

These correspond to two different choices of integration domain for the affine theory, and we will loosely refer to them as two "ensembles" in what follows. In both cases, the fundamental dynamical variable is the affine connection; the metric, when it exists, is always a derived quantity.

Independent of this choice, the code offers two discretizations of the Ricci tensor that can be plugged into the same action:

- A *finite-difference discretization*, in which $R^{\rm lat}_{\mu\nu}(x)$ is computed from central finite differences of the affine connection, based on Eqs. (11.9).
- A *Wilson-loop discretization*, in which an approximate Ricci tensor $\tilde{R}^{\rm lat}_{\mu\nu}(x)$ is reconstructed from gravitational Wilson loops on plaquettes, as described in Appendix F.0.2.

These should be viewed as two regularizations of the same continuum observable; they define the same ensemble in the continuum limit, up to discretization effects. In our simulations, we checked that the obtained results do not depend on this choice, although using the Wilson loops is approximately an order of magnitude slower, due to the larger number of computations needed in each of the updates described below.

*For all numerical results presented in this section we use the torsionless ensemble together with the finite-difference Ricci tensor.* The torsionful ensemble and the Wilson-loop discretization are implemented in the code and can be used for further studies, but they are not employed in the production runs reported below.

### 11.3.2 Local Metropolis updates of the connection

As mentioned before, we work on a periodic hypercubic lattice, with linear extents $(N_1, \ldots, N_D)$, where each $N_i$ counts the number of lattice sites in each dimension, and spacing $\varepsilon$, so that the total number of lattice sites is $N = \prod_\mu N_\mu$. On each site, we store all components of the connection ${\Gamma^\lambda}_{\mu\nu}(x)$ in a flat array, together with auxiliary fields holding the current Ricci tensor $R^{\text{lat}}_{\mu\nu}(x)$ and its determinant.

Configuration updates are generated by a standard Metropolis algorithm acting locally on a single lattice site $x$ at a time. One Monte Carlo sweep consists of visiting all lattice sites once (in a fixed or random order) and, at each site, performing the following steps:

1. *Identify the affected region.* Changing ${\Gamma^\lambda}_{\mu\nu}(x)$ only affects the curvature in a finite neighbourhood of $x$, determined by the finite-difference stencil (or by the set of plaquettes touching $x$ in the Wilson-loop discretization). We precompute the list of lattice sites whose Ricci tensor must be recomputed when $\Gamma$ at $x$ is changed.

2. *Compute the old local action.* For all affected sites we read the current Ricci tensors $R^{\text{lat}}_{\mu\nu}$ and compute their contributions to $S^{\text{lat}}_{\text{Edd}}$, summing over those sites only. This defines a local "old action" $S^{\text{old}}_{\text{loc}}$.

3. *Gaussian proposal for the connection.* At the chosen site $x$ we propose a new value of the connection by adding a Gaussian random perturbation to all components at once,

$$ {\Gamma^\lambda}_{\mu\nu}(x) \longrightarrow {\Gamma^\lambda}_{\mu\nu}(x) + \delta{\Gamma^\lambda}_{\mu\nu}(x), \tag{11.25} $$

where each $\delta\Gamma^{\lambda}{}_{\mu\nu}(x)$ is drawn independently from a normal distribution with zero mean and variance $\sigma^2$. In the torsionless ensemble, the proposal is immediately projected onto the symmetric sector by enforcing $\Gamma^{\lambda}{}_{\mu\nu}(x) = \Gamma^{\lambda}{}_{\nu\mu}(x)$ after the update. In the torsionful case, the code instead works with the full set of components and imposes the projective gauge condition (11.15).

4. *Compute the new local action.* Using the updated connection at $x$, we recompute the Ricci tensor on all affected sites and sum their contributions to obtain a new local action $S^{\text{new}}_{\text{loc}}$.

5. *Metropolis test.* The proposed change is accepted with probability

$$P_{\text{acc}} = \min\Big\{1, \exp\Big[-(S^{\text{new}}_{\text{loc}} - S^{\text{old}}_{\text{loc}})\Big]\Big\}, \tag{11.26}$$

which is the usual Metropolis rule, guaranteeing detailed balance. If the move is accepted, we keep the new $\Gamma^{\lambda}{}_{\mu\nu}(x)$ and overwrite the stored Ricci tensors in the affected region with their new values; otherwise, we restore the old connection at $x$ and discard the trial Ricci tensors.

The step size $\sigma$ of the Gaussian proposal is tuned during the simulation to keep the acceptance rate in a reasonable range, more precisely between 30% and 60%, using a simple feedback rule based on the measured acceptance in each sweep. Once a satisfactory acceptance rate is reached, $\sigma$ is kept fixed for the remainder of the run. If the acceptance rate drops below 30%, then $\sigma$ is reduced by a factor of 1.2, and if the acceptance rate increases above 60%, then $\sigma$ is increased by a factor of 1.2. At least in the test case analyzed later on in this work, this proved to be an efficient mechanism to keep the acceptance rate after thermalization close to 35%. We checked that updating each component individually increases the acceptance rate, but also increases simulation time substantially, so we compromised in this local, more abrupt proposal for all components at once. This local update scheme exploits the locality of the action and is sufficient for the exploratory simulations presented here; more sophisticated algorithms (such as hybrid Monte Carlo) could be implemented within the same framework in future studies.

### 11.3.3 General analysis strategy and observables

Ultimately, in dimensions $D > 2$ one would like to characterize the continuum limit of the theory through observables. Observables should be invariant under the symmetries of the theory, and thus they must be diffeomorphism invariant.

In the current ensembles studied in $D = 2$, we do not constrain the connection to be such that it gives rise to an invertible Ricci tensor, nor such that it has a definite signature to define Euclidean distances. As stressed in Sec. 11.1, the emergent metric obtained from the affine variables is not automatically well defined on every configuration: as the formalism is now in $D = 2$, the path integral integrates over all connections compatible with the chosen torsion and gauge conditions, without imposing further constraints on it. This means that we are not guaranteed that the Ricci tensor can define a meaningful metric in the sense given by (11.6).

This unconstrained connection ensemble therefore prevents us from measuring geometric characteristics such as Hausdorff or spectral dimensions, in the spirit of other lattice approaches to quantum gravity [59, 61]. Nevertheless, there are other easily accessible observables whose expectation value can be computed in this setting.

A simple and easily accessible choice is to monitor local quantities constructed from contractions of the Riemann tensor, since this tensor is readily available in lattice simulations. In particular, constructing diffeomorphism-invariant quantities without a notion of a metric that allows for raising and lowering indices reduces the number of possibilities. Nevertheless, there are those that do not require raising and lowering of indices, and are easily accessible in this setting. The path taken here to build observables is to compute integrals of tensor densities. Luckily, in any dimension $D$, we can build volume measures from (0, 2) tensors built purely from the affine connection, as we will describe below.

A canonical choice for such an observable that *would be* proportional to the physical volume of the system *when a metric can be defined* from (11.6), is

$$V_R \equiv \int d^D x \sqrt{|\det\big(R_{(\mu\nu)}(x)\big)|}, \tag{11.27}$$

where the integral over the manifold is replaced by a discrete sum in the lattice simulations.

Another possibility, given the ensembles we work with, is to build another volume measure from a tensor called the homothetic tensor

$$Q_{\mu\nu} \equiv R^{\alpha}{}_{\alpha\mu\nu} \,, \tag{11.28}$$

an antisymmetric tensor, which is often called the "curl of the connection". This tensor is zero in Riemannian geometry, and colloquially, it is said to measure how much the affine connection fails to preserve volume under parallel transport, and shows quantitative deviations from Riemannian manifolds, see [291]. Because it is skew-symmetric, its determinant is always greater than or equal to zero, and from it, we can integrate the density and obtain

$$V_Q \equiv \int d^D x \sqrt{\det(Q_{\mu\nu}(x))} \,, \tag{11.29}$$

where again, the integral over the manifold is replaced by a discrete sum in the lattice simulations.

Notice that any quantity that is built from (11.27) and (11.29) by just multiplying with a constant scalar is also an observable. In the lattice simulations, we will compute in each path integral configuration the quantities

$$\overline{V_R^{\mathrm{lat}}} \equiv \frac{1}{N} \sum_x \varepsilon^D \sqrt{\left| \det\left( R^{\mathrm{lat}}_{(\mu\nu)}(x) \right) \right|} \qquad \overline{V_Q^{\mathrm{lat}}} \equiv \frac{1}{N} \sum_x \varepsilon^D \sqrt{\det\left( Q^{\mathrm{lat}}_{\mu\nu}(x) \right)} \,, \tag{11.30}$$

with $N$ the number of lattice sites, and where $\overline{V_R^{\mathrm{lat}}}$ and $\overline{V_Q^{\mathrm{lat}}}$ could be intuitively interpreted as the average physical volume, in the sense that the Ricci is proportional to the emergent metric in suitable scenarios (11.6), and the average volumetric deviation from Riemannian geometry of each hypercube of the lattice, respectively.

It is worth highlighting that these observables are independent of the labeling of the vertices, as well as from particular lattice directions around a particular point, and are thus free from the problem of discrete lattice symmetries explained before.

On top of the previously constructed observables, it is interesting to look at another order parameter. As previously mentioned, the ensembles are not constrained to give rise to symmetric Ricci tensors with fixed signature, and it is thus interesting to study whether the system has a particular preference for the signature of the resulting tensors in the simulations. In particular, an easily accessible quantity to look at is the fraction of points in the lattice that have $\det\left(R^{\text{lat}}_{(\mu\nu)}(x)\right) < 0$, meaning that at least one of its eigenvalues is negative

$$f_{d_R<0} \equiv \frac{1}{N} || \{ x \,/\, det(R^{\text{lat}}_{(\mu\nu)}(x)) < 0 \} || \,. \tag{11.31}$$

In the toy model studied in this work, where $D = 2$, the fraction $f_{d_R<0}$ measures the percentage of points in the manifold that have one negative and one positive eigenvalue. Since changes in the signature imply a change of signature of one eigenvalue, and that therefore it is zero in some parts of the manifold, this fraction also provides information about how the symmetric Ricci tensor is non-invertible in some parts of the manifold. In a more constrained ensemble, one should integrate over connections that give rise to an invertible symmetric part of the Ricci tensor, providing a well-defined metric tensor.

In the two-dimensional study of Sec. 11.4 we use the observables (11.30) and the order parameter (11.31) as thermalization indicators, and we compute their expectation values as a function of the bare coupling of Eddington's action.

### 11.3.4 Units

The affine connection has units of inverse length in all dimensions. It is interesting to see that the uniform lattice spacing $\varepsilon$ can be absorbed in a new dimensionless connection variable

$$\tilde{\Gamma}^{\sigma}_{\mu\nu} \equiv \varepsilon \, \Gamma^{\sigma}_{\mu\nu}. \tag{11.32}$$

The transformation to these dimensionless variables has a trivial path integral Jacobian, and will be used for simulations. Notice that when using these variables, the lattice spacing disappears from all the relevant quantities that are used in this work, obtained by integrating over the entire manifold, such as the action, or the observables (11.30). For other quantities that involve lower-dimensional integrals, such as some suitable measure of distance that will be discussed in the appendix, the lattice spacing appears again. Still, since in this work we will restrict the analysis to the simple quantities defined in (11.30), the lattice spacing will play no role. This transformation can effectively be thought of as setting $\varepsilon = 1$, which is what we will do in the following.

## 11.4 Two-dimensional results

As discussed in Sec. 11.1, in $D = 2$ the metric cannot be reconstructed from the affine variables, but the purely affine action itself has a smooth and non-trivial $D \to 2$ limit. Besides, the two-dimensional case is the one with the lowest number of connection components, an appealing feature for testing simulation aspects. This makes two dimensions an attractive testing ground: the theory remains dynamically non-trivial due to its non-vanishing action, yet any direct comparison to metric formulations of gravity is impossible by construction. Before describing the numerical experiments, let us make this $D \to 2$ limit explicit.

### 11.4.1 Action in the $D \to 2^+$ limit

In the pure gravity case, the continuum Eddington action in $D$ dimensions can be written as

$$S_{\text{Edd}}[\Gamma] \;=\; \alpha_D(\Lambda, G) \int d^D x \sqrt{\left| \det R_{(\mu\nu)}(x) \right|}\,, \tag{11.33}$$

with

$$\alpha_D(\Lambda, G) \;\equiv\; \frac{(D-1)(D-2)^{\frac{D-2}{2}}\, 2\Lambda}{(16\pi G)^{\frac{D+2}{2}}} \left| \frac{2\Lambda}{16\pi G} \right|^{-\frac{D}{2}}. \tag{11.34}$$

For definiteness, take $\Lambda > 0$ so the absolute values can be dropped. The $(D-2)^{\frac{(D-2)}{2}}$ has a finite, real value when approaching $D \to 2^+$, thus, in two dimensions, the Eddington reduces to

$$S_{\rm Edd}^{(2D)}[\,\Gamma\,] \;=\; \alpha \int d^2x \,\sqrt{\big|\det R_{(\mu\nu)}(x)\big|}\,, \qquad \alpha \equiv \alpha_2(\Lambda, G) = \frac{1}{16\pi G}. \tag{11.35}$$

The dependence on $\Lambda$ has dropped out completely in this limit; only $G$ survives in the overall prefactor.

On the lattice, the continuum integral is replaced by a sum,

$$S_{\rm Edd}^{(2D)\,\rm lat}[\,\Gamma\,] \;=\; \alpha \sum_x \varepsilon^2 \sqrt{\big|\det R^{\rm lat}_{(\mu\nu)}(x)\big|}\,, \tag{11.36}$$

with $R^{\rm lat}_{(\mu\nu)}(x)$ given by the finite-difference discretization introduced in Sec. 11.2 (or its Wilson-loop improvement). Strictly speaking, the local "volume element" entering the action is $\varepsilon^2\sqrt{|\det R^{\rm lat}_{(\mu\nu)}(x)|}$; in the simulations reported below, we fix $\varepsilon = 1$, basically absorbing the lattice spacing into the connection variables as described in (11.32).

11.4 It is convenient to treat the prefactor $\alpha$ in front of the sum as a tunable bare coupling in the statistical ensemble, and we explore the statistical ensemble defined by the Boltzmann weight

$$Z(\alpha) \;=\; \int \mathcal{D}\Gamma\, e^{-S_{\rm Edd}^{(2D)\,\rm lat}[\,\Gamma\,]}. \tag{11.37}$$

Here $\alpha$ appears as a dimensionless parameter controlling how strongly configurations with large local curvature determinant are suppressed; in practice, we probe several values of $\alpha$ in a reasonable range, treating it as a bare coupling.

### 11.4.2 Simulation setup in two dimensions

While the provided code supports simulations in $D$ dimensions, and clearly the most physically interesting case would be the four-dimensional one, the computational cost scales at least with the size of the simulated manifold, which for $N$ lattice points in a given direction is equal to $N^D$. On top of that, at this stage, where we are interested in a proof of concept implementation, and where we do not constrain the ensemble of affine connections to generate fixed signature emergent metrics, we consider this exercise unnecessary, and we restrict the analysis in this work to the simplest possible case, $D = 2$.

All simulations reported here are performed on periodic hypercubic lattices of sizes $N = 50 \times 50$, $N = 100 \times 100$ and $N = 150 \times 150$ in $D = 2$, with $\varepsilon = 1$. The dynamical variable is the affine connection $\Gamma^{\lambda}{}_{\mu\nu}(x)$ at each site, updated with the local Gaussian Metropolis algorithm described in Sec. 11.3. The local Gaussian standard deviation is always initialized at $\sigma = 1$, and once the system is thermalized it typically drops to around $\sigma \simeq 0.12$ for the smaller values of $\alpha$ explored in this work, and $\sigma \simeq 0.08$ for the larger values of $\alpha$ explored in this work, with an acceptance ratio that stabilizes between 33% to 37%.

We use the finite-difference discretization of the Ricci tensor and work in the torsionless ensemble, enforcing symmetry in the lower indices after each local update.

For each value of the coupling $\alpha$, we generate a Markov chain starting from the "cold" configuration

$$\Gamma^{\lambda}{}_{\mu\nu}(x) = 0 \qquad \text{for all sites } x, \tag{11.38}$$

projected onto the torsionless sector. We checked that starting from a "hot" configuration, where each component at each site was sampled from a Gaussian distribution, did not change the results. We then perform a large number of full-lattice sweeps. By studying the convergence of the observables (11.30) and the order parameter (11.31), we established that the first 10000 to 15000 sweeps can be regarded as thermalization, depending on the values of $\alpha$ used here. Therefore, for safety, we always discard the first 25 000 sweeps as thermalization, while the rest of the sweeps (up to 250 000 in total) are used for measurements. Configurations are recorded every 100 sweeps, allowing every vertex of the manifold to be updated 100 times between measurements, a number of sweeps considered sufficient at this exploratory stage to reduce correlations between successive measurements, giving a total of 2000 measured configurations for each value of $\alpha$. As a cross-check for this number of sweeps between measurements, we determined that measuring every 75, 150, and every 200 sweeps gives the same results within error bars. The code exports the values of the connection at each site for each configuration, and has the option to export the discretized Ricci tensor as well.

### 11.4.3 Numerical results

In $D = 2$ it is impossible to relate this affine theory to General relativity, since (11.6) breaks down and therefore no GR metric tensor exists. However, as discussed before, we can still measure the observables (11.30) that provide different notions of manifold volume and deviations from Riemannian geometry.

The Monte Carlo data are processed as follows. At a fixed value of $\alpha$, for each saved configuration $c$ in the Markov chain, we first compute a certain observable or order parameter, generally denoted as $\mathcal{O}(c)$ on that configuration, and later approximate the path integral expectation value as

$$\langle \mathcal{O} \rangle \approx \frac{1}{N_{\text{cfg}}} \sum_{c=1}^{N_{\text{cfg}}} \mathcal{O}(c) \,, \tag{11.39}$$

where $c$ labels the saved configurations, and $N_{\text{cfg}}$ is the number of stored configurations at that value of $\alpha$. In this work, we used $N_{\text{cfg}} = 2000$ for all the studied values of $\alpha$.

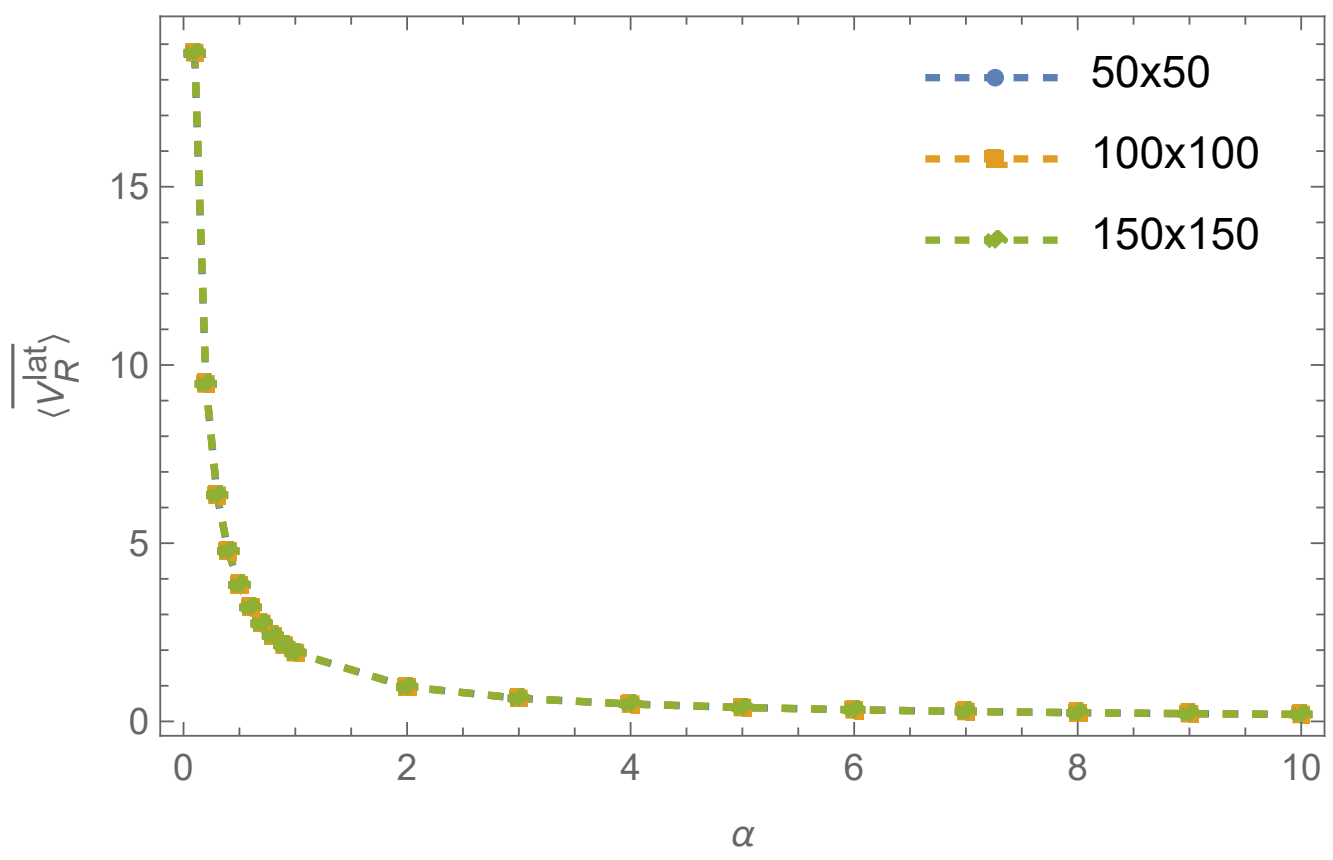

**Figure 11.1.:** Expectation value of the observable $\overline{V_R^{\text{lat}}}$ over 2000 sampled path integral configurations from the Monte Carlo simulations, as a function of the bare parameter $\alpha$, for three different lattice sizes. The results are overlapping monotonically decreasing curves, with no signs of a phase transition. Statistical error bars are below 0.01% of the reported values, and are too small to be seen.

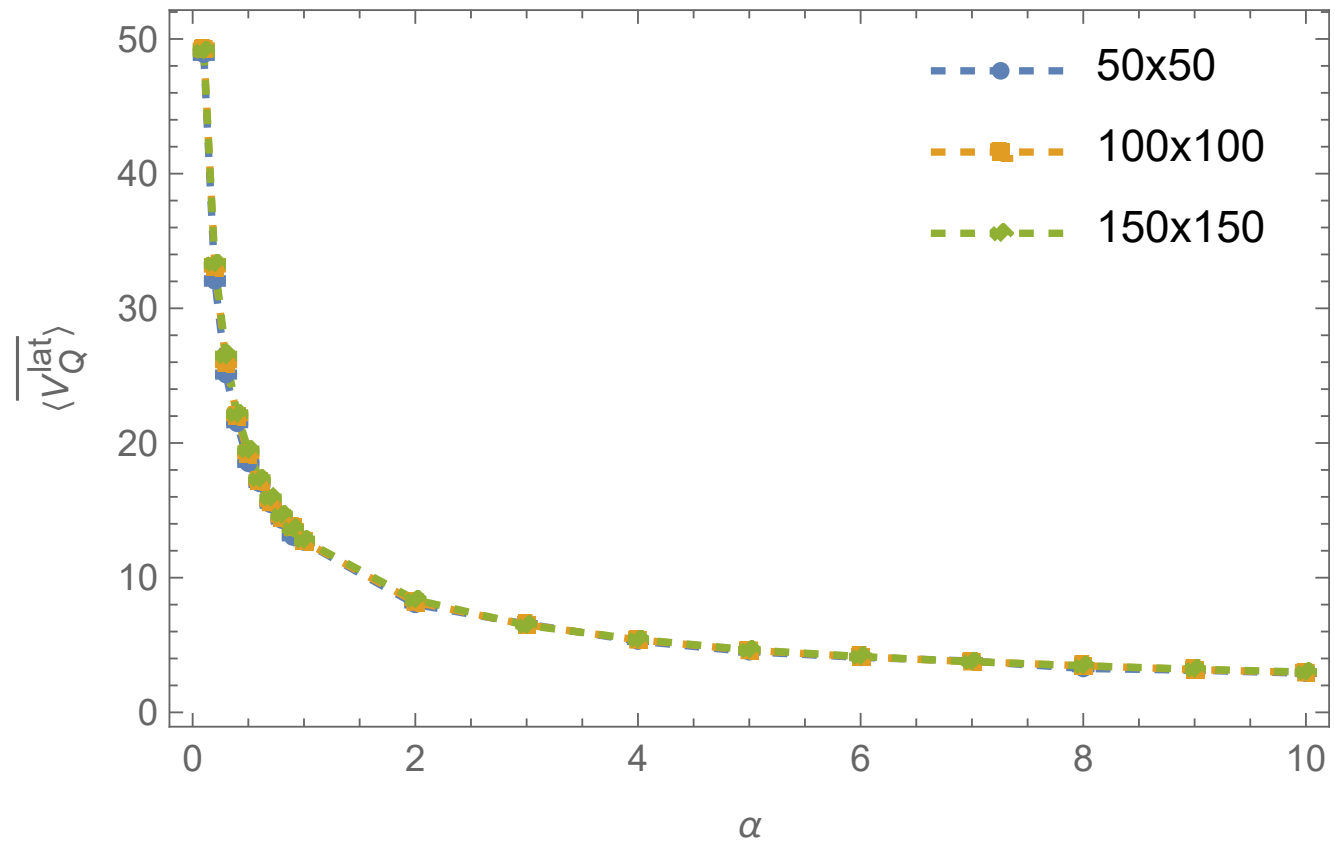

**Figure 11.2.:** Expectation value of the observable $\overline{V_Q^{\text{lat}}}$ over 2000 sampled path integral configurations from the Monte Carlo simulations, as a function of the bare parameter $\alpha$, for three different lattice sizes. The results are overlapping monotonically decreasing curves, with no signs of a phase transition. Statistical error bars are below 0.01% of the reported values, and are too small to be seen.

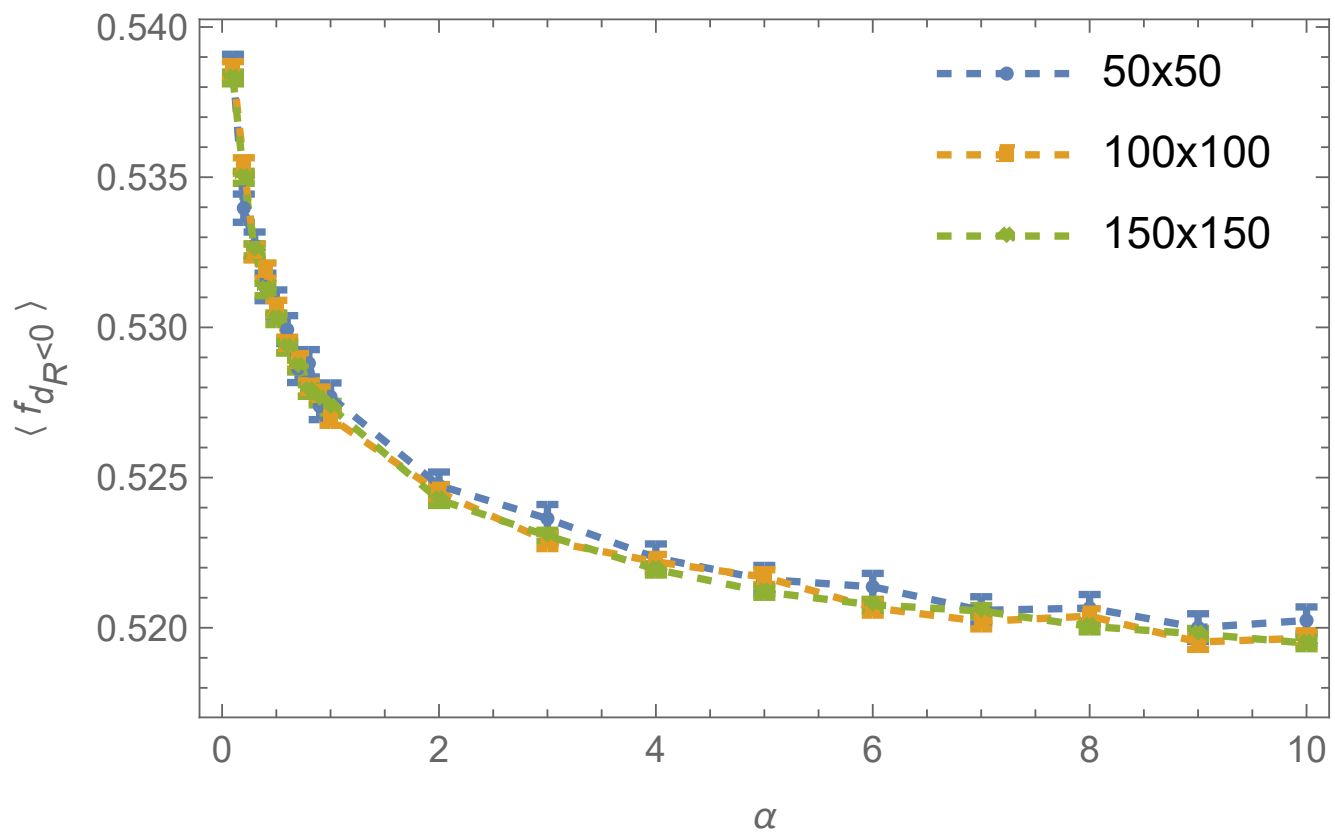

**Figure 11.3.:** Expectation value of $f_{d_R<0}$ over 2000 sampled path integral configurations from the Monte Carlo simulations, as a function of the bare parameter $\alpha$, for three different lattice sizes. The results are overlapping monotonically decreasing curves, with no signs of a phase transition, but always above 0.5, meaning that the configurations have a majority of points with a symmetric part of the Ricci tensor with $(-,+)$ signature, instead of $(-,-)$ or $(+,+)$.

For each of the lattice sizes $N = 50 \times 50$, $N = 100 \times 100$ and $N = 150 \times 150$, we studied a range of $\alpha$ values that covers two orders of magnitude, specifically $\alpha \in \{0.1, 0.2, \ldots, 0.9, 1, 2, \ldots, 10\}$. Within this range, we computed the expectation value of the observables $\overline{V_R^{\text{lat}}}$ and $\overline{V_Q^{\text{lat}}}$ defined in (11.30), and the order parameter $f_{d_R<0}$ defined in (11.31). The numerical results of each of them can be seen in Figs. 11.1, 11.2 and 11.3 respectively.

Within the explored range of couplings, both observables $\overline{V_R^{\text{lat}}}$ and $\overline{V_Q^{\text{lat}}}$ and the order parameter $f_{d_R<0}$ vary smoothly and monotonically with $\alpha$. We do *not* see any clear sign of non-monotonic behaviour or singular features that could be interpreted as a phase transition in terms of the measured quantities. If a non-trivial phase structure exists in the two-dimensional model, it does not show up through the measured geometrical properties on the lattices and coupling range used here.

There is, nevertheless, an interesting feature to be highlighted. Despite not having imposed a constraint on the ensemble of connections, such that there is a preferred signature of the symmetric part of the Ricci tensor, it seems that the system thermalizes on configurations that have approximately between 52% and 54% of their domain covered by a tensor $R_{(\mu\nu)}$ with signature $(-, +)$. It remains to be seen in future work if a more sophisticated ensemble, perhaps with further constraints on the affine connection, changes this small preference for signature average.

All results of the observables (11.30) and the order parameter (11.31) overlap within statistical error bars, meaning that we do not observe any non-trivial scaling within the explored range of lattice sizes, and the factor $\frac{1}{N}$ in the definition of each quantity is sufficient to overlap all curves.

Given the special role of two dimensions in the classical theory, the present results should be viewed as a proof of principle rather than as a test of continuum General Relativity. They show that the purely affine Eddington-type action leads to a well-defined lattice ensemble whose basic properties can be studied numerically, and that the coupling $\alpha$ provides a clean handle to scan through different regimes of the model. In higher dimensions, where the metric can be reconstructed, one can build more intuitive geometric observables (such as scaling dimensions) and use similar Monte Carlo techniques to investigate the continuum limit and possible phase structure of the theory. The absence of such a metric in the present two-dimensional study is therefore not the endpoint of the construction, but points directly towards the next questions: whether a fixed-signature sector can be sampled configuration by configuration, whether physical time and volume can be reconstructed, and whether the resulting geometry presents a recognizable semiclassical behaviour.

The main lesson from these two-dimensional simulations is that the purely affine lattice theory defined by $S_{\rm Edd}^{(2D)\,\rm lat}$ behaves as a perfectly ordinary statistical model of a high-dimensional field, even though no metric can be reconstructed in $D = 2$. The action prefactor $\alpha$ monotonically controls the typical scale of the diagnostic observables built from the Riemann tensor, and the simple local Gaussian Metropolis dynamics is able to explore the configuration space without apparent instabilities or pathologies. Having established this computational proof of principle, we now turn to a setting in which the questions left open by the two-dimensional model can be addressed more directly.

## 11.5 Four-dimensional affine minisuperspace and emergent geometry

The two-dimensional simulations presented above provide a proof of concept for the lattice implementation, but they do not allow us to reconstruct an emergent metric and make contact with General Relativity. To address this limitation, we are currently studying a four-dimensional homogeneous and isotropic reduction of the affine ensemble, or in other words, a "minisuperspace" approximation of theory space. While this minisuperspace setting does not yet amount to an unrestricted four-dimensional lattice simulation, it allows us to impose a fixed-signature sector, reconstruct the metric configuration by configuration, explore the theory's first genuinely geometric observables, and understand some of the difficulties that will arise when the symmetry restrictions are relaxed.

We consider a reduced theory space, constrained to the symmetry assumptions of spatial homogeneity and isotropy. If these symmetry constraints are made at the level of the connection, and one assumes a torsionless connection, then one is left with only 3 degrees of freedom, parametrized by three functions $A(t)$, $B(t)$ and $C(t)$,

$$\Gamma^0{}_{00} = A(t), \qquad \Gamma^0{}_{ij} = B(t)\gamma_{ij}, \qquad \Gamma^i{}_{0j} = \Gamma^i{}_{j0} = C(t)\delta^i_j, \tag{11.40}$$

where $t$ is coordinate time, and $\gamma_{ij}$ is the assumed metric of the purely spatial sector of the emergent metric, in this case chosen to be that of the unit three-sphere. The symmetric Ricci tensor is block diagonal,

$$R_{00} = F(t), \qquad R_{ij} = G(t)\gamma_{ij}, \tag{11.41}$$

with

$$F = 3[-\dot{C} + C(A - C)], \qquad G = 2 + \dot{B} + B(A + C). \tag{11.42}$$

As discussed before, we would like to constrain the ensemble of connection configurations of the path integral to give rise to well defined metrics in each configuration, and thus, we will impose the system to remain during simulation time in the Euclidean fixed-signature sector, meaning $F > 0$ and $G > 0$ (for positive cosmological constant), giving a well defined euclidean metric at every path integral configuration, that will allow us to compute geometric observables across the entire simulation. In this situation, the Ricci-emergent metric takes the FLRW form

$$ds^2_\Gamma = \lambda F(t)dt^2 + \lambda G(t)d\Omega_3^2 \equiv N^2(t)dt^2 + a^2(t)d\Omega_3^2, \tag{11.43}$$

where $\lambda = 1/\Lambda$ in four dimensions, or in tensor notation

$$g_{\mu\nu} = \frac{1}{\Lambda}R_{(\mu\nu)}, \tag{11.44}$$

with its corresponding inverse, simply denoted as the symmetric Ricci tensor with indices up

$$R^{(\mu\alpha)}R_{(\alpha\nu)} = \delta^\mu{}_\nu, \qquad g^{\mu\nu} = \Lambda R^{(\mu\nu)}. \tag{11.45}$$

In this way, every accepted affine configuration has positive signature, and defines its own lapse, scale factor, proper time $d\tau = N(t)dt$, and spatial-volume profile $V(\tau) \propto a^3(\tau)$. While there are multiple ways of achieving this positive signature constraint, in this case, during Monte Carlo simulations, we simply reject configurations that present a different signature for the symmetric Ricci tensor. This has led to approximately 47% rejected updates during simulation time. Due to the simplicity of the system, this is not yet a problem, but more sophisticated methods to preserve a particular signature should be explored if one reduces the symmetry constraints.

As mentioned before, the determinant action by itself has flat directions: it controls the total volume given by the determinant $FG^3$, but not all deformations of $F$, $G$, or of the connection that leave the integral of the square root of this determinant unchanged. In the case of two dimensions, we did not care about these, since we studied simply the total volume as a function of the bare coupling, and thus flat directions cancel out in the expectation value computations. Nevertheless, the Monte Carlo simulations were slower due to sometimes being stuck in such constant volume spaces, and this certainly requires fixing if one is to take this programme seriously. There are, of course, multiple mechanisms to lift flat directions of path integrals, which in general come in the form of regulators that must be removed to obtain physical results. For example, one can add mass terms to each of the quantized fields, as is done in [292], and later take the limit of zero mass numerically. In our case, since we do not want to break the covariance of the action itself, adding such a mass term for the connection fields is not so simple, since contractions of the connection with itself do not create scalar invariants. Nevertheless, a mass term of the difference of two connections, which would be a tensor, can easily be constructed.

In the present implementation, we lift these directions by choosing a fixed de Sitter background connection $\bar{\Gamma}^{\lambda}{}_{\mu\nu}$ and defining the tensor

$$C^{\lambda}{}_{\mu\nu} = \Gamma^{\lambda}{}_{\mu\nu} - \bar{\Gamma}^{\lambda}{}_{\mu\nu}. \tag{11.46}$$

We then simulate the action

$$S_{sim} = S_{Edd} + S_{reg}, \qquad S_{reg}[\Gamma, \bar{\Gamma}] = \frac{m^2}{2} \int d^4x \, \sqrt{g} \, C^{\mu\nu\beta} C_{\mu\nu\beta}. \tag{11.47}$$

Here, the indices of $C^{\lambda}{}_{\mu\nu}$ are contracted using the Ricci-emergent metric, as stated before. The fixed background $\bar{\Gamma}^{\lambda}{}_{\mu\nu}$ is chosen to be the Levi-Civita connection of Euclidean de Sitter space, expressed in the same coordinate system used to compute the finite differences entering Eddington's action. This is the natural choice in the present setting: under the assumptions of spatial homogeneity and isotropy, in vacuum and with a positive cosmological constant, de Sitter space is the available classical solution. Thus, within this reduced theory space, choosing a fixed de Sitter background amounts to choosing the Levi-Civita connection singled out by the classical equations. Nevertheless, the regulator remains defined as the difference between the sampled affine connection and this fixed background connection.

This mass term allows us to compute expectation values without the flat directions of Eddington's action driving the system away during simulation time, since flat directions that keep Eddington's action constant but have different connections are eventually exponentially suppressed by the added regulator. The regulator is obviously not interpreted as part of the fundamental affine theory, but as a mathematical artifact that must eventually be removed. The procedure followed here is therefore to compute expectation values at successively decreasing values of the regulator mass $m$, approach the limit $m \to 0$ numerically, and assess the stability and convergence of the measured quantities.

### 11.5.1 Simulation setup and preliminary numerical results

The simulations use a one-dimensional periodic time lattice with $N$ sites and lattice spacing $\varepsilon = 1$. The spatial slices are closed and have the geometry of the unit three-sphere, $k = 1$, with spatial dimension $n = 3$, so the reduced model is four-dimensional. Periodic central finite differences are used for all time derivatives. We work in the Euclidean sector, and a proposed update is assigned zero statistical weight whenever it violates $F_i > 0$ or $G_i > 0$ at any affected site. The connection functions $A_i$, $B_i$ and $C_i$ are updated with a local Gaussian Metropolis proposal. The proposal width is adapted during thermalization and subsequently kept fixed during measurements.

After integrating over the unit three-sphere, whose volume is

$$V_{S^3} = 2\pi^2, \tag{11.48}$$

the lattice Eddington term used by the code is

$$S_{Edd}^{lat} = \alpha\,\varepsilon \sum_{i=1}^{N} \sqrt{F_i G_i^3}. \tag{11.49}$$

Here the factor $V_{S^3}$ has already been absorbed into the effective one-dimensional coupling,

$$\alpha = V_{S^3}\alpha_4 = 2\pi^2 \frac{3}{16\pi G\Lambda} = \frac{3\pi}{8G\Lambda}. \tag{11.50}$$

All production runs shown here use $\alpha = 1$. Thus, the overall multiplicative coefficient of the reduced sum in Eq. (11.49) is one; equivalently, before performing the $S^3$ integral, $\alpha_4 = 1/(2\pi^2)$ in these conventions. The code uses the same sphere-volume convention for the regulator parameter $m$. If the $S^3$ factor is instead kept explicit, the corresponding sphere-volume-independent parameter is $\bar{m} = m/\sqrt{2\pi^2}$.

The spatial volume is reconstructed without normalizing away its overall scale. In the lattice units used in the simulations, in which $a_i^2 = G_i$ and $N_i^2 = F_i$, it is

$$V_3(t_i) = 2\pi^2 a_i^3 = 2\pi^2 G_i^{3/2}, \tag{11.51}$$

and its integral over emergent proper time gives the four-volume,

$$V_4 = \int d\tau\, V_3(\tau) = 2\pi^2 \varepsilon \sum_{i=1}^{N} \sqrt{F_i G_i^3}. \tag{11.52}$$

The simulations do not fix this total volume. Rather, it fluctuates and is weighted by the Eddington action, while the complete unnormalized spatial-volume profile is retained in the analysis.

We simulated the five lattice sizes

$$N \in \{50, 100, 200, 350, 550\}, \tag{11.53}$$

and the twelve regulator masses

$$m \in \{1, 0.7, 0.5, 0.3, 0.2, 0.15, 0.1, 0.07, 0.05, 0.03, 0.015, 0.01\}. \tag{11.54}$$

For each of the resulting 60 parameter combinations, we combined two independent Markov chains. Each chain was preceded by 50 000 thermalization sweeps, and stored configurations were separated by 3789 full-lattice sweeps. The first chain contributed 80 011 configurations and the second 120 000, giving

$$N_{cfg} = 200\,011 \tag{11.55}$$

stored configurations for every pair $(N, m)$, or 12 000 660 stored configurations over the full scan. Autocorrelations are accounted for in the analysis; consequently, the effective sample sizes (ESS), quoted in the figures, are smaller than these raw configuration counts and vary across the parameter scan.

The mass values where chosen such that one crosses the "regulator-dominated" region towards the Eddington-dominated sector of the action, such that at large values of the mass, the average value of the action is mostly given by the regulator term, but at low masses, the total action used for the simulation is mostly given by the purely Eddington term, and thus, we numerically reduce the effect of the regulator and explore the $m \to 0$ limit. In this case, we did not reach that limit exactly, but we covered a substantial part of the spectrum, as can be seen in Table 11.1.

| $m$ | $\langle S_{\rm Edd} \rangle$ | $\langle S_{\rm reg} \rangle$ | $\langle S_{\rm tot} \rangle$ | $S_{\rm Edd}/S_{\rm tot}$ | $S_{\rm reg}/S_{\rm tot}$ | dominant |
|---|---|---|---|---|---|---|
| 1.0 | 362.45 | 763.97 | 1126.43 | 32.2% | 67.8% | $S_{\rm reg}$ |
| 0.7 | 334.38 | 614.15 | 948.52 | 35.3% | 64.7% | $S_{\rm reg}$ |
| 0.5 | 342.03 | 492.56 | 834.59 | 41.0% | 59.0% | $S_{\rm reg}$ |
| 0.3 | 363.79 | 365.97 | 729.76 | 49.9% | 50.1% | $S_{\rm reg}$ |
| 0.2 | 379.41 | 295.96 | 675.37 | 56.2% | 43.8% | $S_{\rm Edd}$ |
| 0.15 | 387.98 | 256.70 | 644.67 | 60.2% | 39.8% | $S_{\rm Edd}$ |
| 0.1 | 397.06 | 200.93 | 598.00 | 66.4% | 33.6% | $S_{\rm Edd}$ |
| 0.07 | 399.19 | 167.34 | 566.53 | 70.5% | 29.5% | $S_{\rm Edd}$ |
| 0.05 | 397.10 | 137.40 | 534.50 | 74.3% | 25.7% | $S_{\rm Edd}$ |
| 0.03 | 387.11 | 87.92 | 475.03 | 81.5% | 18.5% | $S_{\rm Edd}$ |
| 0.015 | 377.05 | 45.52 | 422.56 | 89.2% | 10.8% | $S_{\rm Edd}$ |
| 0.01 | 373.99 | 31.91 | 405.89 | 92.1% | 7.9% | $S_{\rm Edd}$ |

**Table 11.1.:** Ensemble-averaged split of the Euclidean action $S = S_{\rm Edd} + S_{\rm reg}$ at $N = 550$ as the regulator mass $m$ is reduced, averaged over all configurations per mass. $S_{\rm Edd}$ is the Eddington term; $S_{\rm reg}$ is the de Sitter pinning regulator. Dominance passes from $S_{\rm reg}$ to $S_{\rm Edd}$ at $m \approx 0.3$.

The first results of this ongoing analysis can be seen in Figs. 11.4, 11.5 and 11.6. Since the time direction is periodic, every configuration is first translated by an integer number of lattice sites so that the circular centroid of its scale-factor profile lies at a common reference site. This translation is an exact symmetry of the periodic lattice. The unnormalized spatial-volume and lapse profiles are then averaged using the same shifts, and the

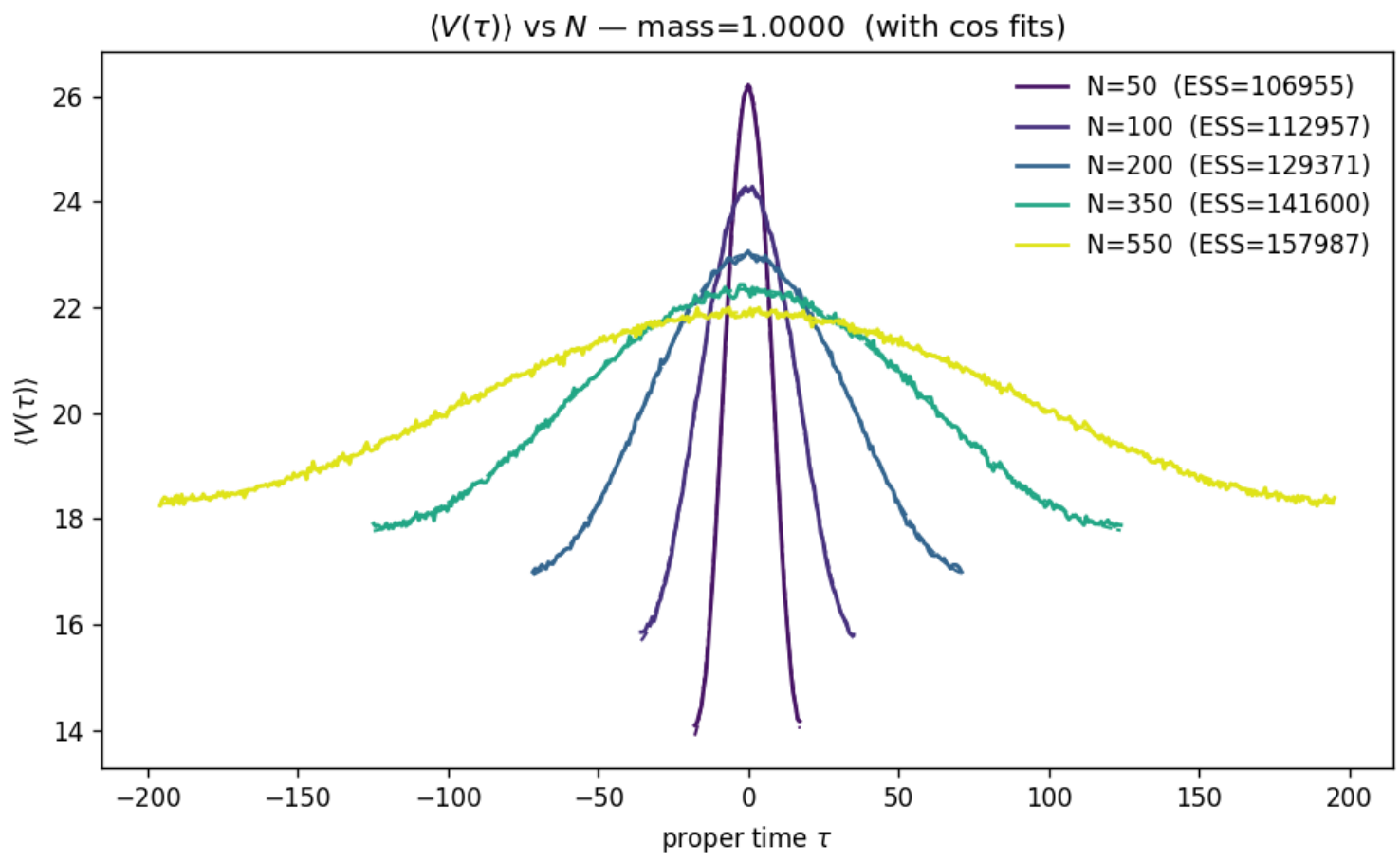

**Figure 11.4.:** Preliminary expectation value of the spatial-volume profile $V(\tau) \propto a^3(\tau)$, reconstructed using the proper time of the emergent metric, for several lattice sizes at regulator mass $m = 1$. The solid curves show the ensemble averages and the dashed curves the de Sitter-motivated fits $\langle V(\tau)\rangle = c_1 \cos^3(2\pi\tau/T) + c_2$, where $T$ is the proper-time extension. The profiles are centred configuration by configuration before taking the expectation value.

11.5

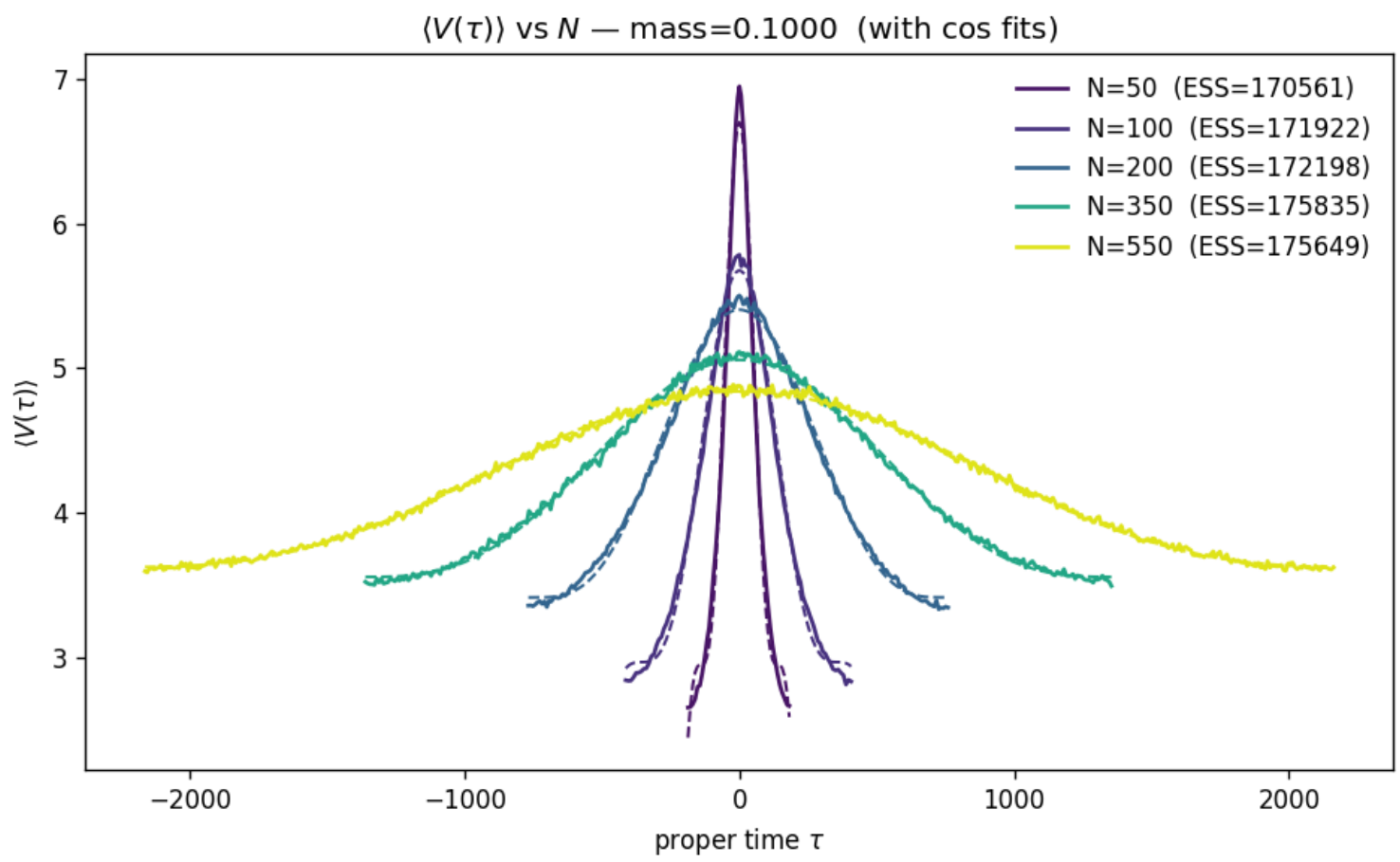

**Figure 11.5.:** Preliminary expectation value of the reconstructed spatial-volume profile for several lattice sizes at regulator mass $m = 0.1$. As in Fig. 11.4, the solid curves are ensemble averages and the dashed curves are fits to Eq. (11.56).

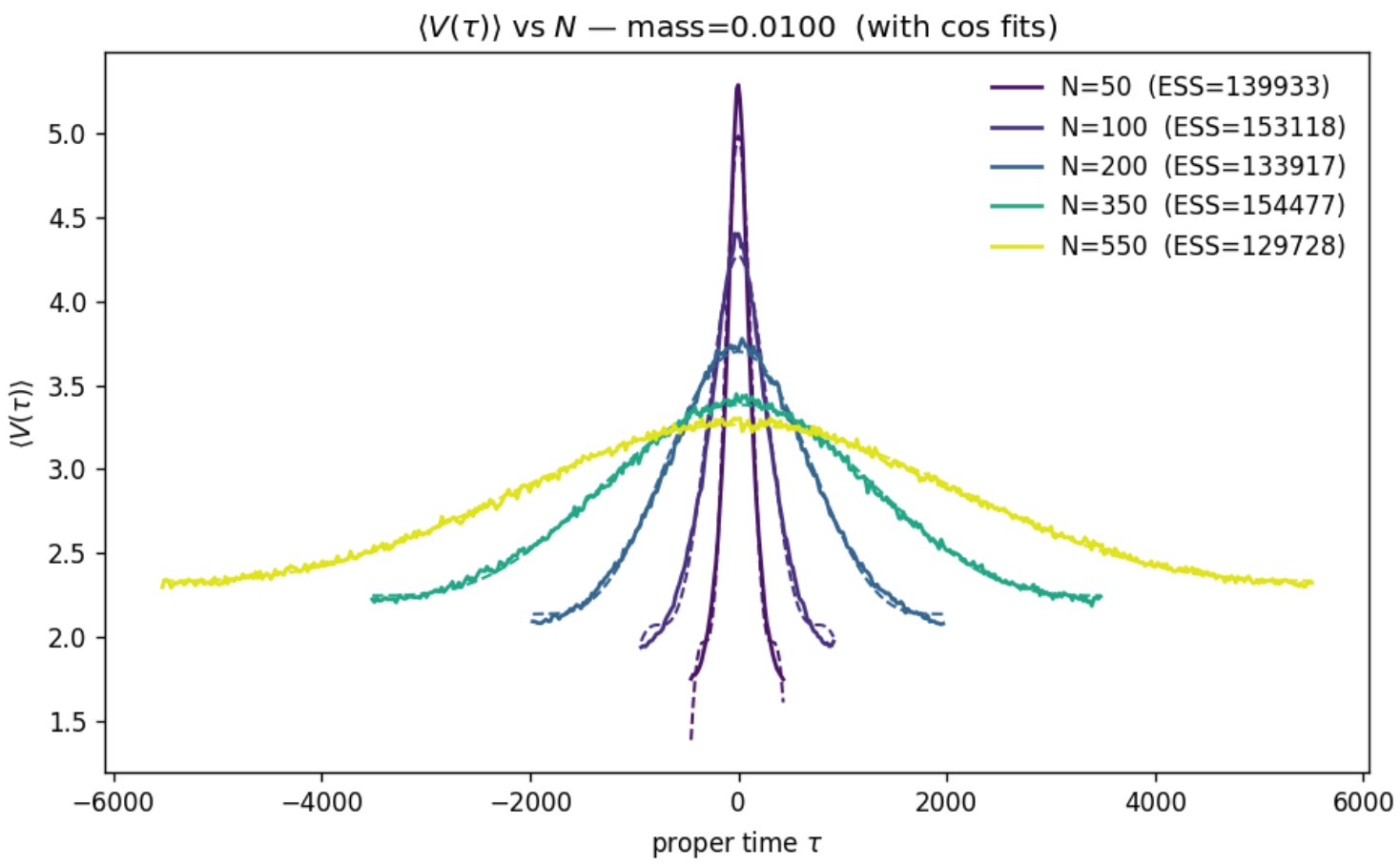

**Figure 11.6.:** Preliminary expectation value of the reconstructed spatial-volume profile for several lattice sizes at regulator mass $m = 0.01$. Together with Figs. 11.4 and 11.5, this shows the size study at three successively decreasing regulator masses. The solid curves are ensemble averages and the dashed curves are fits to Eq. (11.56).

11.5

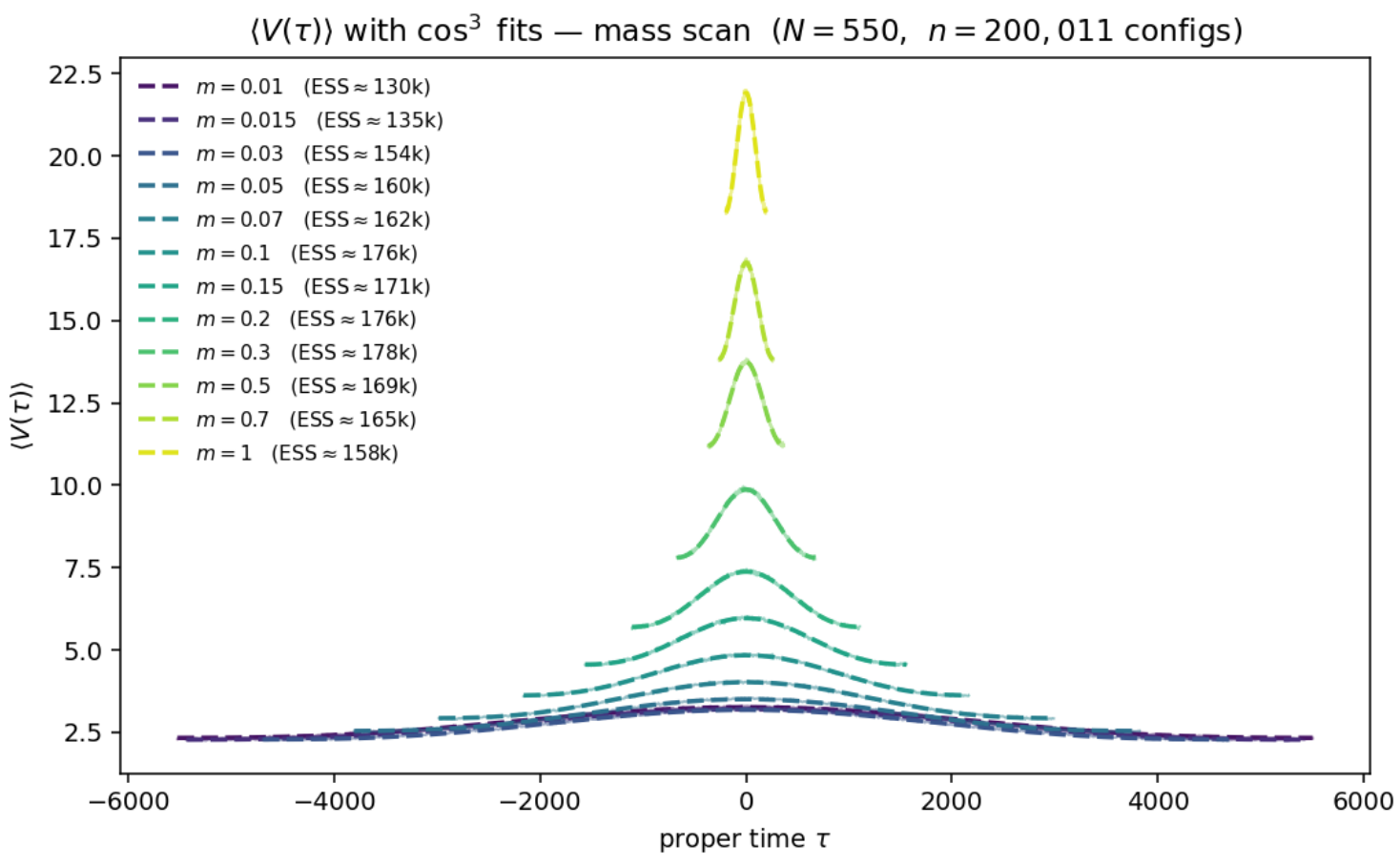

**Figure 11.7.:** Preliminary mass scan of the reconstructed spatial-volume profile at the largest simulated size, $N = 550$. The curves show how the expectation value and its $\cos^3$ fit change as the regulator mass is decreased. This comparison makes the regulator dependence of the proper-time extension and volume scale explicit and provides the starting point for the numerical study of the $m \to 0$ limit.

proper-time grid is reconstructed from the aligned mean lapse. This is closely analogous to the centre-of-volume alignment used in CDT [71], but retains the complete volume scale of every configuration. For a Euclidean de Sitter geometry, the spatial volume follows a $\cos^3$ profile. We therefore fit the reconstructed profiles with the explicit form

$$\langle V(\tau)\rangle = c_1 \cos^3\left(\frac{2\pi\tau}{T}\right) + c_2, \tag{11.56}$$

where $T$ is the proper-time extension of the fitted geometry. We perform this comparison for several lattice sizes and for decreasing regulator masses $m = 1$, $m = 0.1$ and $m = 0.01$. The same qualitative shape is recovered throughout the three scans, while increasing the lattice size broadens the profile in proper time. This is an encouraging indication that the ensemble explores a de Sitter-like ground state.

The comparison between the three masses is equally important. Since the regulator favours the Einstein sector, the physically relevant test is not only the quality of the fit at a finite value of $m$, but how the reconstructed geometry changes as the regulator is decreased. Figure 11.7 shows this change directly at our largest size, $N = 550$.

The proper-time extension increases and the scale of the profile changes substantially as $m$ is lowered. These results should therefore be understood as a first study of the approach towards $m \to 0$, and not yet as the recovery of a regulator-independent continuum ground state. They also indicate which quantities must be followed in the next stage of the analysis: the fitted proper-time extension $T$, the volume normalization and offset, and their scaling with both $N$ and $m$.

It is important to stress that the volume profile shown above is not, by itself, a fully diffeomorphism-invariant observable. The function $V(\tau)$ assigns a spatial volume to a particular value of the proper-time coordinate, and therefore relies on the homogeneous and isotropic foliation, on the reconstruction of $\tau$, and on the prescription used to centre the configurations before averaging.

It is nevertheless a very useful diagnostic in the present symmetry-reduced setting: it allows us to compare the dominant geometry with the Euclidean de Sitter profile and to follow how that geometry changes with $N$ and $m$. Its role is thus similar to that of a gauge-fixed or relationally defined intermediate quantity, rather than that of an observable that could be carried unchanged to an arbitrary affine configuration.

This is precisely why the observable programme must move from profiles expressed in a chosen coordinate to quantities defined intrinsically by the emergent geometry. Once a metric is reconstructed on every configuration, one can label pairs of points by their physical geodesic distance and construct curvature correlators such as the Kretschmann–Kretschmann correlator, that we will show below. One can likewise study the volume of geodesic balls or spheres as a function of their physical radius, Hausdorff and spectral dimensions, diffusion observables, curvature distributions, and higher-point correlators of scalars constructed from the full Riemann tensor. These quantities do not refer to a preferred coordinate label, and can therefore probe the geometry in a form that remains meaningful when the minisuperspace restriction is eventually relaxed.

Some of these intrinsic observables, in particular geodesic-volume profiles and Hausdorff or spectral dimensions, also have counterparts in CDT. The difference is that the present affine formalism combines access to an explicit ensemble of quantum geometries with the smooth tensor calculus available on each reconstructed configuration. CDT provides the former but, on its piecewise-flat geometries, does not provide the same direct access to local smooth curvature tensors and their arbitrary contractions. The usual FRG treatment provides the latter tensorial language at the level of effective fields and correlation functions, but does not sample individual geometries from a specified microscopic path integral. Purely affine lattice gravity therefore offers the possibility of studying both kinds of information within the same non-perturbative construction. The Kretschmann correlator is the first example considered here, but it is intended as the beginning of this broader observable programme.

The construction of observables in this setting has an interesting peculiarity. Since the metric is defined algebraically from the symmetric Ricci tensor, the Ricci scalar constructed from this same tensor is fixed identically, and likewise

$$g_{\mu\nu} = \frac{1}{\Lambda} R_{(\mu\nu)} \quad \to \quad R = g^{\mu\nu} R_{(\mu\nu)} = 4\Lambda \quad \to \quad R_{\mu\nu} R^{\mu\nu} = 4\Lambda^2. \quad (11.57)$$

Their connected correlation functions of the Ricci scalar and the Squared Ricci tensor are therefore trivial. This does not mean that all curvature information is trivial, but that these particular contractions partly repeat the algebraic definition of the emergent metric. This is a distinctive feature of this formalism. In other formalisms, like CDT, such correlators, if one were able to compute them exactly, would not be expected to be trivial. In this setting, where the quantized variables are the connection components, and not the metric, one has this distinctive feature, which could eventually be used to distinguish between the observables of this approach with respect to other theories.

A non-trivial scalar can instead be constructed from the full Riemann tensor of the emergent geometry, such as the Kretschmann scalar

$$K(x) = R_{\mu\nu\rho\sigma}(x)R^{\mu\nu\rho\sigma}(x). \tag{11.58}$$

To define a coordinate-independent two-point function, we subtract its volume-weighted mean, $K_c(x) = K(x) - \overline{K}$, and bin pairs of points by their physical geodesic distance $\ell = d_g(x, y)$,

$$C_{KK}(\ell) = \left\langle \frac{\int dV_x dV_y \, K_c(x)K_c(y) \, \delta(d_g(x,y) - \ell)}{\int dV_x \, K_c^2(x)} \right\rangle . \tag{11.59}$$

In the homogeneous and isotropic reduction, the spatial dependence reduces to the angular separation $\chi$ on $S^3$. The geodesic joining two points lies in a two-dimensional submanifold with metric $ds^2 = d\tau^2 + a^2(\tau)d\chi^2$. Its angular momentum is conserved, and we solve the resulting one-dimensional boundary-value problem numerically for each pair of endpoints. We then accumulate the volume-weighted products $K_c(x)K_c(y)$ in bins of the obtained geodesic distance and finally average over the Monte Carlo ensemble.

Figure 11.8 shows a preliminary example of this observable, where error bars are not included since they have not been computed yet, and therefore a detailed analysis cannot be performed, but at least a nontrivial polynomial decay is observed in the studied domain. The correlator decreases with physical distance and displays an encouraging agreement among the larger lattice sizes. A complete interpretation requires a systematic study of its dependence on the regulator, lattice spacing, volume, binning procedure, and correlations among the sampled configurations. Nevertheless, it already illustrates why the four-dimensional setting is qualitatively richer than the two-dimensional proof of principle: once an emergent metric exists, the affine configurations can be probed through physical volumes, geodesic distances, and non-trivial curvature correlations. The same mass-removal analysis initiated with the volume profiles must therefore be carried over to these correlators, in order to determine which properties represent the affine ensemble itself and which remain controlled by the regulator.

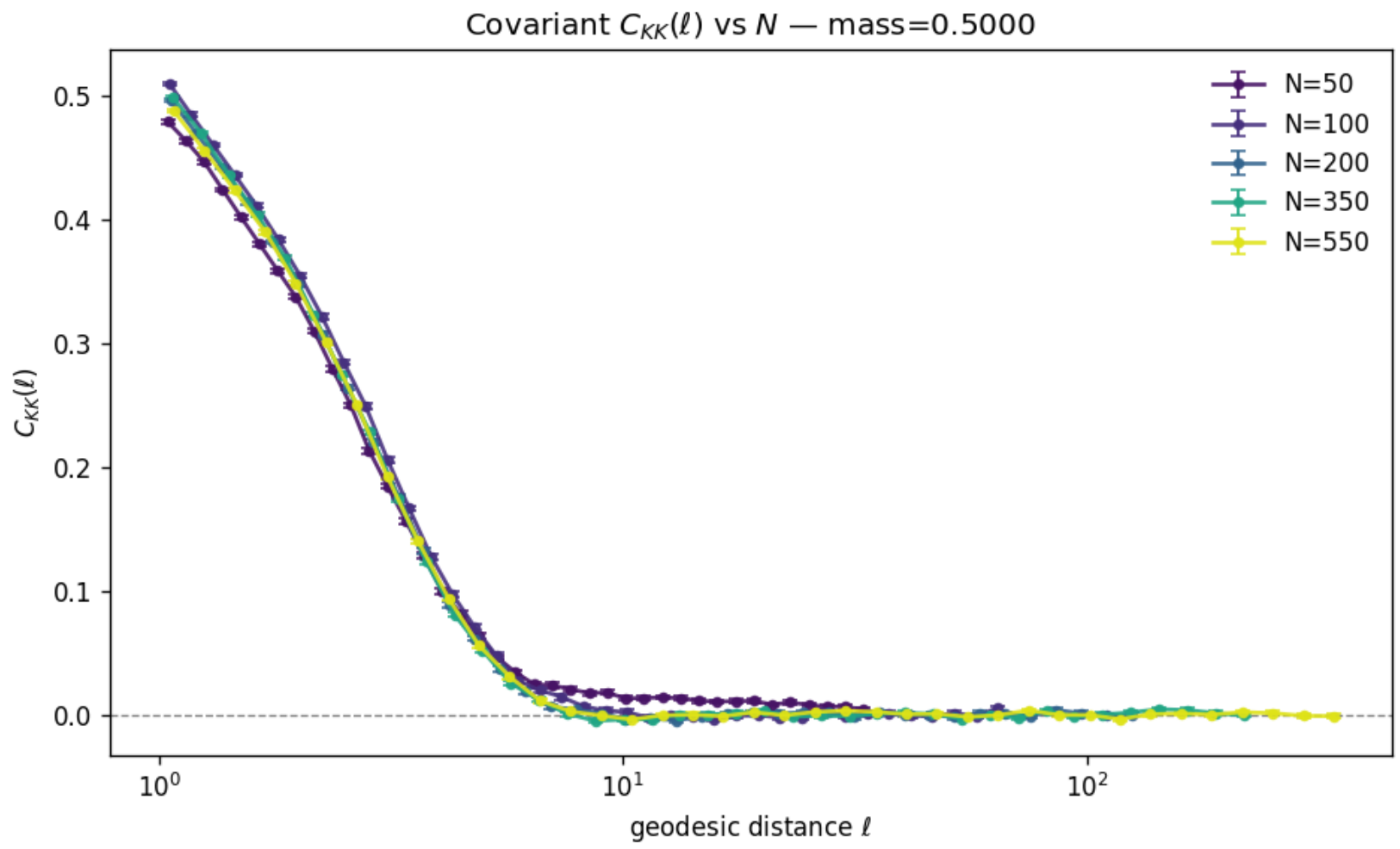

**Figure 11.8.:** Preliminary connected Kretschmann–Kretschmann correlator, normalized at the first reliable distance bin and binned by the geodesic distance $\ell$ of the emergent metric, for several lattice sizes at regulator mass $m = 0.5$.

The current calculations therefore represent an intermediate step between the unrestricted two-dimensional lattice implementation and the desired full four-dimensional simulations. They provide a controlled setting in which the regulator problem, the reconstruction of the metric, and the definition of observables can be studied explicitly. The next steps are to complete the regulator-removal analysis, extend the set of physical observables, and determine which features survive when the homogeneous and isotropic restriction is relaxed.

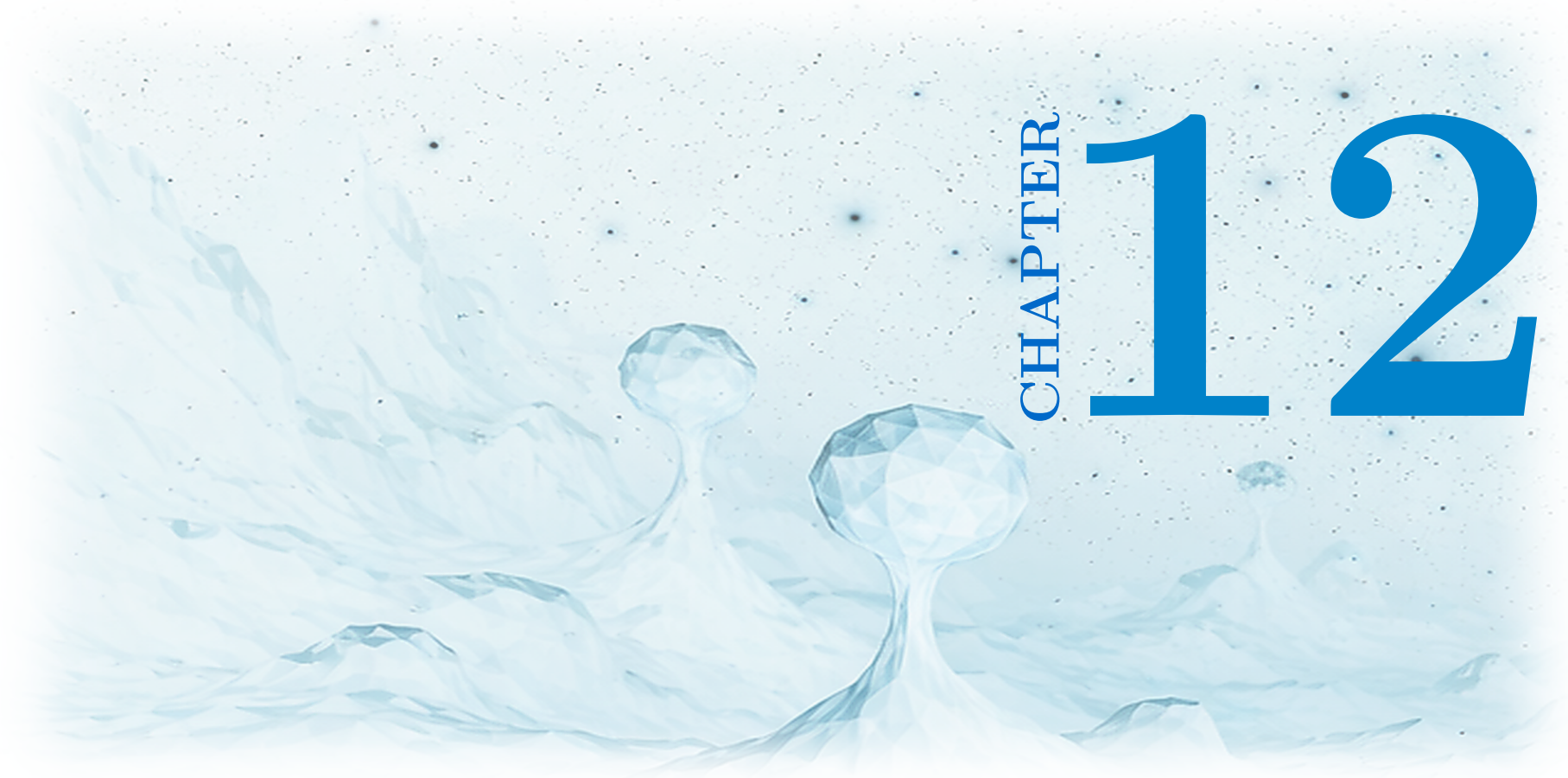

CHAPTER 12

# Summary part IV

The central objective of this part of the thesis was to take the first steps towards developing and exploring a new non-perturbative lattice framework for the quantization of gravity, in which the affine connection is taken as the fundamental field and the metric, when it exists, is an emergent quantity. Instead of asking only how General Relativity should be quantized, this part was guided by a more foundational question: what is it that we are choosing to quantize when we say “Quantum Gravity”?

**Chapter 11 - Lattice Quantum Affine Gravity.** Starting from the Palatini formulation of general relativity with matter, we recalled how, in dimensions higher than two and under mild assumptions, the metric can be eliminated algebraically in favour of the symmetric Ricci tensor, leading to an Eddington-type purely affine action that is classically equivalent to general relativity. This action has a definite sign, being positive for a positive cosmological constant, making a statistical treatment natural using its action as a Boltzmann weight. On a fixed hypercubic lattice, we discretized this action and constructed both torsionful and torsionless ensembles.

We complemented this construction with an explicit Monte Carlo implementation. The accompanying C++ code simulates the purely affine lattice action in arbitrary dimension using a standard local Metropolis algorithm, and offers both torsionful and torsionless ensembles, as well as different possible discretizations of the Ricci tensor. The first numerical study focused on the simplest non-trivial case, a two-dimensional torsionless ensemble with a finite-difference discretization of the Ricci tensor. Even though no emergent metric exists in two dimensions, the purely affine action has a smooth limit and can be used to define a statistical model of the connection. We explored the ensemble as a function of the only bare coupling of the theory, and monitored the expectation value of two diffeomorphism-invariant quantities related to volume measures constructed from the Riemann tensor. Their expectation values vary smoothly and monotonically with the bare coupling. Our data does not show any phase transition, but establishes that the model is numerically well-behaved, presenting neither divergences nor runaway behaviours in the explored ranges. This first study should therefore be understood as the computational proof of principle on which the rest of the programme is built.

The limitations of the two-dimensional ensemble then motivated the second stage of the chapter: a four-dimensional, closed homogeneous and isotropic reduction of the affine ensemble. In this setting, we explicitly restrict every accepted configuration to the sector $F > 0$ and $G > 0$. Consequently, a Euclidean FLRW metric exists configuration by configuration, and its lapse and scale factor define proper time, spatial volume, physical distance and curvature observables. Thus, the absence of a fixed-signature emergent metric is a limitation of the unrestricted two-dimensional study, but is precisely one of the issues treated in the four-dimensional minisuperspace construction.

The determinant action has flat directions, which we regulate by penalizing the tensorial difference between the dynamical connection and a fixed de Sitter background connection. This background is the Levi-Civita connection of the classical solution selected by the closed homogeneous and isotropic vacuum sector with positive cosmological constant. The regulator is not interpreted as part of the fundamental affine action. We simulated five lattice sizes and twelve masses, using 200 011 stored configurations for each pair $(N, m)$ after combining two independently thermalized chains. The figures emphasize $m = 1$, $m = 0.1$ and $m = 0.01$, for which the spatial-volume profiles are well described by the de Sitter-motivated form $\langle V(\tau)\rangle = c_1 \cos^3(2\pi\tau/T) + c_2$, where $T$ is the proper-time extension. Their comparison at the largest size, $N = 550$, also makes clear that the geometry still changes as the regulator is decreased. This gives an encouraging first indication of de Sitter-like behaviour, while at the same time identifying regulator removal and finite-size scaling as the central tests still to be completed.

Once the emergent metric is available, one can also move beyond the diagnostic quantities of the two-dimensional study. The volume profile is useful for identifying the dominant semiclassical geometry, but depends on the chosen foliation and centring prescription and is therefore not by itself a fully diffeomorphism-invariant observable. The Ricci scalar and Ricci-tensor square, on the other hand, are fixed by the algebraic relation used to define the metric and therefore have trivial connected correlators. The full Riemann tensor still contains non-trivial information, which we probe through the connected Kretschmann–Kretschmann correlator binned by physical geodesic distance. More generally, the reconstructed metric makes it possible to study geodesic-ball and geodesic-sphere volumes, Hausdorff and spectral dimensions, diffusion observables, curvature distributions, and higher-point curvature correlators. This illustrates the qualitative step achieved in four dimensions: the connection ensemble can now be studied through intrinsic properties of the geometry that it produces. The next analysis must determine how these observables vary with $m$, $N$, and the chosen background regulator.

The developed framework and simulation code open several directions for further work. The immediate task is to complete the regulator-removal analysis and establish the finite-size and finite-spacing scaling of the volume profiles and curvature observables. Beyond this, one must progressively relax the homogeneous and isotropic restriction, study the dependence on the fixed background used at finite regulator mass, and determine whether a fixed-signature, metric-compatible sector can still be sampled efficiently in a less constrained four-dimensional ensemble. The formulation remains statistical, and the question of how, or whether, a Lorentzian, unitary quantum theory can be recovered is still open. These are the questions naturally exposed by the present construction, rather than prerequisites that have prevented the first four-dimensional exploration from being performed.

In a more refined ensemble, if a suitable Lorentzian continuation exists, as well as an appropriate phase transition of the lattice simulations, the quanta of this purely affine theory would be excitations of the connection rather than of the metric, suggesting a “connecton” picture of gravitational degrees of freedom, closer in spirit to the other fundamental interactions. While the classical equivalence of the purely affine theory with GR implies that the propagating degrees of freedom in each formulation are the same, a spin 2 massless particle, in this formulation, has a different origin than a metric formulation. Whether this speculative picture can be made precise, and whether the lattice framework developed here can serve as a good starting point for such a construction, are questions we will further explore in future work.

So far, we have shown that purely affine gravity, long regarded mainly as a classical reformulation, can be brought into the non-perturbative lattice arena in a concrete and computable way, and that its basic statistical properties can be explored with standard Monte Carlo techniques. We believe that this work is a first step into a rich research line in the quest for the long-standing problem of the quantization of gravity.

This affine detour has been the most personal part of this thesis. It began not from the certainty that the affine connection must provide the correct fundamental description of quantum gravity, but from the conviction that some of the most interesting paths in physics begin by questioning an assumption that has become almost invisible through familiarity. In this case, that assumption is that the metric may not necessarily be the object that must be quantized. The framework developed here is still at an early stage, and many conceptual and numerical difficulties remain ahead. Nevertheless, it has already transformed a personal intuition into a concrete construction: a well-defined statistical ensemble, a working numerical implementation, and the first indications that familiar geometry may emerge from a quantum theory in which the metric is not fundamental. My hope is that these results mark the beginning, rather than the conclusion, of this programme. There is now a wide landscape to explore, from regulator removal and genuinely four-dimensional simulations to new observables, continuum limits, and the possible emergence of Lorentzian spacetime. Whether this path ultimately leads to a complete theory or instead teaches us where the affine viewpoint reaches its limits, I believe it asks a question worth pursuing. For me, opening a concrete route along which that question can be investigated is already one of the most meaningful outcomes of this journey.

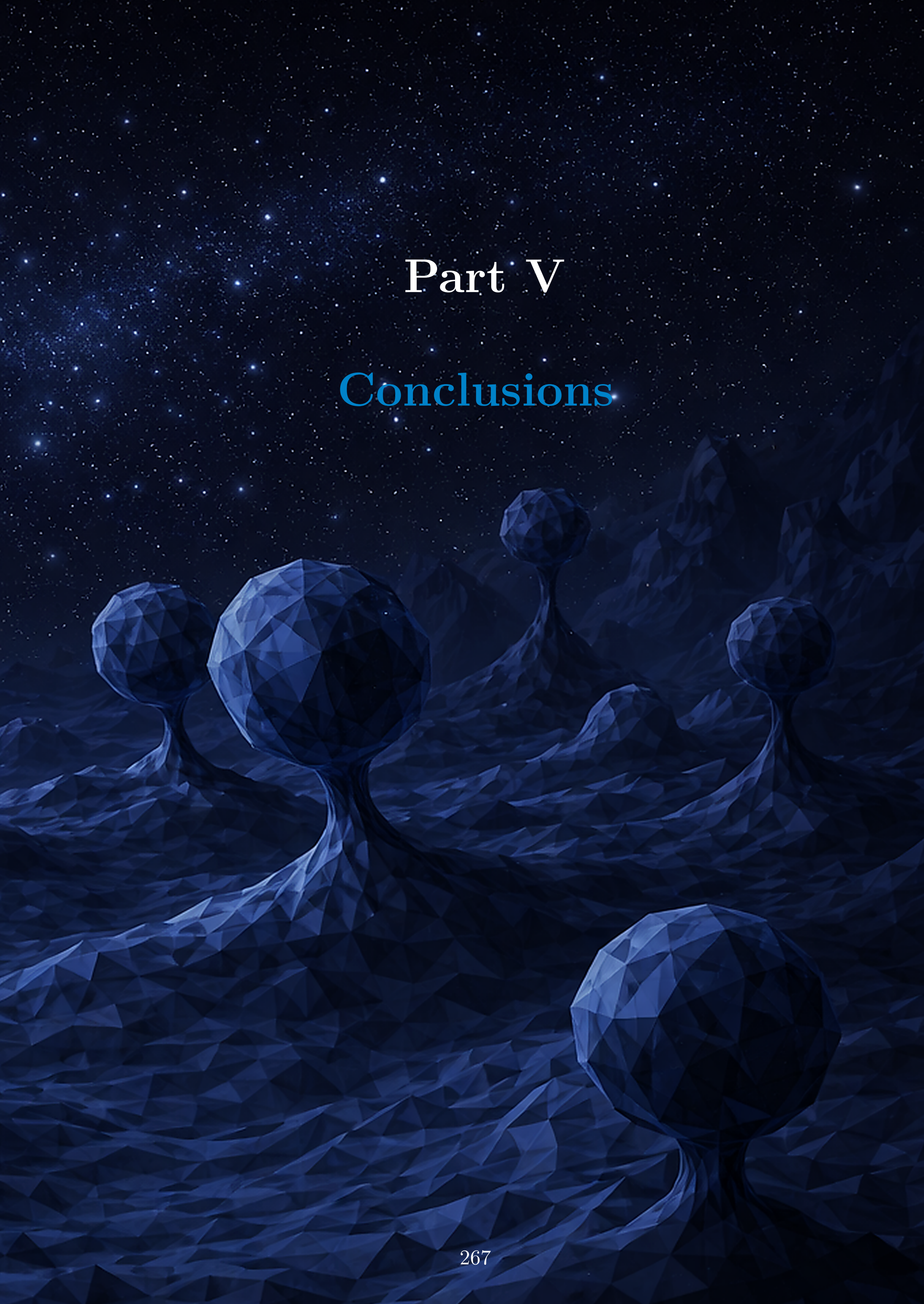

# Part V

# Conclusions

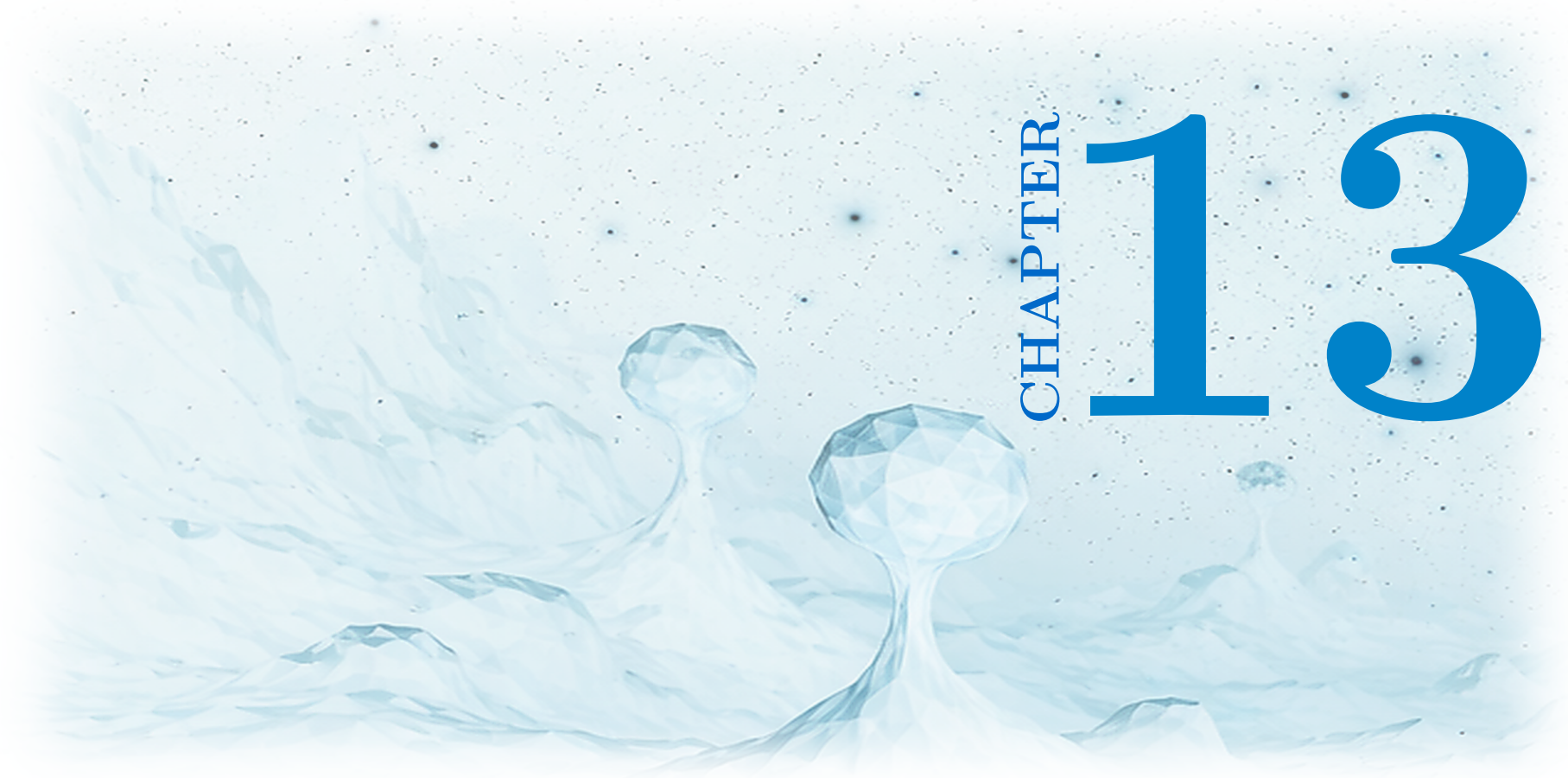

# CHAPTER 13

# The culmination of this journey

## 13.1 A short recap of a long walk

This journey began with a tension between two main pillars of modern physics: General Relativity and Quantum Mechanics. At the shortest distances, where quantum fluctuations are not small ripples on top of a fixed background, but where the background itself becomes part of the quantum dynamics, one question becomes unavoidable: what does it mean to define a quantum theory *of spacetime*?

Part I set the stage by emphasizing one guiding principle of this thesis: a conservative attitude towards that question. In the absence of direct experimental guidance at Planckian scales, progress depends on pushing as far as possible with the tools that have proven reliable elsewhere in quantum field theory: path integrals, renormalization, and non-perturbative control. The standard perturbative approach to quantum gravity is not merely technically challenging; it is mathematically inconsistent as a definition of a predictive theory, because the perturbative expansion around a fixed background does not close on a finite set of couplings. The natural lesson is not that gravity is "incompatible with quantum mechanics", but that the perturbative approach is not the right one. This is why non-perturbative avenues are pursued in this thesis.

A second guiding principle, just as important as the conservative one, is that understanding a physical system often requires learning to look at it from multiple complementary angles. This is both a methodological choice and a personal one: when the object of study is spacetime itself, restricting oneself to a single language or approach risks mistaking the language for the physics. This multifaceted viewpoint is the reason the thesis covers broad ground. It also explains its organization: each part is opened by a concrete question, and each question naturally selects a different set of tools.

Part II approached quantum gravity from a lattice/computational perspective, using Causal Dynamical Triangulations (CDT) as a concrete, conservative, non-perturbative definition of the gravitational path integral. The driving question there was: if gravity is defined as a sum over geometries, what do typical quantum geometries look like, and how can we measure geometric properties when the familiar language of smooth tensors is not available? The core emphasis was therefore on *observables*. The work in this part developed and implemented several probes designed to test quantum geometries in a background-independent, strongly fluctuating setting: an effective notion of topology based on coarse-graining and homology; multiscale measures of statistical homogeneity as non-perturbative analogues of translation symmetry diagnostics; and generalized curvature observables formulated to remain meaningful in discrete quantum geometries. These tools were tested in two-dimensional CDT, not because two dimensions are the target of quantum gravity, but because it provides a non-trivial laboratory where one can build intuition, test definitions together with their robustness and interpretations, and learn what kinds of obstacles appear already in this simplified setting to later explore more realistic ones.

Part III switched viewpoint and asked a complementary question: given that any lattice simulation has access only to a finite window of short-distance scales, how can one connect the physics of spacetime at Planckian scales, to infrared, measurable physics? Here, the non-perturbative functional renormalization group (FRG) became the main tool to study Quantum Gravity. Within the Asymptotic Safety framework, ultraviolet (UV) completion is phrased as the existence of a non-trivial quantum scale-invariant regime, and predictivity of the quantum theory of gravity is tied to the dimensionality and structure of the UV critical surface. The contributions of this part were therefore aimed at making predictivity practical in complicated theory spaces, such as the ones encountered in Quantum Gravity. We introduced and exploited RG flow-invariant structures as a way to organize

RG flow trajectories; we developed a computational strategy to extract these structures efficiently, including a machine-learning implementation (the Flow Invariant Neural Network); and we explored functional scalar-tensor flows as an arena where effective interaction potentials can *emerge* from UV consistency conditions rather than being imposed by hand. The early universe dynamics served as a concrete illustration of the general framework of the predictivity of Asymptotically Safe Quantum Gravity: a UV-complete Quantum Field Theory of gravity can constrain, and in the studied cases, even select, infrared (IR) physics, compatible with observational constraints.

Part IV was explicitly framed as a personal detour to a more foundational layer of the same story. After exploring two non-perturbative, metric-based viewpoints, I asked a question that sits one step before: what do we choose to quantize when we say "quantum gravity"? The detour investigated a conservative alternative in which the affine connection is treated as the fundamental variable and the metric (when it exists) is a derived object. The first result was a concrete and computable non-perturbative lattice framework for the statistical partition function of purely affine gravity, together with an explicit Monte Carlo implementation and exploratory numerical results in two dimensions. This provided a proof of principle, but also made its own limitation clear: in two dimensions no non-degenerate metric can be reconstructed from the symmetric Ricci tensor. This limitation led naturally to the four-dimensional homogeneous and isotropic construction developed afterwards. There, we restrict the ensemble to a positive-definite sector and reconstruct a Euclidean FLRW metric configuration by configuration. The preliminary volume profiles, studied across several sizes and decreasing regulator masses, show a de Sitter-like behaviour, while curvature correlators provide a first example of the non-trivial observables that can be defined using physical distances in the emergent geometry. The point of this part was not to claim a complete framework to quantize gravity, but to show how a small change of perspective on General Relativity leads from a well-defined affine statistical ensemble to concrete questions about the emergence of semiclassical four-dimensional geometry.

Taken together, one can see that the thesis does not present quantum gravity as a one-way road with a single solution. It presents it as a landscape that must be approached through multiple facets: by defining the microscopic arena (lattice path integrals), by agnostically exploring a vast space of possible theories (renormalization), by asking geometric questions within a regime where standard tools do not work (observables), and by a willingness to question what is truly fundamental (alternative quantization variables). If there is a common narrative thread, it is the idea that progress in quantum gravity will be driven not only by bold new principles and assumptions, but also by careful, rigorous construction of well-defined non-perturbative frameworks, of meaningful probes, and of bridges between UV definitions and IR phenomenology.

## 13.2 Open questions and next steps

I believe that a work is measured not only by what it completes, but also by what it enables. The results presented here open several directions for further work. Importantly, these open questions are not detached from the thesis; they are the natural continuation of the choices made throughout it: to remain conservative in assumptions, to be explicit about what the available tools can and cannot decide, and to build technical machinery that can be exported to richer settings.

A first class of open questions concerns the CDT observable program developed in Part II. The tools introduced there were designed to be background-independent and intrinsically scale-dependent, but their full relevance ultimately depends on applying them to the physically motivated setting in CDT: four spacetime dimensions. Extending the effective topology diagnostics, statistical homogeneity measures, and generalized curvature observables beyond the two-dimensional testing ground is therefore a direct next step. The central technical and conceptual task is to turn these probes into instruments that can be used systematically in the study of emergence: to determine at which scales, and in which phases, if any, smooth manifold-like behaviour, effective symmetries, and other physically desirable properties appear, and to what extent those properties are stable under changes of the probing scale and coarse-graining procedure.

A second class of open problems concerns the FRG and Asymptotic Safety perspective developed in Part III. The results there illustrate how UV consistency conditions can shape IR physics, but they are obtained within specific approximations and computational schemes. An obvious next step is therefore a robustness test: analyzing the qualitative inflationary picture under variations of truncation, regulator choice, gauge fixing, background and splitting choice, and signature choice. This is not of merely technical interest; it is the condition under which statements about emergence and predictivity become trustworthy beyond the specific setup in which they were first observed, and should be studied with the rigour it deserves.

At the same time, the methodological tools developed in this thesis are particularly well-suited to push the predictivity of the program forward, precisely where brute-force exploration becomes impractical. Flow invariants provide a compact way of encoding the organizing relevant directions of theory space, and the FINN strategy provides a new, potentially scalable, numerical basis to determine predictions from UV completions. A natural continuation is to test these tools in larger truncations and in theory spaces that are closer to more realistic matter-gravity systems, and to sharpen the connection between UV complete frameworks and phenomenological targets. The scalar-tensor explorations should be read in this spirit: as a proof of principle that quantum gravity motivated UV constraints can do more than accommodate effective physics; they can restrict it and even predict it. Extending our developed framework to broader sectors and to additional phenomenological questions would be a concrete and well-motivated direction for future work.

A third class of open questions concerns the purely affine lattice framework introduced in Part IV. The most basic one is the same that governs any lattice definition of a quantum field theory: does the model admit a controlled continuum limit? Establishing this requires more than just larger simulations; it requires a systematic study of finite-size and finite-spacing effects, improved update algorithms to better explore configuration space, and a clearer map of the ensemble's structure. The unrestricted two-dimensional study did not impose configuration-by-configuration conditions ensuring the existence of a fixed-signature emergent metric. Rather than leaving this only as a future possibility, the four-dimensional minisuperspace construction takes the first concrete step towards treating it: the ensemble is restricted to a positive-definite homogeneous and isotropic sector, and a Euclidean emergent metric is reconstructed for every accepted configuration. The flat directions of the affine action are controlled by a mass regulator defined relative to a fixed de Sitter background connection. This is the Levi-Civita connection of the classical solution selected by the homogeneous and isotropic vacuum sector with positive cosmological constant. The central question is now whether the observed de Sitter-like behaviour, the fitted proper-time spatial volume extension and the geometric correlators stabilize as the regulator is removed and the lattice size is increased. Extending the mass-removal analysis to observables based on physical geodesic dis-tance, progressively relaxing the minisuperspace restriction, and testing the dependence on the fixed background at finite regulator mass are the next concrete steps. In parallel, there are conceptual issues intrinsic to the affine viewpoint: the framework remains statistical, and the question of how, or whether, a Lorentzian, unitary quantum theory can be recovered remains open. On another note, the perturbative renormalization of this formulation of quantum gravity has, surprisingly, not been studied yet, and I intend to do so in the near future. I believe this is a concrete and very interesting immediate research direction that must be analyzed, just like the perturbative renormalization of the standard formulation of General Relativity was studied decades ago. It still amazes me that this other side of the coin, or facet of general relativity, more in line with the title of my thesis, has not yet been investigated.

These open problems share a common lesson that is worth stating explicitly: the aim of this thesis was not to "solve quantum gravity in one go". It aimed to develop tools, perspectives, and concrete non-perturbative constructions that clarify what questions can be asked sharply, what can be computed, and what is still missing. In a field where the destination is still uncertain, building reliable instruments and walkable bridges is already substantive progress. This work contributes such instruments in the lattice setting (observables), such bridges in the RG setting (flow invariants and scalable reconstruction), and an explicit alternative arena where foundational assumptions could be tested (purely affine lattice gravity).

## 13.3 Final words

Gravity is the interaction that most directly shapes our intuition: it is the quiet background that grounds our everyday experience, and it is the dynamical geometry that forces us to rethink what we mean by distance, time, and causality. It is, therefore, not surprising that trying to quantize it pushes us into questions that feel simultaneously counterintuitive and very technical. Working on this thesis has reinforced, for me, that the hardest parts of quantum gravity are not just the difficulty of calculations, but also the ability to ask, and properly formulate the right questions in a setting where the familiar structures in which the standard training of a physicist fails.

If there is a personal conclusion I draw from this journey, it is that progress in this subject should come from a careful balance. On one side, one must be conservative enough to build controlled frameworks and to demand well-defined computations. On the other hand, one must be flexible enough to change one's viewpoint when a language stops being useful, and humble enough to question if what one has been taking as fundamental is indeed fundamental. This is the balance that guided my choices during this PhD: moving between lattice path integrals and renormalization group flows, as well as between geometries sampled configuration by configuration and effective descriptions valid energy by energy, as well as finally stepping four years back to ask whether the most basic assumption made to progress is the most natural starting point.

This thesis ends, therefore, not with a claim of closure, but with a clearer map of what has been explored and what remains open. The story is unfinished, but to be fair, almost all the stories in physics are somewhat eternal. Nevertheless, what we achieved is not to be diminished: the tools developed here aim to make future steps sharper, more computable, and more concrete. If this work helps the community to probe quantum spacetime more directly, to connect Planckian definitions to infrared questions more systematically, or to consider alternatives with a clearer computational footing, then it will have served its purpose.

To those who paved the way, and to those making sure the journey continues, I can only add one final invitation that mimics the beginning of this story: let us keep exploring the many facets of spacetime, from Planckian to cosmological scales, with a never-ending curiosity.

# Appendices

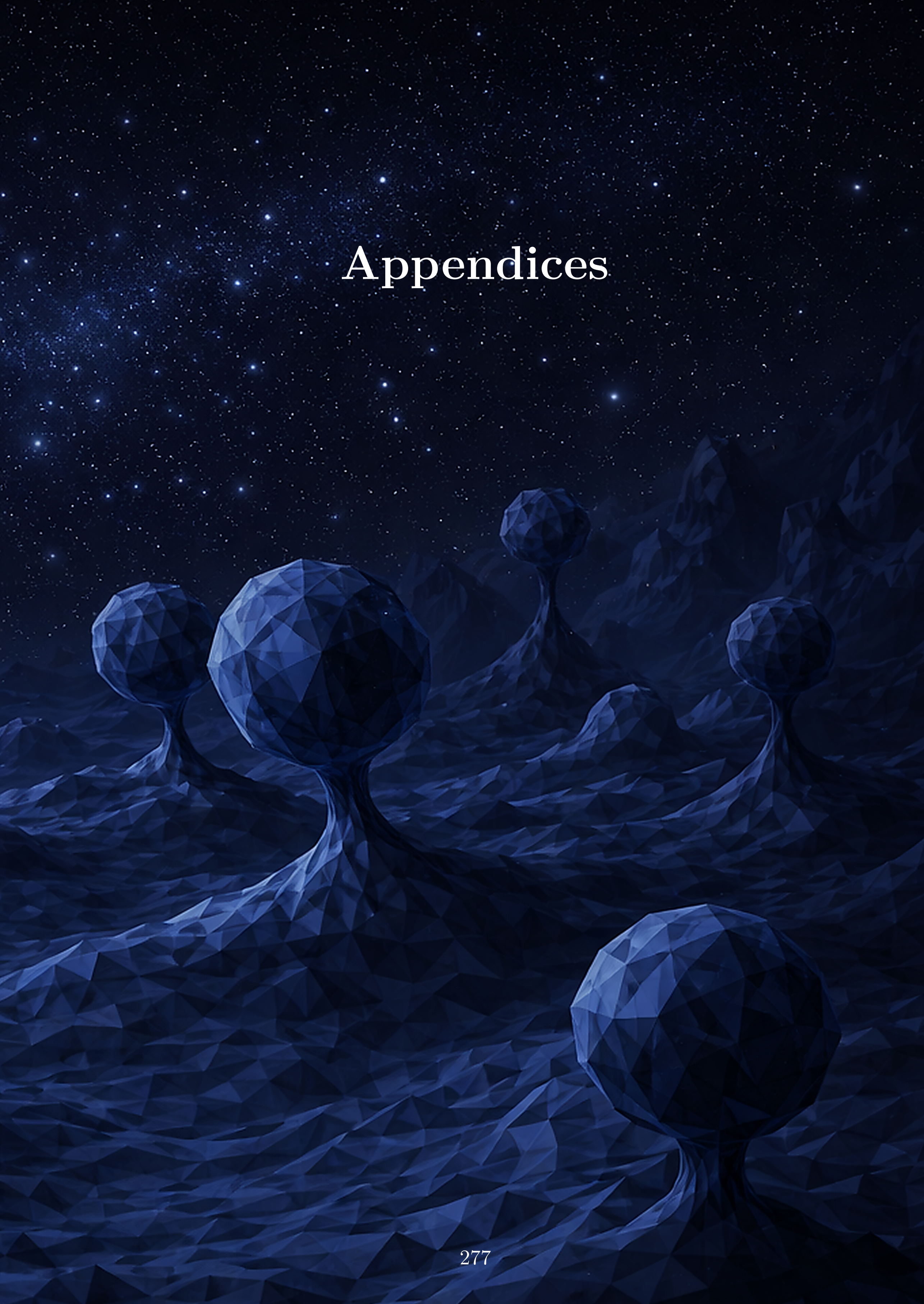

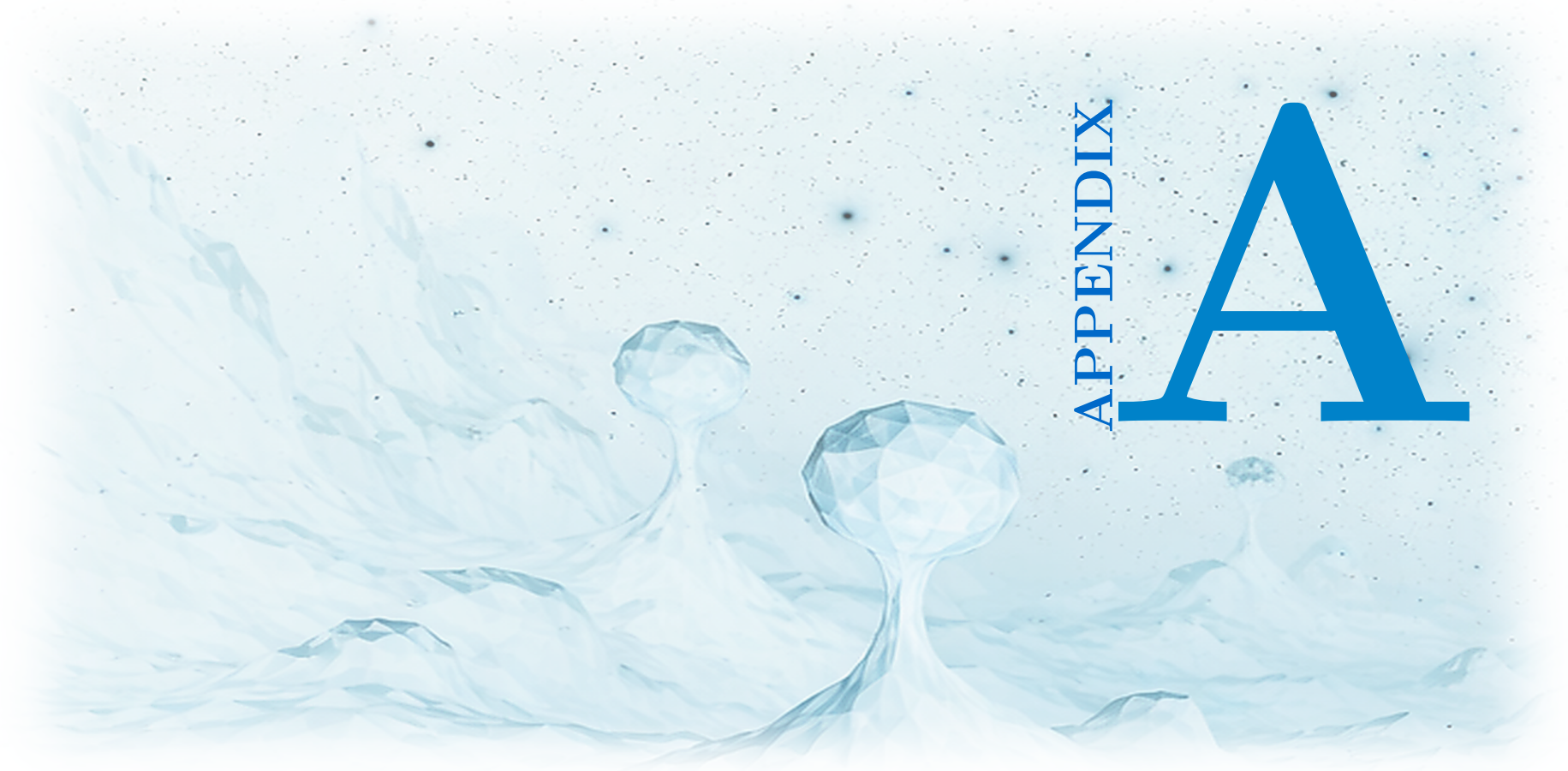

# Monte Carlo Simulations for CDT

In this appendix, we explain in more detail how a Markov chain Monte Carlo (MCMC) algorithm can be implemented to estimate the Euclidean expectation values of observables in 2D CDT. This section will closely follow [293], as the code used in Chapter 4 was an extension of the code developed by J. Brunekreef. For a more detailed explanation than the one you will find here, we refer the reader to his PhD thesis [293], where there is an extended explanation of the algorithm and the computational technicalities encountered in the implementation.

To approximate the expectation values of observables (2.7) by an average over a sample of path integral configurations, one must generate such geometries with a probability distribution of the form

$$P(T) = \frac{1}{Z} e^{-S_R^E[T]} \,. \tag{A.1}$$

Markov chain Monte Carlo methods allow us to approximate this probability distribution by performing a random walk in the space of all possible triangulations $T$. By sampling configurations within this random walk that are sufficiently different from each other (uncorrelated), one obtains a set of independent triangulations. If the different triangulations are generated using Markov chain sampling, then in this setting one is guaranteed to have a sample that follows the probability distribution (A.1). In practice, we work with labeled triangulations, so each vertex has a unique index, and links and triangles are ordered tuples of vertices. This means that the actual probability distribution one converges to in MCMC simulations of CDT is

$$P_l(T) = \frac{1}{Z}\frac{1}{N_0[T]!}e^{-S_R^E[T]}\,, \tag{A.2}$$

where there is a combinatorial factor associated with the number $N_0[T]$ of vertices in $T$.

When performing a local modification in a geometry $T$ that makes it transition to a geometry $T'$, a transition with associated probability $p(T \to T')$, one must check that the move satisfies *detailed balance*. This is in order to guarantee convergence to (A.2). The detailed balance condition is such that

$$P_l(T)\,p(T \to T') = P_l(T')\,p(T' \to T)\,. \tag{A.3}$$

This condition is met by proposing local geometrical updates with a selection probability $g(T \to T')$, and an acceptance probability $A(T \to T')$, thus setting $p(T \to T') = g(T \to T')\,A(T \to T')$. The probability $g(T \to T')$ depends on the type of local move, and the geometry $T$, but one can define the acceptance ratio to be

$$A(T \to T') = min(1, \frac{g(T' \to T)P_l(T')}{g(T \to T')P_l(T)}) \tag{A.4}$$

$$A(T' \to T) = min(1, \frac{g(T \to T')P_l(T)}{g(T' \to T)P_l(T')})\,, \tag{A.5}$$

A.0 which is exactly the so-called *Metropolis-Hastings* algorithm [294, 295].

In order to obtain a set of independent configurations, one performs a sequence of many local updates to the geometry, called Pachner moves [113]. These updates consist of taking a triangulation, such as Fig. 2.1, and performing local deformations of the types *add/delete* and *flip*, depicted in Fig. A.1.

The *add* move consists of randomly selecting a triangle $t$ with uniform probability over all the triangles in the triangulation $T$. Once this is done, one finds its unique neighbour sharing a space-like edge, called $tv$. In the case of the upper part of Fig. A.1, these are $t = (0\,1\,2)$ and $tv = (0\,3\,1)$. Then, one adds a new vertex ($v = (4)$ in the upper part of Fig. A.1) on the space-like edge, and two new time-like links $l1 = (4\,2)$ and $l2 = (3\,4)$, while updating the list of connections of vertices 2 and 3, effectively adding two triangles to $T$. This move does not require any constraint on the nature of the selected triangle $t$. The *delete* move consists of selecting uniformly a vertex $v$ among all the vertices with coordination number 4. Once this vertex is selected, one removes it from $T$, together with its two associated time-like links. In the upper part of Fig. A.1 this corresponds to identifying vertex $v = (4)$, and removing it together with the edges $l1 = (4\,2)$ and $l2 = (3\,4)$, while updating the connectivity of the vertices 0, 1, 2, 3, effectively removing two triangles from $T$.

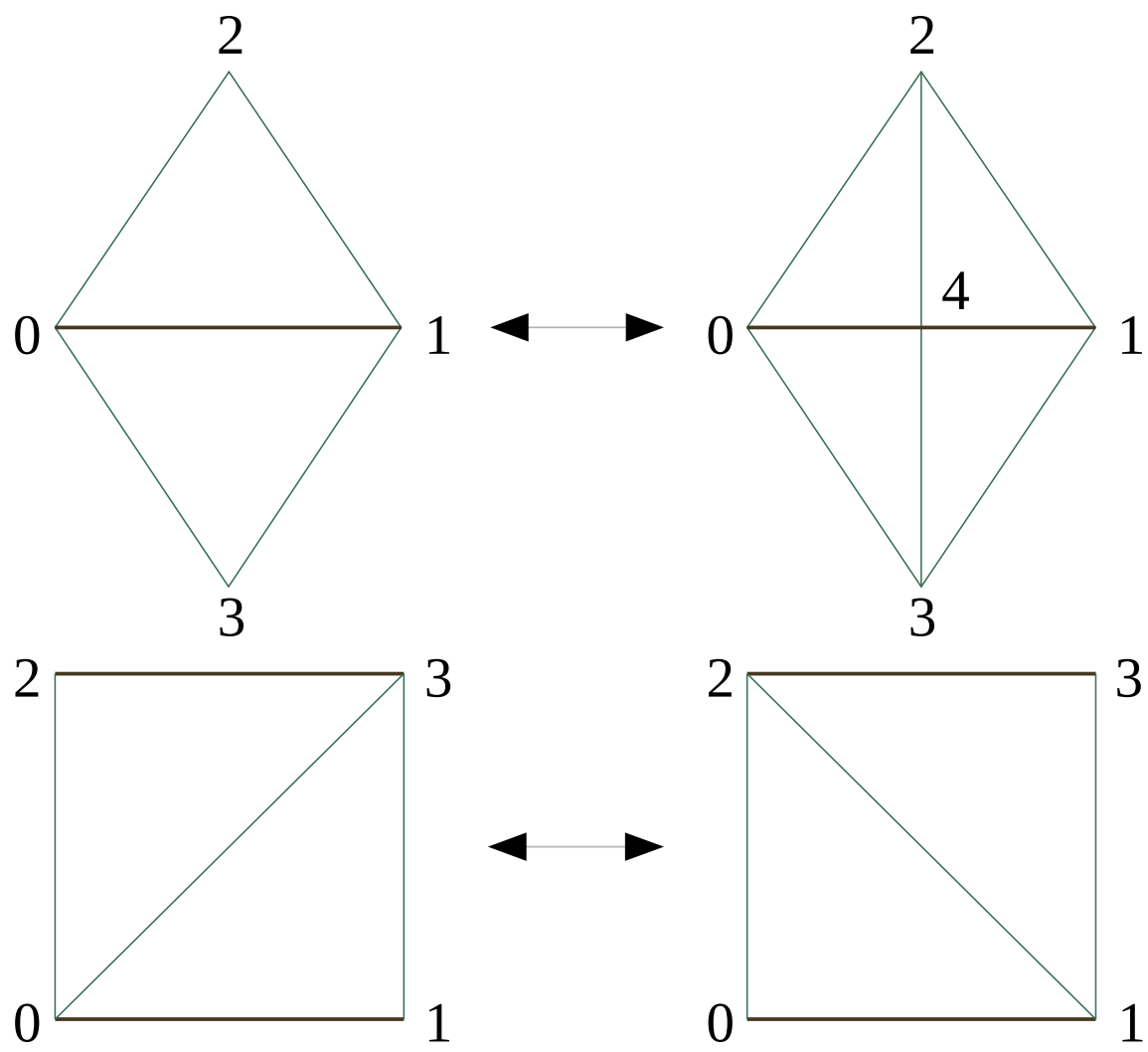

**Figure A.1.:** The two basic geometric local updates one can perform in 2D CDT. On the top are the *add/delete* moves, and on the bottom the *flip* move. The vertices are labeled, and therefore these moves make actual geometrical changes. Figure taken from [293].

The *flip* move is slightly more complicated. It requires keeping track of the "ordering" of the vertices composing triangles, effectively distinguishing between "up" and "down" triangles as described in Chapter 2. While keeping track of the type of triangles, one searches for a pair of neighbouring "up" and "down" triangles that share a time-like edge. In the lower part of Fig. A.1, this consists of finding the pair of triangles $tl = (0\,3\,2)$ and $tr = (0\,1\,3)$ and identifying the *flip* candidate link $l = (0\,3)$. In practice, this is implemented by always keeping track of the type of triangles, either "up" or "down", and keeping track of the triangles that have a neighbour of the opposite type. Then, one samples uniformly among all the links $l$ that are shared by these types of pairs of triangles, and exchanges the extremes of this link between the two triangles, as is shown in the lower part of Fig. A.1. This move, while changing the geometry because of the update of the connectivity of the vertices composing the selected pair of triangles, does not add any triangle to $T$. The *flip* is its own inverse, so there is no other extra move to discuss.

With this set of moves, and using the *Metropolis-Hastings* algorithm, one can mutate configurations into others, and get an ensemble of geometries, used to approximate expectation values by averages over this ensemble. One basic code for such implementation can be found on [296].

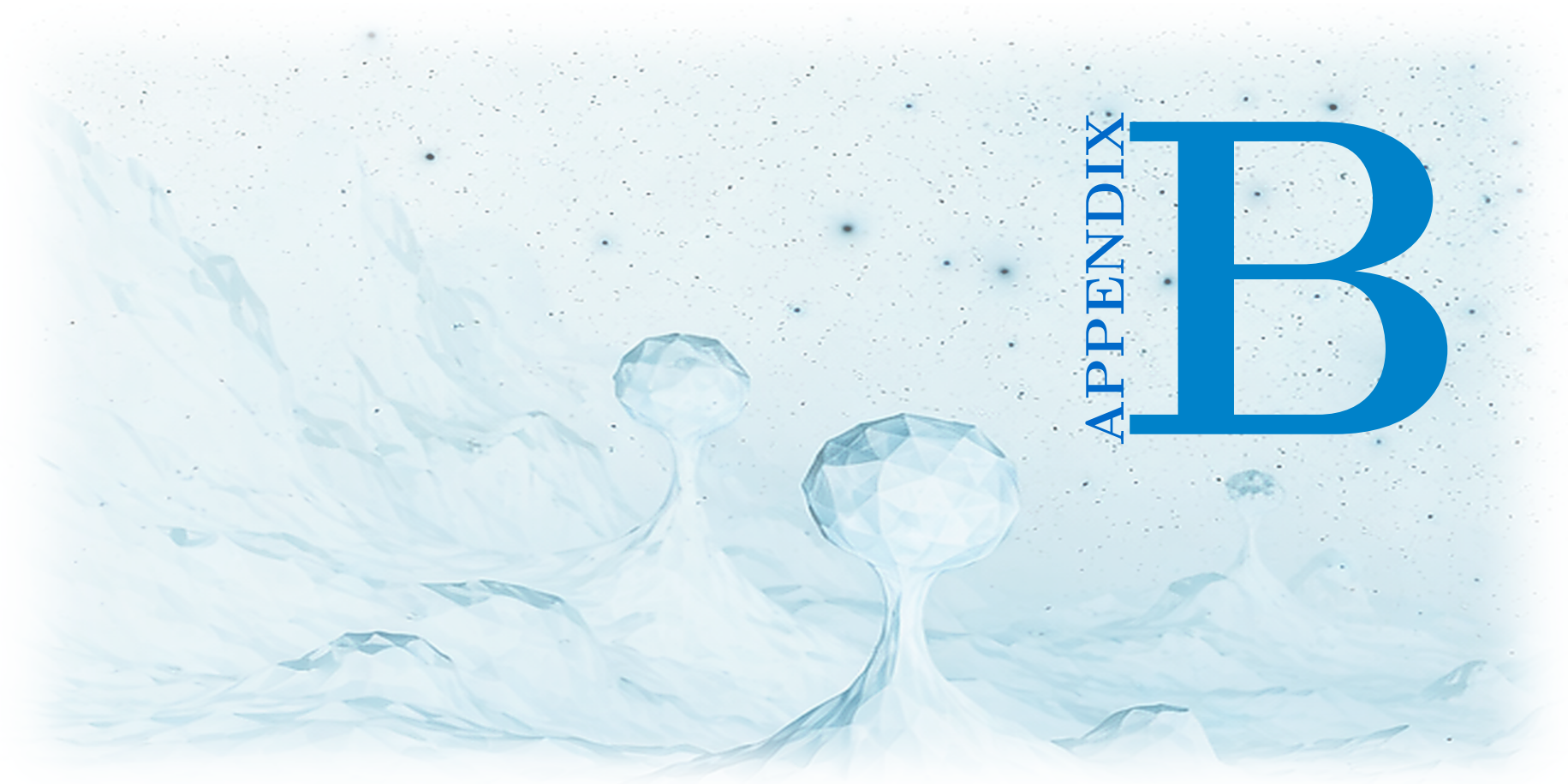

# TDA technicalities for CDT

## B.1 Appendix TDA 1

In this appendix, we provide some details about how we construct the Voronoi decomposition of a triangulation $T$ whose dual is a coarse-grained Delaunay triangulation $T_\delta$ of a given resolution $\delta$. Our starting point is the original triangulation $T$, where each triangle of $T$ has already been coloured and decorated according to the prescription outlined in Sec. 3.3.2 and illustrated in Fig. 3.3. We separately keep track of type-1 and type-2 triangles, to make sure that during the construction of the Voronoi decomposition every piece of boundary segment and every triple intersection of such segments has been taken into account.

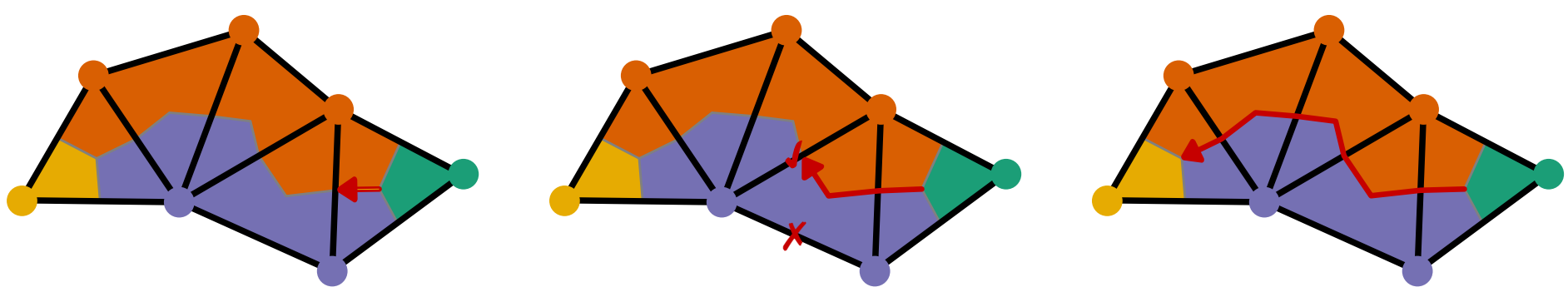

**Figure B.1.:** Moving through a sequence of triangles to identify a boundary segment between two triple points belonging to the Voronoi decomposition.

Our algorithm proceeds by considering each type-2 triangle in turn, and establishing the three nearest triple points that can be reached from the triple point at the centre of this triangle by following boundaries between differently coloured regions on the triangulation $T$. This determines the three boundary segments of the Voronoi decomposition meeting at the initial triple point.

The process of finding a nearest neighbour[1] is illustrated by Fig. B.1. Starting from a triple vertex (left figure), we choose to walk along one of the three boundaries to the centre of the next triangle, in this case by crossing the triangle edge between the violet and the orange vertex of the triangle. There are then two possibilities: if the second triangle is of type 2, we have found the other triple point where the boundary segment ends; we then go back to the first triangle and explore the next direction, say, anticlockwise. If the second triangle is of type 1 (as is the case here, central figure), we next cross the triangle edge, which again has a violet vertex on the left (in the direction indicated by the arrow) and an orange vertex on the right. Such an edge always exists, since the third vertex of the triangle has to be either violet or orange. We then repeat the same step by continuing to the next triangle, always crossing the triangle edge with the violet vertex on the left and the orange one on the right, until the process ends when we encounter a type-2 triangle with a triple point (right figure). Each time we traverse a type-1 triangle, we remove it from the list, to keep track of which such triangles have been visited.

B.1 In this manner we exhaust all type-2 triangles. It can then happen that not all type-1 triangles have been visited yet. They must be parts of closed boundary segments without any triple points, like the ones we met earlier in Fig. 3.9. One proceeds by picking such a triangle from the list of remaining type-1 triangles and following the boundary between the differently coloured regions by traversing subsequent triangles, analogous to what is shown in Fig. B.1. Eventually this path must close on itself. This process is repeated until no type-1 triangles are left. In this way we guarantee that all boundaries have been visited. The result is the Voronoi decomposition, whose cells, boundary segments and triple points we associate with the dual vertices, edges and triangles respectively of the Delaunay triangulation, as described in Sec. 3.3.3.

[1] We use a data structure where, together with each triangle, we store the labels of its neighbouring triangles, such that finding a neighbour is simply a look-up in constant time.

## B.2 Appendix TDA 2

This appendix provides some details of our procedure to adjust a Delaunay triangulation such that it provides a well-defined input to GUDHI's computation of Betti numbers, without changing its homology. This input requires that each $p$-simplex is given in the form of a list $\{v_0, v_1, \ldots, v_p\}$ of $p+1$ vertices. It cannot handle a situation where distinct edges share the same end points $\{u, v\}$ or distinct triangles share the same corner points $\{u, v, w\}$. Both of these cases can arise during the coarse-graining procedure of Sec. 3.3, as was explained there.

To eliminate these irregularities systematically, we proceed in two steps. First, we identify all vertex pairs of the Delaunay triangulation which are connected by more than one edge. For each such vertex pair $\{u, v\}$, we insert an additional vertex in the middle of all but one of the edges connecting $u$ and $v$, such that there are two edges instead of one. At the same time, the triangles on either side of the edge are subdivided accordingly, such that there are four triangles instead of two, as illustrated by Fig. B.2, left. (If the edge was a loose edge, there are no triangles to be subdivided.) It can in principle happen that a newly created edge is part of a loop of length two, in which case the same procedure has to be applied to it also.

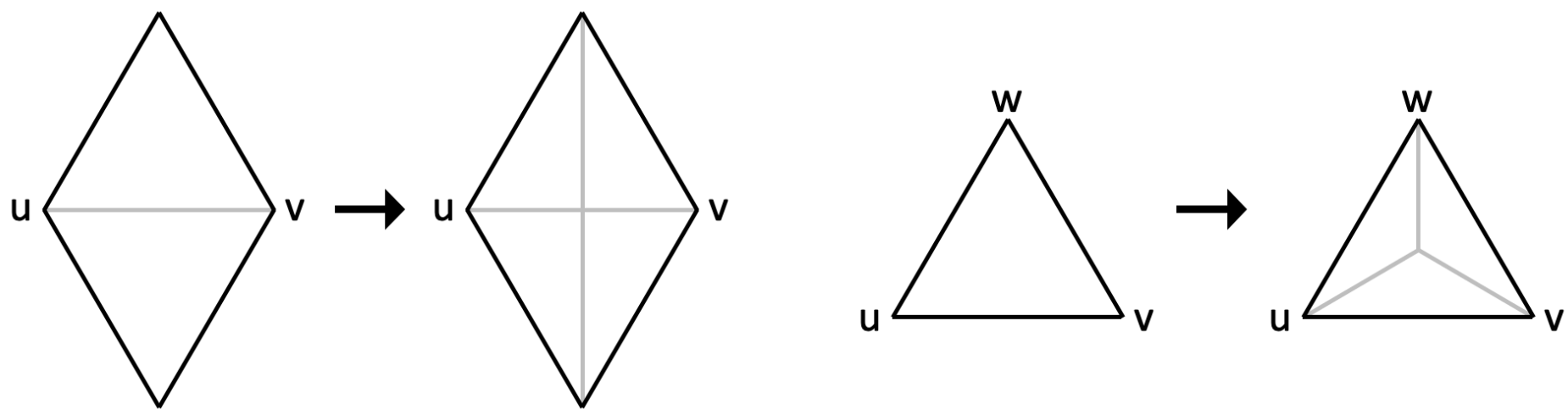

**Figure B.2.:** Removing irregularities from Delaunay triangulations: multiple edges sharing the same two vertices $u$ and $v$ are subdivided (left); multiple triangles sharing the same three vertices $u$, $v$ and $w$ are subdivided (right).

After all loops of length two have been eliminated, we identify all vertex triples of the Delaunay triangulation which are shared by more than one triangle. For each such triple $\{u, v, w\}$, we insert an additional vertex in the middle of all but one of the triangles with corner points $u$, $v$ and $w$, such that there are three triangles instead of one, as illustrated by Fig. B.2, right.

Note that the two operations depicted here coincide with local Monte Carlo moves implemented in two-dimensional CDT and EDT respectively (see, for example, [70, 89]), and do not change the connectivity of the triangulations. The final result is a two-dimensional simplicial complex, from which we prepare the input for GUDHI in the form of three lists:

- a list of vertices: $\{v_0, v_1, \ldots, v_{N_0}\}$,
- a list of edges: $\{\{u_0, v_0\}, \{u_1, v_1\}, \ldots\}$, and
- a list of triangles: $\{\{u_0, v_0, w_0\}, \{u_1, v_1, w_1\}, \ldots\}$.

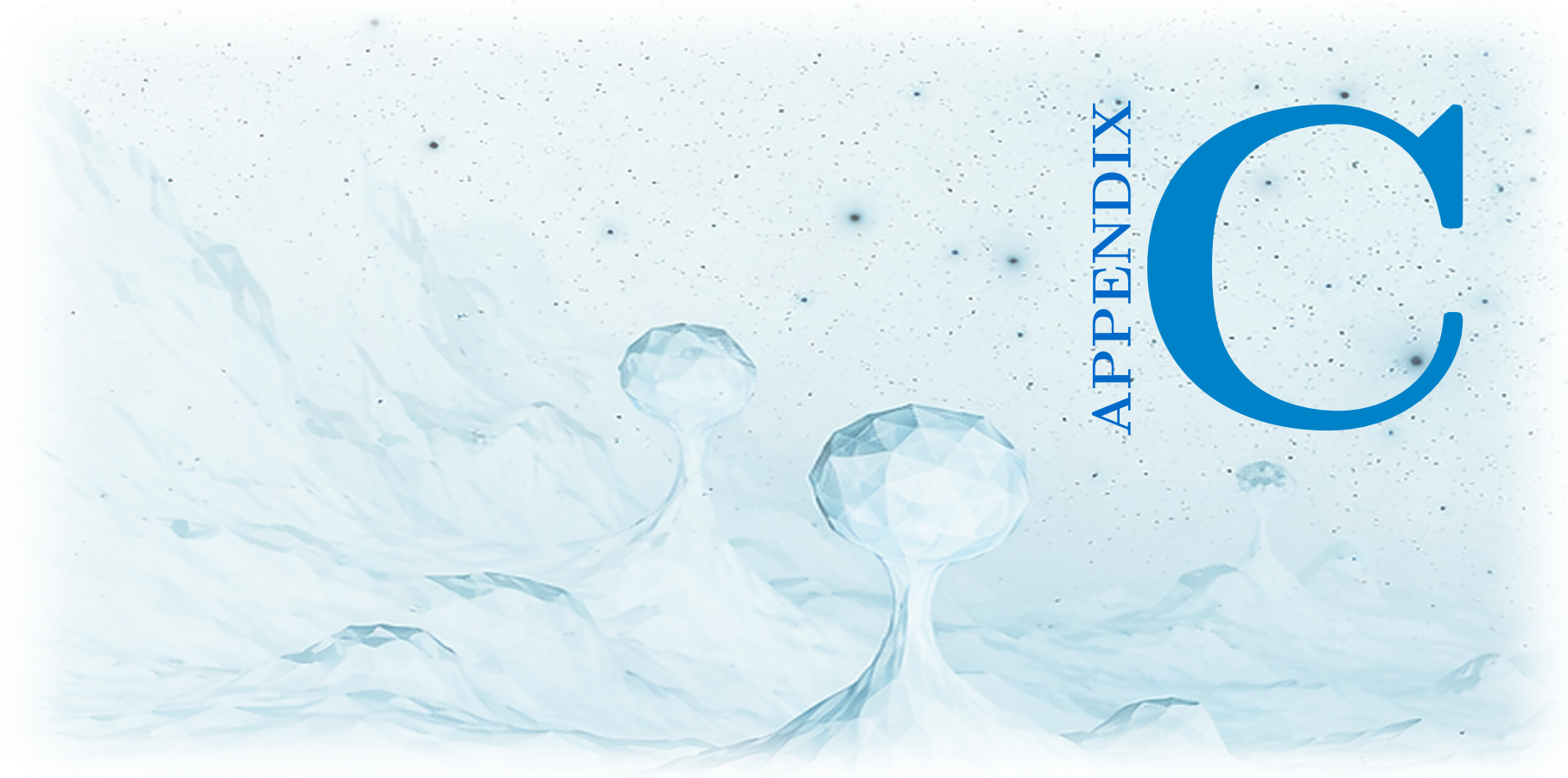

# Multi-scale homogeneity

In chapter 4, we commented on the difference between our homogeneity measures and those previously introduced in [117] and applied there to CDT quantum gravity in three dimensions. The primary difference is the presence of an explicit averaging (over geodesic balls) and of an associated averaging scale (the ball radius $\delta$) in our set-up. This averaging scale is distinct from the scale intrinsically associated with a geometric quantity whose variance and thus homogeneity is considered, the scale we have called $r$ in the main body of the thesis. To illustrate qualitatively that the $\delta$- and $r$-dependence of a homogeneity measure are in general different, we will consider the homogeneity of the quantity $Q(x;r)=\mathrm{v}ol(\mathcal{S}_x(r))$, where the corresponding averaged quantity $Q(x,\delta;r)$ is obtained by averaging the shell volumes $\mathrm{v}ol(\mathcal{S}_x(r))$ over $\delta$-balls according to the prescription (4.1). This is similar to the volumetric homogeneity measure in [117], which was based on the volume of geodesic balls $B_x(r)$ rather than the volume of geodesic shells.[1]

Fig. C.1 shows the expectation value $\langle\mathcal{H}^{rel}_{Q(r)}(\delta)\rangle$ of the relative homogeneity measure for $Q(r)=\mathrm{v}ol(\mathcal{S}(r))$, evaluated on two-dimensional CDT geometries with $N_2=80k$ and $t_{tot}=82$, based on data from 500 sampled configurations $T$. The range for both parameters is $[1,20]$. The data points for fixed $r=1$ (the front sequence of dark blue dots in Fig. C.1) coincide

[1] In our notation, the volumetric measure of [117] corresponds to (the square of) our homogeneity measure $\mathcal{H}_Q(\delta=0)$, where $Q(x;r)=\mathrm{v}ol(B_x(r))$.

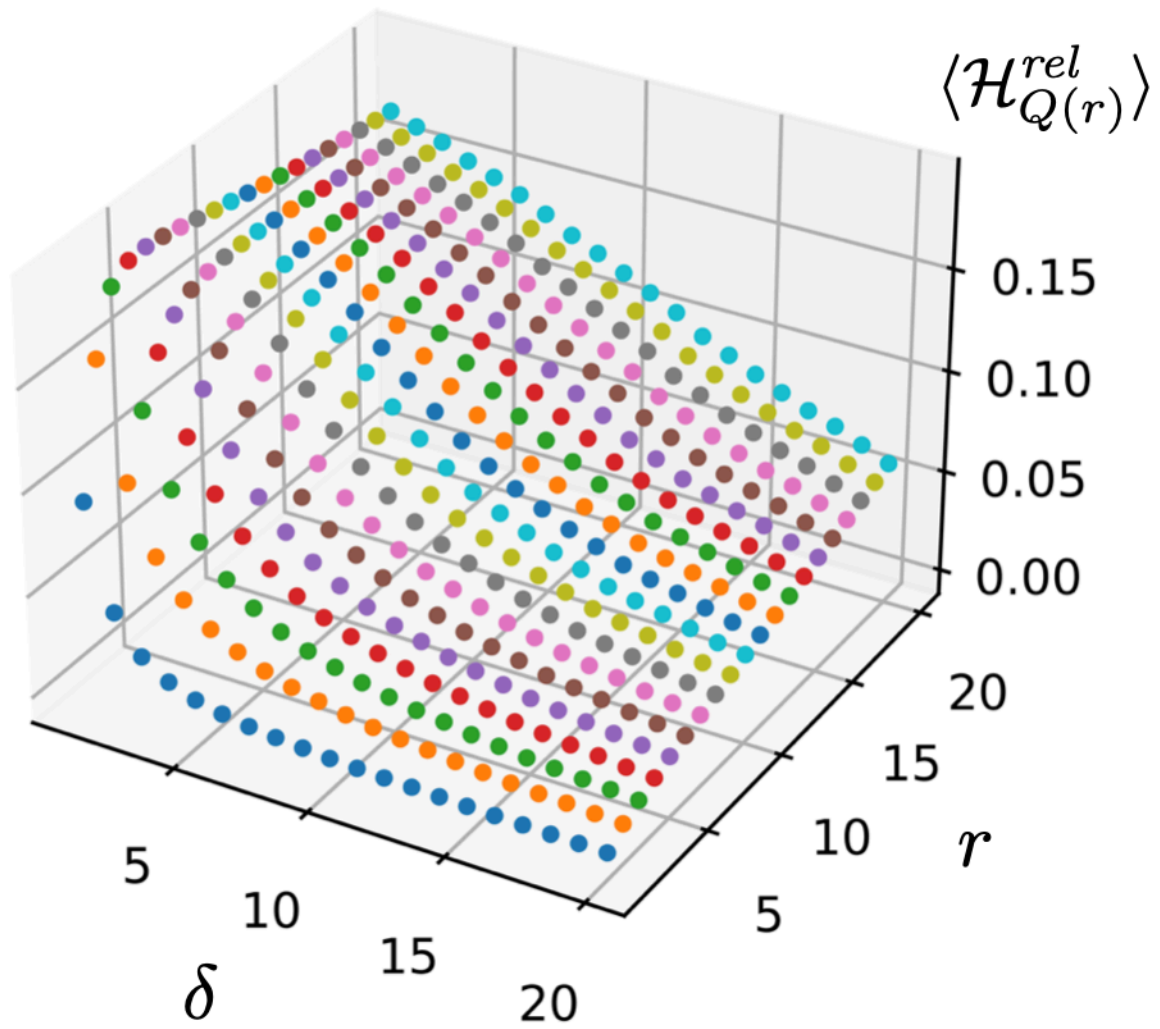

**Figure C.1.:** Expectation value $\langle \mathcal{H}^{rel}_{Q(r)}(\delta) \rangle$ of the relative homogeneity measure for $Q(r) = vol(\mathcal{S}(r))$, the volume of spheres of radius $r$, as a function of $r$ and the averaging scale $\delta$. Measurements taken on an ensemble of two-dimensional CDT configurations of volume $N_2 = 80k$ and time extension $t_{tot} = 82$.

C.0

with those of the relative homogeneity measure for the coordination number of Sec. 4.1.2, since the discrete length of a spherical shell of radius 1 at a vertex $x$ is always equal to the coordination number of $x$. They display the familiar rapid fall-off as a function of the averaging scale $\delta$. Examining other sequences of data points at fixed $r$, they are also monotonically decreasing as a function of $\delta$, but their decline becomes slower with increasing $r$. Note that from the perspective on averaging taken in the present work, one would usually consider only the region $\delta \gtrsim r$, such that the averaging scales $\delta$ are larger than the linear scale $r$ characterizing the quantity that is being averaged. What comes closest to the situation studied in [117] in our case are the data points for fixed minimal averaging scale (here given by $\delta = 1$) as a function of $r$. Even if one disregards the initial region $r \leq 5$ to avoid discretization artefacts, $\langle \mathcal{H}^{rel}_{Q(r)}(\delta = 1) \rangle$ as a function of $r$ behaves very differently from $\langle \mathcal{H}^{rel}_{Q(r=1)}(\delta) \rangle$ as a function of $\delta$, as we claimed above. If one takes the former as a measure of relative homogeneity, it decreases beyond $r = 4$, but at a rather slow rate. At $r = 20$ it is still above 15% and its behaviour is therefore not particularly homogeneous in the range we have investigated.

APPENDIX
D

# $\mathcal{SC}_i(||\mathcal{S}_p^r||)$ in constant curvature manifolds

In the case where the underlying space is a D-sphere of radius $R$, one can compute exactly the functional form of $\mathcal{SC}_i(||\mathcal{S}_p^r||)$ as a function of $r$, $R$ and $D$. This case is interesting since it could be used for direct comparison of $\mathcal{SC}_i(||\mathcal{S}_p^r||)$ of a general metric space with that of a D-sphere.

On a D-sphere, one can compute the surface $||\mathcal{S}_p^r||$ of a geodesic sphere of radius $r$ on top of it

$$||\mathcal{S}_p^r|| = 2\pi A_D R^{D-1} \sin^{D-1}\left(\frac{r}{R}\right), \tag{D.1}$$

where $A_D$ is a dimensional dependent constant of the form

$$A_D = \prod_{i=2}^{D-1} (\int_0^{\pi} \sin^{D-i}(x) dx). \tag{D.2}$$

Using this, one can compute exactly all the generalized curvature definitions

$$\mathcal{SC}_1(||\mathcal{S}_p^r||) = (D-1)\,(\frac{r}{R})\left(\cot\left(\frac{r}{R}\right) - (\frac{r}{R})\csc^2(\frac{r}{R})\right), \tag{D.3}$$

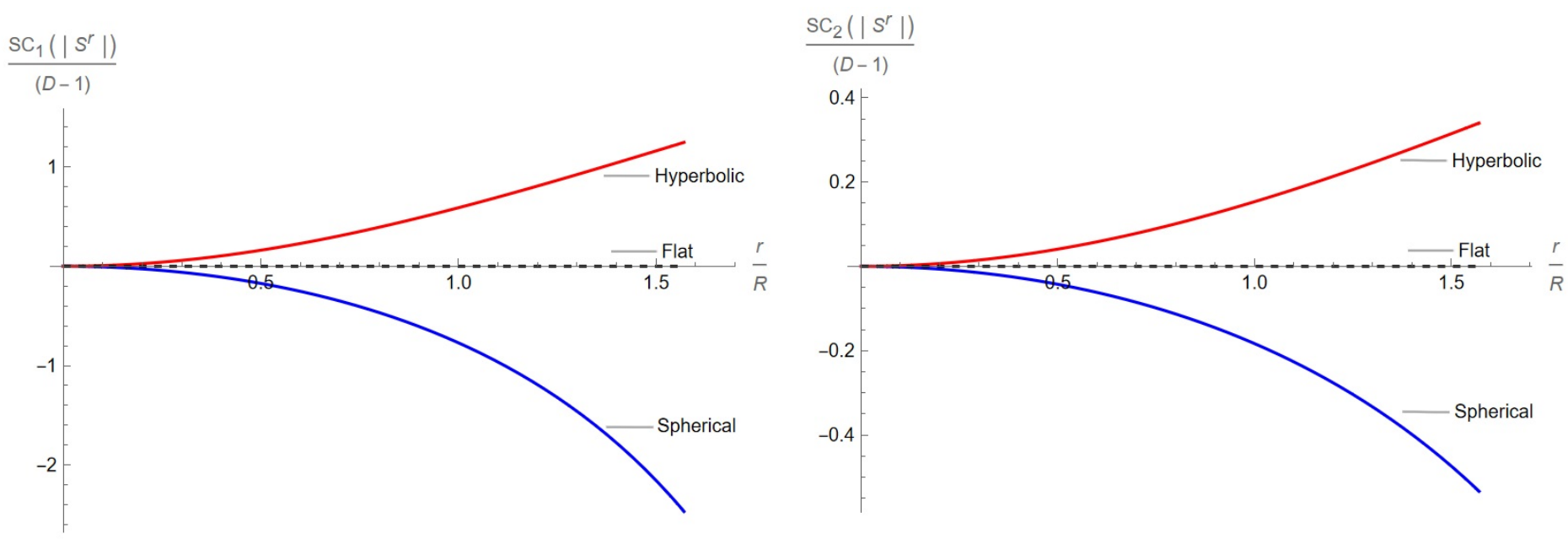

**Figure D.1.:** $\mathcal{SC}_1(||\mathcal{S}_p^r||)$ and $\mathcal{SC}_2(||\mathcal{S}_p^r||)$ for a D-Sphere of scalar curvature $\frac{D(D-1)}{R^2}$, a hyperbolic D-space of scalar curvature $\frac{D(1-D)}{R^2}$, and Euclidean D-dimensional space.

$$\mathcal{SC}_2(||\mathcal{S}_p^r||) = (D-1)\left(\frac{\left(\frac{r}{R}\right)\left(3\cot\left(\frac{r}{R}\right)-2i\right)-6\log\left(1-e^{-2i\left(\frac{r}{R}\right)}\right)}{3} - \frac{2i\,\mathrm{Li}_2\left(e^{-2i\left(\frac{r}{R}\right)}\right)}{\left(\frac{r}{R}\right)} + \frac{\zeta(3)-\mathrm{Li}_3\left(e^{-2i\left(\frac{r}{R}\right)}\right)}{\left(\frac{r}{R}\right)^2}\right), \tag{D.4}$$

where $\mathrm{Li}_n$ is the Polylogarithm.

Notice that the dimensional dependence of each of the $\mathcal{SC}_i(||\mathcal{S}_p^r||)$ appears to be just a multiplicative factor. This allows for a simple comparison with fractal like geometries, like the ones appearing in 2D Euclidean dynamical triangulations, where the ground states shares some properties with a 5-dimensional sphere.

Now, consider a hyperbolic $D$-space with scalar curvature equal to $\frac{D(1-D)}{R^2}$, where we define the radius $R$ to be such that this holds. In this case, one can also compute the surface of a sphere $||\mathcal{S}_p^r||$, where one obtains

$$||\mathcal{S}_p^r|| = 2\pi A_D R^{D-1}\sinh^{D-1}\left(\frac{r}{R}\right), \tag{D.5}$$

where it differs from the sphere case since there is a hyperbolic sine function instead of a sine function.

Using this, one can compute exactly all the generalized curvature definitions

$$\mathcal{SC}_1(||\mathcal{S}_p^r||) = (D-1)\,(\frac{r}{R})\,(\coth\left(\frac{r}{R}\right) - (\frac{r}{R})\,\text{csch}^2\left(\frac{r}{R}\right)), \tag{D.6}$$

$$\begin{aligned}\mathcal{SC}_2(||\mathcal{S}_p^r||) = (D-1)\Bigg((\frac{r}{R})(\frac{2}{3} + \coth\left(\frac{r}{R}\right)) - 2\log\left(e^{2(\frac{r}{R})} - 1\right) - 2i\pi \\ - \frac{2\,\text{Li}_2\left(e^{2(\frac{r}{R})}\right)}{\left(\frac{r}{R}\right)} + \frac{\text{Li}_3\left(e^{2(\frac{r}{R})}\right) - \zeta(3)}{\left(\frac{r}{R}\right)^2}\Bigg)\end{aligned} \tag{D.7}$$

Again, one can see that the dimension only enters as a multiplicative factor. For the sake of completeness, we include a plot for $\frac{\mathcal{SC}_1(||\mathcal{S}_p^r||)}{(D-1)}$ and $\frac{\mathcal{SC}_2(||\mathcal{S}_p^r||)}{(D-1)}$ as a function of $\frac{r}{R}$ in Fig. D.1.

One might be interested in which $\mathcal{SC}_i(||\mathcal{S}_p^r||)$ propagates smaller errors to the estimations of the effective radius $R$ obtained from global fittings or quadratic approximations. This can be done by computing

$$(\frac{\delta R}{R})_i := \delta\mathcal{SC}_i(||\mathcal{S}_p^r||)(R\,\frac{\partial\mathcal{SC}_i(||\mathcal{S}_p^r||)}{\partial R})^{-1}, \tag{D.8}$$

where one can notice that using a quadratic approximation of each $\mathcal{SC}_i$ one gets $(\frac{\delta R}{R})_i < 0.03$ for $r < \frac{3}{4}R$, in both $\mathcal{SC}_i$. Furthermore, one can see that for fixed $\delta\mathcal{SC}_i(||\mathcal{S}_p^r||)$, $(\frac{\delta R}{R})_2 < (\frac{\delta R}{R})_1$, showing that $\mathcal{SC}_2(||\mathcal{S}_p^r||)$ is likely to provide more robust predictions, as was seen in our results.

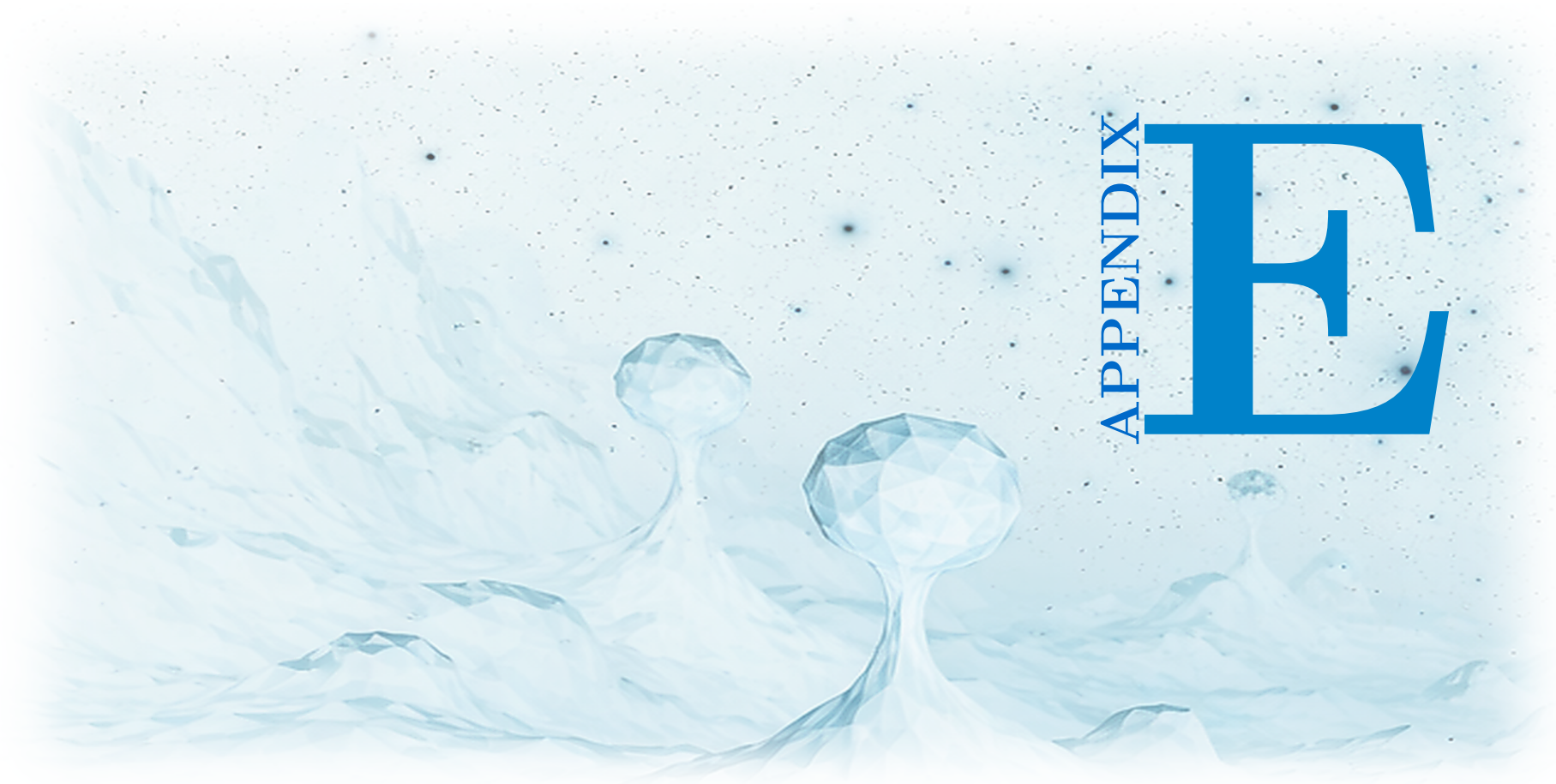

# Boundary conditions and coordinate transformations in scalar-tensor cosmology

As was mentioned in chapter 9, the results obtained in Figs. 9.5, and therefore in 9.6 and 9.7 are subject to a particular choice for the free parameters $m_0^f$ and $m_0^v$. Different choices for these parameters lead to different outcomes, and their effect on the obtained potentials is non-trivial. In Fig. E.1 and E.2, you can observe the result of changing these parameters to different values on the predicted potentials $\mathsf{f}$ and $\mathsf{v}$.

One can see that the effect of changing the values of the parameters $m_0^f$ and $m_0^v$ seems to be mostly an overall multiplication factor, since the shape of these potentials are very similar so each others. We do not have a proof of this conjecture, but it would be interesting to determine of this holds on a more formal setting than just looking at the results for some different values of $m_0^f$ and $m_0^v$.

Not all the analysed values give rise to an effective potential in the Einstein frame that is compatible with observations, although, one can find other combinations apart from the ones showed in this thesis that still work. The resulting effective potentials have different forms though, so their impact on other predictions, like primordial black holes, preheating and reheating can be different.

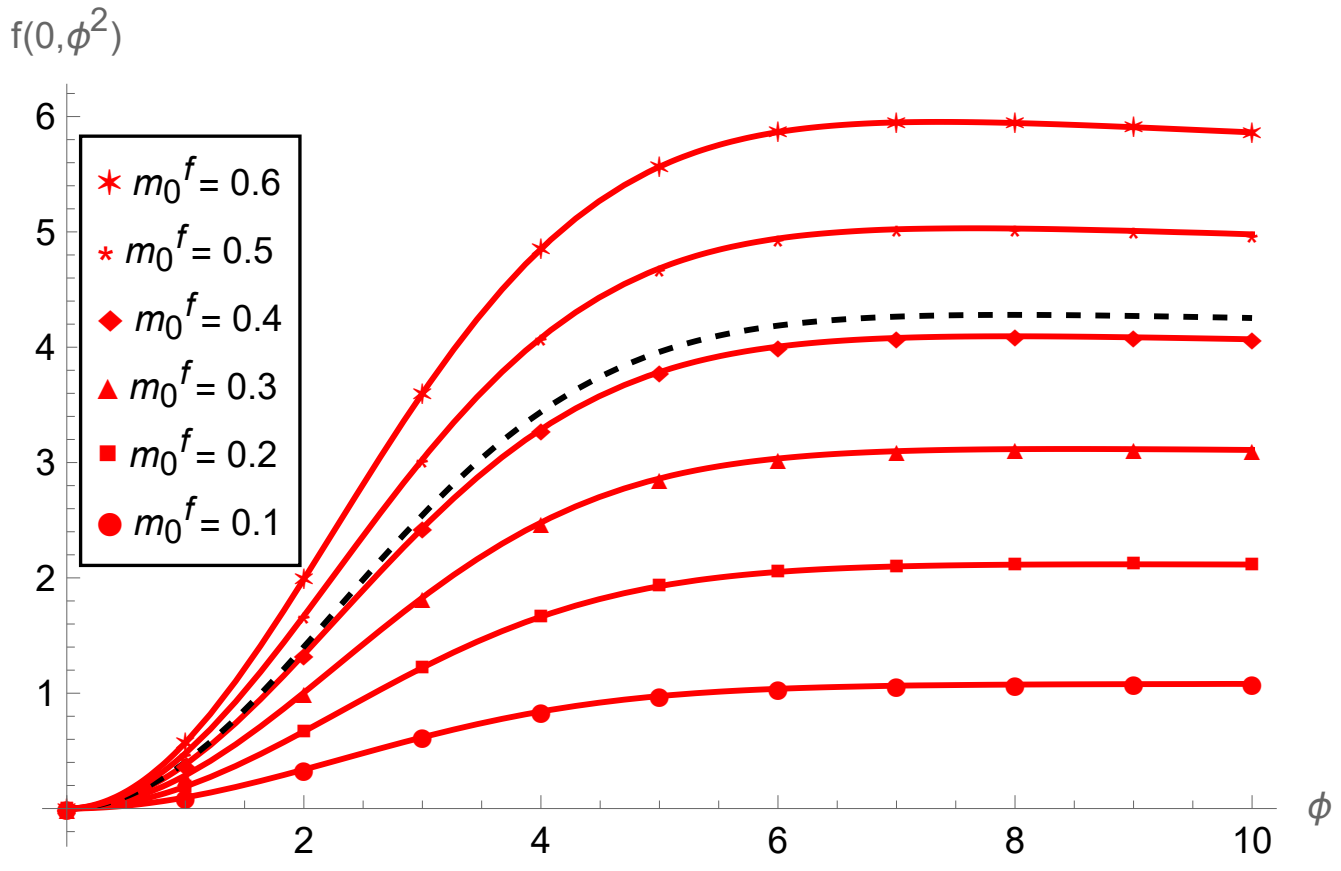

**Figure E.1.:** Effective potential $f$ in the limit of vanishing regulator ($k = 0$) for different choices of the boundary condition parameter $m_0^f$, and fixing $m_0^v = 2.75 \times 10^{-10}$. With dashed lines is the reference potential used in this thesis, corresponding to $m_0^f = 0.42$.

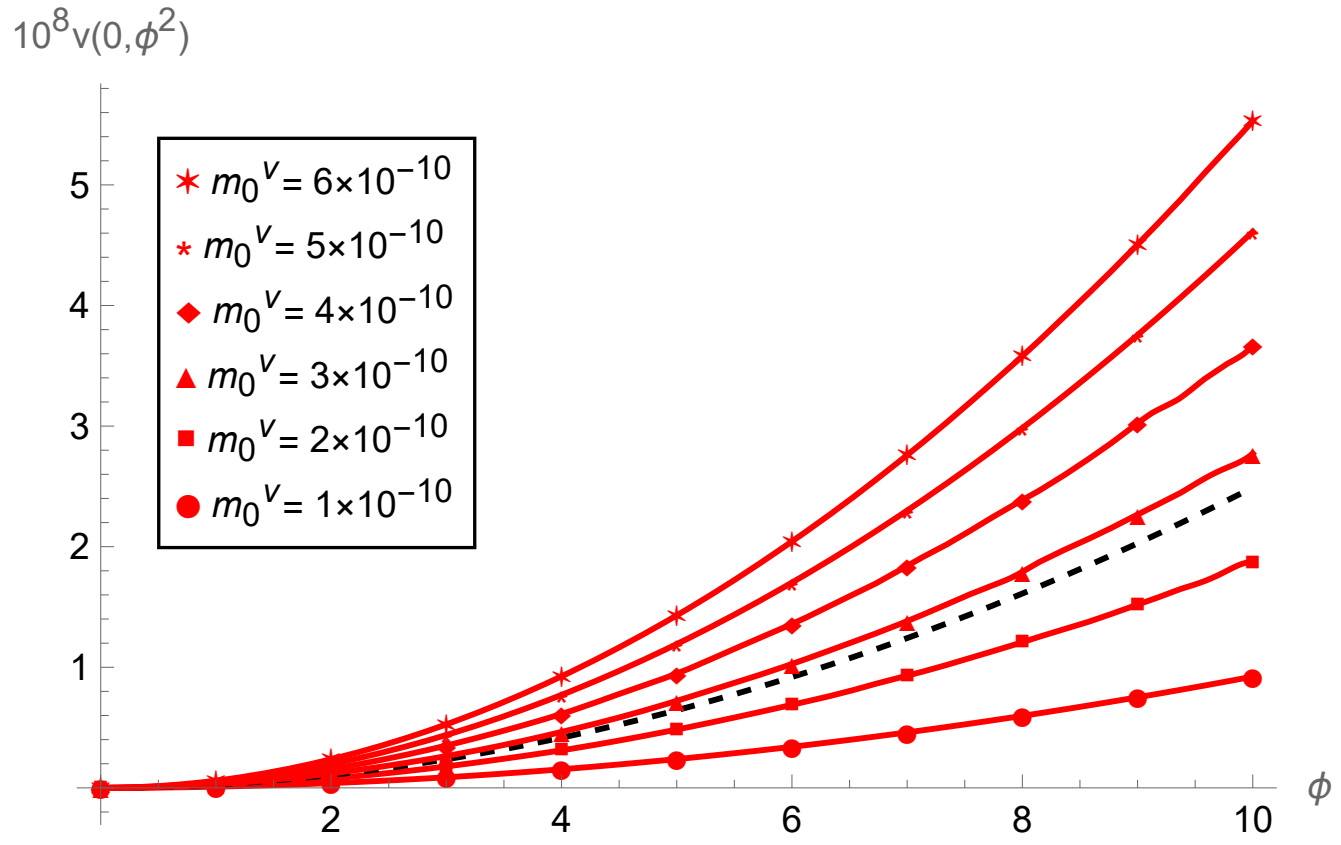

**Figure E.2.:** Effective potential $v$ in the limit of vanishing regulator ($k = 0$) for different choices of the boundary condition parameter $m_0^v$, and fixing $m_0^f = 0.42$. With dashed lines is the reference potential used in this thesis, corresponding to $m_0^v = 2.75 \times 10^{-10}$.

E.0

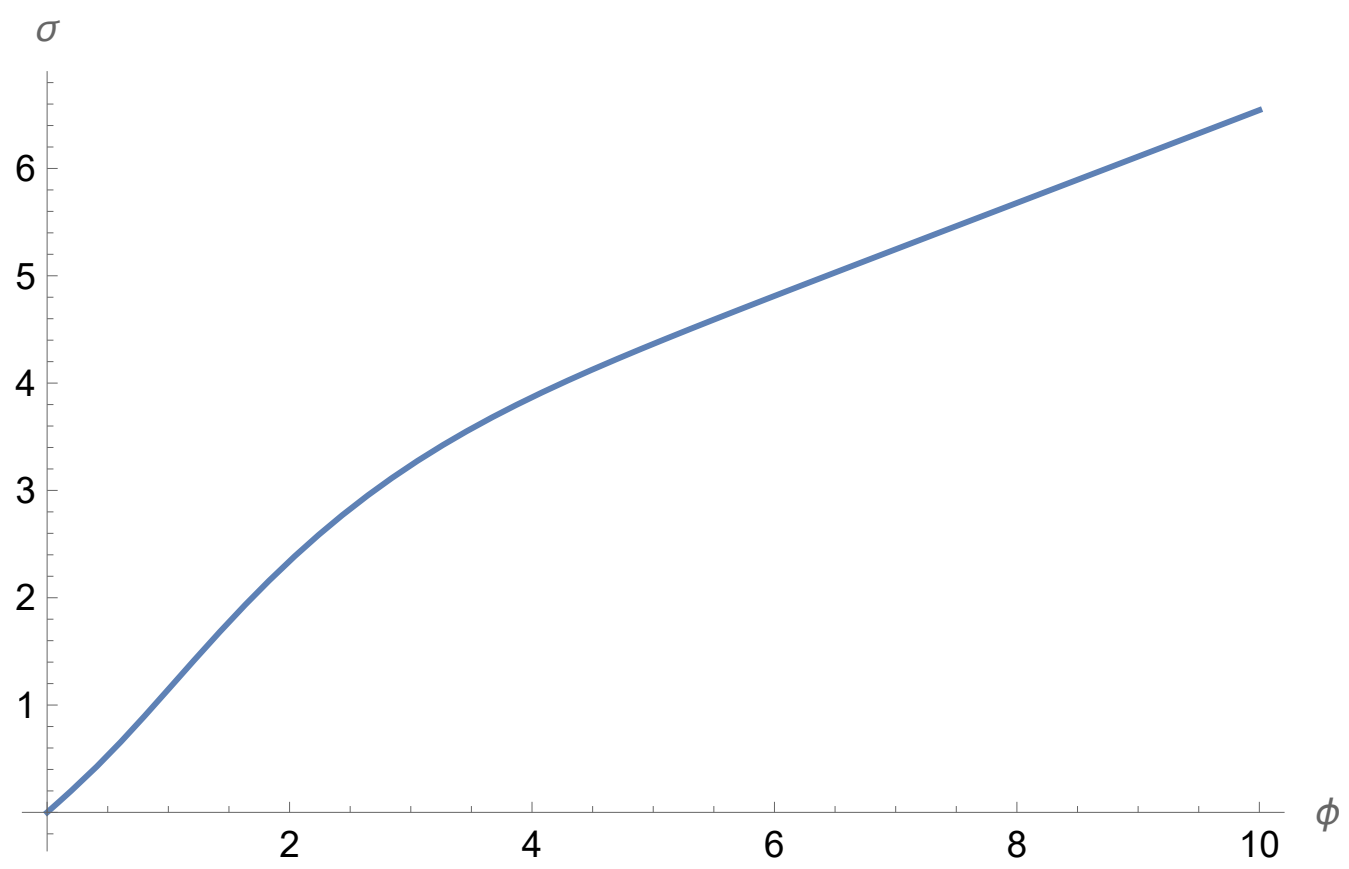

**Figure E.3.:** Numerical result of integrating (E.2), to obtain the functional relation between field variables in the Einstein and Jordan frames at $k = 0$.

## E.1 Weyl variable transformation

The Weyl transformation to go from the Jordan frame to the Einstein frame is

$$g^E_{\mu\nu} \equiv (1 + \mathsf{f}(g, \phi^2))g_{\mu\nu}. \tag{E.1}$$

As shown in the body of the thesis, the relation between the two scalar fields satisfies, for a given value of $k$, the equation

$$(\frac{\partial\sigma}{\partial\phi})^2 = \frac{1}{(1 + \mathsf{f}(g, \phi^2))} + \frac{3\,(\mathsf{f}'(g, \phi^2))^2}{(1 + \mathsf{f}(g, \phi^2))^2}. \tag{E.2}$$

This relation can be solved numerically in our setting, and at $k = 0$ the obtained result can be seen in Fig. E.3. This relation can obviously be mapped into the inflation potential as a function of the Weyl transformed field, which modifies the picture shown in Fig. 9.6. This result can be seen in Fig. E.4.

For slow-roll inflation, this gives the one-to-one correspondence at $k = 0$, of the initial and final values of the inflaton field

$$\phi_i = 1.55 \to \sigma_i \simeq 1.8 \qquad\qquad \phi_f = 0.25 \to \sigma_f \simeq 0.26\,. \tag{E.3}$$

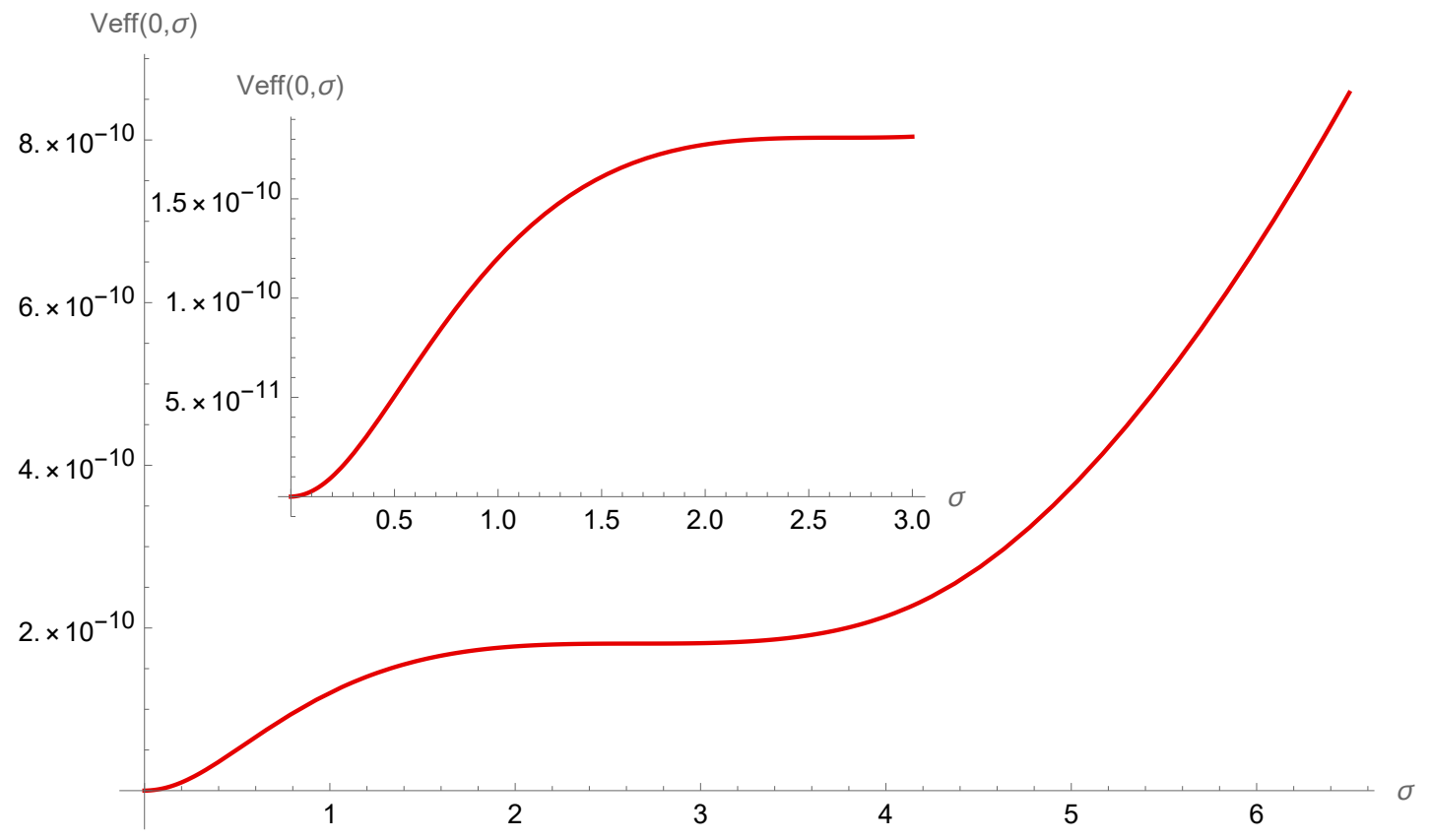

**Figure E.4.:** Effective potential for the inflaton field expressed in terms of the Weyl transformed field variable $\sigma$. The zoomed-in region corresponds to the region of inflationary cosmology in this variable.

| **Einstein Frame** | **Jordan Frame** |
|---|---|
| $\epsilon = \frac{1}{8\pi}\frac{1}{2}\left(\frac{V'_{\text{eff}}(\sigma)}{V_{\text{eff}}(\sigma)}\right)^2$ | $\epsilon = \frac{1}{8\pi}\frac{1}{2}\left(\frac{V'_{\text{eff}}(\phi)}{V_{\text{eff}}(\phi)}\right)^2\left(\frac{\partial\sigma}{\partial\phi}\right)^{-2}$ |
| $\eta = \frac{1}{8\pi}\left(\frac{V''_{\text{eff}}(\sigma)}{V_{\text{eff}}(\sigma)}\right)$ | $\eta = \frac{1}{8\pi}\left[\frac{V''_{\text{eff}}(\phi)}{V_{\text{eff}}(\phi)} - \frac{V'_{\text{eff}}(\phi)}{V_{\text{eff}}(\phi)}\frac{\frac{\partial^2\sigma}{\partial\phi^2}}{\frac{\partial\sigma}{\partial\phi}}\right]\left(\frac{\partial\sigma}{\partial\phi}\right)^{-2}$ |
| $n_s = 1 - 6\epsilon + 2\eta$ | $n_s = 1 - 6\epsilon + 2\eta$ |
| $r = 16\epsilon$ | $r = 16\epsilon$ |
| $A_s = \frac{1}{4}\frac{V_{\text{eff}}}{24\pi^2\epsilon}$ | $A_s = \frac{1}{4}\frac{V_{\text{eff}}}{24\pi^2\epsilon}$ |
| $N_{ef} = 8\pi\int_{\sigma_i}^{\sigma_f}\frac{d\sigma}{\sqrt{2\epsilon}}$ | $N_{ef} = 8\pi\int_{\phi_i}^{\phi_f}\frac{d\phi}{\sqrt{2\epsilon}}\frac{\partial\sigma}{\partial\phi}$ |

**Table E.1.:** Expression of slow-roll inflationary observables in both frames.

The relation between the slow-roll observables of inflation in these two frames can be seen in table E.1.

E.1

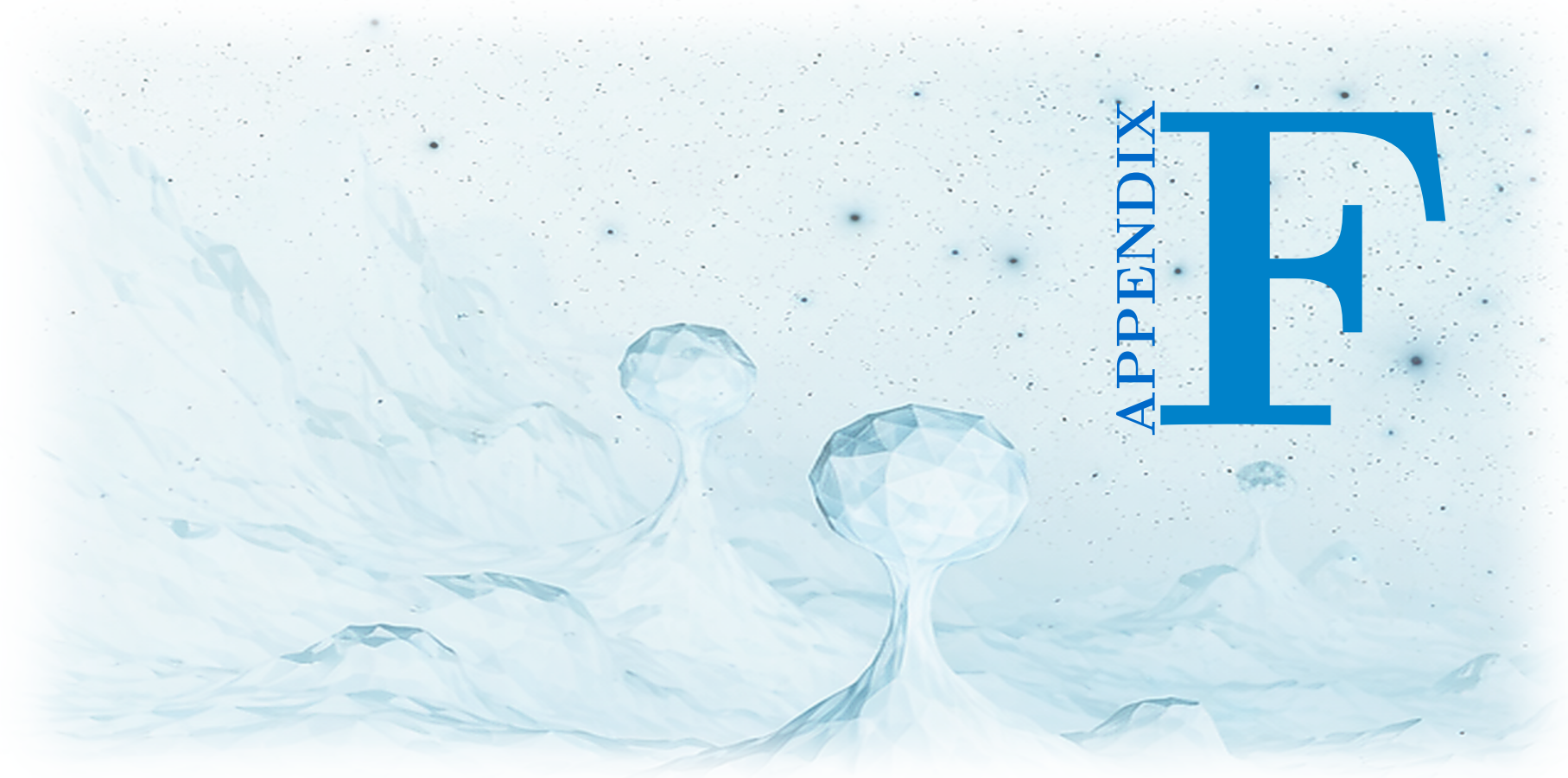

APPENDIX F

# Physical lengths and gravitational Wilson loops in affine gravity

### F.0.1 Physical lengths

While the constructed hypercubic lattice has coordinate length $\varepsilon$, the physical length of an elementary link of the lattice $l = (x,\, x + \varepsilon\hat{\mu})$ with extremes $x$ and $x + \varepsilon\hat{\mu}$ is, in the purely affine formulation, supposed to be given by the emergent metric constructed from the symmetric part of the Ricci tensor as discussed in chapter. 11. Whenever the algebraic relation between the metric and the symmetric Ricci tensor is well defined and yields an invertible tensor with the desired signature, one can assign a physical length and volume to the lattice elements. For example, the square length of an elementary link of the lattice $l = (x,\, x + \varepsilon\hat{\mu})$ can be defined as (no sum over $\mu$)

$$L^2_{phys}(l = (x,\, x + \varepsilon\hat{\mu})) = \varepsilon^2 \left(\frac{D-2}{2\Lambda}\right) \frac{1}{2} \left(R^{lat}_{(\mu\mu)}(x) + R^{lat}_{(\mu\mu)}(x + \varepsilon\hat{\mu})\right) , \tag{F.1}$$

where a mid-point approximation was used. From this length, one could construct geodesic distances, and from it, covariant correlators, effective scaling dimensions, and many other geometrical properties.

If the metric is well defined, then one could also, for example, compute the total physical volume of the lattice using this induced metric,

$$V_{phys} = |\frac{D-2}{2\Lambda}|^{\frac{D}{2}} \sum_x \varepsilon^D \sqrt{|\det R^{\rm lat}_{(\mu\nu)}(x)|}\,. \tag{F.2}$$

In particular, a Renormalization Group step in this setting could be done by increasing the lattice size and decreasing the lattice spacing, while simultaneously keeping fixed the physical volume $V_{phys}$, or other physical lengths. This was not explored in this work, but must be done in future work to explore the continuum limit of this approach.

### F.0.2 Ricci curvature from Wilson loops

For completeness, we now describe a more geometric discretization of the curvature, based on parallel transport around plaquettes (gravitational Wilson loops [56]). This construction is closer in spirit to lattice gauge theory and provides an approximation to the continuum curvature that is manifestly built from holonomies. Its main drawback is that it requires repeated matrix exponentiations or accurate approximations thereof.

Consider an elementary square of the lattice in the domain of a chart of $M$, of side $\varepsilon$ in the $(\mu,\nu)$ plane, based at site $x$.

On this elementary square, or "plaquette" from now on, we will consider the parallel transport operator around a counterclockwise loop, as depicted in Fig. F.1. Let $U_\mu(x)$ be the operator that transports vectors along $x \to x + \varepsilon\hat{\mu}$:

$$V^\alpha(x+\varepsilon\hat{\mu}) \;=\; \left(U_\mu(x)\right)^\alpha{}_\beta\, V^\beta(x), \qquad U_\mu(x) = \mathcal{P}\exp\left(-\int_0^\varepsilon ds\, \bar{\Gamma}_\mu(x+s\hat{\mu})\right). \tag{F.3}$$

where $\Gamma^\alpha{}_{\beta\mu}(x)$ is the affine connection and we defined the matrix $(\bar{\Gamma}_\mu(x))^\alpha{}_\beta \equiv \Gamma^\alpha{}_{\beta\mu}(x)$ be the connection matrix. To set an example of how one gets curvature from these variables, expanding to $\mathcal{O}(\varepsilon^2)$, one gets for the first two

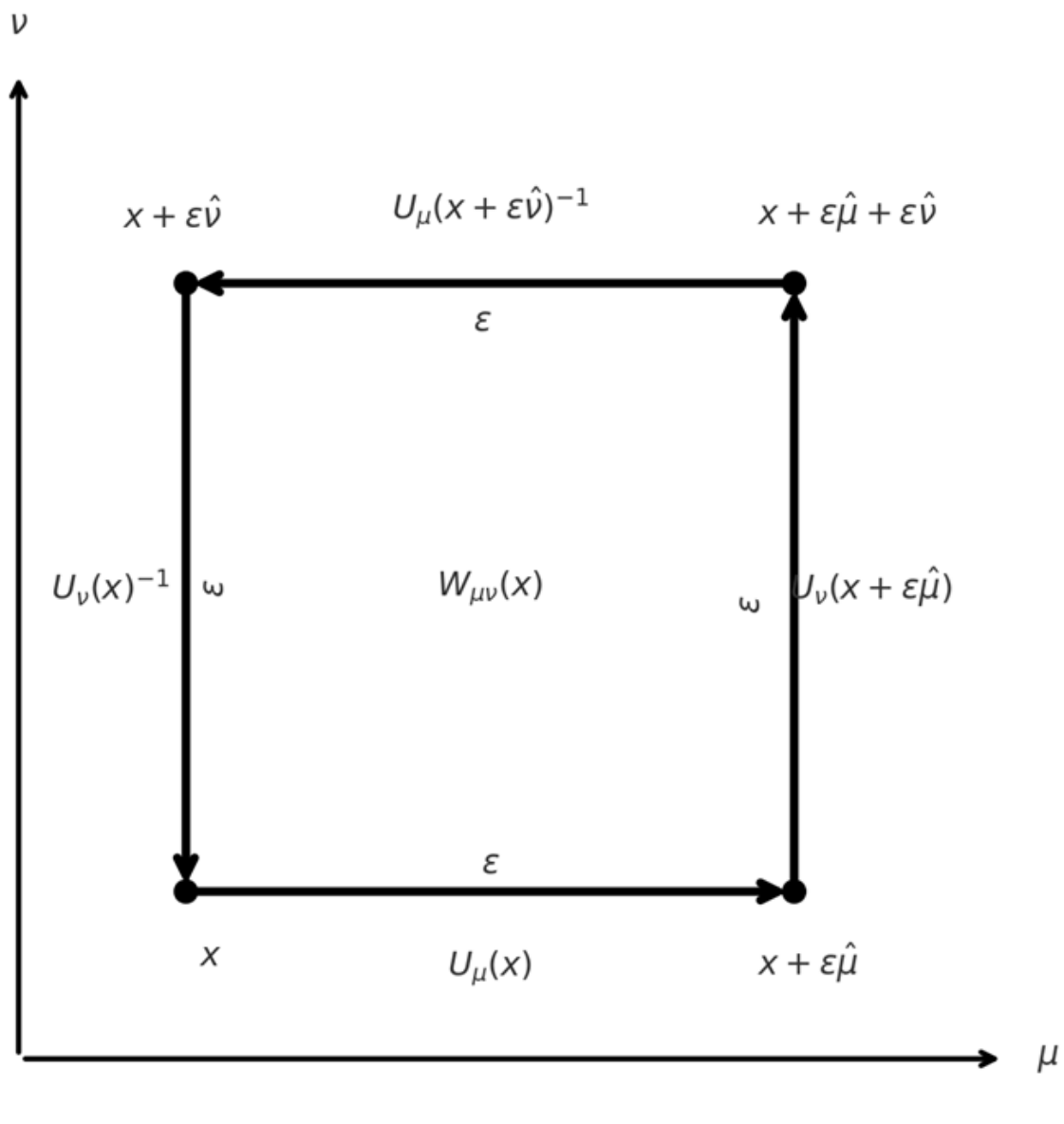

**Figure F.1.:** Wilson loop on a square plaquette with edge length $\varepsilon$.

elements of the loop (here $\mu$ and $\nu$ are not summed over)

$$(U_\mu(x))^\lambda{}_\rho = \delta^\lambda{}_\rho - \varepsilon\,\Gamma^\lambda{}_{\rho\mu}(x) + \frac{\varepsilon^2}{2}\Big(\partial_\mu\Gamma^\lambda{}_{\rho\mu}(x) + \Gamma^\lambda{}_{\kappa\mu}(x)\,\Gamma^\kappa{}_{\rho\mu}(x)\Big) + \mathcal{O}(\varepsilon^3), \tag{F.4}$$

$$(U_\nu(x+\varepsilon\hat{\mu}))^\lambda{}_\gamma = \delta^\lambda{}_\gamma - \varepsilon\Big(\Gamma^\lambda{}_{\gamma\nu}(x) + \varepsilon\,\partial_\mu\Gamma^\lambda{}_{\gamma\nu}(x)\Big) + \frac{\varepsilon^2}{2}\Big(\partial_\nu\Gamma^\lambda{}_{\gamma\nu}(x) + \Gamma^\lambda{}_{\beta\nu}\Gamma^\beta{}_{\gamma\nu}\Big) + \mathcal{O}(\varepsilon^3), \tag{F.5}$$

and similarly for the inverse matrices on the paths on the top and left edges of the loop in Fig. F.1

$$(U_\mu^{-1})^\lambda{}_\rho = \delta^\lambda{}_\rho + \varepsilon\,\Gamma^\lambda{}_{\rho\mu} + \frac{\varepsilon^2}{2}\Big(\Gamma^\lambda{}_{\kappa\mu}\,\Gamma^\kappa{}_{\rho\mu} - \partial_\mu\Gamma^\lambda{}_{\rho\mu}\Big) + \mathcal{O}(\varepsilon^3)\,. \tag{F.6}$$

From the oriented plaquette loop described in Fig. F.1, we define the gravitational Wilson loop operator (here $\mu$ and $\nu$ are not summed over)

$$(W_{\mu\nu}(x))^{\alpha}{}_{\beta} \;=\; (U_\mu(x))^{\alpha}{}_{\kappa}\,(U_\nu(x+\varepsilon\hat{\mu}))^{\kappa}{}_{\delta}\,(U_\mu^{-1}(x+\varepsilon\hat{\nu}))^{\delta}{}_{\phi}\,(U_\nu^{-1}(x))^{\phi}{}_{\beta}. \tag{F.7}$$

Multiplying the four expanded links and keeping all $\mathcal{O}(\varepsilon^2)$ terms, the $\mathcal{O}(\varepsilon)$ pieces cancel pairwise and one obtains

$$\big(W_{\mu\nu}(x)\big)^{\alpha}{}_{\beta} = \delta^{\alpha}{}_{\beta} \;-\; \varepsilon^2\, R^{\alpha}{}_{\beta\mu\nu}(x) \;+\; \mathcal{O}(\varepsilon^3), \tag{F.8}$$

where the Riemann tensor was identified. Denoting the inverse of the Wilson loop $W_{\mu\nu}(x)$ as $W_{\mu\nu}^{-1}(x) = W_{\nu\mu}(x)$, since it is just the reverse direction loop, we get the expression at the lowest order in $\varepsilon$

$$R^{\alpha}{}_{\beta\mu\nu}(x) \;\simeq\; -\,\frac{1}{2\,\varepsilon^2}\big(\,W_{\mu\nu}(x) - W_{\nu\mu}(x)\,\big)^{\alpha}{}_{\beta} \;+\; \mathcal{O}(\varepsilon) \tag{F.9}$$

Which is the hypercubic lattice approximation of the curvature tensor on a plaquette $(\mu, \nu)$. A more sophisticated approximation could use a clover average of the four plaquettes touching $x$ in the $(\mu, \nu)$ plane, but for now this approximation suffices.

We are now in a situation where we can compute the Ricci tensor from Wilson loops. The previous setup of an hypercubic lattice allows us to reconstruct the Ricci tensor from the Wilson loops in an approximation up to the lowest order in the plaquette size

F.0

$$R_{\rho\nu}(x) \;=\; R^{\alpha}{}_{\rho\alpha\nu} \longrightarrow R^{\text{lat,W}}_{\rho\nu}(x) \;\simeq\; -\frac{1}{2\,\varepsilon^2}\sum_{\alpha}\big(\,W_{\alpha\nu}(x) - W_{\nu\alpha}(x)\,\big)^{\alpha}{}_{\rho}\,. \tag{F.10}$$

Since exponentiation of matrices to construct the link variables is computationally expensive, we will resort to a simple approximation of each link variable. We define ($\mu$ is not summed over)

$$(\tilde{U}_\mu(x))^\lambda{}_\rho \equiv \delta^\lambda{}_\rho - \varepsilon\,\Gamma^\lambda{}_{\rho\mu}(x) + \frac{\varepsilon^2}{2}\Big(\Gamma^\lambda{}_{\kappa\mu}(x)\,\Gamma^\kappa{}_{\rho\mu}(x) + \frac{1}{\varepsilon}\Big(\Gamma^\lambda{}_{\rho\mu}(x+\varepsilon\hat{\mu}) - \Gamma^\lambda{}_{\rho\mu}(x)\Big)\Big), \tag{F.11}$$

$$(\tilde{U}_\mu^{-1}(x))^\lambda{}_\rho \equiv \delta^\lambda{}_\rho + \varepsilon\,\Gamma^\lambda{}_{\rho\mu}(x) + \frac{\varepsilon^2}{2}\Big(\Gamma^\lambda{}_{\kappa\mu}(x)\,\Gamma^\kappa{}_{\rho\mu}(x) - \frac{1}{\varepsilon}\Big(\Gamma^\lambda{}_{\rho\mu}(x+\varepsilon\hat{\mu}) - \Gamma^\lambda{}_{\rho\mu}(x)\Big)\Big). \tag{F.12}$$

which are accurate expressions for the link variables up to order $\mathcal{O}(\varepsilon^3)$. With these variables, we define the approximated Wilson loop

$$(\tilde{W}_{\mu\nu}(x))^\alpha{}_\beta \;\equiv\; (\tilde{U}_\mu(x))^\alpha{}_\kappa\,(\tilde{U}_\nu(x+\varepsilon\hat{\mu}))^\kappa{}_\delta\,(\tilde{U}_\mu^{-1}(x+\varepsilon\hat{\nu}))^\delta{}_\phi\,(\tilde{U}_\nu^{-1}(x))^\phi{}_\beta. \tag{F.13}$$

and the approximated lattice Ricci tensor

$$\tilde{R}^{\mathrm{lat,W}}_{\rho\nu}(x) \;\simeq\; -\frac{1}{2\,\varepsilon^2}\sum_\alpha \big(\,\tilde{W}_{\alpha\nu}(x) - \tilde{W}_{\nu\alpha}(x)\,\big)^\alpha{}_\rho\,. \tag{F.14}$$

In the code provided in this manuscript, this $\tilde{R}^{\mathrm{lat,W}}_{\rho\nu}(x)$ is used as an optional improved replacement of the finite-difference Ricci tensor $R^{\mathrm{lat}}_{\rho\nu}(x)$ in the lattice action and in geometric observables.

F.0

APPENDIX
G

# Research Data Management

The data and code generated in this thesis have been managed in accordance with the Research Data Management (RDM) policy of the Institute for Mathematics, Astrophysics, and Particle Physics (IMAPP).[1] Where possible, published data and the software required to reuse it are made available through public repositories. When this is impractical (e.g., large raw outputs), we retain the raw data on institutional storage and provide sufficient information to regenerate or request it, in line with the IMAPP policy.

- **Chapters 1–2, 6–7, 10, and 12–13:** These chapters are theoretical and conceptual and do not present standalone research datasets.
- **Chapters 3–5 (and Appendices A–C):** The computational work consists of Markov chain Monte Carlo ensembles of triangulations, coarse-grained/Delaunay triangulations, and homology-related structures, and the numerical values underlying the figures and tables.
    - *Simulation software:* Triangulation ensembles were generated using established/open simulation software, such as Python, C++, and Rust. The implementation for the EDT and CDT models used in Chapter 3 is open-source `https://gitlab.com/dynamical-triangulation/dyntri-rs.`. The observables im-

[1] `https://www.ru.nl/sites/default/files/2025-02/imapp_rdm_policy-3.pdf`

plementation in Chapter 4 builds on earlier CDT simulation code developed by J. Brunekreef, further explained in [293], as acknowledged in the thesis.

  - *Raw data retention:* Full raw triangulation samples and large intermediate objects are retained on managed institutional storage with associated run logs (parameter files, random seeds, code versions). Due to their volume, these raw outputs are not mirrored in publicly available sources; they can be made available upon reasonable request, subject to storage/transfer constraints and any third-party licensing conditions.

- **Chapters 8–9 (and Appendix E):** This part produces derived/published numerical outputs (e.g. trained-model predictions, sampled solution data for figures) and log data (training configuration, optimizer settings, seeds, software versions).

  - *Supplementary code and notebooks:* The Mathematica implementation and related supplementary calculations used for the numerical solutions and model variations discussed in Chapter 9 are provided in the public repository `https://github.com/silvaagu/Supplementary-Calculations-EIPFASG`.

  - *Reproducibility:* Numerical solvers and trained models were generated using established/open simulation software, such as Python, and Mathematica. The guidelines and code provided in [6] includes the full set of input parameters, boundary-condition choices, sample points, and trained network details, sufficient to reproduce the reported plots.

- **Chapter 11 (and Appendix F):** The computational work consists of Markov chain Monte Carlo ensembles of affine manifolds, derived observables, and the numerical values underlying the figures and tables.

  - *Simulation software:* The C++ Monte Carlo code implementing the lattice affine-gravity ensembles is publicly available at `https://github.com/silvaagu/LatticeQuantumAffineGravityMonteCarlo`.

  - *Raw data retention:* Stored, per-configuration exports (e.g., full connection fields/Ricci tensors) are large, and therefore, not publicly available; these are retained on institutional storage together with run logs and can be regenerated using the public code and recorded run parameters. Stored data can be made available upon reasonable request, subject to storage/transfer constraints and any third-party licensing conditions.

# Summary

Gravity is a name we physicists associate with a single fundamental interaction of our Universe, but in practice, it plays many different roles. It grounds us on Earth, organizes the motion of planets and stars, shapes galaxies and black holes, and becomes inseparable from the questions we ask about the origin and evolution of the Universe. According to Einstein's theory of General Relativity, gravity is not simply a force acting between objects. It is a manifestation of spacetime itself: matter shapes the geometry of spacetime, while this geometry determines how matter moves.

General Relativity has been extraordinarily successful. Nevertheless, most physicists do not believe that it is the final description of spacetime. It remains a classical theory, while all other known fundamental interactions are described quantum mechanically. At sufficiently short distances and high energies, where quantum fluctuations can no longer be ignored, spacetime itself is expected to participate in the quantum nature of our Universe.

This brings us to one of the most difficult questions in theoretical physics: what does it actually mean to quantize gravity?

A quantum theory of gravity is often expected to solve an ambitious collection of problems, ranging from singularities inside black holes to the origin of the Universe. It is also common to encounter speculative pictures of Planckian spacetime involving microscopic wormholes, topology changes, fundamental discreteness, additional dimensions, or new symmetries. These possibilities may contain part of the answer, but direct experiments at the relevant scales are currently unavailable. We therefore have little observational guidance about which new microscopic ingredients, if any, should be introduced.

The perspective followed in this thesis is deliberately conservative. Before changing the fundamental ingredients of General Relativity or Quantum Field Theory, we ask how far one can proceed using tools that are already known to work elsewhere: path integrals, lattice regularizations, renormalization, and non-perturbative computational methods. The aim is not to provide a single final model resolving every problem associated with Quantum Gravity. Instead, the thesis explores several complementary questions about what a quantum spacetime is, how its properties can be measured, and how its microscopic behaviour may constrain physics at much larger scales. This general philosophy and the basic language used throughout the thesis are introduced in Chapter 1.

This thesis develops along two main directions, followed by a final, more foundational detour. The first uses Causal Dynamical Triangulations to study the geometries appearing in a non-perturbative gravitational path integral. The second uses the Functional Renormalization Group and Asymptotic Safety to study the scale dependence of effective theories and their possible infrared predictions. One viewpoint gives access to quantum geometries configuration by configuration; the other follows effective descriptions energy scale by energy scale. The final part then steps back and asks whether the metric itself must be the fundamental variable that is quantized.

**Quantum geometry and the need for observables.**

In the path-integral formulation of Quantum Mechanics, a system is described by summing over all possible configurations connecting its initial and final states. Applying this idea to gravity means summing over possible geometries, since the metric itself is the dynamical field of General Relativity. Formally, one would like to define a path integral over spacetime metrics and use it to compute expectation values of geometric observables.

This formal expression is difficult to control. The space of possible geometries is highly complicated, the metric contains gauge redundancies associated with coordinate transformations, and the Lorentzian path integral is strongly oscillatory. Moreover, the standard perturbative quantization of General Relativity does not produce a fundamental predictive theory. In the perturbative approach, at progressively higher orders new ultraviolet divergences require additional curvature terms and new independent couplings. This failure of perturbation theory does not necessarily imply that gravity cannot be quantized. It may simply mean that perturbation theory is not the correct tool when spacetime is quantized.

Causal Dynamical Triangulations, introduced in Chapter 2, provides a concrete non-perturbative implementation of the gravitational path integral. Smooth geometries are replaced by simplicial manifolds built by gluing together flat building blocks. The individual simplices are simple, while curvature is encoded in the way they are connected. The lattice is used as a regulator rather than being assumed to represent a fundamental physical discreteness. A continuum theory should ultimately be obtained by removing this regulator near a suitable quantum scale-invariant regime.

A distinctive feature of CDT is that its geometries preserve a causal slicing. This makes it possible to define a geometric analogue of a Wick rotation and map the oscillatory gravitational path integral to a statistical system with real weights. The resulting ensemble can then be explored using Monte Carlo simulations.

This gives us a computable sum over geometries, but it comes with some caveats. Typical configurations are not smooth spacetimes with small perturbations around a classical solution. They are highly fluctuating and may possess fractal properties. Standard tensor calculus is therefore not directly available. There are no preferred coordinates and, in general, no smooth Riemann tensor that can simply be evaluated on every configuration.

The physical content of a quantum theory is carried by its observables. In CDT, these observables must be constructed using the structures that remain available, such as connectivity, distances, volumes, and diffusion processes. At the same time, they must have a meaningful continuum interpretation; otherwise, they would describe only the lattice regulator. The first main contribution of this thesis is the development of such observables for topology, homogeneity, and curvature.

**Topology, homogeneity, and curvature.**

In Chapter 3, we introduce a scale-dependent notion of effective topology. Although the microscopic topology of CDT configurations is fixed, the topology visible at a finite resolution need not be. Thin necks or small structures may become unresolved under coarse-graining, producing effective pinchings or bubble-like regions.

To capture this behaviour, we select approximately uniformly separated vertices and use them to construct a Voronoi decomposition of the triangulation. Its dual defines a coarser Delaunay triangulation. Changing the separation between sampled vertices produces effective geometries at different resolutions. Their topology is characterized through Betti numbers, which count connected components, independent non-contractible loops, and higher-dimensional enclosed regions.

The construction was tested in two-dimensional Lorentzian and Euclidean models of quantum gravity. In CDT, the effective topology remains stable over a broad range of local coarse-graining scales. Changes arise only when the coarse-graining scale becomes comparable to the global size of the compact geometry. We therefore find no evidence for local topology-changing structures in the range studied.

The Euclidean model behaves very differently. Its effective geometries develop local pinchings and a hierarchy of bubble-like regions connected through narrow necks. These structures reproduce the familiar picture of mother and baby universes in two-dimensional Euclidean quantum gravity. Moreover, their volume distribution allows us to recover the known string-susceptibility behaviour of the model. This provided a non-trivial validation that the construction detects genuine universal properties rather than arbitrary features of the coarse-graining procedure.

The next question addressed here, is whether a quantum geometry can be called homogeneous. In a smooth classical spacetime, homogeneity is usually formulated through continuous symmetries. These are not directly available on an irregular triangulation. In Chapter 4, we therefore introduce an operational definition: a scalar quantity is averaged inside geodesic balls centered at different points, and the variation of these averages measures how strongly the geometry depends on the chosen position.

This construction makes clear that homogeneity is not a unique property of a geometry in isolation. It depends on what is measured and at which scale. We applied the measure to the coordination number, the deficit angle, and a local Hausdorff dimension in two-dimensional CDT. The coordination number becomes highly homogeneous, but it is primarily a lattice quantity. The deficit angle, which is related to curvature, retains a larger degree of inhomogeneity. The local Hausdorff dimension, by contrast, is compatible with becoming homogeneous in the continuum limit.

A quantum geometry may therefore be homogeneous with respect to its effective dimension while remaining inhomogeneous with respect to its curvature. This is not an inconsistency. It reflects the physical meaning of homogeneity: two observers can agree on one measured property of the geometry while finding different values of another.

Curvature presents a related but more fundamental difficulty. In General Relativity, it is defined through derivatives of the metric. Generic metric spaces and triangulations do not possess the smooth structure required for this construction. Nevertheless, curvature affects other quantities that remain meaningful, such as the growth of geodesic spheres and the return probability of a diffusion process.

In Chapter 5, we use these properties to construct generalized scalar curvatures. On a smooth manifold, geodesic volumes and diffusion probabilities contain short-distance corrections proportional to the Ricci scalar. Our definitions combine derivatives and integrals in such a way that unknown leading normalization factors are removed, leaving a scale-dependent curvature quantity that can be evaluated without coordinates or tensors.

We tested the constructions on triangulations approximating a flat periodic plane and a two-sphere. After separating short-distance lattice artefacts from large-distance finite-size effects, the observables reproduce the expected flat and positively curved behaviour. We then applied them to the quantum geometries of two-dimensional CDT. In the range of scales where discretization effects are under control, the ensemble-averaged generalized curvatures are compatible with zero.

This does not mean that individual CDT configurations resemble ordinary flat planes. They remain highly irregular. The result means that the specific coarse-grained curvature observables introduced here have expectation values compatible with ground-state flatness. More importantly, these quantities can be used to construct curvature correlation functions, providing a possible way to study how classical curvature and its correlations emerge in higher-dimensional quantum spacetime.

Together, these results show how familiar geometric questions can be reformulated when coordinates and smooth tensors are unavailable. The two-dimensional models act as controlled testing grounds, but the principal target is four-dimensional CDT. There, the same tools may determine whether stable topology, approximate cosmological homogeneity, and classical curvature emerge from Planckian quantum geometries.

**Renormalization-group flows and predictivity.**

Lattice simulations provide direct information about quantum geometries, but they necessarily probe a limited interval of scales. Computational restrictions impose both a minimum lattice spacing and a finite total volume. The second part of this thesis therefore asks a complementary question:

how can a microscopic quantum theory constrain effective physics in the infrared?

The basic framework is introduced in Chapter 7. The Renormalization Group describes how a physical theory changes when it is examined at different energy scales. In the Functional Renormalization Group, this information is encoded in a scale-dependent Effective Average Action. At high momentum scales, only short-wavelength fluctuations have been incorporated. As the scale is lowered, longer-wavelength fluctuations are gradually included, until the full quantum effective action is reached.

Asymptotic Safety proposes that gravity may remain well defined at arbitrarily high energies because its renormalization-group flow approaches a non-Gaussian fixed point. At such a point, the dimensionless couplings become scale independent. The set of trajectories reaching the fixed point defines its ultraviolet critical surface.

This surface is central to the predictive power of the theory. Its relevant directions correspond to parameters that must be fixed by observations. Irrelevant couplings are instead determined by the requirement that the theory originates from the ultraviolet fixed point. If only finitely many directions are relevant, a finite number of measurements determines an otherwise infinite collection of interactions.

The conventional way to explore the ultraviolet critical surface is to integrate many individual trajectories from the neighborhood of the fixed point towards the infrared. This becomes inefficient as the dimension of theory space grows. In Chapter 8, we introduce flow invariants as a global alternative.

A flow invariant is a function of the couplings that remains constant along renormalization-group trajectories. It satisfies a linear partial differential equation determined by the beta functions. With suitable conditions near the fixed point, the common zero set of several invariants reconstructs the ultraviolet critical surface. Instead of sampling one trajectory after another, the method aims to encode the entire predictive surface in a small collection of functions.

We first validated the construction in simple dynamical systems and in a three-dimensional scalar-field model. To extend it to more complicated theory spaces, we developed the Flow Invariant Neural Network, or FINN. This is a physics-informed neural network trained directly from the flow-invariant differential equation and the known fixed-point behaviour, rather than from a large precomputed dataset.

The method was tested in a four-dimensional scalar-tensor model. In this case, one can find an ultraviolet fixed point which has four relevant directions, while two additional couplings are predicted by ultraviolet consistency. The critical surface reconstructed with the FINNs agrees with conventional shooting calculations at approximately the percent level in the region studied.

This result is a proof of concept rather than a final phenomenological prediction. The truncation is limited, and the existence of a complete set of regular flow invariants is not guaranteed in arbitrary systems. Nevertheless, it provides a potentially scalable method for determining the infrared predictions of ultraviolet-complete theories when direct trajectory-by-trajectory exploration becomes impractical.

**An inflaton potential emerging from the ultraviolet.**

The final chapter applies the renormalization-group viewpoint to early-Universe cosmology. Inflation is a period of accelerated expansion proposed to explain the large-scale uniformity of the Universe and the origin of primordial density fluctuations. In many models, it is driven by a scalar field whose potential is chosen to produce successful cosmological evolution.

In Chapter 9, we ask whether the inflaton potential can instead emerge from the renormalization-group flow of an asymptotically safe scalar–tensor theory, rather than being chosen to satisfy observations. Instead of approximating the scalar interactions with a short polynomial, we retain the full field dependence of the scalar potential and the non-minimal coupling between the scalar field and the scalar curvature.

The ultraviolet starting point is a shift-symmetric non-Gaussian fixed point, where the scalar potential is flat. As the theory flows towards the infrared, this symmetry is broken dynamically, and non-trivial scalar interactions emerge. The functional flow equations require more information than the usual counting of relevant fixed-point directions suggests. After imposing ultraviolet regularity, the solutions remain labelled by two mass-like parameters.

For suitable values of these parameters, the infrared theory develops an Einstein-frame scalar potential with a broad, approximately flat region. The resulting slow-roll evolution produces about sixty-six e-folds of inflation and values of the scalar spectral index, tensor-to-scalar ratio, and fluctuation amplitude compatible with observational constraints.

The important point is not merely that the potential supports inflation. Many potentials do. Here, its functional form is generated by the flow from an ultraviolet fixed point instead of being postulated from the beginning. The evolution also suggests a possible mechanism for preparing the initial state: a close-to-scale-invariant beginning of the universe that develops a scale-dependent minimum of the potential, may guide the scalar field towards the region from which inflation begins.

This interpretation remains conditional. It assumes a relation between renormalization-group scale evolution and cosmological time evolution, and the calculation relies on a particular truncation, regulator, gauge, background approximation, and Euclidean formulation. Testing the stability of the result under improvements of these ingredients is therefore essential.

**What should be quantized in gravity?**

The final part of this thesis returns to a question that precedes the choice of any particular non-perturbative method: what should be regarded as fundamental in gravity? Both CDT and the Functional Renormalization Group usually begin with the metric. We instead explore a purely affine formulation, classically equivalent to General Relativity under suitable conditions, in which the affine connection is fundamental and the metric, when it exists, is reconstructed from it. A particularly appealing feature is that, for a positive cosmological constant, the resulting Eddington-type action has a definite sign and can be used to define a statistical lattice ensemble.

We first constructed and simulated this ensemble in two dimensions as a proof of principle. Although no non-degenerate metric can be reconstructed there, the model provides a concrete and numerically well-behaved non-perturbative path integral for affine connections. We then moved to a four-dimensional homogeneous and isotropic reduction, where a positive-definite emergent metric can be required configuration by configuration. A fixed de Sitter background connection regulates flat directions while its strength is progressively decreased. The reconstructed spatial-volume profiles display an encouraging de Sitter-like shape across several lattice sizes and regulator masses. Curvature correlations defined through physical geodesic distance illustrate how this framework can combine a statistical ensemble of quantum geometries with the smooth tensor calculus unavailable on generic triangulations. These results remain preliminary, but they open a concrete route towards unrestricted four-dimensional affine simulations and a broader programme of intrinsic geometric observables.

**Bringing the facets together.**

The perspectives studied in this thesis enter Quantum Gravity from different directions. CDT begins with a regulated gravitational path integral and produces individual fluctuating geometries. The Functional Renormalization Group begins with effective actions and follows their dependence on scale. Purely affine lattice gravity returns to the prior question of which field should be quantized and explores a path integral in which geometry emerges from a connection.

Neither approach currently provides a complete theory of Quantum Gravity. CDT still requires stronger control over its four-dimensional continuum limit and the relation between its statistical construction and Lorentzian quantum physics. Functional Renormalization Group calculations rely on truncations and other mathematical choices whose robustness must be checked. A direct relation between the microscopic geometries of the lattice path integral and the effective couplings of the continuum renormalization-group description also remains to be established.

Their complementarity is nevertheless clear. A microscopic construction without suitable observables cannot tell us what its quantum spacetime looks like. An ultraviolet fixed point without control over its outgoing trajectories cannot tell us which low-energy world it predicts. Progress requires both well-defined non-perturbative theories and reliable methods for extracting their physical content.

The main contributions of this thesis are therefore tools and concrete constructions. In the lattice setting, we developed scale-dependent observables for effective topology, homogeneity, and generalized scalar curvature. In the renormalization-group setting, we introduced flow invariants and their neural-network implementation as a method to reconstruct predictive ultraviolet critical surfaces, and showed how a viable inflaton potential may emerge from ultraviolet consistency. Finally, we constructed a non-perturbative lattice formulation of purely affine gravity and began exploring the emergence of four-dimensional geometry within it.

The open questions and possible continuations are discussed in Chapter 13. The geometric observables should be applied to four-dimensional CDT, while flow invariants and FINNs should be tested in larger gravity–matter systems. The inflationary scenario must be examined under changes of truncation and other approximations. In the affine setting, the regulator-removal and scaling analyses must be completed before the homogeneous and isotropic restriction can be relaxed towards a full four-dimensional ensemble.

This thesis began with a curiosity about what spacetime becomes when its quantum nature can no longer be ignored. It does not end with one final microscopic picture or with a solution to every problem commonly assigned to Quantum Gravity. Instead, it provides more concrete questions and a set of instruments with which to approach them. How can topology be measured at finite resolution? What does it mean for a fluctuating geometry to be homogeneous? Can curvature be defined without coordinates or tensors? How can the predictions of an ultraviolet fixed point be extracted efficiently? Can cosmological interactions emerge from quantum-gravitational consistency conditions? And should the metric itself be fundamental, or can geometry emerge from a quantum connection?

The story remains unfinished, as was to be expected. However, in a field where even formulating the correct questions can be difficult, constructing reliable observables and bridges between scales already represents substantive progress. There are still many facets of gravity left to explore, from Planckian geometries to cosmological scales.

# Samenvatting

Zwaartekracht is een naam die wij natuurkundigen associëren met één fundamentele wisselwerking in ons heelal, maar in de praktijk vervult zij veel verschillende rollen. Zij houdt ons op aarde, organiseert de beweging van planeten en sterren, geeft vorm aan sterrenstelsels en zwarte gaten, en is onlosmakelijk verbonden met de vragen die wij stellen over de oorsprong en evolutie van het heelal. Volgens Einsteins algemene relativiteitstheorie is zwaartekracht niet eenvoudigweg een kracht die tussen objecten werkt. Zij is een manifestatie van de ruimtetijd zelf: materie vormt de geometrie van de ruimtetijd, terwijl die geometrie bepaalt hoe materie beweegt.

De algemene relativiteitstheorie is buitengewoon succesvol gebleken. Toch geloven de meeste natuurkundigen niet dat zij de uiteindelijke beschrijving van de ruimtetijd is. Het blijft een klassieke theorie, terwijl alle andere bekende fundamentele wisselwerkingen kwantummechanisch worden beschreven. Op voldoende korte afstanden en bij voldoende hoge energieën, waar kwantumfluctuaties niet langer kunnen worden verwaarloosd, wordt verwacht dat de ruimtetijd zelf deelneemt aan de kwantumaard van ons heelal.

Dit brengt ons bij een van de moeilijkste vragen in de theoretische natuurkunde: wat betekent het eigenlijk om zwaartekracht te kwantiseren?

Van een kwantumtheorie van de zwaartekracht wordt vaak verwacht dat zij een ambitieuze verzameling problemen oplost, van singulariteiten binnen zwarte gaten tot de oorsprong van het heelal. Ook komen speculatieve beelden van ruimtetijd op de Planckschaal vaak voor, met microscopische wormgaten, veranderingen van topologie, fundamentele discretisatie, extra dimensies of nieuwe symmetrieën. Deze mogelijkheden kunnen een deel van het antwoord bevatten, maar directe experimenten op de relevante schalen zijn momenteel niet beschikbaar. We hebben daarom weinig observationele aanwijzingen voor de vraag welke nieuwe microscopische ingrediënten, als die al nodig zijn, moeten worden ingevoerd.

Het perspectief dat in dit proefschrift wordt gevolgd, is bewust behoudend. Voordat we de fundamentele ingrediënten van de algemene relativiteitstheorie of de kwantumveldentheorie veranderen, vragen we hoe ver we kunnen komen met hulpmiddelen waarvan elders al bekend is dat ze werken: padintegralen, roosterregularisaties, renormalisatie en niet-perturbatieve computationele methoden. Het doel is niet om één definitief model te geven dat elk probleem rond kwantumzwaartekracht oplost. In plaats daarvan onderzoekt dit proefschrift verschillende complementaire vragen over wat een kwantumruimtetijd is, hoe haar eigenschappen kunnen worden gemeten en hoe haar microscopische gedrag de fysica op veel grotere schalen kan beperken. Deze algemene filosofie en de basistaal die in het hele proefschrift wordt gebruikt, worden ingevoerd in Hoofdstuk 1.

Dit proefschrift ontwikkelt zich langs twee hoofdrichtingen, gevolgd door een laatste, meer fundamentele omweg. De eerste gebruikt Causal Dynamical Triangulations om de geometrieën te bestuderen die voorkomen in een niet-perturbatieve gravitatiepadintegraal. De tweede gebruikt de Functional Renormalization Group en Asymptotic Safety om de schaalafhankelijkheid van effectieve theorieën en hun mogelijke infraroodvoorspellingen te onderzoeken. Het ene gezichtspunt geeft toegang tot kwantumgeometrieën, configuratie voor configuratie; het andere volgt effectieve beschrijvingen, energieschaal voor energieschaal. Het laatste deel doet vervolgens een stap terug en vraagt of de metriek zelf noodzakelijkerwijs de fundamentele variabele is die moet worden gekwantiseerd.

**Kwantumgeometrie en de behoefte aan observabelen.**

In de padintegraalformulering van de kwantummechanica wordt een systeem beschreven door over alle mogelijke configuraties te sommeren die zijn begin- en eindtoestand verbinden. Als we dit idee op zwaartekracht toepassen, moeten we over mogelijke geometrieën sommeren, aangezien de metriek zelf het dynamische veld van de algemene relativiteitstheorie is. Formeel zouden we een padintegraal over ruimtetijdmetrieken willen definiëren en die gebruiken om verwachtingswaarden van geometrische observabelen te berekenen.

Deze formele uitdrukking is moeilijk onder controle te krijgen. De ruimte van mogelijke geometrieën is uiterst ingewikkeld, de metriek bevat gauge-redundanties die samenhangen met coördinatentransformaties, en de Lorentziaanse padintegraal is sterk oscillerend. Bovendien levert de gebruikelijke perturbatieve kwantisering van de algemene relativiteitstheorie geen fundamentele voorspellende theorie op. In de perturbatieve benadering vereisen nieuwe ultraviolette divergenties bij steeds hogere ordes extra krommingstermen en nieuwe onafhankelijke koppelingen. Dit falen van de perturbatietheorie betekent niet noodzakelijk dat zwaartekracht niet kan worden gekwantiseerd. Het kan eenvoudig betekenen dat perturbatietheorie niet het juiste hulpmiddel is wanneer de ruimtetijd zelf wordt gekwantiseerd.

Causal Dynamical Triangulations, ingevoerd in Hoofdstuk 2, biedt een concrete niet-perturbatieve implementatie van de gravitatiepadintegraal. Gladde geometrieën worden vervangen door simpliciale variëteiten die zijn opgebouwd door vlakke bouwstenen aan elkaar te lijmen. De afzonderlijke simplices zijn eenvoudig, terwijl de kromming is gecodeerd in de manier waarop zij met elkaar zijn verbonden. Het rooster wordt als regulator gebruikt en niet noodzakelijk geïnterpreteerd als een fundamentele fysieke discretisatie. Uiteindelijk moet een continuümtheorie worden verkregen door deze regulator te verwijderen in de buurt van een geschikt kwantumregime met schaalinvariantie.

Een onderscheidend kenmerk van CDT is dat de geometrieën een causale foliatie behouden. Hierdoor kan een geometrisch analogon van een Wick-rotatie worden gedefinieerd en kan de oscillerende gravitatiepadintegraal worden afgebeeld op een statistisch systeem met reële gewichten. Het resulterende ensemble kan vervolgens worden onderzocht met Monte-Carlosimulaties.

Dit geeft ons een berekenbare som over geometrieën, maar brengt ook enkele voorbehouden met zich mee. Typische configuraties zijn geen gladde ruimtetijden met kleine verstoringen rond een klassieke oplossing. Zij fluctueren sterk en kunnen fractale eigenschappen bezitten. De gebruikelijke tensorrekening is daarom niet rechtstreeks beschikbaar. Er zijn geen voorkeurscoördinaten en in het algemeen is er geen gladde Riemann-tensor die eenvoudig op iedere configuratie kan worden geëvalueerd.

De fysieke inhoud van een kwantumtheorie wordt gedragen door haar observabelen. In CDT moeten deze observabelen worden geconstrueerd uit de structuren die wel beschikbaar blijven, zoals connectiviteit, afstanden, volumes en diffusieprocessen. Tegelijkertijd moeten zij een betekenisvolle continuüminterpretatie hebben; anders zouden zij alleen de roosterregulator beschrijven. De eerste belangrijke bijdrage van dit proefschrift is de ontwikkeling van zulke observabelen voor topologie, homogeniteit en kromming.

**Topologie, homogeniteit en kromming.**

In Hoofdstuk 3 voeren we een schaalafhankelijk begrip van effectieve topologie in. Hoewel de microscopische topologie van CDT-configuraties vastligt, hoeft de topologie die bij een eindige resolutie zichtbaar is dat niet te doen. Dunne halzen of kleine structuren kunnen bij coarse-graining onzichtbaar worden, waardoor effectieve insnoeringen of belvormige gebieden ontstaan.

Om dit gedrag te beschrijven, selecteren we punten die ongeveer gelijkmatig over de triangulatie zijn verdeeld en gebruiken we deze om een Voronoi-ontbinding van de triangulatie te construeren. De duale constructie definieert een grovere Delaunay-triangulatie. Door de afstand tussen de geselecteerde punten te veranderen, verkrijgen we effectieve geometrieën met verschillende resoluties. Hun topologie wordt gekarakteriseerd met Betti-getallen, die samenhangende componenten, onafhankelijke niet-contractibele lussen en ingesloten gebieden van hogere dimensie tellen.

De constructie werd getest in tweedimensionale Lorentziaanse en Euclidische modellen van kwantumzwaartekracht. In CDT blijft de effectieve topologie stabiel over een breed bereik van lokale coarse-graining-schalen. Veranderingen treden pas op wanneer de coarse-graining-schaal vergelijkbaar wordt met de globale afmeting van de compacte geometrie. In het onderzochte bereik vinden we daarom geen aanwijzingen voor lokale structuren die van topologie veranderen.

Het Euclidische model gedraagt zich heel anders. De effectieve geometrieën ontwikkelen lokale insnoeringen en een hiërarchie van belvormige gebieden die door nauwe halzen met elkaar zijn verbonden. Deze structuren reproduceren het bekende beeld van moeder- en baby-universa in tweedimensionale Euclidische kwantumzwaartekracht. Bovendien stelt hun volumeverdeling ons in staat het bekende gedrag van de stringsusceptibiliteit van het model terug te vinden. Dit leverde een niet-triviale bevestiging dat de constructie werkelijke universele eigenschappen detecteert, en niet slechts willekeurige kenmerken van de coarse-graining-procedure.

De volgende vraag die we behandelen, is of een kwantumgeometrie homogeen kan worden genoemd. In een gladde klassieke ruimtetijd wordt homogeniteit gewoonlijk geformuleerd met continue symmetrieën. Zulke symmetrieën zijn niet rechtstreeks beschikbaar op een onregelmatige triangulatie. In Hoofdstuk 4 voeren we daarom een operationele definitie in: een scalaire grootheid wordt gemiddeld binnen geodetische bollen met verschillende middelpunten, en de variatie van deze gemiddelden meet hoe sterk de geometrie afhangt van de gekozen positie.

Deze constructie maakt duidelijk dat homogeniteit geen unieke eigenschap is van een geometrie op zichzelf. Zij hangt af van wat wordt gemeten en op welke schaal. We pasten de maat toe op het coördinatiegetal, de deficithoek en een lokale Hausdorff-dimensie in tweedimensionale CDT. Het coördinatiegetal wordt sterk homogeen, maar is voornamelijk een roostergrootheid. De deficithoek, die met kromming samenhangt, behoudt een grotere mate van inhomogeniteit. De lokale Hausdorff-dimensie is daarentegen verenigbaar met homogeniteit in de continuümlimiet.

Een kwantumgeometrie kan dus homogeen zijn met betrekking tot haar effectieve dimensie en tegelijkertijd inhomogeen blijven met betrekking tot haar kromming. Dit is geen inconsistentie. Het weerspiegelt de fysieke betekenis van homogeniteit: twee waarnemers kunnen het eens zijn over één gemeten eigenschap van de geometrie, terwijl zij verschillende waarden van een andere vinden.

Kromming vormt een verwant, maar fundamenteler probleem. In de algemene relativiteitstheorie wordt zij gedefinieerd met afgeleiden van de metriek. Algemene metrische ruimten en triangulaties bezitten niet de gladde structuur die daarvoor nodig is. Toch beïnvloedt kromming andere grootheden die wel betekenisvol blijven, zoals de groei van geodetische bollen en de terugkeerwaarschijnlijkheid van een diffusieproces.

In Hoofdstuk 5 gebruiken we deze eigenschappen om gegeneraliseerde scalaire krommingen te construeren. Op een gladde variëteit bevatten geodetische volumes en diffusiewaarschijnlijkheden kort-afstandscorrecties die evenredig zijn met de Ricci-scalar. Onze definities combineren afgeleiden en integralen zo dat onbekende leidende normalisatiefactoren worden verwijderd. Wat overblijft, is een schaalafhankelijke krommingsgrootheid die zonder coördinaten of tensoren kan worden geëvalueerd.

We testten de constructies op triangulaties die een vlak periodiek vlak en een tweedimensionale bol benaderen. Nadat roosterartefacten op korte afstand waren gescheiden van eindige-grootte-effecten op grote afstand, reproduceerden de observabelen het verwachte gedrag van vlakke en positief gekromde geometrieën. Vervolgens pasten we ze toe op de kwantumgeometrieën van tweedimensionale CDT. In het schaalbereik waarin discretisatie-effecten onder controle zijn, zijn de over het ensemble gemiddelde gegeneraliseerde krommingen verenigbaar met nul.

Dit betekent niet dat afzonderlijke CDT-configuraties op gewone vlakke vlakken lijken. Zij blijven sterk onregelmatig. Het resultaat betekent dat de specifieke coarse-grained krommingsobservabelen die hier zijn ingevoerd verwachtingswaarden hebben die verenigbaar zijn met vlakheid van de grondtoestand. Belangrijker nog is dat deze grootheden kunnen worden gebruikt om krommingscorrelatiefuncties te construeren. Daarmee ontstaat een mogelijke manier om te onderzoeken hoe klassieke kromming en haar correlaties in hoger-dimensionale kwantumruimtetijd tevoorschijn komen.

Gezamenlijk laten deze resultaten zien hoe vertrouwde geometrische vragen kunnen worden geherformuleerd wanneer coördinaten en gladde tensoren niet beschikbaar zijn. De tweedimensionale modellen dienen als gecontroleerde testomgevingen, maar het belangrijkste doel is vierdimensionale CDT. Daar kunnen dezelfde hulpmiddelen bepalen of stabiele topologie, benaderende kosmologische homogeniteit en klassieke kromming voortkomen uit Planckiaanse kwantumgeometrieën.

**Renormalisatiegroepstromen en voorspellend vermogen.**

Roostersimulaties geven directe informatie over kwantumgeometrieën, maar bestrijken noodzakelijkerwijs slechts een beperkt interval van schalen. Computationele beperkingen leggen zowel een minimale roosterafstand als een eindig totaal volume op. Het tweede deel van dit proefschrift stelt daarom een complementaire vraag: hoe kan een microscopische kwantumtheorie de effectieve fysica in het infrarood beperken?

Het basiskader wordt ingevoerd in Hoofdstuk 7. De renormalisatiegroep beschrijft hoe een fysische theorie verandert wanneer zij op verschillende energieschalen wordt onderzocht. In de Functional Renormalization Group wordt deze informatie gecodeerd in een schaalafhankelijke effectieve gemiddelde actie. Op hoge momentschalen zijn alleen fluctuaties met korte golflengte opgenomen. Wanneer de schaal wordt verlaagd, worden geleidelijk fluctuaties met langere golflengte toegevoegd, totdat de volledige kwantumeffectieve actie wordt bereikt.

Asymptotic Safety stelt voor dat zwaartekracht bij willekeurig hoge energieën goed gedefinieerd kan blijven doordat haar renormalisatiegroepstroom een niet-Gaussiaans vast punt nadert. Op zo'n punt worden de dimensieloze koppelingen schaalonafhankelijk. De verzameling trajecten die het vaste punt bereiken, definieert het ultraviolette kritische oppervlak.

Dit oppervlak staat centraal in het voorspellend vermogen van de theorie. De relevante richtingen komen overeen met parameters die door waarnemingen moeten worden vastgelegd. Irrelevante koppelingen worden daarentegen bepaald door de eis dat de theorie uit het ultraviolette vaste punt voortkomt. Als slechts eindig veel richtingen relevant zijn, bepaalt een eindig aantal metingen een anders oneindige verzameling wisselwerkingen.

De gebruikelijke manier om het ultraviolette kritische oppervlak te onderzoeken, is veel afzonderlijke trajecten vanuit de omgeving van het vaste punt naar het infrarood te integreren. Dit wordt inefficiënt naarmate de dimensie van de theorieruimte toeneemt. In Hoofdstuk 8 voeren we stroominvarianten in als een globaal alternatief.

Een stroominvariant is een functie van de koppelingen die constant blijft langs renormalisatiegroeptrajecten. Zij voldoet aan een lineaire partiële differentiaalvergelijking die door de bètafuncties wordt bepaald. Met geschikte voorwaarden nabij het vaste punt reconstrueert de gemeenschappelijke nulverzameling van meerdere invarianten het ultraviolette kritische oppervlak. In plaats van het ene traject na het andere te bemonsteren, probeert de methode het volledige voorspellende oppervlak in een kleine verzameling functies te coderen.

We valideerden de constructie eerst in eenvoudige dynamische systemen en in een driedimensionaal scalair-veldmodel. Om haar uit te breiden naar ingewikkeldere theorieruimten ontwikkelden we het Flow Invariant Neural Network, of FINN. Dit is een physics-informed neuraal netwerk dat rechtstreeks wordt getraind met de differentiaalvergelijking voor de stroominvariant en het bekende gedrag van het vaste punt, in plaats van met een grote vooraf berekende dataset.

De methode werd getest in een vierdimensionaal scalair–tensorieel model. In dit geval kan een ultraviolet vast punt worden gevonden met vier relevante richtingen, terwijl twee extra koppelingen door ultraviolette consistentie worden voorspeld. Het met FINNs gereconstrueerde kritische oppervlak komt in het onderzochte gebied op ongeveer procentniveau overeen met conventionele shooting-berekeningen.

Dit resultaat is een proof of concept en geen definitieve fenomenologische voorspelling. De truncatie is beperkt en het bestaan van een volledige verzameling regelmatige stroominvarianten is in algemene systemen niet gegarandeerd. Niettemin biedt de methode een potentieel schaalbare manier om de infraroodvoorspellingen van ultraviolet complete theorieën te bepalen wanneer directe verkenning traject voor traject onpraktisch wordt.

**Een inflatonpotentiaal die uit het ultraviolet voortkomt.**

Het laatste hoofdstuk past het renormalisatiegroepperspectief toe op de kosmologie van het vroege heelal. Inflatie is een periode van versnelde expansie die is voorgesteld om de grootschalige uniformiteit van het heelal en de oorsprong van primordiale dichtheidsfluctuaties te verklaren. In veel modellen wordt inflatie aangedreven door een scalair veld waarvan de potentiaal zo wordt gekozen dat zij een succesvolle kosmologische evolutie oplevert.

In Hoofdstuk 9 vragen we of de inflatonpotentiaal in plaats daarvan kan voortkomen uit de renormalisatiegroepstroom van een asymptotisch veilig scalair–tensorieel model, in plaats van te worden gekozen om aan waarnemingen te voldoen. In plaats van de scalaire interacties met een korte polynoom te benaderen, behouden we de volledige veldafhankelijkheid van de scalaire potentiaal en van de niet-minimale koppeling tussen het scalaire veld en de scalaire kromming.

Het ultraviolette beginpunt is een verschuivingssymmetrisch niet Gaussiaans vast punt, waar de scalaire potentiaal vlak is. Wanneer de theorie naar het infrarood stroomt, wordt deze symmetrie dynamisch gebroken en ontstaan niet-triviale scalaire interacties. De functionele stroomvergelijkingen vereisen meer informatie dan de gebruikelijke telling van relevante richtingen van het vaste punt doet vermoeden. Nadat ultraviolette regulariteit is opgelegd, worden de oplossingen nog steeds gelabeld door twee massa-achtige parameters.

Voor geschikte waarden van deze parameters ontwikkelt de infraroodtheorie in het Einstein-frame een scalaire potentiaal met een breed, ongeveer vlak gebied. De resulterende slow-roll-evolutie levert ongeveer zesenzestig e-folds inflatie op, evenals waarden voor de scalaire spectrale index, de tensor-scalarverhouding en de fluctuatieamplitude die verenigbaar zijn met observationele beperkingen.

Het belangrijke punt is niet alleen dat de potentiaal inflatie ondersteunt. Veel potentialen doen dat. Hier wordt haar functionele vorm gegenereerd door de stroom vanaf een ultraviolet vast punt, in plaats van vanaf het begin te worden gepostuleerd. De evolutie suggereert ook een mogelijk mechanisme om de begintoestand voor te bereiden: een begin van het heelal dat bijna schaalinvariant is en een schaalafhankelijk minimum van de potentiaal ontwikkelt, kan het scalaire veld naar het gebied leiden van waaruit inflatie begint.

Deze interpretatie blijft voorwaardelijk. Zij veronderstelt een relatie tussen de evolutie van de renormalisatiegroepschaal en de kosmologische tijdsevolutie, en de berekening berust op een specifieke truncatie, regulator, gaugekeuze, achtergrondbenadering en Euclidische formulering. Het is daarom essentieel om de stabiliteit van het resultaat te testen wanneer deze ingrediënten worden verbeterd.

**Wat moet in de zwaartekracht worden gekwantiseerd?**

Het laatste deel van dit proefschrift keert terug naar een vraag die voorafgaat aan de keuze van een bepaalde niet-perturbatieve methode: wat moet in de zwaartekracht als fundamenteel worden beschouwd? Zowel CDT als de Functional Renormalization Group nemen gewoonlijk de metriek als uitgangspunt. Wij onderzoeken daarentegen een zuiver affiene formulering, die onder geschikte voorwaarden klassiek equivalent is aan de algemene relativiteitstheorie, maar waarin de affiene connectie fundamenteel is en de metriek, wanneer zij bestaat, daaruit wordt gereconstrueerd. Een bijzonder aantrekkelijke eigenschap is dat de resulterende actie van het Eddington-type voor een positieve kosmologische constante een bepaald teken heeft en kan worden gebruikt om een statistisch roosterensemble te definiëren.

We construeerden en simuleerden dit ensemble eerst in twee dimensies als een bewijs van principe. Hoewel daar geen niet-gedegenereerde metriek kan worden gereconstrueerd, levert het model een concrete en numeriek goed gedragen niet-perturbatieve padintegraal voor affiene connecties. Vervolgens bestudeerden we een vierdimensionale homogene en isotrope reductie, waarin voor iedere configuratie een positief-definiete emergente metriek kan worden geëist. Een vaste de Sitter-achtergrondconnectie regulariseert vlakke richtingen, terwijl de sterkte van deze regulator stapsgewijs wordt verlaagd. De gereconstrueerde ruimtelijke volumeprofielen vertonen voor verschillende roostergroottes en regulatormassa's een bemoedigende de Sitter-achtige vorm. Krommingscorrelaties, gedefinieerd via fysieke geodetische afstand, illustreren hoe dit raamwerk een statistisch ensemble van kwantumgeome-trieën kan combineren met gladde tensorrekening. Deze resultaten zijn voorlopig, maar openen een concrete route naar algemenere vierdimensionale affiene simulaties en een breder programma van intrinsieke geometrische observabelen.

**De verschillende facetten samengebracht.**

De perspectieven in dit proefschrift benaderen kwantumzwaartekracht vanuit verschillende richtingen. CDT begint met een geregulariseerde gravitatiepadintegraal en produceert afzonderlijke fluctuerende geometrieën. De Functional Renormalization Group begint met effectieve acties en volgt hun schaalafhankelijkheid. Zuiver affiene roosterzwaartekracht keert terug naar de voorafgaande vraag welk veld moet worden gekwantiseerd en onderzoekt een padintegraal waarin geometrie uit een connectie voortkomt.

Geen van beide benaderingen levert momenteel een volledige theorie van kwantumzwaartekracht. Voor CDT is nog meer controle nodig over de vierdimensionale continuümlimiet en over de relatie tussen de statistische constructie en Lorentziaanse kwantumfysica. Berekeningen met de Functional Renormalization Group berusten op truncaties en andere wiskundige keuzes waarvan de robuustheid moet worden gecontroleerd. Ook moet nog een directe relatie worden vastgesteld tussen de microscopische geometrieën van de roosterpadintegraal en de effectieve koppelingen van de continue renormalisatiegroepbeschrijving.

Hun complementariteit is niettemin duidelijk. Een microscopische constructie zonder geschikte observabelen kan ons niet vertellen hoe haar kwantumruimtetijd eruitziet. Een ultraviolet vast punt zonder controle over de uitgaande trajecten kan ons niet vertellen welke laag-energetische wereld het voorspelt. Vooruitgang vereist zowel goed gedefinieerde niet-perturbatieve theorieën als betrouwbare methoden om hun fysieke inhoud te verkrijgen.

De belangrijkste bijdragen van dit proefschrift zijn daarom hulpmiddelen en concrete constructies. In de roostercontext ontwikkelden we schaalafhankelijke observabelen voor effectieve topologie, homogeniteit en gegeneraliseerde scalaire kromming. In de context van de renormalisatiegroep introduceerden we stroominvarianten en hun neurale netwerkimplementatie als methode om voorspellende ultraviolette kritische oppervlakken te reconstrueren, en lieten we zien hoe een levensvatbare inflatonpotentiaal uit ultraviolette consistentie kan voortkomen. Ten slotte con-strueerden we een niet-perturbatieve roosterformulering van zuiver affiene zwaartekracht en begonnen we daarin het ontstaan van vierdimensionale geometrie te onderzoeken.

De open vragen en mogelijke vervolgstappen worden besproken in Hoofdstuk 13. De geometrische observabelen moeten worden toegepast op vierdimensionale CDT, terwijl stroominvarianten en FINNs in grotere zwaartekracht–materiesystemen moeten worden getest. Het inflatiescenario moet onder veranderingen van truncatie en andere benaderingen worden onderzocht. In de affiene formulering moeten de verwijdering van de regulator en de schaalanalyse worden voltooid voordat de homogene en isotrope beperking kan worden losgelaten op weg naar een volledig vierdimensionaal ensemble.

Dit proefschrift begon met nieuwsgierigheid naar wat ruimtetijd wordt wanneer haar kwantumaard niet langer kan worden genegeerd. Het eindigt niet met één definitief microscopisch beeld of met een oplossing voor elk probleem dat doorgaans aan kwantumzwaartekracht wordt toegeschreven. In plaats daarvan levert het concretere vragen en een verzameling instrumenten om die vragen te benaderen. Hoe kan topologie bij eindige resolutie worden gemeten? Wat betekent het dat een fluctuerende geometrie homogeen is? Kan kromming zonder coördinaten of tensoren worden gedefinieerd? Hoe kunnen de voorspellingen van een ultraviolet vast punt efficiënt worden afgeleid? Kunnen kosmologische interacties voortkomen uit kwantumgravitationele consistentievoorwaarden? En moet de metriek zelf fundamenteel zijn, of kan geometrie uit een kwantumconnectie voortkomen?

Het verhaal blijft onaf, zoals te verwachten was. Maar in een vakgebied waarin zelfs het formuleren van de juiste vragen moeilijk kan zijn, vormt het construeren van betrouwbare observabelen en bruggen tussen schalen al wezenlijke vooruitgang. Er zijn nog veel facetten van zwaartekracht te onderzoeken, van Planckiaanse geometrieën tot het kosmologische heelal.

# Resumen

La gravedad es el nombre que los físicos asociamos con una única interacción fundamental de nuestro Universo, pero en la práctica cumple muchos roles diferentes. Nos mantiene pegados a la Tierra, organiza el movimiento de los planetas y las estrellas, da forma a las galaxias y a los agujeros negros, y se vuelve inseparable de las preguntas que nos hacemos sobre el origen y la evolución del Universo. Según la teoría de la Relatividad General de Einstein, la gravedad no es simplemente una fuerza que actúa entre objetos. Es una manifestación del propio espaciotiempo: la materia moldea la geometría del espaciotiempo, mientras que esta geometría, a su vez, determina cómo se mueve la materia.

La Relatividad General ha sido extraordinariamente exitosa. Sin embargo, la mayoría de los físicos no cree que sea la descripción final del espaciotiempo. Sigue siendo una teoría clásica, mientras que todas las demás interacciones fundamentales conocidas se describen cuánticamente. A distancias suficientemente cortas y energías suficientemente altas, donde las fluctuaciones cuánticas ya no pueden ignorarse, se espera que el propio espaciotiempo participe de la naturaleza cuántica de nuestro Universo.

Esto nos lleva a una de las preguntas más difíciles de la física teórica: ¿qué significa realmente cuantizar la gravedad?

A menudo se espera que una teoría cuántica de la gravedad resuelva un conjunto ambicioso de problemas, que van desde las singularidades en el interior de los agujeros negros hasta el origen del Universo. También es frecuente encontrar imágenes especulativas del espaciotiempo planckiano que involucran agujeros de gusano microscópicos, cambios de topología, una discreción fundamental, dimensiones adicionales o nuevas simetrías. Estas posibilidades pueden contener parte de la respuesta, pero en la actualidad no existen experimentos directos en las escalas relevantes. Por lo tanto, contamos con muy poca guía observacional acerca de qué nuevos ingredientes microscópicos, si es que alguno, deberían introducirse.

La perspectiva seguida en esta tesis es deliberadamente conservadora. Antes de modificar los ingredientes fundamentales de la Relatividad General o de la Teoría Cuántica de Campos, nos preguntamos hasta dónde se puede avanzar utilizando herramientas que ya sabemos que funcionan en otros contextos: integrales de camino, regularizaciones reticulares, renormalización y métodos computacionales no perturbativos. El objetivo no es proporcionar un único modelo final que resuelva todos los problemas asociados con la Gravedad Cuántica. En cambio, la tesis explora varias preguntas complementarias acerca de qué es un espaciotiempo cuántico, cómo pueden medirse sus propiedades y cómo su comportamiento microscópico puede restringir la física a escalas mucho mayores. Esta filosofía general y el lenguaje básico utilizado a lo largo de la tesis se introducen en el Capítulo 1.

Esta tesis se desarrolla en dos direcciones principales, seguidas por un último desvío de carácter más fundamental. La primera utiliza las Triangulaciones Dinámicas Causales para estudiar las geometrías que aparecen en una integral de camino gravitatoria no perturbativa. La segunda utiliza el Grupo de Renormalización Funcional y la Seguridad Asintótica para estudiar la dependencia con la escala de las teorías efectivas y sus posibles predicciones infrarrojas. Un punto de vista permite acceder a las geometrías cuánticas configuración por configuración; el otro sigue las descripciones efectivas escala de energía por escala de energía. La parte final da entonces un paso atrás y pregunta si la propia métrica debe ser necesariamente la variable fundamental que se cuantiza.

### Geometría cuántica y necesidad de observables.

En la formulación de integral de caminos de la Mecánica Cuántica, un sistema se describe sumando sobre todas las configuraciones posibles que conectan sus estados inicial y final. Aplicar esta idea a la gravedad significa sumar sobre geometrías posibles, ya que la propia métrica es el campo dinámico de la Relatividad General. Formalmente, querríamos definir una integral de caminos sobre métricas del espaciotiempo y utilizarla para calcular valores esperados de observables geométricos.

Esta expresión formal es difícil de controlar. El espacio de geometrías posibles es sumamente complicado, la métrica contiene redundancias de gauge asociadas con transformaciones de coordenadas y la integral de caminos lorentziana es fuertemente oscilatoria. Además, la cuantización perturbativa estándar de la Relatividad General no produce una teoría fundamental predictiva. En el enfoque perturbativo, a órdenes progresivamente más altos aparecen nuevas divergencias ultravioletas que requieren términos de curvatura adicionales y nuevos acoplamientos independientes.

Este fracaso de la teoría de perturbaciones no implica necesariamente que la gravedad no pueda cuantizarse. Puede simplemente significar que la teoría de perturbaciones no es la herramienta correcta cuando el espaciotiempo está cuantizado.

Las Triangulaciones Dinámicas Causales, introducidas en el Capítulo 2, proporcionan una implementación concreta y no perturbativa de la integral de caminos gravitatoria. Las geometrías suaves se reemplazan por variedades simpliciales construidas pegando bloques planos. Los símplices individuales son simples, mientras que la curvatura queda codificada en la forma en que se conectan. El retículo se utiliza como regulador, en lugar de suponerse que representa una discreción física fundamental. En última instancia, una teoría continua debería obtenerse eliminando este regulador cerca de un régimen cuántico adecuado e invariante de escala.

Una característica distintiva de CDT es que sus geometrías preservan una foliación causal. Esto permite definir un análogo geométrico de una rotación de Wick y mapear la integral de caminos gravitatoria oscilatoria a un sistema estadístico con pesos reales. El ensamble resultante puede entonces explorarse mediante simulaciones de Monte Carlo.

Esto nos proporciona una suma computable sobre geometrías, pero viene acompañado de algunas salvedades. Las configuraciones típicas no son espaciotiempos suaves con pequeñas perturbaciones alrededor de una solución clásica. Presentan grandes fluctuaciones y pueden poseer propiedades fractales. Por lo tanto, el cálculo tensorial estándar no está directamente disponible. No existen coordenadas preferidas y, en general, no hay un tensor de Riemann suave que pueda evaluarse sin más en cada configuración.

El contenido físico de una teoría cuántica está dado por sus observables. En CDT, estos observables deben construirse utilizando las estructuras que siguen disponibles, como la conectividad, las distancias, los volúmenes y los procesos de difusión. Al mismo tiempo, deben tener una interpretación continua significativa; de lo contrario, describirían únicamente el regulador reticular. La primera contribución principal de esta tesis es el desarrollo de observables de este tipo para la topología, la homogeneidad y la curvatura.

#### Topología, homogeneidad y curvatura.

En el Capítulo 3, introducimos una noción dependiente de la escala de topología efectiva. Aunque la topología microscópica de las configuraciones de CDT está fijada, la topología visible a una resolución finita no tiene por qué estarlo. Cuellos delgados o estructuras pequeñas pueden dejar de resolverse bajo un proceso de coarse-graining, produciendo estrangulamientos efectivos o regiones con forma de burbuja.

Para capturar este comportamiento, seleccionamos vértices separados de manera aproximadamente uniforme y los utilizamos para construir una descomposición de Voronoi de la triangulación. Su dual define una triangulación de Delaunay más gruesa. Al cambiar la separación entre los vértices muestreados se obtienen geometrías efectivas a distintas resoluciones. Su topología se caracteriza mediante los números de Betti, que cuentan componentes conexas, lazos independientes no contraíbles y regiones encerradas de dimensión superior.

La construcción se puso a prueba en modelos lorentzianos y euclídeos bidimensionales de gravedad cuántica. En CDT, la topología efectiva permanece estable a lo largo de un amplio rango de escalas locales de coarse-graining. Los cambios aparecen solamente cuando la escala de coarse-graining se vuelve comparable con el tamaño global de la geometría compacta. Por lo tanto, no encontramos evidencia de estructuras locales con cambio de topología en el rango estudiado.

El modelo euclídeo se comporta de manera muy diferente. Sus geometrías efectivas desarrollan estrangulamientos locales y una jerarquía de regiones con forma de burbuja conectadas mediante cuellos estrechos. Estas estructuras reproducen la imagen familiar de universos madre y universos bebé en la gravedad cuántica euclídea bidimensional. Además, su distribución de volúmenes nos permite recuperar el comportamiento conocido de la susceptibilidad de cuerdas del modelo. Esto proporcionó una validación no trivial de que la construcción detecta propiedades universales genuinas, y no características arbitrarias del procedimiento de coarse-graining.

La siguiente pregunta abordada aquí es si una geometría cuántica puede considerarse homogénea. En un espaciotiempo clásico suave, la homogeneidad suele formularse mediante simetrías continuas. Estas no están directamente disponibles en una triangulación irregular. En el Capítulo 4, introducimos por lo tanto una definición operacional: se promedia una cantidad escalar dentro de bolas geodésicas centradas en distintos puntos, y la variación de esos promedios mide cuánto depende la geometría de la posición elegida.

Esta construcción deja claro que la homogeneidad no es una propiedad única de una geometría considerada de manera aislada. Depende de qué se mide y a qué escala. Aplicamos la medida al número de coordinación, al ángulo de déficit y a una dimensión de Hausdorff local en CDT bidimensional. El número de coordinación se vuelve altamente homogéneo, pero es principalmente una cantidad reticular. El ángulo de déficit, relacionado con la curvatura, conserva un grado mayor de inhomogeneidad. En cambio, la dimensión de Hausdorff local es compatible con volverse homogénea en el límite continuo.

Por lo tanto, una geometría cuántica puede ser homogénea con respecto a su dimensión efectiva y, al mismo tiempo, permanecer inhomogénea con respecto a su curvatura. Esto no es una inconsistencia. Refleja el significado físico de la homogeneidad: dos observadores pueden coincidir en una propiedad medida de la geometría y encontrar valores distintos de otra.

La curvatura presenta una dificultad relacionada, pero más fundamental. En la Relatividad General se define mediante derivadas de la métrica. Los espacios métricos genéricos y las triangulaciones no poseen la estructura suave requerida para esta construcción. Sin embargo, la curvatura afecta otras cantidades que siguen siendo significativas, como el crecimiento de las esferas geodésicas y la probabilidad de retorno de un proceso de difusión.

En el Capítulo 5, utilizamos estas propiedades para construir curvaturas escalares generalizadas. En una variedad suave, los volúmenes geodésicos y las probabilidades de difusión contienen correcciones a distancias cortas proporcionales al escalar de Ricci. Nuestras definiciones combinan derivadas e integrales de forma tal que se eliminan los factores de normalización dominantes desconocidos, dejando una cantidad de curvatura dependiente de la escala que puede evaluarse sin coordenadas ni tensores.

Probamos las construcciones sobre triangulaciones que aproximan un plano periódico y una esfera bidimensional. Después de separar los artefactos reticulares de distancias cortas de los efectos de tamaño finito a grandes distancias, los observables reproducen el comportamiento esperado para geometrías planas y con curvatura positiva. Luego los aplicamos a las geometrías cuánticas de CDT bidimensional. En el rango de escalas donde los efectos de discretización están bajo control, las curvaturas generalizadas promediadas sobre el ensamble son compatibles con cero.

Esto no significa que las configuraciones individuales de CDT se parezcan a planos ordinarios. Siguen siendo altamente irregulares. El resultado significa que los observables de curvatura coarse-grained específicos introducidos aquí tienen valores esperados compatibles con la planitud del estado fundamental. Más importante aún, estas cantidades pueden utilizarse para construir funciones de correlación de curvatura, proporcionando una posible forma de estudiar cómo emergen la curvatura clásica y sus correlaciones en un espaciotiempo cuántico de dimensión superior.

En conjunto, estos resultados muestran cómo pueden reformularse preguntas geométricas familiares cuando no están disponibles ni coordenadas ni tensores suaves. Los modelos bidimensionales actúan como bancos de prueba controlados, pero el objetivo principal es CDT en cuatro dimensiones. Allí, las mismas herramientas podrían determinar si una topología estable, una homogeneidad cosmológica aproximada y una curvatura clásica emergen a partir de geometrías cuánticas planckianas.

**Flujos del grupo de renormalización y predictividad.**

Las simulaciones reticulares proporcionan información directa sobre las geometrías cuánticas, pero necesariamente exploran un intervalo limitado de escalas. Las restricciones computacionales imponen tanto un espaciado reticular mínimo como un volumen total finito. Por lo tanto, la segunda parte de esta tesis plantea una pregunta complementaria: ¿cómo puede una teoría cuántica microscópica restringir la física efectiva en el infrarrojo?

El marco básico se introduce en el Capítulo 7. El Grupo de Renormalización describe cómo cambia una teoría física cuando se la examina a distintas escalas de energía. En el Grupo de Renormalización Funcional, esta información queda codificada en una Acción Efectiva Promedio dependiente de la escala. A escalas de momento altas, solo se han incorporado fluctuaciones de longitud de onda corta. A medida que la escala disminuye, se incluyen gradualmente fluctuaciones de longitud de onda mayor, hasta alcanzar la acción efectiva cuántica completa.

La Seguridad Asintótica propone que la gravedad puede permanecer bien definida a energías arbitrariamente altas porque su flujo del grupo de renormalización se aproxima a un punto fijo no gaussiano. En ese punto, los acoplamientos adimensionales se vuelven independientes de la escala. El conjunto de trayectorias que alcanzan el punto fijo define su superficie crítica ultravioleta.

Esta superficie es central para el poder predictivo de la teoría. Sus direcciones relevantes corresponden a parámetros que deben fijarse mediante observaciones. Los acoplamientos irrelevantes, en cambio, quedan determinados por el requisito de que la teoría se origine en el punto fijo ultravioleta. Si solo un número finito de direcciones es relevante, una cantidad finita de mediciones determina una colección de interacciones que, de otro modo, sería infinita.

La forma convencional de explorar la superficie crítica ultravioleta consiste en integrar muchas trayectorias individuales desde una vecindad del punto fijo hacia el infrarrojo. Esto se vuelve ineficiente a medida que aumenta la dimensión del espacio de teorías. En el Capítulo 8, introducimos los invariantes de flujo como una alternativa global.

Un invariante de flujo es una función de los acoplamientos que permanece constante a lo largo de las trayectorias del grupo de renormalización. Satisface una ecuación diferencial parcial lineal determinada por las funciones beta. Con condiciones adecuadas cerca del punto fijo, el conjunto común de ceros de varios invariantes reconstruye la superficie crítica ultravioleta. En lugar de muestrear una trayectoria tras otra, el método busca codificar toda la superficie predictiva en una pequeña colección de funciones.

Primero validamos la construcción en sistemas dinámicos simples y en un modelo de campo escalar tridimensional. Para extenderla a espacios de teorías más complicados, desarrollamos la Flow Invariant Neural Network, o FINN. Se trata de una red neuronal informada por la física, entrenada directamente a partir de la ecuación diferencial de los invariantes de flujo y del comportamiento conocido del punto fijo, en lugar de utilizar un gran conjunto de datos precalculados.

El método se puso a prueba en un modelo escalar–tensorial de cuatro dimensiones. En este caso, puede encontrarse un punto fijo ultravioleta con cuatro direcciones relevantes, mientras que dos acoplamientos adicionales quedan predichos por la consistencia ultravioleta. La superficie crítica reconstruida mediante las FINN concuerda con los cálculos convencionales por métodos de disparo al nivel de aproximadamente un uno por ciento en la región estudiada.

Este resultado es una prueba de concepto, no una predicción fenomenológica final. La truncación es limitada, y la existencia de un conjunto completo de invariantes de flujo regulares no está garantizada en sistemas arbitrarios.

Sin embargo, proporciona un método potencialmente escalable para determinar las predicciones infrarrojas de teorías ultravioletamente completas cuando la exploración directa trayectoria por trayectoria se vuelve poco práctica.

**Un potencial de inflatón emergente del ultravioleta.**

El capítulo final aplica el punto de vista del grupo de renormalización a la cosmología del Universo temprano. La inflación es un período de expansión acelerada propuesto para explicar la uniformidad a gran escala del Universo y el origen de las fluctuaciones primordiales de densidad. En muchos modelos, está impulsada por un campo escalar cuyo potencial se elige para producir una evolución cosmológica exitosa.

En el Capítulo 9, preguntamos si el potencial del inflatón puede emerger, en cambio, del flujo del grupo de renormalización de una teoría escalar–tensorial asintóticamente segura, en lugar de elegirse para satisfacer las observaciones. En vez de aproximar las interacciones escalares mediante un polinomio corto, conservamos la dependencia funcional completa del potencial escalar y del acoplamiento no mínimo entre el campo escalar y la curvatura escalar.

El punto de partida ultravioleta es un punto fijo no gaussiano con simetría de desplazamiento, donde el potencial escalar es plano. A medida que la teoría fluye hacia el infrarrojo, esta simetría se rompe dinámicamente y emergen interacciones escalares no triviales. Las ecuaciones funcionales de flujo requieren más información de la que sugiere el conteo habitual de direcciones relevantes del punto fijo. Después de imponer regularidad ultravioleta, las soluciones siguen estando etiquetadas por dos parámetros de tipo masa.

Para valores adecuados de estos parámetros, la teoría infrarroja desarrolla un potencial escalar en el marco de Einstein con una región amplia y aproximadamente plana. La evolución slow-roll resultante produce alrededor de sesenta y seis e-folds de inflación y valores del índice espectral escalar, de la razón tensor-escalar y de la amplitud de las fluctuaciones compatibles con las restricciones observacionales.

El punto importante no es simplemente que el potencial permita inflación. Muchos potenciales lo hacen. Aquí, su forma funcional es generada por el flujo desde un punto fijo ultravioleta, en lugar de ser postulada desde el comienzo. La evolución también sugiere un posible mecanismo para preparar el estado inicial: un comienzo del Universo cercano a la invariancia de escala, que desarrolla un mínimo del potencial dependiente de la escala, puede guiar el campo escalar hacia la región desde la cual comienza la inflación.

Esta interpretación sigue siendo condicional. Supone una relación entre la evolución de la escala del grupo de renormalización y la evolución temporal cosmológica, y el cálculo depende de una truncación, un regulador, una elección de gauge, una aproximación de fondo y una formulación euclídea particulares. Por lo tanto, es esencial poner a prueba la estabilidad del resultado bajo mejoras de estos ingredientes.

### ¿Qué debería cuantizarse en gravedad?

La parte final de esta tesis vuelve a una pregunta que precede a la elección de cualquier método no perturbativo particular: ¿qué debería considerarse fundamental en gravedad? Tanto CDT como el Grupo de Renormalización Funcional suelen comenzar con la métrica. Aquí exploramos, en cambio, una formulación puramente afín, clásicamente equivalente a la Relatividad General bajo condiciones adecuadas, en la cual la conexión afín es fundamental y la métrica, cuando existe, se reconstruye a partir de ella. Una propiedad particularmente atractiva es que, para una constante cosmológica positiva, la acción resultante de tipo Eddington tiene signo definido y puede utilizarse para construir un ensamble estadístico reticular.

Primero construimos y simulamos este ensamble en dos dimensiones como prueba de principio. Aunque allí no puede reconstruirse una métrica no degenerada, el modelo proporciona una integral de camino no perturbativa concreta y numéricamente bien comportada para conexiones afines. Luego pasamos a una reducción homogénea e isótropa en cuatro dimensiones, donde puede exigirse una métrica emergente definida positiva configuración por configuración. Una conexión de fondo fija de de Sitter regulariza las direcciones planas, mientras su intensidad se disminuye progresivamente. Los perfiles reconstruidos de volumen espacial muestran una forma de tipo de Sitter alentadora para varios tamaños de retículo y masas del regulador. Las correlaciones de curvatura definidas mediante la distancia geodésica física ilustran cómo este marco puede combinar un ensamble estadístico de geometrías cuánticas con el cálculo tensorial suave. Estos resultados son todavía preliminares, pero abren una ruta concreta hacia simulaciones afines más generales en cuatro dimensiones y hacia un programa más amplio de observables geométricos intrínsecos.

**Reuniendo las distintas facetas.**

Las perspectivas estudiadas en esta tesis abordan la Gravedad Cuántica desde direcciones diferentes. CDT comienza con una integral de caminos gravitatoria regularizada y produce geometrías individuales fluctuantes. El Grupo de Renormalización Funcional comienza con acciones efectivas y sigue su dependencia con la escala. La gravedad reticular puramente afín vuelve a la pregunta previa acerca de qué campo debe cuantizarse y explora una integral de camino en la cual la geometría emerge de una conexión.

Ninguno de los dos enfoques proporciona actualmente una teoría completa de la Gravedad Cuántica. CDT todavía requiere un mayor control sobre su límite continuo en cuatro dimensiones y sobre la relación entre su construcción estadística y la física cuántica lorentziana. Los cálculos del Grupo de Renormalización Funcional dependen de truncaciones y de otras elecciones matemáticas cuya robustez debe comprobarse. También queda por establecer una relación directa entre las geometrías microscópicas de la integral de caminos reticular y los acoplamientos efectivos de la descripción continua mediante el grupo de renormalización.

Sin embargo, su complementariedad es clara. Una construcción microscópica sin observables adecuados no puede decirnos cómo es su espaciotiempo cuántico. Un punto fijo ultravioleta sin control sobre las trayectorias que salen de él no puede decirnos qué mundo de bajas energías predice. El progreso requiere tanto teorías no perturbativas bien definidas como métodos confiables para extraer su contenido físico.

Las contribuciones principales de esta tesis son, por lo tanto, herramientas y construcciones concretas. En el contexto reticular, desarrollamos observables dependientes de la escala para la topología efectiva, la homogeneidad y la curvatura escalar generalizada. En el contexto del grupo de renormalización, introdujimos los invariantes de flujo y su implementación mediante redes neuronales como método para reconstruir superficies críticas ultravioletas predictivas, y mostramos cómo un potencial de inflatón viable puede emerger de la consistencia ultravioleta. Finalmente, construimos una formulación reticular no perturbativa de la gravedad puramente afín y comenzamos a explorar en ella la emergencia de geometría cuatridimensional.

Las preguntas abiertas y las posibles continuaciones se discuten en el Capítulo 13. Los observables geométricos deberían aplicarse a CDT en cuatro dimensiones, mientras que los invariantes de flujo y las FINN deberían ponerse a prueba en sistemas gravedad–materia más grandes. El escenario inflacionario debe examinarse bajo cambios de truncación y otras aproximaciones. En la formulación afín, deben completarse la eliminación del regulador y el análisis de escala antes de relajar la restricción homogénea e isótropa hacia un ensamble plenamente cuatridimensional.

Esta tesis comenzó con una curiosidad acerca de en qué se convierte el espaciotiempo cuando su naturaleza cuántica ya no puede ignorarse. No termina con una única imagen microscópica final ni con una solución a todos los problemas comúnmente asignados a la Gravedad Cuántica. En cambio, proporciona preguntas más concretas y un conjunto de instrumentos con los cuales abordarlas. ¿Cómo puede medirse la topología a resolución finita? ¿Qué significa que una geometría fluctuante sea homogénea? ¿Puede definirse la curvatura sin coordenadas ni tensores? ¿Cómo pueden extraerse eficientemente las predicciones de un punto fijo ultravioleta? ¿Pueden las interacciones cosmológicas emerger de condiciones de consistencia gravitacional cuántica? ¿Y debe la propia métrica ser fundamental, o puede la geometría emerger de una conexión cuántica?

La historia sigue inconclusa, como era de esperarse. Sin embargo, en un campo donde incluso formular las preguntas correctas puede ser difícil, construir observables confiables y puentes entre escalas ya representa un progreso sustantivo. Todavía quedan muchas facetas de la gravedad por explorar, desde las geometrías planckianas hasta escalas cosmológicas.

# Acknowledgments

It is difficult to find a story about an adventure that someone undertakes entirely alone. This is not one of those stories and, in fact, it is probably as close to the opposite as one can get.

Although I am aware that my perception of time is somewhat relative, not only in a personal sense, but in a very real physical sense, even so, I find it incredible that I started this adventure of trying to understand the universe approximately a decade ago! The seemingly inevitable arrow of time and the irreversibility of its flow never cease to amaze me and, at times, frighten me. Deep down, it still feels like a very recent event when I began to obsess over understanding the things around me, and over the possibility of studying and understanding the underlying order of the mechanisms that determine the properties of this possibly infinitely complex Universe in which I live. Even so, a decade has passed, and this marks the end of a beautiful and long story arc of which I am very proud, not only because of what I have achieved, but also because I found and was accompanied by people who were, are, and will continue to be genuinely important to me, and because of all the events that make up this adventure I call life. This section is dedicated to some of the people who made it possible for me to be here today.

As some readers may know, there are groups in which the order of the factors does not affect the product. In this particular group of people that I want to mention, the order in which I include them in this section does not necessarily reflect their importance. If they are here, it is because, in one way or another, they have a place within the chaos that is my mind and, if they know me well, then they will know that the mere fact that I have managed to make this section even minimally coherent is an achievement in itself.

## For "the university people"

Given that this is a doctoral thesis, I think it is only appropriate to begin with those who are directly responsible for the fact that, from today onwards, people will have to call me doctor. The main person, and the one directly to blame for this, is my supervisor: Renate. It feels like only yesterday that I was sending you an email translated from Spanish into English by Google, asking you for the opportunity to do a PhD under your supervision, full of curiosity and excitement and, obviously, several weeks after the deadline by which you had asked to receive such requests. Even so, you gave me a chance, and this doctoral adventure began. Four years of deep and stimulating conversations, both productive and recreational. There were even attempts to make it through the borders of highly guarded places in the middle of the night! It is impossible to summarize everything about these years, but there is one thing I am sure of: I had the chance to meet someone as special as you, who became a fundamental pillar of what I have achieved today, and I want to thank you for everything you gave me. Thank you!

Of course, I also have to mention my second supervisor, Frank. Although it was initially in nobody's plans that you would have to deal with a character like me, when the time came for my curiosity to get the better of me and for a complete stranger like me to walk into your office and propose strange ideas for us to discuss, you opened your door in a way that few people would have. A large part of the work in this thesis I owe to you, and I want to thank you for your company, supervision, and collaboration throughout all these years.

Four years of doing a PhD generally come with colleagues, people with whom one shares an office, or simply shares working spaces in general. In turn, those colleagues can end up becoming excellent friends and, in this case, that is exactly what happened.

Cris, my great friend and colleague. We went through everything: learning, travelling, laughter, tears, and so many other things. Not only did we share academic interests and even manage to turn our ideas into publications, but we also forged a friendship unlike almost any other I have had. I think I was very lucky to find someone whom I admire, and with whom I have been able to connect and share so many intellectual and personal aspects that it almost seems suspicious that chance placed us in the same office at the same time. Thank you for everything we have been through and for everything still to come, brother.

Jesse, what can I say? You were there from beginning to end. You saw me and stood by me through both the highs and the lows. We learned, collaborated, celebrated, laughed, and much more. It was a pleasure to walk this path together with you, and it is a pleasure to be able to call you not only my paranymph, but also my friend.

Ari and Patri, because yes, the two of you cannot be separated, I want to tell you that you made me very happy at many moments during this journey. So many meals and beers shared, discovering new countries, deep conversations, support when it was needed, and much more. Thank you for sharing all of this with me.

Marc, I know you do not like cheesy things, so I am going to make this as much your style as possible. I admire you in many ways, and I am grateful for all your support, your company, and all the moments we shared, both at work and socially, every one of which I enjoyed. As I said before, it is a pleasure to be able to call you not only my paranymph, but also my friend.

To my favorite Italians, Luca and Antonio, and my favorite French, Marion. I do not think I need to tell you that your presence at the university made my PhD far better than I could have imagined. Thank you, my friends, for the trips, memories, laughter, drinks, dinners, and conversations we have shared. I truly hope our friendship continues in the future.

To my padawans Tom and Juan, and to my other office mates, Jian, David, and Jesse D.: I have to tell you that you made those eight daily hours at the office (if anyone actually believes I worked that much) a pleasure. I hope our paths cross again in the future so that we can once again enjoy those board games, meals, nights out and, God forbid, work again!

Obviously, there were many other people around my working environment who became a fundamental part of the groups of friends that made this whole adventure, in a country so far away from my own, such a pleasure. In particular, Luca F., Ibai, Vu, and Patrick, thank you for everything! I hope time brings us together again!

## Para “los de afuera de la universidad”

Si bien no creo que, en todos los casos que menciono, la distinción de “gente del trabajo” y “gente de afuera del trabajo” sea aplicable, sí que me gustaría destacar a algunas personas que nunca estuvieron relacionadas con la universidad, pero que hicieron de todo este viaje una experiencia única.

Tres décadas no vienen solas, ni se hacen solas. En mi caso, soy muy afortunado de tener gente que aprecio, valoro, y quiero en mi vida. Gente que por accidente, azar o intención, formó y forma parte de mi vida, y quiero que así siga siendo.

A mis “amigos de Europa”, Santi, Andrea, Rahul y Ampi, les quiero agradecer por haber hecho de Nijmegen un lugar único. Ya probablemente estemos todos cansados de contarle a la gente cómo se armó este grupo, el grupo “random”, pero ¡qué lindo grupo! Que sigan nuestros viajes, nuestras experiencias y nuestra compañía en todo lo que se viene.

A mis amigos santafesinos que siempre me hicieron el aguante en todas, recibiéndome cada año que volvía después de alguna aventura distinta como si nada hubiera cambiado desde que era adolescente, pero especialmente durante estos año del doctorado. Sin ellos, jamás hubiera podido recargar baterias para cada año académico. A Joaco, Ovi, Migue, Lichi, Ale, Lucho, Juli, Dido, Maia, Vale y muchos más, quiero decirles gracias por formar parte de tantas etapas de mi vida. Innumerables experiencias compartidas, incontables juntadas y conversaciones, risas y alegrías.

## Para mi familia

Hay gente que dice que “la familia no se elige”. Si pudiera, yo volvería a elegir a la mía. “Mis logros” no son míos, son de todos nosotros. Hoy brindo por ustedes, mis cimientos y quienes forman parte de mí en cada lugar al que voy y en cada momento de mi vida.

A mis abuelos, Tata, Tati, Belo y Beli. A ellos, que siempre me tuvieron en un pedestal, mucho más alto del que merezco, y que siempre se maravillaron con cada cosa que hacía, aunque fuera minúscula, hoy les dedico lo que yo considero mis logros. Lamentablemente, en lo que hoy defino como mi presente, ellos no están, pero, dado que la definición de tiempo presente es relativa y el presente de un observador puede ser tanto el pasado como el futuro de otro, en un sentido muy real, ellos siguen vivos. Tal vez no en mi línea de mundo, pero sí en otras, y eso me reconforta. Gracias por todo.

A Fran y a Guille. Mis hermanos, esos que compartieron todo conmigo desde el principio. Ustedes conocieron cada faceta de mí y yo las de ustedes. Tenemos un vínculo único, sin el cual estoy seguro de que hoy no sería la persona que soy. Como bien saben, mi limitado uso del lenguaje no me permite expresarles lo mucho que agradezco ser su hermano. Aunque sé que a sus ojos soy idiota más del 99 por ciento de mi tiempo, quiero decirles gracias por quererme, por creer en mí y por apoyarme, a pesar de ese 99 por ciento que tuvieron que tolerar durante décadas de sus vidas. Los quiero mucho.

A papá y a mamá, que ayudaron a crear mis caminos, me guiaron a través de todos ellos y apoyaron literalmente cada uno de mis pasos de forma incondicional, desde mis primeros pasos físicos hasta los otros “pasos” que me llevaron a ser quien soy hoy y a estar donde estoy. A pesar del dolor que les causaba ver a su pajarito volar por ahí, entre lágrimas de orgullo, lograban abrazarme, apoyarme y ayudarme a tomar cada una de las decisiones que tomé. A pesar de las miles de cagadas que me mandé, desafiando y yendo a contramano de sus consejos, me amaron incondicionalmente y estuvieron para mí en literal y absolutamente todos los momentos en los que los necesité, y mucho más que solo lo necesario. Nada de esto hubiera sido posible sin ustedes, y poca gente tiene el lujo de tener padres como ustedes. Hoy puedo decir que estoy orgulloso de que sean mis papás; los amo hasta el infinito y más allá, y espero poder seguir haciéndolos sentir orgullosos en lo que siga de esta aventura.

### Para mi compañera

Para Anto, que obviamente se merece un apartado más largo que esta tesis. A vos, que me seguiste en mis ideas estrafalarias de viajar por el mundo y conocer otras formas de vivir, que te animaste a dejar todo por seguir a un loco como yo. Te volviste mi pilar fundamental y, sinceramente, dudo que merezca a alguien como vos, alguien tan hermosa, inteligente, graciosa, empática, paciente, madre de nuestros dos michijos, una profesional excelente, emprendedora, aventurera, alguien única. Pero acá estoy, un ser terriblemente afortunado de tenerte a mi lado. Este doctorado habría sido literalmente imposible sin vos, y te merecés todo y más de lo que puedo darte.

Sos mi pareja, pero mucho más que eso: sos mi compañera de vida. Entendés y conocés cada una de mis capas, y el hecho de que me ames a pesar de eso me vuela la cabeza. Cuando me mirás, me ves, y yo siento que al revés también es así. Creo que lo que tenemos es algo único. No puedo estar más agradecido de que hayamos experimentado juntos todo lo que vivimos, y solo puedo rogarle al universo que lo que sigue, siga siendo a la par tuya. Sigamos jugando a la vida juntos, porque con vos todo es mejor. Gracias por todo. Te amo infinito no numerable.

# About the author

Agustín Silva was born on the 5th of July, 1996, in Santa Fe de la Vera Cruz, Argentina. In his early years, he succeeded in a competitive career in sprint canoeing: between 2012 and 2014, he was a federated athlete, and a member of the Argentine national team, achieving, among other results, the 18*th* place at the U23–Junior World Championships, in Szeged, Hungary, as well as South American and national champion titles.

In 2015, he began his university studies in his hometown, completing courses of the first three years, of Applied Mathematics and Electrical Engineering, at Universidad Nacional del Litoral and at Universidad Tecnológica Nacional. In 2018, he joined the Instituto Balseiro, in San Carlos de Bariloche, Argentina, where he pursued an academically intensive program with a highly selective intake, under a scholarship awarded by Comisión Nacional de Energía Atómica. There, he obtained his Licentiate's and Master's degrees in Physics, with a special focus on quantum field-theoretic aspects of the presence of boundaries or time-dependent background spacetimes.

In a search for a better understanding of spacetime at a fundamental level, in 2022 he started his PhD at Radboud University (IMAPP) in Nijmegen, the Netherlands, under the supervision of Prof. Dr. Renate Loll. His doctoral research focused on non-perturbative approaches to quantum gravity, in particular, Asymptotic Safety and Causal Dynamical Triangulations. But his work and research interests covered a vast range of topics within gravitational physics, including black hole physics, early universe cosmology, and alternative approaches to quantize gravity.

His research has been disseminated through several publications and conference presentations across Europe and South America, and he has contributed to teaching and supervision at Radboud University, serving as a teaching assistant in multiple Master's and Bachelor's level courses.

In 2026, he will continue the research lines built in this thesis, in a never-ending pursuit of new challenges that spark his curiosity.